\documentclass[11pt, a4paper, headsepline, footsepline]{scrbook}

\usepackage[latin1]{inputenc}
\usepackage{longtable}
\usepackage[hang, small, it, bf]{caption}
\usepackage[dvips]{graphicx}
\usepackage{amsfonts}
\usepackage{amsmath}
\usepackage{mathrsfs}
\usepackage{epsfig}
\usepackage[clearempty]{titlesec}
\usepackage{booktabs}
\usepackage{hhline}
\usepackage{array}
\usepackage{subfigure}
\usepackage{float}
\usepackage{floatflt}
\usepackage{graphicx}
\usepackage{dcolumn}
\usepackage{bm}
\usepackage{mathrsfs} 
\usepackage{amssymb}
\usepackage{amstext}
\usepackage{amsgen}
\usepackage{amsbsy}
\usepackage{amsopn}
\usepackage{cite}
\usepackage{theorem}
\usepackage{color}
\usepackage{maybemath}
\usepackage{url}
\usepackage{wasysym}
\usepackage{multirow}
\usepackage{multicol}
\usepackage{remreset}
\usepackage{numprint}
\usepackage{hyperref}

\newlength{\plotwidth}
\definecolor{uniblue}{rgb}{0.08627,0.21569,0.46275}

\makeatletter
\@addtoreset{figure}{chapter}
\@addtoreset{table}{chapter}

\DeclareFontFamily{U}{euc}{}
\DeclareFontShape{U}{euc}{m}{n}{<-6>eurm5<6-8>eurm7<8->eurm10}{}
\DeclareSymbolFont{AMSc}{U}{euc}{m}{n}
\DeclareMathSymbol{\umu}{\mathord}{AMSc}{"16} 

\newcommand{\dzero}      {D\O}

\newcommand{\ttbar}      {\mbox{$t\bar{t}$}}

\newcommand{\met}        {\mbox{$\not \hspace{-0.1cm} E_{\rm{T}}$}} 	
\newcommand{\notpartial}        {\mbox{$\not\!\!D$}}
\newcommand{\pt}         {\mbox{$p_{\rm{T}}$}}
\newcommand{\Et}         {\mbox{$E_{\rm{T}}$}}

\newcommand{\zee}        {\mbox{$\maybebm{Z/\gamma^* \rightarrow e e}$}}
\newcommand{\wenu}        {\mbox{$\maybebm{W^{\pm} \rightarrow e \nu_e}$}}
\newcommand{\zmumu}      {\mbox{$\maybebm{Z/\gamma^* \rightarrow \mu \mu}$}}

\newcommand{\Nl}       {\mbox{$N_{\rm{loose}}$}}
\newcommand{\Nt}       {\mbox{$N_{\rm{tight}}$}}
\newcommand{\Ns}       {\mbox{$N^{\rm{sig}}$}}
\newcommand{\Nf}       {\mbox{$N^{\rm{fake}}$}}
\newcommand{\epss}       {\mbox{$\varepsilon^{\rm{sig}}$}}
\newcommand{\epsf}       {\mbox{$\varepsilon^{\rm{fake}}$}}
\newcommand{\MTW}{\mbox{$m_{\rm{T}}(W)$}~}
\newcommand{\herwig}{\textsc{herwig}}
\newcommand{\jimmy}{\textsc{jimmy}~}
\newcommand{\powheg}{\textsc{powheg}~}
\newcommand{\pythia}{\textsc{pythia}~}
\newcommand{\alpgen}{\textsc{alpgen}~}

\newcommand{\mcnlo}{\textsc{mc@nlo}}
\newcommand{\sherpa}{\textsc{sherpa}~}
\newcommand{\acermc}{\textsc{acermc}~}
\newcommand{\hathor}{\textsc{Hathor}~}
\newcommand{\phm}       {\phantom{-}}
\newcommand{\pho}       {\phantom{0}}

\DeclareMathOperator{\tr}{Tr}

\DeclareRobustCommand{\Cpp}
{\valign{\vfil\hbox{##}\vfil\cr 
   \textsf{C\kern-.1em}\cr
   $\hbox{\fontsize{\ssf@size}{0}\textbf{+\kern-0.05em+}}$\cr}%
}

\npdecimalsign{\ensuremath{.}}
\begin{document}


  %
%
\begin{titlepage}
  \vspace*{3cm}
  \begin{center}
     {\Large \bf
    Measurement of the charge asymmetry\\[0.2cm]in top quark pair production in $\boldsymbol{p}\boldsymbol{p}$ collision data\\[0.25cm]at
$\sqrt{\boldsymbol{s}}=7$\,TeV using the ATLAS detector}\\[2cm]
    {\Large \bf
      Dissertation \\[2cm]
      }
      
      {\Large
      zur Erlangung des mathematisch-naturwissenschaftlichen Doktorgrades \\
       "Doctor rerum naturalium" \\
      der Georg-August-Universit\"at G\"ottingen \\[2cm]
    }
    {\Large
      vorgelegt von \\[2mm]
      Fabian Kohn  \\[2mm]
      aus Northeim \\[2cm]
    }
    {\Large
      G\"ottingen, 2012 \\
    }

\vfil
  \end{center}
\newpage
\thispagestyle{empty}
\vspace*{20cm}
\noindent Referent: Prof. Dr. Arnulf Quadt\\
\noindent Korreferent: Dr. Carsten Hensel\\[0.5cm]

\noindent Tag der m\"undlichen Pr\"ufung: 07.03.2012

\end{titlepage}

  %
%
\begin{titlepage}
  \vspace*{3cm}
  \begin{center}
    {\Large \bf
    Measurement of the charge asymmetry\\[0.2cm]in top quark pair production in $\boldsymbol{p}\boldsymbol{p}$ collision data\\[0.25cm]at
$\sqrt{\boldsymbol{s}}=7$\,TeV using the ATLAS detector}\\[1.5cm]
    {\Large
      by \\[2mm]
      Fabian Kohn \\[1.5cm]
}
{
  \parbox[t]{140mm}{A measurement of the charge asymmetry in the production of top quark pairs in the semileptonic decay channel has been performed. A dataset corresponding to an integrated luminosity of 1.04\,fb$^{-1}$, obtained at a centre-of-mass energy of $\sqrt{s} = 7$\,TeV with the ATLAS experiment at the LHC, was used. After performing a selection of events with one isolated lepton, at least four jets and missing transverse energy, a kinematic fit was performed to reconstruct the \ttbar~event topology.

The charge asymmetry was determined using the differential distribution of the reconstructed observable $|y_t| - |y_{\bar{t}}|$, where $y_t$ and $y_{\bar{t}}$ denote the top and antitop quark rapidities, respectively. An unfolding procedure was applied to correct for detector acceptance and resolution effects and to obtain the corresponding distribution at parton level. The total charge asymmetry after unfolding was measured to be
\begin{eqnarray*}
  A_C^{\text{unf}} = -0.018 \pm 0.028\,\text{(stat.)} \pm 0.023\,\text{(syst.)}
\end{eqnarray*}
in agreement with the Standard Model prediction.

In addition, a simultaneous unfolding in $|y_t| - |y_{\bar{t}}|$ and the invariant \ttbar~mass, $M_{t \bar{t}}$, was performed.}
}
\vfil
  \end{center}

  \begin{figure}[b]
    \begin{minipage}{\textwidth}
        \begin{raggedright}
        \begin{tabular}{@{}l@{}}
                Post address:  \\
                Friedrich-Hund Platz 1 \\
                37077 G\"ottingen  \\
                Germany      \\
        \end{tabular}
        \end{raggedright}
        \hfill        
       \begin{raggedleft}
        \begin{tabular}{@{}r@{}}
            II.Physik-UniG\"o-Diss-2012/02  \\
            II.~Physikalisches Institut    \\          
            Georg-August-Universit\"at G\"ottingen      \\
            February 2012                   \\             
        \end{tabular}
        \end{raggedleft}
{
\vspace*{-4.5cm}
\begin{center}
\end{center}
}
    \end{minipage}
    \vspace*{5mm}
  \end{figure}
  \newpage
\thispagestyle{empty}
\vspace*{14cm}
  \begin{flushright}
  \textsc{Not only is the universe stranger than we imagine,\\
  it is stranger than we can imagine.}\\
  {\small -- \textit{Sir Arthur Stanley Eddington}}
  \end{flushright}  
\end{titlepage}

%
  
  \pagenumbering{roman}

  \tableofcontents
  \clearpage
  
  \pagenumbering{arabic}
  \setcounter{page}{1}
  
  \chapter{Introduction}
The field of Elementary Particle Physics is concerned with understanding the most fundamental building blocks of Nature and their interactions. The Standard Model of Particle Physics\cite{Glashow:1961tr, Weinberg:1967tq, Salam, Politzer:1973fx,Gross:1973id, Fritzsch:1972jv} is one of the most successful and thorough theories in physics, its predictions being in astonishing agreement with the observed phenomena to highest precision.
Regardless of its success in explaining the interactions of all fundamental particles, the Standard Model is not without shortcomings. The fine-tuning of radiative corrections to the Higgs boson mass\cite{PW1964132, PhysRevLett.13.508}, the strong evidence for the existence of Dark Matter and Dark Energy, and a missing mechanism to describe gravity within the framework of a quantum field theory are only a few of the remaining ambiguities.

According to current knowledge, there exist six quarks in Nature: the up, down, strange, charm, bottom and the top quark along with the six known leptons (electron, muon and tau together with their corresponding neutrinos). Quarks and leptons are grouped into three subsets or {\em generations} of a quark and lepton doublet each.

The top quark is the heaviest known elementary particle and has been the focus of studies for several decades, from indirect searches using electroweak precision data to its discovery by the CDF\cite{PhysRevLett.74.2626} and \dzero\cite{PhysRevLett.74.2422} experiments at the Tevatron in 1995, the observation of the single top quark\cite{PhysRevLett.103.092001,PhysRevLett.103.092002} in 2009, to newest precision measurements of its properties at the Tevatron and the LHC.
Its unique characteristics, namely the large mass of about $173\,\text{GeV}$\cite{Lancaster:2011wr} and its short lifetime, provide the opportunity to perform precise measurements of electroweak interactions. Due to the affinity of its mass to the electroweak scale and the potential link to the vacuum expectation value of the Higgs field, the top quark also allows indirect constraints of the Higgs mass in combination with precision measurements of the $W$ boson mass. Furthermore, since the top quark is the only quark which has a decay width larger than the hadronisation scale, it does not form hadronic bound states. As a result, top quark properties, such as its spin, are accessible without being obscured by the process of hadronisation.

At hadron colliders, such as the Tevatron or the Large Hadron Collider, top quark pairs are mainly produced via the strong interaction, either via gluon-gluon fusion or via quark-antiquark annihilation:
\begin{eqnarray}
  q + \bar{q} & \rightarrow & t + \bar{t}\text{,} \nonumber \\
  g + g       & \rightarrow & t + \bar{t}\text{.} \nonumber
\end{eqnarray}

In the Born approximation, these production mechanisms are entirely symmetric under the exchange of the final state top and antitop quark. Consequently, there is no angular discrimination between the top and antitop quark and the resulting predicted differential distributions are identical for both particles.

In the Standard Model, an asymmetry in the production of top quark pairs arises due to radiative corrections from virtual and real gluon emission if higher order corrections are taken into account. These higher order corrections introduce interferences between amplitudes which are odd under the exchange of the final state quark and antiquark. Interference terms of final state and initial state gluon bremsstrahlung, and of higher order amplitudes with Born level amplitudes contribute to an overall imbalance of the differential distributions of the final state top quark and antitop quark. An additional small contribution originates from the interference of different amplitudes in quark-gluon scattering:
\begin{eqnarray}
  g + q & \rightarrow & t + \bar{t} + q'\text{.} \nonumber
\end{eqnarray}

The measurement of the charge asymmetry in top quark pair production provides the opportunity to verify perturbative Quantum Chromodynamics and consequently, the Standard Model. Moreover, a similar effect is predicted and observed in Quantum Electrodynamics\cite{Berends1981237,Berends:1982dy,Berends1973381}, where radiative corrections lead to an asymmetry in the electroweak production of fermion-antifermion pairs. Since this effect has been studied to very high precision, the verification of its counterpart in Quantum Chromodynamics would be yet another confirmation of the Standard Model and its predictions.

Furthermore, potential new physics, in particular theories involving the breaking of electroweak symmetry, could lead to deviations from the Standard Model expectation due to large anomalous couplings to the top quark predicted in numerous theoretical models. As a matter of fact, current independent measurements performed by CDF\cite{PhysRevLett.101.202001,CDFCONF-10584} and \dzero\cite{PhysRevLett.100.142002,Abazov:2011rq} suggest a possible discrepancy between the predicted and observed charge asymmetry in proton-antiproton collisions. This effect is observed in particular for high invariant $\ttbar$ masses and high $\ttbar$ rapidity differences, which is supported by several models predicting physics beyond the Standard Model.

Since the predicted charge asymmetry is small at hadron colliders due to the probabilistic nature of the initial state parton kinematics, precise knowledge of the detection mechanisms, sophisticated analysis methods and detailed understanding of potential systematic effects are crucial to accomplish such a measurement. This is in particular true for the Large Hadron Collider, where a high centre-of-mass energy and a symmetric hadronic initial state ($pp$) make this measurement even more difficult. The increased fraction of top quark pairs produced via (charge symmetric) gluon-gluon fusion lead to a dilution of the measured asymmetry. In addition, there is no preferred initial state quark direction in proton-proton collisions and hence no resulting forward-backward asymmetry which could be measured directly as it is the case at the Tevatron. Consequently, a new analysis concept and new observables have to be considered to perform this measurement under the conditions of the LHC.

The charge asymmetry has been measured in top quark pair production at the Tevatron by both the CDF\cite{PhysRevLett.101.202001} and \dzero\cite{PhysRevLett.100.142002} collaborations and preliminary results have also been shown by the CMS\cite{CMS-PAS-TOP-11-014} experiment at the LHC. This thesis describes a measurement of the charge asymmetry in top quark pair production which has been performed with the ATLAS experiment for the first time in 2011\cite{CONFNote}.

This document is organised as follows: \mbox{Chapter \ref{Theory}} gives an introduction to the theoretical aspects of the Standard Model and theories beyond, paying special attention to the top quark and the charge asymmetry in top quark pair production. \mbox{Chapter \ref{Experiment}} covers technical aspects of the LHC and the ATLAS detector. Definition and description of the objects taken into account for the described analysis and the trigger strategy is given in \mbox{Chapter \ref{Objects}}, followed by a summary of the event selection performed to increase the fraction of relevant signal events with respect to various background processes in \mbox{Chapter \ref{Selection}}. A summary of the data and Monte Carlo samples used, and a description of data driven methods to determine the background contributions from $W$+jets and QCD multijets, are given in \mbox{Chapter \ref{Samples}}. A detailed explanation of the reconstruction method used to obtain parton level information of the top and antitop quark based on measured quantities follows in \mbox{Chapter \ref{Reconstruction}}. An unfolding approach performed to account for detector acceptance and resolution is described in \mbox{Chapter \ref{Unfolding}}, followed by \mbox{Chapter \ref{Systematics}}, covering relevant systematic uncertainties affecting the analysis. Finally, the results of the analysis are presented in \mbox{Chapter \ref{Results}} and a summary of this thesis is given in \mbox{Chapter \ref{Summary}}.

  \chapter{The Top Quark Charge Asymmetry in Context of the Standard Model}
\label{Theory}
\section{The Standard Model}
\begin{figure}[h!tb]
  \begin{centering}
    \includegraphics[width = 8 cm]{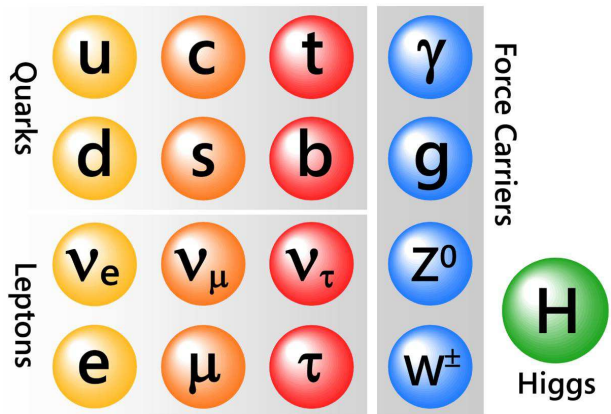}
    \vspace{-0.2 cm}
    \caption[\quad Summary of elementary particles in Nature]{Summary of elementary particles. Aside from the quarks (upper left box) and leptons (lower left box), the gauge bosons (right vertical box) and the hypothetical Higgs boson are shown.}
    \label{fig:SM}
  \end{centering}
\end{figure}

In Nature, all observed matter consists of building blocks which are considered to be elementary, the leptons and quarks, shown schematically in \mbox{Figure \ref{fig:SM}}. They are classified into three generations or families with increasing order of quark masses.
Leptons and quarks obey Fermi-Dirac statistics and hence carry a non-integer spin. Leptons are described by their quantum numbers, electric charge ($Q$), the third component of the weak isospin ($T_3$), and the weak hypercharge ($Y_{\rm{W}} = 2(Q - T_3)$), summarised in \mbox{Table \ref{tab:leptons}}.
\begin{table}[h!tb]
  \small  
  \renewcommand{\arraystretch}{1.2}
  \centering
    \begin{tabular}{|c||c|c|c|c|}
      \hline
      \multicolumn{5}{|c|}{\bf{Leptons (spin $\maybebm{s = \frac{1}{2}}$)}} 																     \cr
      \hline
      \bf{Flavour}  & \bf{Mass [MeV]}         & \bf{$\maybebm{Q}$} & \bf{$\maybebm{T_3}$} & \bf{$\maybebm{Y_{\rm{W}}}$}  \cr
      \hline \hline
      $\nu_e$       &  $\;\,<2 \cdot 10^{-3}$ & $\phm0\phm$  & $\phm\frac{1}{2}\phm$  &  $-1$     \cr 
      $e$           &  $\phantom{23}0.511$    & $-1\phm$     & $-\frac{1}{2}\phm$     &  $-1$     \cr 
      \hline
      $\nu_{\mu}$   &  $<0.19\phantom{^{-5}}$ & $\phm0\phm$  & $\phm\frac{1}{2}\phm$   &   $-1$    \cr 
      $\mu$         &  $105.658$              & $-1\phm$     & $-\frac{1}{2}\phm$    &  $-1$     \cr 
      \hline
      $\nu_{\tau}$  &  $<18.2\phantom{^{-5}}$ & $\phm0\phm$  & $\phm\frac{1}{2}\phm$    &   $-1$   \cr 
      $\tau$        &  $1776.82 \pm 0.16$     & $-1\phm$     & $-\frac{1}{2}\phm$   &   $-1$    \cr 
      \hline      
    \end{tabular}
    \vspace{-0.1 cm}
    \caption[\quad Leptons and their properties and quantum numbers]{Leptons and their properties and quantum numbers, ordered by generation\cite{ReviewParticlePhysics}. Where no uncertainty on the mass is given, it is negligible at the given precision.}
    \label{tab:leptons}
  \renewcommand{\arraystretch}{1.0}
\end{table}

\noindent
Similarly, quarks are assigned the quantum numbers charge ($Q$), the third component of the weak isospin ($T_3$), and hypercharge ($Y = 2(Q - T_3)$), describing qualities of related hadronic bound states. A summary of these properties can be found in \mbox{Table \ref{tab:quarks}}.
\begin{table}[h!tb]
  \small 
  \renewcommand{\arraystretch}{1.2}
  \centering
    \begin{tabular}{|c||c|c|c|c|}
      \hline
      \multicolumn{5}{|c|}{\bf{Quarks (spin $\maybebm{s = \frac{1}{2}}$)}} \cr
      \hline
      \bf{Flavour}      & \bf{Mass [MeV]}                & \bf{$\maybebm{Q}$} & \bf{$\maybebm{T_3}$} & \bf{$\maybebm{Y}$} \cr
      \hline \hline
      $u$               & $3.0$ to $7.0$                 & $\phm\frac{2}{3}\phm$ & $\phm\frac{1}{2}\phm$  & $\phm\frac{4}{3}\phm$             \cr 
      $d$ 			    & $1.5$ to $3.0$                 & $-\frac{1}{3}\phm$    & $-\frac{1}{2}\phm$   & $\phm\frac{1}{3}\phm$             \cr
      \hline
      $c$ 			    & $1250 \pm 90$                  & $\phm\frac{2}{3}\phm$ & $\phm\frac{1}{2}\phm$  & $\phm\frac{4}{3}\phm$             \cr
      $s$ 			    & $\phantom{12}95 \pm 25$        & $-\frac{1}{3}\phm$    & $-\frac{1}{2}\phm$  & $-\frac{2}{3}\phm$              \cr
      \hline
      $t$               & $(173.2 \pm 0.9) \cdot 10^{3}$\cite{Lancaster:2011wr} & $\phm\frac{2}{3}\phm$ & $\phm\frac{1}{2}\phm$   & $\phm\frac{4}{3}\phm$   \cr
      $b$ 			    & $(4.20 \pm 0.07) \cdot 10^{3}$ & $-\frac{1}{3}\phm$  & $-\frac{1}{2}\phm$  & $-\frac{2}{3}\phm$           \cr
      \hline      
    \end{tabular}
  \vspace{-0.1 cm}
  \caption[\quad Quarks and their properties]{Quarks and their properties and quantum numbers, ordered by generation\cite{ReviewParticlePhysics}.}
  \label{tab:quarks}
  \renewcommand{\arraystretch}{1.0}
\end{table}
Furthermore, all quarks carry a colour charge, denoted by a respective quantum number which can take the values red, green and blue.

The first generation is constituted by the up~($u$) and down~($d$) quark doublet (the building blocks of protons and neutrons) alongside the electron~($e$) and the \mbox{electron-neutrino~($\nu_e$)}. The second generation contains the charm~($c$), strange~($s$) quarks, the muon~($\mu$) and muon-neutrino~($\nu_{\mu}$). Finally, the top~($t$) and bottom~($b$) quark compose the third generation, together with the tau~($\tau$) and the tau-neutrino~($\nu_{\tau}$) in the lepton sector. The top quark assumes a quite distinct role among the other quarks due to its large mass of $(173.2 \pm 0.9)\,\text{GeV}$\cite{Lancaster:2011wr}.

Within the Standard Model, particle interactions are described by a quantum field theory consistent with both quantum mechanics and special relativity, combining the electroweak theory and Quantum Chromodynamics into a structure denoted by the gauge symmetry group \mbox{$\text{SU(3)}_C \times \text{SU(2)}_L \times \text{U(1)}_Y$}. This structure describes colour charge ($C$), weak isospin ($L$) and hypercharge ($Y$) gauge groups. The underlying gauge theory is non-Abelian due to the non-commutative nature of the SU(3) and SU(2) field strength tensors.

Given the matter and gauge fields, and requiring local gauge invariance and renormalisability, the Standard Model Lagrangian can be constructed as
\begin{eqnarray}
\mathcal{L}_{\text{SM}} & = & \mathcal{L}_{\text{SU(3)}} + \mathcal{L}_{\text{SU(2)} \times \text{U(1)}} \\
                        & = & \mathcal{L}_{\text{SU(3)}}^{\text{Gauge}} + \mathcal{L}_{\text{SU(3)}}^{\text{Matter\phantom{g}\!\!}} + \mathcal{L}_{\text{SU(2)} \times \text{U(1)}}^{\text{Gauge}} + \mathcal{L}_{\text{SU(2)} \times \text{U(1)}}^{\text{Matter\phantom{g}}} + \mathcal{L}_{\text{SU(2)} \times \text{U(1)}}^{\text{Higgs}} + \mathcal{L}_{\text{SU(2)} \times \text{U(1)}}^{\text{Yukawa\phantom{g}}}.
\end{eqnarray}

The first term, describing strong gauge interactions, is given by
\begin{equation}
\mathcal{L}_{\text{SU(3)}}^{\text{Gauge}} = \frac{1}{2 g_{S}^{2}} \tr{G^{\mu\nu} G_{\mu\nu}},
\end{equation}
where $G^{\mu\nu}$ is the gluon field strength tensor and $g_{S}$ is the strong gauge coupling constant. Strong interactions are described by perturbative Quantum Chromodynamics (QCD).

\noindent
The matter term of the SU(3) Lagrangian contains the gauge covariant derivatives of the quarks:
\begin{equation}
\mathcal{L}_{\text{SU(3)}}^{\text{Matter\phantom{g}\!\!}} = i \bar{q}_{i \alpha} \notpartial_{\beta}^{\alpha} q_{i}^{\beta},
\end{equation}
where $\alpha , \beta \in [1,2,3]$ are the quark colour indices. A summation over the quark flavour index $i$ is implied and it is
\begin{equation}
\notpartial_{\mu \beta}^{\alpha} = \partial_{\mu} \delta_{\beta}^{\alpha} + i g_{S} G_{\mu \beta}^{\alpha}.
\end{equation}

The first term of the electroweak Lagrangian describes the corresponding gauge interactions of the electroweak theory:
\begin{equation}
\mathcal{L}_{\text{SU(2)} \times \text{U(1)}}^{\text{Gauge}} = \frac{1}{2 g^{2}} \tr{W^{\mu\nu} W_{\mu\nu}} - \frac{1}{4 g'^{\,2}} B^{\mu\nu} B_{\mu\nu},
\end{equation}
with the weak isospin and hypercharge gauge field strength tensors $W^{\mu\nu}$ and $B^{\mu\nu}$, respectively.

The SU(2) and U(1) gauge couplings are represented by the constants $g$ and $g'$. Through mixing of the $B$ and $W_3$ fields, the photon and $Z$ boson are generated:
\begin{equation}
\begin{pmatrix}
\gamma \\
Z
\end{pmatrix}
=
\begin{pmatrix}
\phm \cos \theta_W & \sin \theta_W \\
- \sin \theta_W & \cos \theta_W
\end{pmatrix} 
\begin{pmatrix}
B \\
W_3
\end{pmatrix},
\end{equation}
where $\theta_W$ denotes the weak mixing angle. Similarly, the $W^{\pm}$ bosons are generated through mixing of the $W_1$ and $W_2$ fields.

\noindent
The matter term of the electroweak Lagrangian contains the kinetic energy terms from the fermions and their gauge field interactions:
\begin{equation}
\mathcal{L}_{\text{SU(2)} \times \text{U(1)}}^{\text{Matter\phantom{g}}} = i \bar{q}_{L}^{i} \notpartial q_{L}^{i} + i \bar{u}_{R}^{i} \notpartial u_{R}^{i} + i \bar{d}_{R}^{i} \notpartial d_{R}^{i} + i \bar{l}_{L}^{i} \notpartial l_{L}^{i} + i \bar{e}_{R}^{i} \notpartial e_{R}^{i}.
\end{equation}
A summation over the index $i$ is implied. The indices $L$ and $R$ refer to the left and right-handed chiral projections 
\begin{equation*}
\psi_L = (1-\gamma_5) \frac{\psi}{2} \quad \text{and} \quad \psi_R = (1+\gamma_5) \frac{\psi}{2}.
\end{equation*}
The left-handed quark and lepton fields are represented by the SU(2) doublets
\begin{equation}
q_{L}^{i} = {u^i \choose d^i}_L \quad \text{and} \quad l_{L}^{i} = {\nu^i \choose e^i}_L
\end{equation}
while the right-handed fields are represented by the singlets $u_{R}^{i}$, $d_{R}^{i}$ and $e_{R}^{i}$. The gauge covariant terms, describing the electroweak gauge interactions of the fermions, are given by
\begin{eqnarray}
D_{\mu} q_{L}^{i} & = & (\partial_{\mu} + \frac{i g}{2} \tau W_{\mu} + i \frac{g'}{6} B_{\mu}) q_{L}^{i}, \notag\\
D_{\mu} l_{L}^{i} & = & (\partial_{\mu} + \frac{i g}{2} \tau W_{\mu} - i \frac{g'}{2} B_{\mu}) l_{L}^{i}, \notag \\
D_{\mu} u_{R}^{i} & = & (\partial_{\mu} + i \frac{2}{3} g' B_{\mu}) u_{R}^{i}, \\
D_{\mu} d_{R}^{i} & = & (\partial_{\mu} - i \frac{g'}{3} B_{\mu}) d_{R}^{i}, \notag\\
D_{\mu} e_{R}^{i} & = & (\partial_{\mu} - i g' B_{\mu}) e_{R}^{i}. \notag
\end{eqnarray}

\begin{figure}[h!tb]
  \begin{centering}
    \includegraphics[width = 6 cm]{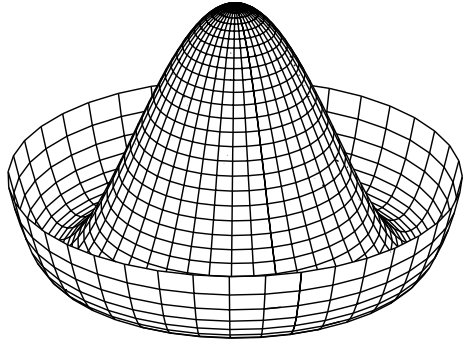}
    \vspace{-0.2 cm}
    \caption[\quad The Higgs potential]{The Higgs potential\footnotemark. Electroweak symmetry breaking induces a non-zero vacuum expectation value in the minima of the Higgs potential, leading to effective masses for the gauge bosons and fermions.}
    \label{fig:higgspot}
  \end{centering}
\end{figure}

Note that there are no mass terms for the fermions in either the SU(3) or the \mbox{SU(2) $\times$ U(1)} gauge theory since such terms are forbidden by the gauge invariance of the Standard Model. A Dirac mass term for a fermion field is not invariant under a chiral transformation and hence would violate the requirement of gauge invariance and renormalisability. However, since weak interactions are observed to be short ranged, the gauge bosons must obtain non-vanishing masses through a different mechanism. Both gauge boson and fermion masses are generated by spontaneous symmetry breaking of the Higgs term in the Standard Model Lagrangian,
\begin{equation}
\mathcal{L}_{\text{SU(2)} \times \text{U(1)}}^{\text{Higgs}} = (D^{\mu} \phi)^{\dag} D_{\mu} \phi + \mu^2 \phi^{\dag} \phi - \lambda (\phi^{\dag} \phi)^2 ,
\end{equation}
containing the kinetic energy of the Higgs field, which is represented by the complex scalar field
\begin{equation}
\phi = {\phi^{+} \choose \phi^{0}}.
\end{equation}
The first term describes the Higgs field interactions with the gauge fields and the latter two denote the Higgs potential, shown in \mbox{Figure \ref{fig:higgspot}}.

The final term in the Standard Model Lagrangian describes the Yukawa interactions of the fermions with the Higgs field:
\begin{equation}
\mathcal{L}_{\text{SU(2)} \times \text{U(1)}}^{\text{Yukawa\phantom{g}}} = - \Gamma_{u}^{ij} \bar{q}_{L}^{i} \epsilon \phi^{\ast} u_R^j - \Gamma_{d}^{ij} \bar{q}_{L}^{i} \epsilon \phi d_R^j - \Gamma_{e}^{ij} \bar{l}_{L}^{i} \epsilon \phi e_R^j + \text{h.c.},
\end{equation}
where $\epsilon = i \sigma_2$ denotes the two dimensional total antisymmetric tensor which ensures electrical neutrality of the individual Yukawa terms and $\sigma_2$ represents the Pauli matrix
\begin{equation} 
\sigma_2 = \left( 
\begin{array}{cc} 
0 & -i \\ 
i & \phm 0 
\end{array} \right). 
\end{equation}
Furthermore, the Yukawa couplings $\Gamma_u$, $\Gamma_d$ and $\Gamma_e$ denote the respective complex $3 \times 3$ matrices in generation space, describing the interactions between the Higgs doublet and the different fermion flavours. As they are not required to be diagonal, a mixing amongst the three generations is allowed.
\footnotetext{Image taken from wikipedia (public domain).}

The electroweak symmetry is spontaneously broken by acquisition of a non-zero vacuum expectation value of the neutral Higgs field component
\begin{equation}
\langle \phi^0 \rangle = \frac{\mu}{\sqrt{2 \lambda}} \equiv \frac{v}{\sqrt{2}}
\end{equation}
which, consequently, generates masses $M_Z$ and $M_W$ for the electroweak gauge bosons through unitarity gauge, and $M_f$ for the fermions from gauge invariant Yukawa couplings $\Gamma_f$ of the Higgs boson to fermions:
\begin{eqnarray}
M_Z & = & \frac{1}{2} v g, \\
M_W & = & \frac{1}{2} v \sqrt{g^2 + g'^{\,2}}, \\
M_f & = & \Gamma_f \frac{v}{\sqrt{2}}.
\end{eqnarray}
A summary of the gauge bosons and their properties can be found in \mbox{Table \ref{tab:bosons}}.

\begin{table}[h!tb]
\begin{center}
\small 
\renewcommand{\arraystretch}{1.2}
    \begin{tabular}{|cl||c|c|c|}
      \hline
      \multicolumn{5}{|c|}{\bf{Bosons (integer spin)}} \cr
      \hline
      \multicolumn{2}{|c||}{\bf{Particle}}  & \bf{Mass [GeV]}  & \bf{Charge} & \bf{Spin}  \cr
      \hline \hline
      $(\gamma)$ & Electromagnetic force      &  $0$      & $\phm0$  & $1$     \cr 
      \hline
      $(g)$ & Strong force           &  $0$      & $\phm0$ & $1$        \cr 
      \hline
      $(W^{-})$ & \multirow{3}{*}{Weak force}     &  $80.403 \pm 0.029$        & $-1$ & $1$       \cr 
      $(W^{+})$ &                           &  $80.403 \pm 0.029$        & $+1$ & $1$      \cr 
      $(Z^{0})$ &                           &  $91.188 \pm 0.003$      & $\phm0$ & $1$      \cr       
      \hline      
      $(H)$ & Mass (hypoth.)                &  $116 - 127\,$(95\,\% C.L.) \cite{ATLAS-CONF-2011-163,CMS-PAS-HIG-11-032}   & $\phm0$ & $0$      \cr     
      \hline 
    \end{tabular}
  \vspace{-0.1 cm}
    \caption[\quad Gauge bosons and their properties]{Gauge bosons and their properties and quantum numbers\cite{ReviewParticlePhysics}.}
    \label{tab:bosons}
  \renewcommand{\arraystretch}{1.0}
  \end{center}    
\end{table}

\section{The Top Quark}
In the following, the production and decay of top quark pairs and singly produced top quarks ({\em single tops}) within the Standard Model will be discussed in detail. Furthermore, an overview of important properties of the top quark and their measurement will be given. In particular, the charge asymmetry in the production of top quarks pairs within the Standard Model and in theories beyond will be covered.

\subsection{Top Quark Production at Hadron Colliders}
\label{sec:topprod}
\begin{figure}[h!tb]
  \begin{centering}
    \includegraphics[width = 14 cm]{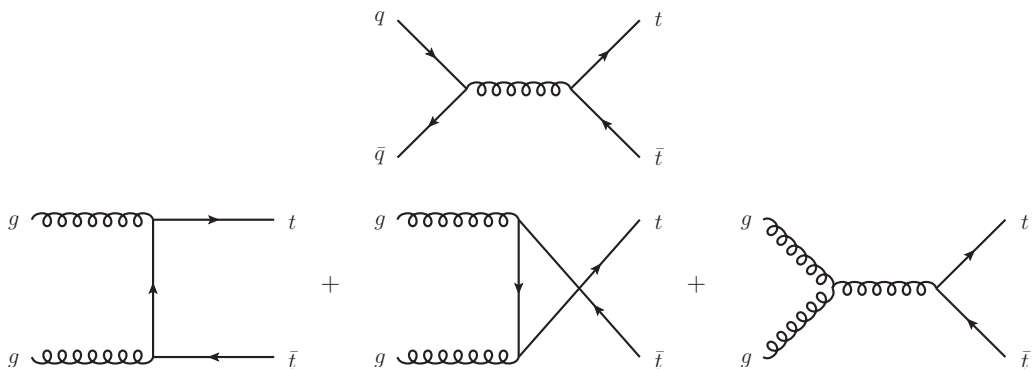}
    \caption[\quad Lowest order diagrams contributing to top quark pair production at hadron colliders]{Lowest order diagrams contributing to top quark pair production at hadron colliders. Top quarks are produced via strong interaction, either in quark-antiquark annihilation (top) or gluon-gluon fusion (bottom).}
    \label{fig:ttbarprod}
  \end{centering}
\end{figure}

At hadron colliders, \ttbar~pairs are mainly produced through strong interactions described by perturbative QCD. Interactions between the quark and gluon constituents of the colliding hadrons (either protons or antiprotons) participate in a hard scattering process and produce a top quark and an antitop quark in the final state. At Born level approximation, top quark pairs can be produced via gluon-gluon fusion ($gg$) or via the annihilation of quark-antiquark pairs ($q\bar{q}$). The relevant leading order Feynman diagrams for the contributing processes are shown in \mbox{Figure \ref{fig:ttbarprod}}.

Due to the fact that hadrons are composite particles, consisting of partons with unknown fractions $x$ of the initial hadron momenta, the initial state of the parton interaction is not precisely known. However, hadron interactions in $pp$ and $p\bar{p}$ collisions can be described by separating the partonic reactions into a short distance and a long distance contribution.

The long distance part can be factorised into longitudinal parton momentum distribution functions (PDFs) $f_i (x_i , \mu_{F}^{2})$, where $\mu_{F}^{2}$ denotes an (arbitrary) factorisation scale describing the separation of the long and short distance contributions. An additional renormalisation scale $\mu_{R}^{2}$ is introduced to account for higher order corrections, where ultraviolet divergent terms may emerge and a renormalisation approach can be used to absorb such divergences into corresponding counter terms. Both scales $\mu_{F}^{2}$ and $\mu_{R}^{2}$ are commonly chosen to correspond to the momentum transfer $\mu_{F}^{2} = \mu_{R}^{2} = Q^2$. Furthermore, for the calculation and simulation of top quark processes, $Q^2$ is typically chosen such that $\mu_{F} = \mu_{R} = Q = m_t$ corresponds to the top pole mass $m_t$ and the associated scale variation dependency is studied.

\begin{figure}[h!tb]
  \begin{centering}
    \mbox{
      \subfigure[$e^{+}p$]{
        \scalebox{0.45}{\includegraphics[width=\textwidth]{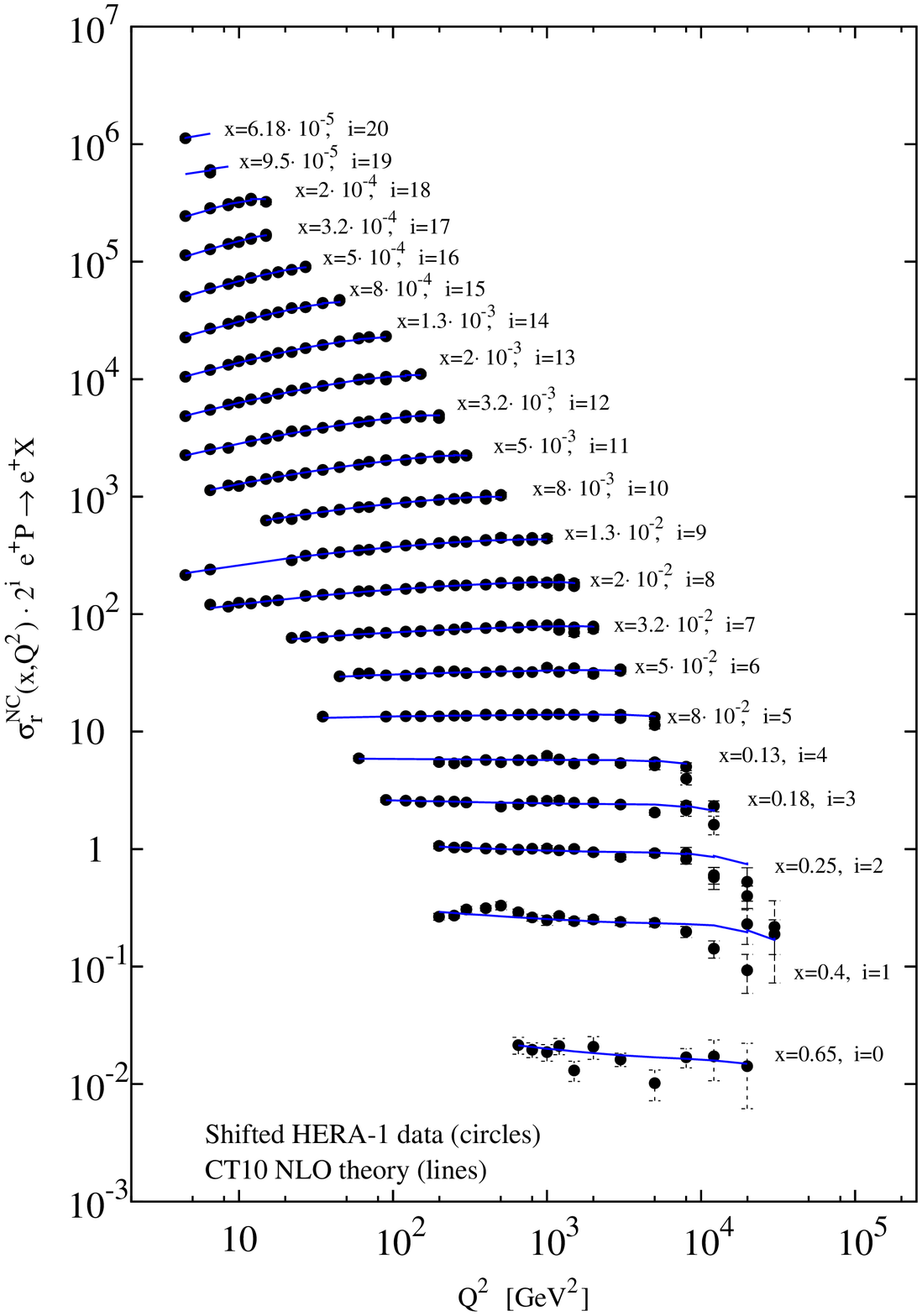}}
      }
      \quad
      \subfigure[$e^{-}p$]{
        \scalebox{0.45}{\includegraphics[width=\textwidth]{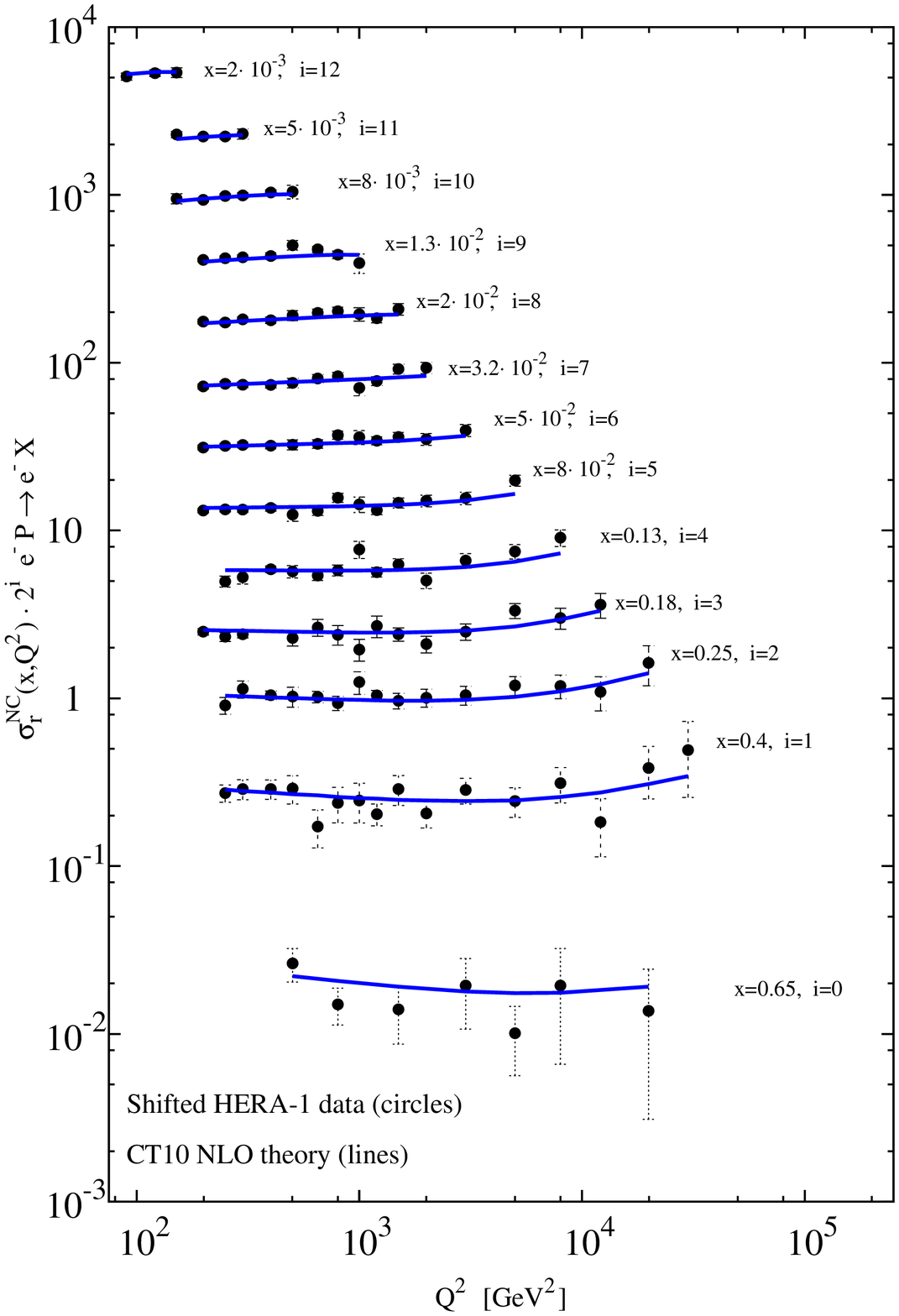}}
      }
    }
    \vspace{-0.2 cm}
    \caption[\quad Comparison of CTEQ10 NLO predictions]{Comparison of CTEQ10 NLO predictions\cite{Lai:2010vv} for reduced cross-sections in $e^{+}p$ (left) and $e^{-}p$ (right) neutral-current deep inelastic scattering experiments from combined HERA-1 data\cite{Hera:2009wt}, with correlated systematic shifts included.}
    \label{fig:HERAPDFs}
  \end{centering}
\end{figure}

The PDFs represent the probability distribution of observing a parton of type $i$ at a given scale $\mu_{F}^{2}$ with a longitudinal parton momentum fraction $x_i$. Since these probabilities cannot be universally derived from QCD, they have to be provided from experimental studies of the proton structure, mostly from deep inelastic lepton-proton scattering experiments at the H1\cite{Abt:1993cb,Adloff:1997sc,Aid:1996au,Adloff:2000qk} and ZEUS\cite{Derrick:1993fta,Derrick:1996hn,Breitweg:1997ff,Chekanov:2001qu} experiments at the HERA electron-proton collider. As an example, the $e^{+}p$ and $e^{-}p$ production cross-sections measured in deep-inelastic scattering experiments at HERA can be found in \mbox{Figure \ref{fig:HERAPDFs}}\cite{Hera:2009wt} in comparison to the CTEQ10 PDF next-to-leading order (NLO) prediction\cite{Lai:2010vv}.

The short distance term arises from the hard scattering process of the respective partons, denoted by the partonic cross-section for partons $i$ and $j$, $\sigma_{ij}$. This contribution is characterised by high momentum transfer. Hence, it is not dependent on the incoming hadron type or the respective wave functions and can be described by perturbative QCD, as indicated by the leading order diagrams in \mbox{Figure \ref{fig:ttbarprod}}.

At a given centre-of-mass energy $\sqrt{s}$ and for a top mass parameter $m_t$, the total top quark pair production cross-section can be calculated from the short distance and long distance terms as
\begin{equation}
\sigma_{t \bar{t}} \left( \sqrt{s}, m_t \right) = \sum_{i,j} \iint \text{d}x_i \text{d}x_j f_i (x_i , Q^2) f_j (x_j , Q^2) \times \sigma_{ij \rightarrow t \bar{t}} \left( \rho, m_t^2 , x_i , x_j , \alpha_s (Q^2), Q^2 \right),
\end{equation}
where the summation is performed over all permutations of $i,j = \lbrace q, \bar{q}, g \rbrace$. The PDFs of the initial state protons are denoted by $f_i (x_i , Q^2)$ and $f_j (x_j , Q^2)$, respectively and the parameter $\rho$ is given by
\begin{equation}
\rho = \frac{4 m_t^2}{\sqrt{x_i x_j s}} = \frac{4 m_t^2}{\sqrt{\hat{s}}},
\end{equation}
where $x_i x_j s \equiv \hat{s}$ denotes the effective centre-of-mass energy in the partonic reaction.

\begin{figure}[h!tb]
  \begin{centering}
    \mbox{
      \subfigure[$\mu = 5\,\text{GeV}$]{
        \scalebox{0.35}{\includegraphics[width=\textwidth]{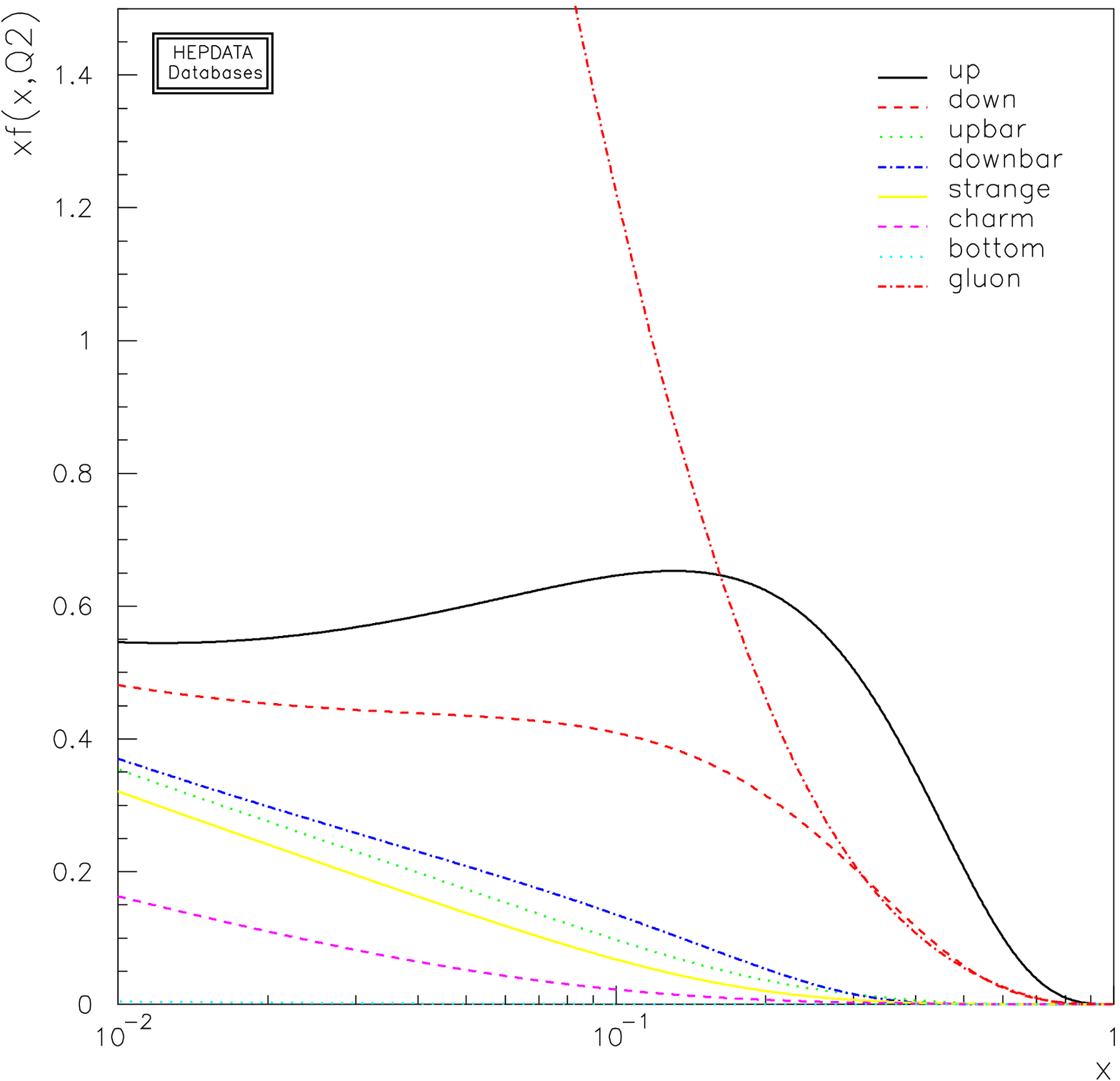}}
      }
      \quad
      \subfigure[$\mu = m_t$]{
        \scalebox{0.35}{\includegraphics[width=\textwidth]{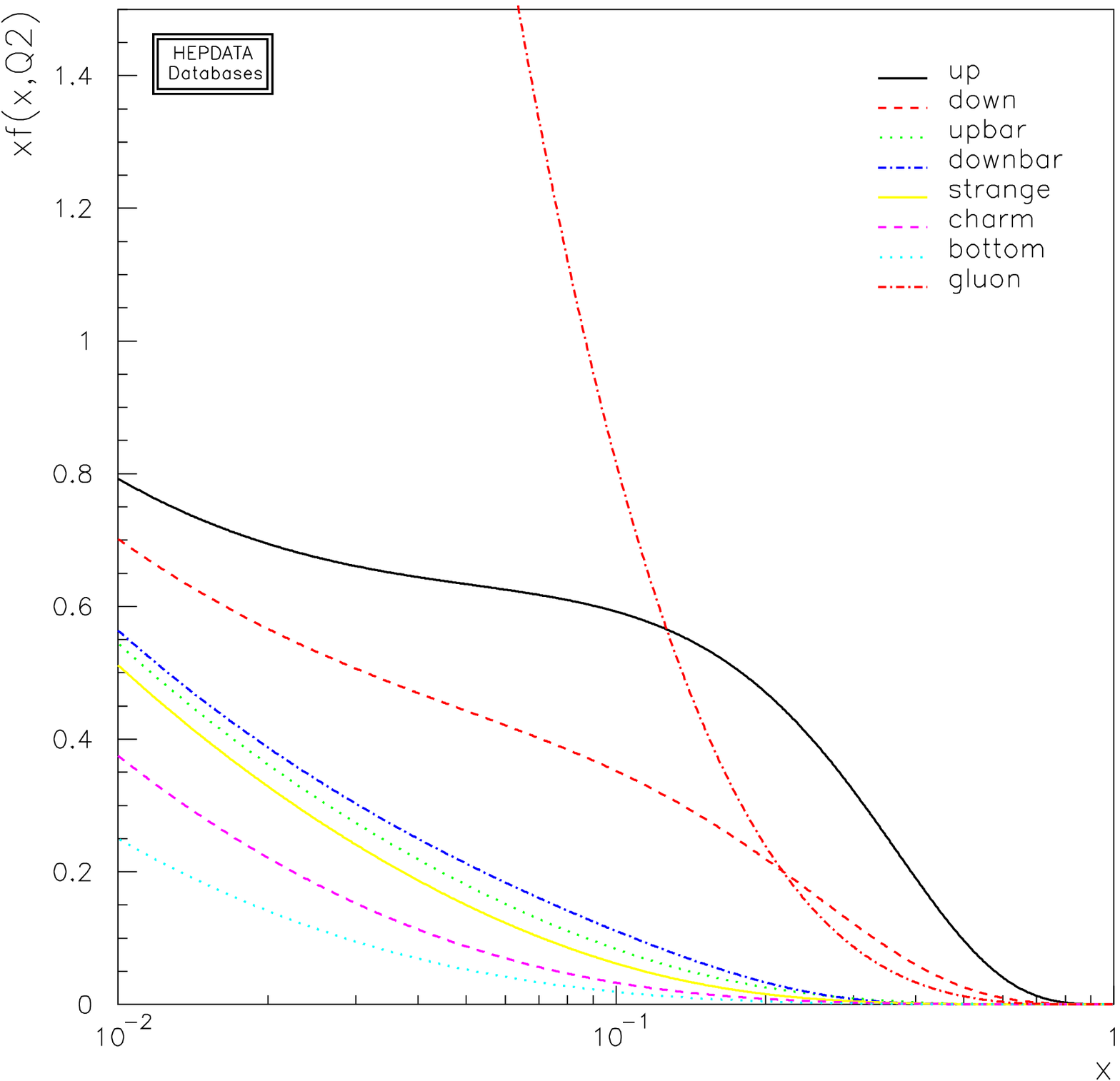}}
      }
    }
    \vspace{-0.2 cm}
    \caption[\quad Parton distribution functions at different momentum transfers]{CTEQ10 parton distribution functions\cite{Lai:2010vv} at different momentum transfers for gluons and different quark/antiquark flavours. Shown are the PDF sets for $\mu = 5\,\text{GeV}$ (left) and $\mu = m_t$ (right).}
    \label{fig:PDFs}
  \end{centering}
\end{figure}

The probability of a parton $i$ to be carrying a momentum fraction of $x_i$ decreases significantly with rising $x_i$, as can be seen in \mbox{Figure \ref{fig:PDFs}}, where two PDFs from the CTEQ10 PDF set\cite{Lai:2010vv} are shown as an example.
The PDFs have been evaluated at scales $\mu = 5\,\text{GeV}$ and $\mu = m_t$, respectively, where $\mu \equiv Q$.

The minimal energy carried by the two incoming partons to produce a top quark pair at the threshold (i.e. at rest), is given by
\begin{equation}
\sqrt{x_i x_j s} \geq 2m_t
\end{equation}
and hence, assuming both partons carrying the same momentum fractions as an approximation, $x_i \approx x_j \equiv x$:
\begin{equation}
x \approx \frac{2m_t}{\sqrt{s}}.
\end{equation}
This corresponds to a typical value of $x \approx 0.05$ at the LHC for a centre-of-mass energy of $\sqrt{s} = 7\,\text{TeV}$. As shown in \mbox{Figure \ref{fig:PDFs}}, the gluon PDFs dominate significantly over any other parton in the corresponding range of $x$. Consequently, the production of top quark pairs at the LHC is dominated by gluon-gluon fusion. At the Tevatron (where the typical value of $x$ is of the order of 0.2) the production of top quarks is dominated by quark-antiquark annihilation processes, in particular involving up and down valence quarks. Since the centre-of-mass energy at the LHC is significantly higher, top quark pairs are typically produced above the threshold, but still within the gluon-gluon fusion dominated range of the PDFs.

The total \ttbar~cross-section at the LHC is predicted in an approximate next-to-next-to-leading order (NNLO) calculation to be $165_{-16}^{+11}\,\text{pb}$\cite{PhysRevD.78.034003,Langenfeld:2009tc,Beneke2010483} for a centre-of-mass energy of $\sqrt{s} = 7\,\text{TeV}$ and $m_t = 172.5\,$GeV. Preliminary measurements have been performed at both ATLAS and CMS, yielding 
\begin{eqnarray}
  \sigma_{t \bar{t}}\,(\sqrt{s} = 7\,\text{TeV}) & = & 179.0_{-9.7}^{+9.8}\text{\,(stat.+syst.)} \pm 6.6\text{\,(lumi.)}\,\text{pb} \text{\cite{ATLAS-CONF-2011-121}}, \\
  \sigma_{t \bar{t}}\,(\sqrt{s} = 7\,\text{TeV}) & = & 165.8 \pm 2.2\text{\,(stat.)} \pm 10.6\text{\,(syst.)} \pm 7.8\text{\,(lumi.)}\,\text{pb} \text{\cite{CMS-PAS-TOP-11-024}},
\end{eqnarray}
respectively. Both measurements are in agreement with the Standard Model prediction.

\begin{figure}[h!tb]
  \begin{centering}
    \includegraphics[width = 6 cm]{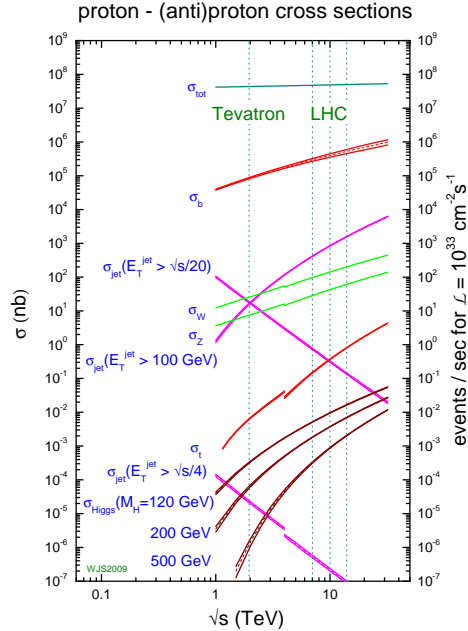}
    \vspace{-0.2 cm}
    \caption[\quad QCD predictions for hard-scattering cross-sections at the Tevatron and the LHC]{QCD predictions for hard-scattering cross-sections at the Tevatron and the LHC\cite{Catani:2000xk}. The top quark pair production cross-section is denoted by $\sigma_t$. The discontinuities in the different curves denote the change from $p\bar{p}$ collisions at the Tevatron to $pp$ collisions at the LHC. For the LHC, different centre-of-mass energies are highlighted by three vertical dashed lines, where the leftmost line corresponds to a centre-of-mass energy of $\sqrt{s} = 7\,\text{TeV}$.}
    \label{fig:SMxsects}
  \end{centering}
\end{figure}

This cross-section is several orders of magnitude lower than, for example, the SM $Z$ and $W$ boson production cross-sections or the inclusive QCD multijet production cross-section at comparable values of $Q^2$. This can be seen in \mbox{Figure \ref{fig:SMxsects}}, where the total production cross-sections for several SM processes are shown as a function of centre-of-mass energy of the colliding (anti)protons.
Consequently, a sophisticated real-time selection to identify the relevant final state particles and obtain a good signal to background separation with respect to other SM processes and, more importantly, the dominant QCD multijet background, is crucial for all top quark related measurements at the LHC. Furthermore, an extensive theoretical understanding and modelling of these backgrounds is necessary to facilitate the measurement of top quark properties to highest precision and in order to achieve a sensitivity to potential deviations from the Standard Model expectations.

\subsection{Production of Single Top Quarks}
\label{theory_ST}
\begin{figure}[h!tb]
  \begin{centering}
    \mbox{
      \subfigure[]{
        \scalebox{0.45}{\includegraphics[width=\textwidth]{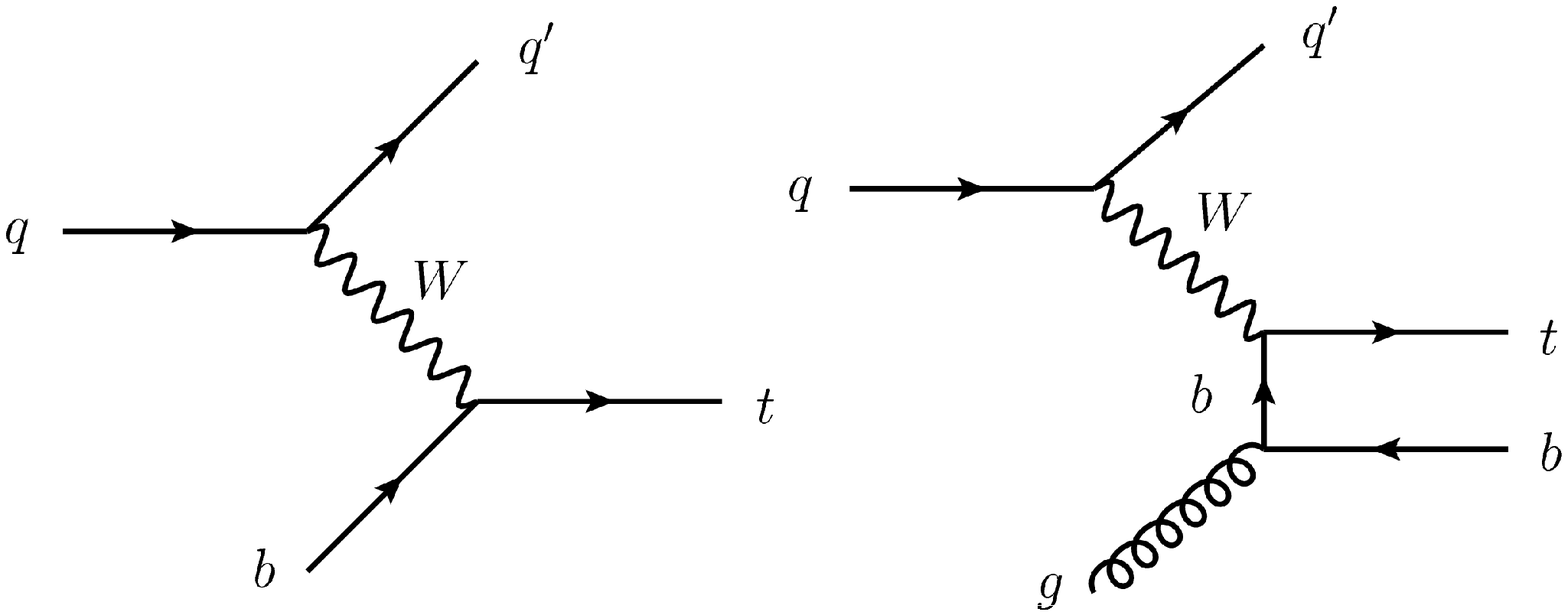}}
      }
      \quad
      \subfigure[]{
        \scalebox{0.21}{\includegraphics[width=\textwidth]{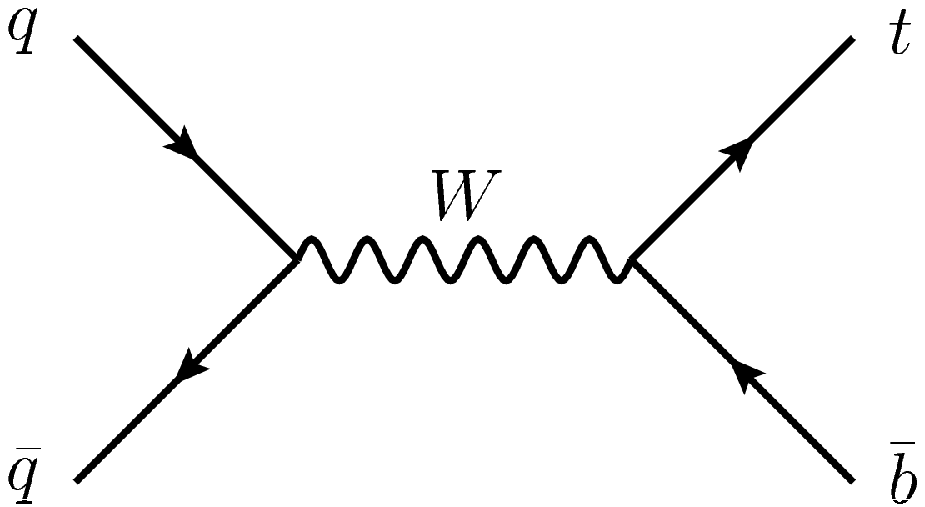}}
      }
      \quad
      \subfigure[]{
        \scalebox{0.21}{\includegraphics[width=\textwidth]{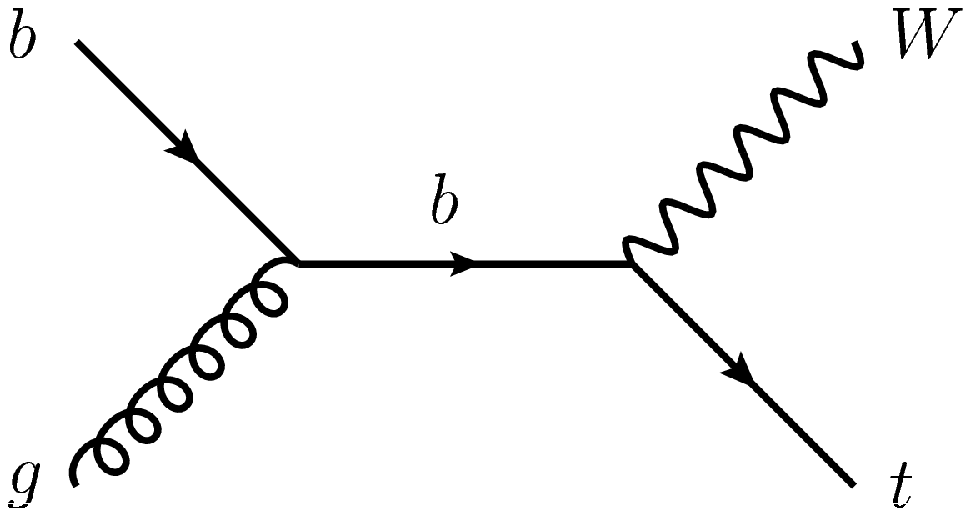}}
      }
    }
    \vspace{-0.2 cm}
    \caption[\quad Electroweak single top production diagrams]{Electroweak single top production diagrams via $W$-gluon fusion (a), exchange of an off-shell $W^{\ast}$ (b) and $Wt$ production (c).}
    \label{fig:stprod}
  \end{centering}
\end{figure}

In addition to the production of top quark pairs via the strong interaction, single top quarks can be produced in electroweak charged current interactions. Three mechanisms for this production exist, as shown in \mbox{Figure \ref{fig:stprod}}. Single top quarks can emerge in the fusion of $W$ bosons and gluons ($t$-channel process) similar to the production of heavy flavour quarks in deep-inelastic scattering via charged current interactions. In addition, they can be produced by annihilation of quark-antiquark pairs ($s$-channel process) and exchange of an off-shell $W^{\ast}$, or by $Wt$ production in quark-gluon interactions.
The corresponding production cross-sections have been approximated at next-to-next-to-leading-order to be 
\begin{eqnarray*}
\sigma^{\text{t-ch}\phantom{Wt}}_{t}\!\!\!\!\!\!\,(\sqrt{s} = 7\,\text{TeV}) & = & 64.57_{-2.01}^{+2.71}\,\text{pb} \text{\cite{st_tchan_pred}}, \\
\sigma^{\text{s-ch}\phantom{Wt}}_{t}\!\!\!\!\!\!\,(\sqrt{s} = 7\,\text{TeV}) & = & \pho4.63_{-0.17}^{+0.19}\,\text{pb} \text{\cite{st_schan_pred}}, \\
\sigma^{Wt\phantom{\text{t-ch}}}_{t}\!\!\!\!\!\!\,(\sqrt{s} = 7\,\text{TeV}) & = & 15.74_{-1.08}^{+1.06}\,\text{pb} \text{\cite{st_tchan_pred}},
\end{eqnarray*}
for $m_t = 172.5\,$GeV in the $t$-channel, in the $s$-channel, and for $Wt$ production, respectively. Since in all production channels a top charged current is involved, the respective cross-sections behave as
\begin{equation*}
\sigma_{t} \propto |V_{tb}|^2 g^2,
\end{equation*}
where $|V_{tb}|$ denotes the relative probability that the top quark decays into a bottom quark via the exchange of a $W$ boson. This is described by the Cabbibo-Kobayashi-Maskawa (CKM) matrix $V_{\text{CKM}}$\cite{ReviewParticlePhysics} which summarises the relative transition probabilities in charged weak interactions. The magnitudes of the CKM matrix elements, here denoted by the matrix $M_{\text{CKM}}$, are
\begin{eqnarray}
M_{\text{CKM}} & = &
\begin{pmatrix}
|V_{ud}| & |V_{us}| & |V_{ub}| \\
|V_{cd}| & |V_{cs}| & |V_{cb}| \\
|V_{td}| & |V_{ts}| & |V_{tb}|
\end{pmatrix} \notag \\
 & = & 
\renewcommand{\arraystretch}{1.2}
\begin{pmatrix}
0.97428 \pm 0.00015 & 0.2253 \pm 0.0007 & 0.00347^{+0.00016}_{-0.00012} \\
0.2252 \pm 0.0007 & 0.97345^{+0.00015}_{-0.00016} & 0.0410^{+0.0011}_{-0.0007} \\
0.00862^{+0.00026}_{-0.00020} & 0.0403^{+0.0011}_{-0.0007} & 0.999152^{+0.000030}_{-0.000045}
\end{pmatrix}.
\renewcommand{\arraystretch}{1.0}
\end{eqnarray}
As a consequence of this dependency, measurements of the single top quark production cross-section provide an implicit sensitivity to the CKM matrix element $V_{tb}$. At the LHC, single tops are predominantly produced via the $t$-channel interaction, followed by $Wt$ production, due to the large initial state gluon contribution at the LHC centre-of-mass energy, while the $s$-channel process is suppressed.

First direct evidence for single top quarks was found in 2006 by the \dzero~collaboration at the Tevatron\cite{Abazov:2006gd}, followed by its observation\cite{Abazov:2009ii,Aaltonen:2009jj} in 2009 by the \dzero~and CDF experiments. First preliminary measurements of the inclusive single top production cross-section have been conducted at the LHC, yielding
\begin{equation*}
\sigma^{t\text{-ch}}_{t}\,(\sqrt{s} = 7\,\text{TeV}) = 90_{-22}^{+32}\,\text{pb} \text{\cite{ATLAS-CONF-2011-101}}
\end{equation*}
and
\begin{equation*}
\sigma^{t\text{-ch}}_{t}\,(\sqrt{s} = 7\,\text{TeV}) = 83.6 \pm 29.8\text{\,(stat.+syst.)} \pm 3.3\text{\,(lumi.)} \,\text{pb} \text{\cite{PhysRevLett.107.091802}}
\end{equation*}
in the $t$-channel as measured by ATLAS and CMS, respectively. The $Wt$ production cross-section has been measured at CMS to be
\begin{equation*}
\sigma^{Wt}_{t}\,(\sqrt{s} = 7\,\text{TeV}) = 22_{-7}^{+9}\text{\,(stat.+syst.)} \,\text{pb} \text{\cite{CMS-PAS-TOP-11-022}}
\end{equation*}
and a limit has been set by a corresponding measurement at ATLAS, corresponding to
\begin{equation*}
\sigma^{Wt}_{t}\,(\sqrt{s} = 7\,\text{TeV}) < 39\,\text{pb} \text{\cite{ATLAS-CONF-2011-104}}
\end{equation*}
at 95\,\% C.L. Furthermore, searches for single top quarks produced in $s$-channel interactions have been conducted at ATLAS, limiting the corresponding production cross-section to
\begin{equation*}
\sigma^{s\text{-ch}}_{t}\,(\sqrt{s} = 7\,\text{TeV}) < 26.5 \,\text{pb} \text{\cite{ATLAS-CONF-2011-118}}
\end{equation*}
at 95\,\% C.L. All measurements are in agreement with Standard Model predictions.

\subsection{Top Quark Decay}
Due to the magnitude of the CKM matrix element $V_{tb}$ being close to unity, the top quark decays in almost $100\,\%$ of the cases via electroweak charged current interaction into a $b$ quark and $W$ boson, which then in turn decays either leptonically into a charged lepton and the corresponding (anti)neutrino or hadronically into a quark-antiquark pair. At leading order, the Standard Model prediction for the total decay width of the top quark, $\Gamma_t^0$, is given by
\begin{equation}
\Gamma_t^0 = \frac{G_F m_t^3}{8 \pi \sqrt{2}} |V_{tb}|^2,
\end{equation}
where $G_F$ denotes the Fermi coupling constant
\begin{equation}
G_F = \frac{\sqrt{2}}{8} \frac{g^2}{M_W^2}.
\end{equation}

Taking into account higher order corrections at next-to-leading order, the total top quark decay width becomes
\begin{equation}
\Gamma_t = \Gamma_t^0 \left( 1 - \frac{M_W^2}{m_t^2} \right)^2 \left( 1 + 2 \frac{M_W^2}{m_t^2} \right)^2 \left[ 1 - \frac{2 \alpha_s}{3 \pi} \left( \frac{2 \pi^2}{3} - \frac{5}{2} \right) \right],
\end{equation}
where terms of order $m_b^2 / m_t^2$ and $(\alpha_s / \pi) M_W^2 / m_t^2$ have been neglected. At a top mass of $170\,\text{GeV}$ and $\alpha_s$ evaluated at the $Z$ scale, this yields an approximate predicted decay width of
\begin{equation}
\Gamma_t \approx 1.3\,\text{GeV},
\end{equation}
and a corresponding mean lifetime of
\begin{equation}
\tau_t \approx 0.5 \cdot 10^{-24}\,\text{s},
\end{equation}
which is significantly lower than the time scale corresponding to the strong hadronisation scale $\Lambda_{\text{QCD}} \approx 250\,\text{MeV}$. Hence, the top quark decays before being able to form hadronic bound states such as the \ttbar-quarkonium. Consequently, the top quark spin/polarisation properties are preserved in its decay and are transferred to the decay products.

Since the top quark decays almost exclusively into a $b$ quark and $W$ boson, the resulting final state decay channels are well defined and can be separated into three cases, characterised by the final state particles:
\begin{itemize}
  \item {\bf Full hadronic final state (alljets):} Both $W$ bosons from the \ttbar~pair further decay into quarks, leading to a total amount of six quarks including the $b$ quarks from the initial top and antitop decays.
  \item {\bf Semileptonic final state (lepton+jets):} One $W$ boson from the \ttbar~pair decays into quarks, while the second one decays leptonically, leading to a total of four quarks including the $b$ quarks from the initial top and antitop decays, and one charged lepton and its corresponding (anti-)neutrino.
  \item {\bf Dileptonic final state (dilepton):} Both $W$ bosons from the \ttbar~decay into a charged lepton and the corresponding (anti-)neutrino, respectively. In addition, two remaining $b$ quarks from the top and antitop decays are produced.
\end{itemize}
If the charged lepton is a $tau$ in the semileptonic or dileptonic channel, either a muon or electron and the corresponding (anti-)neutrino, or further quarks from the hadronic decay of the $tau$ lepton are produced.

\begin{figure}[h!tb]
  \begin{centering}
    \mbox{
      \subfigure{
        \scalebox{0.35}{\includegraphics[width=\textwidth]{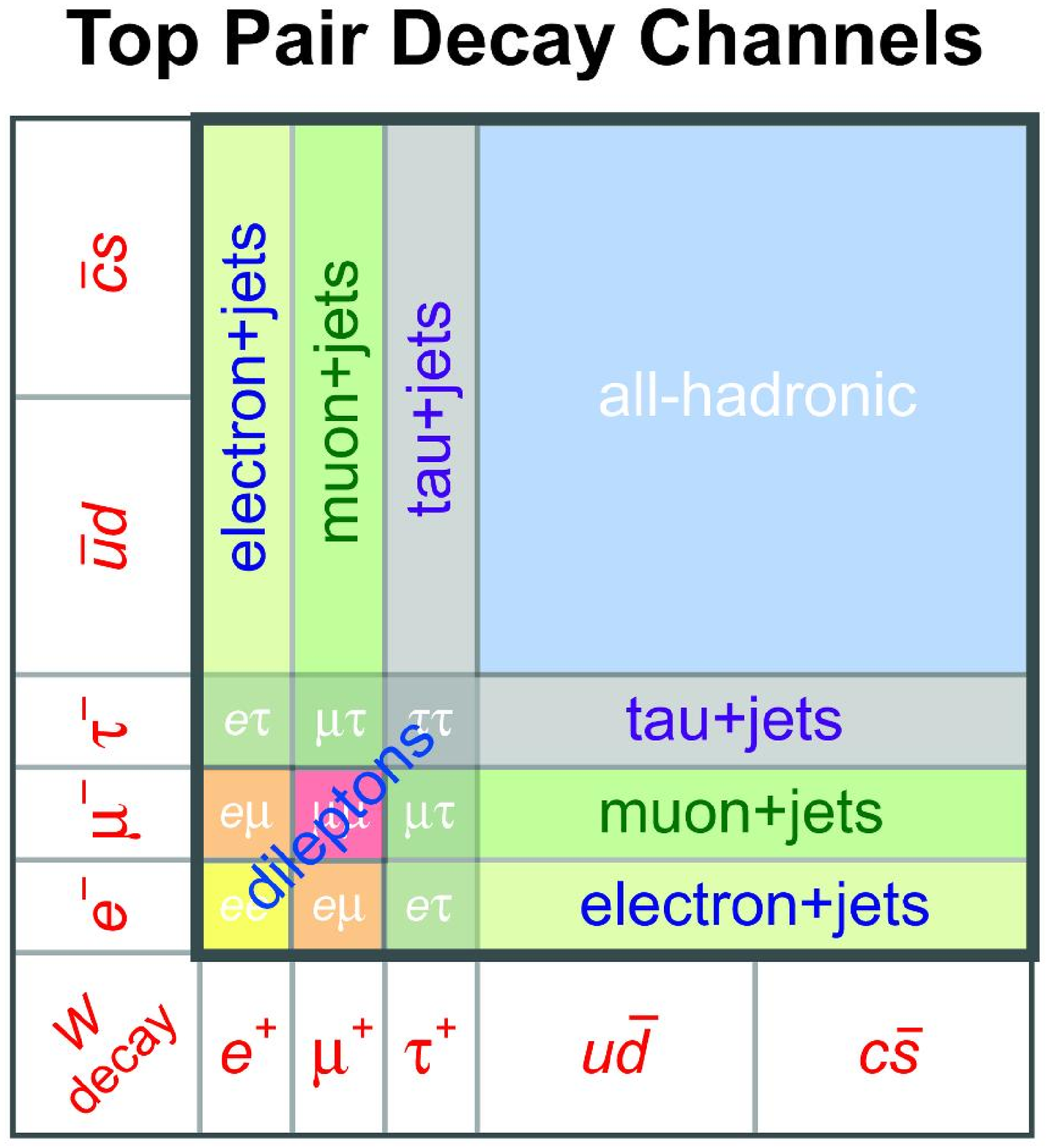}}
      }
      \quad
      \subfigure{
        \scalebox{0.473}{\includegraphics[width=\textwidth]{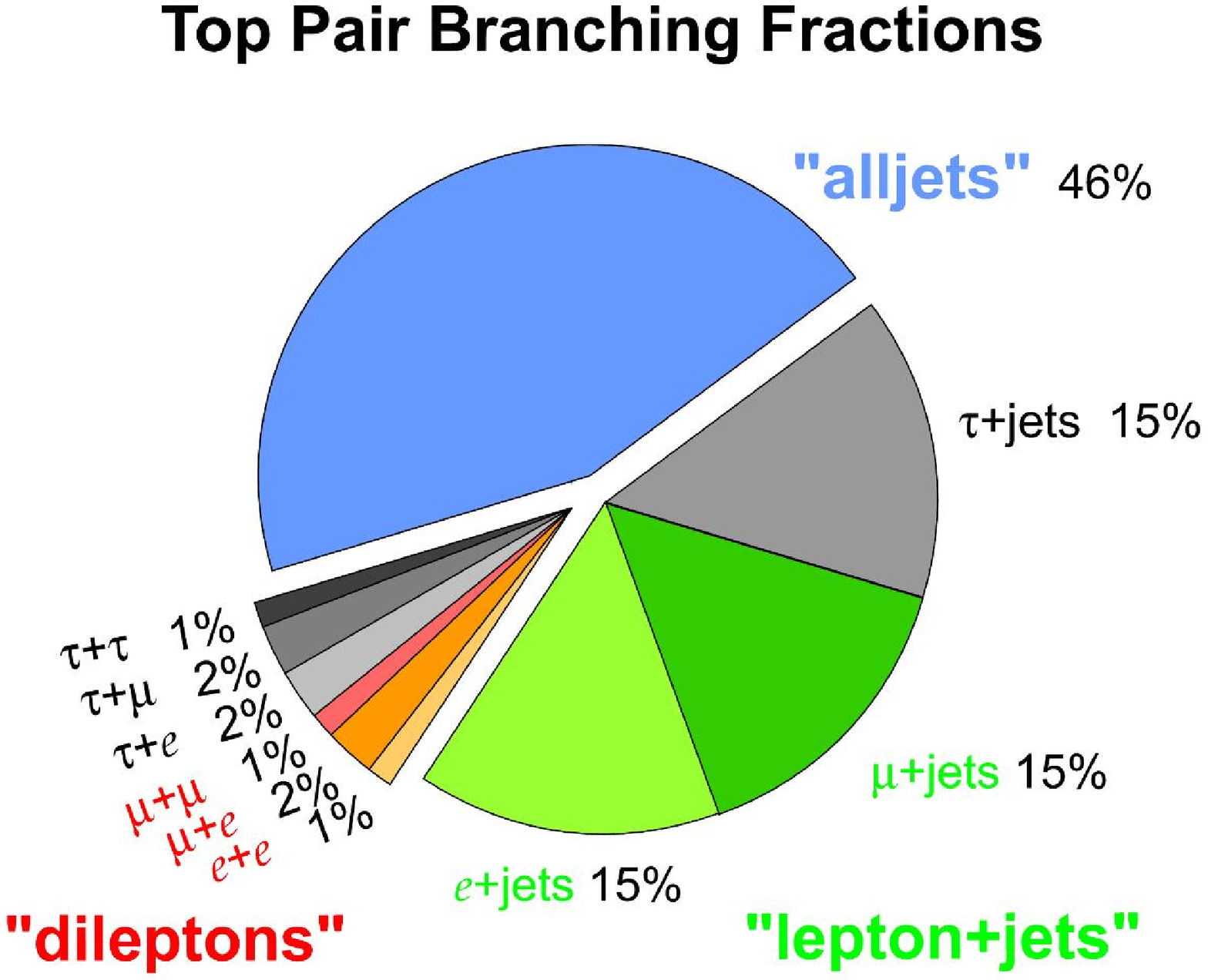}}
      }
    }
    \vspace{-0.2 cm}
    \caption[\quad Top quark pair decay channels and branching fractions]{Top quark pair decay channels (left) and branching fractions (right)\footnotemark.}
    \label{fig:ttbarfrac}
  \end{centering}
\end{figure}

The respective $W$ branching ratios (BR) at leading order (LO) can be found in \mbox{Table \ref{tab:topbfracs}}.
\begin{table}[h!tb]
\begin{center}
\small 
\renewcommand{\arraystretch}{1.2}
    \begin{tabular}{|l||c|c|}
      \hline
      Decay mode & LO BR & Measured BR \cr
      \hline \hline
      $W^{\pm} \rightarrow q\bar{q}$           & $\frac{6}{9}$ & $(67.60 \pm 0.27)\,\%$  \cr
      \hline
      $W^{\pm} \rightarrow e^{\pm}\nu_e$         & $\frac{1}{9}$ & $(10.75 \pm 0.13)\,\%$  \cr
      $W^{\pm} \rightarrow \mu^{\pm}\nu_{\mu}$   & $\frac{1}{9}$ & $(10.57 \pm 0.15)\,\%$  \cr
      $W^{\pm} \rightarrow \tau^{\pm}\nu_{\tau}$ & $\frac{1}{9}$ & $(11.25 \pm 0.20)\,\%$  \cr
      \hline 
    \end{tabular}
  \vspace{-0.1 cm}
    \caption[\quad Theoretical and measured $W$ branching ratios]{Theoretical (LO) and measured $W$ branching ratios\cite{ReviewParticlePhysics}.}
    \label{tab:topbfracs}
  \renewcommand{\arraystretch}{1.0}
  \end{center}    
\end{table}
Taking these branching fractions into account, possible \ttbar~final states and their approximate relative probabilities are shown schematically in \mbox{Figure \ref{fig:ttbarfrac}}.

\subsection{Top Quark Properties}
Several properties of the top quark have been studied in collider experiments such as the Tevatron and the LHC. Amongst them, the top quark mass has been determined with a relative uncertainty of only $0.5\,\%$\cite{Lancaster:2011wr} by combining the most recent measurements from \dzero~and CDF. This combination constitutes the most precise (in relative terms) mass measurement of any quark so far. Since the top quark does not form hadronic bound states, the top quark mass $m_t$ is defined as the pole mass in this context and is measured to be
\begin{equation}
m_t = 173.18 \pm 0.56\,\text{(stat.)} \pm 0.75\,\text{(syst.)}.
\end{equation}

In addition, measurements to exclude an exotic top quark carrying a charge of $4e/3$\footnote{Here, $e$ denotes the electron charge.}, which is predicted to be $2e/3$ in the Standard Model, have been performed. This model has been excluded with approximately $95\,\%$ C.L.\cite{Aaltonen:2010js} and $90\,\%$ C.L.\cite{Abazov:2006vd}, respectively, in independent measurements at CDF and \dzero. A corresponding measurement at ATLAS excludes a top quark charge of $4e/3$ at more than five standard deviations\cite{ATLAS-CONF-2011-141}.
\footnotetext{Images taken from http://www-d0.fnal.gov/Run2Physics/top/.}

Since the top quark does not form bound hadronic states due to its small lifetime, it provides the unique opportunity to measure quark properties which are usually concealed by hadronisation, such as spin correlations between quark-antiquark pairs produced at hadron colliders. Since all other quarks depolarise due to QCD interactions before fragmentation, it is not possible to gain knowledge about the spin from the final state, while the final state particles in the top quark decay preserve significant amount of information about the spins of the initial top and antitop quarks to potentially allow their measurement. First studies at the Tevatron using the angular distributions of the final state particles indicate a correlation strength $C$ of 
\begin{equation}
C = 0.57 \pm 0.31\,\text{(stat.+syst.)} \text{\cite{Abazov:2011ka}}
\end{equation}
in the beam basis, compatible with the next-to-leading order Standard Model prediction of \mbox{$C_{\text{SM}} = 0.78_{-0.04}^{+0.03}$}\cite{Bernreuther:2004jv} and excluding a non-correlation hypothesis at $97.7\,\%$ C.L. Similar studies at ATLAS indicate a correlation strength $A_{\text{helicity}}$ in the helicity basis of
\begin{equation}
A_{\text{helicity}} = 0.34_{-0.11}^{+0.15}\,\text{(stat.+syst.)} \text{\cite{ATLAS-CONF-2011-117}},
\end{equation}
which is compatible with the corresponding next-to-leading order Standard Model prediction.

Furthermore, polarisations of $W$ bosons from the top quark decay are predicted to arise due to the different possible helicity states of the produced on-shell $W$ bosons. Consequently, the $W$ helicity has been measured and the obtained left-handed, longitudinal and right-handed polarisations have been found to be consistent with the Standard Model predictions\cite{Abazov:2010jn,ATLAS-CONF-2011-122}.


The charge asymmetry in top quark pair production is key topic of this thesis and will be discussed in detail in the following section.

\section{Charge Asymmetry in Top Quark Pair Production}
As discussed in \mbox{Section \ref{sec:topprod}}, top quark pairs at hadron colliders are produced via gluon-gluon fusion or quark-antiquark annihilation in Born approximation:
\begin{eqnarray}
  q + \bar{q} & \rightarrow & t + \bar{t}\text{,} \nonumber \\
  g + g       & \rightarrow & t + \bar{t}\text{.} \nonumber
\end{eqnarray}
These leading order processes obviously do not discriminate between the final state top and antitop, as can be seen from the respective Born partonic differential cross-sections for the two production mechanisms,
\begin{eqnarray}
  \frac{d \sigma_{q\bar{q} \rightarrow t\bar{t}}}{d \cos{\hat{\theta}}} & = & \alpha_s^2 \frac{\pi \beta}{3 \hat{s} N_C} \left( 1 + c^2 + 4m_t^2 \right), \\
  \frac{d \sigma_{gg \rightarrow t\bar{t}}}{d \cos{\hat{\theta}}} & = & \alpha_s^2 \frac{\pi \beta}{2 \hat{s}} \left( \frac{1}{N_C (1-c^2)} - \frac{3}{16} \right) \times \left(1 + c^2 + 8m_t^2 - \frac{32m_t^4}{1-c^2} \right),
\end{eqnarray}
where $\hat{\theta}$ denotes the polar angle of the top quark with respect to the incoming parton in the centre-of-mass rest frame, $N_C = 3$, $\beta = \sqrt{1-4m_t^2}$ and $c = \beta \cos{\hat{\theta}}$. Consequently, the same holds for the full $pp \rightarrow t \bar{t}$ differential cross-sections.

If, however, the processes are considered at higher order in the Standard Model, where radiative corrections from real or virtual gluon emission are introduced, a significant asymmetry can be generated in the differential \ttbar~cross-section, leading to a charge asymmetry in the production of top quark pairs\cite{Kuhn:1998kw}. This effect is caused by the interference of amplitudes which are odd under the exchange of the final state top quark and antitop quark. 

\begin{figure}[h!tb]
  \begin{centering}
    \mbox{
      \subfigure{
        \scalebox{1.0}{\includegraphics[height=2cm]{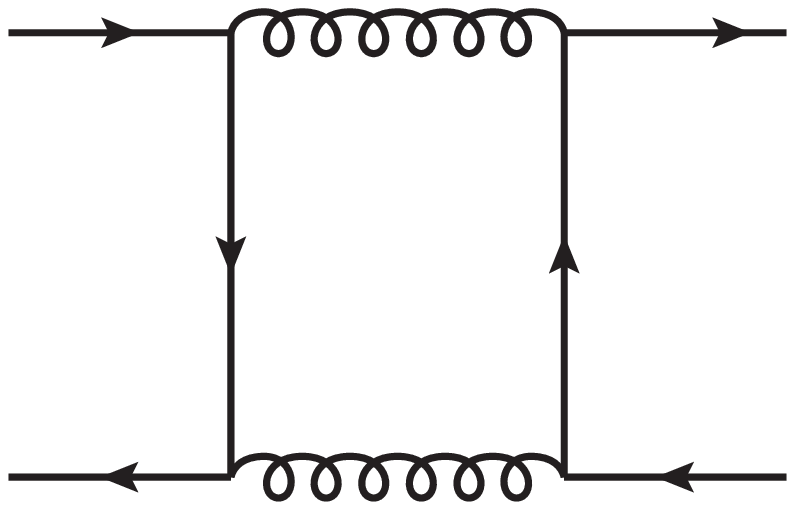}}
      }
      \quad
      \subfigure{
        \scalebox{1.0}{\includegraphics[height=2cm]{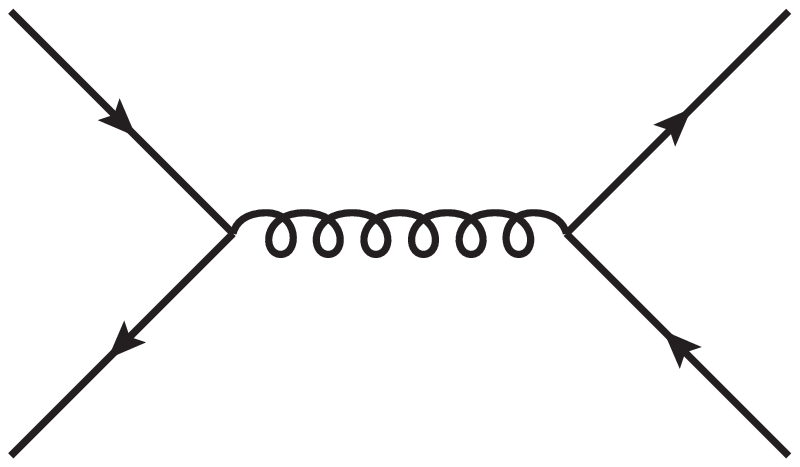}}
      }
    }
    \vspace{-0.2 cm}
    \caption[\quad Box and Born diagrams contributing to the production of top quark pairs through quark-antiquark annihilation]{Box (left) and Born (right) diagrams contributing to the production of top quark pairs through quark-antiquark annihilation.}
    \label{fig:boxborn}
  \end{centering}
\end{figure}

The dominant contribution to an overall charge asymmetry stems from interference between the leading order amplitude for quark-antiquark annihilation and the corresponding one-loop corrections (box diagrams), creating a positive contribution to the total charge asymmetry. The two diagrams are shown in \mbox{Figure \ref{fig:boxborn}}.

\begin{figure}[h!tb]
  \begin{centering}
    \mbox{
      \subfigure{
        \scalebox{1.0}{\includegraphics[height=2cm]{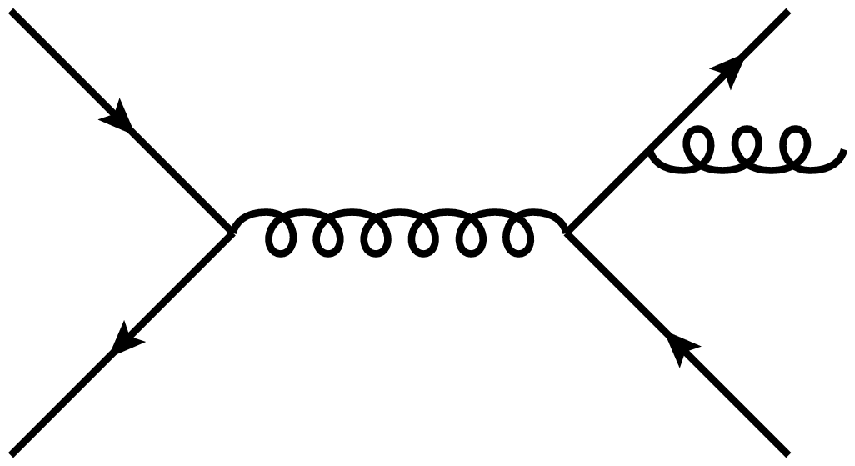}}
      }
      \quad
      \subfigure{
        \scalebox{1.0}{\includegraphics[height=2cm]{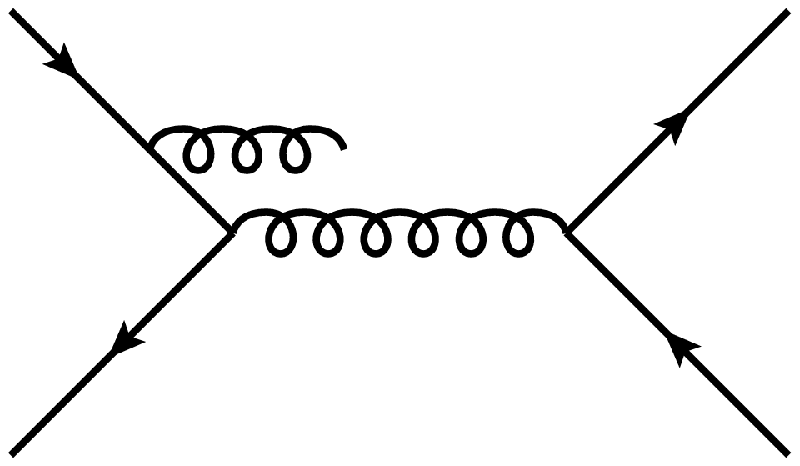}}
      }
    }
    \vspace{-0.2 cm}
    \caption[\quad Final state and initial state bremsstrahlung diagrams contributing to the production of top quark pairs through quark-antiquark annihilation]{Final state (left) and initial state (right) bremsstrahlung diagrams contributing to the production of top quark pairs through quark-antiquark annihilation.}
    \label{fig:bremsstrahlung}
  \end{centering}
\end{figure}

In addition, interferences between initial state and final state gluon bremsstrahlung have to be taken into account, contributing negatively but typically of lower magnitude than the box-Born interference to the overall asymmetry. The respective diagrams are shown in \mbox{Figure \ref{fig:bremsstrahlung}}.

\begin{figure}[h!tb]
  \begin{centering}
    \mbox{
      \subfigure[]{
        \scalebox{1.0}{\includegraphics[height=2cm]{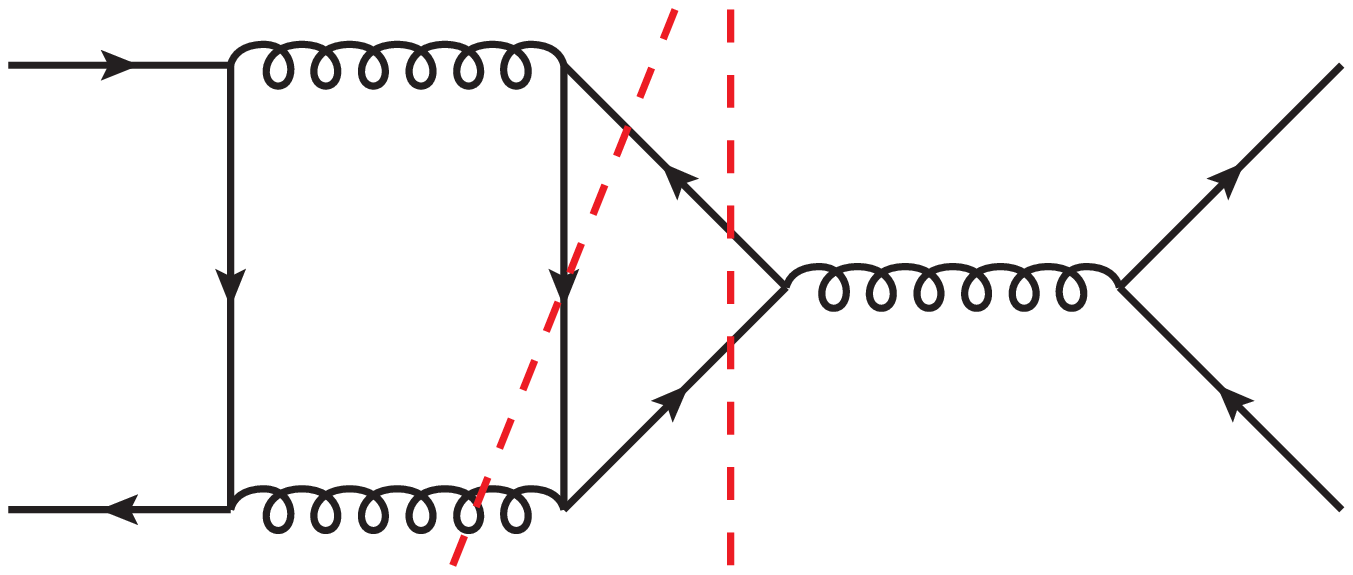}}
      }
      \quad
      \subfigure[]{
        \scalebox{1.0}{\includegraphics[height=2cm]{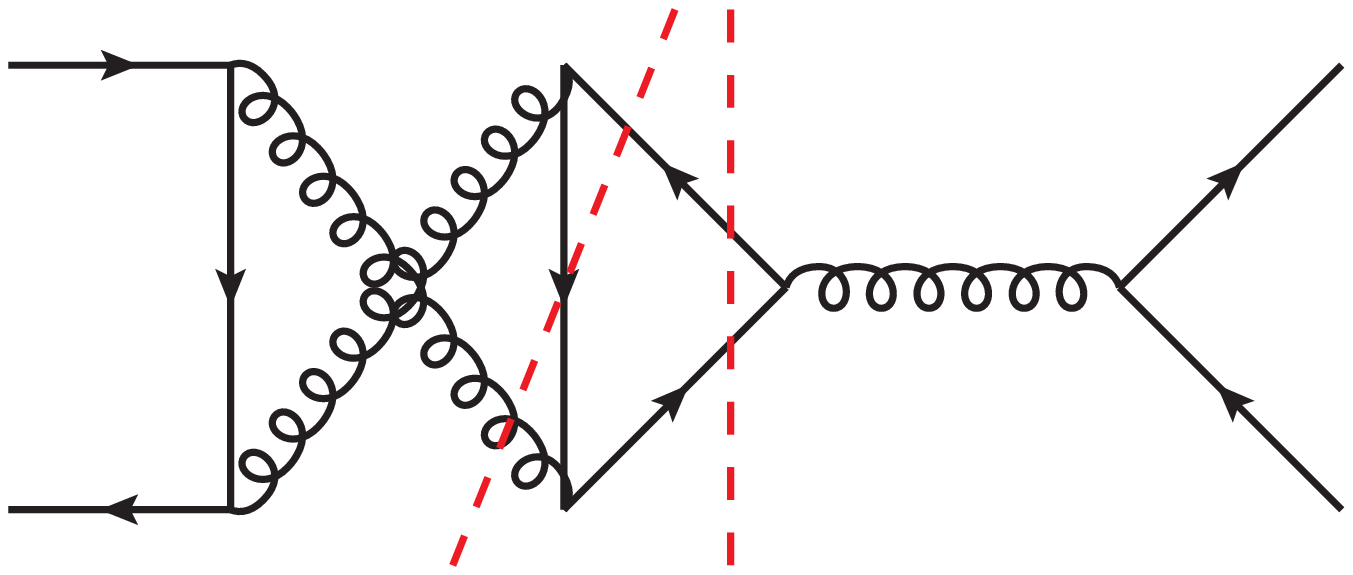}}
      }
    }
    \vspace{-0.2 cm}
    \caption[\quad Cut diagrams contributing to the asymmetry in the production of top quark pairs through quark-antiquark annihilation]{Cut diagrams contributing to the asymmetry in the production of top quark pairs through quark-antiquark annihilation, arising from virtual and real gluon emission. The respective contributions to the cross-section are odd under the exchange of the final state top and antitop quarks.}
    \label{fig:asymdiagrams}
  \end{centering}
\end{figure}

The asymmetry arising from the combination of these contributions can be described by comparing the colour factor terms arising in the differential cross-sections from the two cut diagrams\cite{Cutkosky} shown in \mbox{Figure \ref{fig:asymdiagrams}} after averaging over initial states and summing over final states.
The respective colour factors $C_A$ and $C_B$ for the two diagrams (a) and (b), respectively, can be expressed as\cite{Kuhn:1998kw}
\begin{equation}
  C_A = \frac{1}{N_C^2} \tr{\frac{\lambda^a}{2} \frac{\lambda^b}{2} \frac{\lambda^c}{2}} \tr{\frac{\lambda^a}{2} \frac{\lambda^c}{2} \frac{\lambda^b}{2}} = \frac{1}{16 N_C^2} \left( f_{abc}^2 + d_{abc}^2 \right)\phm\phantom{,}
\end{equation}
and
\begin{equation}
  C_B = \frac{1}{N_C^2} \tr{\frac{\lambda^a}{2} \frac{\lambda^b}{2} \frac{\lambda^c}{2}} \tr{\frac{\lambda^b}{2} \frac{\lambda^c}{2} \frac{\lambda^a}{2}} = \frac{1}{16 N_C^2} \left( -f_{abc}^2 + d_{abc}^2 \right),
\end{equation}
where $f_{abc}^2 = 24$ and $d_{abc}^2 = 40/3$. Consequently, the respective contributions $d\sigma_A$ and $d\sigma_B$ to the cross-section are odd under the exchange of the final state top and antitop quarks:
\begin{equation}
  d\sigma_A \left( t, \bar{t} \right) = - d\sigma_B \left( \bar{t}, t \right).
\end{equation}

\begin{figure}[h!tb]
  \begin{centering}
    \includegraphics[height=5cm]{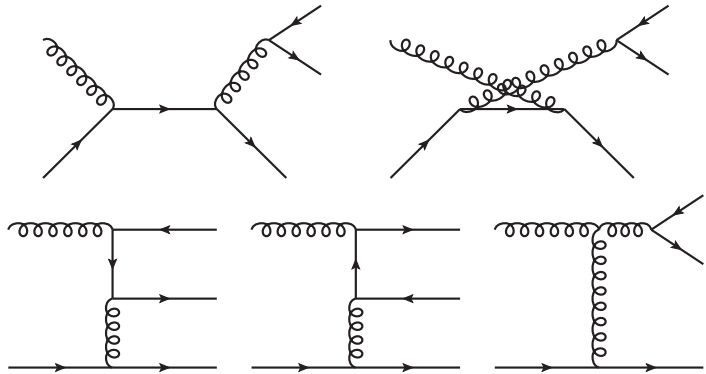}
    \vspace{-0.2 cm}
    \caption[\quad Quark-gluon scattering diagrams contributing to the hadronic production of top quark pairs]{Quark-gluon scattering diagrams contributing to the hadronic production of top quark pairs introduced at order of $\alpha_s^3$.}
    \label{fig:qgscat}
  \end{centering}
\end{figure}

A small contribution is introduced at the order of $\alpha_s^3$ through interferences of different terms in quark-gluon scattering,
\begin{eqnarray}
  g + q & \rightarrow & t + \bar{t} + q'\text{,} \nonumber
\end{eqnarray}
which is shown in \mbox{Figure \ref{fig:qgscat}}.

\begin{figure}[h!tb]
  \begin{centering}
    \includegraphics[width = 6 cm]{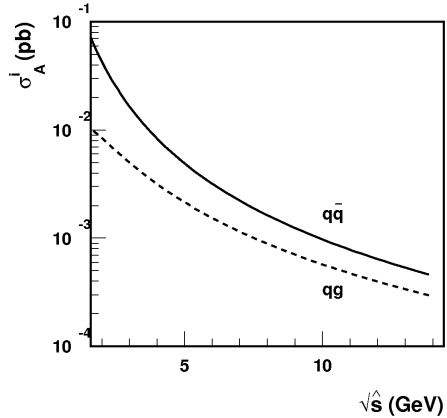}
    \vspace{-0.2 cm}
    \caption[\quad Integrated charge asymmetric contributions in the SM.]{Integrated charge asymmetric parts of the top quark pair production cross-section from quark-antiquark annihilation ($q \bar{q}$) and quark-gluon scattering ($qg$) initiated processes as a function of the partonic centre-of-mass energy\cite{Kuhn:1998kw}.}
    \label{fig:asymvsCM}
  \end{centering}
\end{figure}

The individual charge asymmetric contributions from quark-antiquark annihilation and quark-gluon scattering can be found in \mbox{Figure \ref{fig:asymvsCM}}\cite{Kuhn:1998kw} as a function of the partonic centre-of-mass energy, quantified by the integrated forward-backward contributions
\begin{equation}
\sigma_{A}^i = \int_0^1 \frac{d \sigma_{A}^i}{d \cos{\hat{\theta}}} d \cos{\hat{\theta}} - \int_{-1}^0 \frac{d \sigma_{A}^i}{d \cos{\hat{\theta}}} d \cos{\hat{\theta}},
\end{equation}
where $i$ denotes the two contributions from quark-antiquark annihilation ($q \bar{q}$) and quark-gluon scattering ($qg$).

The resulting overall QCD charge asymmetry $A_{t\bar{t}}$ can be expressed in terms of the asymmetric contributions $\sigma_{\text{A}}$ and symmetric contributions $\sigma_{\text{S}}$ to the total production cross-section as ratio
\begin{equation}
A_{t\bar{t}}^{\text{QCD}} = \frac{\sigma_{\text{A}}}{\sigma_{\text{S}}} = \frac{ \alpha_s^3 \sigma_{\text{A}}^{\text{(1)}} + \alpha_s^4 \sigma_{\text{A}}^{\text{(2)}} + ... } { \alpha_s^2 \sigma_{\text{S}}^{\text{(0)}} + \alpha_s^3 \sigma_{\text{S}}^{\text{(1)}} + ... } = \frac{ \alpha_s^3 \sigma_{\text{A,} q\bar{q}}^{\text{(1)}} + \alpha_s^3 \sigma_{\text{A,} qg}^{\text{(1)}} + \alpha_s^4 \sigma_{\text{A,} q\bar{q}}^{\text{(2)}} + ... } { \alpha_s^2 \sigma_{\text{S}}^{\text{(0)}} + \alpha_s^3 \sigma_{\text{S}}^{\text{(1)}} + ... },
\label{eqn:QCDasym}
\end{equation}
where
\begin{equation}
\sigma_{\text{S}}^{\text{($i$)}} = \sigma_{\text{S,} gg}^{\text{($i$)}} + \sigma_{\text{S,} q\bar{q}}^{\text{($i$)}} + \sigma_{\text{S,} qg}^{\text{($i$)}}
\end{equation}
contains the respective symmetric contributions at a given order $i$. This asymmetry can also be parametrised as
\begin{equation}
A_{t\bar{t}}^{\text{QCD}} = \alpha_s A_{t\bar{t}}^{(0)} + \alpha_s^2 A_{t\bar{t}}^{(1)} + ...
\end{equation}
Higher order corrections arising from QCD which are not taken into account in \mbox{Equation \ref{eqn:QCDasym}}, such as $A_{t\bar{t}}^{(1)}$, can be evaluated in next-to-next-to-leading log approximation using soft gluon resummation techniques\cite{Ahrens:2010zv,Ahrens:2011uf,Kidonakis:2011zn} in order to improve the theoretical predictions on the respective differential cross-section contributions and the associated uncertainties.

\begin{figure}[h!tb]
  \begin{centering}
    \mbox{
      \subfigure[]{
        \scalebox{1.0}{\includegraphics[height=2cm]{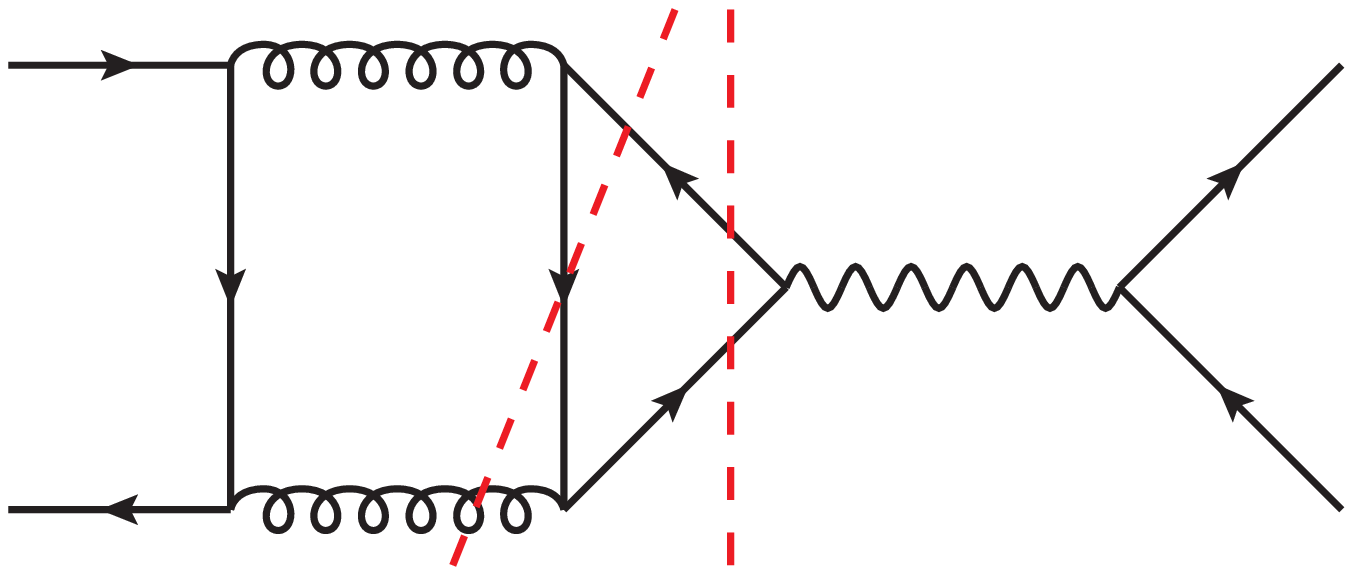}}
        \label{fig:asymdiagrams_weak_a}
      }
      \quad
      \subfigure[]{
        \scalebox{1.0}{\includegraphics[height=2cm]{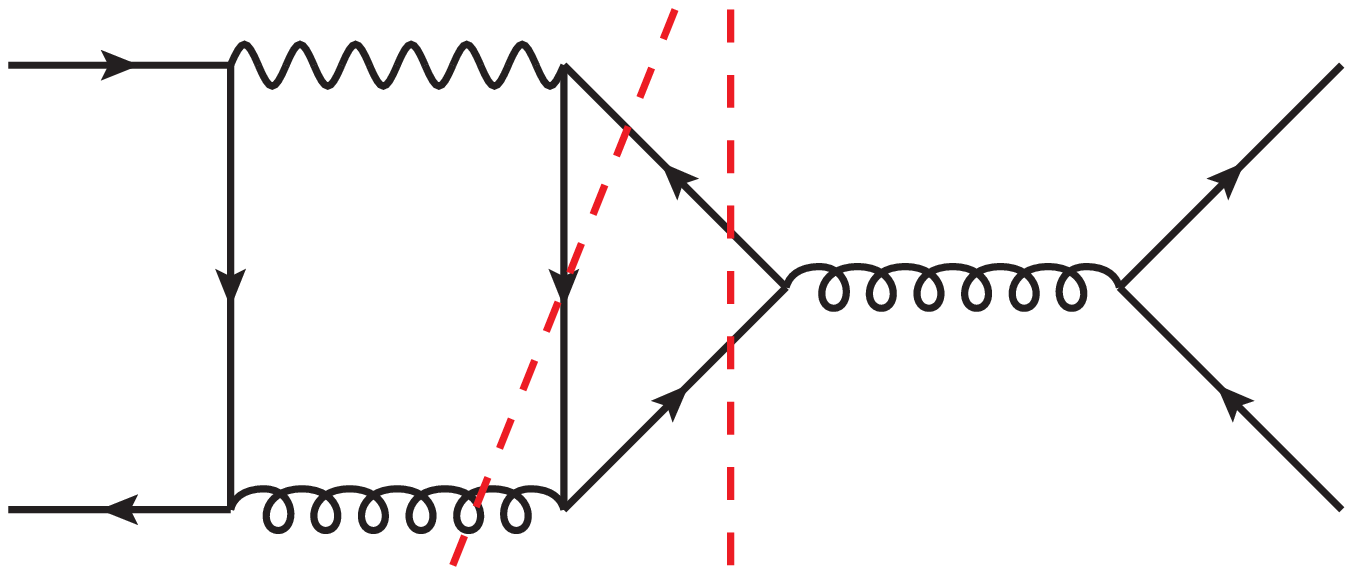}}
        \label{fig:asymdiagrams_weak_b}
      }
    }
    \vspace{-0.2 cm}
    \caption[\quad Mixed QCD-electroweak cut diagrams contributing to the asymmetry in the production of top quark pairs through quark-antiquark annihilation]{Mixed QCD-electroweak cut diagrams contributing to the asymmetry in the production of top quark pairs through quark-antiquark annihilation, arising from interferences between the singlet-state box diagram and initial and final state radiation diagrams with electroweak \ttbar~production (left) and from the interference of the gluon-$\gamma$ and gluon-$Z$ box diagrams with the Born diagram (right).}
  \end{centering}
\end{figure}

In addition to the asymmetry arising from the Standard Model QCD terms, further contributions originate from mixed QCD-electroweak interference terms to the quark-antiquark annihilation process\cite{Kuhn:1998kw,PhysRevD.84.093003}. The \ttbar~colour-singlet configuration of the box diagram in \mbox{Figure \ref{fig:boxborn}} can interfere with the production of top quark pairs through a photon or a $Z$ boson (and similarly for the interference between initial state and final state radiation), as shown in \mbox{Figure \ref{fig:asymdiagrams_weak_a}}.
Furthermore, interferences of the gluon-$\gamma$ and gluon-$Z$ box diagrams with the leading order QCD amplitude as indicated in \mbox{Figure \ref{fig:asymdiagrams_weak_b}} contribute to an additional asymmetry, which together with \mbox{Figure \ref{fig:asymdiagrams_weak_a}} leads to a total increase of asymmetry by a factor of about 1.09\cite{Kuhn:1998kw} with respect to the QCD contributions.

The overall charge asymmetry, including both the QCD and electroweak contributions, can be expressed as
\begin{equation}
A_{t\bar{t}} = \frac{ \alpha^2 \tilde{\sigma}_{\text{A}}^{\text{(0)}} + \alpha_s^3 \sigma_{\text{A}}^{\text{(1)}} + \alpha_s^2 \alpha \tilde{\sigma}_{\text{A}}^{\text{(1)}} + \alpha_s^4 \sigma_{\text{A}}^{\text{(2)}} + ... } { \alpha^2 \tilde{\sigma}_{\text{S}}^{\text{(0)}} + \alpha_s^2 \sigma_{\text{S}}^{\text{(0)}} + \alpha_s^3 \sigma_{\text{S}}^{\text{(1)}} + \alpha_s^2 \alpha \tilde{\sigma}_{\text{S}}^{\text{(1)}} + ... },
\end{equation}
where $\tilde{\sigma}_{\text{A}}^{\text{($i$)}}$ and $\tilde{\sigma}_{\text{S}}^{\text{($i$)}}$ denote the asymmetric and symmetric contributions from QCD-electroweak mixing, respectively.

Note that a similar effect emerges in Quantum Electrodynamics, where interferences at the order of $\alpha^3$ create an asymmetry in the angular distribution of final state particles produced in $e^{+}e^{-}$ collisions due to virtual radiative corrections and soft and hard photon emission\cite{Berends1981237,Berends:1982dy,Berends1973381}.

\subsection{Charge Asymmetry Beyond the Standard Model}
Numerous theoretical models predicting the manifestation of physics beyond the Standard Model (BSM) exist, several of which can have implications on the charge asymmetry observed in the production of top quark pairs at hadron colliders. The most popular models will be explained in the following.

\subsubsection{Colour-octet Gauge Bosons}
In chiral colour models\cite{Pati:1975ze,Hall:1985wz,Frampton:1987dn,Frampton:1987ut,Bagger:1987fz,Frampton:2009rk,Zerwekh:2011wf,wang:2011taa,Haisch:2011up,Tavares:2011zg,Wang:2011hc}, SM colour charge is extended to contain a right-handed and left-handed contribution $\text{SU(3)}_R \times \text{SU(3)}_L$ to reflect the chirality. This extension implies a breaking of the symmetry of the diagonal $\text{SU(3)}_C$ group and the generation of a heavy colour-octet gauge boson, the axigluon. The associated coupling to quarks has a pure axial-vector structure and is of the same magnitude as the QCD coupling.

Alternative models with non-chiral structure imply the existence of massive gauge bosons with pure vector-like couplings to quarks (colorons)\cite{Hill:1991at,Hill:1993hs,Chivukula:1996yr,Choudhury:2007ux} or Kaluza-Klein\cite{Kaluza:1921tu,Klein:1926tv} excited states arising from models including extra dimensions\cite{Randall:1999ee,Dicus:2000hm,Agashe:2006hk,Agashe:2007jb,Lillie:2007yh,Lillie:2007ve,Delaunay:2010dw,Delaunay:2011vv,Djouadi:2011aj}.

Further generalisations include the assumption of different coupling strengths for the different $\text{SU(3)}$ contributions\cite{Cuypers:1990hb,Carone:2008rx,Martynov:2009en,Zerwekh:2009vi,Burdman:2010gr,Bai:2011ed,Barcelo:2011fw,Alvarez:2011hi,AguilarSaavedra:2011ci}, leading to both vector and axial-vector couplings in the interactions of the respective colour-octet resonance $G_{\mu}^{a}$ and quarks. The vector and axial-vector coupling strengths, $g_{V}^{q_i}$ and $g_{A}^{q_i}$, respectively, lead to the following generalised term $\mathcal{L}_{G'}$\cite{Rodrigo:2010gm} in the modified SM Lagrangian:
\begin{equation}
\mathcal{L}_{G'} = g_s t^a \bar{q}_i \left( g_{V}^{q_i} + g_{A}^{q_i} \gamma_{5} \right) \gamma^{\mu} G_{\mu}^{a} q_i.
\end{equation}

The corresponding leading-order cross-section for top quark pair production in the annihilation of quark-antiquark pairs is given by\cite{Ferrario:2008wm,Ferrario:2009ns}
\begin{eqnarray}
\frac{d \sigma_{q\bar{q} \rightarrow t\bar{t}}}{d \cos{\hat{\theta}}} & = & \alpha_s^2 \frac{\pi \beta}{3 \hat{s} N_C} \Biggl( C_{+} + \frac{2 \hat{s} \left( \hat{s} - m_G^2 \right)}{\left( \hat{s} - m_G^2 \right)^2 + m_G^2 \Gamma_G^2} \left[ g_V^q g_V^t \left( C_{+} \right) + 2 g_A^q g_A^t c \right] \\
 & + & \frac{\hat{s}^2}{\left( \hat{s} - m_G^2 \right)^2 + m_G^2 \Gamma_G^2} \left[ \left( (g_V^q)^2 + (g_A^q)^2\right) \times \left( (g_V^t)^2 C_{+} + (g_A^t)^2 C_{-} \right) + 8 g_V^q g_A^q g_V^t g_A^t c\right] \Biggr), \notag
\end{eqnarray}
where $C_{\pm} = 1 + c^2 \pm 4m_t^2$. The vector and axial-vector couplings of the resonances to the light quarks and top quarks are given by the constants $g_V^q$ and $g_A^q$, and by $g_V^t$ and $g_A^t$, respectively.

Compared to the SM cross-section, an additional asymmetric contribution is introduced by terms which are odd in c. Hence, a large positive asymmetry can be generated in models where $g_A^q g_A^t < 0$ or where the term $8 g_V^q g_A^q g_V^t g_A^t c$ is dominant. A negative asymmetry on the other hand can be created in flavour universal models where $g_A^q = g_A^t$.

Models involving a colour-octet resonance typically require the resonance to be off-shell, either heavy\cite{Barcelo:2011fw}, or below the \ttbar~production threshold\cite{AguilarSaavedra:2011ci}, since in the intermediate mass range a distinct excess would emerge in the \ttbar~mass spectrum, which is not observed. Alternatively, a very broad resonance\cite{AguilarSaavedra:2011ci,Barcelo:2011vk} would be concealed due to limited statistics in the tail of the $M_{t \bar{t}}$ distribution.

\subsubsection{Extra Weak Gauge Bosons}
Different theoretical models, such as some Grand Unified Theories (GUTs), topcolor or left-right\cite{Barger:2010mw} models predict the existence of extra weak gauge bosons, such as the $W'$ or $Z'$. Furthermore, these states can appear as Kaluza-Klein excitations in extra dimensional models\cite{Djouadi:2007eg,Djouadi:2009nb}. A sizable contribution to the charge asymmetry can only be introduced in $W'$ or $Z'$ $t$-channel interactions\cite{Cao:2009uz,Jung:2009jz,Cheung:2009ch,Barger:2010mw,Cao:2010zb,Cheung:2011qa,Shelton:2011hq,Barger:2011ih,Shu:2011au,AguilarSaavedra:2011hz,Cao:2011ew,Berger:2011ua,Bhattacherjee:2011nr,Barreto:2011au,Jung:2011zv,AguilarSaavedra:2011zy,Jung:2011ua,Duraisamy:2011pt,Ko:2011vd,Jung:2011id}.  The $s$-channel \ttbar~production through a $Z'$ is suppressed due to the fact that the corresponding amplitudes do not create interferences with the SM amplitude\cite{Rodrigo:2010gm}.

A potentially significant asymmetry can be created by the introduction of flavour violating couplings into the SM Lagrangian by a term $\mathcal{L}_{W'/Z'}$\cite{Rodrigo:2010gm}, such as
\begin{equation}
\mathcal{L}_{W'/Z'} = \bar{t} \left( g_V^{Z'} + g_A^{Z'} \gamma_5 \right) \gamma^{\mu} Z'_{\mu} u + \bar{t} \left( g_V^{W'} + g_A^{W'} \gamma_5 \right) \gamma^{\mu} W'_{\mu} d.
\end{equation}

A real $Z'$ contribution is constrained by the absence of like-sign top quark pair production\cite{AguilarSaavedra:2011zy}, except for very light $Z'$ resonances. Furthermore, for a $Z'$ leading to a sizable positive charge asymmetry, a large corresponding $Z'$ coupling has to be assumed and a corresponding excess in the tail of the \ttbar~mass distribution would be generated. Such excess is not observed, however. Furthermore, $W'$ and $Z'$ left-handed couplings are disfavoured by precision measurements in $B$ hadron systems\cite{AguilarSaavedra:2011ug}.

\subsubsection{Coloured Scalars}
In addition to gauge bosons, neutral or charged coloured scalars\cite{Jung:2009pi} can be introduced in particular in the presence of larger gauge groups such as SU(5) or SO(10)\cite{Patel:2011eh}. These occur primarily in GUT models close to the unification scale. However, some of the coloured scalar states can manifest at lower energy scales\cite{Perez:2008ry,Dorsner:2009mq}, for example due to gauge coupling unification.

Similar to extra weak gauge bosons, \ttbar~production via coloured scalars in the $s$-channel is not affected by any charge asymmetric contributions due to the absence of interferences with the SM amplitudes\cite{Rodrigo:2010gm}. Consequently, only $t$-channel flavour-violating couplings can introduce a significant asymmetry in the production cross-section, such as the exchange of scalar colour singlets\cite{Shu:2009xf,Cao:2009uz,Blum:2011fa}, triplets\cite{Dorsner:2009mq,Shu:2009xf,Arhrib:2009hu,Cao:2009uz,Dorsner:2010cu,Babu:2011yw,AguilarSaavedra:2011ug,Ligeti:2011vt,Grinstein:2011dz}, sextets\cite{Shu:2009xf,AguilarSaavedra:2011ug,Grinstein:2011yv,Grinstein:2011dz} and octets\cite{Dorsner:2009mq,Shu:2009xf}. Furthermore, a light scalar contribution is disfavoured due to the implied existence of a highly constrained new top quark decay channel ($t \rightarrow S' u$)\cite{ReviewParticlePhysics,Amsler:2008zzb}.

A generalised contribution from a coloured scalar SU(2) doublet $S'$ to the SM Lagrangian can be described by an additional term $\mathcal{L}_{S'}$\cite{Rodrigo:2010gm,Shu:2009xf}, given by
\begin{equation}
\mathcal{L}_{S'} = t^a \bar{t} \left( g_S + g_P \gamma_5 \right) \phi^a u,
\end{equation}
where $g_S$ and $g_P$ denote the scalar and pseudoscalar coupling constants, respectively. The resulting asymmetry $y_{S'}$ created in the exchange is hence given by
\begin{equation}
y_{S'} = \sqrt{g_S^2 + g_P^2}.
\end{equation}
This contribution is typically negative for a heavy coloured scalar. However, due to the destructive interference of the scalar contribution with the SM, a positive overall asymmetry can be generated.

\subsection{Top Quark Charge asymmetry at Hadron Colliders}
\label{chap:topasymHadColl}
\begin{figure}[h!tb]
  \begin{centering}
    \includegraphics[width = 8 cm]{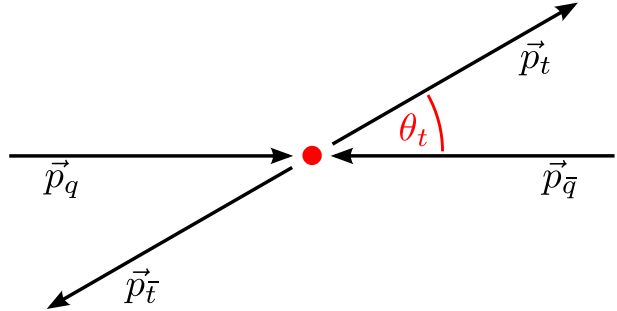}
    \vspace{-0.2 cm}
    \caption[\quad Top quark pair production kinematics]{Top quark pair production kinematics in quark-antiquark annihilation. The initial state partons and their momenta $\vec{p}_q$ and $\vec{p}_{\bar{q}}$ and the produced top quarks and their momenta $\vec{p}_t$ and $\vec{p}_{\bar{t}}$, respectively, are shown. In addition, the production angle of the top quark $\theta_t$ is shown.}
    \label{fig:theta_t}
  \end{centering}
\end{figure}

In order to quantify a potential charge asymmetry created in the Standard Model or BSM models, the natural choice of observable would be the production angle $\theta_t$ of the top/antitop quarks with respect to the incoming partons from the hard scattering process, as depicted in \mbox{Figure \ref{fig:theta_t}}.
The corresponding differential charge asymmetry $A(\cos{\theta_t})$ at the partonic level is given by
\begin{equation}
A(\cos{\theta_t}) = \frac{N_t(\cos{\theta_t}) - N_{\bar{t}}(\cos{\theta_t})}{N_t(\cos{\theta_t}) + N_{\bar{t}}(\cos{\theta_t})},
\label{eqn:A_diff}
\end{equation}
in the $q\bar{q}$ rest frame, where 
\begin{equation}
N_t(\cos{\theta_t}) = \frac{d \sigma}{d \Omega}(\cos{\theta_t})
\end{equation}
and $N_{\bar{t}}(\cos{\theta_t}) = N_t(- \cos{\theta_t})$ due to the symmetry of charge conjugation. Consequently, the integrated charge asymmetry $A$ can be quantified such that
\begin{equation}
A = \frac{N_t(\cos{\theta_t} \geq 0) - N_{\bar{t}}(\cos{\theta_t} \geq 0)}{N_t(\cos{\theta_t} \geq 0) + N_{\bar{t}}(\cos{\theta_t} \geq 0)}.
\label{eqn:A_incl}
\end{equation}

However, in the strong production of top quark pairs at hadron colliders, the production angle as such is not accessible experimentally due to the fact that the initial state of the partonic reaction is of probabilistic nature. Since the available information is limited to the hadronic initial state ($pp$ or $p\bar{p}$) and the PDFs of the protons and/or antiprotons, respectively, different methods to measure the charge asymmetry, making use only of the final state information of the hadronic collision, must be employed.

At non-symmetric hadron colliders such as the Tevatron, where protons are brought to collision with antiprotons, a charge asymmetry in the production of top quark pairs as introduced in \mbox{Equation \ref{eqn:A_diff}}, observed in the \ttbar~rest frame, corresponds directly to an equal-sized geometric forward-backward asymmetry, $A_{\text{FB}}$, since $N_t(y) = N_{\bar{t}}(- y)$. Since this quantity is experimentally accessible in a direct way due to the fact that the initial directions of the proton and antiproton are known, the measurement of the underlying charge asymmetry in the laboratory frame is possible. 

At $pp$ colliders such as the LHC, no forward-backward asymmetry is visible in the laboratory frame due to the intrinsic charge conjugation symmetry of the initial state collisions. However, since top quarks are preferentially emitted in the direction of the incoming parton, and quarks in the proton on average carry a larger momentum fraction than antiquarks, an excess of top quarks in the forward and backward regions is expected in the laboratory frame. Consequently, different widths of the corresponding rapidity distributions of top quarks and antitop quarks, and hence, the respective decay products, are predicted. The underlying charge asymmetry in the \ttbar~rest frame can be extracted either from the final state particles directly or by performing a kinematic reconstruction of the \ttbar~decay signature.

Since an asymmetry can solely be created from the quark-antiquark annihilation contribution to the top quark pair production cross-section, the total charge asymmetry in both $pp$ and $p\bar{p}$ collisions can be significantly diluted due to the (symmetric) gluon-gluon fusion contribution. The magnitude of the overall asymmetry depends strongly on the centre-of-mass energy since the fraction of soft gluons in the proton/antiproton PDFs and hence the probability of gluon interactions in the partonic reaction increases with rising hadron momentum. In addition, the top quark pair production cross-section in $pp$ collisions shows a higher contribution from gluon-gluon fusion since interactions of sea quarks are dominant. This differs from $p\bar{p}$ collisions, where potential interactions of valence (anti)quarks from the colliding (anti)protons lead to an increased contribution from quark-antiquark annihilation in the overall cross-section.

In order to quantify the charge asymmetry at the LHC, a proper observable, taking into account the potential differences in the rapidity distributions of the top and antitop quark in the laboratory frame has to be chosen. Different frame-invariant variables based on rapidity or pseudorapidity differences of the final state top and antitop quarks are typically used to measure the asymmetry. 

Potential observables to quantify the inclusive charge asymmetry in the \ttbar~rest frame
\begin{equation}
A_{t\bar{t}} = \frac{N^{+} - N^{-}}{N^{+} + N^{-}}
\end{equation}
based on rapidities and pseudorapidities of the final state top and antitop quarks include parametrisations such as
\begin{equation}
\begin{array}{lcl}
N^{+} = N(\eta_t - \eta_{\bar{t}} \geq 0)      & , & N^{-} = N(\eta_t - \eta_{\bar{t}} \leq 0); \\
N^{+} = N(y_t - y_{\bar{t}} \geq 0)      & , & N^{-} = N(y_t - y_{\bar{t}} \leq 0)\text{\cite{PhysRevLett.101.202001,Aaltonen:2011kc,Abazov:2011rq}}; \\
N^{+} = N(|\eta_t| - |\eta_{\bar{t}}| \geq 0)  & , & N^{-} = N(|\eta_t| - |\eta_{\bar{t}}| \leq 0)\text{\cite{CMS-PAS-TOP-11-014}}; \\
N^{+} = N(|y_t| - |y_{\bar{t}}| \geq 0)  & , & N^{-} = N(|y_t| - |y_{\bar{t}}| \leq 0)\text{\cite{CONFNote}}.
\end{array}
\end{equation}
Since 
\begin{equation}
\Delta y = y_t - y_{\bar{t}} = 2 y_t^{t\bar{t}}
\end{equation}
and correspondingly for the pseudorapidities of the final state top and antitop quarks, it follows that
\begin{equation}
A_{t\bar{t}} = \frac{N(\Delta y \geq 0) - N(\Delta y \leq 0)}{N(\Delta y \geq 0) + N(\Delta y \leq 0)}.
\end{equation}
At the Tevatron, the predicted charge asymmetry within the Standard Model for the given observable $A_{t\bar{t}}$, evaluated for a centre-of-mass energy of $1.96\,\text{TeV}$, is
\begin{equation}
A_{t\bar{t}} = 0.087 \pm 0.010 \text{\cite{Kuhn:2011ri}}.
\end{equation}

In the following analysis, the charge asymmetry will be quantified using a parametrisation based on the difference of absolute rapidities of the top and antitop quarks, defined as
\begin{equation}
A_C = \frac{N(|y_t| - |y_{\bar{t}}| \geq 0) - N(|y_t| - |y_{\bar{t}}| \leq 0)}{N(|y_t| - |y_{\bar{t}}| \geq 0) + N(|y_t| - |y_{\bar{t}}| \leq 0)}.
\label{eqn:myobservable}
\end{equation}

At the LHC, the predicted charge asymmetry within the Standard Model for the given observable $A_C$, evaluated for a centre-of-mass energy of $7\,\text{TeV}$, is
\begin{equation}
A_C = 0.0115 \pm 0.0006 \text{\cite{Kuhn:2011ri}}.
\end{equation}

\begin{figure}[h!tb]
  \begin{centering}
    \mbox{
      \subfigure{
        \scalebox{0.4}{\includegraphics[width=\textwidth]{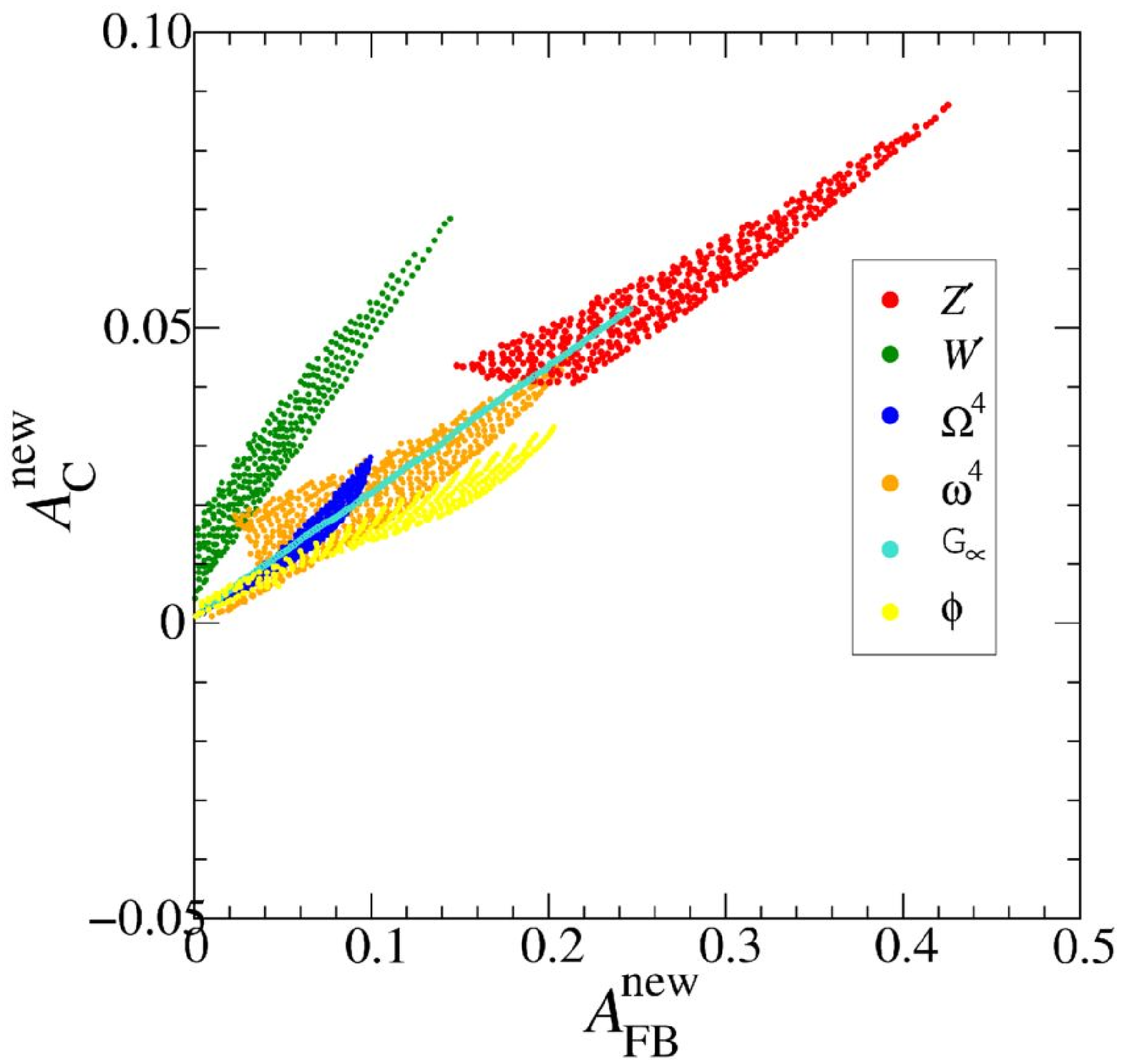}}
      }
      \quad
      \subfigure{
        \scalebox{0.4}{\includegraphics[width=\textwidth]{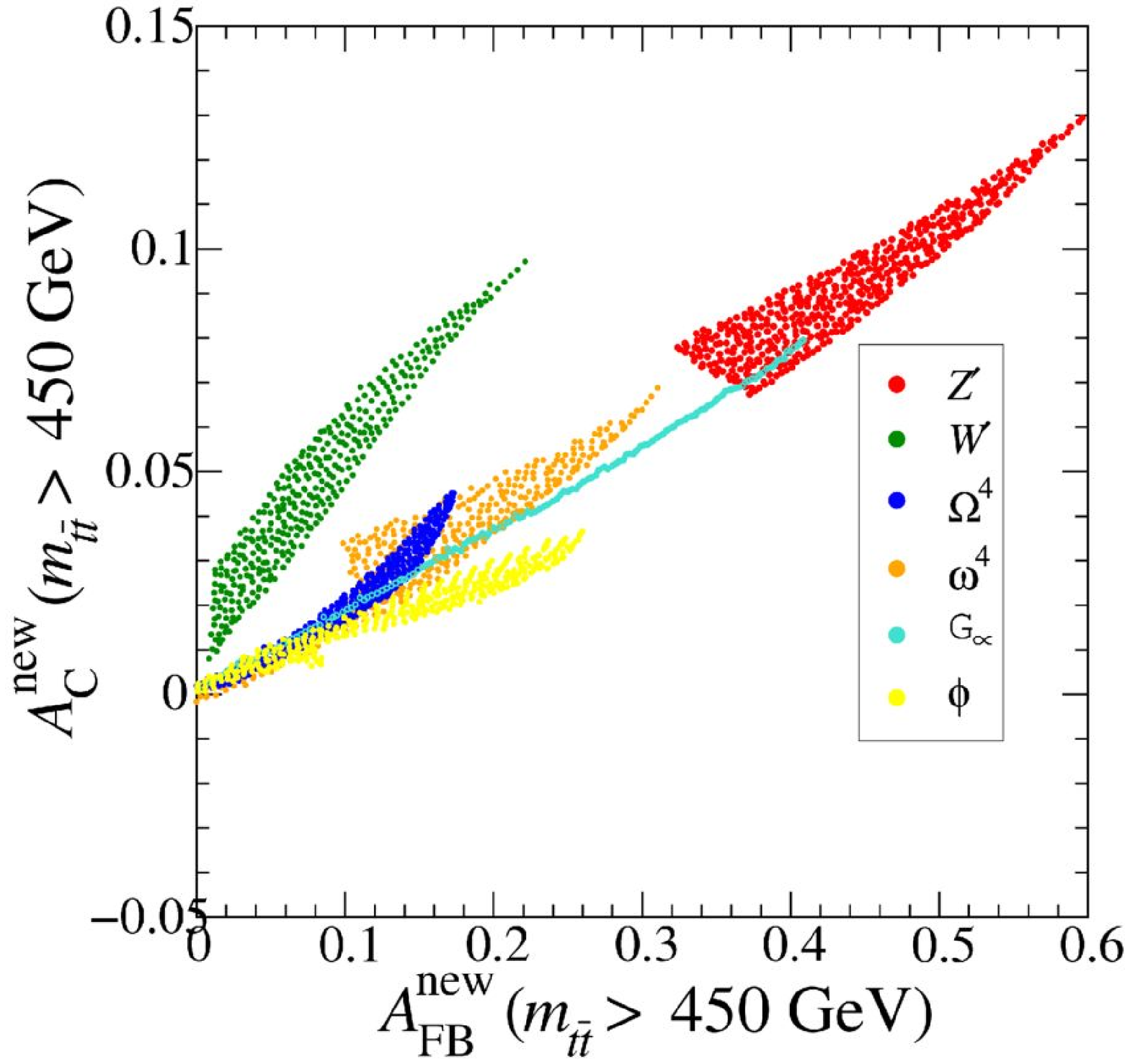}}
      }
    }
    \vspace{-0.2 cm}
    \caption[\quad Predicted charge asymmetries at the Tevatron and LHC]{Predicted charge asymmetries at the Tevatron and LHC for various BSM models\cite{AguilarSaavedra:2011ug,AguilarSaavedra:2011hz}. The inclusive charge asymmetry originating from new physics, $A_C^{\text{new}}$, at the LHC vs. the forward-backward asymmetry $A_{\text{FB}}^{\text{new}}$ at the Tevatron (left) is shown. Furthermore, the identical predictions in a high invariant mass region where $M_{t \bar{t}} > 450\,\text{GeV}$ (right) for the different models in the created phase space is shown. The Standard Model prediction corresponds to $A_{\text{FB}}^{\text{new}} = 0$ and $A_C^{\text{new}} = 0$, respectively.}
    \label{fig:AsymTeVLHC}
  \end{centering}
\end{figure}

A summary of the predicted charge asymmetries for Tevatron and LHC measurements for various BSM models can be found in \mbox{Figure \ref{fig:AsymTeVLHC}}. Potential regions in the phase space of inclusive charge asymmetry from new physics, indicated by the variable $A_C^{\text{new}}$ at the LHC plotted against the associated forward-backward asymmetry $A_{\text{FB}}^{\text{new}}$ at the Tevatron, are highlighted\footnote{Note that $A_{\text{FB}}^{\text{new}}$ and $A_C^{\text{new}}$ denote only the respective contributions to the overall charge asymmetry and forward-backward asymmetry originating from the corresponding BSM model. The predicted Standard Model contribution in the respective variables is subtracted (and hence corresponds to $A_{\text{FB}}^{\text{new}} = 0$ and $A_C^{\text{new}} = 0$, respectively).}. Different generalised predictions and parametrisations\cite{AguilarSaavedra:2011ug,AguilarSaavedra:2011hz} of $Z'$ and $W'$ models, scalar triplet ($\omega^4$) and sextet ($\Omega^4$) models, a generalised colour-octet resonance model ($\mathcal{G}_{\mu}$) and a colour-singlet Higgs-like isodoublet $\phi$ are shown. The same model predictions are shown for a high \ttbar~invariant mass region.

\begin{figure}[h!tb]
  \begin{centering}
    \includegraphics[width = 6 cm]{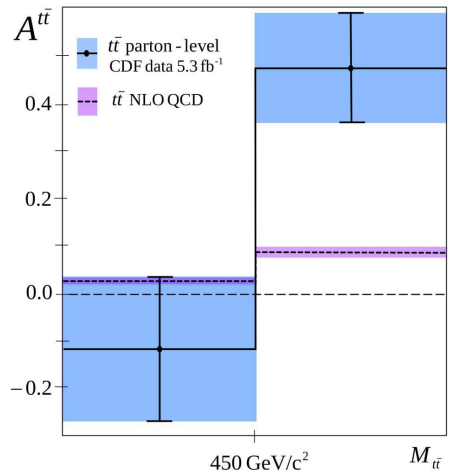}
    \vspace{-0.2 cm}
    \caption[\quad Summary of high $t \bar{t}$ invariant mass result at CDF]{Summary of results obtained in a $t \bar{t}$ invariant mass dependent measurement of the charge asymmetry at CDF\cite{Aaltonen:2011kc}. A deviation of $3.4 \sigma$ is observed in the high $M_{t \bar{t}}$ region.}
    \label{fig:summaryTevatron}
  \end{centering}
\end{figure}

Depending on the observed asymmetries at the Tevatron and the LHC, the exclusion of different theories can be possible. Recent measurements of the charge asymmetry by the CDF and \dzero~collaborations at the Tevatron indicate large partonic asymmetries $A_{t \bar{t}}$ in the \ttbar~rest frame of
\begin{equation}
A_{t \bar{t}} = 0.201 \pm 0.065\,\text{(stat.)} \pm 0.018\,\text{(syst.)} \text{\cite{CDFCONF-10584}}
\end{equation}
in the combination of the dileptonic and semileptonic decay channel, indicating a $2.9 \sigma$ excess, and
\begin{equation}
A_{t \bar{t}} = 0.196 \pm 0.065\,\text{(stat.+syst.)} \text{\cite{Abazov:2011rq}},
\end{equation}
respectively, indicating a $1.9 \sigma$ excess above the Standard Model prediction. 
Furthermore, larger deviations for high $t \bar{t}$ invariant masses\cite{Aaltonen:2011kc} and for high rapidity differences\cite{CDFCONF-10224} have been observed. In particular, an asymmetry of
\begin{equation}
A_{t \bar{t}} = 0.475 \pm 0.114\,\text{(stat.+syst.)} \text{\cite{Aaltonen:2011kc}}
\end{equation}
for $t \bar{t}$ invariant masses above 450\,GeV has been observed, indicating a $3.4 \sigma$ deviation from the Standard Model prediction, as shown in \mbox{Figure \ref{fig:summaryTevatron}}.

Similar measurements have been performed by the ATLAS and CMS collaborations at the LHC. In previous ATLAS measurements, a charge asymmetry in the variable $A_C$ of
\begin{equation}
A_C = -0.024 \pm 0.016\,\text{(stat.)} \pm 0.023\,\text{(syst.)} \text{\cite{CONFNote}}
\end{equation}
has been measured, while a measurement at CMS using the corresponding observable based on pseudorapidities instead of rapidities indicates a charge asymmetry of
\begin{equation}
A_C = -0.016 \pm 0.030\,\text{(stat.)} _{-0.021}^{+0.026}\,\text{(syst.)} \text{\cite{CMS-PAS-TOP-11-014}},
\end{equation}
in compatibility with the Standard Model prediction.

\vspace{3mm}
\clearpage

  \chapter{Experimental Setup}
\label{Experiment}
This chapter describes the technical details of the Large Hadron Collider and the ATLAS experiment, focusing on the detector subsystems and their properties after a short general overview. In addition, the ATLAS trigger system is explained in more detail concerning technical and functional parameters.

\section{LHC and ATLAS Technical Overview}
\begin{figure}[h!tb]
  \begin{centering}
    \includegraphics[width = 8 cm]{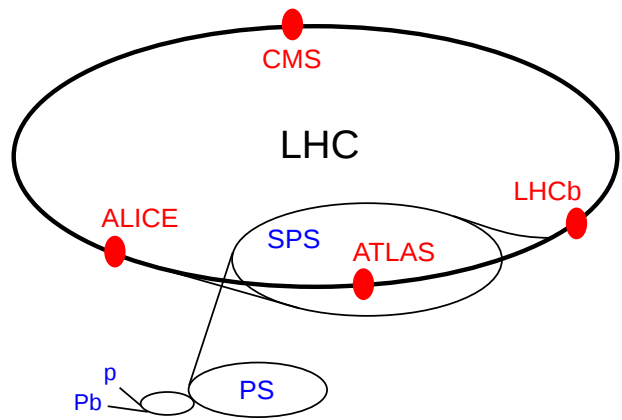}
    \vspace{-0.2 cm}
    \caption[\quad Layout of the LHC accelerator complex]{Layout of the LHC accelerator complex, including four of the LHC experiments\footnotemark.}
    \label{fig:LHC_layout}
  \end{centering}
\end{figure}

The Large Hadron Collider (LHC)\cite{LHCMachine} is situated at CERN, the European Centre for Nuclear Research near Geneva, Switzerland, and is the technologically most advanced particle accelerator so far. It is designed as a proton-proton accelerator with the potential to accelerate and collide heavy ions as well. Its construction started in 1999 and physics operation commenced in November 2009 with the first proton-proton collisions. The collider has been constructed in the former accelerator ring of the Large Electron Positron Collider (LEP), which is about 27~kilometres in circumference and 100 to 125~metres below ground level, partly making use of the already existing infrastructure of the LEP ring as well as two existing caverns for experiments. A schematic view of the accelerator ring is shown in \mbox{Figure \ref{fig:LHC_layout}}.
\footnotetext{Image taken from wikipedia (public domain).}

The accelerator complex incorporates six experiments, two of them being multi-purpose experiments, ATLAS\cite{AtlasExperiment} and CMS\cite{CMSExperiment}, while the other four are designed for more specific fields of research. Having two independently designed multi-purpose detectors is vital for cross-confirmation of any potential discoveries made and naturally allows to combine the results of both experiments. Among the four other detectors, the LHCb\cite{LHCbExperiment} experiment focuses on \mbox{$b$ physics}, while ALICE\cite{AliceExperiment} has been designed for heavy ion physics. TOTEM\cite{TotemExperiment} and LHCf\cite{LHCfExperiment} are both designed to conduct studies of forward physics, including soft and hard diffractive processes and low-$x$ QCD.

Protons are extracted from hydrogen atoms in a duoplasmatron\cite{Brown:1113300} and then accelerated by a linear accelerator (up to 50\,MeV) before being injected into the first circular accelerator, the Proton Synchrotron Booster ($50\,\text{MeV} \rightarrow 1.4\,\text{GeV}$). Afterwards, they are injected into the Proton Synchrotron (PS, $1.4\,\text{GeV} \rightarrow 26\,\text{GeV}$) and subsequently into the Super Proton Synchrotron (SPS, $26\,\text{GeV} \rightarrow 450\,\text{GeV}$), before being transferred into the main LHC ring. The two LHC proton beams deliver an energy of 3.5\,TeV each (7\,TeV at design specifications) and bunches of about $10^{11}$~protons can be brought to collision at a bunch crossing rate (BCR) of 40\,MHz within one of the various detectors.


Collision interactions are typically characterised by the instantaneous luminosity $\mathcal{L}$, which relates the cross-section $\sigma$ of a given process to the corresponding event rate $\dot{N}$, given the experimental acceptance $A$ and measurement efficiency $\varepsilon$:
\begin{equation}
  \mathcal{L} = \frac{\dot{N}}{\sigma A \varepsilon}.
\end{equation}
\noindent
At the design luminosity of $\mathcal{L} = 10^{34}\,\rm{cm}^{-2}\rm{s}^{-1}$ and the given collision parameters, this leads to a total of about 23 proton - proton collisions per bunch crossing on average, which implies an overall interaction rate in the GHz regime.

The ATLAS detector is designed to investigate a wide range of physics processes, including the measurement of well-known Standard Model processes, the search for the Higgs boson, extra dimensions, and particles that could constitute dark matter.

It is 44\,m in length, 25\,m in height and 25\,m in width, with a total weight of about \numprint{7000}~tonnes. As a comparison, CMS, the second LHC multi-purpose experiment, weighs about \numprint{12500}~tonnes while being 21\,m long, 15\,m wide, and 15\,m high. ATLAS features an onion-like structure, which will be discussed in detail in \mbox{Section \ref{sec:ATLASSubSys}}. A summary of the ATLAS and the LHC specifications compared to other accelerator/detector combinations is shown in \mbox{Table
\ref{tab:LHC_comparison}}.

\begin{table}[h!tb]
  \small
  \renewcommand{\arraystretch}{1.5}
  \begin{center}
    \begin{tabular}{|r||c|c|c|c|c|c|}
      \hline
                                      & \bf{Type}                 & \bf{$\maybebm{\sqrt{s}}$}  & \bf{BCR}     &
                                      \bf{$\bm{N_{\rm{CH}}}$} & \bf{Event Size} & \bf{Year} \\ \hline \hline \bf{ATLAS,
                                      LHC (CERN)} & pp                     & $\phantom{14.91}7$\,TeV           & \phantom{.}40 MHz & $\sim 10^8$ & $\sim \numprint{1300}$\,kB & 2009 \\
      \hline
      \bf{CDF/D\O, Tevatron\footnotemark (FNAL)}    & p$\overline{\rm{p}}$   &
      $\phantom{40}1.96$\,TeV & $2.5$ MHz    & $\sim 10^6$ & \phantom{6}$\sim 250$\,kB & 2001 \\
      \hline
      \bf{ZEUS, HERA (DESY)}          & $\rm{e}^{\pm}$p        & $\phantom{1}0.318$\,TeV      & \phantom{.}10 MHz   
      & $\sim 10^5$ & \phantom{6}$\sim 100$\,kB & 1992 \\
      \hline
    \end{tabular}
  \vspace{-0.1 cm}
  \caption[\quad Comparison of LHC/ATLAS to other accelerators/detectors]{Comparison of LHC/ATLAS to other
  accelerators/detectors. The particle types brought to collision, the centre-of-mass energy $\sqrt{s}$, the bunch crossing rate (BCR), the amount of readout channels ($N_{\rm{CH}}$), the average amount of data per event and the year of startup of the accelerator are shown.}
  \label{tab:LHC_comparison}
  \end{center}
  \renewcommand{\arraystretch}{1.0}
  \normalsize
\end{table}
\footnotetext{Values refer to Tevatron Run II.}

As can be seen from the comparison table, the LHC exceeds the bunch crossing rate of the largest particle accelerator up to 2009 (Tevatron in \mbox{Run II}, Fermi National Accelerator Laboratory) by a factor of seven. Furthermore, the amount of detector readout channels used at ATLAS increased by two orders of magnitude with respect to CDF/\dzero, together with the average amount of data per event increasing by about one order of magnitude. This results in a much higher total raw data throughput rate to be processed by the readout and data acquisition system.

As a consequence, a sophisticated trigger system designed for the reduction of data rates is needed to facilitate analyses. This can only be achieved by selecting a subset of potentially relevant events out of the large
amount of collisions that take place at LHC/ATLAS, rejecting a large fraction of raw data.

\section{ATLAS Coordinate System}
The ATLAS coordinate system is a right-handed coordinate frame with the $x$-axis pointing towards the centre of the LHC
ring and the $z$-axis being directed along the beam pipe, while the $y$-axis points upwards (slightly tilted
with respect to the vertical direction ($0.704\,^{\circ} $) due to the general tilt of the LHC tunnel). In this context, the
pseudorapidity can be introduced as
\begin{equation}
  \eta = - \ln \tan \frac{\theta}{2}
\end{equation}
\noindent
with $\theta$ being the polar angle with respect to the positive $y$-axis. For massive objects such as jets, the rapidity is used, given by
\begin{equation}
  y = \frac{1}{2} \ln{ \left( \frac{E + p_z}{E - p_z} \right)}.
\end{equation}
\noindent

In addition, the transverse momentum \pt~of a particle in the detector is defined as the momentum perpendicular to the
$z$-axis:
\begin{equation}
p_{\rm{T}} =  \sqrt{p_x^2 + p_y^2}.
\end{equation}

Furthermore, the azimuthal angle $\phi$ is defined around the beam axis.

\section{The ATLAS Detector Subsystems}
\label{sec:ATLASSubSys}
In order to allow for reliable detection of particles and measurement of their properties, the ATLAS detector requirements include
\begin{itemize}
  \item a good hermiticity with respect to detector acceptance,
  \item a high spatial and timing resolution, in particular to minimise occupancy of individual detector components, to measure \pt~with high resolution and to allow for distinction between different particles,
  \item a low material budget to minimise particle interaction and energy loss with non-active detector materials.
\end{itemize}
ATLAS has a cylindrical shape with layers of detector components arranged in axial succession. Each of these layers is designed to detect different types of particles which are mostly originating from the primary
\begin{figure}[h!tb]
  \begin{centering}
    \includegraphics[width = 15 cm]{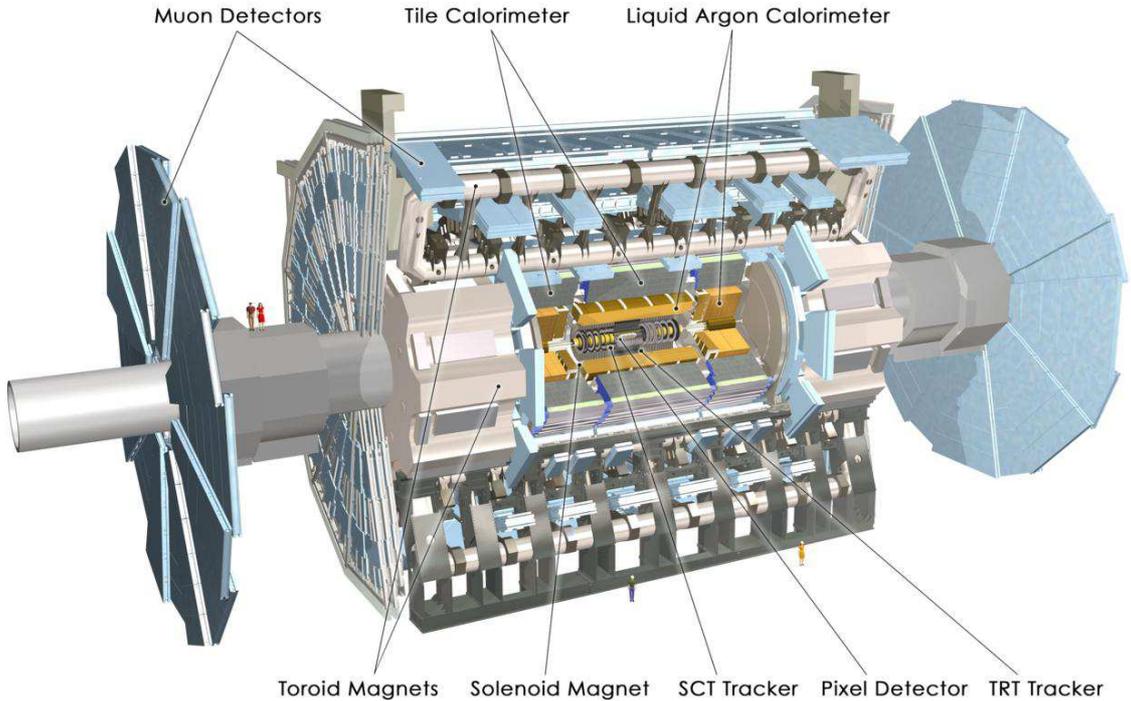}
    \vspace{-0.2 cm}
    \caption[\quad The ATLAS detector subsystems]{The ATLAS detector subsystems\footnotemark.}
    \label{fig:ATLAS_labeled}
  \end{centering}
\end{figure}
interaction point of the proton beams at the centre of ATLAS. As they travel throughout the detector, they can be measured by its successive layers. The different detector subsystems are shown in \mbox{Figure \ref{fig:ATLAS_labeled}} and constitute, from the innermost to the outermost layer, the inner detector, the solenoid magnet, the electromagnetic calorimeter, the hadronic calorimeter, the toroidal magnet, and the muon spectrometer.

The detectors are complementary: charged particles are detected in the innermost layers by their hits in the tracking chambers, where the particle trajectory is bent by the magnetic field of the superconducting solenoid magnet. Using this tracking information, the momentum of the particles can be determined. Around the magnet, the electromagnetic and hadronic calorimeters are designed to measure the energy of particles. These are brought to a stop in the calorimeter by interaction with the detector material\cite{ReviewParticlePhysics} (ionisation), thus depositing all of their energy, which is measured in the calorimeter cells. Finally, the muon chambers allow for additional momentum measurements of muons, which penetrate all other layers of the detector only depositing very little energy in the detector material. The measurements in the muon chambers are performed using the shape of the tracks, which are bent by the magnetic field of the toroidal magnets.
\footnotetext{Image taken from http://www.atlas.ch/.}

\subsection{Inner Detector}
In the following, the inner detector\cite{AtlasExperiment,ATLAS_InDetTDR} is described in more detail. It is situated near the interaction point to allow for high precision measurement of charged particle trajectories. It covers a pseudorapidity range of $|\eta| < 2.5$ and consists of three subsystems, the silicon pixel detector, the semiconductor tracker (SCT) and the transition radiation tracker (TRT). The innermost layers of the inner detector (three in the cylindrical barrel region, three endcap disks on each side of the forward region) constitute the pixel detector, which is designed to measure particle vertices and extract track momenta from the reconstructed particle hits in the detector layers. Due to its close proximity to the primary interaction point, a very high spatial resolution of the pixel detector is required, which is achieved by very small pixel sizes of 50\,$\umu$m~$\times$~400\,$\umu$m and 50\,$\umu$m~$\times$~600\,$\umu$m (about $8 \cdot
10^7$ readout channels in total), with the pixel detector covering a total area of $2.3\,\rm{m}^2$.

Around the pixel detector, the semiconductor tracker (or {\em silicon strip tracker}) measures the momentum of charged particles. It consists of four barrel layers and nine endcap wheels on each side, covering a total area of $61.1\,\rm{m}^2$ and making use of about $6.3 \cdot 10^6$ readout channels.
\begin{figure}[h!tb]
  \begin{centering}
    \includegraphics[width = 13 cm]{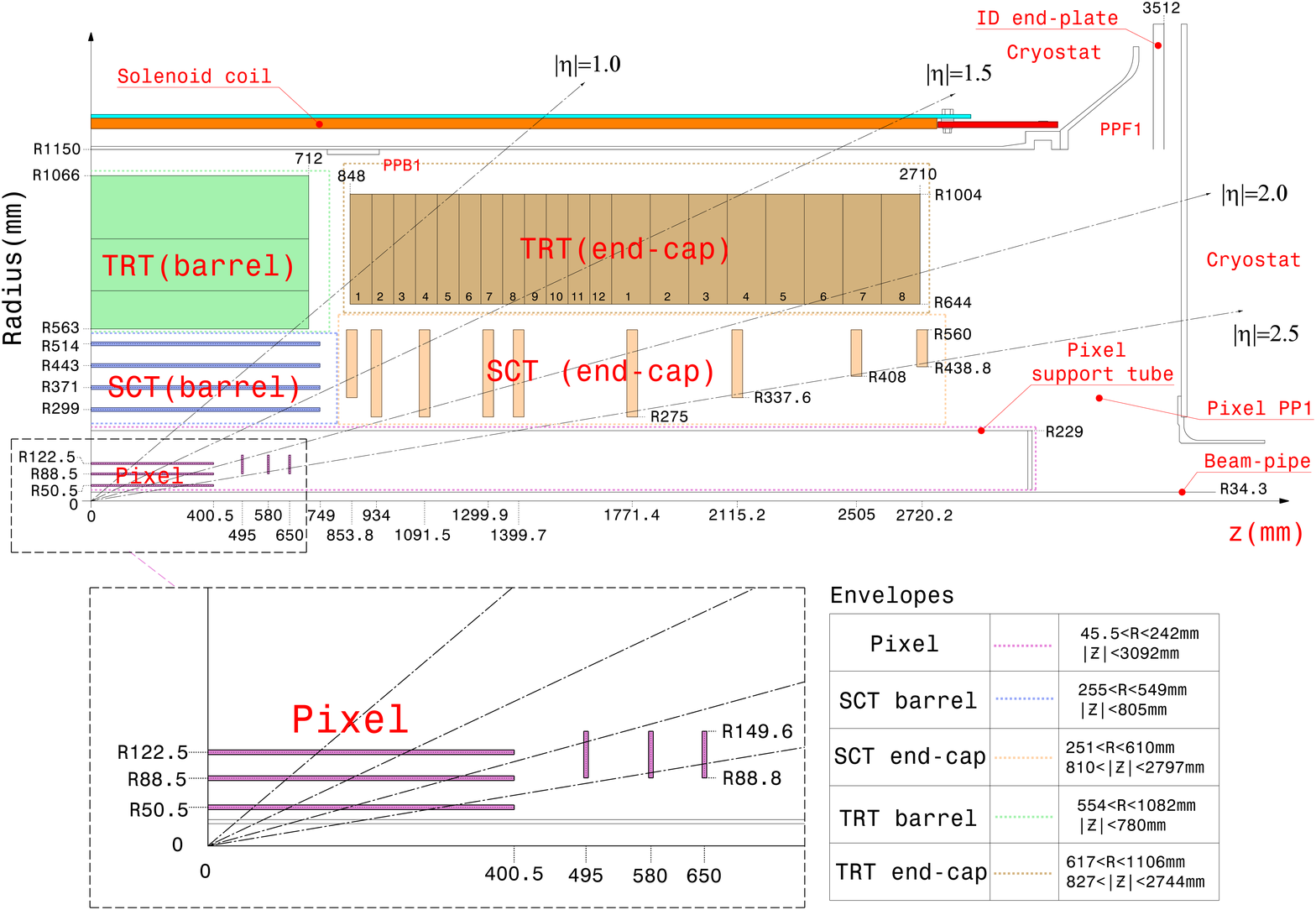}
    \vspace{-0.2 cm}
    \caption[\quad Cross-sectional view of a quarter-section of the ATLAS inner detector]{Cross-sectional view of a quarter-section of the ATLAS inner detector showing each of the major detector elements alongside with its active dimensions and envelopes\cite{AtlasExperiment}.}
    \label{fig:InDet_open}
  \end{centering}
\end{figure}

The outermost part of the inner detector is constituted by the TRT, which consists of straw tubes with a diameter of 4~mm and a maximum length of 80~cm, filled with an ionisable gas. The barrel tubes are divided in two at the centre and read out at each end to reduce occupancy. The ionisation charges created by charged particles travelling through the gas filled tubes are used to enhance track pattern recognition and to improve momentum resolution of the objects identified in the pixel detector and semiconductor tracker with an additional average of 36 hits per track. Furthermore, its function is to distinguish electrons and pions making use of the different amount of transition radiation emitted by these particles when crossing the boundary surface of two media with different dielectric constants (in this case a special radiator foam with a large amount of air bubbles to achieve a maximum material transition surface). The transition radiation tracker has a total of \numprint{351000} readout channels.

The resulting tracking performance of the inner detector subsystems for single particles and particles in jets can be found in \mbox{Table \ref{tab:InDet_acc}}.

\begin{table}[h!tb]
  \small
  \renewcommand{\arraystretch}{1.5}
  \begin{center}
    \begin{tabular}{|l||c|c|c|c|}
      \hline
        \multirow{2}{*}{\bf{Track Parameter}}       & \multicolumn{2}{c|}{\bf{$\maybebm{0.25 < |\eta| < 0.50}$}} &
        \multicolumn{2}{c|}{\bf{$\maybebm{1.50 < |\eta| < 1.75}$}} \cr
                      & \bf{$\maybebm{\sigma_X (\infty)}$} & \bf{$\maybebm{p_X [\text{GeV}]}$}  &
                      \bf{$\maybebm{\sigma_X (\infty)}$} & \bf{$\maybebm{p_X [\text{GeV}]}$}      \cr
      \hline
      \hline
        Momentum ($1 / p_{\rm{T}}$)   & $0.34\,\rm{TeV}^{-1}$ & 44.0  & $0.41\,\rm{TeV}^{-1}$ & 80.0 \cr
      \hline  
        Azimuthal angle ($\phi$)     & $70\,\umu \rm{rad}$ & 39.0 & $92\,\umu \rm{rad}$ & 49.0  \cr
      \hline  
        Polar angle ($\cot{\theta}$) & $0.7 \times 10^{-3}$ & \phantom{0}5.0  & $1.2 \times 10^{-3}$ & 10.0 \cr
      \hline
        Transv. impact parameter ($d_0$) & $10.0\,\umu \rm{m}$ & 14.0  & $12.0\,\umu \rm{m}$ & 20.0  \cr      
      \hline
        Longit. impact parameter ($z_0 \times \sin{\theta}$) & $91.0\,\umu \rm{m}$ & \phantom{0}2.3 
         & $71.0\,\umu \rm{m}$ & \phantom{0}3.7  \cr      
      \hline
    \end{tabular}
  \vspace{-0.1 cm}
  \caption[\quad inner detector performance parameters]{Track parameter resolutions at infinite momentum $\sigma_X
  (\infty)$ and the transverse momentum $p_X$ for which the intrinsic and multiple-scattering contribution equals
  the intrinsic resolution. The momentum and angles correspond to muons, while the impact parameters correspond to
  pions. The values are shown for two $\eta$ regions, one in the barrel inner detector (where the amount of material is
  close to its minimum) and one in the endcap (where the amount of material is close to its maximum)\cite{AtlasExperiment}.}
  \label{tab:InDet_acc}
  \end{center}
  \renewcommand{\arraystretch}{1.0}
  \normalsize
\end{table}

\subsection{Calorimeters}
Around the solenoid magnet (which is described in more detail in \mbox{Chapter \ref{chap:magnets}}), an
electromagnetic (EM) liquid argon sampling calorimeter is used to detect and identify electromagnetically interacting particles and to measure their energy\cite{AtlasExperiment,ATLAS_CaloTDR,ATLAS_TDR1,ATLAS_LArTDR}. It also allows, in combination with the hadronic calorimeters, for reconstruction of hadronic jets and the measurement of the missing energy of an event.

\begin{figure}[h!tb]
  \begin{centering}
    \includegraphics[width = 12 cm]{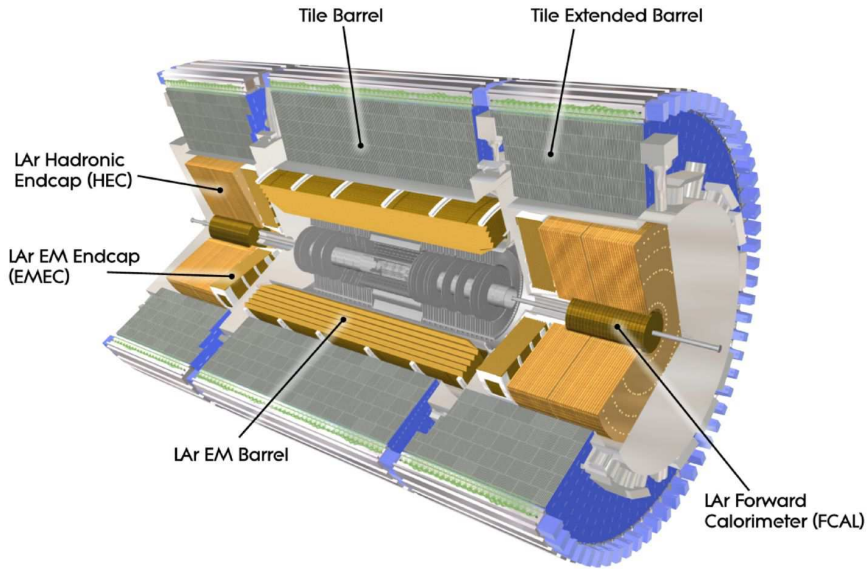}
    \vspace{-0.2 cm}
    \caption[\quad Overall layout of the ATLAS calorimeters]{Overall layout of the ATLAS calorimeters\cite{AtlasExperiment}.}
    \label{fig:Calo_labeled}
  \end{centering}
\end{figure}

The EM calorimeter covers a pseudorapidity range of $|\eta| < 3.2$ and comprises several layers of accordion-shaped
kapton-copper electrodes and lead absorber plates shrouded in stainless steel, with the gaps in between filled with liquid argon at a temperature of 87\,K.
Whenever an electromagnetically interacting particle passes through one of the lead absorber plates, it creates a
particle shower that ionises the liquid argon. Exposed to an electric field, the drifting ionisation charges induce a signal in the electrodes due to capacitive coupling. The resulting signal is sampled
and digitised at 40\,MHz, corresponding to the bunch crossing rate. The EM calorimeter has a total of \numprint{170000}
readout channels, and provides an energy resolution of
\begin{equation*}
\dfrac{\sigma_{\rm{EM}}^{\rm{sp}}}{E} = \dfrac{(10.1 \pm 0.4)\,\%}{\sqrt{E\,[\rm{GeV}]}}~\text{(stochastic)}~\oplus~0.2 \pm 0.1\,\%~\text{(constant)}\text{ \cite{AtlasExperiment}},
\end{equation*}
as measured in particle test beams. The total module thickness in the barrel region corresponds to at least 22 radiation lengths ($X_0$), increasing from $22\,X_0$ to $30\,X_0$ between $|\eta|=0$ and $|\eta|=0.8$ and from $24\,X_0$ to $33\,X_0$ between $|\eta|=0.8$ and $|\eta|=1.3$. In the endcaps, the total thickness is greater than $24\,X_0$ except for $|\eta|<1.475$, increasing from $24\,X_0$ to $38\,X_0$ in the outer wheel ($1.475 < |\eta| < 2.5$) and from
$26\,X_0$ to $36\,X_0$ in the inner wheel ($2.5 < |\eta| < 3.2$). 

In analogy, the hadronic calorimeter (hCAL)\cite{AtlasExperiment,ATLAS_CaloTDR,ATLAS_TDR1} is designed to
measure the energy of hadronic particles that can penetrate the EM calorimeter. This sampling calorimeter consists of iron absorbers for showering which are interleaved with plastic scintillator tiles (thus being referred to as {\em tile calorimeter}) in the barrel part of the detector ($|\eta|<1.7$). The scintillator tiles emit a shower of photons whenever charged particles pass through them due to excitation of the atoms in the scintillating material and subsequent emission of visible or UV photons. These light pulses are carried by optical fibres to photomultiplier tubes and converted to electric signals. The total number of tile calorimeter readout channels is of the order of \numprint{10000}.

Since the scintillating tiles are very sensitive to radiation damage, liquid argon is used as sampling medium together with copper absorbers in the forward endcap regions ($1.5<|\eta|<3.2$) in close proximity to the proton beams. This provides improved radiation hardness in the region of increased particle flux. The total number of channels for both endcaps is \numprint{5632}. For the same reason, a high density copper/tungsten absorber liquid argon forward calorimeter (FCAL) covers the pseudorapidity region $3.1<|\eta|<4.9$, with an additional \numprint{3524} channels for both forward regions together.

In order to estimate the performance of the hadronic calorimeter, test beam studies were conducted, showing an energy resolution of
\begin{equation*}
\dfrac{\sigma_{\pi}^{\rm{HAD}}}{E_{\pi}} = \dfrac{(56.4 \pm 0.4)\,\%}{\sqrt{E\,[\rm{GeV}]}}~\text{(stochastic)}~\oplus~5.5 \pm 0.1\,\%~\text{(constant)}\text{ \cite{AtlasExperiment}}
\end{equation*}
for pions, and a radial depth of approximately 7.4 interactions lengths ($\lambda$) for the tile calorimeter.
The hadronic endcaps show an energy resolution of
\begin{equation*}
\dfrac{\sigma_{e}^{\rm{HEC}}}{E_{e}} = \dfrac{(21.4 \pm 0.1)\,\%}{\sqrt{E\,[\rm{GeV}]}}~\text{(stochastic)}\text{ \cite{AtlasExperiment}}
\end{equation*}
for electrons (the constant term being compatible with zero), and
\begin{equation*}
\dfrac{\sigma_{\pi}^{\rm{HEC}}}{E_{\pi}} = \dfrac{(70.6 \pm 1.5)\,\%}{\sqrt{E\,[\rm{GeV}]}}~\text{(stochastic)}~\oplus~5.8 \pm 0.2\,\%~\text{(constant)}\text{ \cite{AtlasExperiment}}
\end{equation*}
for pions.

The jet energy resolution for the overall calorimeter system is described by the parametrisation
\begin{equation*}
  \dfrac{\sigma_{\rm{jet}}}{E_{\rm{jet}}^{\phantom{2}}} = \sqrt{ \dfrac{a^2}{E_{\rm{jet}}^{\phantom{2}}} +
  \dfrac{b^2}{E_{\rm{jet}}^2} + c^2 }.
\end{equation*}
For central jets in the region $0.2<|\eta|<0.4$, it is $a \approx 60\,\%\,\sqrt{\rm{GeV}}$ (stochastic), $c \approx 3\,\%$ (constant) and the noise term $b$ increases from 0.5\,GeV to 1.5\,GeV from barrel to endcap ranges\cite{AtlasExperiment}.


\subsection{Muon Chambers}
The muon chambers\cite{AtlasExperiment,ATLAS_MTDR,ATLAS_TDR1} are designed to detect muons which are able to
pass all other detector systems depositing only a small amount of energy in the material due to the fact that
muons in the GeV regime are approximately minimum ionising particles.

\begin{figure}[h!tb]
  \begin{centering}
    \includegraphics[width = 7 cm]{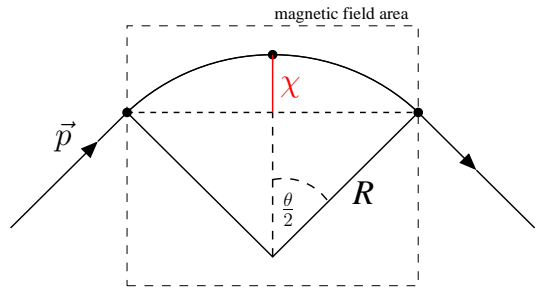}
    \vspace{-0.2 cm}
    \caption[\quad Illustration of the track sagitta]{Illustration of the track sagitta $\chi$ and its geometric
    relation to the bending radius $R$ for a track of a particle with a given momentum $\vec{p}$ travelling through
    a homogeneous magnetic field, shown for the case of three equidistant track hits along the curvature.}
    \label{fig:sagitta}
  \end{centering}
\end{figure}

Being deflected by the magnetic field in the detector, it is possible to determine the muon momentum and
sign of electric charge by measuring its trajectory as it passes through the tracking chambers. The
muon spectrometer is also designed to trigger on these particles, utilising dedicated trigger chambers. The driving performance goal is a
standalone transverse momentum resolution of approximately $10\,\%$~for 1\,TeV tracks, which translates into a sagitta
along the z axis of about $\chi = 500\,\umu$m, to be measured with a resolution of~$\sigma_{\chi}
\leq50\,\umu$m. The sagitta $\chi$~is given by
\begin{equation}
  \chi = R - R \cos{\frac{\theta}{2}},
\end{equation}
\noindent
where $R$ is the radius of the track curvature and $\theta$ is the angle enclosed by the outermost of three equidistant points along the track, as can be seen in \mbox{Figure \ref{fig:sagitta}}.

To achieve high spatial tracking resolution, three layers (stations) of drift chambers (precision chambers) are employed both in the barrel and in the endcap region. In the barrel these chambers are arranged in concentric cylinders, with the radii of the detector layers being at about 5\,m, $7.5$\,m and 10\,m, covering a pseudorapidity range of $|\eta| < 1.0$. Two of these layers are placed near the inner and outer field boundary, while the third is situated within the field volume. The muon momentum is determined from the track sagitta. In this region, exclusively monitored drift tube chambers (MDTs) are used.

Due to the magnet cryostats in the endcap region, however, placing one station within the field is not
possible. Hence it is necessary to rely on a point-angle measurement to determine the track momentum in this detector region (a point in the inner station and an angle in the combined middle-outer stations). The endcap layers are arranged in four concentric discs at 7\,m, 10\,m, 14\,m, and 21-23\,m from the detector origin, covering a pseudorapidity range of $1.0<|\eta|<2.7$. Here, in addition to MDT chambers, cathode strip chambers (CSCs) are used in the innermost layer of the inner station due to radiation hardness requirements.

The only exception to the continuous $\eta$ coverage of the muon chambers is made at $|\eta| < 0.05$~in the $R$-$\phi$ plane to allow for cable and service outlets for the inner detector, the central solenoid and the calorimeters (central gap). Further regions of reduced acceptance can be found at the feet of the detector.
\begin{figure}[h!tb]
  \begin{centering}
    \includegraphics[width = 9 cm]{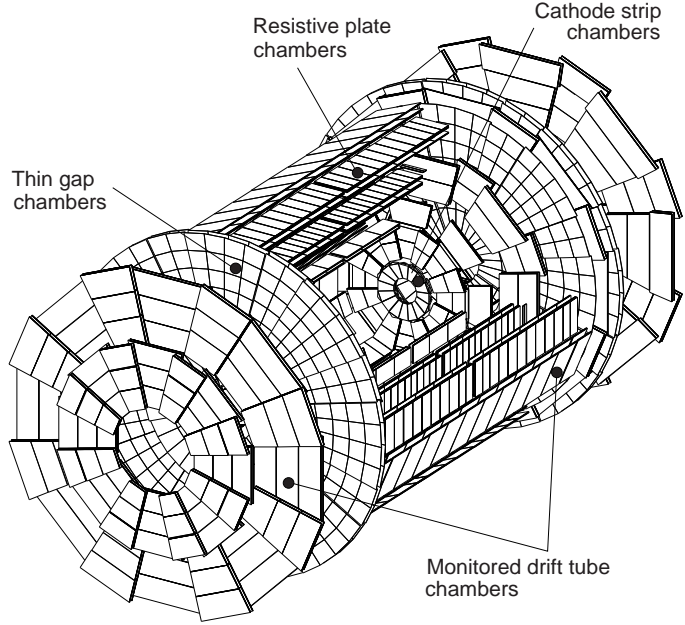}
    \vspace{-0.2 cm}
    \caption[\quad The ATLAS muon system]{The ATLAS muon system\cite{ATLAS_TDR1}. The trigger chambers (RPC, TGC) and the precision
    chambers (MDT, CSC) are shown.}
    \label{fig:Muon_Chambers}
  \end{centering}
\end{figure}

Regardless of the high spatial resolutions, the timing resolution of the precision chambers as shown in \mbox{Table
\ref{tab:muon_chambers}} is too low due to the drift time to ensure differentiation between muons from subsequent bunch
crossings for the trigger (where the scale of the required timing resolution is set by the bunch crossing interval of
25\,ns). Thus it is necessary to employ additional drift chambers with high timing resolution (at cost of spatial
resolution), the trigger chambers. These provide a fast momentum estimate and are primarily used for the trigger (since
their spatial resolution is too low with respect to the precision chambers), which will be described in \mbox{Chapter \ref{chap:triggersys}}.

In the barrel region, two out of three resistive plate chambers (RPCs) are placed directly
in front of and behind the central MDT, while the third is situated directly below or above (according to the
mechanical constraints in the respective region) the outermost precision chamber. The RPCs are furthermore used to
determine the second coordinate for the MDT chambers (in tube wire direction). In the endcaps, three layers of thin gap
chambers (TGCs) are located near the central endcap MDT layer for triggering.

A comparison of the different chamber technologies, necessary to allow for both fast triggering and high precision
measurements, is shown in \mbox{Table \ref{tab:muon_chambers}}.

\begin{table}[h!tb]
  \small
  \renewcommand{\arraystretch}{1.4}
  \begin{center}
    \begin{tabular}{|r||c|c|c|c|}
      \hline
                                                          &  \phantom{s}\bf{MDT}\phantom{s}            & 
                                                          \phantom{s}\bf{CSC}\phantom{s}          & \bf{RPC} & \bf{TGC}
                                                          \cr \hline \hline \bf{$z$/$R$ resolution [$\umu$m]} &  35 ($z$) &  40 ($R$)              &
       $10 \times 10^3$ ($z$) &  $(2-6) \times 10^3$ ($R$) \cr
      \hline
      \bf{\# readout channels}                            &  354\,000   & 
      30\,700 &  373\,000  &  318\,000 \cr
      \hline
      \bf{\# chambers}                                    &  \phantom{23}1\,150     &  \phantom{39\,3}32
      &  \phantom{39\,3}606                &  \phantom{39}3\,588 \cr
      \hline
      \bf{$\phi$ resolution [mm]} &  --        &  \phantom{39\,34}5                   & \phantom{39\,34}10      &
       \phantom{39\,3}3-7 \cr
      \hline
      \bf{timing resolution [ns]} &  --        &  \phantom{39\,34}7                   &  \phantom{3943}1.5      &
       \phantom{39\,334}4 \cr
      \hline
      
    \end{tabular}
  \vspace{-0.1 cm}
  \caption[\quad Comparison of muon chamber technologies]{Comparison of muon chamber technologies. Both the trigger chambers (RPC and TGC) and precision chambers (MDT and CSC) and their respective resolution and timing performance are shown.}
  \label{tab:muon_chambers}
  \end{center}
  \renewcommand{\arraystretch}{1.0}
  \normalsize
\end{table}

\subsection{Magnet System}
\label{chap:magnets}
Outside the inner detector follows a solenoid magnet\cite{AtlasExperiment,ATLAS_SolenoidTDR}, which is used to bend the trajectories of charged particles on their way through the inner detector, making it possible to measure the particle momentum with
\begin{figure}[h!tb]
  \begin{centering}
    \includegraphics[width = 9 cm]{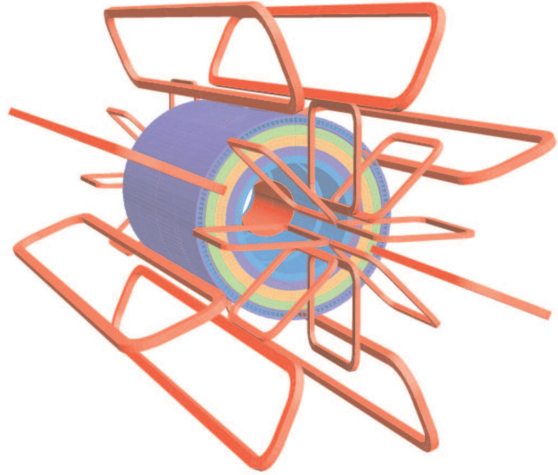}
    \vspace{-0.2 cm}
    \caption[\quad ATLAS solenoid and toroidal magnets]{Schematic view of the ATLAS solenoidal (inner cylinder) and
    toroidal magnets (outer coils)\cite{ATLAS_MTDR}.}
    \label{fig:magnetic_windings}
  \end{centering}
\end{figure}
high resolution. Its axial field strength is about 2\,T (peak $2.6$\,T) using low-temperature superconducting cables cooled down to 1.8\,K with liquid helium during operation with a nominal current flow of \numprint{8000}\,A.

In addition, a large toroidal magnet\cite{AtlasExperiment,ATLAS_ToroidTDR} consisting of eight superconducting coils in the barrel region and eight more at each of the forward regions extends throughout the muon chambers, providing a magnetic field strength of 4\,T (peak $4.7$\,T). The whole toroid system contains over 70\,km of superconducting cable, allowing for a design current of \numprint{20000}\,A with a stored energy of above 1\,GJ. To minimise multiple scattering of the muons, the toroid design incorporates an air core.

Similar to the solenoid magnet, its purpose is to bend the trajectories of muons in order to measure their transverse momentum (in combination with the tracking information from the inner detector).

\section{The ATLAS Trigger System}
\label{chap:triggersys}
The design luminosity of $10^{34}\,\rm{cm}^{-2}\rm{s}^{-1}$, in combination with the bunch crossing rate of 40\,MHz and the amount of protons contained in each single bunch, leads to a proton-proton collision rate in the GHz regime. This corresponds to an extremely high theoretical raw data rate of about 1.5\,PBs$^{-1}$. Being able to store only a fraction of this amount of data on storage media \mbox{($\sim (300-500)$\,MBs$^{-1}$)} and only a small fraction of these collisions being useful for analysis, the ATLAS trigger system\cite{AtlasExperiment,ATLAS_CompTDR,ATLAS_TDR1} has been designed to reduce the initial data rate by several orders of magnitude. \mbox{Figure \ref{fig:SMxsects}} shows the total rates of several physics processes in
comparison to the total interaction rate. As an example, the frequency of the Standard Model \ttbar~production at \mbox{$\sqrt{s}$ = 7\,TeV} is of the order of 1\,mHz, constituting only a small fraction of the total amount of the raw data rate, making the selection of this and other physics processes a crucial task.

\begin{figure}[h!tb]
  \begin{centering}
    \includegraphics[width = 15 cm]{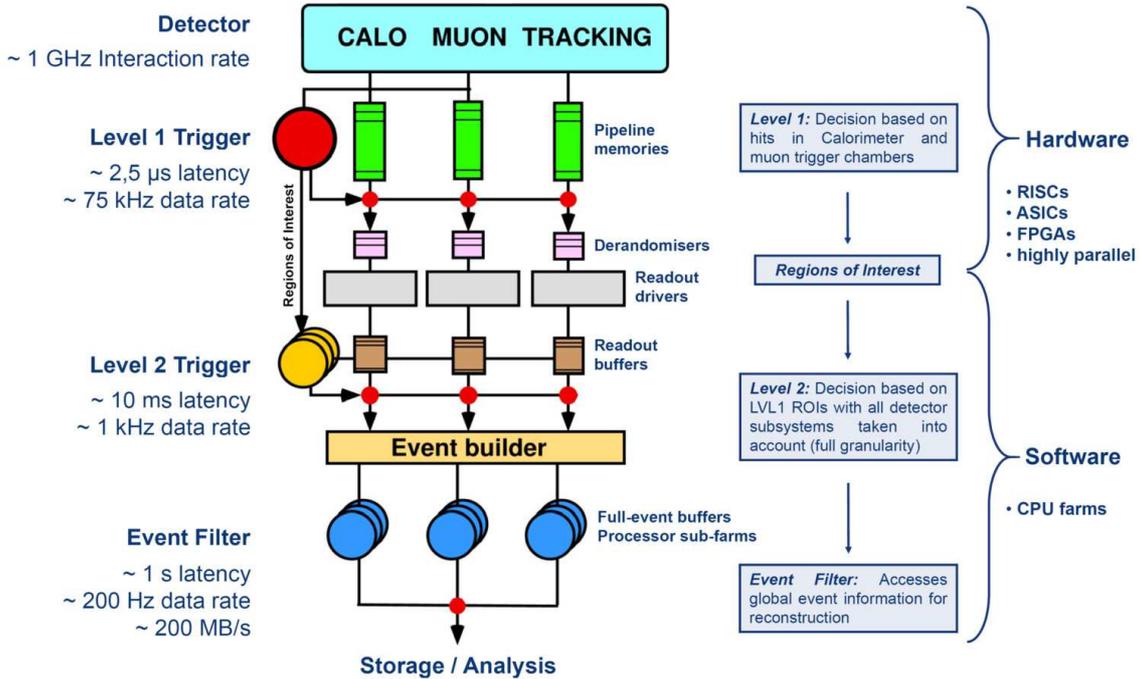}
    \vspace{-0.2 cm}
    \caption[\quad Block diagram of the trigger/DAQ system]{Block diagram of the trigger/DAQ system. On the left side the typical collision and the data equivalent at the different stages of triggering are shown, while in the middle section the different components of the trigger system are shown schematically\cite{ATLAS_TDR1}. The right side of the graphic gives a short summary of the operations and the technologies used at the respective level.}
    \label{fig:TrigLevels}
  \end{centering}
\end{figure}

In order to achieve such a reduction and to select only relevant physics events / processes, ATLAS uses a
three-level trigger system for real-time event selection (with the last two levels being referred to as {\em high-level trigger}), while each trigger level refines the decisions of its predecessor. An overview of the different trigger levels and the global structure is shown in \mbox{Figure \ref{fig:TrigLevels}} and will be discussed in more detail in the following sections.

\subsection{Level 1 Trigger (LVL1)}
The LVL1 trigger is completely hardware-based, where highly specialised components are deployed, including field
programmable gate arrays (FPGAs), application specific integrated circuits (ASICs) and reduced instruction set computing (RISC) chips. Since most of this hardware is integrated directly into the particular detector components in order to reduce material occurrence from cabling and additional readout electronics, the LVL1 trigger system is required to be highly parallelised.

\begin{figure}[h!tb]
  \begin{centering}
    \includegraphics[width = 9 cm]{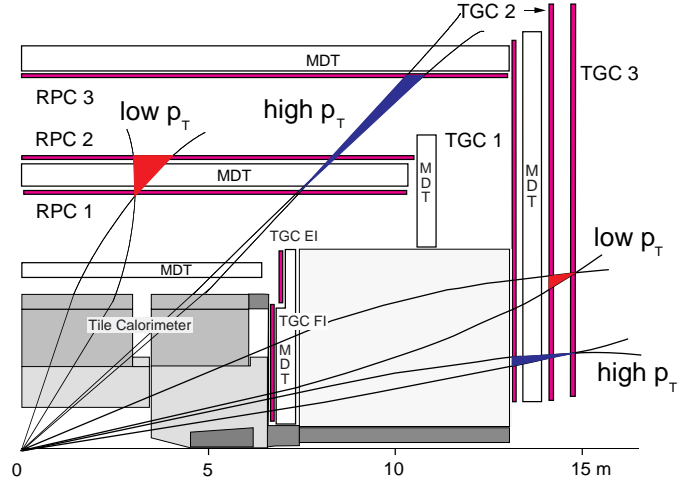}
    \vspace{-0.2 cm}
    \caption[\quad Layout of the muon trigger chambers]{Layout of the muon trigger chambers\cite{ATLAS_MTDR}. A quarter cross-section in the bending plane with typical low \pt~and high \pt~muon tracks and the corresponding coincidence windows in the different layers for barrel (RPC1-3) and endcap (TGC1-3) are shown.}
    \label{fig:TriggerChamberScheme}
  \end{centering}
\end{figure}

Being based on the muon and calorimeter system only, the LVL1 trigger\cite{ATLAS_LVL1TDR} performs an initial selection on the basis of the hits in the muon trigger chambers and calorimeters. In the muon chambers, low-\pt~and high-\pt~muons are identified by measuring the tracks in the trigger chambers (RPCs and TGCs) using coincidence windows for discrimination. Low-\pt~muons have a smaller bending radius, thus allowing for detection by two layers in close proximity to each other (Moreover, track-hit matching can be difficult if the in-plane hit distance in the coincidence layers is too large due to the curvature of the track). In the barrel region, low-\pt~muons are identified by the consecutive layers RPC1 and RPC2 (The number after the chamber type identifies the layer). In contrast, high-\pt~muons produce an almost straight track and therefore the coincidence layers used
should be as separated as possible to allow for the measurement of the track radius. Consequently, high-\pt~muons are measured by the combination of the outermost RPC1 and RPC3 hits in the barrel. In the endcaps, low-\pt~muons are identified by the TGC2 - TGC3 coincidence window, while TGC1 and TGC3 are used to identify high-\pt~tracks. This is shown in \mbox{Figure \ref{fig:TriggerChamberScheme}}, where a quadrant layout of the muon trigger chambers is shown.

Classification of transverse momentum is achieved by using large lookup tables of track hits to find an
estimate for the track momentum. Different exclusive low and high \pt~thresholds are defined (e.g. 10, 11, 20 and 40\,GeV) and can be modified if necessary.

The calorimeter trigger selection is based on low resolution information from all ATLAS calorimeters and is designed to identify high \Et~electrons and photons, hadron jets and the total transverse energy alongside with large missing
transverse energy \met, where this is defined by the sum of all vectored energy depositions $\vec{E}_T$ in the
transversal plane:
\begin{equation}
  \not\!\!\vec{E}_{\rm{T}} = - \sum \vec{E}_{\rm{T}}.
\end{equation}
This definition arises from the fact that the initial transverse momentum of the incoming protons is approximately zero and due to conservation of energy in the transverse plane, the total vector sum of final state transverse energies has to be zero as well. Furthermore, isolation requirements based on calorimeter information are available for the calorimeter trigger on this trigger level.

Since the LVL1 trigger operates synchronously with data taking, the latency for a decision is about~$2.5\,\umu \rm{s}$, which is achieved by making use of pipeline memories despite the bunch crossing and data taking frequency of 40\,MHz. The trigger decision is derived while the information from the sensors is kept in the buffer memory for about 100 bunch crossings. At the end of the latency time, the readout data is rejected or accepted (after leaving the pipeline memory).

Only in the latter case the geometric information of the triggered object is forwarded to the next trigger level as {\em region of interest} (ROI), which includes spatial position ($\eta$ and $\phi$) and \Et~estimates of the embedded objects (e, $\mu$, $\tau$, $\gamma$ and jet candidates) as well as global energy information (\Et~and \met) for further analysis. This way, the data rate is reduced to about (75-100)\,kHz at LVL1.

\subsection{Level 2 Trigger (LVL2)}
After a LVL1 accept, events are read out from the pipeline memories and stored in readout buffers (ROBs) until
being processed by the LVL2 trigger\cite{ATLAS_HLTTDR}. This trigger level uses the ROIs from the previous level to
further reduce the data rate to about 1\,kHz. This is achieved by analysing each ROI in the detector system from which it originated, accessing the data from other detector subsystems in the ROI in addition, including the inner detector tracking information, the full-granularity calorimeter hits and the precision chamber measurements from the muon spectrometer.

With an increased processing latency of 10\,ms and the data rate already reduced at LVL1, the LVL2 trigger allows for more complex algorithms being applied to the trigger objects and the information from the respective ROIs. As a part of the high-level trigger, the LVL2 stage is completely software-based and runs on dedicated computing farms. The trigger decision is performed by an event-driven sequential selection procedure using the detector data. Despite the LVL2 trigger providing more time to conduct the trigger decision than the LVL1 stage, the selection algorithms still have to be kept simple and efficient. Hence, the sequence of the different algorithms / requirements is determined by their complexity, simple ones (with respect to CPU time and memory) are executed first, while each algorithm uses the result of its predecessor ({\em seeded reconstruction}).

\subsection{Event Filter (EF)}
In the final execution level of the trigger, which is software based and runs on CPU farms, the global event is collected from the ROBs. The EF accesses the complete event information and all detector subsystems using full granularity. An event reconstruction at trigger level is possible by using similar algorithms as it is the case for offline reconstruction, accessing calibration and alignment information from databases.

In addition to the reconstruction of the event as a whole, the EF performs extended tasks that are not possible at earlier trigger levels, such as vertex reconstruction, final track fitting and algorithms requiring larger ROIs than available at LVL2 (e.g.~the calculation of \mbox{global \met}).

The latency of the EF is of the order of seconds, after which events passing this final trigger level are stored permanently for further analysis or, in some cases, may be redirected to special storage elements as well (if needed for calibration or alignment exclusively).

\subsection{Trigger Implementation}
The information used to conduct the LVL1 decision is given in terms of multiplicities of {\em trigger items} resembling candidates for physical objects (like electrons, muons, jets,\ldots) detected in the calorimeters or muon trigger chambers which have sufficiently high \pt. These are sent to the central trigger processor (CTP) together with threshold information on global energy sums. The delivered multiplicities are discriminated against corresponding requirements or conditions, leading to logical values for each condition within so-called {\em trigger menus}. Examples for LVL1 trigger items are
\begin{center}
  \begin{tabular}{r l}
    {\bf {\em EM25i}}:  &  electromagnetic, $p_{\rm{T}}>25\,\rm{GeV}$, isolated, \cr
    {\bf {\em MU20}}:   &  muon, $p_{\rm{T}}>20\,\rm{GeV}$.
  \end{tabular}
\end{center}
The final LVL1 event decision is derived from the values of the defined trigger items by applying a logical
{\em OR}. If the event is accepted, the ROIs for all trigger items and the contained trigger elements above
the defined thresholds are delivered to the high-level trigger, where they are used as seeds for further
processing.

The step-by-step execution of trigger algorithms as it is performed on LVL2 and EF level is called a {\em trigger chain}, consisting of different intermediate {\em trigger signatures} (cf.~trigger items on LVL1), where the successive trigger algorithms keep refining these signatures in the course of a trigger chain. Examples for such trigger signatures, as they have been used in this analysis, are
\begin{center}
  \begin{tabular}{r l}
    {\bf {\em EF\_e20\_medium}}:    &  electron, $E_{\rm{T}}>20\,\rm{GeV}$, medium object definition, \cr
    {\bf {\em EF\_mu18}}:           &  muon, $p_{\rm{T}}>18\,\rm{GeV}$.
  \end{tabular}
\end{center}
The individual decisions can also be logically combined to more complex trigger items.

As an additional requirement for the high-level trigger, parallel execution of different trigger chains without
interference is required to be possible in independent {\em slices} (e.g. electron slice and muon slice) to ensure transparency and scalability of the trigger system.

The information of individual high-level trigger signatures is organised in {\em trigger streams}, grouping similar triggers based on the respective purpose, priority, and procedure of processing of the respective events. An inclusive model is chosen to allow for trigger items to be contained in multiple trigger streams in parallel. Raw data or \textsc{physics} streams contain the data events selected for full reconstruction and later analysis and correspond to the respective types of trigger items, such as the \textsc{physics\_Egamma} and \textsc{physics\_Muon} streams which contain events triggered by electron/photon or muon triggers, respectively. Further trigger streams exist for calibration and monitoring purposes, such as the \textsc{Express} stream, which contains a small subset of relevant events for very fast reconstruction in order to allow for real-time monitoring of the data taking and trigger system.

\section{Underlying Event and Pile-Up}
At hadron colliders such as the LHC, a multitude of partons is involved in the hadronic collisions due to the substructure of the colliding particles, while typically only one parton from each of the incoming hadrons is involved in the hard scattering process and has sufficient energy to create high energy/momentum final state particles. Nevertheless, the remaining partons still contribute to the final event signature through low momentum transfer interactions. These additional reactions which occur in parallel to the hard scattering event are denoted {\em underlying event}.

Furthermore, additional contributions to the final event signature arise from the interactions of other protons within the colliding bunches. Despite the fact that the probability of multiple hard scatterings occurring within one bunch crossing is relatively low, the likelihood of soft interactions between the constituent partons originating from additional proton-proton collisions in the same bunch crossing increases significantly with instantaneous luminosity $\mathcal{L}$. In addition to this {\em in-time pile-up} contribution, final state particles from different bunch crossings can lead to additional signatures if the identification of the correct bunch crossing is unsuccessful ({\em out-of-time pile-up}).

Both the underlying event and the pile-up contributions play a significant role in the final state of the observed collisions and hence have to be included in the simulation of signal and background processes to achieve a proper modelling of the data taken. This modelling is in particular difficult due to the fact that both the underlying event and the pile-up cannot fully be described in perturbative QCD, since the particles created in such processes typically have very low momentum/energy. Furthermore, due to time-dependent changes in the LHC environment parameters, the simulation has to be adapted continuously to match the respective setup for a given subset of data taken.
\clearpage

  \chapter{Trigger Strategy and Object Definition}
\label{Objects}
This chapter covers the determination and different approaches of applying trigger and reconstruction efficiencies in analyses. In addition, the object definitions used in this analysis are described.

\section{Data Quality}
\label{chap:DQ}
In order to ensure that only data taken under well defined and stable conditions is taken into account for physics analyses, dedicated online\cite{ATLASOnlineDQ} and offline\cite{ATLASOfflineDQ} monitoring systems ensure data integrity and quality. 

The online data quality monitoring accesses real-time detector status information and makes use of events from the \textsc{Express} trigger stream to provide several low-level quantities and distributions. This allows for a quick response to problems with the LHC beam conditions or the detector that may arise during operation.

The data quality offline monitoring uses a first reconstruction performed in order to identify and record problems in the detector hardware and the data acquisition and processing. All relevant information from the individual detector systems and reconstructed event quantities are combined into a small set of key numbers and distributions to allow for both automatic and manual monitoring.

Information from the online and offline data quality monitoring as well as feedback from the individual shift crews is combined into a database containing LHC beam conditions, detector status and data flow information which can be used to create lists of runs usable for analyses ({\em GoodRunsLists}), containing a set of data taking run and luminosity block (LB) information.

\section{Trigger and Reconstruction Efficiencies}
Trigger and detector acceptance and response are represented by the respective trigger and reconstruction efficiencies, which are typically both estimated in Monte Carlo simulations and measured with data driven methods. However, both the reconstruction and trigger simulation used for Monte Carlo samples typically do not fully reflect the actual conditions due to limitations of the simulation modelling. Hence, discrepancies have to be corrected for in the Monte Carlo. This ensures that the actual detector performance is reflected in the corrected Monte Carlo distributions of relevant physical quantities.

Since for Monte Carlo samples the detector simulation/reconstruction does not retain the information which links objects from the Monte Carlo generator level to reconstructed objects or trigger objects, it can be necessary to perform a matching procedure to identify corresponding objects and their affiliation at the different trigger levels and at offline reconstruction level. This is achieved by a geometric matching in $\Delta R$, defined for example for the matching of objects at reconstruction level to the corresponding generated (true) object as
\begin{eqnarray}
  \Delta R & = & \sqrt{ (\Delta \eta)^2 + (\Delta \phi)^2 } \nonumber \\
           & = & \sqrt{ (\eta_{\text{true}} - \eta_{\text{reco}})^2 + (\phi_{\text{true}} - \phi_{\text{reco}})^2 }
\end{eqnarray}
\noindent
and for the different trigger levels, accordingly.

A positive match of the given objects is denoted by a $\Delta R$~below a predefined threshold. This threshold is typically chosen according to the spatial resolution of the measurement for the respective objects.

Both trigger and reconstruction efficiencies are commonly taken into account in physics analyses using scale factors, relating the respective efficiencies measured with data driven methods to the expected performance from Monte Carlo simulations. Alternatively, in particular trigger efficiencies can be applied by implementing a reweighting approach, taking into account detector and trigger acceptance by direct application of data driven efficiencies in the form of event weights.

In this context, the reconstruction efficiency $\varepsilon_{\text{reco}}$ is typically defined as the fraction of objects which have been identified by a specific reconstruction algorithm,
\begin{equation}
\varepsilon_{\text{reco}} = \frac{N_{\text{reco}}}{N_{\text{candidate}}},
\end{equation}
where $N_{\text{candidate}}$ denotes the amount of reconstruction candidates (e.g. tracks or energy depositions) considered for reconstruction. 

In analogy, the trigger efficiency $\varepsilon_{\text{trig}}$ refers to the fraction of reconstructed objects which have been selected by a given trigger item or chain,
\begin{equation}
\varepsilon_{\text{trig}} = \frac{N_{\text{trig}}}{N_{\text{reco}}}.
\end{equation}
These efficiencies are typically parametrised or binned in kinematic quantities of the given objects, such as \pt~and $\eta$.

\subsection{Measurement of Trigger Efficiencies}
In order to obtain trigger efficiencies from data, a simple Monte Carlo based counting method is not applicable as events rejected by the trigger are typically no longer available offline. Furthermore, it is not desirable to rely on the trigger simulation. To measure trigger efficiencies, there exist several data driven methods, including:
\begin{itemize}
\item \textbf{orthogonal triggers:} the trigger efficiency $\varepsilon_{\rm{A}}$ for a given trigger item A is determined with respect to a different, ideally orthogonal (i.e. completely uncorrelated) trigger B, whose efficiency $\varepsilon_{\rm{B}}$ is known,
\item \textbf{minimum bias datasets:} the trigger efficiency for an arbitrary trigger item is determined on a minimum bias dataset, where only a minimal trigger selection has been applied and thus a trigger efficiency measurement can be performed with minimal bias,
\item \textbf{the Tag \& Probe method:} the trigger efficiency for a given trigger is measured by selecting a specific process from a given data sample (if possible) of which the kinematics are known, allowing for a decoupling of event selection and trigger efficiency determination.
\end{itemize}
\noindent In the following, the Tag \& Probe method is explained in more detail since it is commonly used to determine both trigger and reconstruction efficiencies for physics analyses.

\begin{figure}[h!tb]
  \begin{centering}
    \includegraphics*[width = 7 cm]{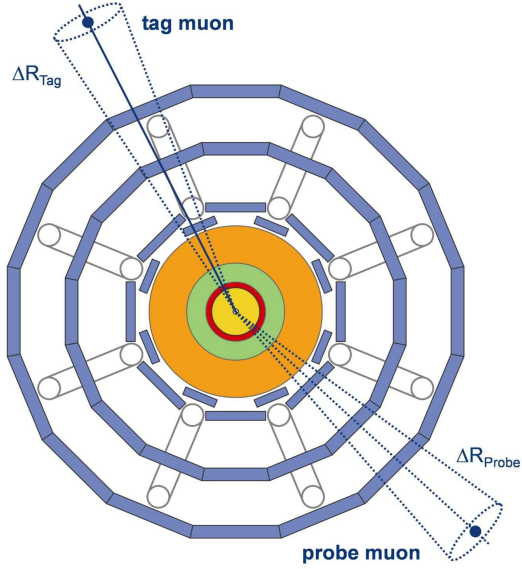}
    \vspace{-0.2 cm}
    \caption[\quad The Tag \& Probe method]{The Tag \& Probe method. An event selection is performed using the tag object, while the actual trigger efficiency is determined only with the probe object to ensure independence of the two processes.}
    \label{fig:tagprobe}
  \end{centering}
\end{figure}

To measure the trigger efficiency from data, it is possible to select a specific process, e.g. \zmumu~events for the determination of muon trigger efficiencies by application of the Tag \& Probe data method, which is shown schematically in \mbox{Figure \ref{fig:tagprobe}}.
The selection is achieved by making use of the fact that if one isolated muon is present in a \zmumu~event, there has to be a second muon in the observed event that is isolated as well, on which the actual efficiency measurement can be performed. The basic concept of this method focuses on decoupling event selection and the actual determination of the trigger efficiency.

Events are selected by first identifying a muon that has triggered the event, the {\em tag
muon}. Moreover, several requirements are applied to the tag muon, e.g.~\pt~and isolation
criteria. To ensure the \zmumu~sample to be as pure as possible, these requirements should be very tight, leading to an enrichment of \zmumu~events.

If the tag muon meets all requirements, a second muon (if present) in the event is selected, the {\em probe muon,}
which must comply to requirements that typically correspond to the selection performed in the analysis for which the obtained efficiencies are used.

In addition, the dimuon invariant mass $M_{\mu \mu}$ has to be sufficiently close to the $Z$ pole mass. If
that is the case, there is a high probability that the probe muon does not originate from a background
process, but constitutes a muon from the \zmumu~signal.

The actual trigger efficiency $\varepsilon_{\rm{TP}}$ for the Tag \& Probe method can be determined by the fraction of probe muons that have been triggered:
\begin{equation}
  \varepsilon_{\rm{TP}} = \frac{\rm{N}_{\mu}^{\rm{probe\,\&\,trigger}}}{\rm{N}_{\mu}^{\rm{probe}}}.
\end{equation}
\noindent

Since the invariant mass constraint directly relates the tag and the probe muon, it is obvious that the geometric correlation can create a bias in the estimated trigger efficiency as well. This can, for example, be the case if one of the two muons produced in the $Z$ decay emerges into an acceptance gap of the detector, while the other muon is properly reconstructed. In the context of the Tag \& Probe method, this muon could pass the tag muon criteria, but since there is no second muon in the event, no efficiency can be determined, the event not being taken into account at all. Hence, the Tag \& Probe method does not cover events where only one of the muons was found, neglecting the corresponding contribution to the overall efficiency. This effect, however, is typically very small.

\subsection{Scale Factors and Event Rejection Approach}
If the data driven efficiency $\varepsilon_{\text{data}}$ for a specific trigger item or trigger chain has been measured and the corresponding Monte Carlo trigger simulation efficiency $\varepsilon_{\text{MC}}$ is known, the respective scale factor $f$ is defined by
\begin{equation}
  f = \frac{\varepsilon_{\text{data}}}{\varepsilon_{\text{MC}}},
\end{equation}
which quantifies the discrepancies between measured and predicted trigger or reconstruction efficiency. Typically, both the underlying efficiencies and scale factors are parametrised according to actual dependencies on geometric and/or analysis related quantities such as transverse momentum or pseudorapidity.

The scale factors are applied by rejecting events on MC which do not pass the trigger criteria of the trigger simulation and applying the scale factors as an event weight based on all relevant objects on MC events passing the trigger simulation. For events requiring exactly one reconstructed object, the event weight equals the scale factor corresponding to the respective object properties. However, if more than one object is required in the final state, the probability $p$ of an event to be triggered is determined by the logical OR of the trigger probabilities of the all $N$ objects taken into account,
\begin{equation}
  p = 1 - \prod_{i=1}^N \left( 1 - \varepsilon_i \right),
\end{equation}
where $\varepsilon_i$ corresponds to the trigger efficiency/probability of the $i$-th object. Correlations can be taken into account by an appropriate parametrisation of the underlying efficiencies and scale factors. The corresponding scale factor for events requiring more than one final state object is given by
\begin{equation}
  \label{eq:sf_multiobject}
  f = \frac{1 - \prod \left( 1 - \varepsilon_i^{\text{data}} \right)}{1 - \prod \left( 1 - \varepsilon_i^{\text{MC}} \right)}.
\end{equation}
Consequently, this scale factor depends on the individual object properties, even if the respective efficiencies $\varepsilon_i^{\text{data}}$ and $\varepsilon_i^{\text{MC}}$ do not, since \mbox{Equation \ref{eq:sf_multiobject}} does not factorise in these quantities. Due to limited statistics, taking into account all possible combinations of parameters for objects $i,j,$ necessary for an accurate representation of scale factors, is not feasible. In particular, the treatment of correlations in the statistical uncertainties of the scale factors becomes increasingly difficult both analytically and computationally.

Furthermore, depending on the magnitude of the underlying efficiencies, a significant amount of MC events is rejected when requiring the trigger simulation to have selected events, leading to a reduction of available MC statistics and practical loss of CPU time spent on the generation of the rejected events. This can be of great importance, in particular if analyses are limited by MC statistics. However, since efficiencies of some of the relevant objects, such as electrons, usually are relatively high ($>95\,\%$) in the used signal regions, the corresponding loss of MC statistics can be negligible for those objects, depending on the amount of available generated events.

\subsection{Reweighting Approach}
\label{chap:TrigReweighting}
As an alternative to the determination and application of scale factors, a trigger reweighting approach, making exclusive use of data driven measurements of trigger efficiencies, can be applied to MC events. If the data driven efficiency $\varepsilon_{\text{data}}$ for a specific trigger item or trigger chain is known, the corresponding event weights can be determined regardless of the amount of required final state objects. The event weight $w$ is calculated directly from the respective parametrisation of the object efficiencies:
\begin{equation}
  w = 1 - \prod_{i=1}^N \left( 1 - \varepsilon_i^{\text{data}} \right),
\end{equation}
which factorises in the trigger (in)efficiencies.

Note that only objects following identical selection criteria as used in the efficiency determination should be taken into account in the event weight calculation to ensure that the respective parametrisation is valid. Furthermore, the choice of a proper parametrisation of the efficiencies is crucial due to the fact that neglecting potential dependencies of the efficiencies can lead to significant mis-modelling of the trigger response on MC. This effect is typically smaller for the scale factor approach, where the weights are typically close to one even if the overall efficiencies significantly differ from one (and hence, the impact of the parametrisation is smaller).

However, the trigger reweighting approach offers several additional advantages over a scale factor approach, most prominently due to the fact that no trigger simulation is involved in either the determination or application of the trigger efficiencies. Since the simulated trigger decision is not taken into account, all MC events are preserved regardless of the magnitude of the underlying trigger efficiencies, retaining the full MC statistics. In addition, efficiencies obtained from data can be applied to MC independent of potential changes or errors in the implementation of the trigger simulation, which would make repeated extraction of scale factors or alternative solutions necessary. Since the trigger requirements for a given data period typically do not change, the obtained efficiencies have to be determined only once for a set of reconstruction algorithm parameters and remain unchanged as well. 

Possible correlations in the statistical uncertainties derived for the object efficiencies can be modelled properly in a relatively simple way due to the fact that all objects obtain efficiencies from the same parametrisation. Hence, a propagation of the statistical uncertainties to the final event weights, event yields or even distributions of object and event quantities is possible. However, since the uncertainties on trigger efficiencies are usually relatively small compared to other sources of uncertainties in physics analyses, the impact is expected to be small when neglecting the respective correlations. Typically, uncertainties on both scale factors and efficiencies are regarded as fully correlated in a conservative approximation.

\section{Objects}
In the following, a summary of objects used in this analysis and their definition will be given. The algorithms used to reconstruct the individual objects and the corresponding performance will be discussed.

\subsection{Jets}
\label{chap:obj:jets}
Jets represent collimated collections of long-lived hadrons, originating mostly from partons after the fragmentation and hadronisation process\cite{Webber:1983if}. Jets were reconstructed by associating objects from the EM calorimeter and hCAL, and tracks reconstructed in the detector into physical jet objects representing the underlying fragmentation and parton shower processes.

Jet reconstruction algorithms are preferably collinear and infra-red safe, i.e. collinear splitting and soft gluon emissions should not change the algorithm response of the final reconstructed jet. In this context, the jet reconstruction was performed using the anti-$k_t$ ($R=0.4$) jet clustering algorithm\cite{jetantikt1,jetantikt2}. Tracks and energy depositions ({\em topological clusters}\cite{AtlasExperiment}) in the calorimeter identified at the electromagnetic energy scale\cite{AtlasExperiment} (EM scale), calibrated to yield a correct energy response for electrons and photons, were associated into combined jets of particles. This was achieved by iterative combination of pairs of objects into proto-jets according to the respective geometric distance of the individual constituents and the proto-jets until a convergent state was reached.

Jet quality criteria were applied to identify jets which do not correspond to physical in-time energy deposits in the calorimeter. Possible sources for such {\em bad jets} are for example hardware problems, calorimeter showers induced by cosmic rays and beam remnants.

The jet reconstruction efficiencies were determined from data with a Tag \& Probe method, using jets from charged tracks in the ID, where the efficiency was defined as the fraction of probe track jets matching a corresponding calorimeter jet. The jet energy resolution $\sigma (E_{\rm{T}}) / E_{\rm{T}}$ for jets in the region $|\eta| < 2.8$ ranged from about 0.2 (for 20\,GeV jets) to 0.05 (for 900\,GeV jets\footnote{Updated plots at {\em https://twiki.cern.ch/twiki/bin/view/AtlasPublic/JetEtmissApproved2011JetResolution}})\cite{ATLAS-CONF-2010-054}. Differences between measured efficiencies and efficiencies obtained in Monte Carlo simulations were taken into account by randomly removing jets from events according to a Gaussian probability distribution corresponding to the jet reconstruction efficiency uncertainty.

In addition, a calibration of the jet energy scale\cite{AtlasExperiment} (JES) was performed to obtain the energy of the final state particle-level jet from the measured energy of the reconstructed jet. The differences in response from the EM calorimeter and hCAL and additional effects such as energy mis-measurements of particles not coming to a stop in the calorimeter were taken into account.

The following requirements were applied to reconstructed jets (calibrated at the electromagnetic + JES (EM+JES) scale unless stated otherwise) after electron/muon overlap removal:
\begin{itemize}
\item Jet transverse momentum: $p_{\rm{T}}(\text{jet}) > 25\,\text{GeV}$,
\item Jet pseudorapidity at the EM scale, including JES $\eta$ correction: $|\eta_{\text{EMscale}}(\text{jet})| < 2.5$.
\end{itemize}

\subsubsection{$b$ Tagging}
Due to the dominant decay of top quarks into a $W$ boson and a $b$ quark, events containing top quarks are characterised by hard $b$ jets, which have the distinction of possessing a long lifetime, a large corresponding $B$ hadron mass and a large branching ratio into leptons. Hence, the identification of $b$ jets and the discrimination against light quark jets can significantly increase the signal to background ratio for top quark decays.

The identification (or {\em tagging}) of $b$ jets is mostly driven by exploiting the increased lifetime with respect to light jets and the associated significant flight path length $l$, leading to secondary vertices and measurable transversal and longitudinal impact parameters $d_0$ and $z_0$ of the final state particles. These denote the spatial distance of closest approach with respect to the associated primary vertex in the transversal respectively longitudinal plane.

For this analysis, the JetFitterCombNN tagging algorithm\cite{ATLAS-CONF-2011-102} was used, which determines a $b$ tag probability/weight $w$ for a given jet according to a Neural Network combination of the weights from the IP3D and JetFitter tagging algorithms\cite{ATLAS-CONF-2011-102}. The IP3D algorithm used the significance of $d_0$ and $z_0$ of each track contained in the respective jet to determine a likelihood corresponding to the $b$ jet probability. The JetFitter algorithm implemented a Kalman Filter\cite{Kalman_1960} to identify a common line of primary vertex and weak $b$ and $c$ hadron decay vertices within the jets. Using the approximated hadron flight path, a $b$ jet likelihood was obtained based on the masses, momenta, flight length and track multiplicities of the reconstructed vertices.

A cut on the $b$ tag weight of $w > 0.35$ was applied to classify $b$ jets, corresponding to an approximate $b$ tagging efficiency of $70\,\%$ in simulated \ttbar~decays and a rejection rate of about 5 for $c$ jets and about 100 for light flavour jets. The performance of the $b$ tagging algorithm was determined on specific data samples and compared to the MC prediction. A set of scale factors, parametrised in jet $p_{\rm{T}}$ was determined and applied to MC events for both the $b$ tagging efficiency and the light jet mis-tag rate to take into account differences between data and MC. The respective event weight was determined as logical OR from the scale factors of all jets taken into account.

\subsection{Muons}
Muons were identified and selected both at online (trigger) and offline (reconstruction) level in accordance with the respective object quality criteria.

Muon reconstruction was performed with the MuId\cite{Benekos:2005uu} algorithm, taking into account the tracks of the ID and the MS and creating a combined reconstructed track. Particle hits identified in the MS chambers were associated to the corresponding track segments in the individual layers of the ID. A combined fit was performed to obtain an optimised reconstructed trajectory from the joint information of both systems and to optimise geometrical and transverse momentum resolution. The MS momentum resolution in $\sigma (p_{\rm{T}}) / p_{\rm{T}}$ ranges from 0.04 (for 20\,GeV muons) to 0.06 (for 200\,GeV muons) in the central detector region and from 0.05 (for 20\,GeV muons) to 0.20 (for 200\,GeV muons) in the forward detector region. The ID momentum resolution in $\sigma (p_{\rm{T}}^{-1})$ ranges from 0.001 (for 20\,GeV muons) to 0.0005 (for 200\,GeV muons) in the central detector region and from 0.004 (for 20\,GeV muons) to 0.005 (for 200\,GeV muons) in the forward detector region\cite{ATLAS-CONF-2011-046}.

In addition, differences in the muon momentum resolution between data and Monte Carlo were corrected for by randomly changing the muon $p_{\rm{T}}$ on Monte Carlo according to a Gaussian resolution function to reflect the resolution observed in data.

Muon trigger and reconstruction efficiencies were measured with data driven methods on \zmumu~events using the Tag \& Probe method and were compared to Monte Carlo simulations. Both the respective efficiencies and scale factors were parametrised in muon transverse momentum, pseudorapidity and azimuthal angle to reflect differences in the detector acceptance and the $p_{\rm{T}}$ dependency for both the trigger and reconstruction efficiencies.

Muons were selected at trigger level by requiring the {\em EF\_mu18} trigger chain (corresponding to a muon with a $p_{\rm{T}} > 18\,$GeV at trigger level), which was seeded at LVL1 by the {\em L1\_MU10} trigger item and at LVL2 by the {\em L2\_mu18} trigger item, to have fired. This requirement was applied to data only, while the trigger efficiency was taken into account for Monte Carlo events by determining an event weight based directly on the trigger efficiency associated with the reconstructed and selected muons. Hence, no explicit (simulated) trigger decision was required. Furthermore, no geometric matching of trigger objects and reconstructed muons has been applied due to a software error resulting in incorrect trigger object representation within the used data and Monte Carlo samples. The resulting mis-modelling of the trigger response has been estimated and was taken into account as an additional systematic uncertainty (c.f. \mbox{Chapter \ref{Systematics}}).

Reconstructed muons from the MuId algorithm were required to be combined muons, i.e. to have a combined ID and MS track passing the respective track criteria. In addition, cuts specific to select semileptonic top quark decays were applied in order to reject muons from heavy and light flavour decays such as $b$ and $c$ hadrons or in-flight decays of kaons and pions. In particular, isolation requirements were applied based on the momentum and energy deposition within a $\Delta R$ cone of size 0.3 around the muon track, $E_{\text{T,cone30}}(\mu)$ and $p_{\text{T,cone30}}(\mu)$, respectively. The following requirements have been used:
\begin{itemize}
\item Muon transverse momentum: $p_{\rm{T}}(\mu) > 20\,\text{GeV}$,
\item Muon pseudorapidity: $|\eta(\mu)| < 2.5$,
\item Jet overlap removal by requiring $\Delta R (\mu,\text{closest reconstructed jet}) > 0.4$, where jets reconstructed by the anti-$k_t$ algorithm with $R=0.4$ were taken into account, calibrated to the EM+JES scale and with $p_{\rm{T}}(\text{jet}) > 20\,\text{GeV}$,
\item Isolation: $E_{\text{T,cone30}}(\mu) < 4\,\text{GeV}$ and $p_{\text{T,cone30}}(\mu) < 4\,\text{GeV}$.
\end{itemize}

\subsection{Electrons}
Similarly to muons, electrons were identified and selected according to the respective electron quality criteria. 

Electrons were reconstructed by matching energy cluster hits (or {\em seeds}) above a threshold of about $3\,\text{GeV}$ in the electromagnetic calorimeter to corresponding extrapolated ID tracks, vetoing tracks from photon conversion pairs. In addition, a matching of the track momentum and cluster energy was performed by application of a cut on $E/p$. Information from the TRT chambers was used to enhance the separation of electron candidates from pions in the reconstruction process. The electron energy resolution was about $(1.2 \pm 0.1\,\text{(stat.)} \pm 0.3\,\text{(syst.)})\,\%$ in the central detector region, about 1.8\,\% in the endcaps, and of the order of 3\,\% in the forward detector region\cite{Aad:2011mk}.

The electron energy scale and resolution were obtained with data driven methods in kinematic regions similar to top quark pair production events, using events from \zee~decays. On data, the energy scale was corrected as a function of electron transverse energy $E_{\rm{T}}(e)$ and cluster pseudorapidity $|\eta_{\text{cluster}}|$. On MC, the electron energy was corrected by applying a randomised Gaussian resolution function to the electron energy, following a similar parametrisation. Statistical and systematic uncertainties on both scale and energy corrections were taken into account and were assigned to the MC.

Electron trigger and reconstruction efficiencies were measured with data driven methods on \zee~and \wenu~events using the Tag \& Probe method and compared to Monte Carlo simulations. The respective electron scale factors were parametrised in electron pseudorapidity $|\eta_{\text{cluster}}|$ (using cluster quantities) to reflect differences in the detector acceptance for both the trigger and reconstruction efficiencies.

Electrons were selected at trigger level by requiring the {\em EF\_e20\_medium} trigger chain (corresponding to an electron with a $E_{\rm{T}} > 20\,$GeV at trigger level, and {\em medium} denoting a specific set of object criteria at trigger level, respectively), which was seeded at LVL1 by the {\em L1\_EM14} trigger item and at LVL2 by the {\em L2\_e20\_medium} trigger item, to have fired. This requirement was applied to both data and MC events. In addition, a geometric matching of reconstructed electrons to the corresponding trigger objects in a given event was performed to ensure that the reconstructed electron had fired the trigger. Data/MC scale factors were applied as event weights to take into account differences in the efficiencies between data and trigger simulation.

In this analysis, reconstructed electrons were required to pass the tight\cite{ATL-PHYS-PUB-2011-006} electron quality requirements. In addition, the following requirements were applied to electron candidates:
\begin{itemize}
\item Electron transverse energy: $E_{\rm{T}}(e) > 25\,\text{GeV}$,
\item Electron cluster pseudorapidity: $|\eta_{\text{cluster}}(e)| < 2.47$,
\item Exclusion of the non-covered transition region $1.37 < |\eta_{\text{cluster}}(e)| < 1.52$,
\item Jet overlap removal by requiring $\Delta R (e,\text{closest reconstructed jet}) > 0.2$, where all jets reconstructed by the anti-$k_t$ algorithm with $R=0.4$ were taken into account,
\item Isolation: $E_{\text{T,cone20}}(e) < 3.5\,\text{GeV}$ after leakage and pile-up correction\cite{ATL-PHYS-PUB-2011-006},
\end{itemize}
where $E_{\text{T,cone20}}(e)$ denotes the energy deposition within a $\Delta R$ cone of size 0.2 around the electron cluster. The electron transverse energy $E_{\rm{T}}(e)$ was determined from the energy deposited in the cluster in the calorimeter $E_{\text{cluster}}(e)$ and the associated track pseudorapidity $\eta_{\text{track}}(e)$,
\begin{equation}
E_{\rm{T}}(e) = \frac{E_{\text{cluster}}(e)}{\cosh{\left( \eta_{\text{track}}(e) \right)}}.
\end{equation}

\subsection{Missing Transverse Energy}
The missing transverse energy \met~is an object-based quantity derived from the topological clusters in the calorimeter calibrated at the EM scale, corrected for the energy scale of the corresponding object associated to the cluster. Objects taken into account were electrons and jets, where jets were separated into high- and low-$p_{\rm{T}}$ (soft) jets. The electron contribution used electrons passing the tight electron quality criteria with $p_{\rm{T}} > 10\,\text{GeV}$, while jets with $p_{\rm{T}} > 20\,\text{GeV}$ were corrected to the EM+JES scale and soft jets with $7\,\text{GeV} < p_{\rm{T}} < 20\,\text{GeV}$ were included at the EM scale.

In addition, muons were included in the definition using the transverse momentum of the corresponding tracks due to the fact that muons typically only deposit small amounts of energy in the calorimeter. Muons reconstructed with the MuId reconstruction algorithm with $|\eta| < 2.5$ were taken into account, distinguishing between isolated (requiring \mbox{$\Delta R (\mu,\text{closest jet}) > 0.3$}, where all jets reconstructed by the anti-$k_t$ algorithm with $R=0.4$ were taken into account) and non-isolated (where the energy deposited in the calorimeter was taken into account in the jet term). Furthermore, muons in detector regions with low acceptance ($|\eta| < 0.1$ and $1.0 < |\eta| < 1.3$) were taken into account making use of the corresponding calorimeter response.

Clusters not associated to any object were included in an additional ({\em CellOut}) term, calibrated at the EM scale, as well as the energy deposition of isolated muons in the calorimeter.

The missing transverse energy terms in the $x$ and $y$ directions were calibrated and combined into an overall missing energy in the respective dimension:
\begin{equation}
\not\!\!E_i = E_i^{\text{electrons}} + E_i^{\text{jets}} + E_i^{\text{softjets}} + E_i^{\text{muons}} + E_i^{\text{CellOut}},
\end{equation}
where $i = \lbrace x,y \rbrace$. Consequently, the respective scalar transverse missing energy \met~is given by
\begin{equation}
\not\!\!E_{\rm{T}} = \sqrt{\not\!\!E_x^2 + \not\!\!E_y^2}.
\end{equation}

The resolution of the $\not \hspace{-0.1cm} E_x$ and $\not \hspace{-0.1cm} E_y$ components ranged from about 2\,GeV (for a total transverse energy of 20\,GeV) to 14\,GeV (for a total transverse energy of 700\,GeV)\cite{Aad:2011re}.

  \chapter{Event Selection}
\label{Selection}
The event selection used for the top charge asymmetry measurement was aimed at maximising the signal contribution from the \ttbar~decay in the semileptonic decay channel. At the same time, the background contribution from different sources was minimised. The two distinct observable final states, muon+jets and electron+jets, were treated independently.

As discussed in \mbox{Chapter \ref{Theory}}, top quark pairs are dominantly produced at the LHC by gluon-gluon fusion. Both the top and antitop quark then decay into a $W$ boson and a $b$ quark in almost 100\,\% of all cases. The $W$ boson decays into two jets or a charged lepton and a neutrino, which can only be measured indirectly as missing transverse energy. Hence, the event selection for the semileptonic decay channel was focused on topologies with at least four reconstructed jets, exactly one isolated lepton (muon or electron) and missing transverse energy.

A preselection of the delivered raw data based on a common {\em GoodRunsList} (c.f. \mbox{Chapter \ref{chap:DQ}}), ensuring stable beam conditions and data quality, has been applied prior to the event selection. This selection included global data quality flags, e.g. requiring stable beams at a centre-of-mass energy of 7\,TeV and the LVL1 central trigger and luminosity measurement to be functional, indicating stable running and data taking conditions of the LHC and ATLAS, respectively. Furthermore, the data quality of the individual detector subsystems has been verified. These criteria correspond to a data quality selection efficiency of 84.1\,\% in the electron+jets channel and 84.3\,\% in the muon+jets channel.

In addition to the baseline selection, which will be described in the following, several corrections have been applied to Monte Carlo samples on an event-by-event basis to account for potential mis-matches in the detector simulation with respect to data, as described in \mbox{Chapter \ref{Objects}}. Muon trigger efficiencies have been taken into account directly by performing a trigger reweighting using trigger efficiencies obtained with a Tag \& Probe method, while discrepancies between the simulation and data for electron and muon reconstruction efficiencies and the electron trigger efficiencies as well as the $b$ tagging efficiencies have been taken into account by applying the corresponding scale factors to the Monte Carlo events. Furthermore, each Monte Carlo event has been assigned a weight according to the average amount of $pp$ interactions per bunch crossing in the respective sample to account for differences in the modelling of pile-up.

The following event selection has been used to enhance the signal to background ratio in the recorded samples. The described selections have been applied to both data and Monte Carlo samples unless stated otherwise:
\begin{itemize}
  \item The electron or muon trigger was required to have fired. The trigger item used was {\em EF\_e20\_medium} in the electron+jets channel for both Monte Carlo and data. In the muon+jets channel, the {\em EF\_mu18} trigger item was required for data, while a trigger reweighting using {\em EF\_mu18} trigger efficiencies obtained with data driven methods was applied to Monte Carlo events, made necessary by a mis-modelling of the trigger simulation in the used Monte Carlo samples. 
  \item A primary vertex with at least five tracks associated to it was required to improve rejection of non-collision background from the underlying event, pile-up and cosmic radiation.
  \item Exactly one isolated lepton (one electron and no muon or vice versa) passing the respective object quality criteria with $E_{\rm{T}}>25\,\rm{GeV}$ (electron+jets) or $p_{\rm{T}}>20\,\rm{GeV}$ (muon+jets), respectively, was required.
  \item The selected lepton was required to match the object of the fired trigger (electron+jets only, since due to a software problem, the muon trigger matching requirement was dropped from the selection).
  \item Any event where a reconstructed electron and muon share a common track was rejected.
  \item Any event containing a {\em bad jet} (c.f. \mbox{Chapter \ref{chap:obj:jets}}) with $p_{\rm{T}}>20\,\rm{GeV}$ at the EM+JES scale was rejected.
  \item A missing transverse energy $\not \hspace{-0.1cm} E_{\rm{T}} > 35\,\rm{GeV}$ (electron+jets) or $\not \hspace{-0.1cm} E_{\rm{T}} > 20\,\rm{GeV}$ (muon+jets) was required.
  \item A cut of $m_{\rm{T}}(W)\footnote{In this context, \MTW denotes the $W$ boson transverse mass, defined as
\begin{equation}
\MTW = \sqrt{2 p_{\rm{T}}^l p_{\rm{T}}^{\nu} \left( 1 - \cos{\left(\phi^l - \phi^{\nu}\right)} \right)},
\end{equation}
where $p_{\rm{T}}^{l/\nu}$ and $\phi^{l/\nu}$ describe the lepton and neutrino transverse momentum and azimuthal angle, respectively. The neutrino information is represented by the measured missing transverse energy, \met.} > 25\,\rm{GeV}$ (electron+jets) or a triangular cut of $\not \hspace{-0.1cm} E_{\rm{T}} + m_{\rm{T}}(W) > 60\,\rm{GeV}$ (muon+jets) was applied in order to suppress the QCD multijet background contribution, since these events typically have low $m_{\rm{T}}(W)$ and low $\not \hspace{-0.1cm} E_{\rm{T}}$.
  \item At least four jets with $p_{\rm{T}}>25\,\rm{GeV}$ passing the jet quality criteria were required.
  \item Any event where a jet was found in an area of LAr calorimeter defects was rejected and electrons which were affected were removed from the respective events, correcting the measured $\not \hspace{-0.1cm} E_{\rm{T}}$ accordingly. For Monte Carlo, a randomised subset of events was dropped according to the relative fraction of the data sample affected by these defects in order to correct for the created mis-match between data and simulation.
  \item At least one jet which has been $b$ tagged using the JetFitterCombinedNN algorithm with a weight $w > 0.35$ (corresponding to an overall $b$ tagging efficiency of about 70\,\% in simulated \ttbar~events) was required to further improve the signal to background ratio.
\end{itemize}

  \chapter{Samples and Process Modelling}
\label{Samples}
This chapter describes the data sample used in this analysis and the Monte Carlo samples generated to simulate the signal contribution and most of the backgrounds. Furthermore, the data driven estimations of the dominant background contributions, $W$+jets and QCD multijet production, are described.

\section{Data Sample}
\label{SamplesData}
A set of ATLAS data taken in the course of the year 2011, corresponding to an integrated luminosity of
\begin{equation}
\int \mathcal{L} \, dt = (1.04 \pm 0.04)\,\text{fb}^{-1}
\end{equation}
has been analysed, after the preselection of the delivered raw data using the corresponding {\em GoodRunsList}. The data has been recorded between March, $22^{\text{rd}}$, 2011 and June, $28^{\text{th}}$, 2011. A peak instantaneous luminosity of about $\mathcal{L} = 1.3 \cdot 10^{33}\,\rm{cm}^{-2}\rm{s}^{-1}$ was reached and a bunch spacing of 50\,ns was used.

\section{Signal and Background Monte Carlo Samples}
\label{SamplesMC}
Several Monte Carlo samples have been generated to facilitate this analysis, including nominal samples for the signal contribution and various background processes as well as several additional samples used in the evaluation of systematic uncertainties. All samples correspond to the {\em mc10b} production commonly performed for all ATLAS analyses using a generalised set of parameters to match the data taking conditions during the time period that was considered.

In particular, the contribution from in-time and out-of-time pile-up was added to all generated Monte Carlo events in the parton showering simulation process after the generation of the initial hard scattering. A fixed configuration of average proton-proton interactions per bunch crossing was used, while the actual data taking conditions with respect to pile-up changed over the course of time. In order to correct for the mis-match between individual data taking periods and the simulated pile-up contribution in the MC, an event-based reweighting was performed, taking into account the expected and observed distribution of the average number of interactions per bunch crossing. This {\em pile-up reweighting} was performed for all MC samples.

The \ttbar~signal process has been simulated using the \mcnlo~generator \cite{FRI-0201} (v3.41) which incorporates the CTEQ6.6 \cite{Nadolsky:2008zw} parton distribution function set and makes use of a next-to-leading order calculation approach for QCD processes. Both the parton showering and fragmentation processes, and the underlying event have been modelled using the \herwig~v6.510 \cite{COR-0001} and \jimmy \cite{JButterworth:1996zw} generators utilising the CTEQ6.6 and AUET1\cite{PUB-2010-014} tunes to match the ATLAS data, respectively. The inclusive \ttbar~cross section has been estimated to approximately next-to-next-to-leading order using the \hathor tool \cite{Aliev:2010} to be $165_{-16}^{+11}\,\text{pb}$\cite{PhysRevD.78.034003,Langenfeld:2009tc,Beneke2010483} for $m_t = 172.5\,$GeV and the MC has been scaled accordingly.
For this analysis, only semileptonic decays of the top quark pairs were considered. The respective cross-section, taking into account the proper branching fractions, was 89.3\,pb, including a $k$-factor of 1.117 to rescale the next-to-leading order perturbative QCD cross section in \mcnlo~to the approximate next-to-next-to-leading order cross-section. The signal sample contained \numprint{15000000} simulated events, corresponding to an integrated luminosity of about $150\,\text{fb}^{-1}$.

The electroweak single top production was simulated using the \mcnlo~and \jimmy generators and the respective cross-sections have been calculated at approximately next-to-next-to-leading order to be $64.57_{-2.01}^{+2.71}$\,pb in the t-channel, $4.63_{-0.17}^{+0.19}$\,pb in the s-channel, and $15.74_{-1.08}^{+1.06}$\,pb for $Wt$ production, as introduced in \mbox{Chapter \ref{theory_ST}}.

The background contribution from the production of heavy gauge bosons with additional jets was modelled using the leading order \alpgen generator \cite{MAN-0301}, interfaced to \herwig~and \jimmy for the purposes of parton shower and hadronisation simulation. The CTEQ6.1\cite{Stump:2003yu} parton distribution functions and the AUET1 tune were employed for proper ATLAS data matching for both the matrix element evaluations and the parton showering. The production of additional partons was taken into account by generating different subsamples with different final state parton multiplicities, where additional partons can be either light ($u$,$d$,$s$) partons (simulated in $W$+jets and $Z/\gamma$+jets light flavour samples) or heavy quarks (simulated in $W$+$c$+jets, $W$+$c\bar{c}$+jets, $W$+$b\bar{b}$+jets, and $Z$+$b\bar{b}$+jets samples, respectively). Since the inclusive $W$+jets and $Z/\gamma$+jets samples included contributions from both light partons and heavy quarks in the matrix element and parton shower simulation, the created overlap in phase space between the inclusive samples and the heavy quark contribution was taken into account by removing double counted events from the respective samples. The production cross-sections of the used \alpgen samples were normalised to the corresponding approximate next-to-next-to-leading order cross-sections using $k$-factors of 1.20 ($W$+jets) and 1.25 ($Z$+jets), respectively. Furthermore, the relative fractions of the individual $W$+jets heavy quark contributions to the overall $W$+jets sample have been determined in data driven methods\cite{AsymPaper,Aad:2010ey}, and were accounted for by applying corresponding scale factors to the individual samples. They have been found to be $1.63 \pm 0.76$ for $W$+$c\bar{c}$+jets and $W$+$b\bar{b}$+jets, and $1.11 \pm 0.35$ for the $W$+$c$+jets contribution, respectively. The $W$+jets light quark contributions were scaled accordingly to conserve the overall predicted cross-section.

Contributions from diboson ($WW$, $WZ$ and $ZZ$) production and decays was simulated using \herwig~at leading order, and the corresponding production cross-sections were normalised to the next-to-next-to leading order predictions, using $k$-factors of 1.48 ($WW$), 1.60 ($WZ$) and 1.30 ($ZZ$), respectively. Each sample was inclusive and has been filtered to include only events containing at least one lepton (electron or muon) with $p_{\rm{T}} > 10\,\text{GeV}$ and $|\eta| < 2.8$ at parton level. The $k$-factors have been determined such that the unfiltered \herwig~cross sections agree with the next-to-next-to-leading order calculations.

For the evaluation of the systematic uncertainties of the various generators and the simulation of hadronisation and fragmentation, alternative samples have been used for the signal contribution and the $Z$+jets background contribution. The \ttbar~production and decay has been simulated with the \powheg generator \cite{FRI-0701}, and the corresponding parton showering and fragmentation processes have been modelled using both \herwig~and \jimmy (as used for the nominal \mcnlo~sample), and \pythia in order to evaluate systematic effects from the parton showering. Furthermore, additional \mcnlo~signal samples using different top mass hypotheses of 170\,GeV and 180\,GeV have been used in order to quantify systematics arising from the uncertainties on the top mass prediction. Finally, several samples using different strengths of initial state radiation (ISR) and final state radiation (FSR) have been generated at leading order using the \acermc generator \cite{KER-0401}, corresponding to different contributions of ISR and FSR based on observations in data\cite{Aad:2009wy}. An alternative inclusive $Z$+jets background modelling was performed using the \sherpa \cite{GLE-0901} generator and the CTEQ6.6 parton distribution functions. Details on the evaluation of systematic uncertainties can be found in \mbox{Chapter \ref{Systematics}}.

All background samples and the samples used for the evaluation of systematic uncertainties corresponded to an integrated luminosity of about $10-30\,\text{fb}^{-1}$ before analysis specific selection.

\section{Data Driven Estimation of the QCD Multijet Contribution}
\label{Fakes}
The identification of top quark pairs decaying semileptonically relies on the identification of one lepton in the final state, carrying a large transverse momentum. Hence, mis-identified leptons ({\em fake} leptons), which can originate from various sources, pose a non-negligible background to the identification of \ttbar~signal events. Potential sources for mis-identified leptons include
\begin{itemize}
\item semileptonic decays of $b$ quarks into $b$ jets containing leptons with mis-identified isolation properties,
\item long lived weakly decaying particles such as $\pi^{\pm}$ or $K$ mesons,
\item $\pi^{0}$ mesons which are mis-identified as electrons,
\item direct photon conversion and reconstruction of electrons produced in the process.
\end{itemize}
These processes are most dominant in regions where the contribution from real leptons are small, most prominently for the background contribution from QCD multijets.

Despite the fact that object and event selection are designed to ensure the suppression of these backgrounds by requiring stringent criteria, the QCD multijet production cross-section is orders of magnitudes higher than the top quark pair production cross-section. Since the simulation of these backgrounds in Monte Carlo simulations is highly difficult and several of the described contributions are detector dependent, data driven methods are necessary to obtain reliable estimates for the fake lepton background contribution from QCD multijet events.

In order to estimate the contribution from QCD multijet fake muons and electrons, a data driven method, the Matrix Method, was applied to data using different control regions for electrons and muons dominated by QCD multijet processes. The Matrix Method allows to statistically separate two contributions of a data sample based on the impact of a defined selection on the respective subsamples.

\begin{figure}[h!tb]
  \begin{centering}
    \includegraphics[width = 6 cm]{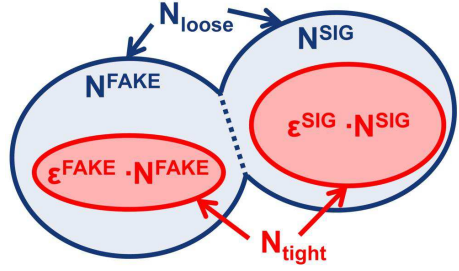}
    \vspace{-0.2 cm}
    \caption[\quad An illustration of the Matrix Method]{An illustration of
    the Matrix Method and the effect of the applied selection on the underlying sample subsets.}
    \label{fig:matrixmethod}
  \end{centering}
\end{figure}

The Matrix Method defines two subsets, \Nl~and \Nt, of the data sample before and after application of a particular requirement applied in the event selection, typically with a large discrimination power between signal (in this case real leptons) and background (in this case fake leptons) contributions. The number of events in the {\em loose} sample, \Nl, is given by the sum of the signal and the background contributions \Ns~and \Nf~in the given sample:
\begin{equation}
  N_{\rm{loose}} = N^{\rm{sig}} + N^{\rm{fake}}.
\end{equation}
After the application of the defined requirement, which is passed by signal events with a probability (or {\em efficiency}) of \epss~and by background events with a probability of \epsf, the number of events \Nt~in the {\em tight} sample follows by taking into account the respective probabilities for the imposed requirement for the signal and background contributions:
\begin{equation}
    N_{\rm{tight}} = \varepsilon^{\rm{sig}} N^{\rm{sig}} + \varepsilon^{\rm{fake}} N^{\rm{fake}}.
\end{equation}
This situation is illustrated in \mbox{Figure \ref{fig:matrixmethod}}.

The linear system of two equations with two unknown variables can be rewritten as a matrix equation:
\begin{equation}
  \left( \begin{array}{c}
    N_{\rm{loose}} \\
    N_{\rm{tight}}
  \end{array} \right)
  =
  \left( \begin{array}{cc} 
    1                   & 1 \\
    \varepsilon^{\rm{sig}} & \varepsilon^{\rm{fake}}                           
  \end{array} \right)
  \left( \begin{array}{c} 
    N^{\rm{sig}} \\
    N^{\rm{fake}}                          
  \end{array} \right)
\end{equation}
Solving the matrix equations yields the signal and background contributions in the data sample prior to the isolation cut:
\begin{eqnarray}
N^{\rm{fake}} & = & \frac{N_{\rm{tight}} - \varepsilon^{\rm{sig}} N_{\rm{loose}}}{\varepsilon^{\rm{fake}} -
      \varepsilon^{\rm{sig}}}, \label{eqn:MM_nfake}\\
      N^{\rm{sig}} & = & \frac{\varepsilon^{\rm{fake}} N_{\rm{loose}} - N_{\rm{tight}}}{\varepsilon^{\rm{fake}} -
      \varepsilon^{\rm{sig}}}.
\end{eqnarray}

In order to determine the background contribution in the signal region (which corresponds to the {\em tight} selection by construction), the fraction of $N^{\rm{fake}}$ in the {\em tight} sample can be calculated using \mbox{Equation \ref{eqn:MM_nfake}} as
\begin{eqnarray}
N_{\rm{tight}}^{\rm{fake}} & = & \varepsilon^{\rm{fake}} N^{\rm{fake}} \notag \\
                           & = & \frac{\varepsilon^{\rm{fake}}}{\varepsilon^{\rm{sig}} - \varepsilon^{\rm{fake}}} \left(\varepsilon^{\rm{sig}} N_{\rm{loose}} - N_{\rm{tight}} \right).
\label{eqn:MM_nfaketight}
\end{eqnarray}
If the selection probabilities \epss~and \epsf~for signal and background, respectively, are sufficiently different, the overall contribution of the QCD multijet background can be used to determine event based weights for the used data sample in order to obtain the distributions of the QCD multijet background contribution in arbitrary variables. This is done by assigning a weight to each data event based on the chosen requirement and the corresponding signal and fake probabilities of the objects taken into account for a given event. If it passes the {\em loose} selection only, the event weight is given by setting $N_{\rm{loose}} = 1$, $N_{\rm{tight}} = 0$ in \mbox{Equation \ref{eqn:MM_nfaketight}}, yielding
\begin{equation}
w_{\text{loose}} = \frac{\varepsilon^{\rm{sig}} \varepsilon^{\rm{fake}}}{\varepsilon^{\rm{sig}} - \varepsilon^{\rm{fake}}}.
\end{equation}
Similarly, if both the {\em loose} and {\em tight} requirements are fulfilled, the event weight is given by setting $N_{\rm{loose}} = 1$, $N_{\rm{tight}} = 1$ in \mbox{Equation \ref{eqn:MM_nfaketight}}:
\begin{equation}
w_{\text{tight}} = \frac{\left( \varepsilon^{\rm{sig}} -1 \right) \varepsilon^{\rm{fake}}}{\varepsilon^{\rm{sig}} - \varepsilon^{\rm{fake}}}.
\end{equation}
This approach allows for a purely data driven estimation of both the normalisation and the shape of the QCD multijet background in semileptonic decays of top quark pairs. The individual parameters that have been used in the estimation for both the muon and electron channels in this analysis will be covered in the following sections. In addition, detailed studies of the performance and stability of the methods used will be demonstrated for the muon+jets channel alongside with an approach to obtain well-defined statistical and systematic uncertainties on the estimate.

\subsection{Muon+jets Channel}
\label{chap:QCDMu}
\begin{figure}[h!tb]
  \begin{centering}
    \includegraphics[width = 10cm]{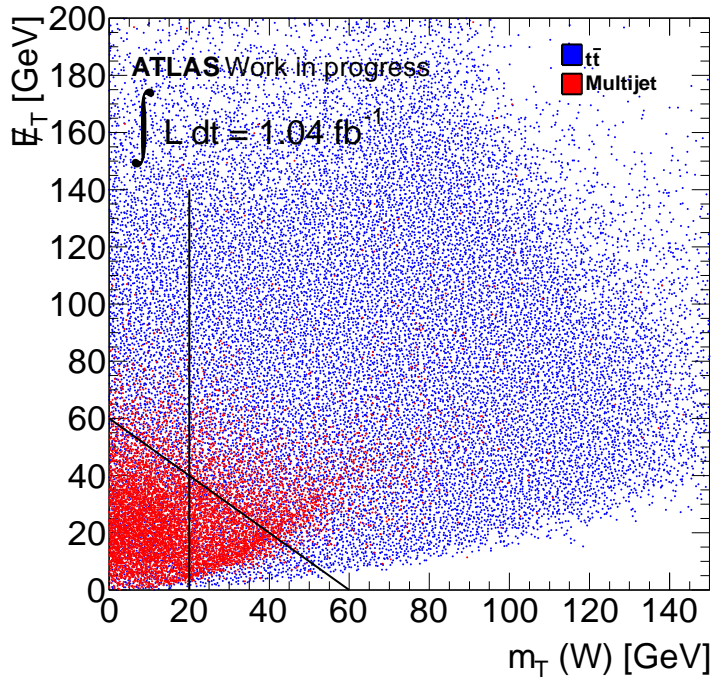}
    \vspace{-0.2 cm}
    \caption[\quad Impact of the QCD multijet suppression requirements]{Impact of the requirement of $m_{\rm{T}}(W) < 20\,\rm{GeV}$ and the triangular cut, \mbox{$\not \hspace{-0.1cm} E_{\rm{T}}$} $+\,m_{\rm{T}}(W) < 60\,\rm{GeV}$, on both the QCD multijet estimate and the \ttbar~signal contribution, taken from Monte Carlo simulations. The imposed cuts are illustrated by the black lines.}
    \label{fig:QCDTriangCut}
  \end{centering}
\end{figure}

In order to estimate the contribution from QCD multijet fake muons, the matrix method was applied to data in the QCD multijet-enriched low-\MTW control region. Furthermore, an inverted triangular cut was imposed to achieve orthogonality to the signal region in the determination of the fake probabilities \epsf~used in the Matrix Method:
\begin{equation*}
 m_{\rm{T}}(W) < 20\,\rm{GeV} \text{ and } \not\!\!E_{\rm{T}} + m_{\rm{T}}(W) < 60\,\rm{GeV}.
\end{equation*}
The impact of the described requirements on the QCD multijet estimate and the simulated \ttbar~signal is shown in \mbox{Figure \ref{fig:QCDTriangCut}}.

The signal probabilities \epss~were determined in the signal region using a \zmumu~Tag \& Probe method in order to select prompt muons from $Z$ decays.

The loose selection was identical to the full selection applied in the signal region (for details on the individual requirements, refer to \mbox{Chapter \ref{Objects}} and \mbox{Chapter \ref{Selection}}), except for the muon isolation. The {\em tight} selection in addition requires isolation criteria based on both momentum and energy depositions around the muon tracks,
\begin{equation*}
  p_{\rm{T,cone30}} < 4.0\,\rm{GeV} \text{ and }  E_{\rm{T,cone30}} < 4.0\,\rm{GeV},
\end{equation*}
corresponding to the full analysis event selection.

The fake probabilities have been obtained separately both with and without explicitly requiring at least one $b$ tagged jet ($\geq 0$ $b$ tags and $\geq 1$ $b$ tags, respectively). In addition, the fake probabilities have been determined requiring at least two $b$ tagged jets, for completeness. For the latter cases, the JetFitCombNN $b$ tagging algorithm with a working point of $w = 0.35$ (corresponding to an overall $b$ tagging efficiency of 70\,\%) has been used in accordance with the signal region event selection. Furthermore, the signal muon contribution from $W$+jets and $Z$+jets in the control region was obtained from Monte Carlo and subtracted to obtain a purer QCD multijet estimation. This contamination was of the order of 1.7\,\% ($\geq 0$ $b$ tags) and 1.8\,\% ($\geq 1$ $b$ tags).

\begin{figure}[h!tb]
  \begin{centering}
    \mbox{
      \subfigure[Signal eff.]{\scalebox{0.3}{\includegraphics[width =
      \textwidth]{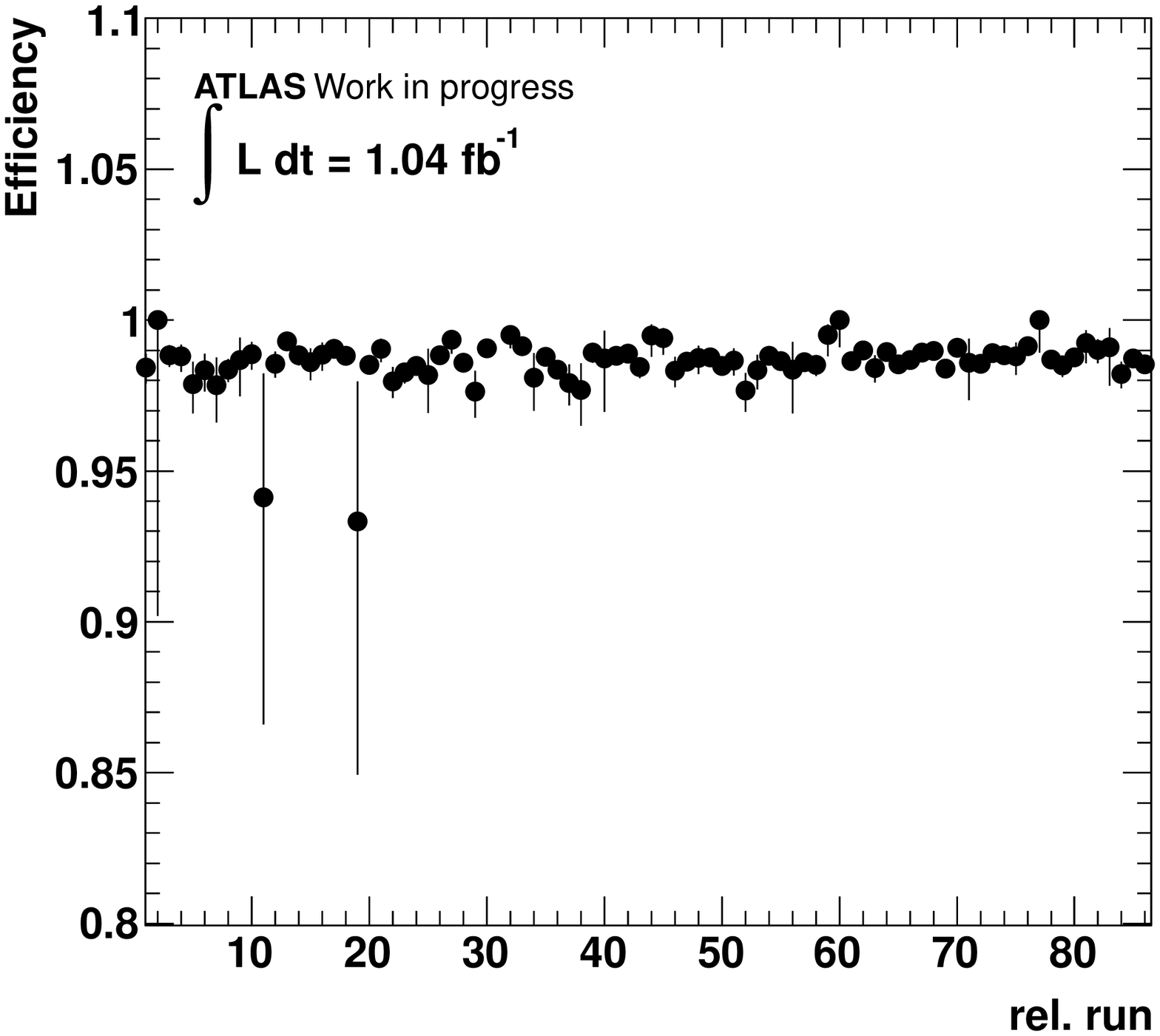}}} \quad
      \subfigure[Fake eff. ($\geq 0$ $b$ tags)]{\scalebox{0.3}{\includegraphics[width =
      \textwidth]{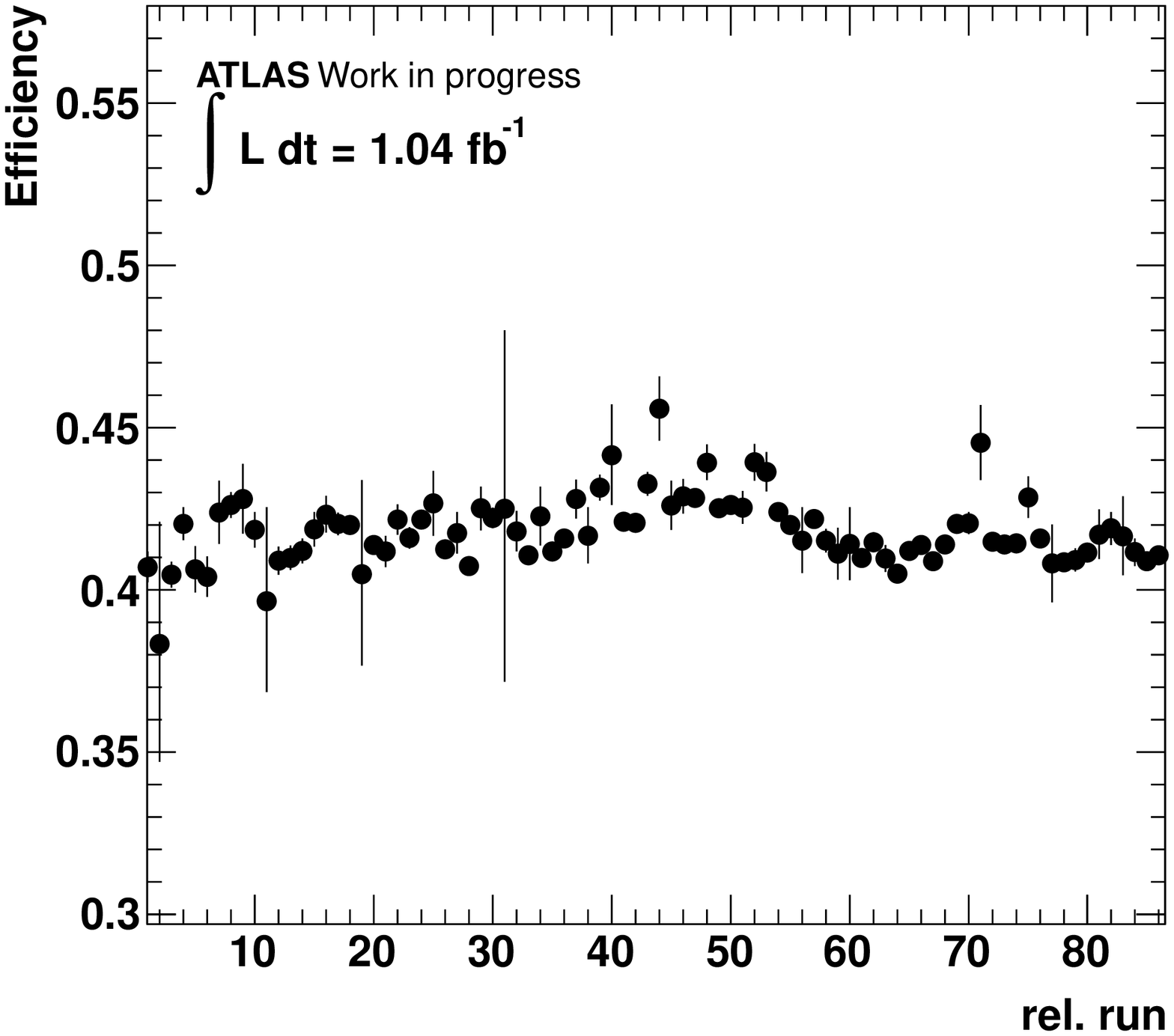}}} \quad
      \subfigure[Fake eff. ($\geq 1$ $b$ tags)]{\scalebox{0.3}{\includegraphics[width =
      \textwidth]{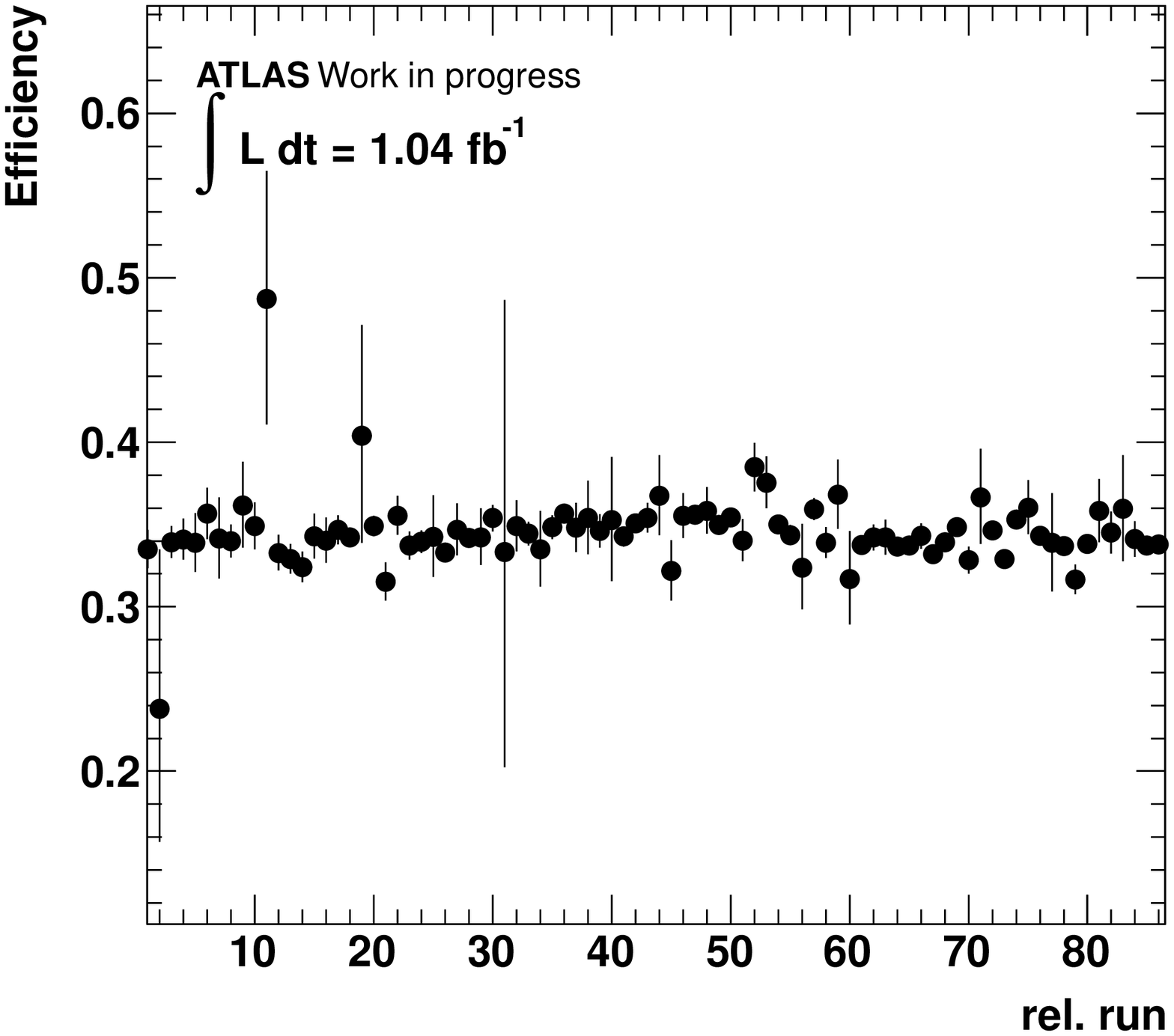}}}
    }
    \caption[\quad Signal and fake probabilities vs. relative run number]{Integrated signal and fake (without and with the requirement of at least one $b$ tagged jet) probabilities, shown as a function of relative run number (full dataset with {\em GoodRunsList} applied). The observed probabilities were stable over the whole data taking period under consideration.}
    \label{fig:eff_vsrun}
  \end{centering}
\end{figure}

In order to verify the stability of the method, the signal and fake (both without and with the requirement of at least one $b$ tagged jet) probabilities are shown as a function of the relative run number (full dataset with {\em GoodRunsList} applied) in \mbox{Figure~\ref{fig:eff_vsrun}}.
As can be seen, neither signal nor fake probabilities showed any significant trend with respect to run number and therefore, to instantaneous luminosity and different pile-up conditions.

\begin{figure}[h!tb]
  \begin{centering}
    \mbox{
      \subfigure[Signal eff.]{\scalebox{0.3}{\includegraphics[width =
      \textwidth]{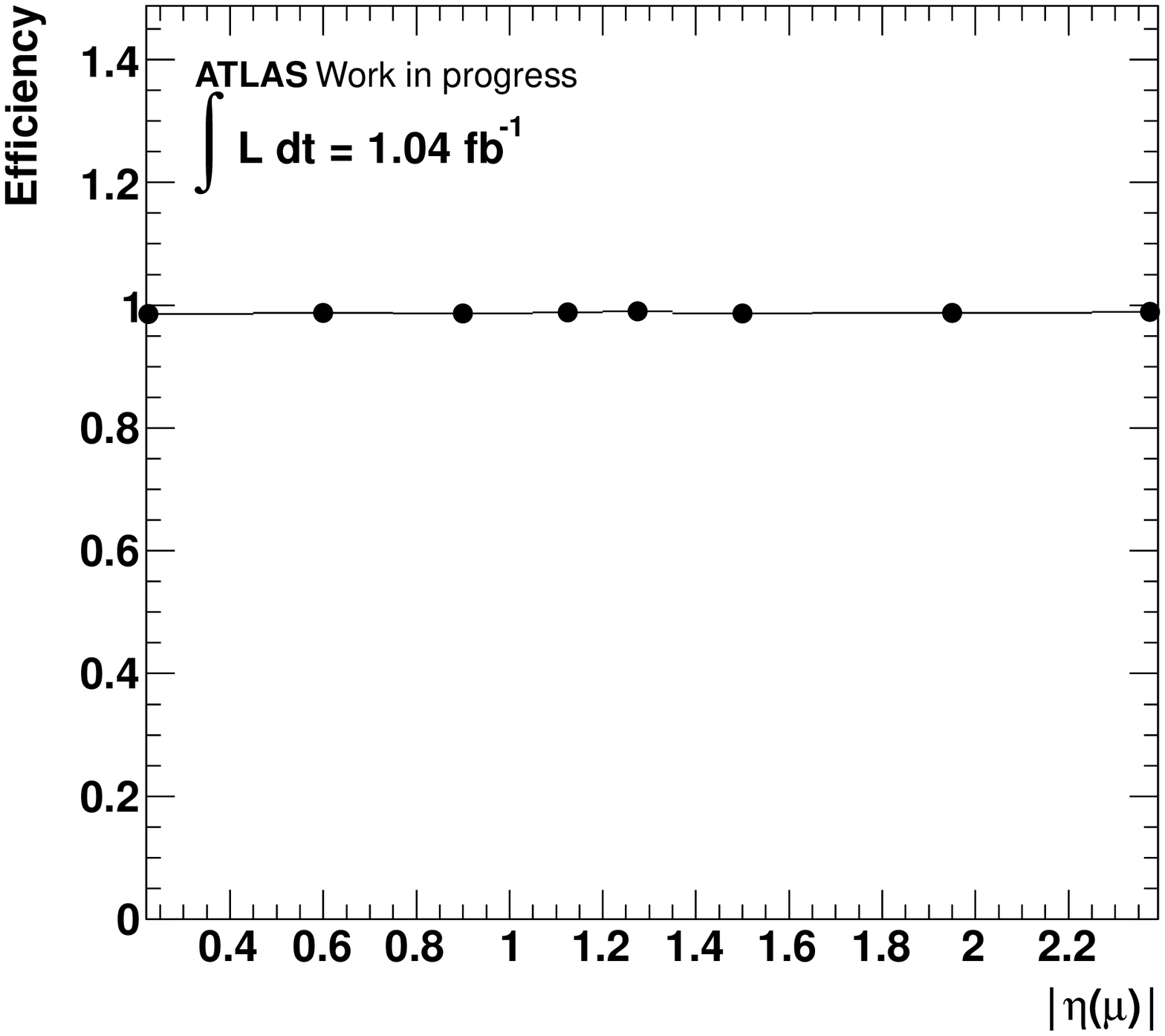}}} \quad
      \subfigure[Fake eff. ($\geq 0$ $b$ tags)]{\scalebox{0.3}{\includegraphics[width =
      \textwidth]{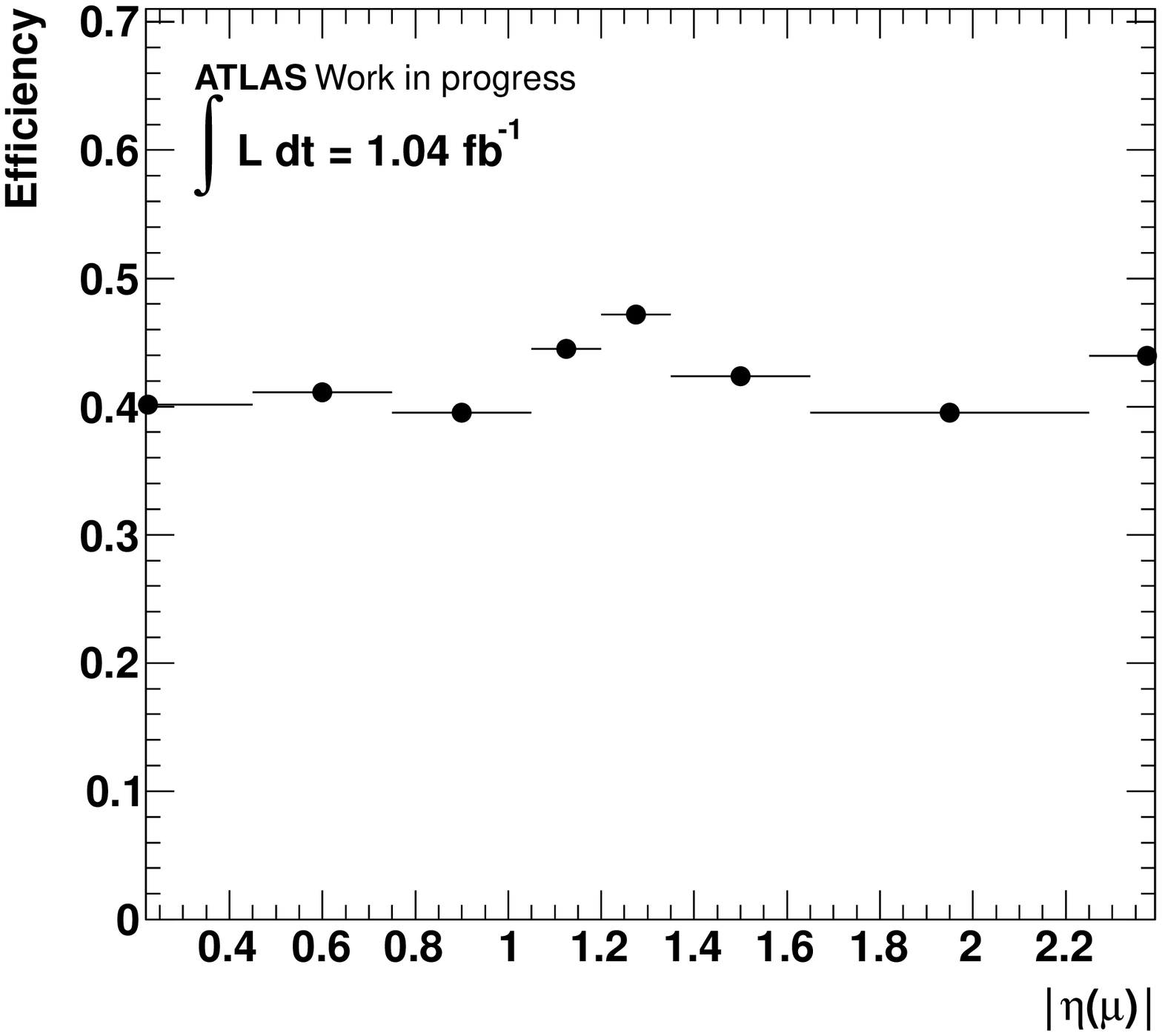}}} \quad
      \subfigure[Fake eff. ($\geq 1$ $b$ tags)]{\scalebox{0.3}{\includegraphics[width =
      \textwidth]{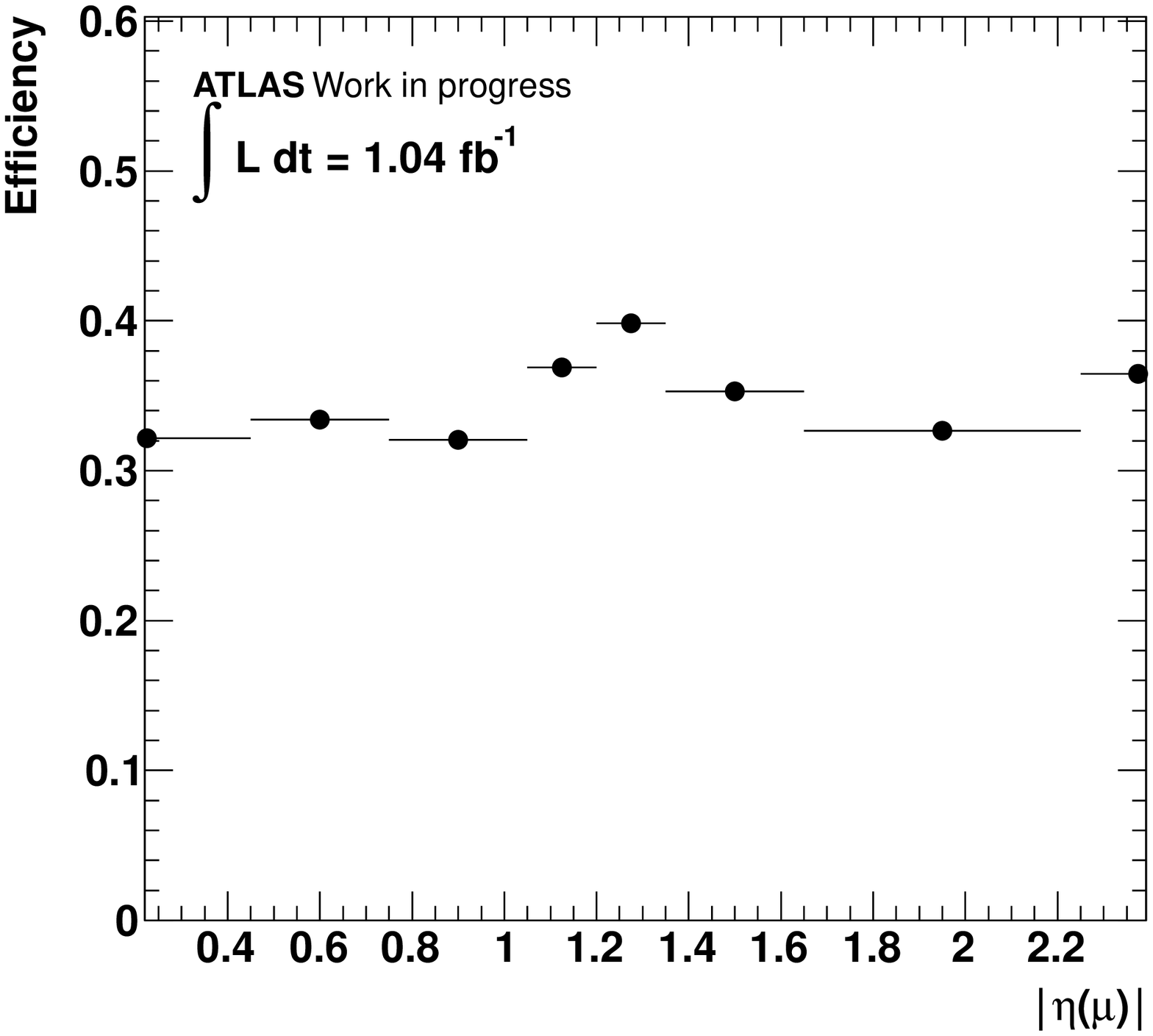}}}
    }
    \mbox{
      \subfigure[Signal eff.]{\scalebox{0.3}{\includegraphics[width =
      \textwidth]{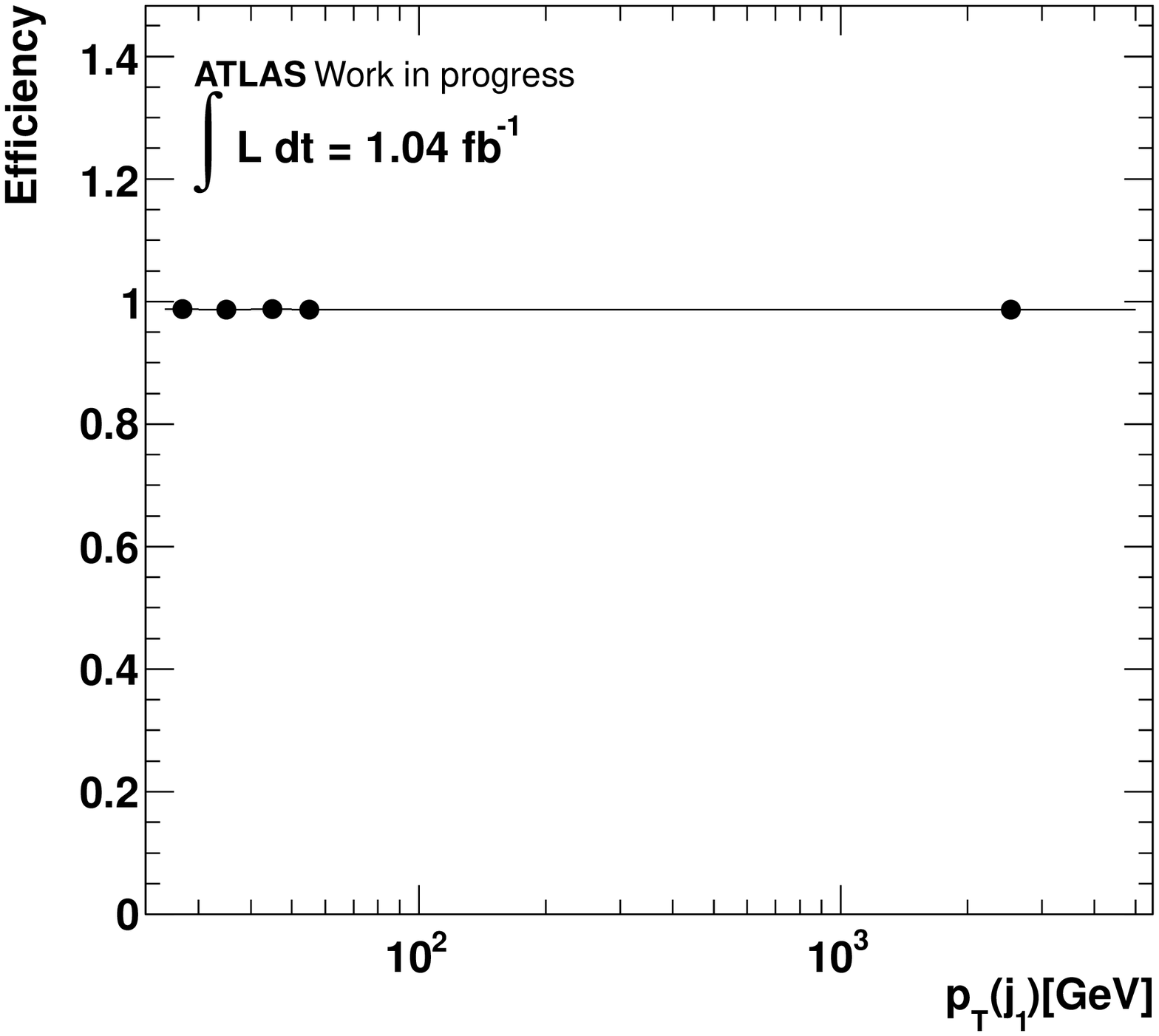}}} \quad
      \subfigure[Fake eff. ($\geq 0$ $b$ tags)]{\scalebox{0.3}{\includegraphics[width =
      \textwidth]{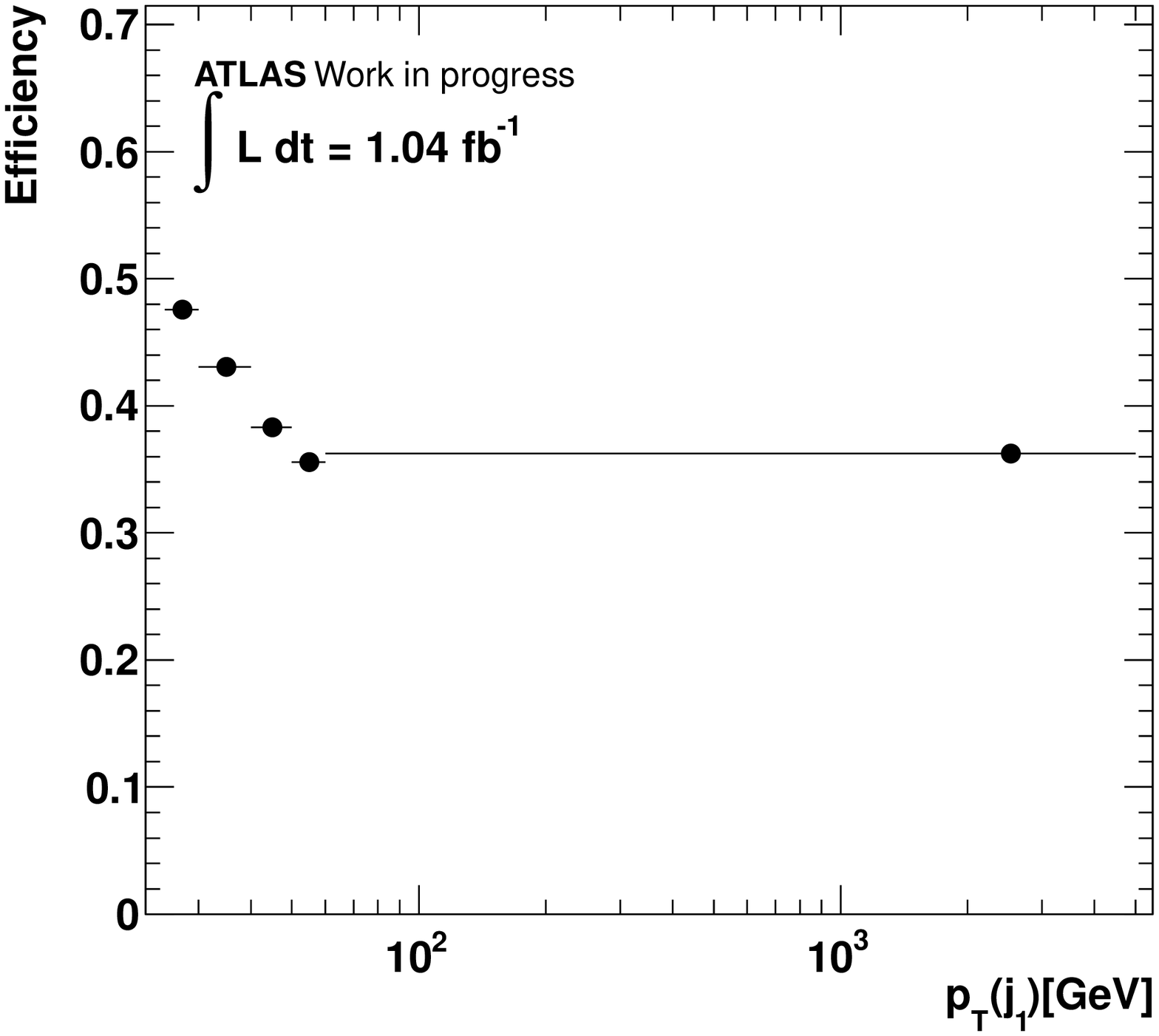}}} \quad
      \subfigure[Fake eff. ($\geq 1$ $b$ tags)]{\scalebox{0.3}{\includegraphics[width =
      \textwidth]{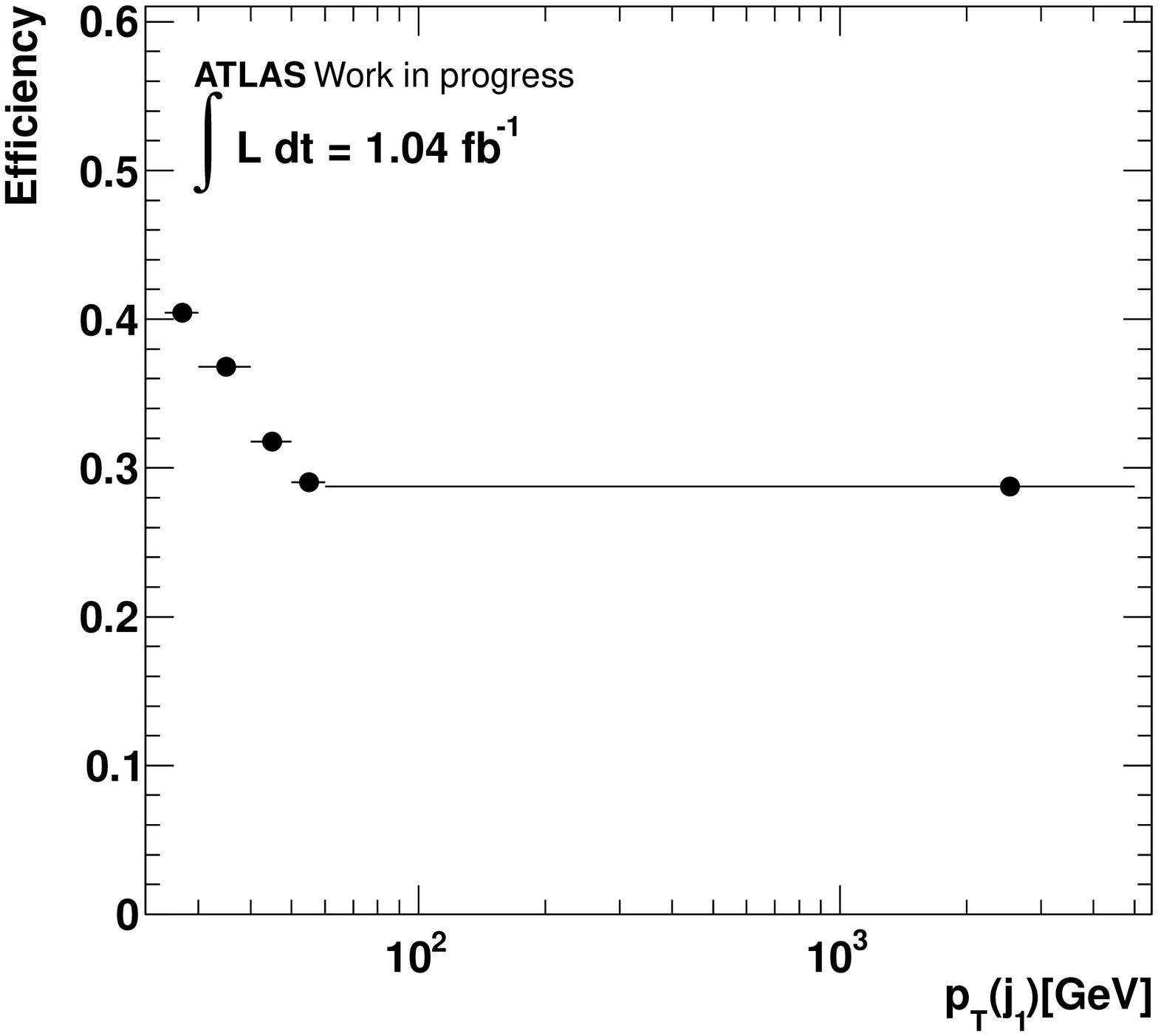}}}
    }
    \caption[\quad Matrix Method probability parametrisation]{Signal and fake probabilities
    as a function of muon $\eta$ and leading jet $p_{\rm{T}}(j_1)$.}
    \label{fig:eff_param}
  \end{centering}
\end{figure}

In order to take into account dependencies on object kinematics, the signal and fake probabilities have been determined as a function of muon pseudorapidity $\eta$ to reflect the dependency on the muon detector acceptance. Furthermore, they have been parametrised as a function of the leading jet transverse momentum $p_{\rm{T}}(j_1)$ in order to take into account the effects of hard jets and hence increased hadronic activity on the muon isolation. The respective projections for muon $\eta$ and $p_{\rm{T}}(j_1)$ can be found in \mbox{Figure \ref{fig:eff_param}}.

The event yields and fractions in the signal region, obtained on a dataset corresponding to $1.04\,\rm{fb}^{-1}$ for different amounts of reconstructed jets required in the event selection ({\em jet bins}) can be found in \mbox{Table~\ref{tab:yields}}.
\begin{table}[htbp]
  \renewcommand{\arraystretch}{1.4}
  \begin{center}
  \begin{tabular}{| l | r | r | r | r | r | r |}
      \hline
      \cline{2-7} \multirow{2}{*}{\centering Jet bin} & \multicolumn{ 2}{c|}{$\geq 0$ $b$ tags} & \multicolumn{ 2}{c|}{$\geq 1$ $b$ tags} & \multicolumn{ 2}{c|}{$\geq 2$ $b$ tags} \\ 
      \cline{2-7}      & Yield & Fraction & Yield & Fraction & Yield & Fraction \\ \hline \hline
        1 jet           &       36820.5 &       5.6\,\%         &       3534.3  &       15.2\,\%        &       0.0     &        \\ \hline
        2 jet           &       14915.9 &       8.9\,\% &       2633.3  &       17.0\,\%        &       106.6   &       9.5\,\% \\ \hline
        3 jet           &       4141.8  &       9.6\,\% &       1003.8  &       11.1\,\%        &       80.5    &       4.0\,\% \\ \hline
        4 jet           &       1109.3  &       8.2\,\% &       348.7   &       6.4\,\% &       34.8    &       1.6\,\% \\ \hline
        4 jet (incl.) \large\phantom{0} \small   &       $1514.9_{-31.9\%}^{+31.8\%}$          &       7.7\,\% &       $519.3_{-36.2\%}^{+36.2\%}$           &       5.6\,\% &       $72.7_{-83.3\%}^{+83.3\%}$            &       1.8\,\% \\ \hline
    \end{tabular}
  \end{center}
  \renewcommand{\arraystretch}{1.0}
  \caption{Event yields and fractions in the signal region for a dataset corresponding to $1.04\,\rm{fb}^{-1}$. Uncertainties correspond to the overall normalisation uncertainty due to statistical uncertainties from the obtained signal and fake probabilities, a second control region, the systematic shift of the $m_{\rm{T}}(W)$ cut and the MC uncertainty on the $W$/$Z$+jets normalisation used in the subtraction of real leptons in the control region. The tagged probabilities correspond to the JetFitCombNN $b$ tagging algorithm with a working point of $w = 0.35$.}
  \label{tab:yields}
\end{table}

The following sources of systematic uncertainty have been taken into account in the estimate and were combined to a single uncertainty on the event weight, which can hence be propagated into a bin-by-bin normalisation and shape uncertainty on an arbitrary variable:
\begin{itemize}
\item \textbf{Statistical uncertainty on the signal and fake probabilities:} Takes into account the uncertainties on both the signal and fake probabilities. The resulting uncertainty on the obtained event weights was evaluated using Gaussian error propagation. Note that the assumption of a symmetric probability distribution function was valid for the signal probabilities as well despite the closeness to unity, owing to the high statistics available from \zmumu~and the resulting small impact on the overall statistical uncertainty.
\item \textbf{Systematic uncertainty from second control region:} A second control region (high $d_0$ significance) was used for the fake probabilities to determine a bin-by-bin systematic discrepancy which was quoted as additional uncertainty on the event weight.
\item \textbf{Systematic uncertainty due to choice of control region cut:} The low transverse $W$ mass control region cut was varied by $5\,$GeV up and down to estimate the impact on the obtained probabilities and event weights.
\item \textbf{Uncertainties on $\boldsymbol{W}$/$\boldsymbol{Z}$+jets Monte Carlo normalisation:} For the 1 jet inclusive bin, the $W$/$Z$+jets Monte Carlo normalisation uncertainties (25\,\%) were used to quantify the effect on the subtraction of real leptons in the control region.
\end{itemize}

The resulting overall normalisation uncertainties are shown in \mbox{Table \ref{tab:yields}} for the signal region ($\geq 4$ jets selection).

\subsection{Electron+jets Channel}
The QCD multijet contribution in the electron+jets channel has been estimated in analogy to the muon+jets channel, following slightly different criteria corresponding to the {\em loose} and {\em tight} selections\cite{AsymPaper}. The signal probabilities used in the Matrix Method were determined in the signal region with a \zee~Tag \& Probe method in order to select prompt electrons from the $Z$ decay.

The fake probabilities have been determined in the QCD multijet-enriched low-\met~control region:
\begin{equation}
  5\,\rm{GeV}~<~\not\!\!E_{\rm{T}}~<~20\,\rm{GeV}. \notag
\end{equation}

Similar to the evaluation in the muon+jets channel, the loose selection was identical to the full selection applied in the signal region, except for the electron isolation criteria, which was modified in the {\em loose} selection:
\begin{eqnarray}
  E_{\rm{T,cone20}} & < & 6.0\,\rm{GeV}, \notag
\end{eqnarray}
while the {\em tight} selection requires
\begin{eqnarray}
  E_{\rm{T,cone20}} & < & 3.5\,\rm{GeV}. \notag
\end{eqnarray}
In addition, slightly less stringent track quality criteria and the corresponding \met~definition were applied to the reconstructed electron candidates in the {\em loose} sample with respect to the {\em tight} sample.

Furthermore, a subtraction of the real lepton contribution in the control region has been performed using the corresponding Monte Carlo predictions, in analogy to the muon+jets channel treatment.

\section{Data Driven Estimation of the $W$+jets Contribution}
\label{chap:WJetsDD}
Due to the fact that the parton density of $u$ quarks in the protons brought to collision in the LHC is on average higher than the parton density of $d$ quarks (which can be observed already at lower momentum transfers, as depicted in \mbox{Figure \ref{fig:PDFs}}), a higher rate of $W^{+}$ than of $W^{-}$ is expected at the LHC. Since the production rates and the asymmetry in the production of $W^{+}$ and $W^{-}$ events has been determined to a higher theoretical precision\cite{Martin:2009iq,Kom:2010mv} than the overall $W$+jets rate at the LHC, the observed $W^{+}/W^{-}$ asymmetry in data can be used to obtain an estimate for the rate of the $W$+jets background contribution in the signal region\cite{AsymPaper}.

Assuming that all processes other than the $W$+jets production are symmetric in the final state lepton charge, the total number of $W$+jets events, $N_W$, can be extracted from the amount of observed data events passing the selection criteria described in \mbox{Chapter \ref{Selection}} (except for the requirement of at least one $b$ tagged jet) with a positively (negatively) charged lepton, given by $D^{+}$ and $D^{-}$, respectively:
\begin{eqnarray}
N_W & = & N_{W^{+}} + N_{W^{-}} \notag \\
    & = & \left( \frac{r_{\text{MC}} + 1}{r_{\text{MC}} - 1} \right) \left( D^{+} - D^{-} \right).
\end{eqnarray}
The fraction
\begin{equation}
r_{\text{MC}} = \frac{\sigma_{pp \rightarrow W^{+}}}{\sigma_{pp \rightarrow W^{-}}}
\end{equation}
has been evaluated on Monte Carlo, based on the same event selection, and has been determined to be $1.56 \pm 0.06$ in the electron+jets channel and $1.65 \pm 0.08$ in the muon+jets channel, respectively. The dominant contributions to the overall uncertainty were due to PDF and jet energy scale uncertainties, and by the uncertainties on the heavy quark contribution fractions (i.e. the relative contributions from $W b\bar{b}$+jets, $W c\bar{c}$+jets and $W c$+jets).

Furthermore, the obtained overall $W$+jets rate $N_W$ has been extrapolated to the full event selection by determining the relative fraction of events passing the requirement of at least one $b$ tagged jet after requiring exactly two reconstructed jets, $f_{2,\geq 1 b \text{ tag}}$, on data, and by determining the ratio $k_{2 \rightarrow \geq 4}$ of the same fraction for the sample where at least four reconstructed jets were required with respect to $f_{2,\geq 1 b \text{ tag}}$, using the $W$+jets Monte Carlo prediction\cite{Aad:2010ey}:
\begin{equation}
N_{W,\geq 1 b \text{ tag}} = N_W \cdot f_{2,\geq 1 b \text{ tag}} \cdot k_{2 \rightarrow \geq 4}.
\end{equation}
The fraction $f_{2,\geq 1 b \text{ tag}}$ has been measured to be $0.063 \pm 0.005$ in the electron+jets channel and $0.068 \pm 0.005$ in the muon+jets channel, including statistical and systematic uncertainties. The extrapolation factor $k_{2 \rightarrow \geq 4}$ has been determined to be $2.52 \pm 0.36$ in the electron+jets channel and $2.35 \pm 0.34$ in the muon+jets channel, respectively. The uncertainties include contributions from the statistical limitation of the used Monte Carlo samples and systematic uncertainties.

  \chapter{Kinematic Event Reconstruction}
\label{Reconstruction}
As described in \mbox{Chapter \ref{Selection}}, the signature of a \ttbar~event in the semileptonic decay channel at leading order is the observation of four reconstructed jets, one isolated lepton and missing transverse energy. A reconstruction of the full \ttbar~final state was performed following a likelihood approach. A probability for the observation of a set of measured quantities under the assumption of a specific model and a set of model parameters was assigned. In this particular case, the model described the \ttbar~decay and the input quantities were the measured energies of the four jets, the measured energy of the lepton, and the missing transverse energy. The fit parameters of the likelihood were the parton energies, the lepton transverse momentum and the three neutrino momentum components. The likelihood was used to assign the measured jets to the decay products of the \ttbar~system. For this study, all permutations with four out of the five leading jets (if exist) were taken into account for the event reconstruction to increase the probability of identifying the proper combination in the presence of additional jets (e.g. from ISR or pile-up). Moreover, the (non-Gaussian) partonic energy resolution (the resolution of the particle jets with respect to the partons) of the final state objects were taken into account through the use of object specific transfer functions in order to obtain the final likelihood:
\begin{eqnarray}   
L & = & \mathcal{B}(\widetilde{E}_{\rm{p,}1}, \widetilde{E}_{\rm{p,}2} | m_W,   \Gamma_W) \cdot \mathcal{B}(\widetilde{E}_{\rm l}, \widetilde{E}_{\nu} | m_W, \Gamma_W) \cdot \nonumber \\ 
           &   & \mathcal{B}(\widetilde{E}_{\rm{p,}1}, \widetilde{E}_{\rm{p,}2},     \widetilde{E}_{\rm{p,}3} | m_t, \Gamma_t) \cdot \mathcal{B}(\widetilde{E}_{\rm l}, \widetilde{E}_{\nu},    \widetilde{E}_{\rm{p,}4} | m_t, \Gamma_t) \cdot \nonumber \\ 
	   &   & \mathcal{W}( \hat{E}_{x}^{miss}| \widetilde{p}_{x, \nu}) \cdot \mathcal{W}(\hat{E}_{y}^{miss} |    \widetilde{p}_{y, \nu}) \cdot \mathcal{W}(\hat{E}_{\rm{lep}} | \widetilde{E}_{\rm{lep}}) \cdot \nonumber \\ 
	   &   & \prod_{\rm{i=1}}^4 \mathcal{W}(\hat{E}_{\rm{jet,}i} | \widetilde{E}_{\rm{p,}i}) \cdot P(\textrm{$b$ tag} ~| ~\textrm{quark}),    
\label{eq:KLF_Likelihood}
\end{eqnarray}
where:
\begin{itemize}  
\item the $\mathcal{B}$s represent the Breit-Wigner parametrisation of the  parton (from which the associated jets originated) energies $\widetilde{E}_{\rm{p,}i}$ and lepton energies $\widetilde{E}_{\rm{lep}}$ with respect to the fitted ones,
\item the $\mathcal{W}$s are the transfer functions associating the measured jets/leptons to the partonic objects, where the mapping functions of the objects are parametrised with a double Gaussian,
\item the $m_W$ and $\Gamma_W$ denote the $W$ boson mass and its decay width. The parameters are fixed to $m_W = 80.4\,\rm{GeV}$ and $\Gamma_W = 2.1\,\rm{GeV}$, respectively,
\item the $m_t$ and $\Gamma_t$ denote the top quark pole mass and its decay width. The parameters were fixed to $m_t = 172.5\,\rm{GeV}$ and $\Gamma_t = 1.5\,\rm{GeV}$, respectively,
\item the $\widetilde{X}$ are the partonic object quantities and $\hat{X}$ their corresponding measured values, 
\item $P(\textrm{$b$ tag} ~| ~\textrm{quark})$ is a $b$ tag probability or rejection efficiency, depending on the quark flavor.
\end{itemize}
The probability $P(\textrm{$b$ tag} ~| ~\textrm{quark})$ was used to take into account the tagging efficiency and rejection rate of the used $b$ tagging algorithms at a specific 
working point. 

\begin{figure}[h!tb]  
\begin{centering}    
\mbox{      \subfigure[muon+jets channel]{\scalebox{0.4}{\includegraphics[width =\textwidth]{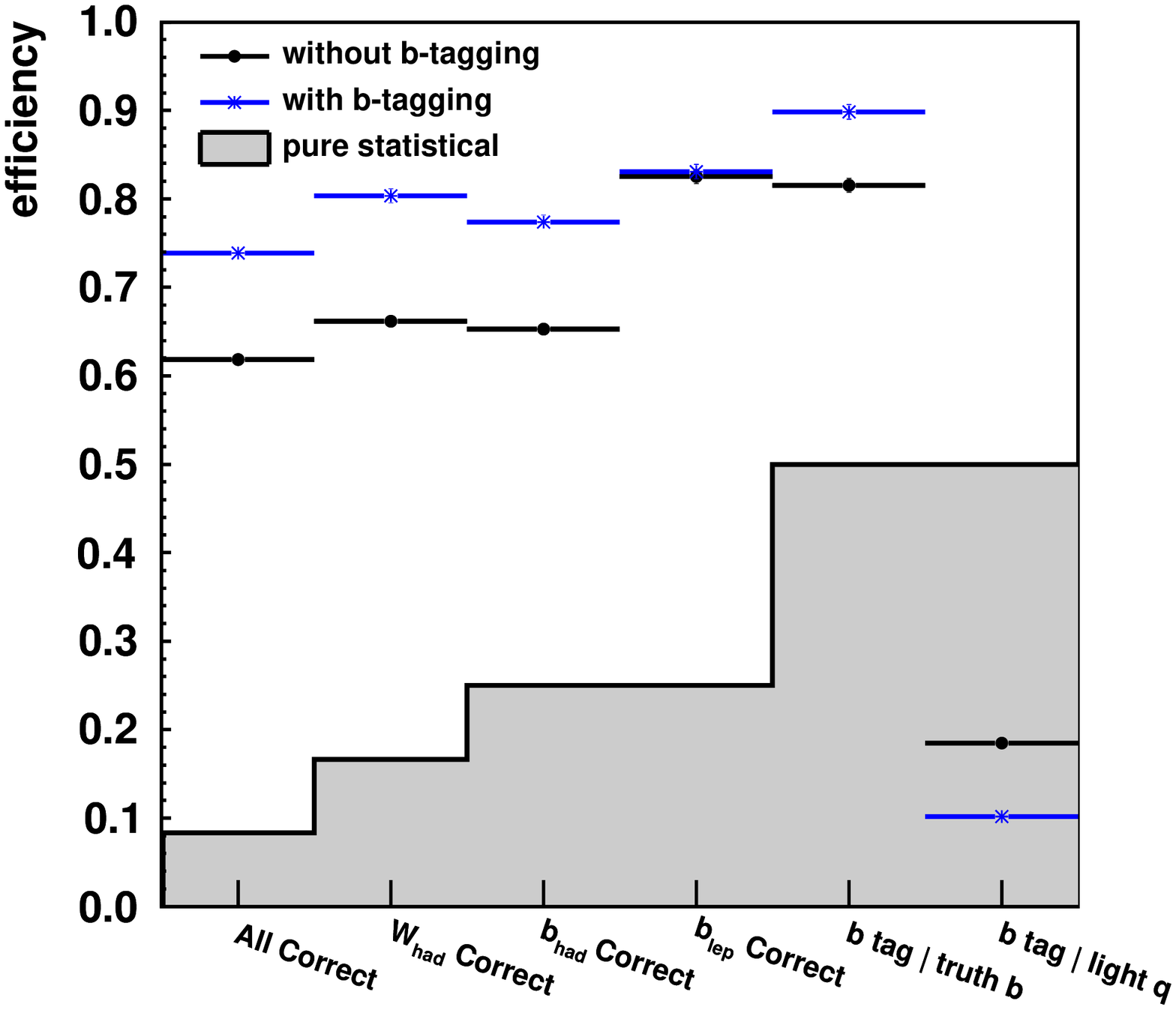}}} \quad      
\subfigure[electron+jets channel]{\scalebox{0.4}{\includegraphics[width = \textwidth]{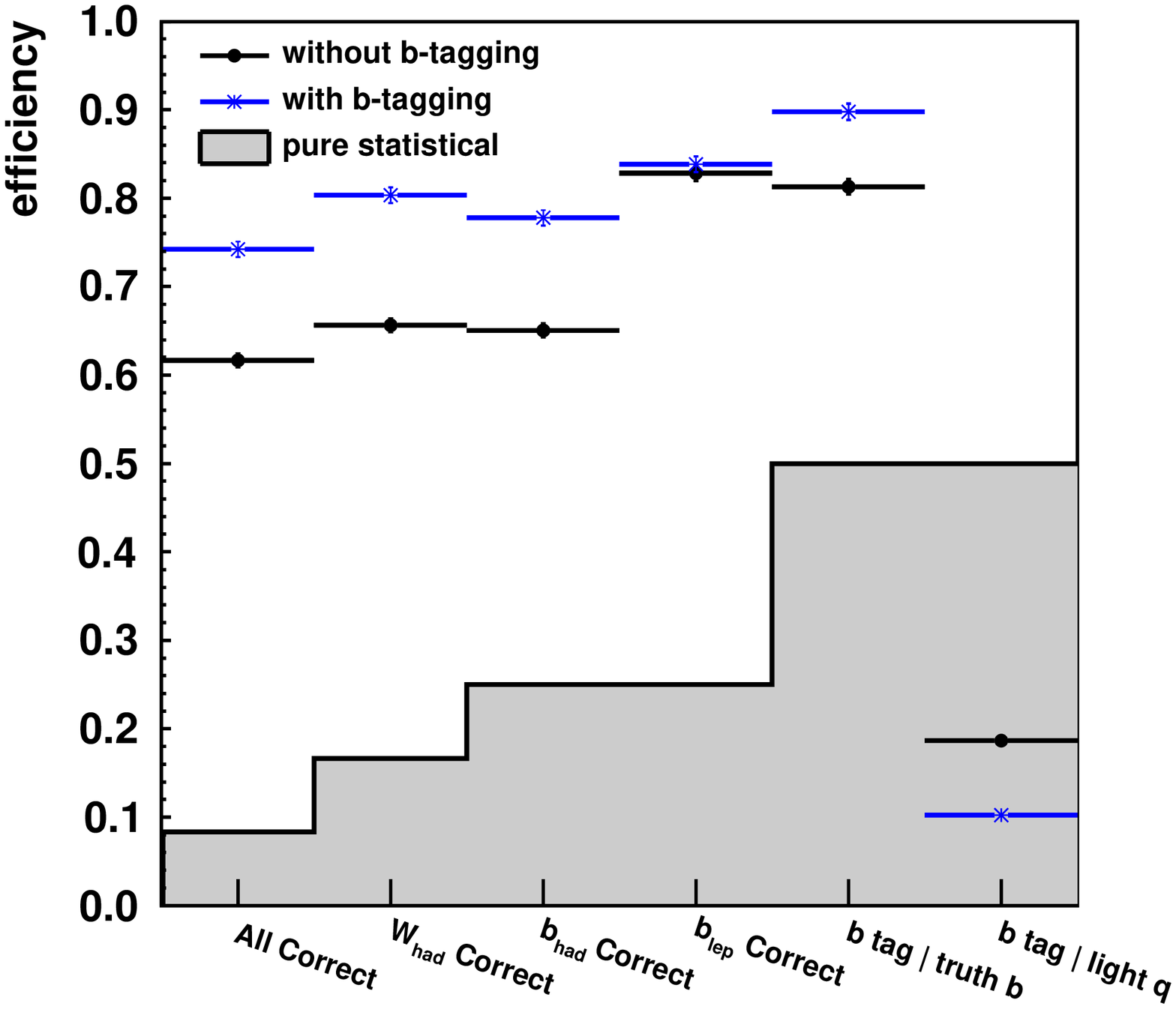}}} }    
\caption[\quad Reconstruction efficiencies in the muon+jets and electron+jets channel]{Reconstruction efficiencies for the muon+jets (left) and 
electron+jets (right) channel. The indicated efficiencies denote the probability of reconstructing the correct (or {\em true}) combination of objects (only matched events taken 
into account). The bars marked \textit{pure statistical} indicate the efficiency which is expected by choosing a combination at random.}   
\label{fig:KLF_Perf}  
\end{centering}
\end{figure}

The most probable event topology hypothesis was chosen by iterating over all possible permutations of reconstructed jets, the lepton and the missing energy and by maximising the logarithmic likelihood, $\log{L}$. The permutation with the highest event probability was used for all further studies. The reconstruction efficiency, obtained in simulations, is shown for both the muon+jets and electron+jets channel in Figure~\ref{fig:KLF_Perf}.

As can be seen, the overall efficiencies for the reconstruction of the correct event topology ({\em All Correct}) in both channels were 62\,\% (74\,\%) with a fixed mass parameter, without (with) $b$ tagging information taken into account. In order to associate the reconstructed objects with the corresponding truth quarks and leptons, a simple $\Delta R$ matching was applied, using cone sizes of $0.3$ for jets and $0.1$ for leptons. An event was considered matched if all truth partons originating from the hard scattering process could successfully be identified with reconstructed jets and the truth lepton was matched to a reconstructed one. For the shown performance evaluation, only events where the four reconstructed jets and the lepton were successfully matched to corresponding truth level objects (contributing positively to the reconstruction efficiency) were taken into account. The matching efficiency on simulated \ttbar~events was about 30\,\% in both channels.

\begin{figure}[h!tb]  
\begin{centering}    
\includegraphics[width = 12cm]{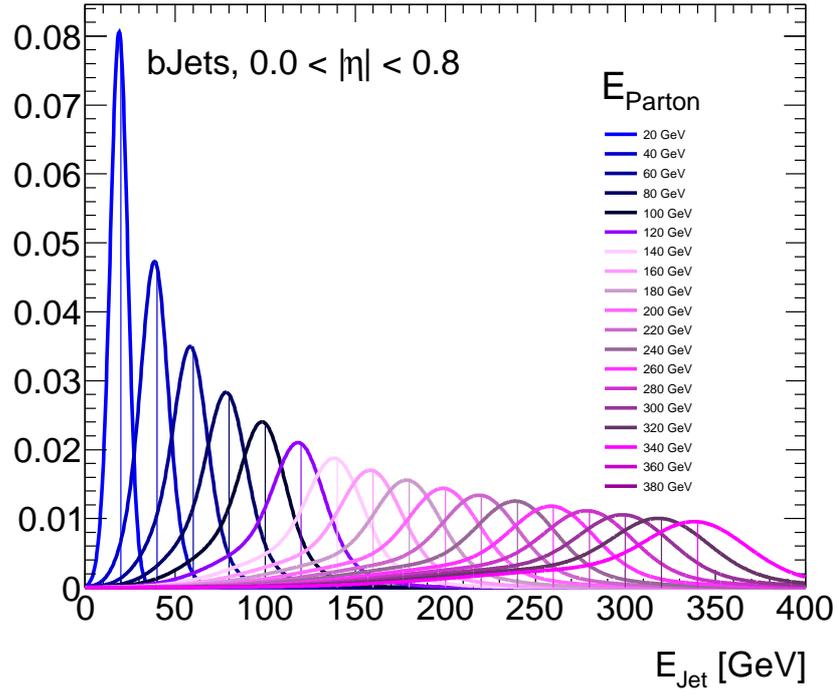} 
\caption[\quad Transfer functions for $b$ jets]{The transfer functions mapping the measured $b$ jets to the corresponding partonic objects, in the range $|\eta|<0.8$, for different energies.}   
\label{fig:TF_KLFa}  
\end{centering}
\end{figure}

\begin{figure}[h!tb]  
\begin{centering}    
\includegraphics[width = 12cm]{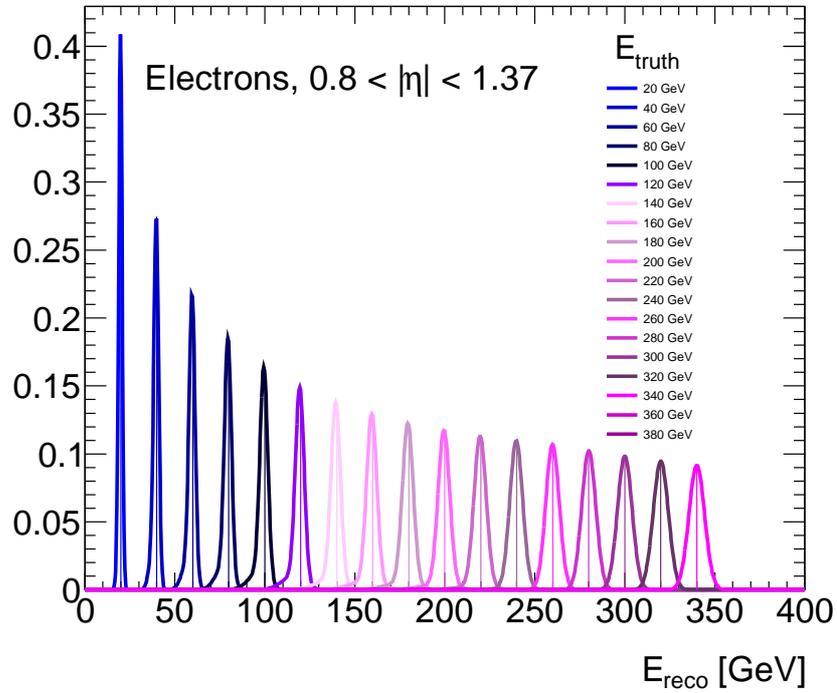} 
\caption[\quad Transfer functions for electrons]{The transfer functions mapping the measured electrons to the corresponding partonic objects, in the range $0.8<|\eta|<1.37$, for different energies.}   
\label{fig:TF_KLFb}  
\end{centering}
\end{figure}

Examples for the transfer functions used in the likelihood can be found in \mbox{Figure~\ref{fig:TF_KLFa}} and \mbox{Figure~\ref{fig:TF_KLFb}}, where the fit functions in different energy regions for $b$ jets in the pseudorapidity range $|\eta|<0.8$ and for electrons in the pseudorapidity range $0.8<|\eta|<1.37$ are shown.

A double Gaussian function was used in the fit of the transfer functions:
\begin{eqnarray}W(E_{\mathrm{true}}, E_{\mathrm{reco}}) = \frac{1}{2 \pi (p_2 + p_3 p_5)} (e^{- \frac{(\Delta E - p_1)^2}{2 p_2^2} } + p_3 e^{- \frac{(\Delta E - p_4)^2}{2p_5^2} } ) ,
\end{eqnarray}
where the parameters $p_1$, $p_2$, $p_3$, $p_4$ and $p_5$ are functions of the true energy of the respective particle and $\Delta E = E_{\mathrm{true}} - E_{\mathrm{reco}}$. 

  \chapter{Unfolding}
\label{Unfolding}
Any measured observable is influenced by imperfections of the used measurement apparatus and procedure itself, such as limited resolution of the detector response, the detector acceptance and possible object and event selections which are applied to the data. Due to these distortions, any measurement of such observable does not fully represent the original (or {\em true}) quantity. Mathematically, the actual measurement can be considered to be a convolution of the true quantity with a function representing the overall detector and selection acceptance and the detector response. 

Let the true quantity be represented by a vector $\smash{\vec{t}}$ (with entries $t_j$ and $j = 1,2,...,n_t$) describing the bin contents of a histogram, and the measured or reconstructed distribution by a corresponding vector $\smash{\vec{k}}$ (with entries $k_i$ and $i = 1,2,...,n_k$), respectively. The underlying detector resolution effects can be described by a transition or {\em response matrix} $R^{\text{res}}$, which contains the individual transition probabilities and hence the migrations between the observed elements of the distribution and the corresponding true values.

\begin{figure}[h!tb]
  \begin{centering}
    \includegraphics[width = 14 cm]{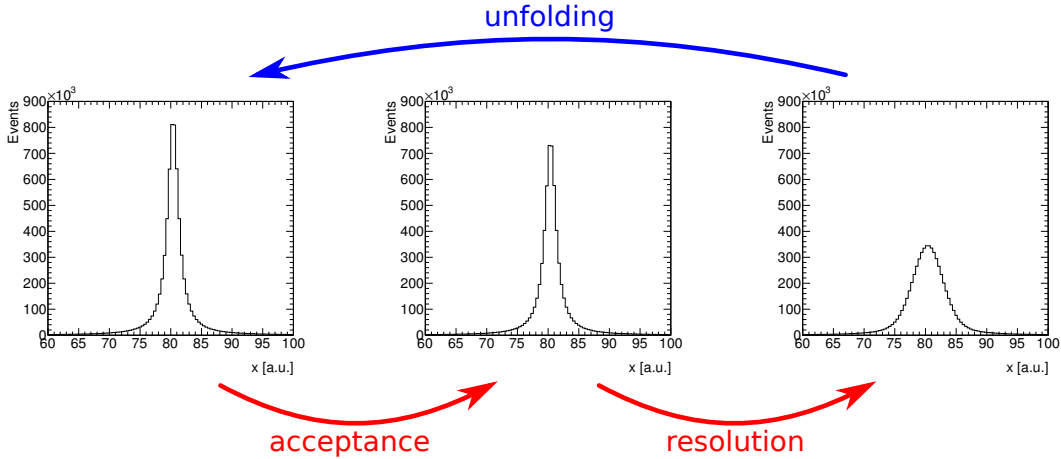}
    \vspace{-0.2 cm}
    \caption[\quad Schematic overview of the measurement and unfolding process]{Schematic overview of the measurement and unfolding process. The true distribution of an arbitrary variable $x$ (left) is affected by acceptance effects (centre) and resolution effects (right) in the measurement process. The unfolding procedure attempts to reverse these effects to obtain the most probable true distribution corresponding to the given measured distribution.}
    \label{fig:unfSchema}
  \end{centering}
\end{figure}

Furthermore, the detector acceptance and applied selection can be quantified by an additional weight factor for each element of $R^{\text{res}}$, taking into account the probability of events from a particular entry of $\smash{\vec{t}}$ being observed at all in the measurement process. This additional correction, which can be described by a second matrix $R^{\text{acc}}$, together with the response matrix describing the resolution effects, yields the overall response matrix $R$:
\begin{equation}
R = R^{\text{acc}} R^{\text{res}},
\end{equation}
denoting the transition probabilities between the observed distribution and the true distribution:
\begin{equation}
\vec{k} = R \vec{t},
\label{eqn:unfeq}
\end{equation}
where
\begin{eqnarray}
R_{ij} & = & P(\text{observed in bin }i~|~\text{expected in bin }j) \\
       & = & P(k_i|t_j).
\end{eqnarray}
The concept of unfolding is illustrated in \mbox{Figure \ref{fig:unfSchema}}, where an example distribution for an arbitrary variable $x$ is shown at the different stages in the measurement process.

In order to find an estimator for the true distribution given the measured distribution, an unfolding method\cite{Cowan:2002Durham} can be applied to correct for the respective acceptance and resolution effects. In this process, the response matrix is derived from an arbitrary reference sample, typically using Monte Carlo simulations. This procedure is called the {\em training} step of the unfolding. The obtained response matrix has to be inverted in order to allow the unfolding of any measured distribution to its corresponding true distribution. Since the response matrix represents the full resolution and acceptance information of the underlying measurement, the unfolding procedure can be performed model-independently, assuming that the detector simulation used in the training sample is sufficiently accurate.

However, in most situations where unfolding is applied, an exact and unique inverse response matrix $R^{-1}$ does not necessarily have to exist, such that
\begin{equation}
  R R^{-1} = I,
\end{equation}
where $I$ is the unity matrix. Hence, approximations are needed to perform the above matrix inversion to acceptable accuracy.

Limited statistics in the reference sample and the resulting statistical fluctuations can lead to additional and inadvertent bin migration effects in the response matrix, which do not represent the underlying resolution and acceptance effects. Consequently, these contributions have to be suppressed in the matrix inversion process, achieved by applying a regularisation procedure in order to limit the propagation of statistical fluctuations into the unfolded distribution or quantity. This regularisation typically involves a cut-off or weight parameter representing the sensitivity of the unfolding approach with respect to short-ranged bin-by-bin changes. Hence, the regularisation can be regarded as a constraint on the smoothness of the response matrix and hence the unfolded distribution.

The obtained approximate inverse matrix $R^{-1}$ is applied to the distribution measured from data, $\smash{\vec{m}}$, and the respective unfolded distribution, $\smash{\vec{u}}$, is obtained as estimator for the true distribution based on the given measurement:
\begin{equation}
\vec{u} = R^{-1} \vec{m}.
\end{equation}

Several procedures have been developed to perform the inversion of the response matrix and the necessary regularisation. These unfolding methods will be briefly explained and evaluated with respect to their value and applicability for the measurement of the charge asymmetry using the observable $A_C$ in the following.
\begin{itemize}
\item \textbf{Bin-by-bin correction} - Neglecting bin migrations, a simple reweighting can be performed by defining a correction factor for each bin of the distribution measured on data based on the true and measured reference distributions\cite{Choudalakis:2011rr}:
\begin{equation}
c_i = \frac{t_i}{k_i}.
\end{equation}
Consequently, the unfolding is performed by application of the respective correction factors to the corresponding bins of the distribution $\smash{\vec{m}}$ obtained from the data measurement to obtain the unfolded distribution $\smash{\vec{u}}$:
\begin{equation}
u_i = c_i \cdot m_i.
\end{equation}
Despite the simplicity of this method, it is prone to biases due to the reference Monte Carlo used in the training step, in particular the shape of the respective distributions. Since a priori no bin migrations are taken into account, it relies strongly on the correct description of the underlying physics and hence does not provide a model-independent approach. Furthermore, ambiguities in the determination of the statistical uncertainties may arise for cases where $k_i > t_i$, where the obtained relative uncertainty can be underestimated, being smaller than the expected uncertainty for an ideal detector (i.e. where $m_i$ would itself be an estimator of $t_i$).

\item \textbf{Singular value decomposition} - This method employs a singular value decomposition (SVD) approach\cite{Hocker:1995kb} in order to perform the inversion of the response matrix in the unfolding procedure. Singular value decomposition of a given real $m \times n$ matrix $R$ involves a factorisation such that
\begin{equation}
R = U S V^T,
\end{equation}
where $U$ and $V$ denote $m \times m$ and $n \times n$ orthogonal matrices, respectively, such that $U U^T = U^T U = I_m$ and $V V^T = V^T V = I_n$ (with $I_m$ and $I_n$ being the corresponding $m \times m$ and $n \times n$ unity matrices). Furthermore, $S$ denotes an $m \times n$ diagonal matrix with non-negative diagonal elements, i.e. $S_{ij} = 0$ for $i \neq j$ and $S_{ii} = s_i \geq 0$. The matrix entries $s_i$ are called singular values of the matrix $R$.

This form can be used in order to decompose the given measured distribution $\smash{\vec{k}}$ and the unknown true distribution $\smash{\vec{t}}$ into a series of orthogonal and normalised functions of the respective $m$ and $n$ classes by performing an appropriate rotation of the respective vectors. Consequently, the given initial system of linear equations as shown in \mbox{Equation \ref{eqn:unfeq}} is reduced to a diagonal system of equations, such that
\begin{equation}
U^T \vec{k} = S V^T \vec{t} \quad\quad \Leftrightarrow \quad\quad \vec{d} = S \vec{z},
\end{equation}
where $\smash{\vec{d}}$ and $\smash{\vec{z}}$ denote the rotated measured and true vectors, respectively. This procedure is in particular effective if the respective singular values $s_i$ are small and the statistical uncertainties on the entries of the measured distribution are large, since in such a case any exact inversion solution is dominated by statistical fluctuations and thus physically meaningless.

By transformation of the given system of linear equations into the form of a weighted least squares minimisation, taking into account measurement uncertainties, a regularisation of the unfolding procedure can be achieved by the addition of a corresponding regularisation or stabilisation term to the expression to be minimised\cite{Blobel:2002pu,Michael1994400,lawson1995solving}. This introduces prior knowledge of the given problem and involves the requirement of minimal curvature of the obtained unfolded solution (i.e. the smoothness of the resulting distribution), eliminating statistical bin-to-bin fluctuations similar to the suppression of high-frequency harmonics in Fourier analysis. A regularisation parameter $\tau$ defines the relative weight of the additional regularisation term with respect to the terms originating from the given system of linear equation in the minimisation:
\begin{equation}
\left( R \vec{t} - \vec{k} \right)^T \left( R \vec{t} - \vec{k} \right) + \tau \left( C \vec{t} \right)^T C \vec{t},
\end{equation}
where $C$ denotes a matrix representing the prior condition on the solution.

\item \textbf{Bayesian iterative unfolding} - A procedure based on an iterative approach to perform the inversion of the response matrix following Bayes' theorem is applied\cite{DAgostini:1994zf}. This approach allows incorporating new knowledge to update a prior probability of observation of a given event\cite{bayes1,bayes2,bayes3} in an iterative procedure.

In order to obtain the inverted response matrix, the posterior probability of obtaining the true distribution $\vec{t}$ given the measured distribution $\vec{k}$ is calculated accordingly, assuming prior knowledge $P_{0}(t_j)$ for the individual components of $\vec{t}$ based on the true distribution obtained in the Monte Carlo training step of the unfolding: 
\begin{equation}
P(t_j|k_i) = \frac{P(k_i|t_j)P_{0}(t_j)}{\sum_{l=1}^{n_t}P(k_i|t_l)P_{0}(t_l)}.
\end{equation}
Note that in this context, the probability $P(k_i|t_j)$ is identical to the transition probability $R_{ij}$ contained in the response matrix.

The obtained posterior probability distribution function is used as a prior in the next iteration, consecutively updating the existing knowledge about the respective probabilities with increasing number of iterations:
\begin{eqnarray}
P_{1}(t_j) & \propto & \sum_{i} P(t_j|k_i) \cdot k_i  \propto \sum_{i} P(k_i|t_j)\cdot P_{0}(t_j) \cdot k_i \\
P_{2}(t_j) & \propto & \sum_{i} P(t_j|k_i) \cdot k_i  \propto \sum_{i} P(k_i|t_j)\cdot P_{1}(t_j) \cdot k_i \\
P_{3}(t_j) & \propto & \sum_{i} P(t_j|k_i) \cdot k_i  \propto \sum_{i} P(k_i|t_j)\cdot P_{2}(t_j) \cdot k_i \\
& \vdots & \notag
\end{eqnarray}

Regularisation of the Bayes iterative unfolding procedure can be achieved naturally by restricting the number of iterations $N_{\text{It}}$ such that the underlying true distribution is recovered within the statistical uncertainties, and bin-to-bin fluctuations which are of purely statistical nature are suppressed. For a large number of iterations, a convergent state is reached, yielding the true, but strongly fluctuating inverse of the response matrix (thus minimising any remaining systematic bias of the unfolded distribution at the cost of larger statistical uncertainties). The number of iterations necessary to reach convergence depends on different conditions, including the choice of binning, the strength of bin migrations in the response matrix (i.e. the magnitude of its off-diagonal elements), and the choice of prior. The optimal choice of $N_{\text{It}}$ is case dependent and must be determined following a well-defined procedure, balancing remaining bias and statistical uncertainty of the obtained result.
\end{itemize}

Due to the fact that the bin-by-bin unfolding cannot account for bin migrations and does not allow for a regularised unfolding procedure, and hence is expected to be heavily model-dependent, this procedure was not eligible for the usage in this analysis. It only allows for a comparison of the measured asymmetry with the Standard Model prediction (unless the unfolding is performed for different model hypotheses), which can be achieved using the measured asymmetry directly as well.

The SVD unfolding approach, despite providing a well defined methodology, cannot perform an unfolding in more than one parameter in its current technical implementations, which would limit the analysis to an inclusive unfolding in a single parameter.

Given the requirements of the measurement, a Bayesian iterative unfolding was performed in this analysis in order to recover the inclusive $|y_t| - |y_{\bar{t}}|$ distribution and the resulting charge asymmetry observable $A_C$ at the parton level. In particular, this approach allowed for a simultaneous unfolding in multiple observables due to the fact that Bayesian unfolding is independent of the ordering of the classes/bins. Since many of the BSM models summarised in \mbox{Chapter \ref{Theory}} predict different magnitudes of the charge asymmetry for low and high $M_{t\bar{t}}$ regions, a simultaneous unfolding of $|y_t| - |y_{\bar{t}}|$ and the invariant \ttbar~mass, $M_{t\bar{t}}$, has been performed, taking into account bin migrations in both dimensions. Semileptonic \ttbar~events generated with the \mcnlo~generator have been used as reference sample to obtain the response matrix based on the detector simulation. Furthermore, a simple extraction of the covariance matrix of the unfolded distributions\cite{Adye:2011p620} was possible.

The SVD unfolding was performed for the inclusive measurement as a cross-check to verify the stability and consistency of the Bayes iterative approach (c.f. \mbox{Appendix \ref{App:SVDUnfolding}}).

The techniques utilised were available in the {\em RooUnfold} package\cite{Adye:2011p620}, which provided simple interfaces and efficient implementations for all three mentioned unfolding methods.

  \chapter{Systematic Uncertainties}
\label{Systematics}
In addition to the statistical uncertainty originating from limited data statistics, there were a multitude of systematic effects that can have an impact on the performed measurement. These effects were studied individually and a corresponding systematic uncertainty on the measurement result was assigned for each contribution.

In order to evaluate the individual effects, the analysis was performed for each systematic under consideration with a modified response matrix and/or background contribution depending on the modelled effect. The changes typically corresponded to an uncertainty of one or more parameters (e.g. a shift of the muon trigger efficiencies according to the respective uncertainties) or an alternate model (e.g. a different MC generator used for the simulation of \ttbar~events). The uncertainty was extracted in each case based on the shift in the measurement central value when changing the parameters accordingly, and was symmetrised.

In order to suppress the statistical component of the obtained uncertainty inherent in the re-evaluation of the central value by changing different parameters of the performed measurement, the requirement of at least one $b$ tagged jet in the event selection was replaced by a reweighting method. The $|y_t| - |y_{\bar{t}}|$ distribution for the Monte Carlo based background contributions ($W$+jets, $Z$+jets, single top and diboson background) was obtained by direct application of the $b$ tag weights to the events passing the nominal event selection without the requirement of at least one $b$ tagged jet. This approach is very similar to the trigger reweighting approach described in \mbox{Chapter \ref{chap:TrigReweighting}}. The same procedure was applied for the simultaneous unfolding in $|y_t| - |y_{\bar{t}}|$ and $M_{t \bar{t}}$. Furthermore, the normalisation of the resulting distributions was modified to match the distribution obtained using the nominal event selection to avoid the introduction of a potential bias in the background subtraction. Control plots showing the agreement of the two approaches can be found in \mbox{Appendix \ref{AppBTagW}}. Since the weighted distributions were compatible within statistical uncertainties with the nominal distributions, no additional systematic uncertainty was assigned. The bin-by-bin statistical uncertainty in the background distribution is reduced by up to 25\,\%.

The following systematics were considered and evaluated for this analysis. All contributions were assigned to the Monte Carlo prediction (signal and/or background, were applicable).
\begin{itemize}  
\item \textbf{QCD multijet uncertainty} - The QCD multijet background was estimated with data driven methods in both the electron and muon channel. Since both the shape and the overall normalisation of the estimation can only be verified to a limited extend, a very conservative systematic uncertainty of 100\,\% was quoted despite the availability of more advanced estimates (c.f. \mbox{Chapter \ref{chap:QCDMu}}), following the recommendations of the performance groups. Results following a less conservative approach are discussed in \mbox{Appendix \ref{App:AltApproach}}. The QCD multijet background normalisation was shifted up and down by 100\% to quantify this uncertainty. Since the QCD multijet background contribution is intrinsically charge symmetric, but enters the background subtraction, the normalisation and not shape was the dominant source of systematic uncertainty.

\item \textbf{Jet energy scale} - The jet energy scale (JES) uncertainty was derived using information from test beam data, LHC collision data and simulation and was taken into account by scaling up and down the energy of all considered jets by \mbox{$1\,\sigma$} of the associated transverse momentum uncertainty based on the jet \pt~and $\eta$\cite{Aad:2011he}. The full event selection and kinematic reconstruction has been re-run with the scaled jets. In addition, the missing transverse energy has been re-evaluated, taking into account the scaled contributions of the jets in $p_{x}$ and $p_{y}$. For jets within the acceptance range, the JES uncertainty varied from about 2.5\,\% for high \pt~jets in the central detector region to about 14\,\% for low \pt~jets in the forward region.

\item \textbf{Pile-up (JES)} - Depending on the instantaneous luminosity and the amount of vertices, an event weight was assigned to match the pile-up contribution in the simulation to data. An additional systematic uncertainty of 5\% (7\%) for  low $p_{\rm{T}}$ jets or 2\% (3\%) for high $p_{\rm{T}}$ jets in the $|\eta|<2.1$ ($2.1<|\eta|<4.5$) region was added to the JES uncertainty in quadrature\cite{ATLAS-CONF-2011-030}.

\item \textbf{$\boldsymbol{b}$ jet energy scale} - In order to account for the difference of the energy scale for $b$ jets with respect to light quark jets, all $b$ jets (i.e.~jets which have a matched truth $b$ quark in simulations) were scaled by an additional fraction ranging from 2.5\,\% in the low $p_{\rm{T}}$ jet region to 0.76\,\% in the high $p_{\rm{T}}$ jet region and added to the JES uncertainty in quadrature\cite{Aad:2011he}. 
 
\item \textbf{Jet reconstruction efficiency} - The jet reconstruction efficiency (JRE) was evaluated by randomly dropping jets from events with a probability of about 2\%\cite{ATLAS-CONF-2010-054}. The resulting difference with respect to the nominal case was symmetrised and quoted as JRE systematic uncertainty. 

\item \textbf{Jet energy resolution} - A smearing of the jet transverse momentum corresponding to a resolution of about 10\,\% was applied to Monte Carlo events as systematic to reflect the difference between the jet energy resolution (JER) observed on data and Monte Carlo\cite{ATLAS-CONF-2010-054}. The resulting discrepancies were symmetrised and quoted as JER systematic uncertainty.

\item \textbf{Muon efficiencies} - In order to account for the trigger and reconstruction efficiencies for muons\cite{ATLAS-CONF-2011-008,ATLAS-CONF-2011-021}, global and object based scale factors and efficiencies were taken into account and a systematic uncertainty was assigned on an event-by-event basis (for global scale factors) or on an object basis. These were combined into an overall muon efficiency uncertainty of the order of 1\,\%. In addition, a one-sided uncertainty of 1.5\,\% (events with 0 or 1 $b$ tagged jet) or 2.2\,\% (events with more than 1 $b$ tagged jet) was assigned to the muon trigger efficiency to account for a mis-modelled and as a consequence discarded trigger object matching algorithm in the Monte Carlo samples.

\item \textbf{Muon scales and resolution} - Since the Monte Carlo muon momentum scales and resolution differed from the ones observed in data, the muon momentum was smeared and a scaling of up to 1.5\,\% was applied on object level to account for this discrepancy\cite{ATLAS-CONF-2011-046}. A systematic uncertainty at the sub-percent level was assigned by scaling up and down both the momentum scaling and smearing by \mbox{1\,$\sigma$} according to the respective uncertainty. In addition, the missing energy was re-evaluated with the modified four-vector information. The full event selection and kinematic reconstruction was performed for the different scales, resulting in a symmetrised systematic uncertainty based on the comparison of the different results of the measurement.

\item \textbf{Electron efficiencies} - In order to account for the trigger and reconstruction efficiencies for electrons, global and object based scale factors were taken into account\cite{Aad:2011mk}. A systematic uncertainty was assigned on an event basis (for global scale factors) or on an object basis, which were combined into an overall electron efficiency uncertainty of the order of 1\,\%.

\item \textbf{Electron scales and resolution} - In order to take into account discrepancies between the electron energy resolution on Monte Carlo and data, a Gaussian smearing procedure was applied to the electron energy for Monte Carlo events to reflect the resolution in data\cite{Aad:2011mk}. In addition, the electron energy in data was corrected to account for a scaling mis-match between data and Monte Carlo. Both systematic uncertainties were of the order of 1\,\% to 2\,\% and were assigned to the Monte Carlo prediction for consistency.

\item \textbf{$\boldsymbol{b}$ tag scale factors} - Due to discrepancies in the $b$ tagging efficiencies and fake rates between data and simulation, all Monte Carlo jets were assigned a specific weight to account for this effect\cite{ATLAS-CONF-2011-089,ATLAS-CONF-2011-102}. The obtained $b$ tag weights for each jet were combined into an event weight by multiplication (corresponding to a logical AND of all jets taken into account). The provided scale factors contained uncertainties which result in small shape variations. In order to determine the deviation in the shapes from the nominal case due to the $b$ tag scaling and to quantify the corresponding systematic uncertainty on the measurement, the resulting samples were shifted up and down by the provided uncertainties. These were of the order of 8\,\%, depending on the jet \pt~and $\eta$.
  
\item \textbf{PDF uncertainty} - For the \mcnlo~signal Monte Carlo, CTEQ6.6 PDFs were utilised to model the incoming partons to the hard scattering process, as described in \mbox{Chapter \ref{SamplesMC}}. The impact of the choice of PDFs was evaluated by varying the eigenvalues of the CTEQ parametrisation\cite{1126-6708-2006-02-032} or by comparison with the respective MRST2001 parametrisation\cite{dur1538}, using the \textsc{Lhapdf} tool\cite{Whalley:2005nh}. Event weights were determined and the variations in the resulting pseudosamples were taken into account as PDF systematic uncertainty.

\item \textbf{LAr defects} - Parts of the LAr calorimeter readout electronics were inoperative during a significant time period of data taking due to a technical problem. Having occurred after the production of the used Monte Carlo samples, it was necessary to correct for the resulting mismatch between data and simulated events at the analysis level. Monte Carlo events were dropped with a probability corresponding to the relative amount of data affected (84.0\,\%) if an electron or a jet entered the region of degraded acceptance (taking into account the isolation requirements). A systematic uncertainty was assigned based on different transverse momentum requirements for the jets taken into account for this procedure after symmetrisation.

\item \textbf{ISR and FSR} - In order to take into account initial and final state radiation, which can introduce additional jets in the observed events, different Monte Carlo samples with varying ISR and FSR contributions\cite{Aad:2009wy} (ISR and FSR contributions scaled up and down independently and in combination) were evaluated by replacing the nominal signal sample in the measurement. The systematic uncertainty was quoted as the maximum relative deviation from the nominal leading order sample observed in these variations and applied to the \mcnlo~prediction. The parameters were varied in a range comparable to those used in the Perugia Soft/Hard tune variations\cite{Skands:2010ak}.

\item \textbf{$\boldsymbol{t} \bar{\boldsymbol{t}}$ modelling} - The impact of using different MC generators for the signal process modelling was studied. In addition to the \mcnlo~generator, the \powheg generator was used for comparison and a symmetrised systematic uncertainty was assigned based on the variations in the measurement results for the alternate modelling.

\item \textbf{Parton shower / fragmentation} - In addition to the matrix element level MC generator, the effect of different showering models was taken into account by comparing the results for the \powheg generator with showering performed by \pythia and by \herwig, and a symmetrised systematic uncertainty was assigned based on the variations in the measurement results for the alternate shower modelling.

\item \textbf{Top mass} - Since the top mass parameter was considered fixed, the uncertainty on the measurement of the mass was taken into account separately. Two different Monte Carlo samples generated with different mass parameters (scaled up and down to 180\,GeV and 170\,GeV, respectively) were used and the observed deviations were linearly interpolated according to the actual uncertainty of 0.5\,\% of the top mass measurement\cite{Lancaster:2011wr} and a symmetrised systematic uncertainty was quoted.

\item \textbf{$\boldsymbol{W}$+jets background uncertainty} - The W+jets background normalisation was estimated with a data driven method and a systematic uncertainty based on the limited statistics available and several systematic contributions to the method have been evaluated (see \mbox{Chapter \ref{chap:WJetsDD}} for details). An overall $W$+jets normalisation uncertainty of 22.4\,\% and 22.7\,\% was obtained in the muon+jets and electron+jets channel, respectively. In addition, the $W$+jets background shape uncertainty has been evaluated by modifying several generator parameters such as the renormalisation scale or the functional form of the factorisation scale compared to the nominal \alpgen parameters, based on the leading jet \pt. A symmetrised systematic uncertainty on the charge asymmetry measurement has been assigned based on the deviations for two different sets of parameters with respect to the nominal results.

\item \textbf{$\boldsymbol{Z}$+jets background uncertainty} - In order to quantify the uncertainty on the $Z$+jets contribution normalisation, a Berends-Giele scaling uncertainty\cite{Berends199132} was calculated, corresponding to an overall normalisation uncertainty of 34\,\%. In addition, the $Z$+jets background was determined independently from both the \alpgen and \sherpa generator to quantify the shape uncertainty, which was quoted based on the symmetrised discrepancy of the results obtained with \sherpa with respect to the nominal case, in which \alpgen samples were used.

\item \textbf{Charge mis-identification} - Since the detector momentum resolution is finite, and the lepton charge was identified by taking into account the bending radius of the particle track, a certain probability for mis-identifying the lepton charge exists, especially for high transverse momentum leptons due to their almost straight trajectories. This probability was evaluated on Monte Carlo and on data to be of the order of 0.2\,\% to 0.5\,\% in the central detector region, and up to 2.5\,\% in the forward region, respectively, in the electron+jets channel\cite{AsymPaper}. In the muon+jets channel, the probability was found to be below 0.003\,\% in all cases. A corresponding symmetrised systematic uncertainty on the measurement of the charge asymmetry was determined.

\item \textbf{$\boldsymbol{b}$ tag charge} - A dependency of the $b$ tag efficiencies on the $b$ quark charge can lead to a bias in the measurement due to the requirement of at least one $b$ tagged jet. Hence, a simple Monte Carlo study on parton level was performed by simulating a difference in the $b$ tag efficiency of 5\,\% between $b$ and $\bar{b}$ quarks. The resulting impact on the charge asymmetry on parton level was studied and the difference with respect to the nominal case (assuming identical tagging efficiencies for $b$ and $\bar{b}$ quarks) was quoted as systematic uncertainty.

\item \textbf{MC generator statistics} - Since the signal Monte Carlo sample entered directly into the unfolding response matrix and statistical fluctuations in the bins of this matrix can have an impact on the unfolding process, an additional ensemble test was performed by fluctuating the obtained nominal response matrix on a bin-by-bin basis following a Gaussian probability distribution (since the statistics from the \mcnlo~sample were very high a Gaussian model could safely be assumed). Uncertainties of the order of 0.3\,\% to 3\,\% have been obtained, depending on the statistics in each bin of the response matrix.

\item \textbf{Unfolding convergence} - Based on the convergence evaluation which was used to determine the optimal amount of regularisation in the unfolding process (for details refer to \mbox{Chapter \ref{chap:results:unfolding}}), a remaining absolute uncertainty of 1\,\permil, corresponding directly to the choice of convergence criterion, was assigned, representing the potential remaining change with respect to further increase in regularisation.

\item \textbf{Unfolding bias} - Closure tests have been performed using ensembles of pseudodata to quantify any remaining bias from the unfolding at the chosen regularisation strengths by obtaining pull distributions for the measured asymmetry, normalised with respect to the respective unfolding statistical uncertainty. The corresponding distributions can be found in \mbox{Figure \ref{fig:pull_dist}} in \mbox{Appendix \ref{AppControlPlots}}. The remaining bias was extracted from the residuals in the pull distributions in the closure tests and was taken into account as an additional systematic uncertainty. The residual bias after unfolding was extracted from the respective pull distributions corresponding to the regularisation used in the individual cases and taken into account as an additional relative uncertainty on the unfolded results, which was of the order of 1\,\% to 11\,\%.

\item \textbf{Other backgrounds} - For the small backgrounds from single top and diboson production, only normalisation uncertainties were considered. For the single top contribution, the uncertainties of the approximate next-to-next-to-leading order prediction (for details, refer to \mbox{Chapter \ref{theory_ST}}) were taken into account (corresponding to an uncertainty of about 11\,\%), while for the diboson production, an overall uncertainty of 5\,\% was assumed.

\item \textbf{Luminosity} - The relative uncertainty on the measurement of the integrated luminosity of the used data sample was estimated to be 3.7\,\%\cite{ATLAS-CONF-2011-116} and was taken into account for the measurement.

\item \textbf{Pile-up} - In order to take into account the difference in pile-up conditions between Monte Carlo and data, an event based reweighting was performed to reflect the distribution of average observed number of interactions per bunch crossing on data. The impact of the pile-up conditions on the measurement before unfolding is shown in detail in \mbox{Chapter \ref{chap:results:unfolding}}. Since no significant pile-up dependency was observed, no additional systematic uncertainty was assigned.

\end{itemize}

  \chapter{Results}
\label{Results}

\section{Event Yields \& Control Plots}
The final expected and observed number of events in both the muon+jets and the electron+jets channel after performing the event selection can be found in \mbox{Table \ref{evtnumbers}}, both without and with the requirement of at least one $b$ tagged jet.
\begin{table}[htbp]
  \begin{center}
  {\small
  \begin{tabular}{| l | rrr | rrr | rrr | rrr |}
      \hline
      {\bf Channel}   & \multicolumn{6}{c|}{{\bf muon+jets}} & \multicolumn{6}{c|}{{\bf electron+jets}} \\
      \hline
      Selection & \multicolumn{3}{c|}{$\geq 0$ $b$ tags} & \multicolumn{3}{c|}{$\geq 1$ $b$ tags} & \multicolumn{3}{c|}{$\geq 0$ $b$ tags} & \multicolumn{3}{c|}{$\geq 1$ $b$ tags}            \\ \hline \hline
      $t\bar{t}$              & 7187   &$\!\!\!\pm\!\!\!$&    588     & 6342   & $\!\!\!\pm\!\!\!$  &    519   &  4832   & $\!\!\!\pm\!\!\!$  &    395     & 4256   & $\!\!\!\pm\!\!\!$  &    348   \\
      Single top              & 458  &$\!\!\!\pm\!\!\!$&    40    & 366  & $\!\!\!\pm\!\!\!$  &    32  &  320  & $\!\!\!\pm\!\!\!$  &    28    & 256  & $\!\!\!\pm\!\!\!$  &    22  \\ \hline
      $W$+jets (data)           & 8617  &$\!\!\!\pm\!\!\!$&    1173    & 1390  & $\!\!\!\pm\!\!\!$  &    311  &  5407  & $\!\!\!\pm\!\!\!$  &    768    & 883  & $\!\!\!\pm\!\!\!$  &    200  \\
      $Z$+jets                  & 940  &$\!\!\!\pm\!\!\!$&    333    & 134  & $\!\!\!\pm\!\!\!$  &    47  &  756  & $\!\!\!\pm\!\!\!$  &    267    & 105  & $\!\!\!\pm\!\!\!$  &    37  \\
      Diboson                 & 134 &$\!\!\!\pm\!\!\!$&    7   & 22 & $\!\!\!\pm\!\!\!$  &    2 &  80 & $\!\!\!\pm\!\!\!$  &    5   & 13 & $\!\!\!\pm\!\!\!$  &    1 \\ \hline
      QCD multijet (data)        & 1515     &$\!\!\!\pm\!\!\!$&    757       & 519     & $\!\!\!\pm\!\!\!$  &    519     &  944     & $\!\!\!\pm\!\!\!$  &    472       & 247     & $\!\!\!\pm\!\!\!$  &    247     \\ \hline \hline
      Total background        & 11663   &$\!\!\!\pm\!\!\!$&    1436     & 2431   & $\!\!\!\pm\!\!\!$  &    608   &  7508   & $\!\!\!\pm\!\!\!$  &    941 & 1504   & $\!\!\!\pm\!\!\!$  &    321   \\ \hline
      Total expected          & 18850   &$\!\!\!\pm\!\!\!$&    1552     & 8774   & $\!\!\!\pm\!\!\!$  &    799   &  12340   & $\!\!\!\pm\!\!\!$  &    1020     & 5760   & $\!\!\!\pm\!\!\!$  &    473   \\ \hline \hline
      Observed                & \multicolumn{3}{r|}{19639}                                                                           & \multicolumn{3}{r|}{9124}                                                                         & \multicolumn{3}{r|}{12096}                                                                            & \multicolumn{3}{r|}{5829}                                                                         \\ \hline
    \end{tabular}
  }
  \end{center}
  \caption{Observed number of data events in comparison to the expected number of Monte Carlo signal events and different background contributions for the event selection, both with and without the requirement of at least one $b$ tagged jet. The QCD multijet and $W$+jets contributions were estimated using data driven methods (c.f. \mbox{Chapter \ref{Samples}}). Uncertainties are statistical and include the respective systematic uncertainties on the normalisation for QCD multijet and $W$+jets and the cross-section uncertainties on all other contributions. For the QCD multijet background, a conservative 50\,\% (100\,\%) overall normalisation uncertainty for the selection without (with) requiring at least one $b$ tagged jet was assumed.
  }
  \label{evtnumbers}
\end{table}

\begin{figure}[!htbp]
  \begin{center}
    \includegraphics[width=\plotwidth]{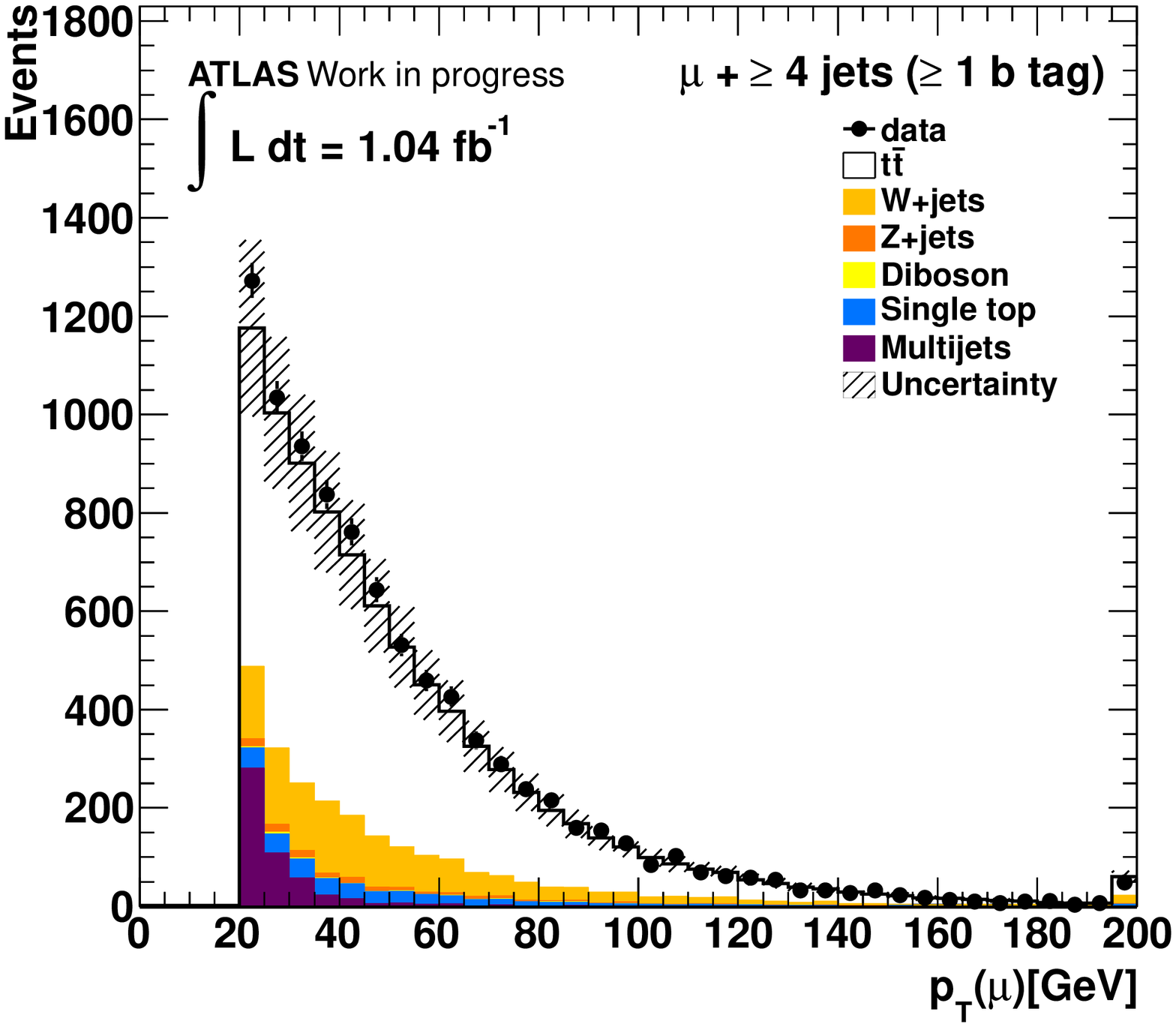}
    \quad\quad
    \includegraphics[width=\plotwidth]{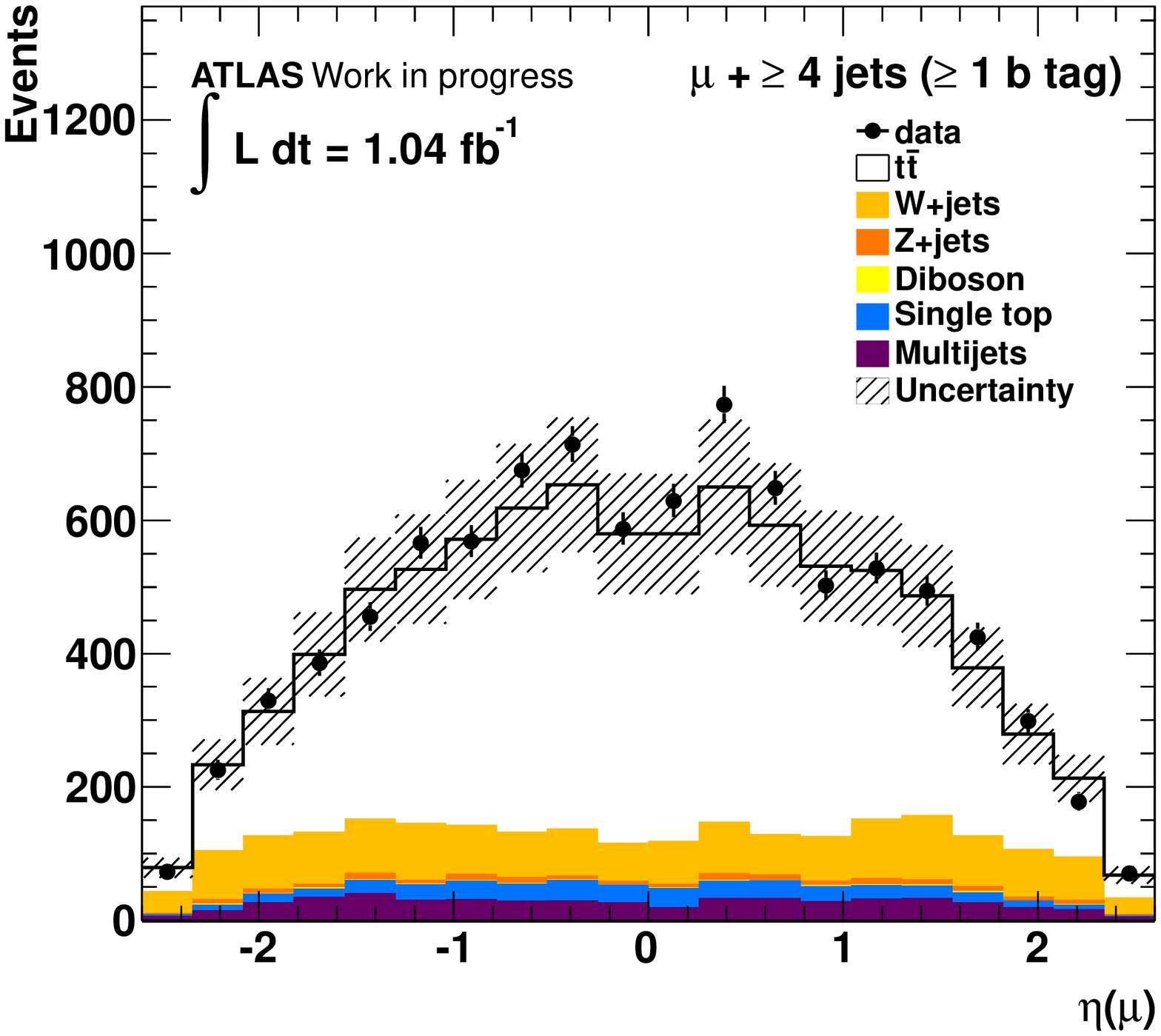}
    \includegraphics[width=\plotwidth]{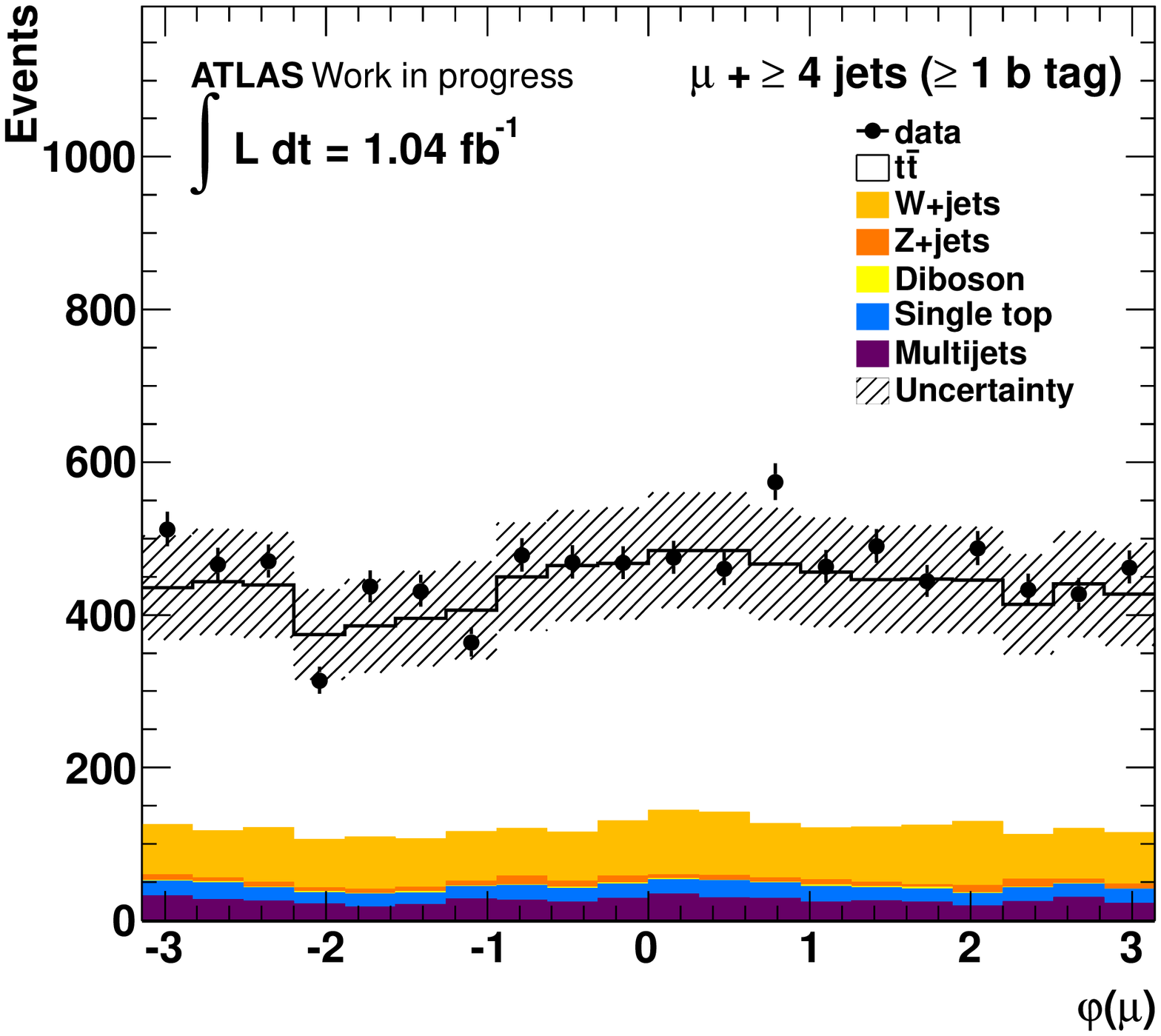}
    \quad\quad
    \includegraphics[width=\plotwidth]{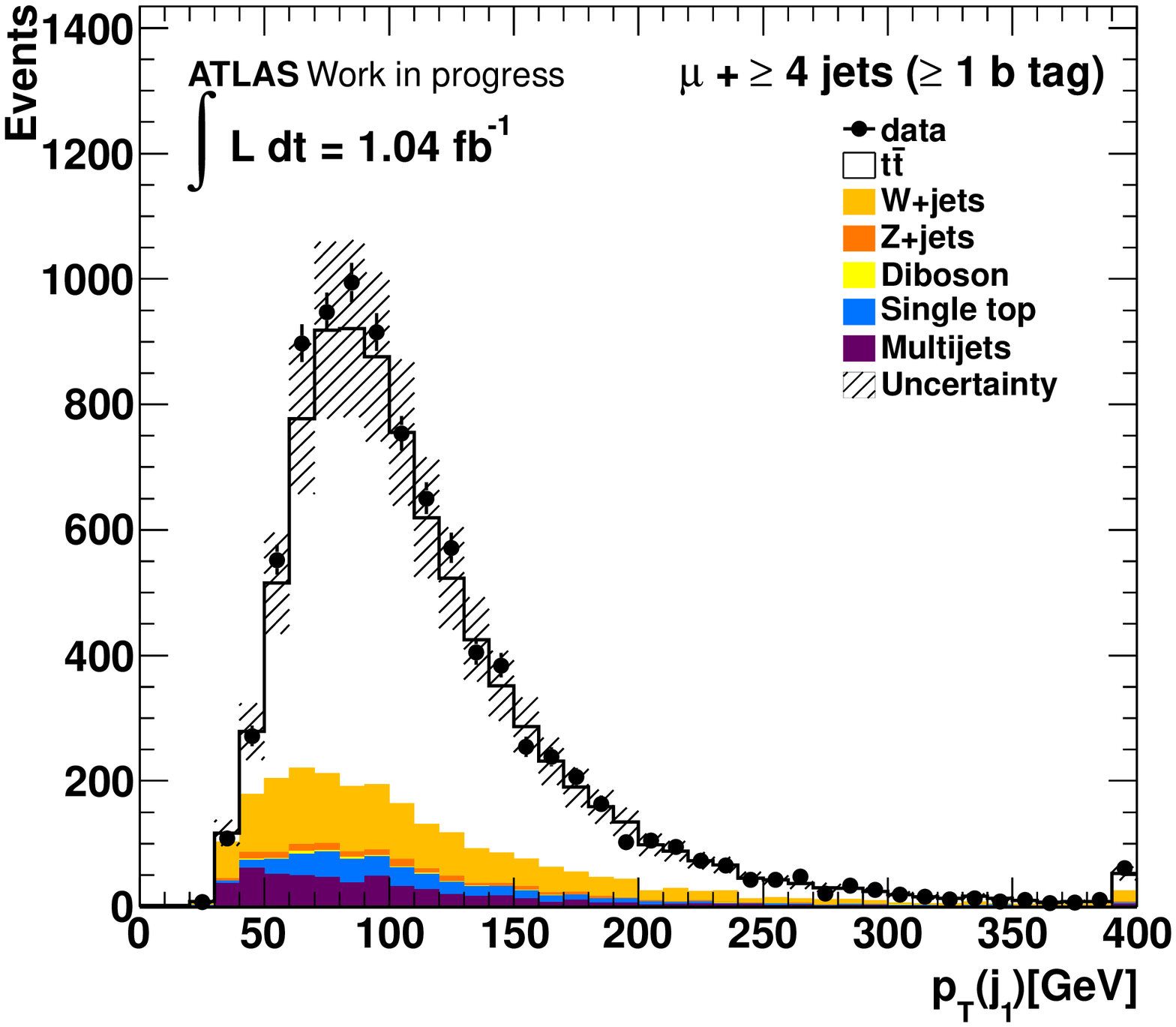}
    \includegraphics[width=\plotwidth]{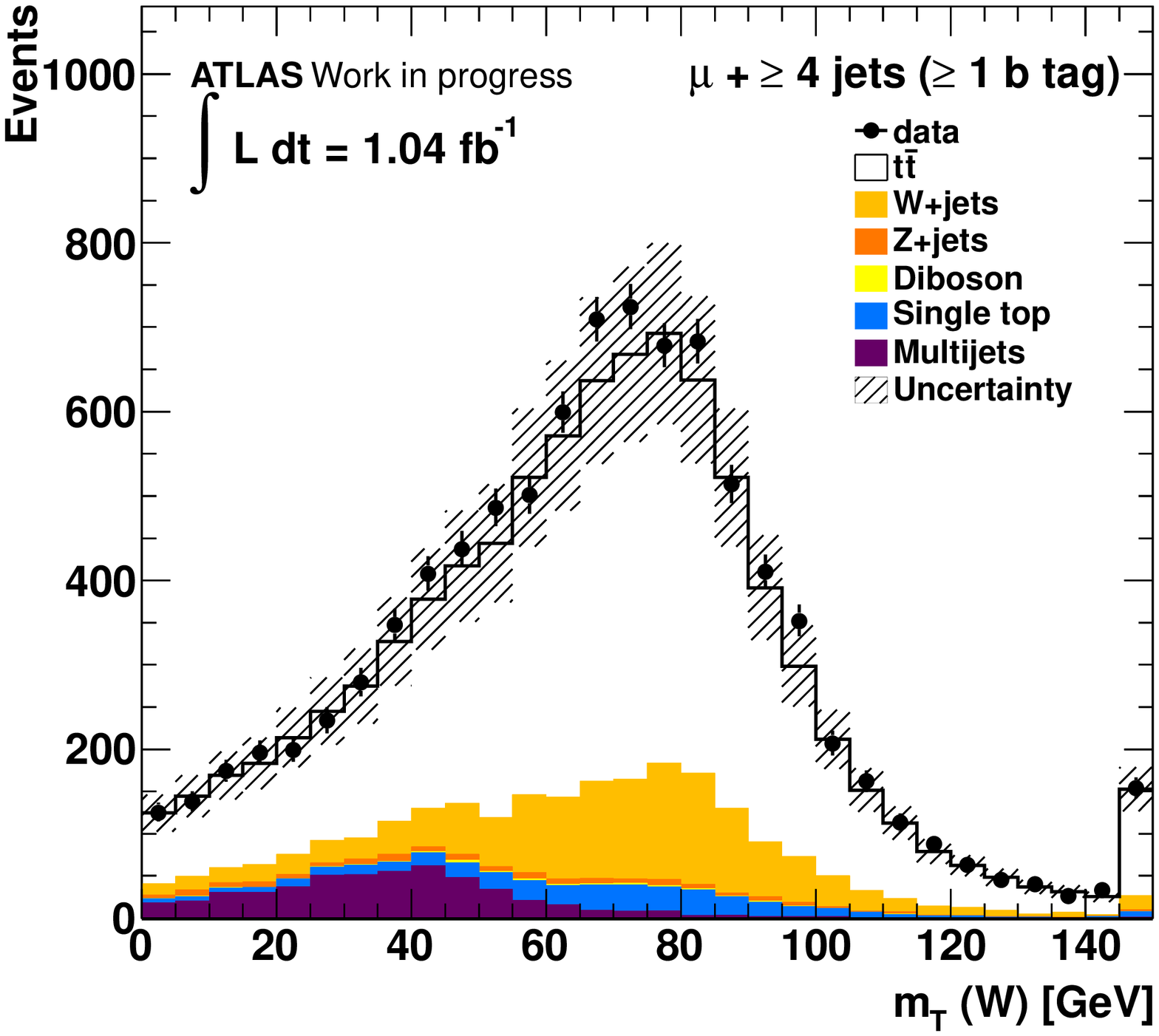}
    \quad\quad
    \includegraphics[width=\plotwidth]{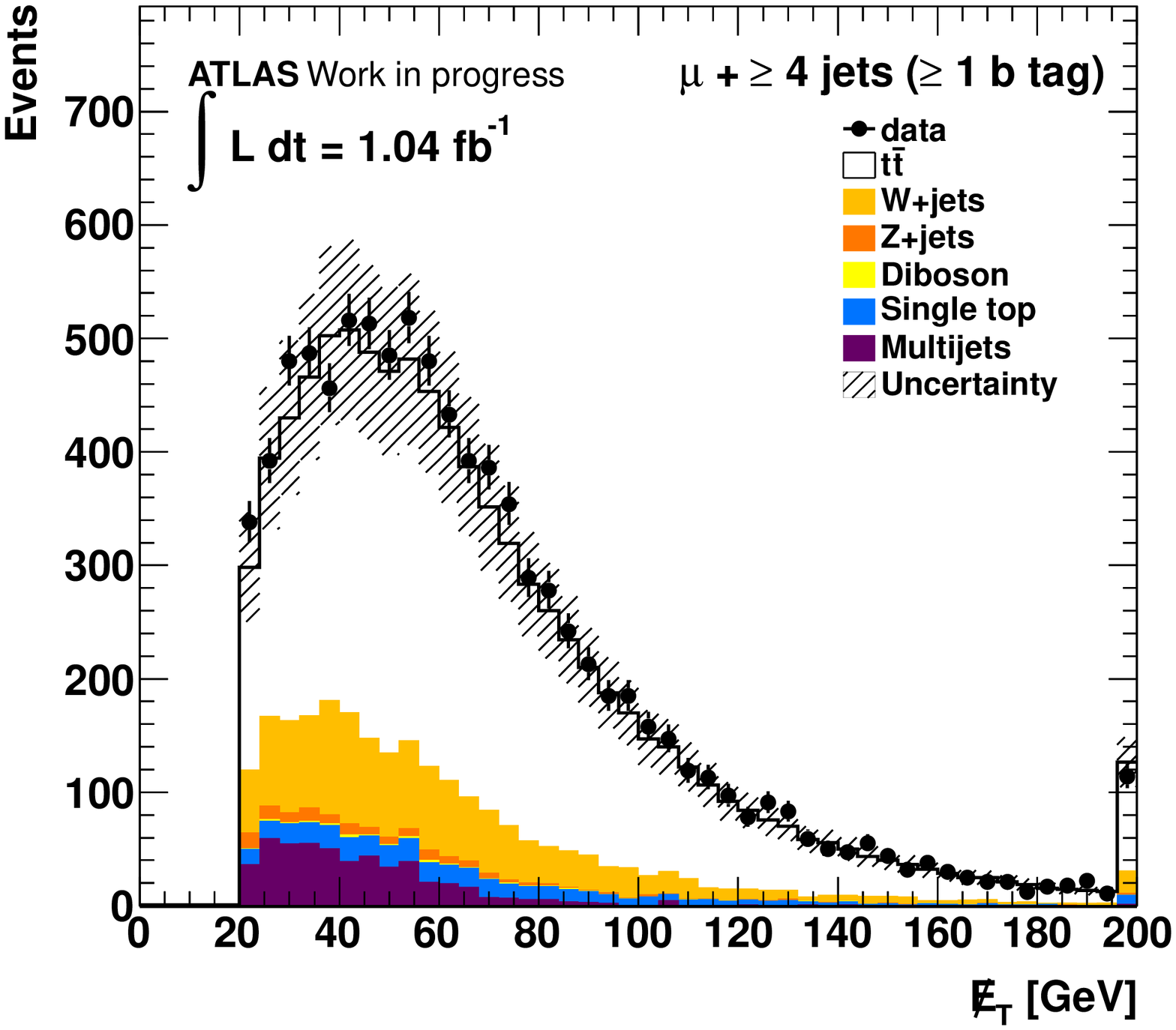}
    
    \vspace{-0.2 cm}
    \caption{Control plots for the muon+jets channel. From the top left to the bottom right, the transverse momentum $p_{\rm{T}}$, the pseudorapidity $\eta$ and the azimuthal angle $\phi$ of the selected muon are shown. Additional plots show the transverse momentum of the leading jet ($p_{\rm{T}}(j_1)$), the $W$ transverse mass $m_{\rm{T}}(W)$ and the transverse missing energy \met. Uncertainties are statistical and for $W$+jets also include systematic uncertainties on normalisation. For the QCD multijet background, a conservative 100\,\% systematic uncertainty was assumed. In addition, the uncertainties on luminosity, jet energy scale, $b$ tag scale factors and \ttbar~cross-section are shown.}
    \label{controlplots_muons_tagged}
  \end{center}
\end{figure}

\begin{figure}[!htbp]
  \begin{center}
    \includegraphics[width=\plotwidth]{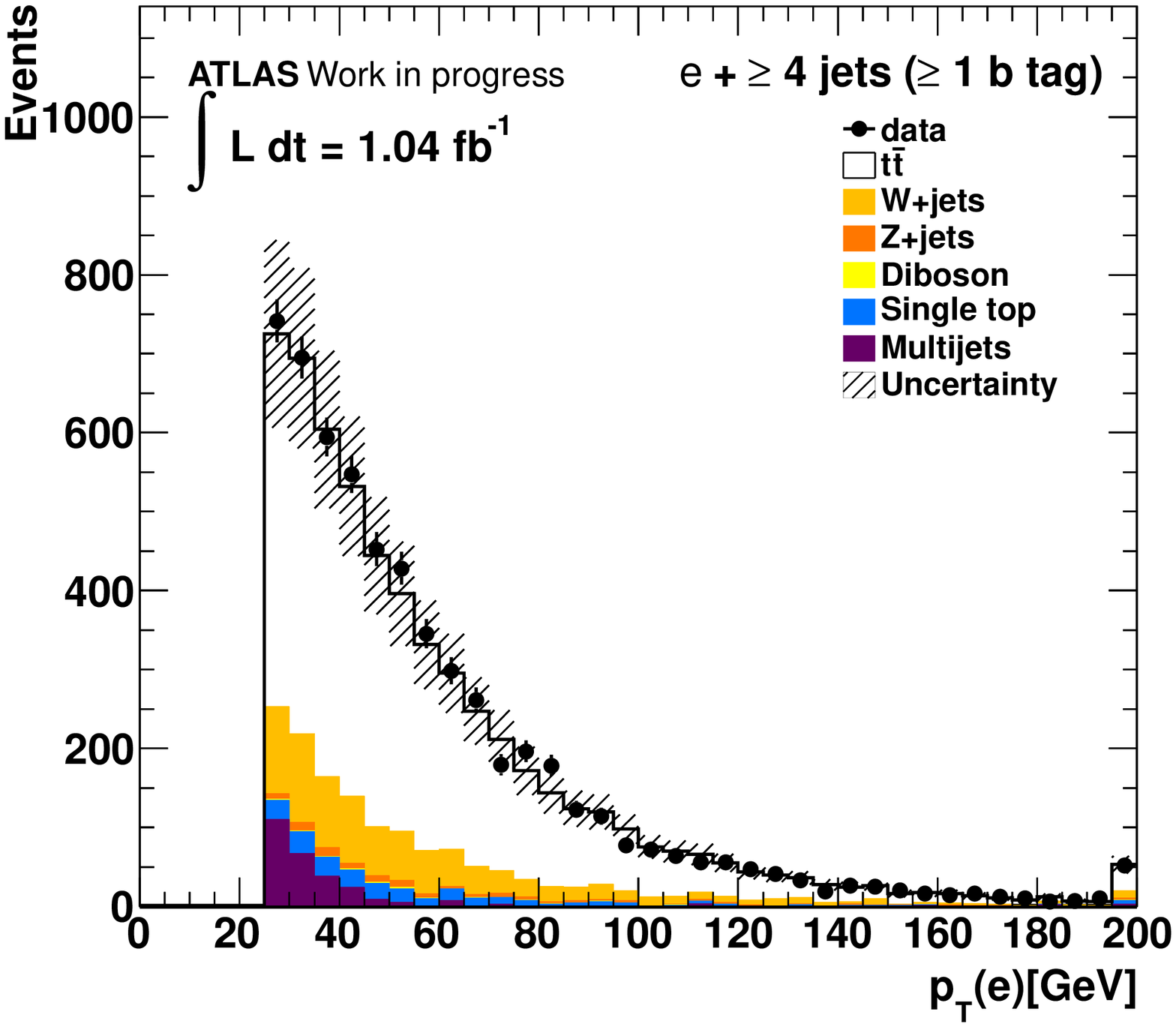}
    \quad\quad
    \includegraphics[width=\plotwidth]{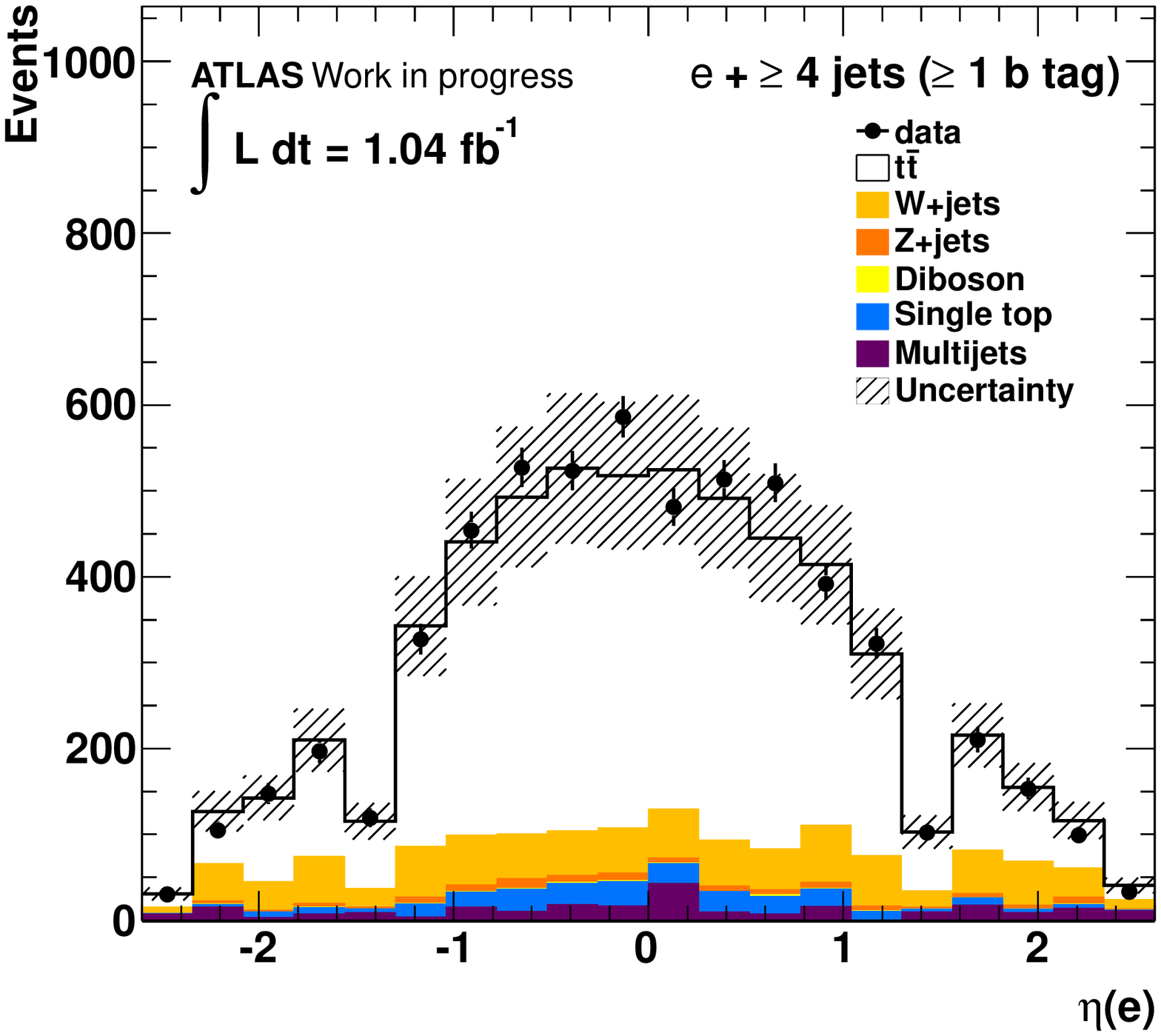}
    \includegraphics[width=\plotwidth]{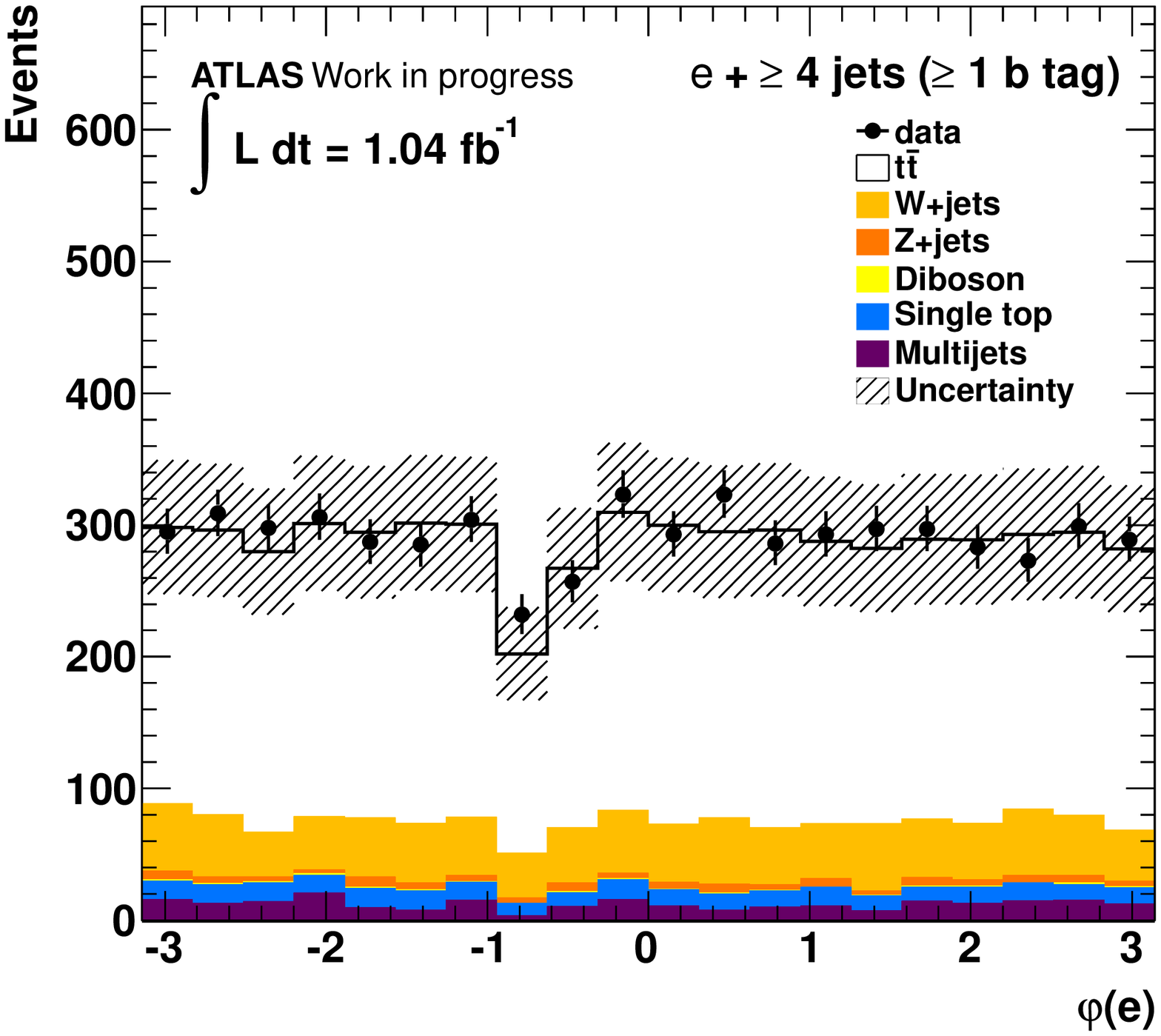}
    \quad\quad
    \includegraphics[width=\plotwidth]{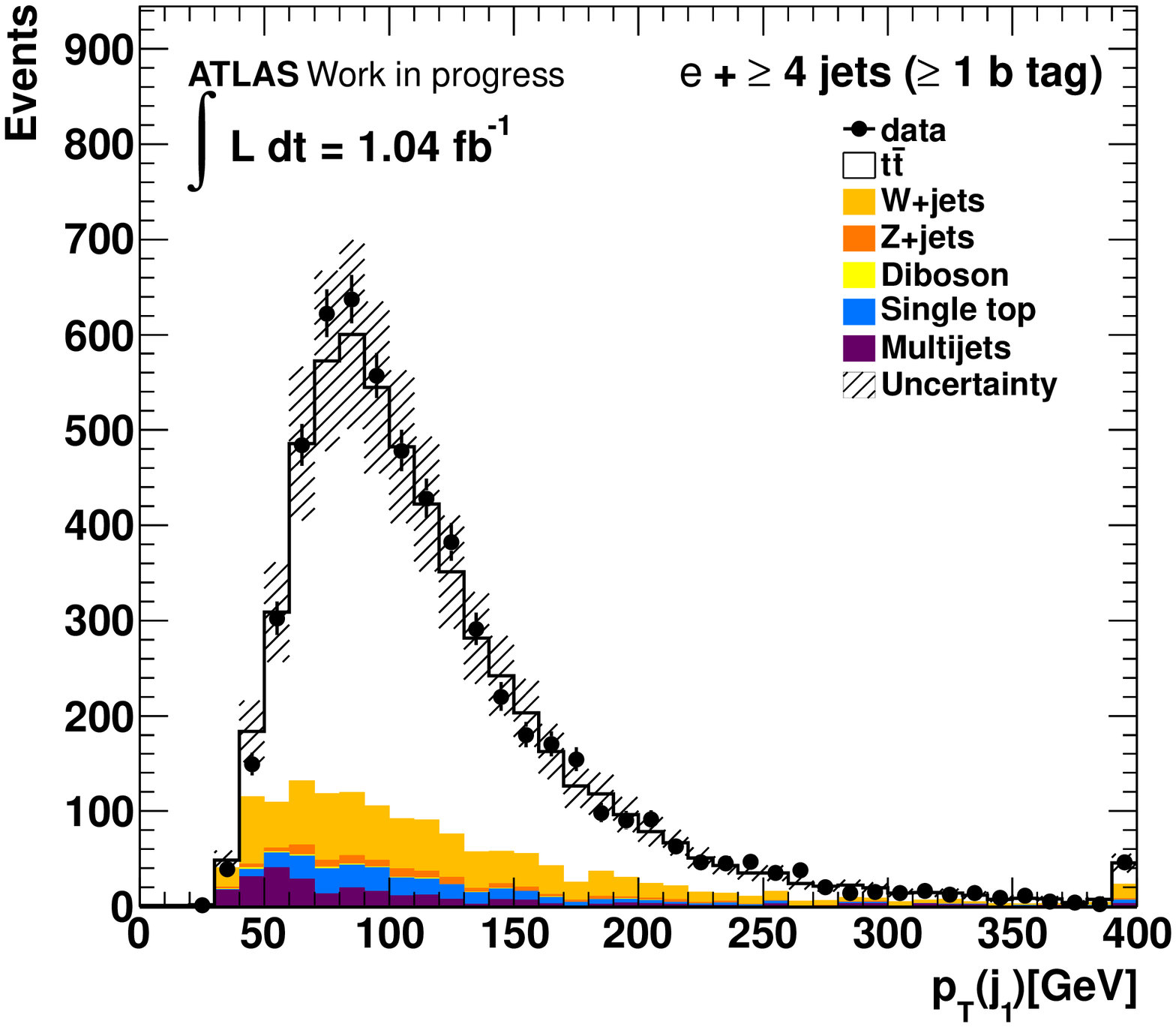}
    \includegraphics[width=\plotwidth]{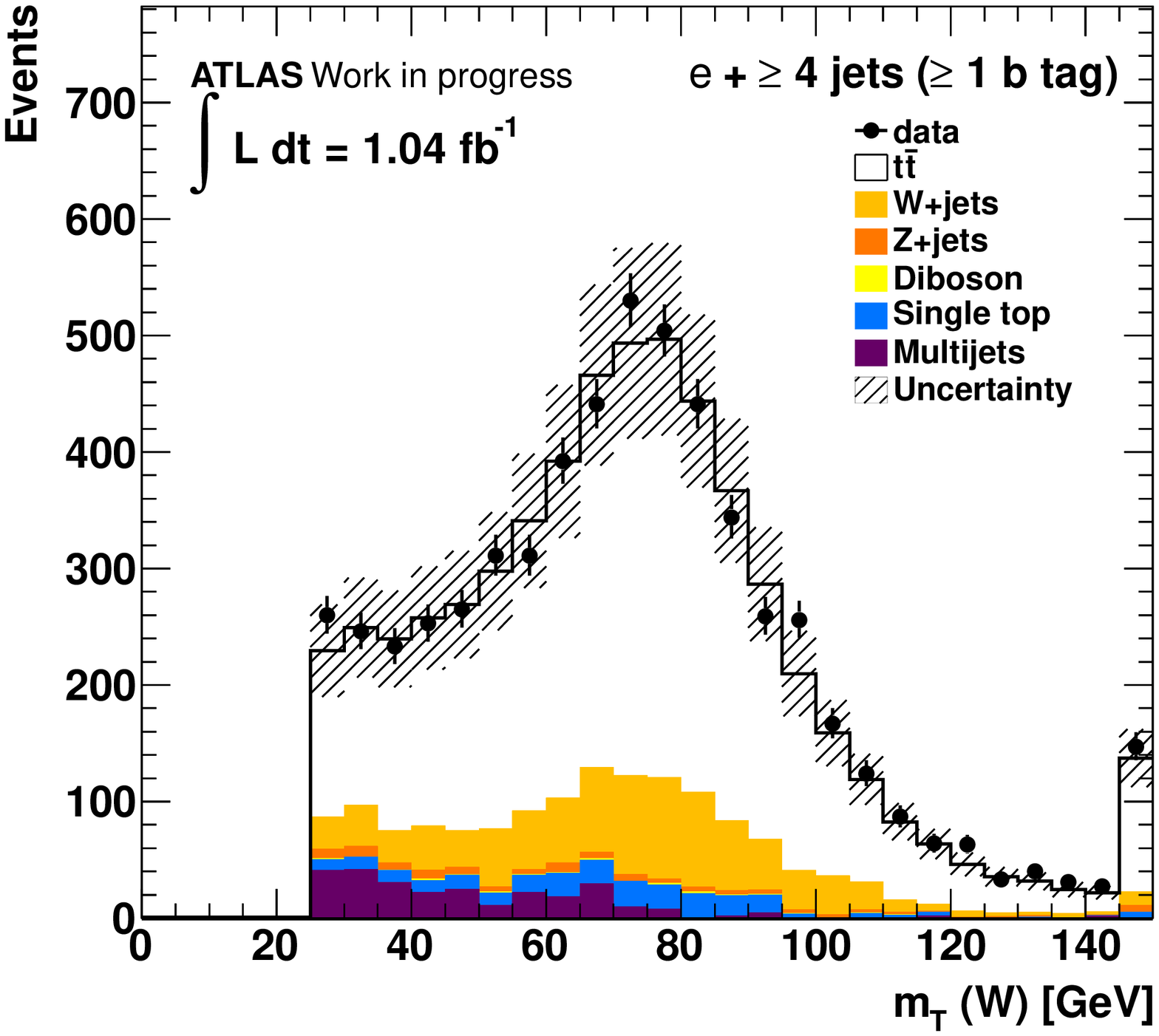}
    \quad\quad
    \includegraphics[width=\plotwidth]{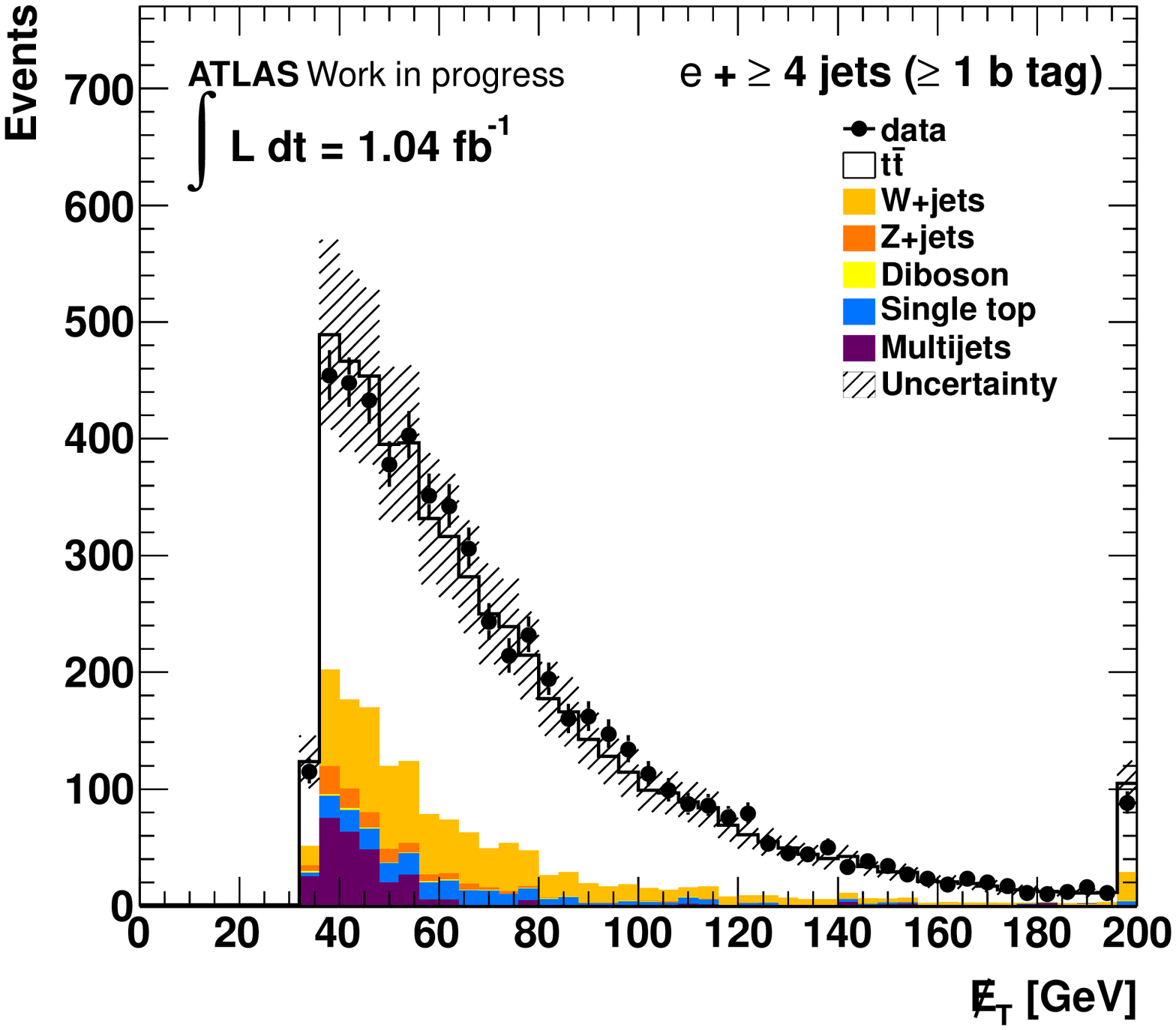}
    
    \vspace{-0.2 cm}
    \caption{Control plots for the electron+jets channel. From the top left to the bottom right, the transverse momentum $p_{\rm{T}}$, the pseudorapidity $\eta$ and the azimuthal angle $\phi$ of the selected electron are shown. Additional plots show the transverse momentum of the leading jet ($p_{\rm{T}}(j_1)$), the $W$ transverse mass $m_{\rm{T}}(W)$ and the transverse missing energy \met. Uncertainties are statistical and for $W$+jets also include systematic uncertainties on normalisation. For the QCD multijet background, a conservative 100\,\% systematic uncertainty was assumed. In addition, the uncertainties on luminosity, jet energy scale, $b$ tag scale factors and \ttbar~cross-section are shown.}
    \label{controlplots_electrons_tagged}
  \end{center}
\end{figure}

Control plots for the full event selection as described in \mbox{Chapter \ref{Selection}}, showing a comparison of the observation and expectation for several object and event quantities, can be found in \mbox{Figure \ref{controlplots_muons_tagged}} and \mbox{Figure \ref{controlplots_electrons_tagged}}, for both the muon+jets and the electron+jets channel, respectively. The uncertainties on the expectation include statistical and the leading systematic uncertainties, that is QCD multijet background and $W$+jets normalisation, luminosity, jet energy scale, $b$ tag scale factors and \ttbar~cross-section uncertainty.

\begin{figure}[!htbp]
  \begin{center}
    \includegraphics[width=\plotwidth]{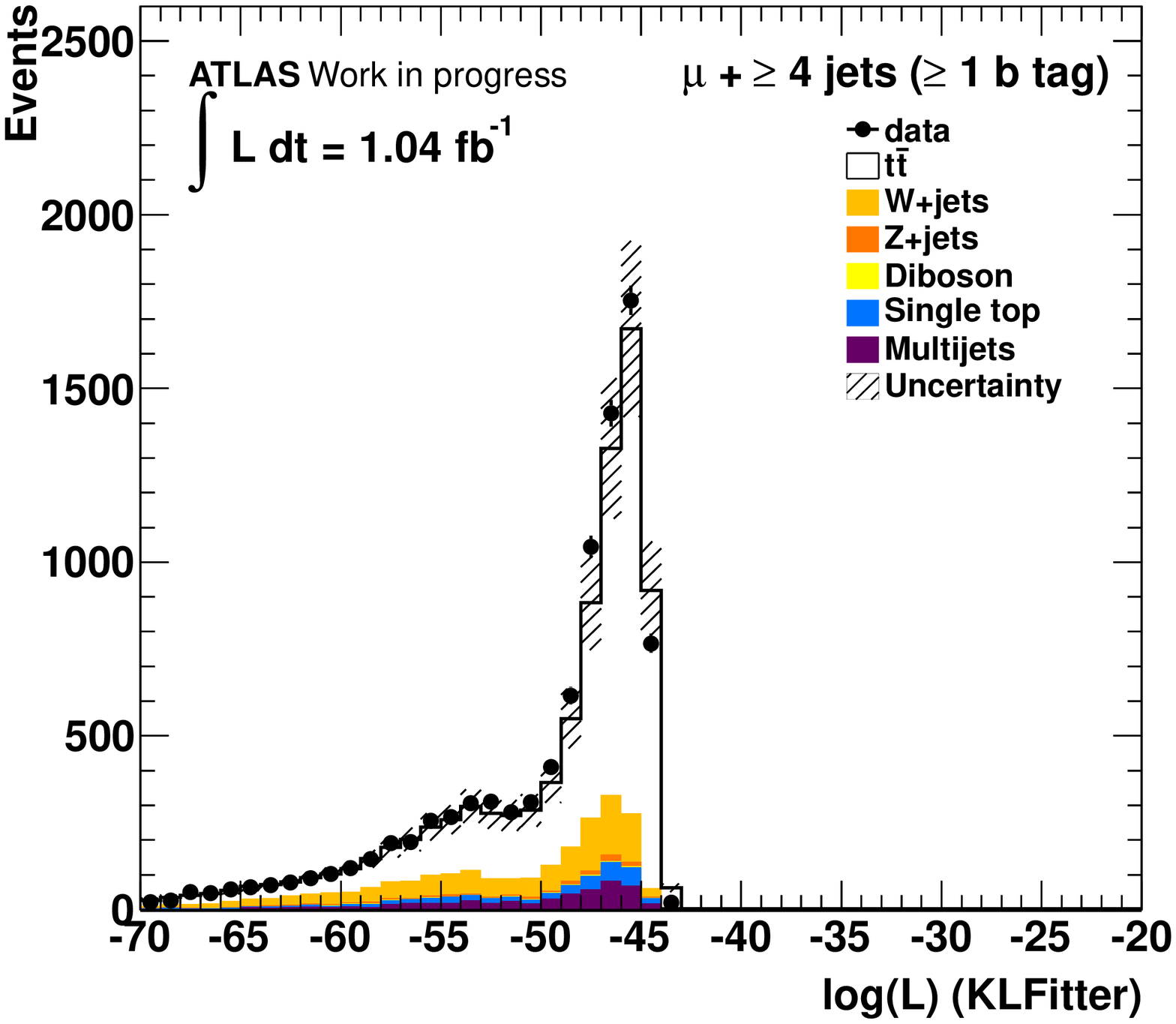}
    \quad\quad
    \includegraphics[width=\plotwidth]{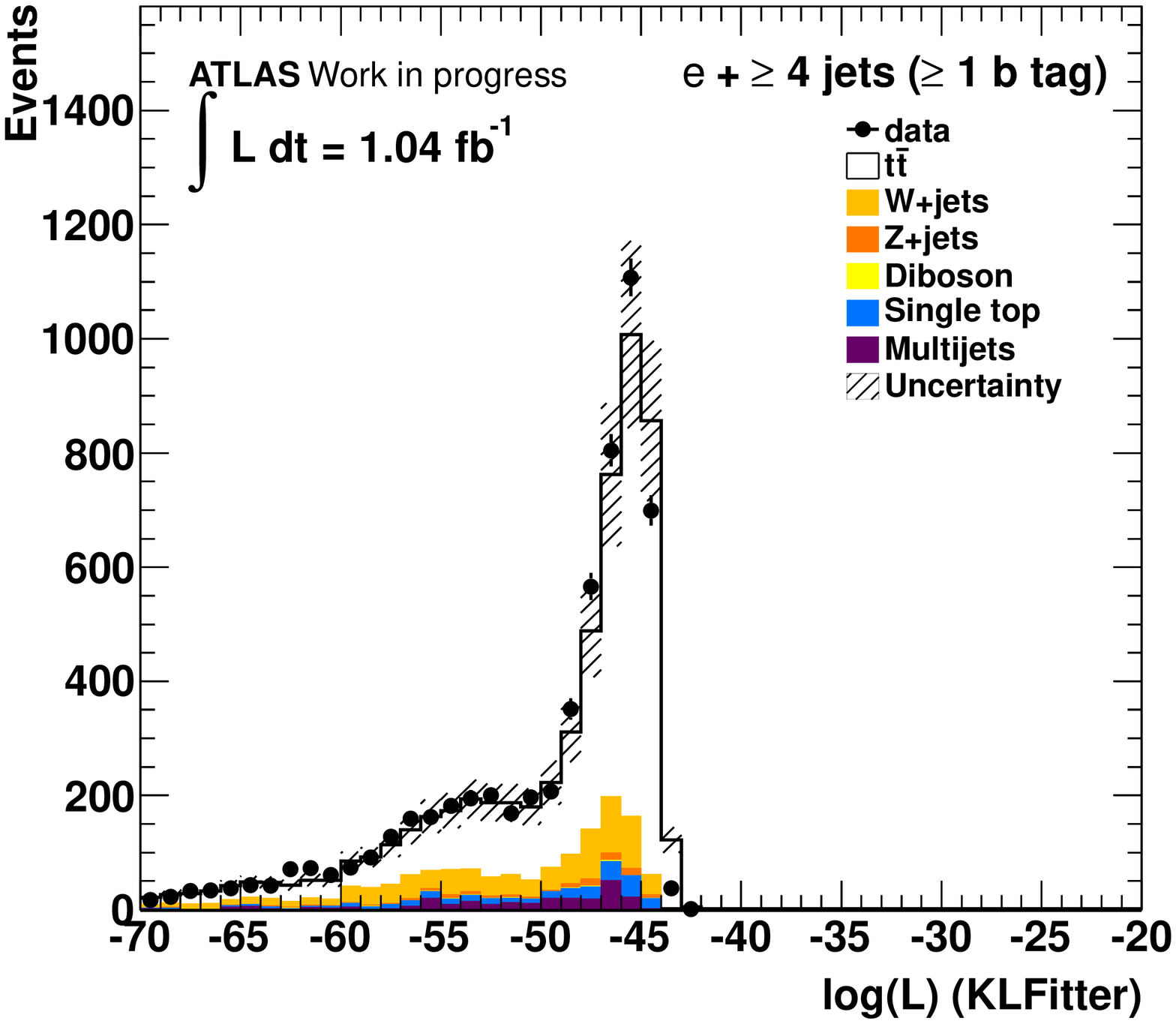}
    \includegraphics[width=\plotwidth]{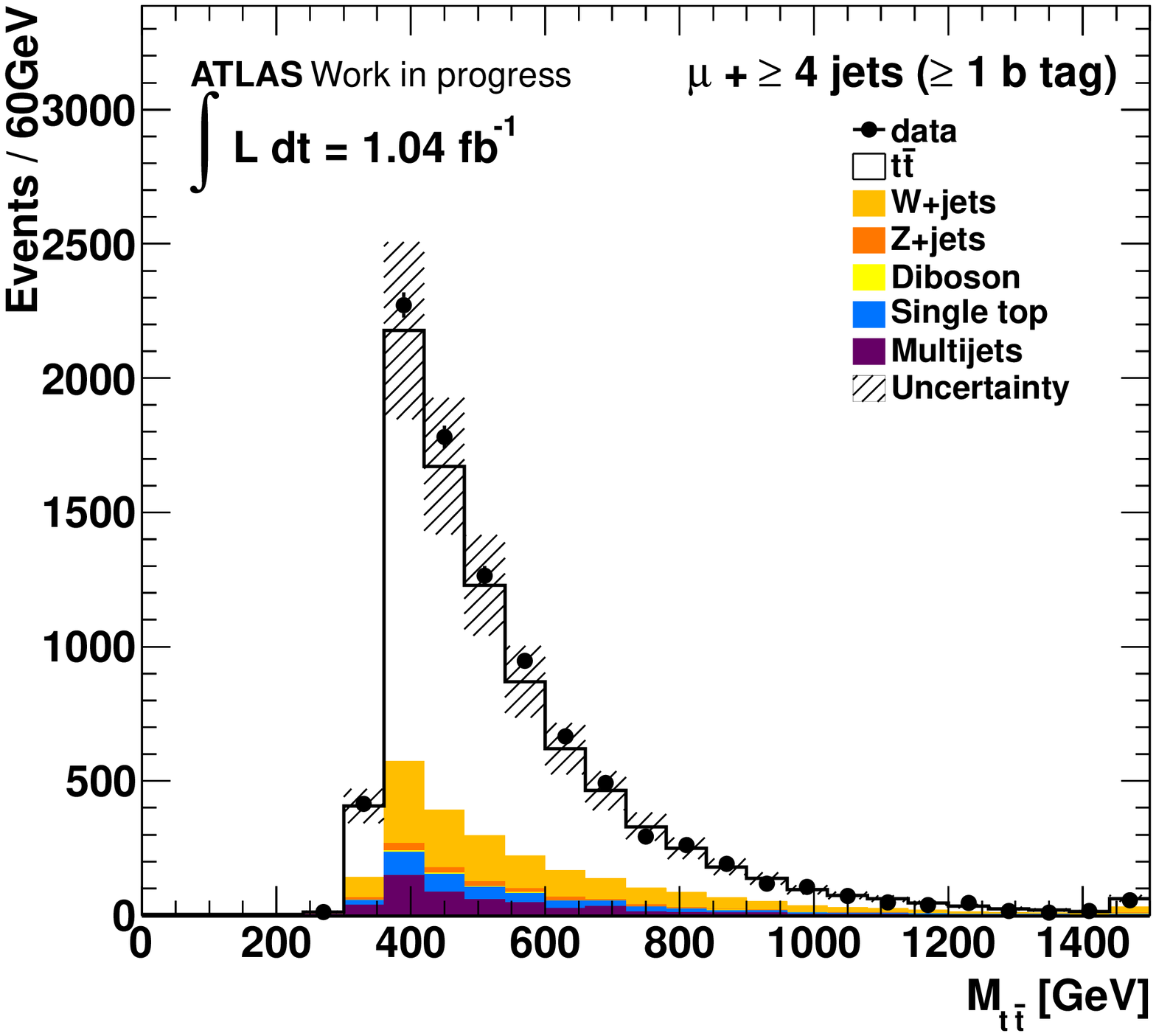}
    \quad\quad
    \includegraphics[width=\plotwidth]{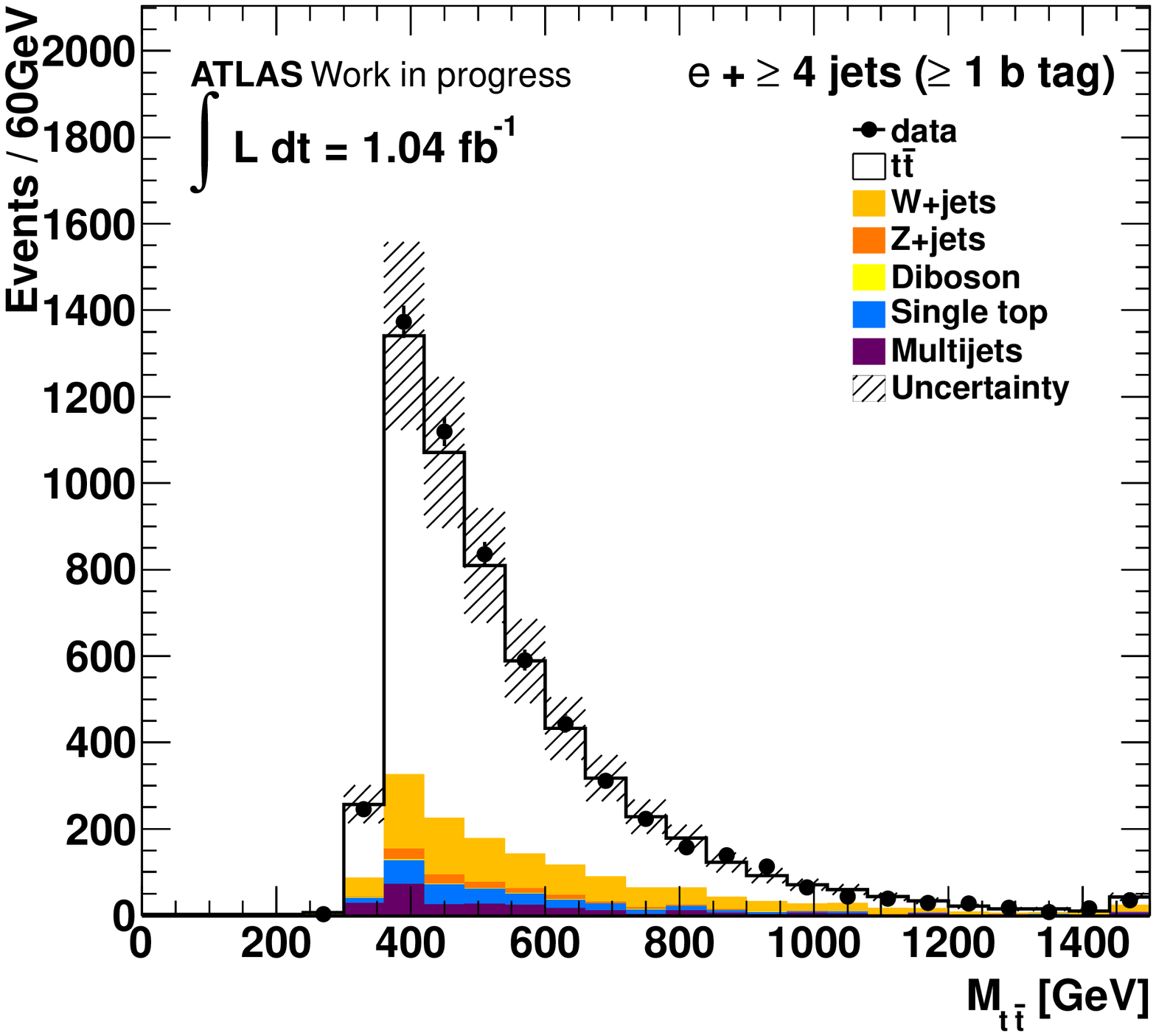}
    \includegraphics[width=\plotwidth]{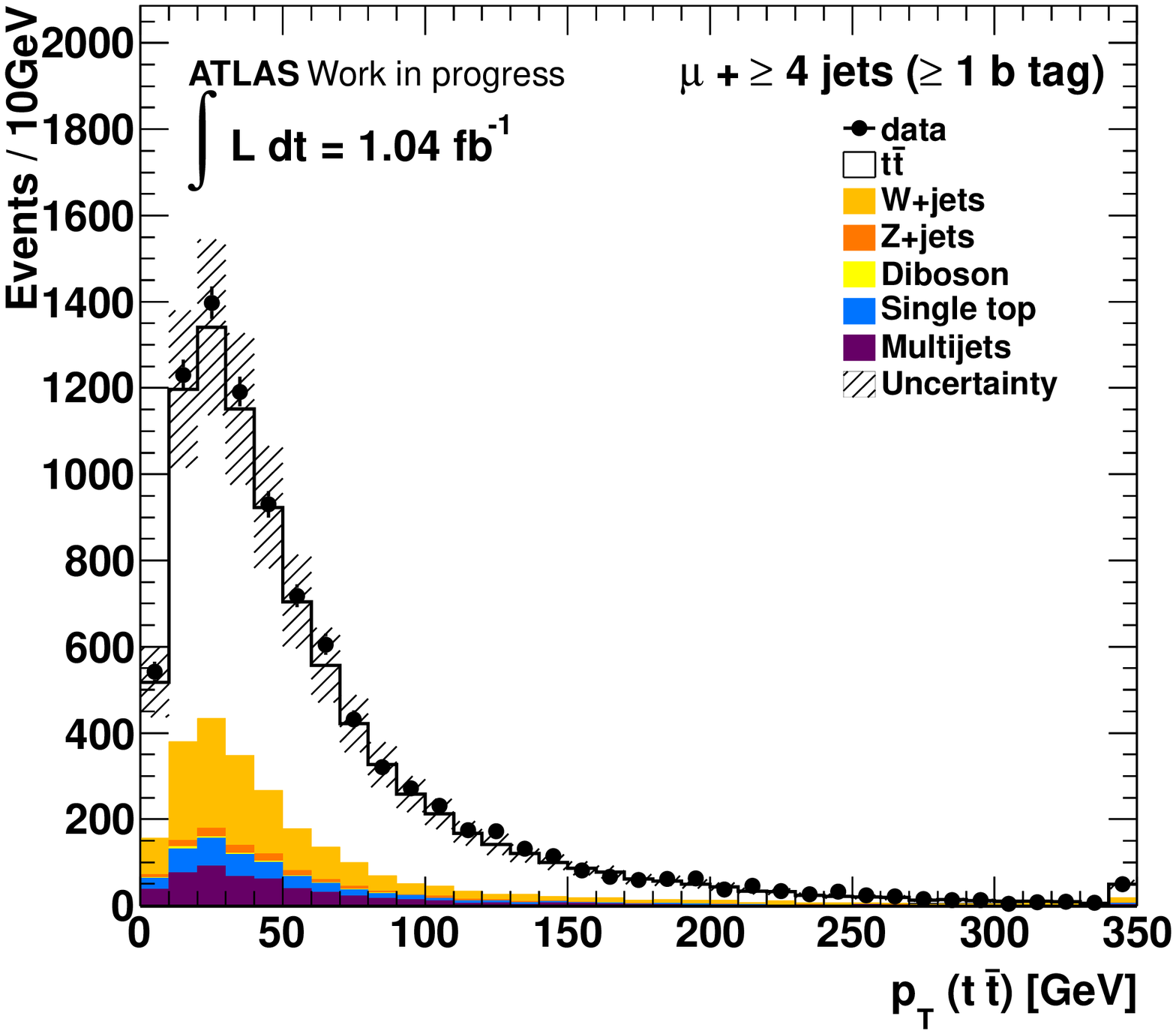}
    \quad\quad
    \includegraphics[width=\plotwidth]{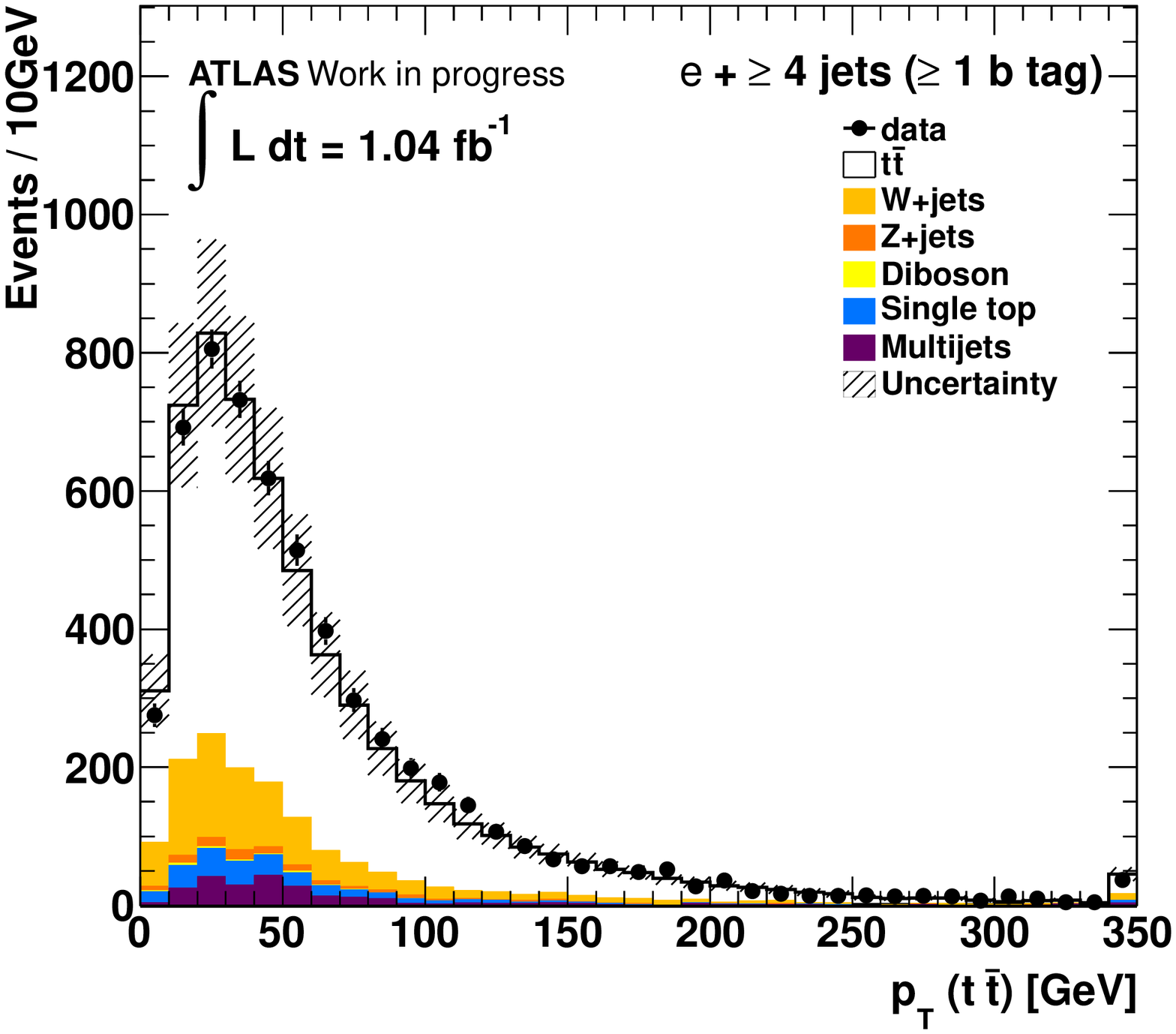}
    
    \vspace{-0.2 cm}
    \caption{Control plots for the \ttbar~event reconstruction, on the left for the muon+jets, on the right for the electron+jets channel. The top row shows the logarithmic likelihood $\log{L}$ of the kinematic fit, followed by the invariant mass $M_{t \bar{t}}$ and the transverse momentum $p_{\rm{T}}$ of the \ttbar~system. Uncertainties are statistical and for $W$+jets also include systematic uncertainties on normalisation. For the QCD multijet background, a conservative 100\,\% systematic uncertainty was assumed. In addition, the uncertainties on luminosity, jet energy scale, $b$ tag scale factors and \ttbar~cross-section are shown.}
    \label{fig:LogL_KLF}
  \end{center}
\end{figure}

Due to the tighter definition of selection criteria in the electron+jets channel, the number of events in the electron+jets channel was significantly lower than it was the case for the muon+jets channel. This was necessary in order to reduce the contribution from the increased number of electron fakes originating from the QCD multijet background with respect to the muon channel, where the expected fake rate was significantly lower. The overall agreement between Monte Carlo prediction and data was very good in both channels. Additional control plots for the same quantities without the explicit requirement of at least one $b$ tagged jet can be found in \mbox{Figure \ref{controlplots_muons_pretag}} and \mbox{Figure \ref{controlplots_electrons_pretag}} of \mbox{Appendix \ref{AppControlPlots}} for completeness.

In addition to the control plots showing basic object and event kinematics, several more complex quantities based on the kinematic event reconstruction, as described in \mbox{Chapter \ref{Reconstruction}}, can be found in \mbox{Figure~\ref{fig:LogL_KLF}}. The respective distributions for the logarithmic likelihood of the kinematic fit, and the invariant mass and transverse momentum of the reconstructed \ttbar~system in both the muon+jets and the electron+jets channel are shown. The agreement between data and prediction is very good, indicating a proper modelling of the \ttbar~signal and background kinematics in the various Monte Carlo samples and data driven background estimates.

\section{Measurement of the Charge Asymmetry}
Events passing the described event selection were taken into account to determine the differential and integrated charge asymmetry based on the distribution of $|y_t| - |y_{\bar{t}}|$. A subtraction of the predicted background contributions was performed on the distributions measured on data to obtain an estimate for the \ttbar~signal contribution only.

\begin{figure}[!htbp]
  \begin{center}
    \includegraphics[width=\plotwidth]{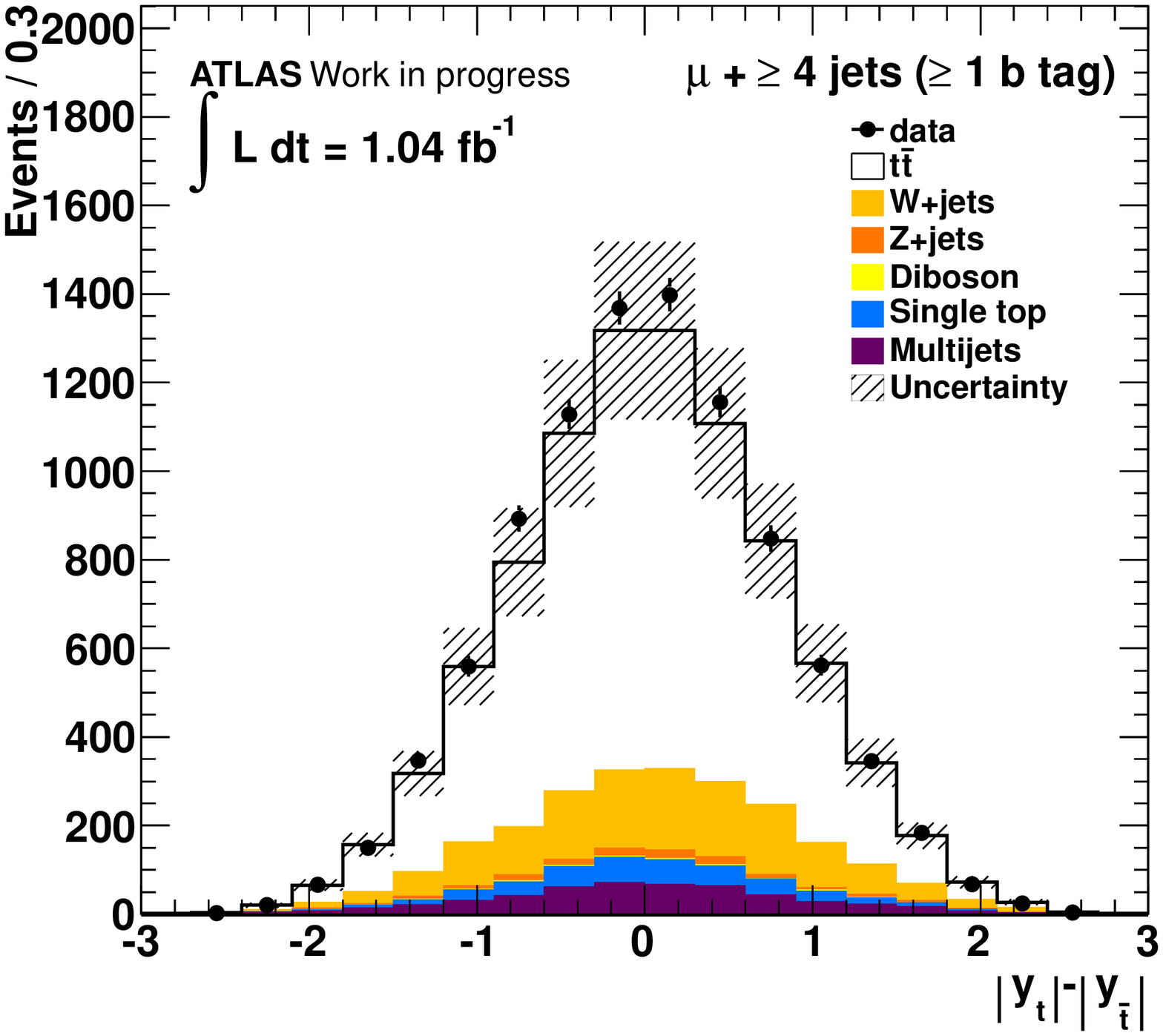}
    \quad\quad
    \includegraphics[width=\plotwidth]{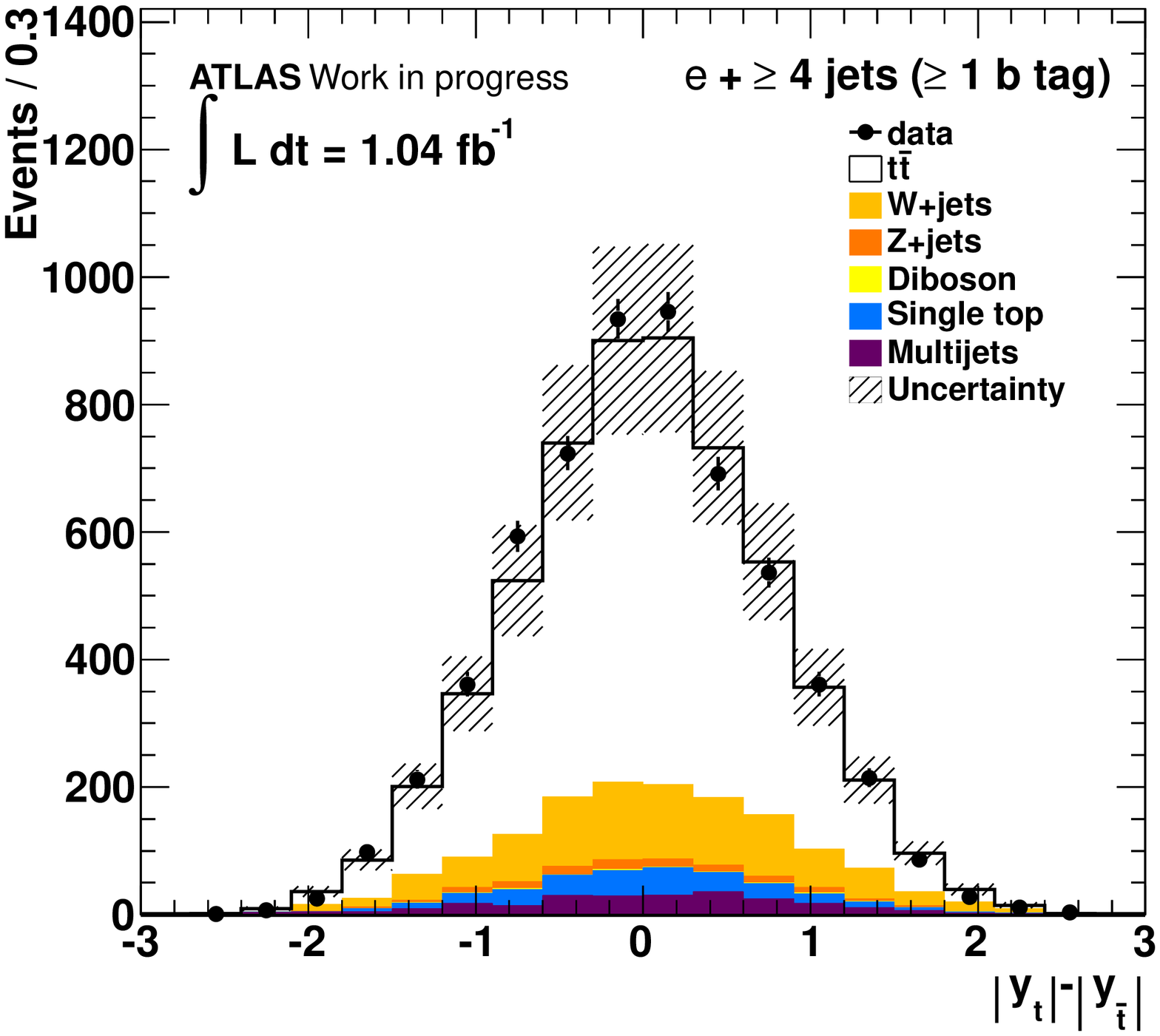}
    \includegraphics[width=\plotwidth]{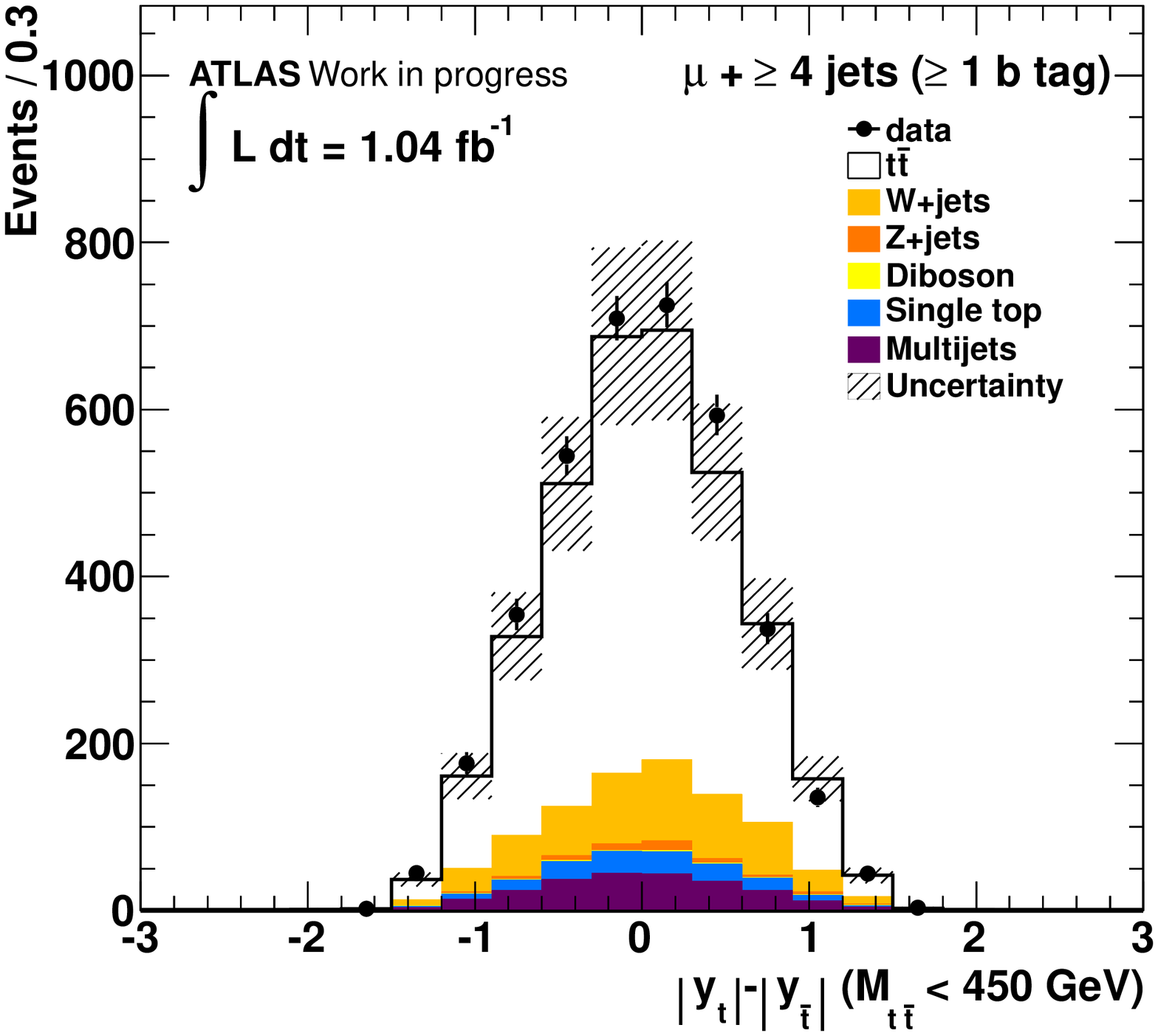}
    \quad\quad
    \includegraphics[width=\plotwidth]{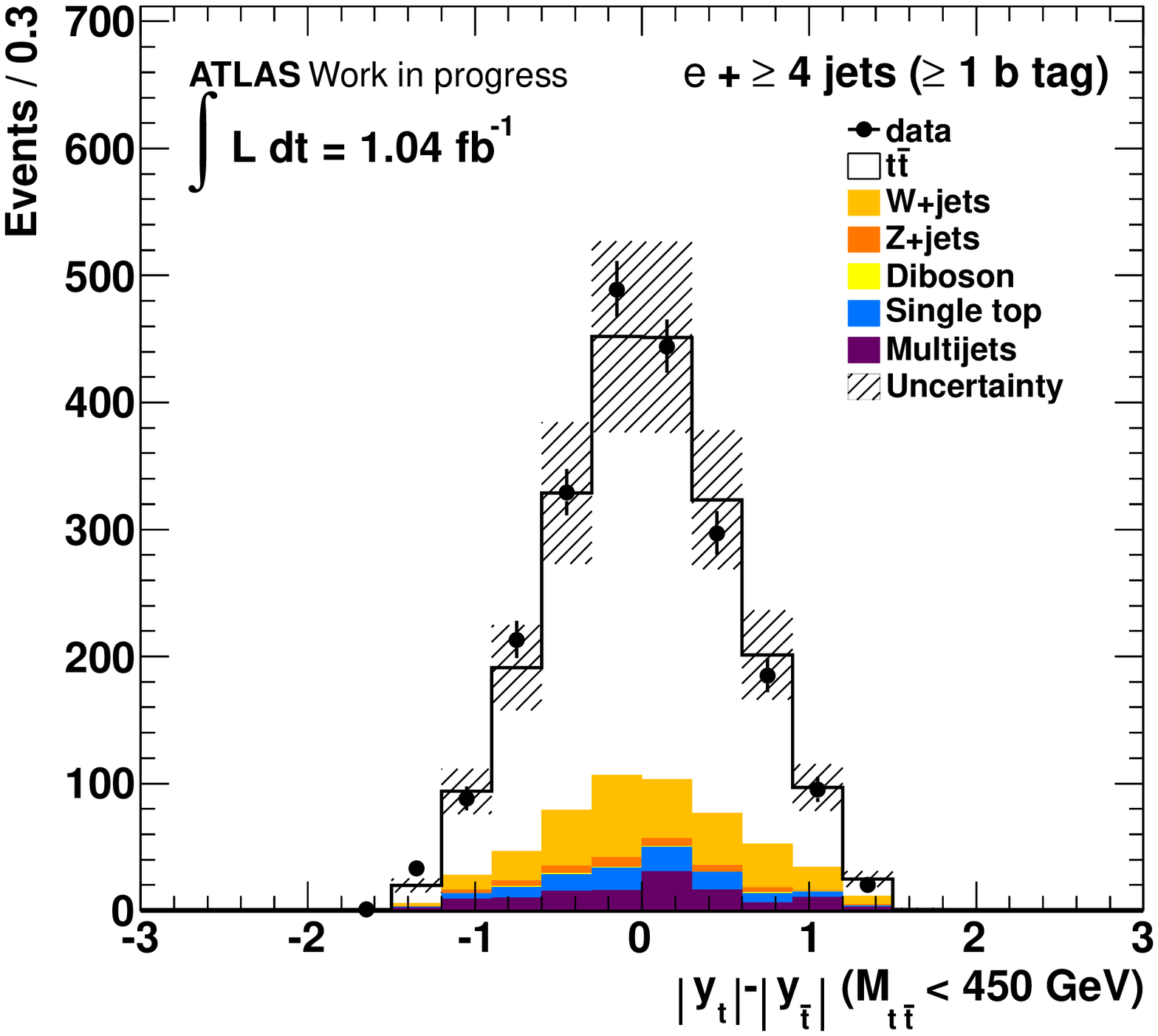}
    \includegraphics[width=\plotwidth]{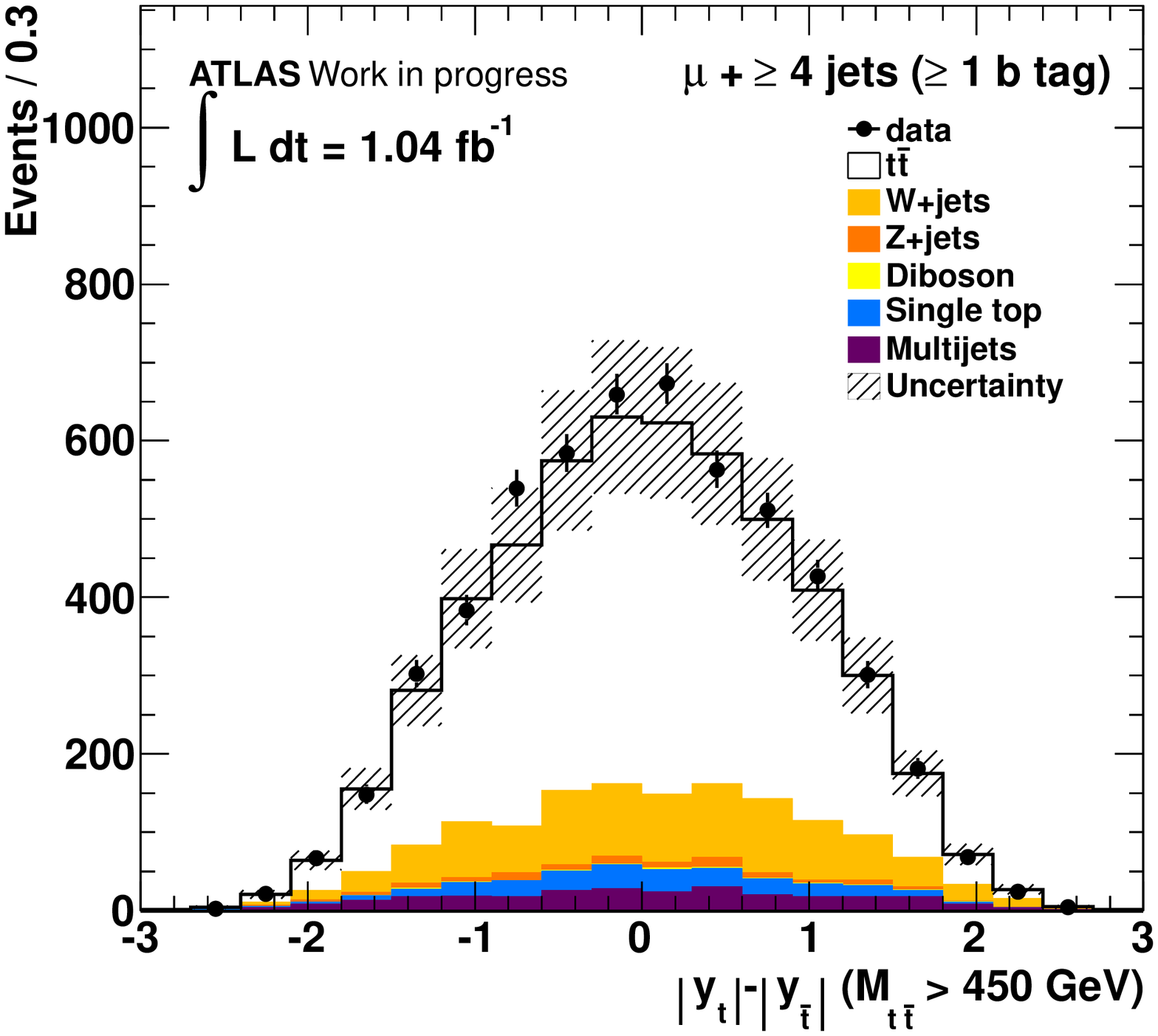}
    \quad\quad
    \includegraphics[width=\plotwidth]{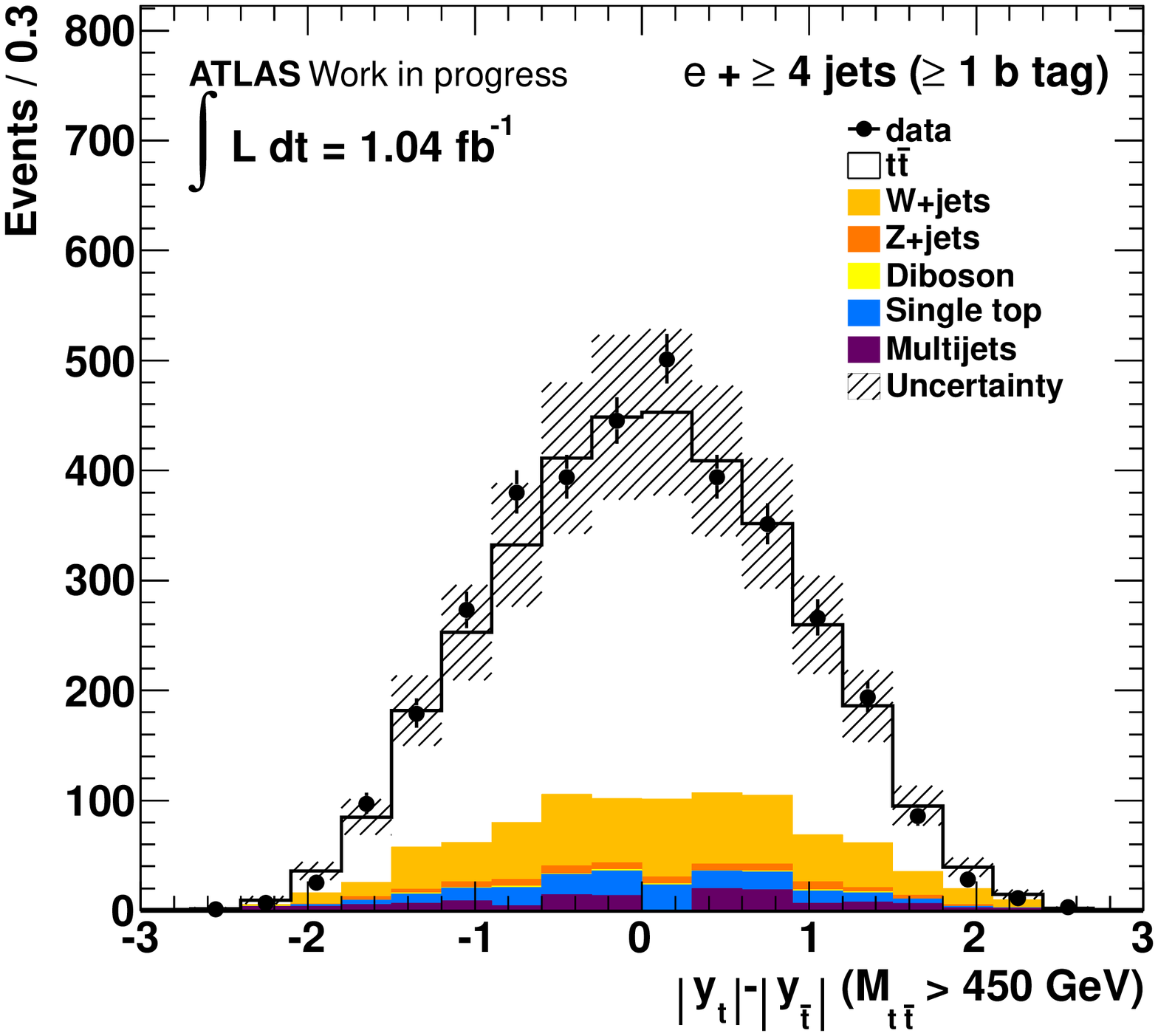}
    
    \vspace{-0.2 cm}
    \caption{Distributions for $|y_t| - |y_{\bar{t}}|$, on the left for the muon+jets, on the right for the electron+jets channel. The top row shows the distribution for the inclusive measurement, while the lower rows show the corresponding distributions requiring $M_{t\bar{t}} < 450$\,GeV and $M_{t\bar{t}} > 450$\,GeV, respectively. Uncertainties are statistical and for $W$+jets also include systematic uncertainties on normalisation. For the QCD multijet background, a conservative 100\,\% systematic uncertainty was assumed. In addition, the uncertainties on luminosity, jet energy scale, $b$ tag scale factors and \ttbar~cross-section are shown.}
    \label{fig:DeltaAbsYPlots}
  \end{center}
\end{figure}

The corresponding distributions are shown in \mbox{Figure \ref{fig:DeltaAbsYPlots}} for both the inclusive measurement of the $|y_t| - |y_{\bar{t}}|$ distribution and the corresponding measurement for the two cases of \mbox{$M_{t \bar{t}} < 450$\,GeV} and \mbox{$M_{t \bar{t}} > 450$\,GeV}. For the $M_{t \bar{t}}$ dependent measurement, an additional requirement on the event reconstruction logarithmic likelihood of $\log{L} > -52$ (see \mbox{Figure~\ref{fig:LogL_KLF}}) has been applied in order to improve the resolution in the \ttbar~invariant mass. The respective relative resolution in $M_{t \bar{t}}$ for both channels before and after the application of the additional requirement on $\log{L}$ can be found in \mbox{Figure \ref{fig:ttbarres_before}} and \mbox{Figure \ref{fig:ttbarres_after}} in \mbox{Appendix \ref{AppControlPlots}}. The relative resolution improves from 28.8\,\% to 18.4\,\% in the muon+jets channel and from 28.8\,\% to 17.7\,\% in the electron+jets channel.

The resulting charge asymmetries for the observable $A_C^{\text{reco}}$ obtained after background subtraction can be found in \mbox{Table \ref{table:ChAsymm}} for both the muon+jets channel and the electron+jets channel. The observed results are shown alongside the predicted \ttbar~charge asymmetries obtained with \mcnlo~for comparison. The uncertainties on the prediction correspond to the limited Monte Carlo statistics in the used sample after the applied selection.
\begin{table}[!htbp]
\begin{center}
\begin{tabular}{|l|c|c|}
\hline
Measurement                & Observed                                                     & Predicted (\mcnlo) \\
\hline
\hline
\multicolumn{3}{|c|}{Muon+jets Channel} \\
\hline
Inclusive                  & $-0.010 \pm 0.015\,\text{(stat.)} \pm 0.008\,\text{(syst.)}$ & $\phantom{-}0.0021 \pm 0.0011\,\text{(stat.)}$ \\
$M_{t\bar{t}} < 450$\,GeV  & $-0.012 \pm 0.024\,\text{(stat.)} \pm 0.007\,\text{(syst.)}$ & $-0.0043 \pm 0.0018\,\text{(stat.)}$ \\
$M_{t\bar{t}} > 450$\,GeV  & $-0.011 \pm 0.022\,\text{(stat.)} \pm 0.007\,\text{(syst.)}$ & $\phantom{-}0.0064 \pm 0.0015\,\text{(stat.)}$ \\
\hline
\hline
\multicolumn{3}{|c|}{Electron+jets Channel} \\
\hline
Inclusive                  & $-0.034 \pm 0.019\,\text{(stat.)} \pm 0.010\,\text{(syst.)}$ & $-0.0003 \pm 0.0014\,\text{(stat.)}$ \\
$M_{t\bar{t}} < 450$\,GeV  & $-0.060 \pm 0.033\,\text{(stat.)} \pm 0.019\,\text{(syst.)}$ & $\phantom{-}0.0007 \pm 0.0022\,\text{(stat.)}$ \\
$M_{t\bar{t}} > 450$\,GeV  & $-0.032 \pm 0.028\,\text{(stat.)} \pm 0.015\,\text{(syst.)}$ & $-0.0010 \pm 0.0017\,\text{(stat.)}$ \\
\hline
\end{tabular}
\caption{Measured values of the charge asymmetry observable $A_C^{\text{reco}}$ for the muon+jets and electron+jets channel after subtraction of the various background contributions. The results for the inclusive measurement and the respective measurements for $M_{t \bar{t}} < 450$\,GeV and $M_{t \bar{t}} > 450$\,GeV are shown together with the \mcnlo~predictions.}
\label{table:ChAsymm}
\end{center}
\end{table}
As can be seen in the table, the obtained results were compatible with the \mcnlo~Standard Model prediction in all cases within the statistical and systematic uncertainties. However, in the electron+jets channel, a tendency towards more negative asymmetries was observed, in particular for the measurement of $A_C^{\text{reco}}$ for $M_{t \bar{t}} < 450$\,GeV. A breakdown of the individual systematic uncertainties taken into account can be found in \mbox{Table \ref{Tab:SystematicsMeas}} and \mbox{Table \ref{Tab:SystematicsMeas2D}} of \mbox{Appendix \ref{App:Systematics}} for completeness.

\begin{figure}[h!tb]
  \begin{centering}
    \mbox{
      \includegraphics[width=\plotwidth]{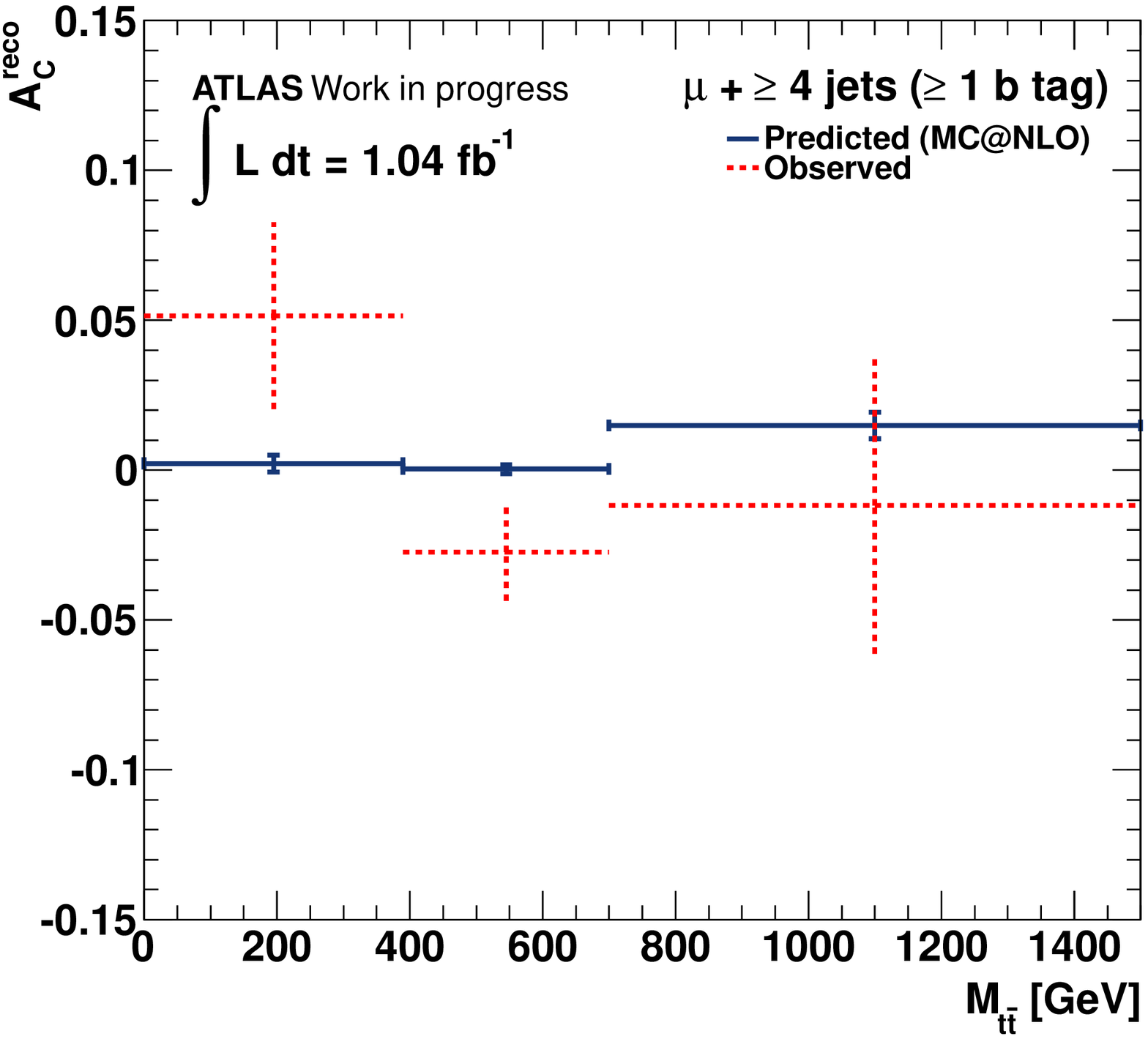}
      \quad\quad
      \includegraphics[width=\plotwidth]{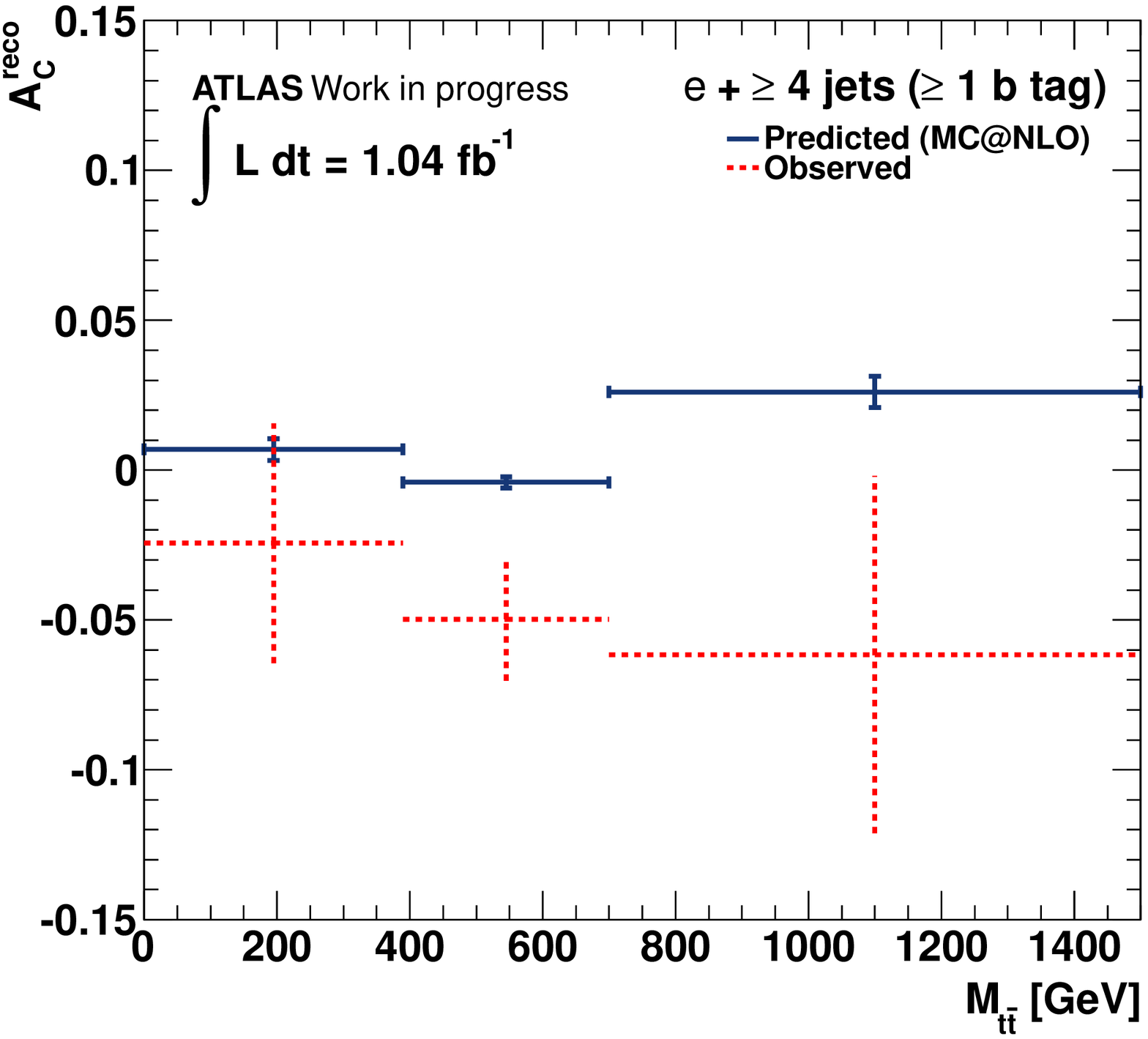}
    }
    
    \vspace{-0.2 cm}
    \caption[\quad Integrated charge asymmetry as a function of $M_{t \bar{t}}$]{Integrated charge asymmetry $A_C^{\text{reco}}$ after background subtraction as a function of $M_{t \bar{t}}$. The observed asymmetries (red, dashed) and the asymmetries predicted by \mcnlo~(blue) are shown. Uncertainties are statistical only.}
    \label{fig:ACrecovsmttbar}
  \end{centering}
\end{figure}

In addition, the integrated charge asymmetry $A_C^{\text{reco}}$ after background subtraction as a function of the invariant \ttbar~mass, $M_{t \bar{t}}$ is shown in \mbox{Figure \ref{fig:ACrecovsmttbar}} together with the Standard Model prediction obtained from simulated \ttbar~events.
The observed asymmetry values show no statistically significant deviation from the Standard Model prediction. Since, in addition, no systematic uncertainties were included, the observed discrepancies are negligible.

\section{Unfolding}
\label{chap:results:unfolding}
In order to perform the Bayes iterative unfolding and to correct the measured differential and integrated charge asymmetry for detector resolution and acceptance effects, the corresponding response matrix has been obtained using the full set of \ttbar~signal Monte Carlo events generated with \mcnlo~(\numprint{15000000} simulated events). A unified binning for the true and reconstructed distributions (and consequently for the response matrices) in the variable $|y_t| - |y_{\bar{t}}|$ has been used, employing six bins with variable width to ensure sufficient statistics in the tails of the respective distributions. Bin edges at $\{ -3.0, -1.2, -0.6, 0.0, 0.6, 1.2, 3.0 \}$ for the inclusive unfolding and $\{ -3.00, -0.96, -0.48, 0.0, 0.48, 0.96, 3.00 \}$ for the simultaneous unfolding in $|y_t| - |y_{\bar{t}}|$ and $M_{t \bar{t}}$ were chosen.

Furthermore, for the simultaneous unfolding in $|y_t| - |y_{\bar{t}}|$ and $M_{t \bar{t}}$, an additional requirement on the event reconstruction logarithmic likelihood of $\log{L} > -52$ (see \mbox{Figure~\ref{fig:LogL_KLF}}) has been applied in order to improve the resolution of the \ttbar~invariant mass in the same way as for the measurement of the charge asymmetry before unfolding.

\begin{figure}[h!tb]
  \begin{centering}
    \mbox{
      \includegraphics[width=\plotwidth]{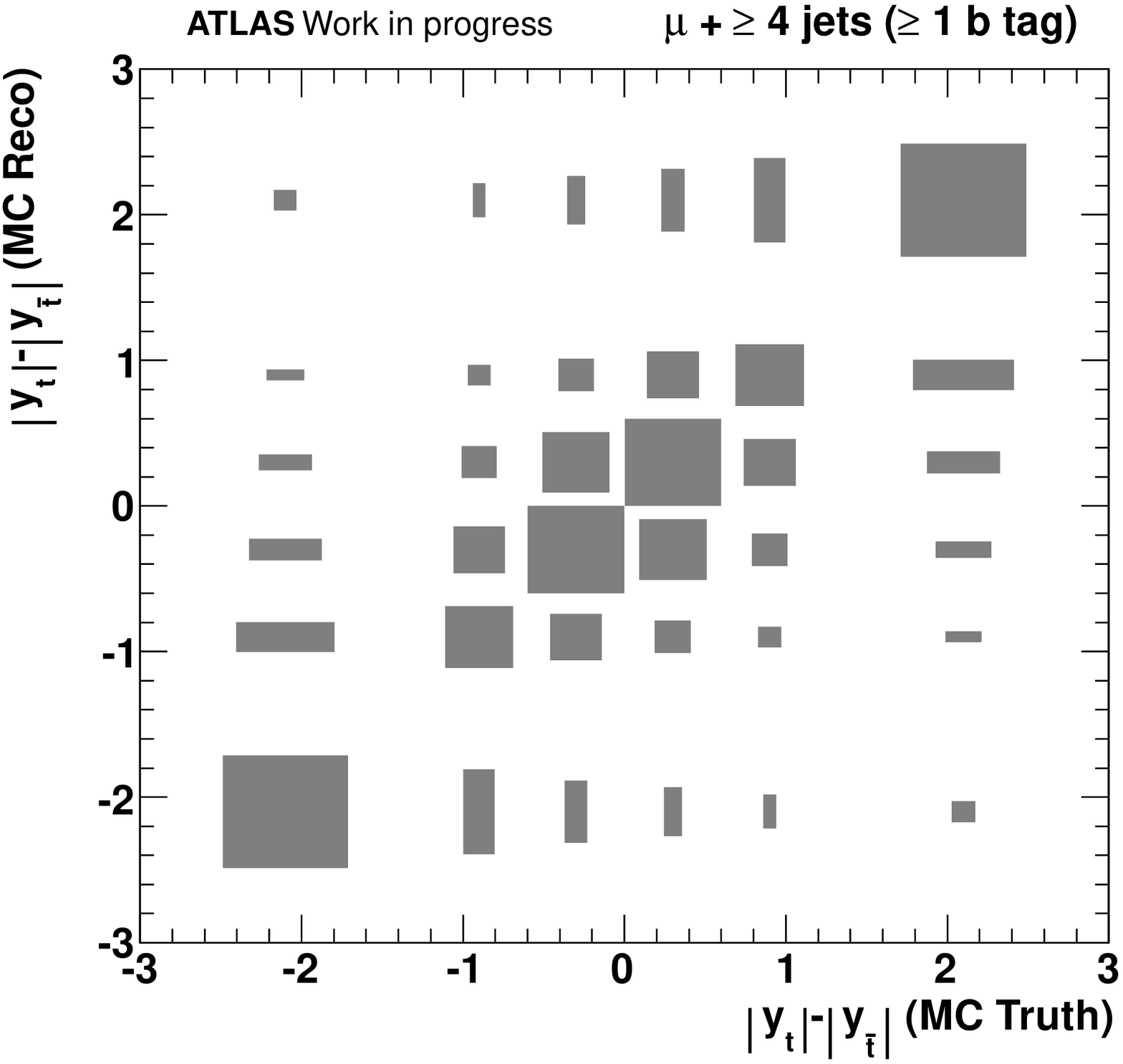}
      \quad\quad
      \includegraphics[width=\plotwidth]{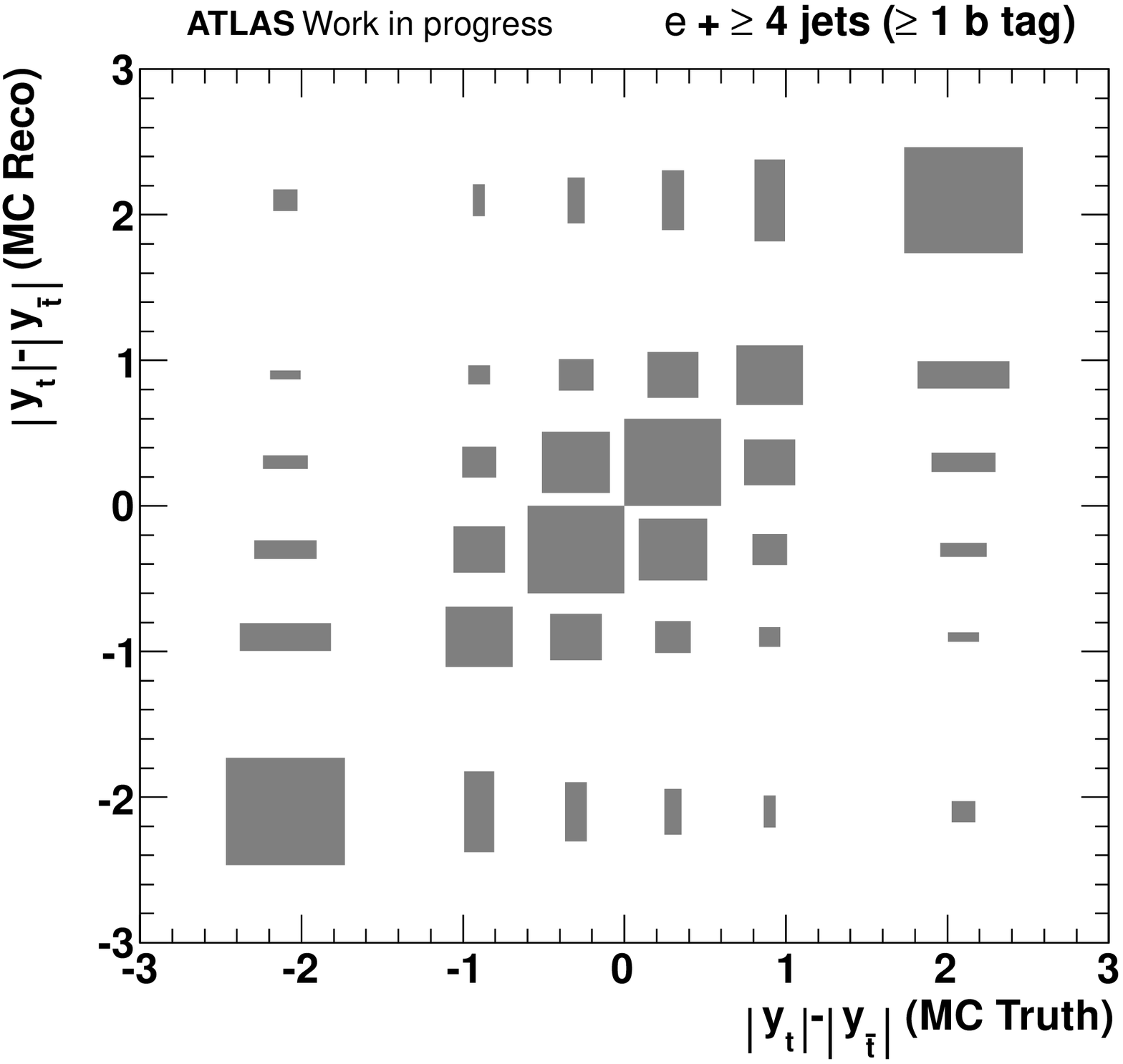}
    }
    
    \vspace{-0.2 cm}
    \caption[\quad Inclusive unfolding response matrices]{Unfolding response matrices for the inclusive unfolding of the charge asymmetry distribution $|y_t| - |y_{\bar{t}}|$. The bin migration probability corresponds to the box sizes displayed in the response matrix, and is shown independently for the muon+jets channel (left) and the electron+jets channel (right).}
    \label{fig:unfolding:responsematrix}
  \end{centering}
\end{figure}

\begin{figure}[h!tb]
  \begin{centering}
    \mbox{
      \includegraphics[width=\plotwidth]{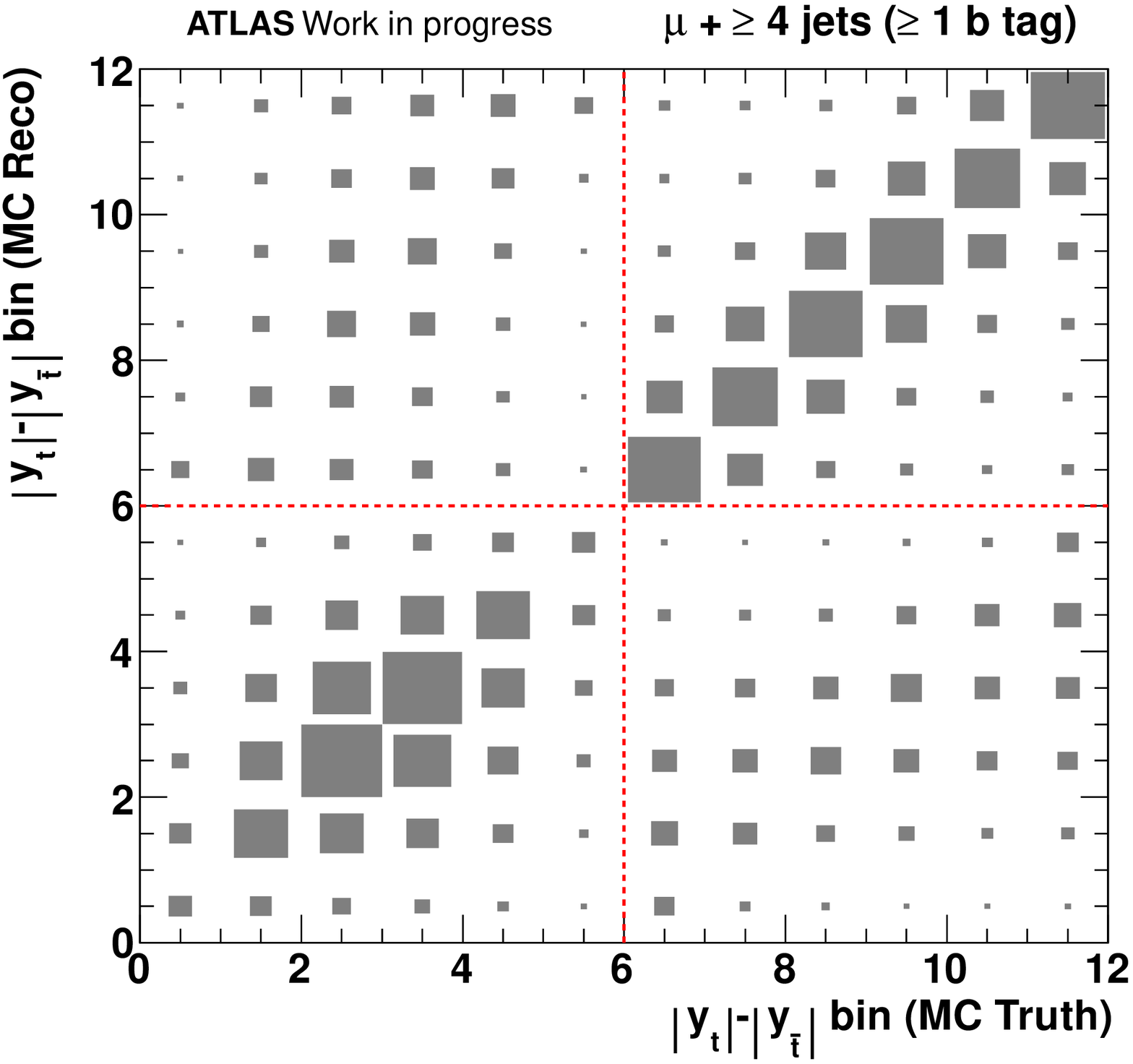}
      \quad\quad
      \includegraphics[width=\plotwidth]{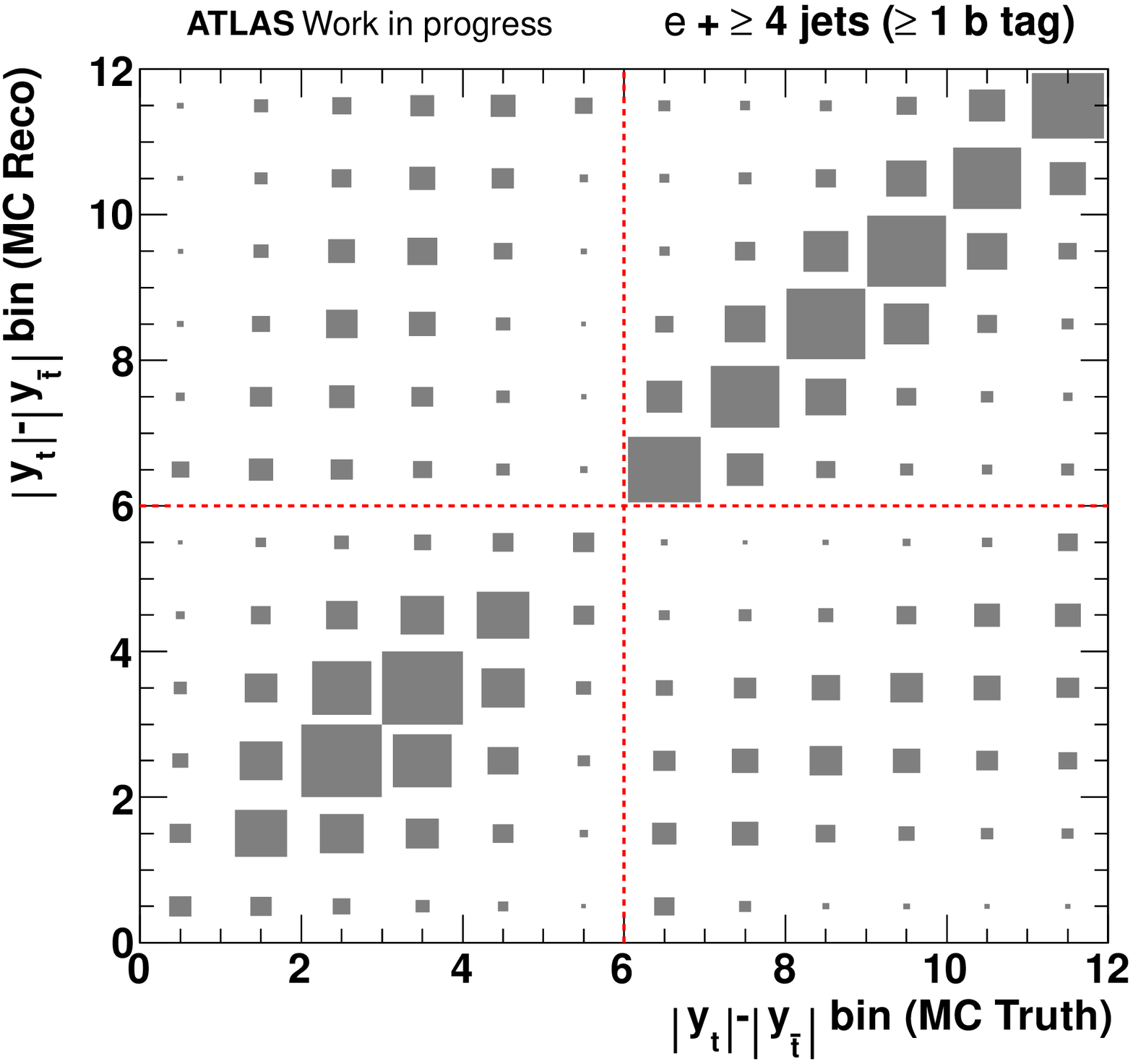}
    }
    
    \vspace{-0.2 cm}
    \caption[\quad Unfolding response matrices]{Unfolding response matrices for the simultaneous unfolding of the distribution in $|y_t| - |y_{\bar{t}}|$ and $M_{t \bar{t}}$ showing the discrete bin correspondence of the truth distribution with respect to the distribution obtained after event selection and reconstruction. The bin migration probability corresponds to the box sizes displayed in the response matrix, and is shown independently for the muon+jets channel (left) and the electron+jets channel (right). The four quadrants represent the respective transition probabilities for the true and reconstructed $|y_t| - |y_{\bar{t}}|$ distributions in the low and high $M_{t \bar{t}}$ regions of phase space.}
    \label{fig:unfolding:responsematrix:2D}
  \end{centering}
\end{figure}

The response matrix representation used in the inclusive measurement of the distribution of $|y_t| - |y_{\bar{t}}|$ and of $A_C^{\text{unf}}$ is shown in \mbox{Figure \ref{fig:unfolding:responsematrix}} for both the muon+jets and electron+jets channel.
Similarly, the corresponding matrices for the simultaneous unfolding in $|y_t| - |y_{\bar{t}}|$ and $M_{t \bar{t}}$ can be found in \mbox{Figure \ref{fig:unfolding:responsematrix:2D}}. The transition probability information of the corresponding two-dimensional distributions (with $2 \times 6$ bins) was encoded in the matrix by linearisation of the respective histograms, concatenating the bins from the two $M_{t \bar{t}}$ regions into a single one dimensional histogram (with $1 \times 12$ bins). Hence, the associated response matrix contained $2 \times 2$ quadrants, corresponding to the transition probabilities for the true and reconstructed $|y_t| - |y_{\bar{t}}|$ distributions in the low and high $M_{t \bar{t}}$ regions of phase space, respectively.

A closure test has been performed to verify that the Bayesian iterative unfolding approach can be used to recover an arbitrary asymmetry present in the true distribution. The full \mcnlo~\ttbar~signal sample was used as a basis to verify the correct unfolding response for different artificially injected truth level asymmetries. This was achieved by performing an event-by-event reweighting of the \ttbar~signal sample to asymmetries of -10\,\%, -5\,\%, 0\,\%, 5\,\% and 10\,\% in the variable $A_C^{\text{true}}$ by systematically increasing the weights of events with $|y_t| - |y_{\bar{t}}| < 0$ and decreasing the weights of events with $|y_t| - |y_{\bar{t}}| > 0$ by the same amount.

Pseudoexperiments, using statistically independent sets of pseudodata corresponding to the statistics expected in the used data sample (after background subtraction), were conducted based on a Poissonian fluctuation of the respective distribution in $|y_t| - |y_{\bar{t}}|$, taking into account the additional event weights determined in the asymmetry reweighting. Ensemble tests were performed to confirm the linearity of the unfolding response in the true value of the chosen charge asymmetry observable $A_C^{\text{true}}$. Furthermore, the dependency of the obtained results on the regularisation of the unfolding procedure was studied. 

\begin{figure}[!htbp]
  \begin{center}
    \includegraphics[width=\plotwidth]{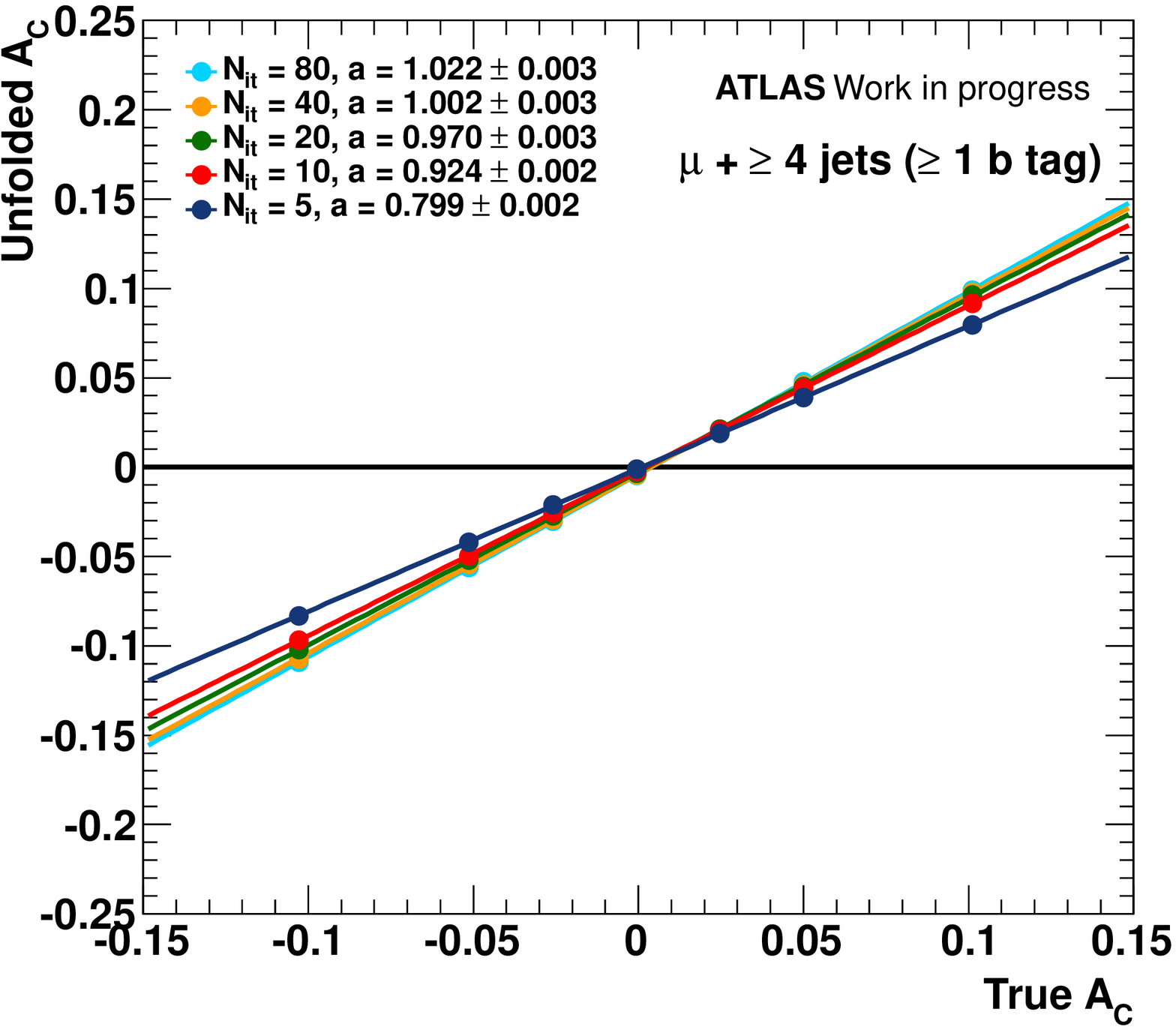}
    \quad\quad
    \includegraphics[width=\plotwidth]{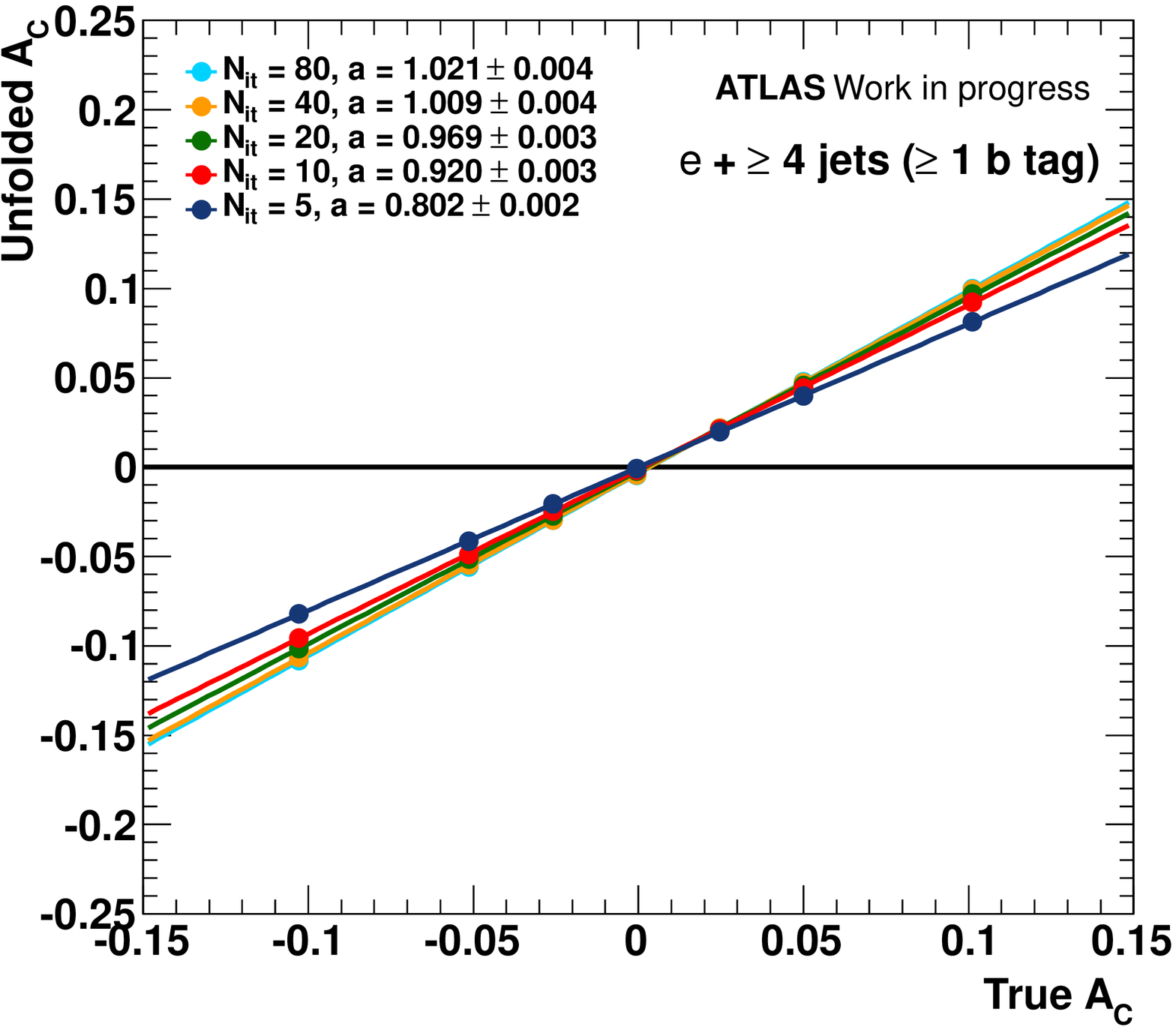}
    \includegraphics[width=\plotwidth]{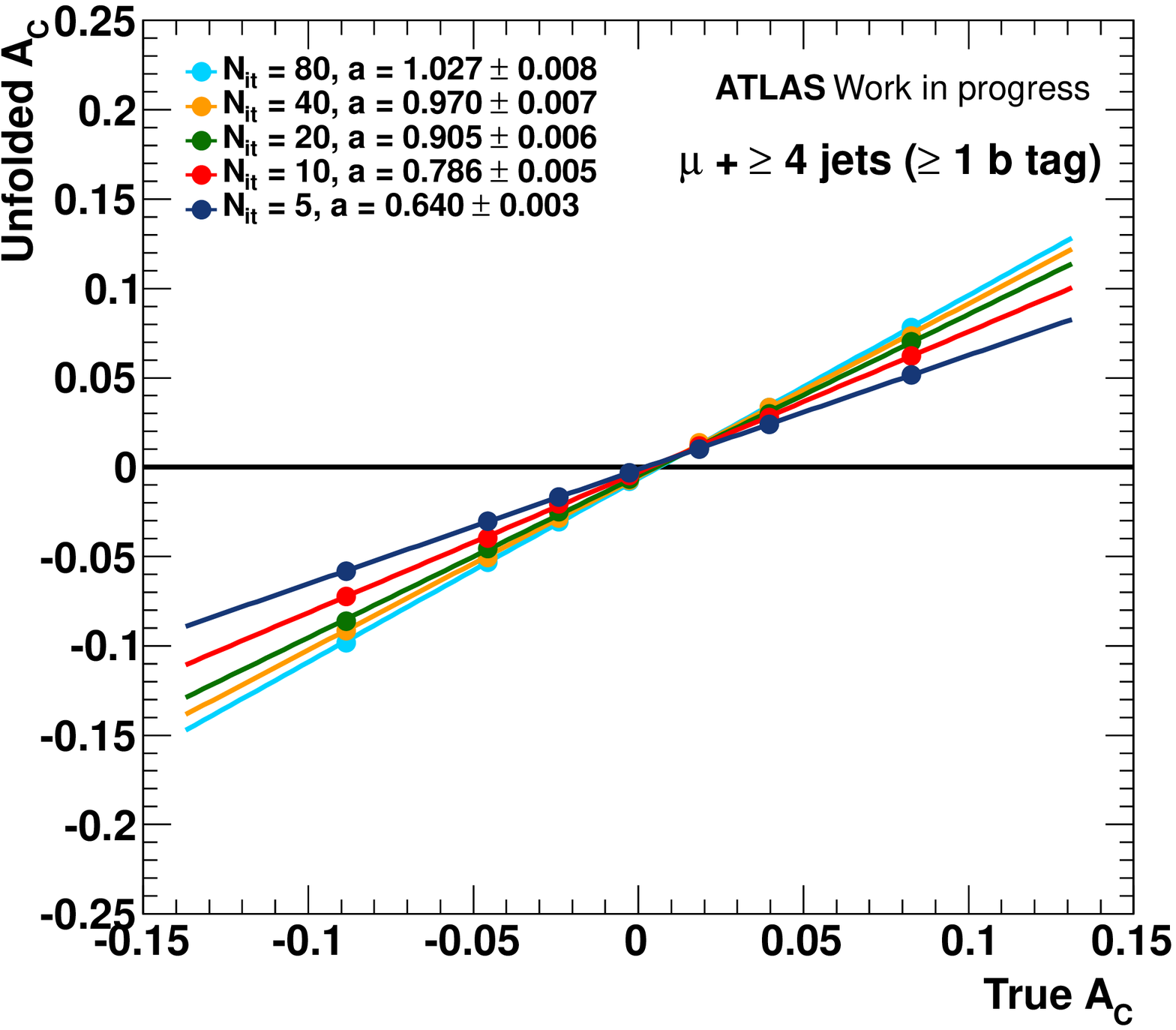}
    \quad\quad
    \includegraphics[width=\plotwidth]{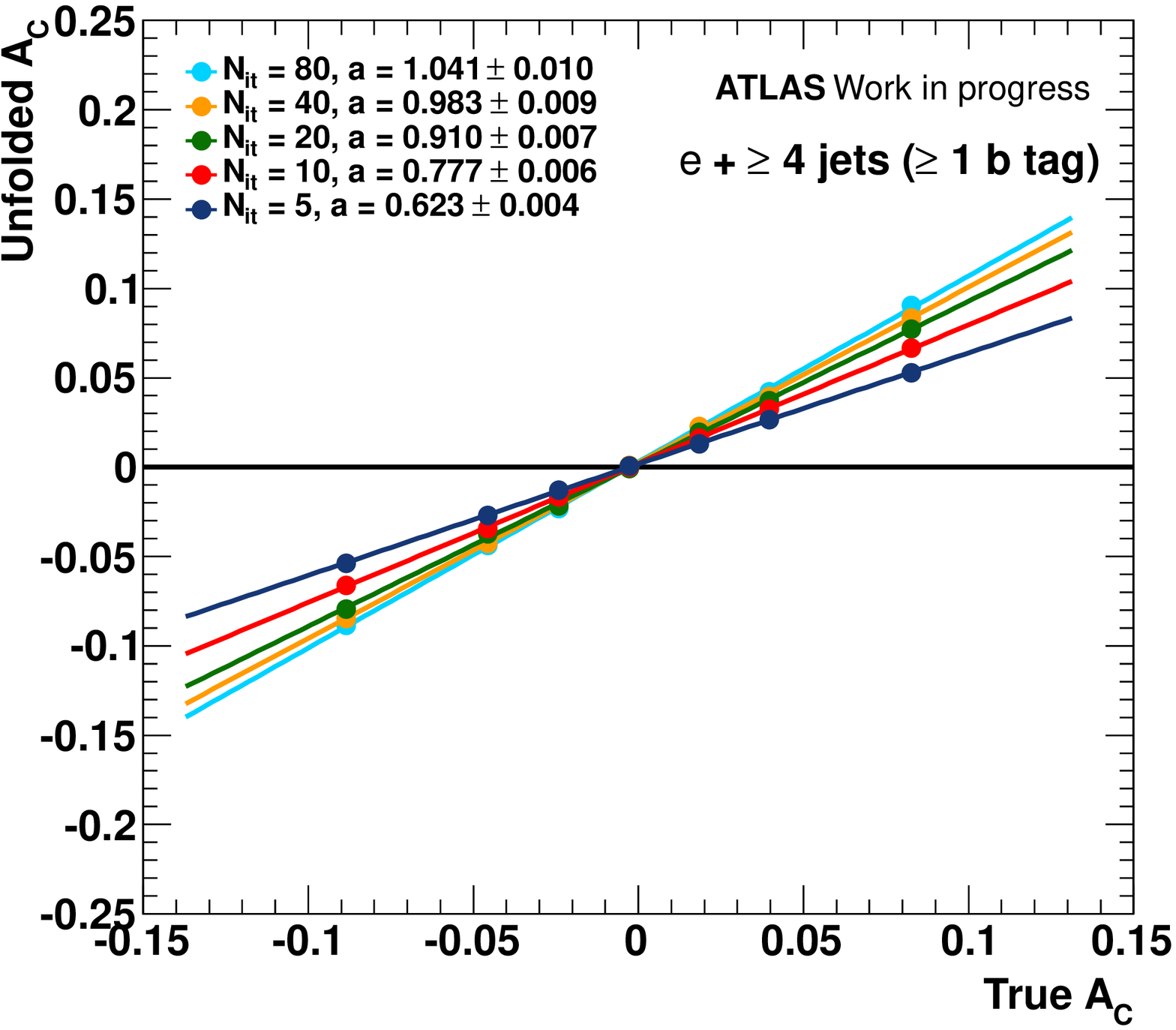}
    \includegraphics[width=\plotwidth]{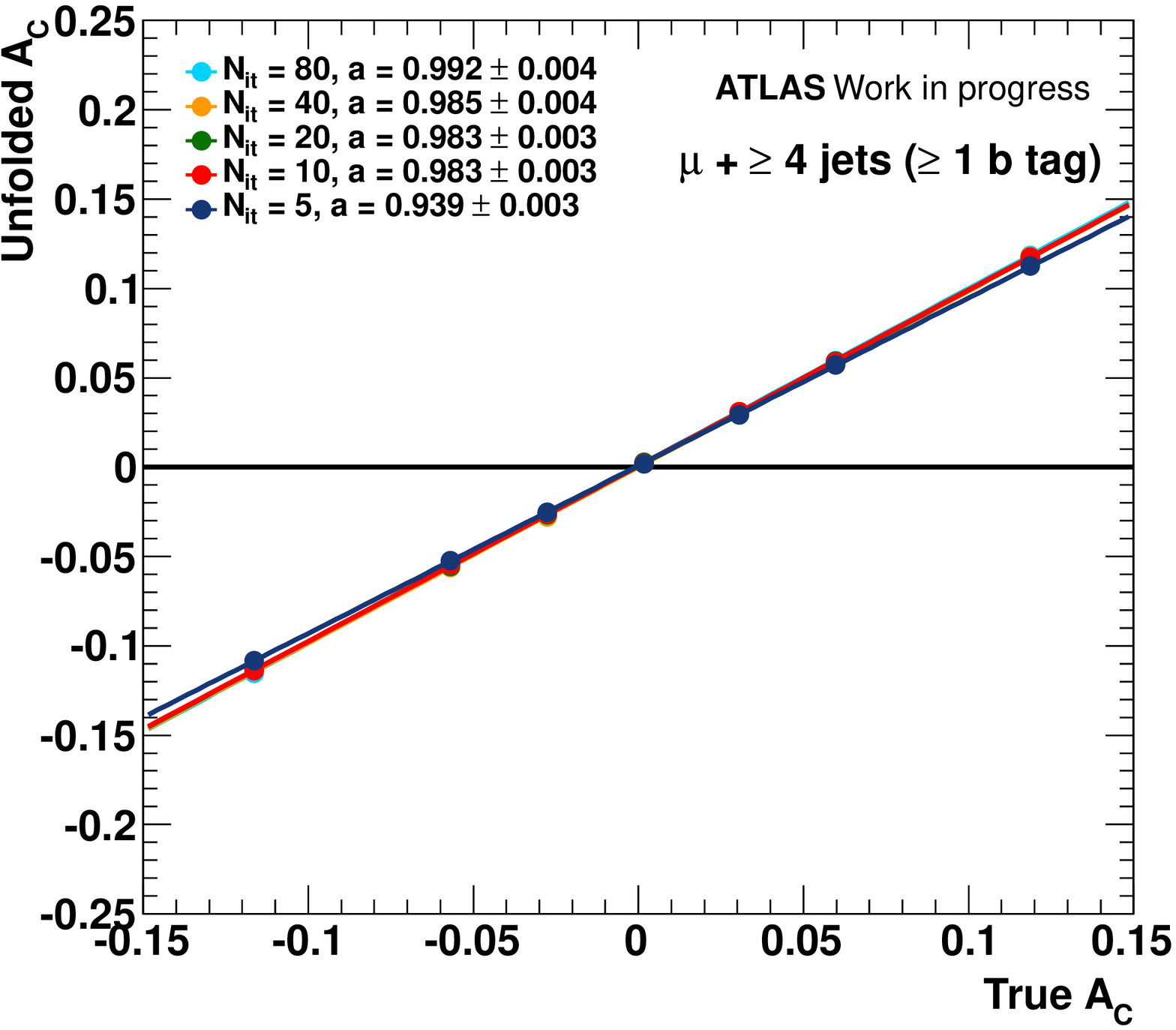}
    \quad\quad
    \includegraphics[width=\plotwidth]{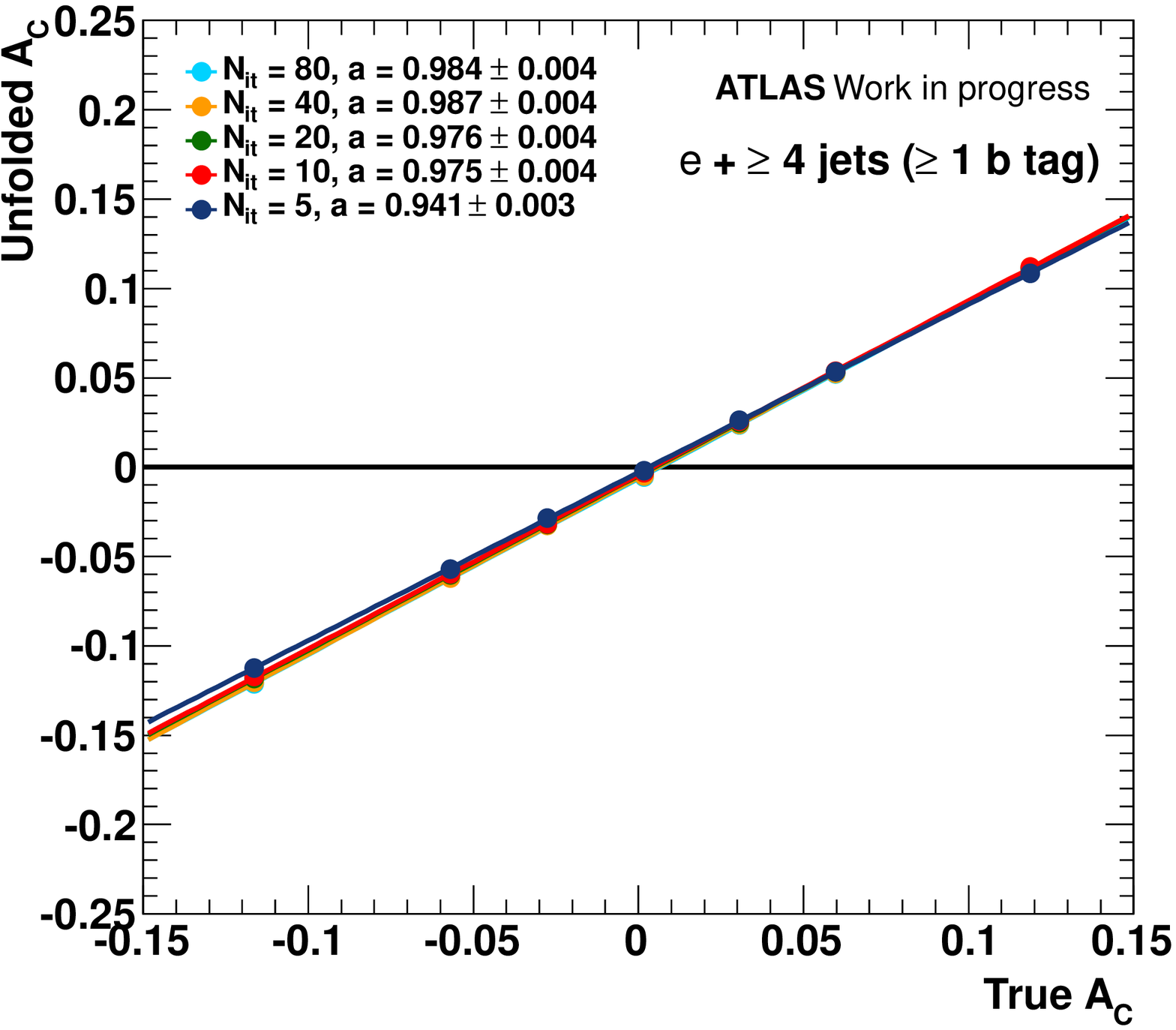}
    
    \vspace{-0.2 cm}
    \caption{The obtained overall inclusive asymmetry after unfolding as a function of injected true asymmetry $A_C^{\text{true}}$ for different regularisation parameters $N_{\text{It}}$ for both the muon+jets channel (left) and electron+jets channel (right). The top row shows the respective distribution for the inclusive measurement, while the lower rows show the corresponding distributions for $M_{t\bar{t}} < 450$\,GeV and $M_{t\bar{t}} > 450$\,GeV, respectively. For the simultaneous unfolding in $|y_t| - |y_{\bar{t}}|$ and $M_{t \bar{t}}$, a cut on the event reconstruction likelihood $\log{L}$ was applied to improve the $M_{t\bar{t}}$ resolution of the selected events.}
    \label{fig:unfolding:calib}
  \end{center}
\end{figure}

The average unfolded value of $A_C^{\text{unf}}$ obtained from the sets of pseudodata as a function of the injected true value of $A_C^{\text{true}}$ can be found for both the inclusive unfolding and the simultaneous unfolding in $|y_t| - |y_{\bar{t}}|$ and $M_{t \bar{t}}$ in \mbox{Figure \ref{fig:unfolding:calib}} for different regularisation strengths, using $N_{\text{It}} = 5$, 10, 20, 40 and 80 iterations in the Bayesian unfolding.

A straight line fit using a parametrisation of the form
\begin{equation}
A_C^{\text{unf}} = a \cdot A_C^{\text{true}} + b
\end{equation}
has been performed, where $a$ and $b$ denote the slope and offset parameters, respectively. The procedure has been repeated for each of the individual choices of $N_{\text{It}}$ to obtain the calibration curves in order to verify the linearity and determine the respective slopes and offsets of the fit. These provided a measure for the average remaining bias in the unfolded differential and integrated asymmetry after unfolding for different regularisation strengths. As can be seen in the fits, a slope close to one could be achieved in all cases, indicating a proper average correspondence of the unfolded asymmetry to the injected true value, independent of the strength of the injected asymmetry.

In addition, the expected statistical uncertainty on $A_C^{\text{unf}}$ after the unfolding procedure is shown in \mbox{Figure \ref{fig:unfolding:calibuncert}} in \mbox{Appendix \ref{AppControlPlots}} as a function of $N_{\text{It}}$. As expected, the uncertainty increases with the number of iterations used in the regularisation due to the increasing sensitivity to bin-to-bin statistical fluctuations in the inversion process of the response matrix. The expected uncertainties are independent of the injected value of $A_C^{\text{true}}$.

In order to achieve a convergent state in the iterative unfolding process, the dependency of the unfolded result as a function of the number of iterations has been studied with respect to a defined convergence criterion. For individual ensembles of pseudodata, the unfolding procedure was considered to be converged if the absolute change in the unfolded asymmetry in terms of $A_C^{\text{unf}}$, $\Delta A_C^{\text{unf}}$, for a given amount of iterations $N_{\text{It}} = i$ with respect to the previous amount of iterations $N_{\text{It}} = i-1$ was lower than 1\,\permil, i.e. if
\begin{equation}
\Delta A_C^{\text{unf}} = |A_C^{\text{unf}}(N_{\text{It}} = i) - A_C^{\text{unf}}(N_{\text{It}} = i-1)| < 0.001.
\end{equation}
\mbox{Figure \ref{fig:conv1}} in \mbox{Appendix \ref{AppControlPlots}} shows the percentage of ensembles which have reached the defined convergence criterion as a function of regularisation strength, parametrised by the amount of iterations $N_{\text{It}}$. At $N_{\text{It}} = 40$ (inclusive unfolding), and $N_{\text{It}} = 80$ (simultaneous unfolding in $|y_t| - |y_{\bar{t}}|$ and $M_{t\bar{t}}$) iterations, all ensembles of pseudodata were ensured to have reached convergence for all injected value of $A_C^{\text{true}}$. Hence, these regularisation strengths were chosen for the inclusive unfolding and the simultaneous unfolding in $|y_t| - |y_{\bar{t}}|$ and $M_{t\bar{t}}$, respectively. An additional systematic uncertainty of 1\,\permil, corresponding directly to the choice of convergence criterion was assigned to the unfolded value of $A_C^{\text{unf}}$ (c.f. \mbox{Chapter \ref{Systematics}}).

In addition, a further cross-check was performed to verify that the unfolding procedure reached a convergent state by determining the average standard deviation of the variable $\Delta A_C^{\text{unf}}$ for the used set of ensembles as a function of $N_{\text{It}}$, which shows a monotonous falling behaviour in all cases as expected. The corresponding additional control plots can be found in \mbox{Figure \ref{fig:conv2}} in \mbox{Appendix \ref{AppControlPlots}} for completeness.

Despite the large amount of iterations necessary to reach convergence, in particular for the simultaneous unfolding in $|y_t| - |y_{\bar{t}}|$ and $M_{t \bar{t}}$ and the associated expected statistical uncertainties, this approach allowed performance of the unfolding following a well-defined and model-independent procedure. Furthermore, the remaining expected bias in the unfolding procedure was minimised, as can be seen in the calibration curves in \mbox{Figure \ref{fig:unfolding:calib}} for the used choices of $N_{It} = 40$ and $N_{It} = 80$, respectively. The corresponding slope and offset parameters extracted from the straight line fit in the calibration can be found in \mbox{Table \ref{tab:calibfactors}} for completeness.
\begin{table}[h!tdp]
\begin{center}
\begin{tabular}{|l|c|c|c|c|}
\hline
           \multicolumn{1}{|c|}{} & \multicolumn{2}{c|}{electron+jets channel} & \multicolumn{2}{c|}{muon+jets channel} \\
\cline{2-5}\multicolumn{1}{|c|}{Unfolding} & $a$ (slope) & $b$ (offset) & $a$ (slope) & $b$ (offset) \\
\hline\hline
Inclusive & \phantom{-}1.009 $\pm$ 0.004 & -0.0032 $\pm$ 0.0002 & \phantom{-}1.002 $\pm$ 0.003 & -0.0037 $\pm$ 0.0002 \\
\hline
$M_{t\bar{t}} < 450\,\text{GeV}$ & \phantom{-}1.041 $\pm$ 0.001 & \phantom{-}0.0030 $\pm$ 0.0005 & \phantom{-}1.027 $\pm$ 0.008 & -0.0064 $\pm$ 0.0004 \\
\hline
$M_{t\bar{t}} > 450\,\text{GeV}$ & \phantom{-}0.984 $\pm$ 0.004 & -0.0063 $\pm$ 0.0003 & \phantom{-}0.992 $\pm$ 0.004 & \phantom{-}0.0007 $\pm$ 0.0003 \\
\hline
\end{tabular}
\caption{Slopes and offsets from the linear fit in the unfolding calibration. The parameters were obtained for a linear fit of the average unfolded value of $A_C^{\text{unf}}$ as a function of the true $A_C^{\text{true}}$ value, obtained from sets of pseudoexperiments for 40 (inclusive unfolding) and 80 (simultaneous unfolding in $|y_t| - |y_{\bar{t}}|$ and $M_{t \bar{t}}$) iterations, respectively. For the simultaneous unfolding in $|y_t| - |y_{\bar{t}}|$ and $M_{t \bar{t}}$, a cut on the event reconstruction likelihood $\log{L}$ was applied to improve the $M_{t\bar{t}}$ resolution of the selected events. The shown fit parameter uncertainties are statistical only.}
\label{tab:calibfactors}
\end{center}
\end{table}
Additional closure tests have been performed using ensembles of pseudodata to quantify any remaining bias from the unfolding at the chosen regularisation strengths and a corresponding systematic uncertainty was assigned to the unfolded result accordingly (c.f. \mbox{Chapter \ref{Systematics}}), while no correction of the unfolded results for the obtained calibration was performed.

The unfolded integrated asymmetries in $A_C^{\text{unf}}$ are shown in \mbox{Table \ref{tab:unfolding}} for both the inclusive unfolding and for the simultaneous unfolding in $|y_t| - |y_{\bar{t}}|$ and $M_{t \bar{t}}$, using 40 and 80 iterations in the unfolding process, respectively.
\begin{table}[!htbp]
\begin{center}
\begin{tabular}{|l|c|c|}
\hline
Unfolding                  & Observed                                                     & Predicted (\mcnlo) \\
\hline
\hline
\multicolumn{3}{|c|}{Muon+jets Channel} \\
\hline
Inclusive                  & $-0.002 \pm 0.036\,\text{(stat.)} \pm 0.023\,\text{(syst.)}$ & $0.0056 \pm 0.0003\,\text{(stat.)}$ \\
$M_{t\bar{t}} < 450$\,GeV  & $-0.002 \pm 0.084\,\text{(stat.)} \pm 0.049\,\text{(syst.)}$ & $0.0024 \pm 0.0004\,\text{(stat.)}$ \\
$M_{t\bar{t}} > 450$\,GeV  & $-0.003 \pm 0.045\,\text{(stat.)} \pm 0.034\,\text{(syst.)}$ & $0.0086 \pm 0.0004\,\text{(stat.)}$ \\
\hline
\hline
\multicolumn{3}{|c|}{Electron+jets Channel} \\
\hline
Inclusive                  & $-0.047 \pm 0.045\,\text{(stat.)} \pm 0.028\,\text{(syst.)}$ & $0.0056 \pm 0.0003\,\text{(stat.)}$ \\
$M_{t\bar{t}} < 450$\,GeV  & $-0.196 \pm 0.119\,\text{(stat.)} \pm 0.091\,\text{(syst.)}$ & $0.0024 \pm 0.0004\,\text{(stat.)}$ \\
$M_{t\bar{t}} > 450$\,GeV  & $-0.016 \pm 0.055\,\text{(stat.)} \pm 0.035\,\text{(syst.)}$ & $0.0086 \pm 0.0004\,\text{(stat.)}$ \\
\hline
\end{tabular}
\caption{Unfolded values of the charge asymmetry observable $A_C^{\text{unf}}$ for the muon+jets and electron+jets channel. The results for the inclusive measurement and the respective results for the simultaneous unfolding in $|y_t| - |y_{\bar{t}}|$ and $M_{t \bar{t}}$ for $M_{t \bar{t}} < 450$\,GeV and $M_{t \bar{t}} > 450$\,GeV are shown. For the simultaneous unfolding in $|y_t| - |y_{\bar{t}}|$ and $M_{t \bar{t}}$, a cut on the event reconstruction likelihood $\log{L}$ was applied to improve the $M_{t\bar{t}}$ resolution of the selected events. Furthermore, the respective \mcnlo~predictions are shown.}
\label{tab:unfolding}
\end{center}
\end{table}
\begin{figure}[!htbp]
  \begin{center}
    \includegraphics[width=6.3cm]{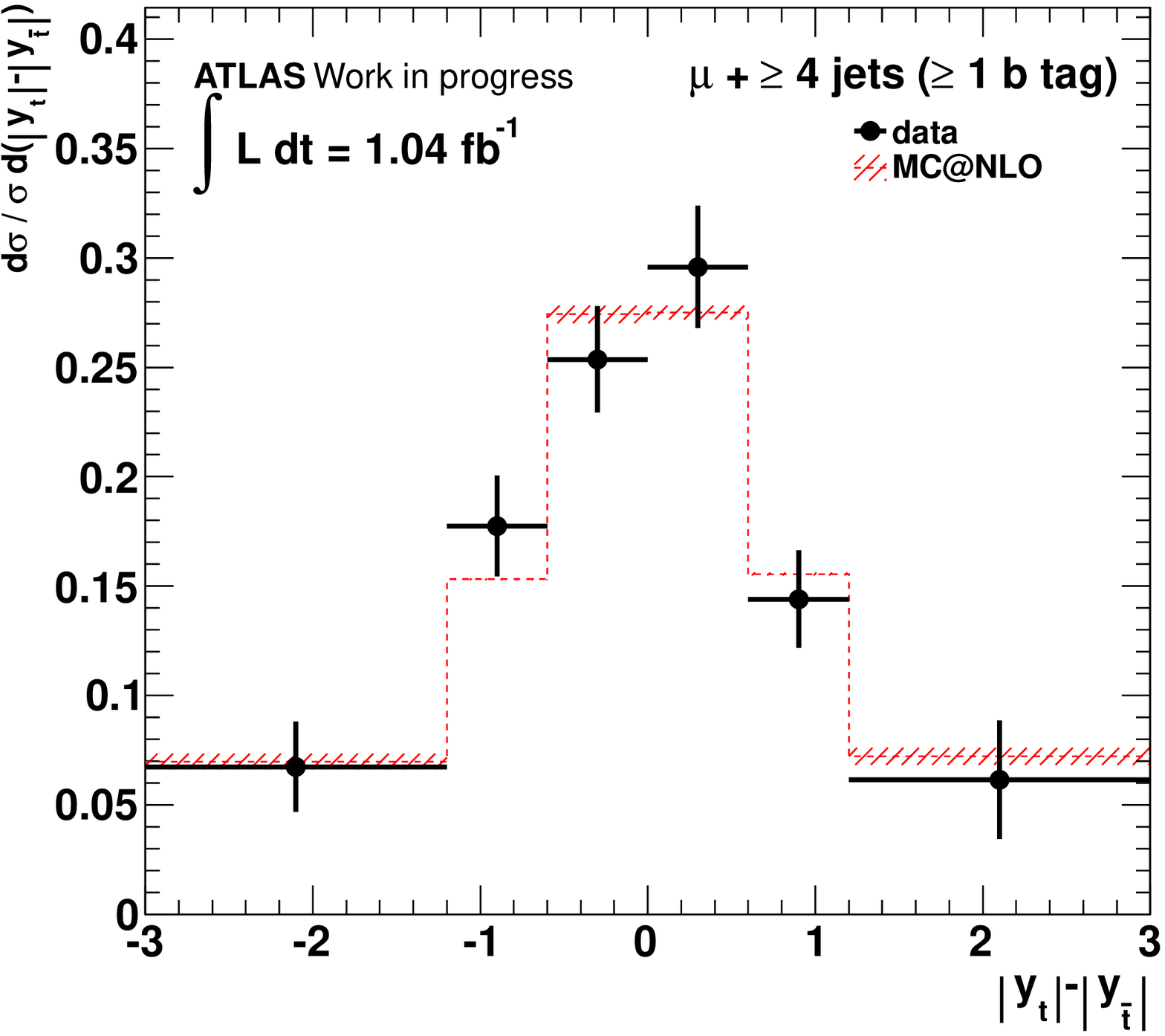}
    \quad\quad
    \includegraphics[width=6.3cm]{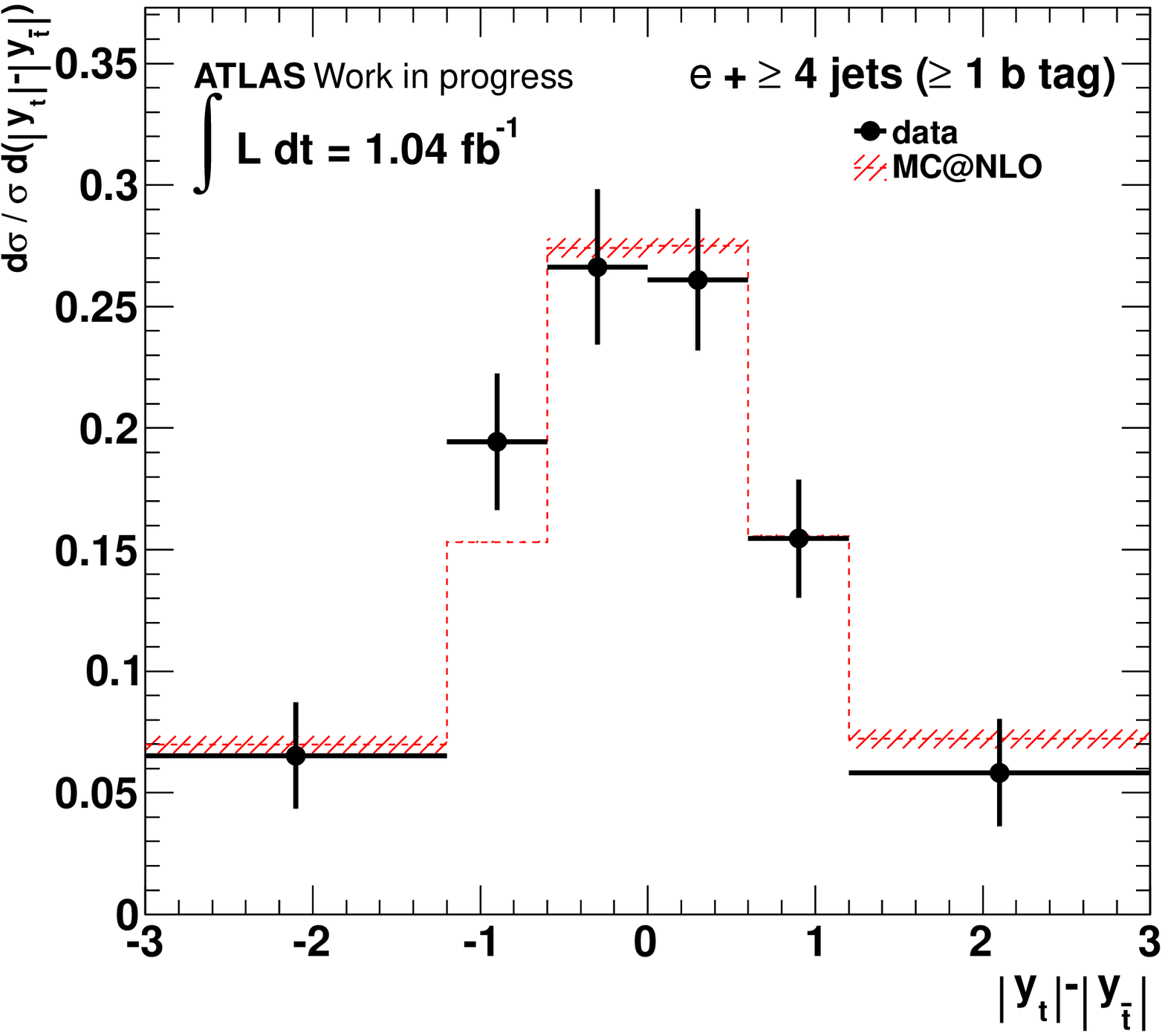}
    \includegraphics[width=6.3cm]{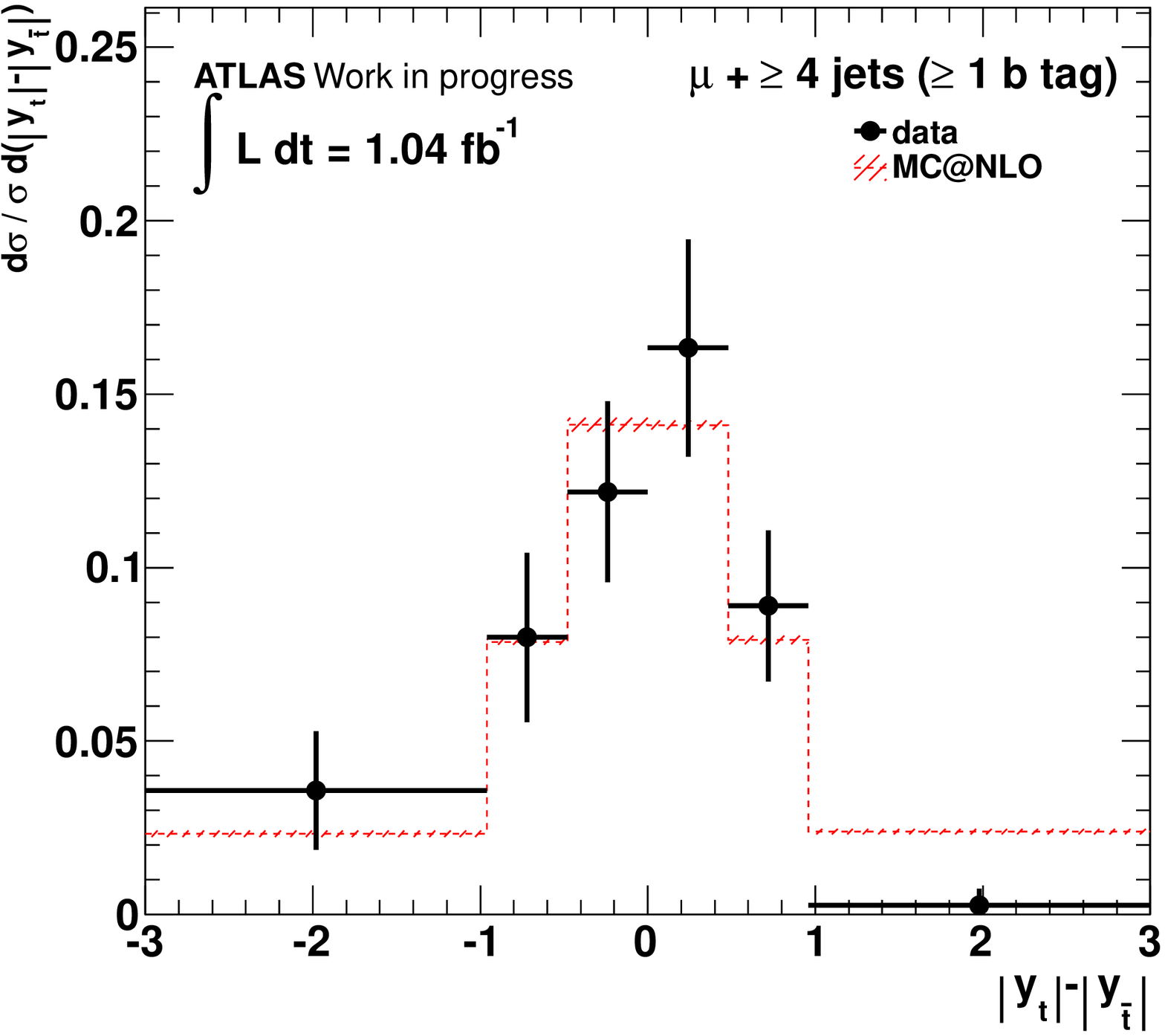}
    \quad\quad
    \includegraphics[width=6.3cm]{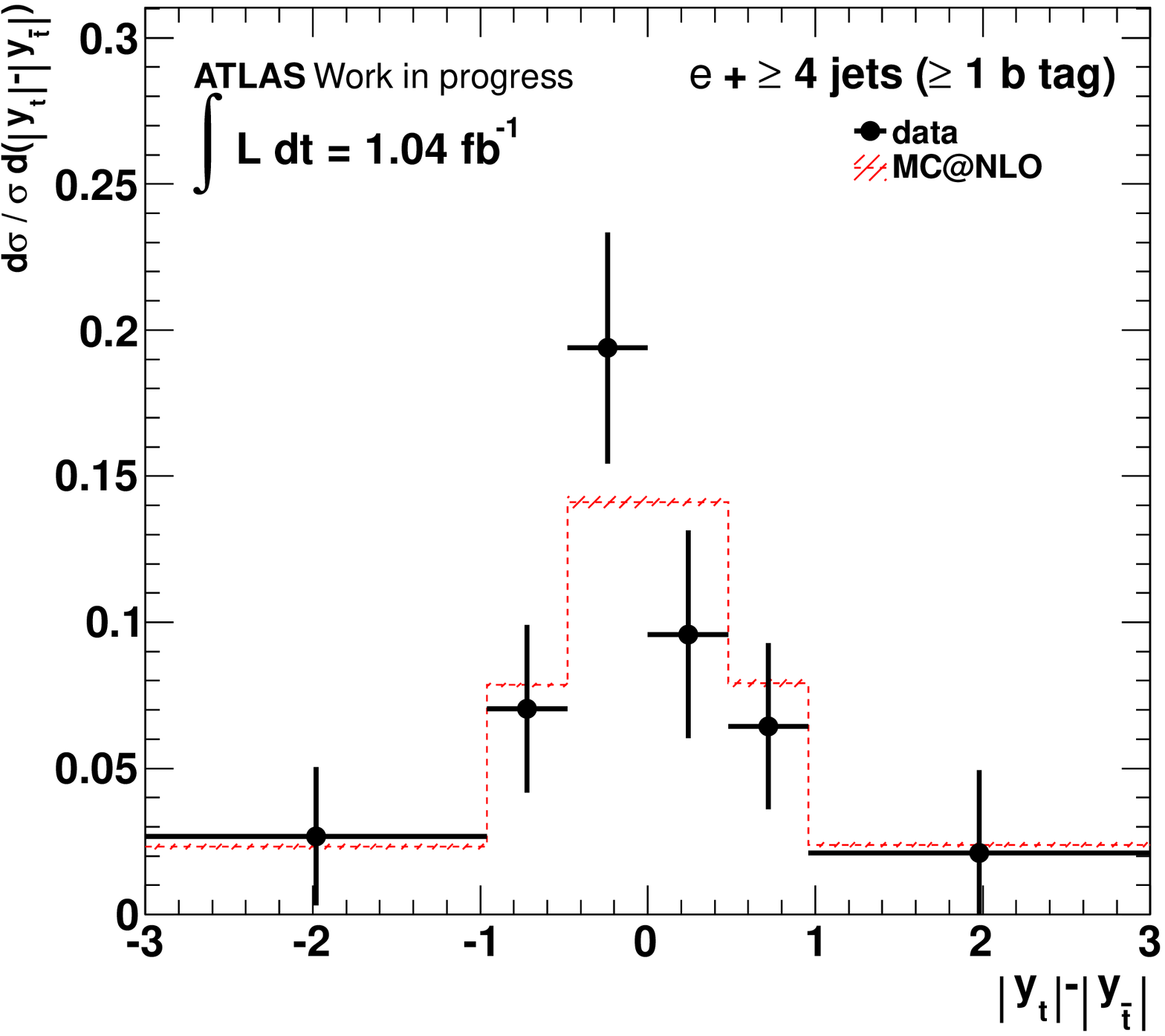}
    \includegraphics[width=6.3cm]{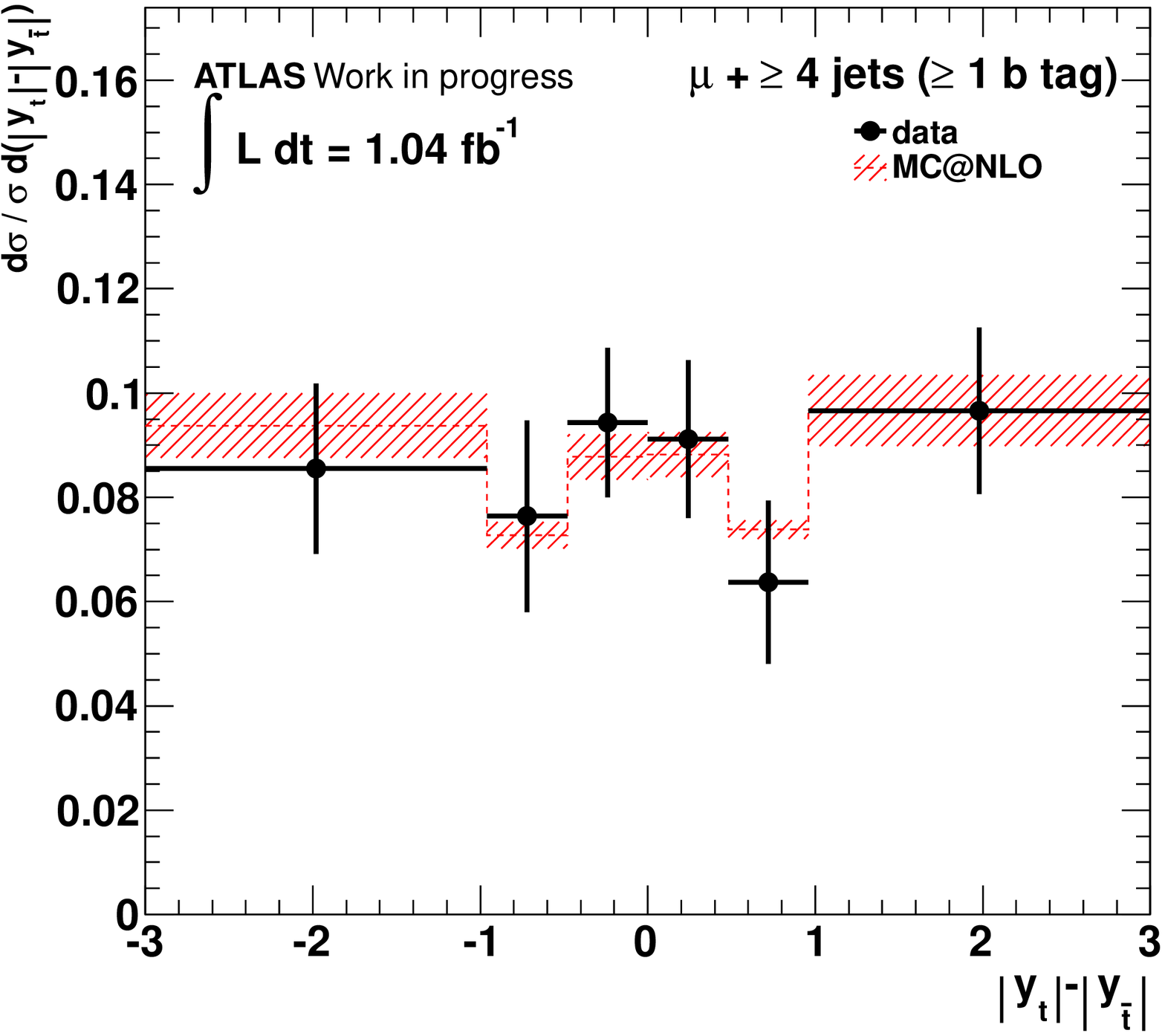}
    \quad\quad
    \includegraphics[width=6.3cm]{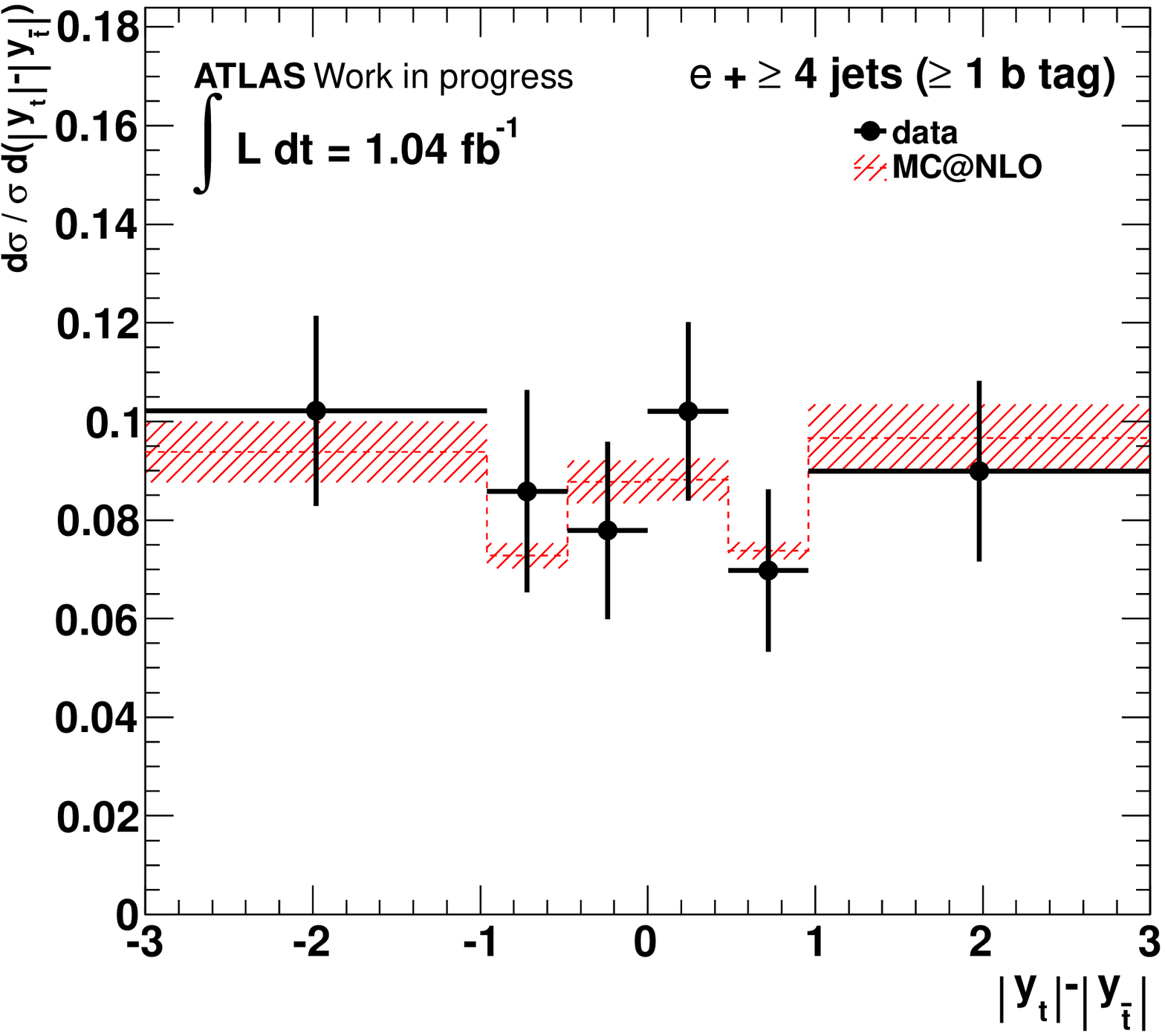}
    
    \vspace{0.2 cm}
    \caption{Unfolded distribution of $|y_t|-|y_{\bar{t}}|$, normalised to unity for both the muon+jets channel (left) and electron+jets channel (right). The top row shows the distributions for the inclusive measurement, while the lower rows show the corresponding distributions for $M_{t\bar{t}} < 450$\,GeV and $M_{t\bar{t}} > 450$\,GeV, respectively. For the simultaneous unfolding in $|y_t| - |y_{\bar{t}}|$ and $M_{t \bar{t}}$, a cut on the event reconstruction likelihood $\log{L}$ was applied to improve the $M_{t\bar{t}}$ resolution of the selected events. The uncertainties include both statistical and systematic shape components.}
    \label{fig:unfolding:data}
  \end{center}
\end{figure}
In addition, \mbox{Figure \ref{fig:unfolding:data}} shows the obtained distributions after unfolding for both the inclusive unfolding and the simultaneous unfolding in $|y_t| - |y_{\bar{t}}|$ and $M_{t \bar{t}}$. The unfolded distributions have been normalised to unity and the shape uncertainties obtained from all systematic effects described in \mbox{Chapter \ref{Systematics}} and the bin-by-bin statistical uncertainties have been included.

\begin{figure}[h!tb]
  \begin{centering}
    \mbox{
      \includegraphics[width=\plotwidth]{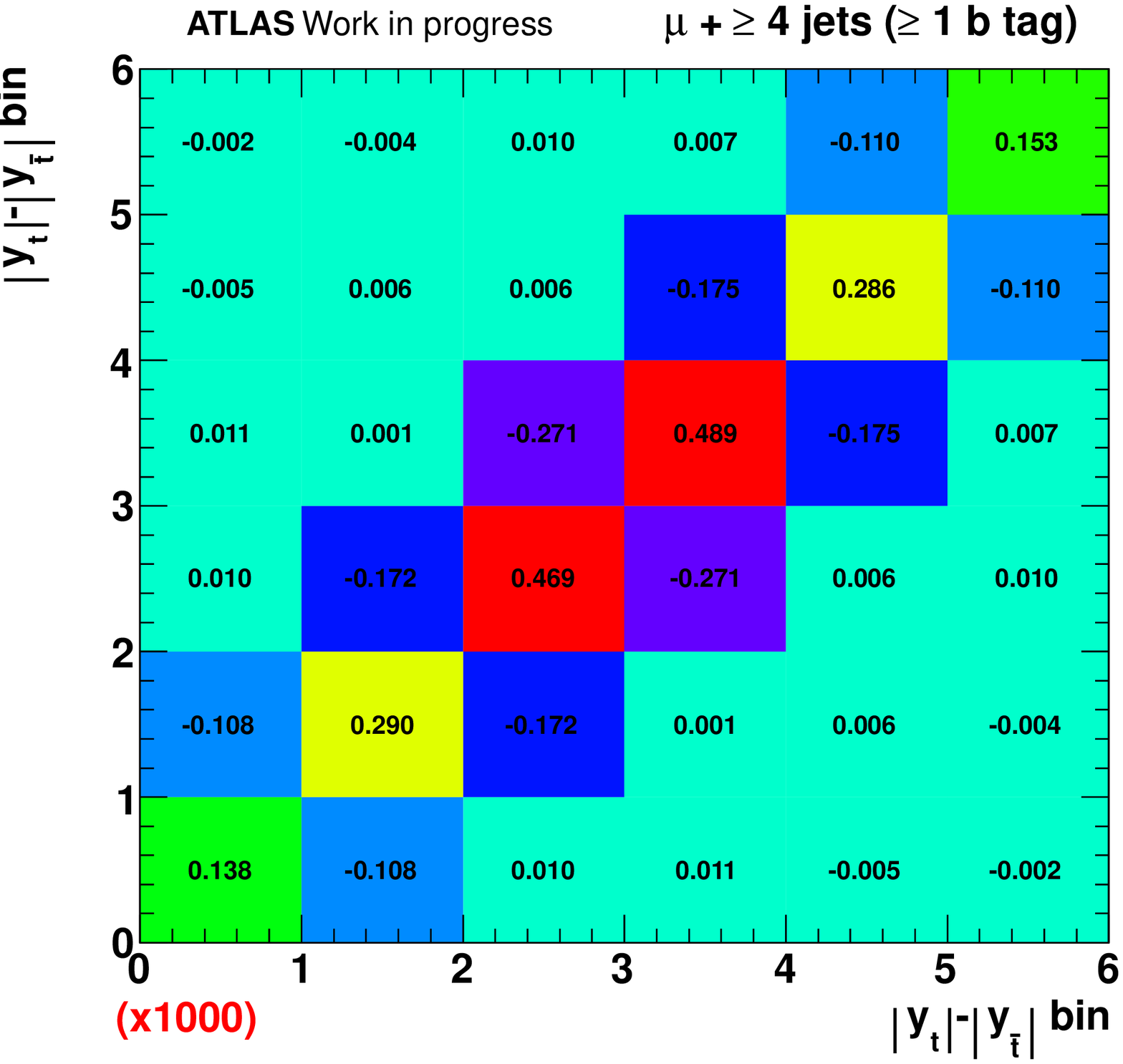}
      \quad\quad
      \includegraphics[width=\plotwidth]{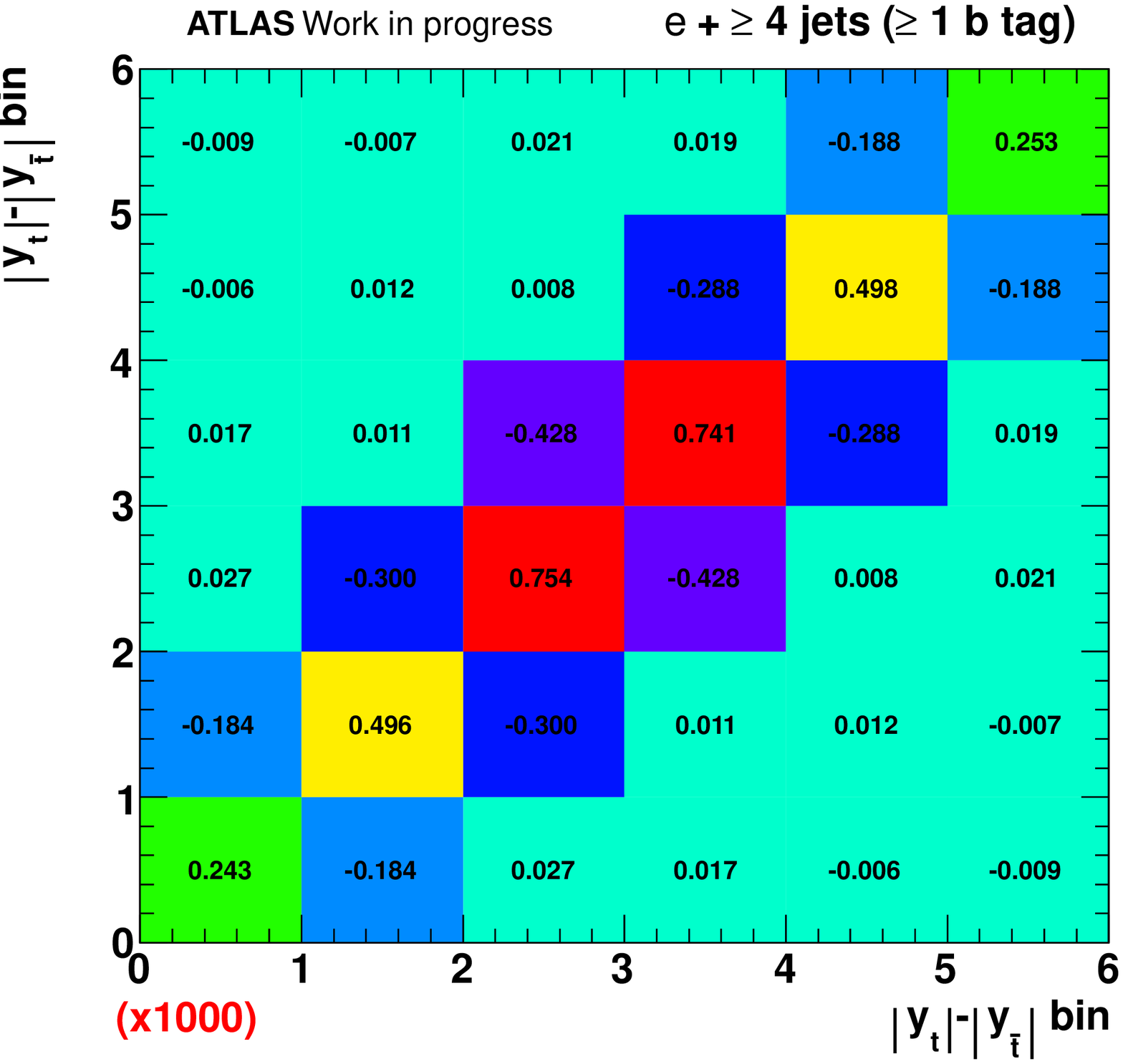}
    }
    
    \vspace{-0.2 cm}
    \caption[\quad Inclusive unfolding covariance matrices]{Covariance matrices corresponding to the unfolded $|y_t|-|y_{\bar{t}}|$ distribution for the muon+jets channel (left) and the electron+jets channel (right) for the inclusive unfolding. The numbers inside the boxes represent the values and the sign of the correlation among the different bins and have been scaled by a factor of 1000 for readability.}
    \label{fig:unfolding:covar}
  \end{centering}
\end{figure}

\begin{figure}[h!tb]
  \begin{centering}
    \mbox{
      \includegraphics[width=\plotwidth]{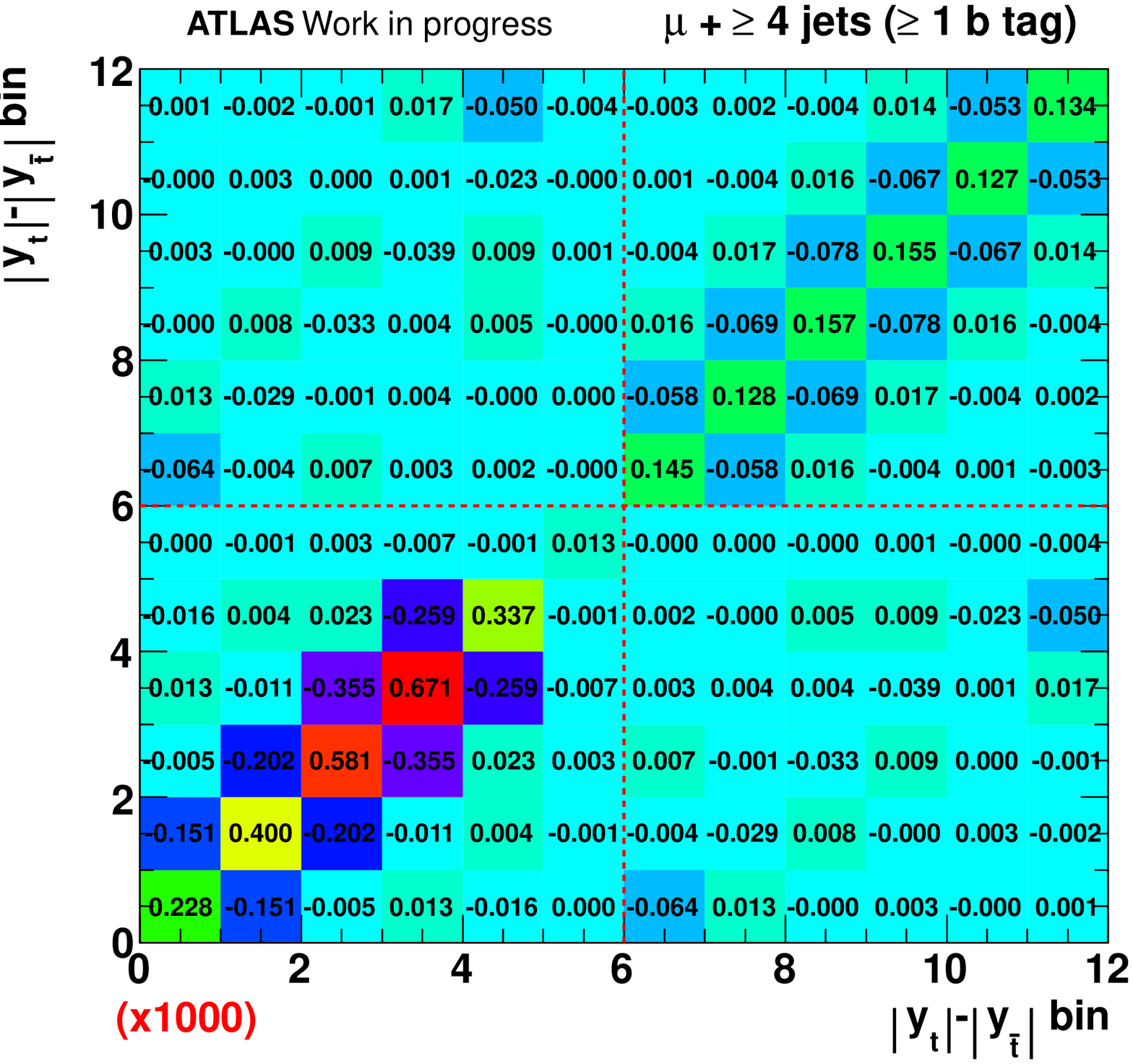}
      \quad\quad
      \includegraphics[width=\plotwidth]{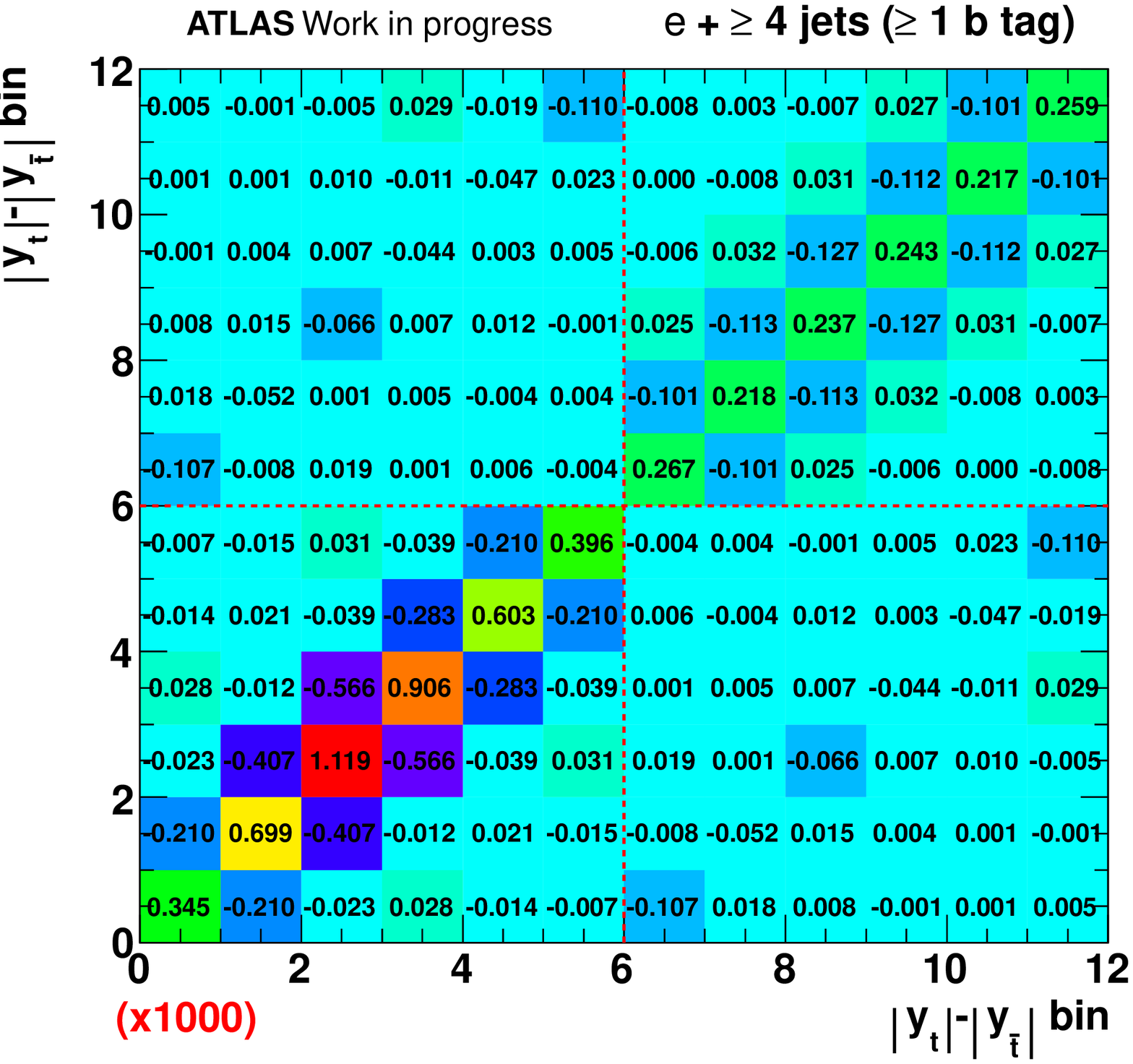}
    }
    
    \vspace{-0.2 cm}
    \caption[\quad Unfolding covariance matrices dependent on $M_{t \bar{t}}$)]{Covariance matrices corresponding to the unfolded $|y_t|-|y_{\bar{t}}|$ distribution for the muon+jets channel (left) and the electron+jets channel (right) for the simultaneous unfolding in $|y_t| - |y_{\bar{t}}|$ and $M_{t \bar{t}}$. The numbers inside the boxes represent the values and the sign of the correlation among the different bins and have been scaled by a factor of 1000 for readability.}
    \label{fig:unfolding:covar2D}
  \end{centering}
\end{figure}

Furthermore, the covariance matrices corresponding to the unfolded $|y_t|-|y_{\bar{t}}|$ distributions are shown in \mbox{Figure \ref{fig:unfolding:covar}} and \ref{fig:unfolding:covar2D} for the inclusive unfolding and the simultaneous unfolding in $|y_t| - |y_{\bar{t}}|$ and $M_{t \bar{t}}$, respectively.

A summarised list of all systematics and their contribution to the overall systematic uncertainties can be found in \mbox{Table \ref{Tab:SystematicsUnf}} for the inclusive unfolding and in \mbox{Table \ref{Tab:SystematicsUnf2D}} for the simultaneous unfolding in $|y_t| - |y_{\bar{t}}|$ and $M_{t \bar{t}}$.
\begin{table}[htbp]
\begin{center}
{\small
\begin{tabular}{|l|r|r|}
\cline{2-3}\multicolumn{1}{c|}{} & \multicolumn{ 2}{c|}{Absolute systematic uncertainty} \\
\cline{2-3}\multicolumn{1}{c|}{} & Muon Channel & Electron Channel \\
\hline
QCD multijet  & 0.001 & 0.011 \\ 
\hline
Jet energy scale  & 0.006 & 0.012 \\ 
$b$ tag jet energy scale  & 0.001 & 0.001 \\ 
Pile-up jet energy scale  & 0.001 & 0.002 \\ 
\hline
Jet reco efficiency  & 0.003 & 0.001 \\ 
Jet energy resolution  & 0.006 & 0.001 \\ 
\hline
Muon efficiencies  & 0.001 & (n.a.) \\ 
Muon scales / resolution  & $<$ 0.001 & $<$ 0.001 \\ 
Electron efficiencies  & (n.a.) & 0.001 \\ 
Electron scales / resolution  & 0.001 & 0.002 \\ 
\hline
$b$ tag scale factors  & 0.002 & 0.004 \\ 
PDF uncertainty  & $<$ 0.001 & $<$ 0.001 \\ 
LAr hole uncertainty  & 0.004 & 0.001 \\ 
\hline
ISR and FSR  & 0.010 & 0.009 \\ 
\ttbar~modelling  & (0.005) 0.011 & 0.011 \\ 
Parton shower / fragmentation  & (0.004) 0.010 & 0.010 \\ 
Top mass  & 0.007 & (0.002) 0.007 \\ 
\hline
$W$+jets normalisation  & 0.005 & 0.007 \\ 
$W$+jets shape  & $<$ 0.001 & 0.003 \\ 
$Z$+jets normalisation  & $<$ 0.001 & $<$ 0.001 \\ 
$Z$+jets shape  & 0.001 & 0.005 \\ 
\hline
Single top  & $<$ 0.001 & $<$ 0.001 \\ 
Diboson  & $<$ 0.001 & $<$ 0.001 \\ 
Charge mis-identification  & $<$ 0.001 & $<$ 0.001 \\ 
$b$ tag charge  & 0.001 & 0.001 \\ 
\hline
MC statistics  & 0.005 & 0.006 \\
Unfolding convergence  & 0.001 & 0.001 \\
Unfolding bias  & $<$ 0.001 & 0.004 \\
\hline
Luminosity   & 0.001 & 0.001 \\
\hline
\hline
Combined & 0.023 & 0.028 \\
\hline
\end{tabular}
}
\end{center}
\caption{List of all systematic uncertainties taken into account for the unfolding procedure in the measurement of the top charge asymmetry. The numbers in brackets denote the uncertainties before using the larger uncertainty of both channels as conservative estimate.}
\label{Tab:SystematicsUnf}
\end{table}

\begin{table}[htbp]
\begin{center}
{\small
\begin{tabular}{|l|r|r|r|r|}
\cline{2-5}\multicolumn{1}{c|}{} & \multicolumn{ 4}{c|}{Absolute systematic uncertainty} \\ \cline{2-5}\multicolumn{1}{c|}{} & \multicolumn{ 2}{c|}{Muon Channel} & \multicolumn{ 2}{c|}{Electron Channel} \\ 
\cline{2-5}\multicolumn{1}{c|}{} & $M_{t\bar{t}} < 450\,\text{GeV}$ & $M_{t\bar{t}} > 450\,\text{GeV}$ & $M_{t\bar{t}} < 450\,\text{GeV}$ & $M_{t\bar{t}} > 450\,\text{GeV}$ \\
\hline
QCD multijet  & 0.018 & 0.004 & 0.054 & 0.015 \\ 
\hline
Jet energy scale  & 0.013 & 0.010 & 0.034 & 0.006 \\ 
$b$ tag jet energy scale  & 0.001 & 0.001 & 0.003 & 0.001 \\ 
Pile-up jet energy scale  & 0.002 & 0.002 & 0.004 & 0.002 \\ 
\hline
Jet reco efficiency  & 0.002 & 0.002 & 0.011 & 0.001 \\ 
Jet energy resolution  & 0.016 & 0.001 & 0.036 & 0.002 \\ 
\hline
Muon efficiencies  & $<$ 0.001 & 0.001 & (n.a.) & (n.a.) \\ 
Muon scales / resolution  & 0.002 & 0.001 & 0.001 & 0.001 \\ 
Electron efficiencies  & (n.a.) & (n.a.) & 0.002 & $<$ 0.001 \\ 
Electron scales / resolution  & $<$ 0.001 & 0.001 & 0.007 & 0.001 \\ 
\hline
$b$ tag scale factors  & 0.001 & 0.003 & 0.012 & 0.002 \\ 
PDF uncertainty  & 0.001 & 0.001 & 0.001 & 0.001 \\ 
LAr hole uncertainty  & 0.008 & 0.001 & 0.008 & 0.003 \\ 
\hline
ISR and FSR  & 0.027 & 0.024 & (0.014) 0.019 & 0.019 \\ 
\ttbar~modelling  & 0.018 & (0.001) 0.018 & 0.032 & 0.015 \\ 
Parton shower / fragm.  & (0.002) 0.006 & 0.006 & (0.005) 0.006 & 0.006 \\ 
Top mass  & 0.019 & (0.002) 0.019 & (0.009) 0.011 & 0.011 \\ 
\hline
$W$+jets normalisation  & $<$ 0.001 & 0.006 & 0.014 & 0.004 \\ 
$W$+jets shape  & $<$ 0.001 & $<$ 0.001 & 0.002 & 0.001 \\ 
$Z$+jets normalisation  & $<$ 0.001 & $<$ 0.001 & 0.003 & 0.001 \\ 
$Z$+jets shape  & 0.004 & 0.001 & 0.017 & 0.009 \\ 
\hline
Single top  & $<$ 0.001 & $<$ 0.001 & 0.002 & $<$ 0.001 \\ 
Diboson  & $<$ 0.001 & $<$ 0.001 & $<$ 0.001 & $<$ 0.001 \\ 
Charge mis-identification  & $<$ 0.001 & $<$ 0.001 & $<$ 0.001 & $<$ 0.001 \\ 
$b$ tag charge  & 0.001 & 0.001 & 0.001 & 0.001 \\ 
\hline
MC statistics  & 0.012 & 0.007 & 0.017 & 0.009 \\
Unfolding convergence  & 0.001 & 0.001 & 0.001 & 0.001 \\
Unfolding bias  & $<$ 0.001 & $<$ 0.001 & 0.006 & 0.002 \\
\hline
Luminosity  & $<$ 0.001 & 0.001 & 0.003 & $<$ 0.001 \\
\hline
\hline
Combined & 0.049 & 0.034 & 0.091 & 0.035 \\
\hline
\end{tabular}
}
\end{center}
\caption{List of all systematic uncertainties taken into account for the unfolding procedure in the measurement of the top charge asymmetry. The numbers in brackets denote the uncertainties before using the larger uncertainty of both channels as conservative estimate.}
\label{Tab:SystematicsUnf2D}
\end{table}

All systematic uncertainties based on a replacement of the unfolding matrix which were of the same order of magnitude or lower than the respective MC statistics uncertainty could not be resolved to full extent due to the inherent fluctuations from limited statistics in the response matrix. In those cases, the larger of the systematic uncertainties in both channels was used for the final systematic uncertainty on the unfolded charge asymmetry. The resulting total systematic uncertainties were 0.023 in the muon+jets channel and 0.028 in the electron+jets channel for the inclusive unfolding. For the simultaneous unfolding in $|y_t|-|y_{\bar{t}}|$ and $M_{t\bar{t}}$ the resulting total systematic uncertainties were 0.049 ($M_{t\bar{t}} < 450\,\text{GeV}$) and 0.034 ($M_{t\bar{t}} > 450\,\text{GeV}$) in the muon+jets channel, and 0.091 ($M_{t\bar{t}} < 450\,\text{GeV}$) and 0.035 ($M_{t\bar{t}} > 450\,\text{GeV}$) in the electron+jets channel, respectively. 

The overall uncertainty in the individual channels was dominated by the statistical uncertainty. The systematic uncertainties were dominated by the contributions from ISR/FSR, top mass and jet energy resolution in the muon+jets channel, and by the uncertainties originating from QCD multijet background, \ttbar~modelling and parton shower / fragmentation in the electron+jets channel. As described in \mbox{Chapter \ref{Systematics}}, the QCD multijet background contribution has been estimated very conservatively, assuming a 100\,\% normalisation uncertainty. 
Most of the other large contributions can be traced to the available Monte Carlo statistics in the used samples. Since only \numprint{3000000} (\ttbar~modelling, parton shower / fragmentation), or \numprint{1000000} (top mass, ISR/FSR) Monte Carlo events were available for the respective samples used in the evaluation of the systematic uncertainties (as opposed to \numprint{15000000} for the nominal signal sample), the statistical component in the evaluation of the response matrix uncertainty and the unfolding procedure was larger by factors of two to four with respect to the nominal case.

In addition to the evaluation of the systematic uncertainties in the unfolding process, the effect of pile-up on the measured quantity before the unfolding was studied in order to ensure the stability of the method for different pile-up conditions. \mbox{Figure \ref{fig:pile-up}} in \mbox{Appendix \ref{AppControlPlots}} shows the measured integrated asymmetry before background subtraction, $A_C^{\text{data}}$, as a function of the number of primary vertices and of the bunch timing, i.e. at which relative position in the respective bunch the $pp$ collision corresponding to the respective event occurred. As there was no statistically significant dependence of the measured asymmetry on the bunch timing or on the number of primary vertices in either channel, no additional systematic uncertainty due to pile-up was assigned.

Both the differential and the integrated asymmetries after unfolding were in agreement with the \mcnlo~Standard Model prediction within the estimated uncertainties. The provided differential distributions alongside with the respective covariance matrices can directly be put into context with analogue measurements at other experiments and theoretical predictions.

\section{Combination}
\label{chap:results:comb}
A best linear unbiased estimator (\textsc{Blue}) method\cite{Lyons:1988, Valassi:2003} has been used to combine the results from the muon+jets and electron+jets channel after unfolding, taking into account systematic uncertainties and the associated correlations\footnote{Note that since the exact correlations are unknown for most of the systematic uncertainties contributing to the overall result, correlation coefficients between the muon+jets and electron+jets channel of either zero or one have been assumed.}. All systematic uncertainties have been considered to be fully correlated between the muon+jets and the electron+jets channel except for the contributions from the Monte Carlo statistics of the response matrix, LAr defects (since the treatment of electrons does not affect the muon+jets channel), QCD multijet and $W$+jets normalisation (since both have been determined with data driven methods based on orthogonal datasets), charge mis-identification, and the systematic uncertainty from unfolding convergence and remaining bias. No correlation was assumed for the statistical uncertainties.

For the inclusive unfolding, a combined value of
\begin{equation*}
A_C^{\text{unf}} = -0.018 \pm 0.028\,\text{(stat.)} \pm 0.023\,\text{(syst.)}
\end{equation*}
was obtained, where the relative weight of the muon+jets channel result was 63.9\,\%.

The combination of the results for the simultaneous unfolding in $|y_t| - |y_{\bar{t}}|$ and $M_{t \bar{t}}$ yielded
\begin{eqnarray*}
A_C^{\text{unf}} (M_{t\bar{t}} < 450\,\text{GeV}) & = & -0.053 \pm 0.070\,\text{(stat.)} \pm 0.054\,\text{(syst.)},\\
A_C^{\text{unf}} (M_{t\bar{t}} > 450\,\text{GeV}) & = & -0.008 \pm 0.035\,\text{(stat.)} \pm 0.032\,\text{(syst.)},
\end{eqnarray*}
where the relative weights of the muon+jets channel were 73.7\,\% and 59.5\,\%, respectively. Note that the combined systematic uncertainty was slightly lower than the systematic uncertainties in both the muon+jets and the electron+jets channel. This effect is inherent to the \textsc{Blue} uncertainty propagation and is due to the assumed correlations of the individual contributions, which can be regarded as additional prior information in the propagation of the uncertainties.

The combined results were compatible with the \mcnlo~Standard Model expectation (c.f. \mbox{Table \ref{tab:unfolding}}) within the estimated uncertainties, not indicating any significant deviation.

  \chapter{Summary \& Conclusion}
\label{Summary}
A measurement of the charge asymmetry in the production of top quark pairs at the ATLAS experiment was performed, using a dataset corresponding to an integrated luminosity of 1.04\,fb$^{-1}$ taken over the course of 2011 at a centre-of-mass energy of $\sqrt{s}$ = 7\,TeV.

An object and event selection was employed in the lepton+jets decay channel in order to identify events with a signature corresponding to a semileptonic decay of a \ttbar~pair, given by one isolated lepton (muon or electron) with large transverse momentum, at least four reconstructed jets and large missing transverse momentum. The selection was furthermore chosen such that various background contributions, including the production of single top quarks, heavy gauge bosons in association with jets, the contribution from diboson production and fake leptons predominantly produced in QCD multijet events, were reduced. A set of background contributions was estimated using Monte Carlo simulations, while the $W$+jets and QCD multijet normalisation was determined using data driven methods.

A kinematic fit was performed in order to reconstruct \ttbar~events based on the measured objects after application of the selection, yielding the most probable object kinematics at the parton level under the assumption of a semileptonic top quark decay event topology.

The reconstructed objects were used to obtain the differential distribution of the difference of absolute rapidities of the reconstructed top and antitop, $|y_t| - |y_{\bar{t}}|$. A subtraction of the various background contributions was performed in order to obtain the integrated charge asymmetry of the \ttbar~signal contribution at reconstruction level, $A_C^{\text{reco}}$. Furthermore, $A_C^{\text{reco}}$ was determined as a function of the invariant \ttbar~mass, $M_{t \bar{t}}$.

\begin{figure}[h!tb]
  \begin{centering}
    \includegraphics[width = \plotwidth]{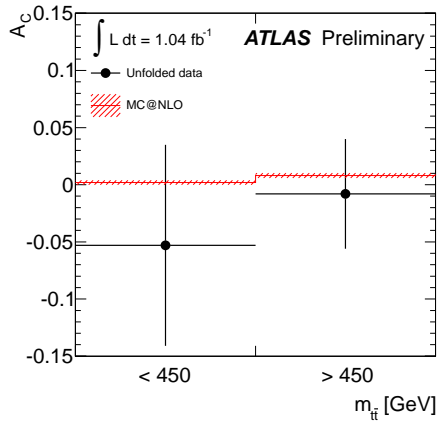}
    \vspace{-0.2 cm}
    \caption[\quad Unfolded asymmetries in two regions of $M_{t \bar{t}}$]{Unfolded asymmetries in two regions of $M_{t \bar{t}}$ compared to the prediction from \mcnlo\cite{AsymPaper}. The error bands on the \mcnlo~prediction include uncertainties from parton distribution functions and renormalisation and factorisation scales.}
    \label{fig:HighMttbarSummary}
  \end{centering}
\end{figure}

An unfolding procedure was applied to the reconstructed $|y_t| - |y_{\bar{t}}|$ distribution in order to correct for resolution and acceptance effects and to obtain the corresponding distribution at parton level. A Bayesian iterative unfolding procedure was used and cross-checks and calibrations were performed to verify the linearity, convergence and stability of the approach. The unfolding was performed both for the inclusive $|y_t| - |y_{\bar{t}}|$ distribution and in addition simultaneously in $|y_t| - |y_{\bar{t}}|$ and $M_{t \bar{t}}$. A combination of obtained results in the muon+jets and electron+jets channel was performed using the \textsc{Blue} method, yielding a combined integrated charge asymmetry after unfolding of
\begin{equation}
A_C^{\text{unf}} = -0.018 \pm 0.028\,\text{(stat.)} \pm 0.023\,\text{(syst.)},
\end{equation}
while the \mcnlo~prediction was $0.0056 \pm 0.0003\,\text{(stat.)}$. For the simultaneous unfolding in $|y_t| - |y_{\bar{t}}|$ and $M_{t \bar{t}}$, combined values of
\begin{eqnarray*}
A_C^{\text{unf}} (M_{t\bar{t}} < 450\,\text{GeV}) & = & -0.053 \pm 0.070\,\text{(stat.)} \pm 0.054\,\text{(syst.)},\\
A_C^{\text{unf}} (M_{t\bar{t}} > 450\,\text{GeV}) & = & -0.008 \pm 0.035\,\text{(stat.)} \pm 0.032\,\text{(syst.)}
\end{eqnarray*}
were obtained. The \mcnlo~predictions were $0.0024 \pm 0.0004\,\text{(stat.)}$ and $0.0086 \pm 0.0004\,\text{(stat.)}$, respectively. A summary of the results obtained in the simultaneous unfolding in $|y_t| - |y_{\bar{t}}|$ and $M_{t \bar{t}}$ can be found in \mbox{Figure~\ref{fig:HighMttbarSummary}}.

The obtained results were in agreement with the Standard Model prediction. However, both for the measured asymmetry after reconstruction and the corresponding unfolded values a tendency to more negative integrated asymmetries was observed in the electron channel. This effect is most dominant for the simultaneous unfolding in the region where $M_{t\bar{t}} < 450\,\text{GeV}$. Nevertheless, the discrepancy is covered by the estimated overall uncertainties including systematics.

The estimated uncertainties were dominated by the statistical uncertainty, in particular for the simultaneous unfolding in $|y_t| - |y_{\bar{t}}|$ and $M_{t \bar{t}}$. The largest contributions to the systematic uncertainties originated from ISR/FSR, top mass and jet energy resolution in the muon+jets channel, and from the uncertainties from QCD multijet background, \ttbar~modelling and parton shower/fragmentation in the electron+jets channel. Most of these contributions, except for the QCD multijet background uncertainty, can be attributed to large parts to the statistical component inherent in the evaluation of the systematic uncertainties, which involves a replacement of the unfolding response matrix with a corresponding matrix obtained from a different sample. Since only a fraction of events of about 10\,\% with respect to the nominal \ttbar~sample was available in those cases, statistical fluctuations in the response matrix dominated over the actual systematic shift to be evaluated. This also explains the discrepancies observed in the comparison of the individual contributions between the muon+jets and electron+jets channel.
Furthermore, the QCD multijet normalisation uncertainty was conservatively estimated to be 100\,\% following the recommendations of the performance groups, which led to a large contribution from this uncertainty, in particular in the electron+jets channel.

\begin{figure}[h!tb]
  \begin{centering}
    \mbox{
      \subfigure{
        \scalebox{0.4}{\includegraphics[width=\textwidth]{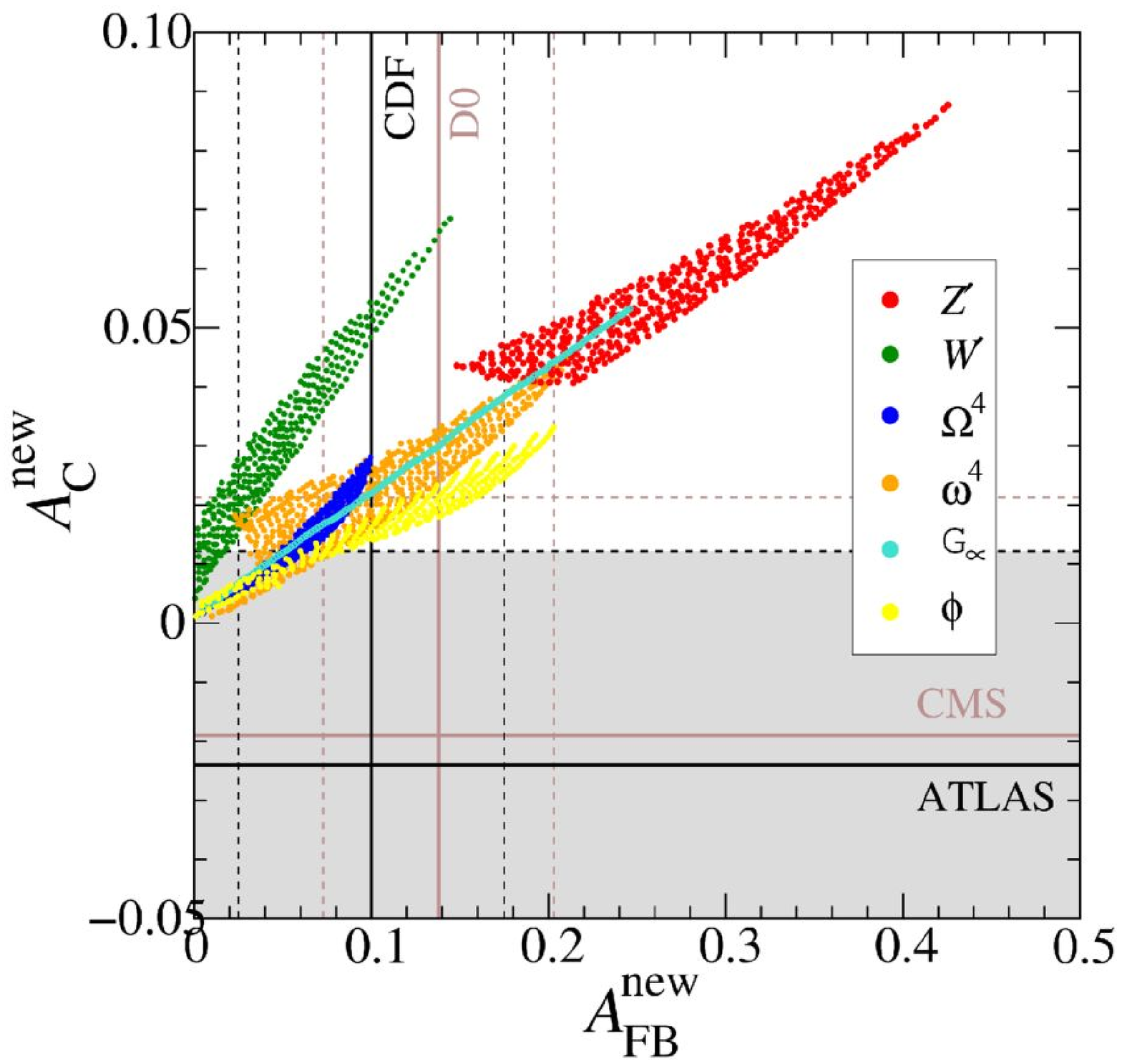}}
      }
      \quad
      \subfigure{
        \scalebox{0.4}{\includegraphics[width=\textwidth]{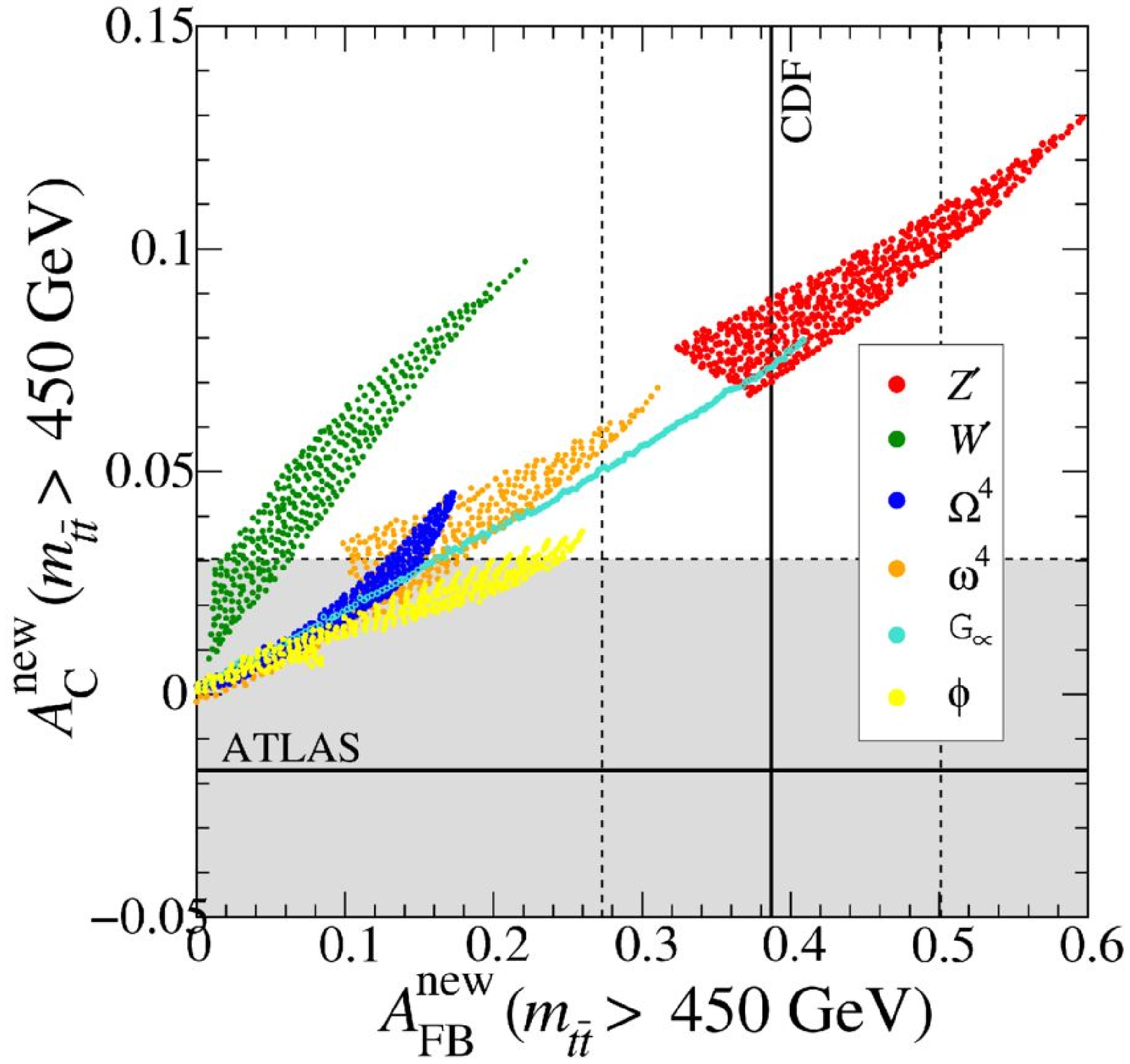}}
      }
    }
    \vspace{-0.2 cm}
    \caption[\quad Measurement results and predicted charge asymmetries at the Tevatron and LHC]{Measurement results and predicted charge asymmetries at the Tevatron and LHC for various BSM models\cite{AguilarSaavedra:2011ug,AguilarSaavedra:2011hz}. The inclusive charge asymmetry from new physics $A_C^{\text{new}}$ at the LHC vs. corresponding the forward-backward asymmetry $A_{\text{FB}}^{\text{new}}$ at the Tevatron (left) and the identical predictions in a high invariant mass region where $M_{t \bar{t}} > 450\,\text{GeV}$ (right) for the different models in the created phase space is shown. The labelled lines indicate the central values of the results measured at different experiments, while the dashed lines indicate the $1\,\sigma$ uncertainty regions (gray area denotes the ATLAS measurement uncertainty). The Standard Model prediction corresponds to $A_{\text{FB}}^{\text{new}} = 0$ and $A_C^{\text{new}} = 0$, respectively.}
    \label{fig:AsymTeVLHCComp}
  \end{centering}
\end{figure}

A summary of the obtained results alongside with recent results obtained by \dzero, CDF and CMS in comparison to several theoretical models beyond the Standard Model is shown in \mbox{Figure \ref{fig:AsymTeVLHCComp}}. Potential regions in the phase space of inclusive charge asymmetry from new physics, indicated by the variable $A_C^{\text{new}}$ at the LHC plotted against the associated forward-backward asymmetry $A_{\text{FB}}^{\text{new}}$ at the Tevatron, are highlighted (c.f. \mbox{Chapter \ref{chap:topasymHadColl}}).
The measurement performed in this analysis together with the results from other experiments already puts pressure on several of the proposed models. This is most prominent for $Z'$ models, which are disfavoured at $2\,\sigma$ to $3\,\sigma$ by this measurement and a similar measurement by CMS, while being favoured in the high invariant mass region by CDF. Furthermore, disagreements between favoured regions from the measurements at the Tevatron and at the LHC are observed, in particular for the high invariant \ttbar~mass region. These measurements of the charge asymmetry at hadron colliders provide the first step towards a better understanding of Quantum Chromodynamics and possible extensions of the Standard Model.

Despite the limitation of the measurement by the statistical uncertainty, in particular for the simultaneous unfolding in $|y_t| - |y_{\bar{t}}|$ and $M_{t \bar{t}}$, an increased dataset from the ATLAS experiment will quickly reduce this uncertainty, significantly increasing the sensitivity. This will allow to make much more precise statements especially in regions where an increased asymmetry is expected, making it possible to fully exclude several of the available BSM models with sufficient significance. Furthermore, non-excluded models could be further constrained in their respective parameters.

Further improvements can be achieved for the systematic uncertainties, where methods will evolve and in particular existing estimates can be replaced by more precise studies and methods. As an example, the QCD multijet systematic uncertainty has been evaluated in \mbox{Appendix \ref{App:AltApproach}} using more advanced methods, as described in \mbox{Chapter \ref{chap:QCDMu}}. However, one of the most prominent sources of uncertainty is the limited statistics of the used Monte Carlo samples, in particular for the evaluation of the systematic uncertainties. These could be reduced significantly if larger samples would be available. 

Optimisations in the unfolding procedure could lead to further reductions of the statistical uncertainties. Further studies of the unfolding procedure and the associated statistical and systematic uncertainties are discussed in \mbox{Appendix \ref{App:AltApproach}}.

Finally, an additional cross-check has been performed for the inclusive charge asymmetry, comparing the obtained results to an SVD unfolding approach, using otherwise identical parameters. Comparable values have been obtained.

  \cleardoublepage
  \begingroup
    \let\cleardoublepage\relax
    \begin{appendix}
\label{Appendix}

\chapter{Additional Control Plots}
\label{AppControlPlots}
\begin{figure}[h!tb]
  \begin{centering}
    \mbox{
      \includegraphics[width=\plotwidth]{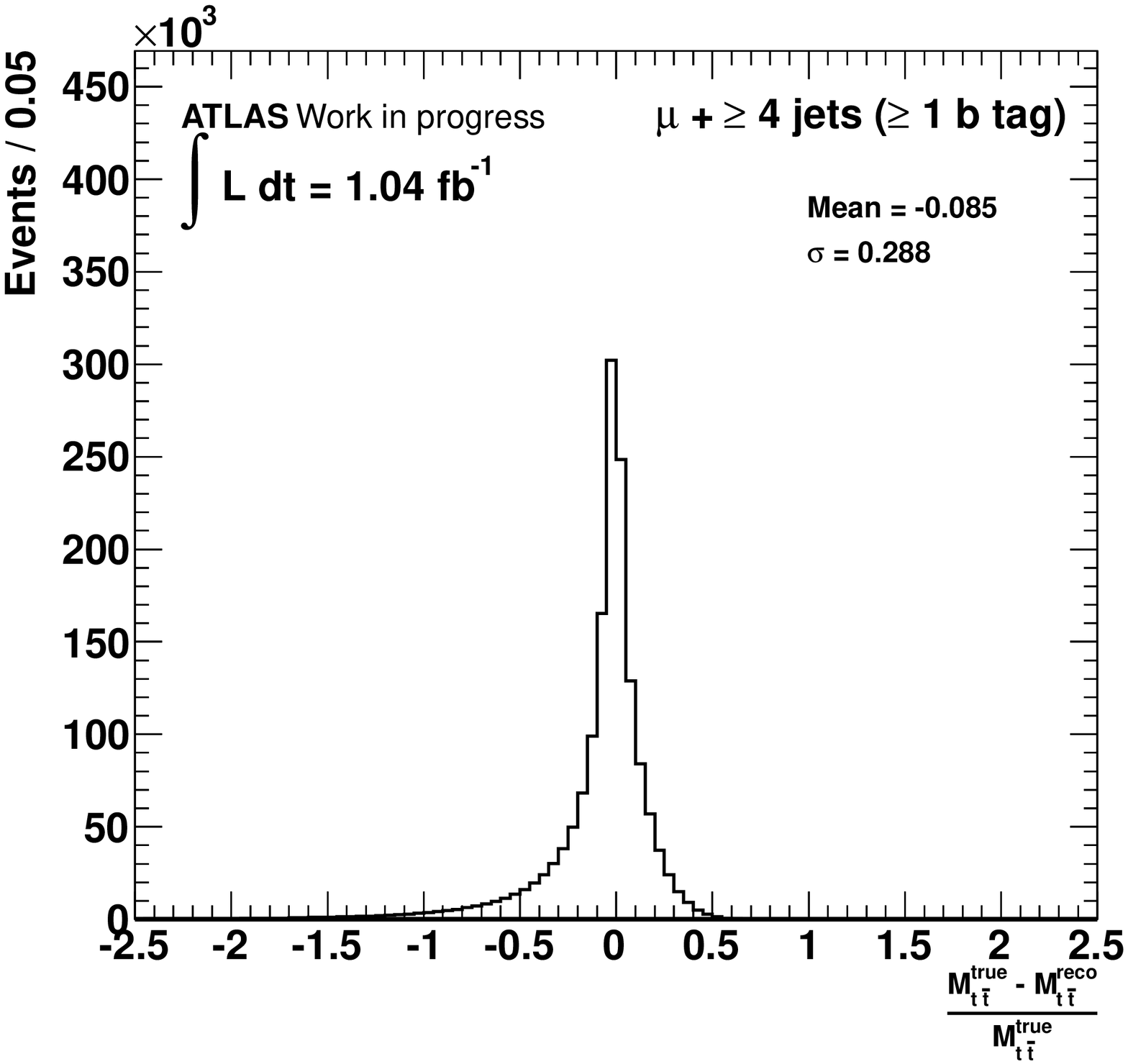}
      \quad\quad
      \includegraphics[width=\plotwidth]{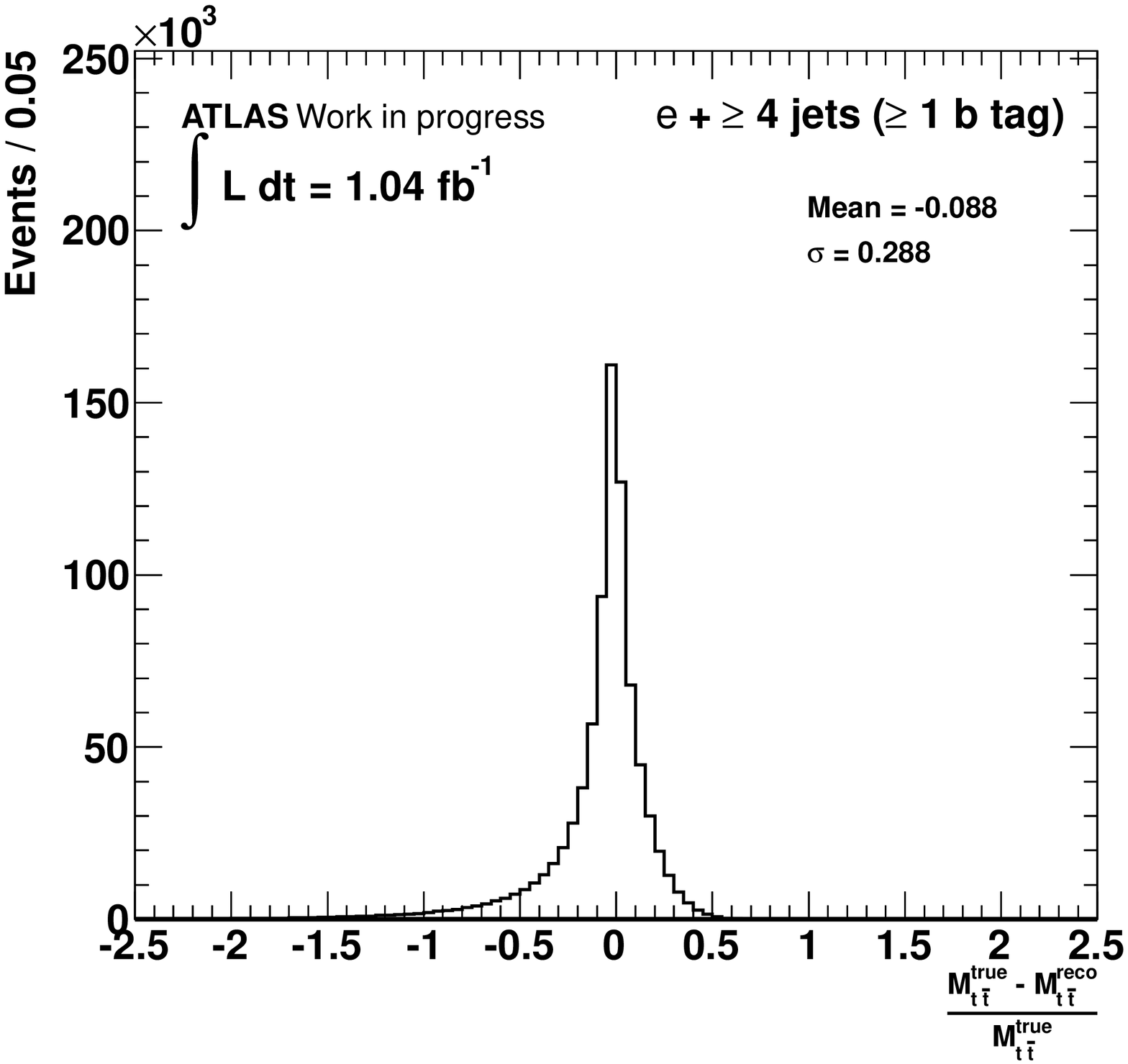}
    }
    
    \vspace{-0.2 cm}
    \caption[\quad $M_{t\bar{t}}$ resolution before $\log{L}$ cut]{Relative resolution of $M_{t\bar{t}}$ for the kinematic event reconstruction without any requirement on the reconstruction logarithmic likelihood $\log{L}$ for the muon+jets channel (left) and the electron+jets channel (right).}
    \label{fig:ttbarres_before}
  \end{centering}
\end{figure}
\begin{figure}[h!tb]
  \begin{centering}
    \mbox{
      \includegraphics[width=\plotwidth]{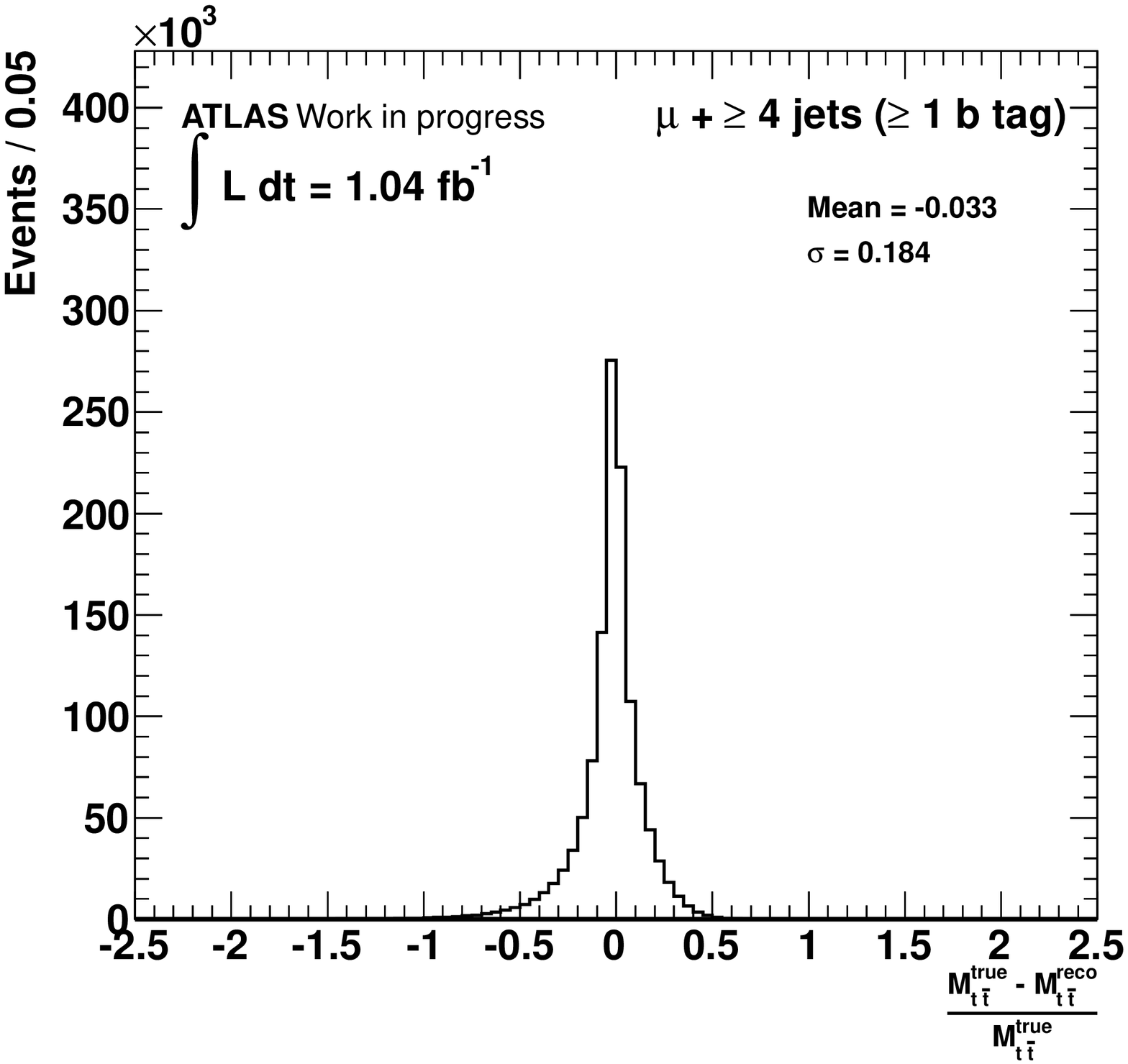}
      \quad\quad
      \includegraphics[width=\plotwidth]{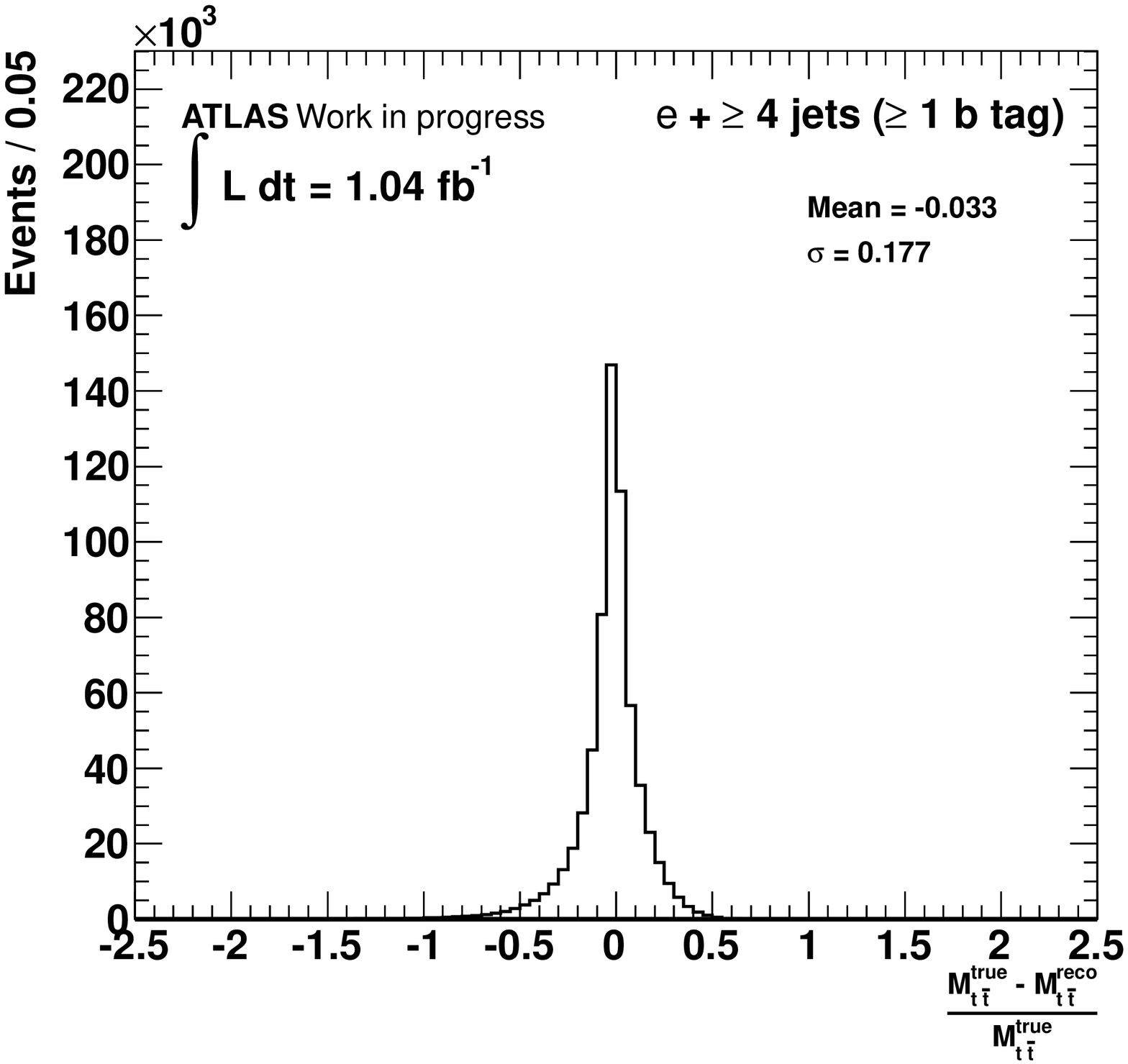}
    }
    
    \vspace{-0.2 cm}
    \caption[\quad $M_{t\bar{t}}$ resolution after $\log{L}$ cut]{Relative resolution of $M_{t\bar{t}}$ for the kinematic event reconstruction after application of a requirement on the reconstruction logarithmic likelihood of $\log{L} > -52$. The relative resolution improves from 28.8\,\% to 18.4\,\% in the muon+jets channel (left) and from 28.8\,\% to 17.7\,\% in the electron+jets channel (right).}
    \label{fig:ttbarres_after}
  \end{centering}
\end{figure}

\begin{figure}[!htbp]
  \begin{center}
    \includegraphics[width=\plotwidth]{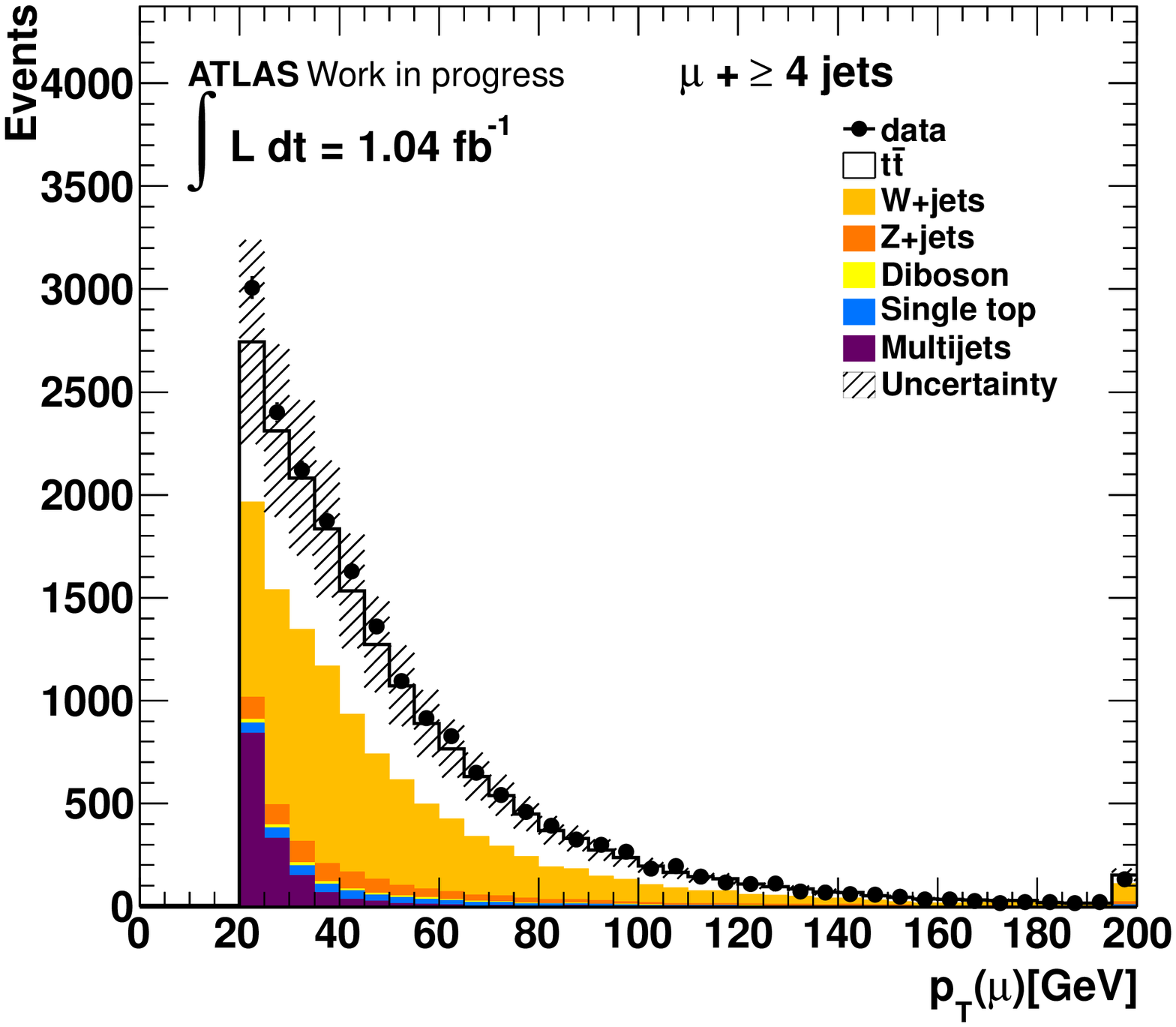}
    \quad\quad
    \includegraphics[width=\plotwidth]{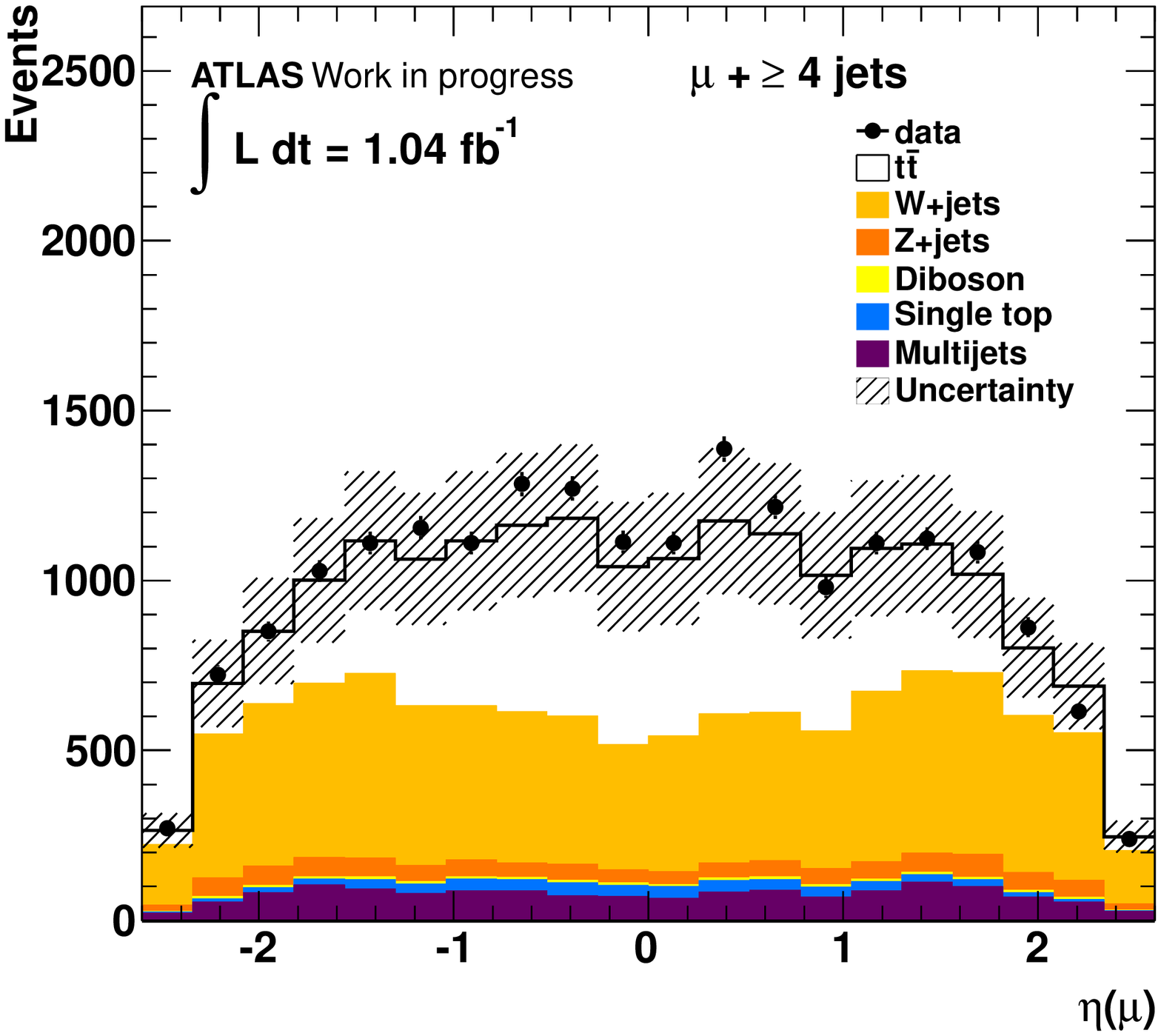}
    \includegraphics[width=\plotwidth]{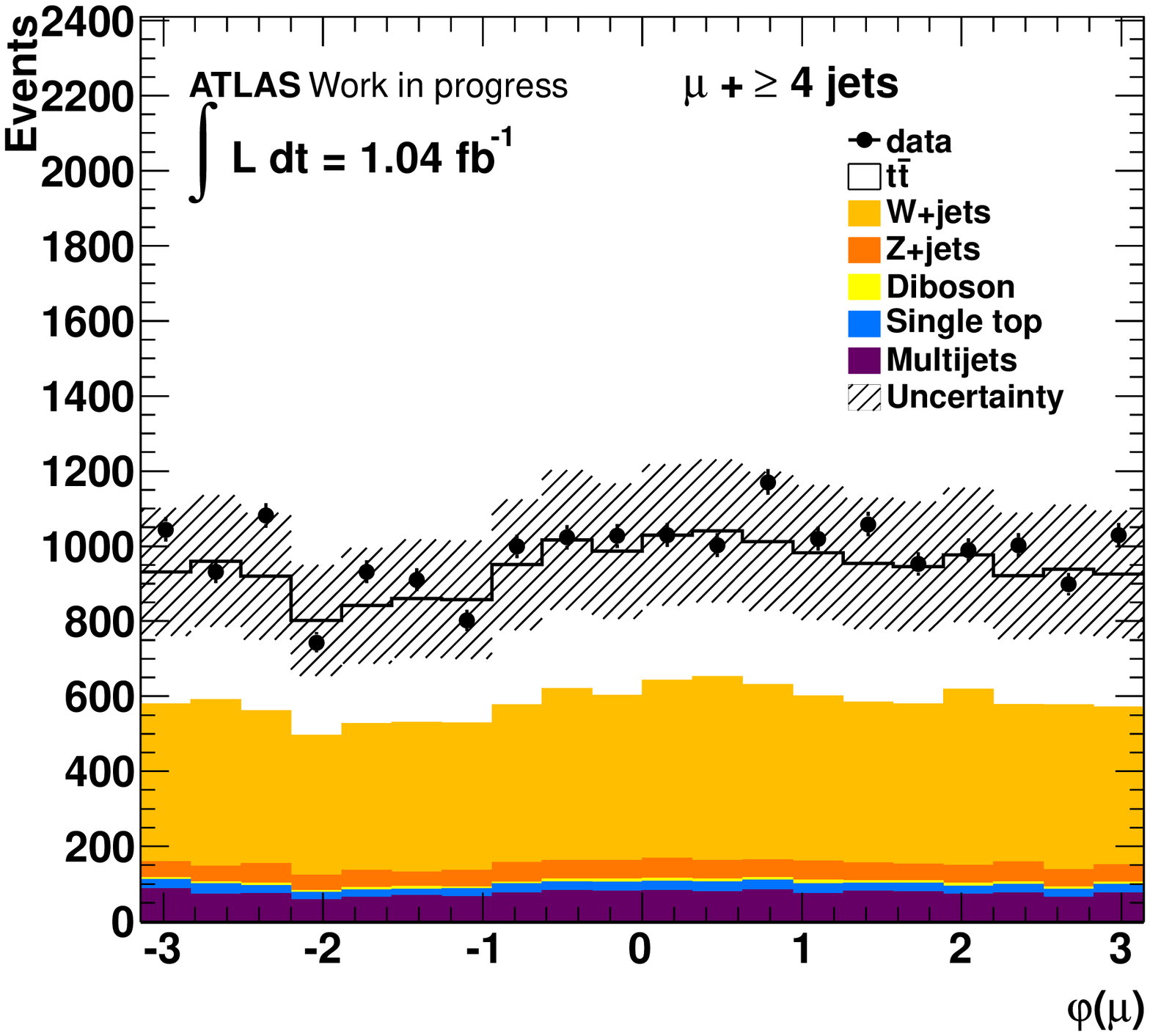}
    \quad\quad
    \includegraphics[width=\plotwidth]{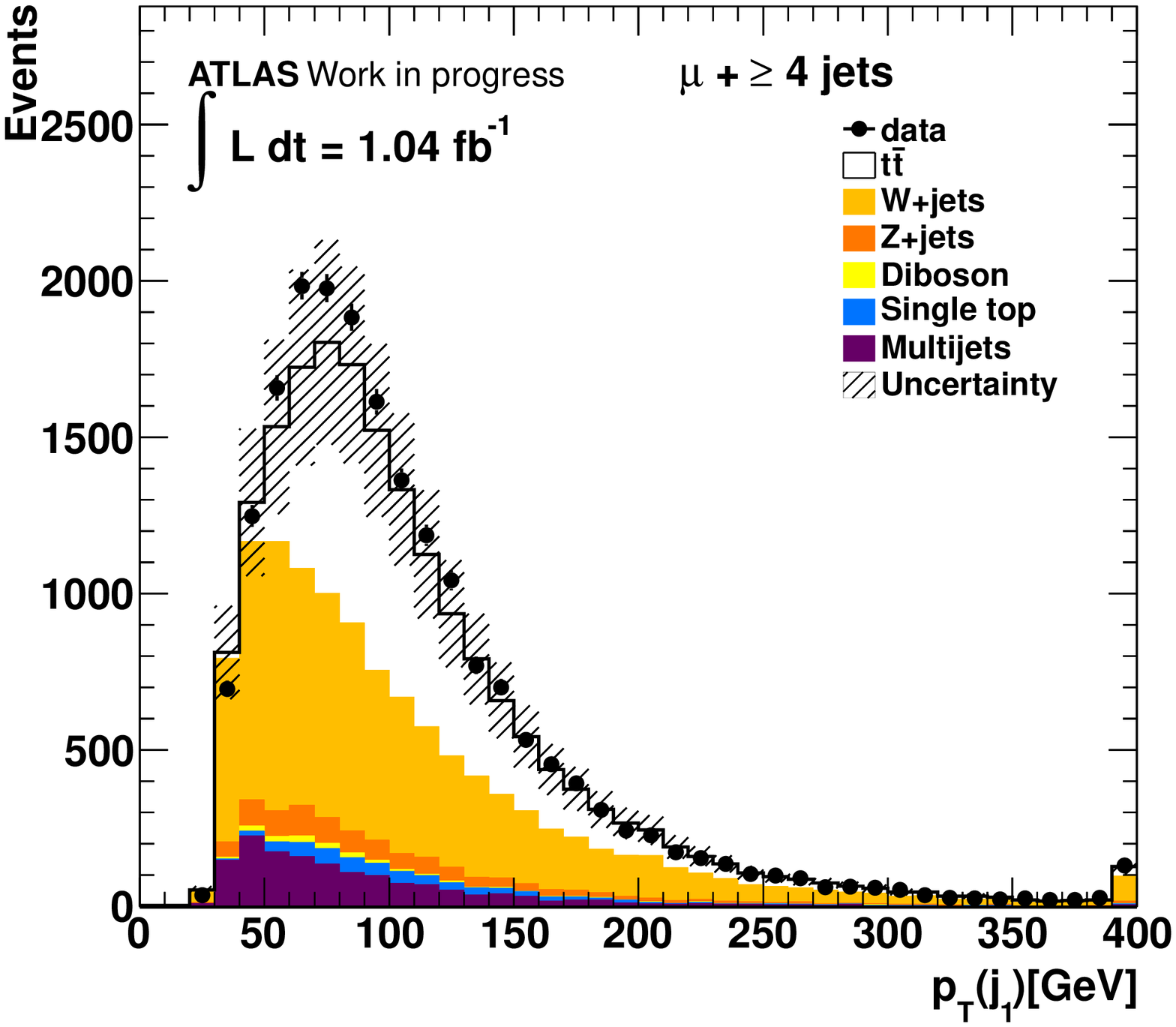}
    \includegraphics[width=\plotwidth]{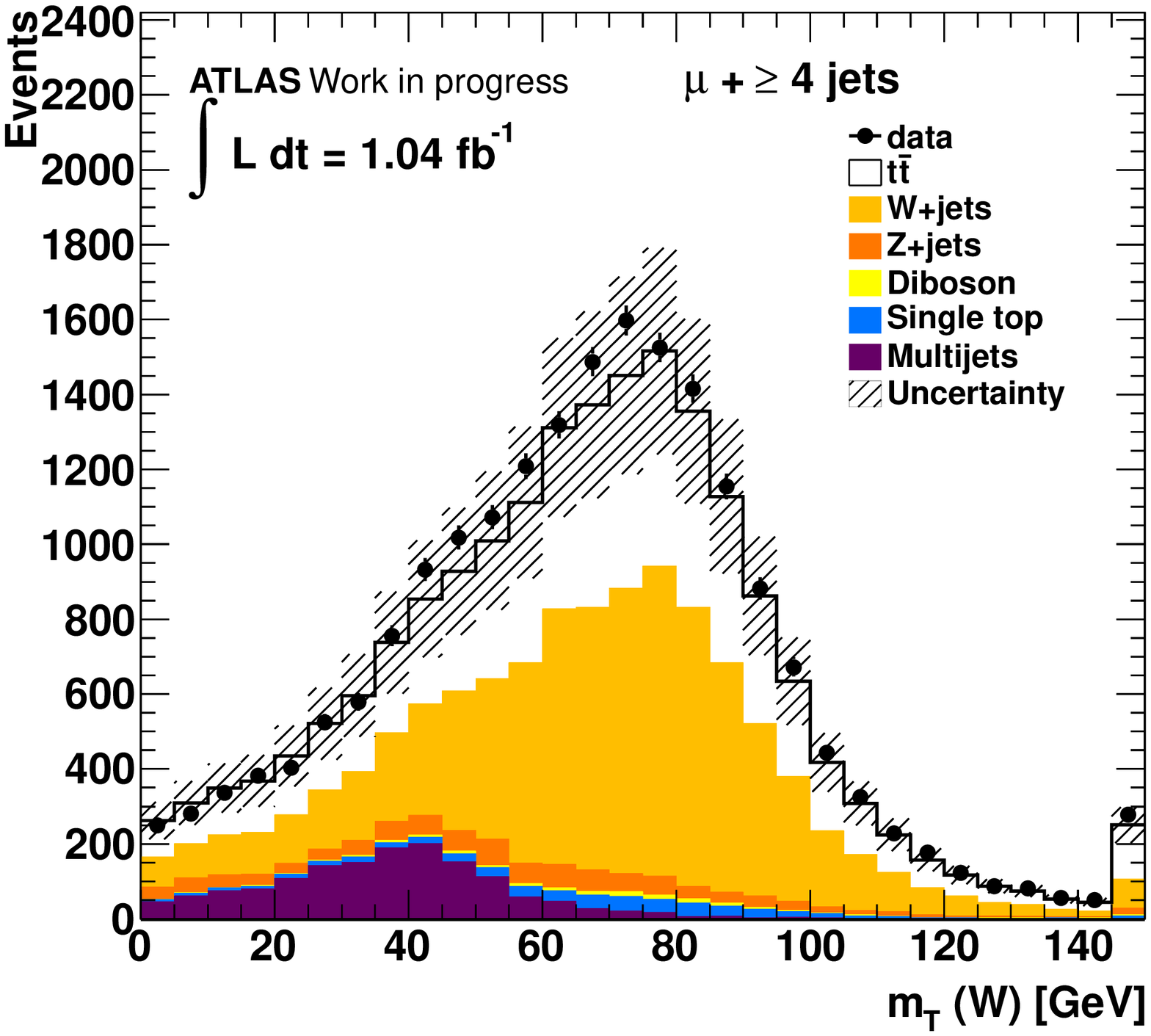}
    \quad\quad
    \includegraphics[width=\plotwidth]{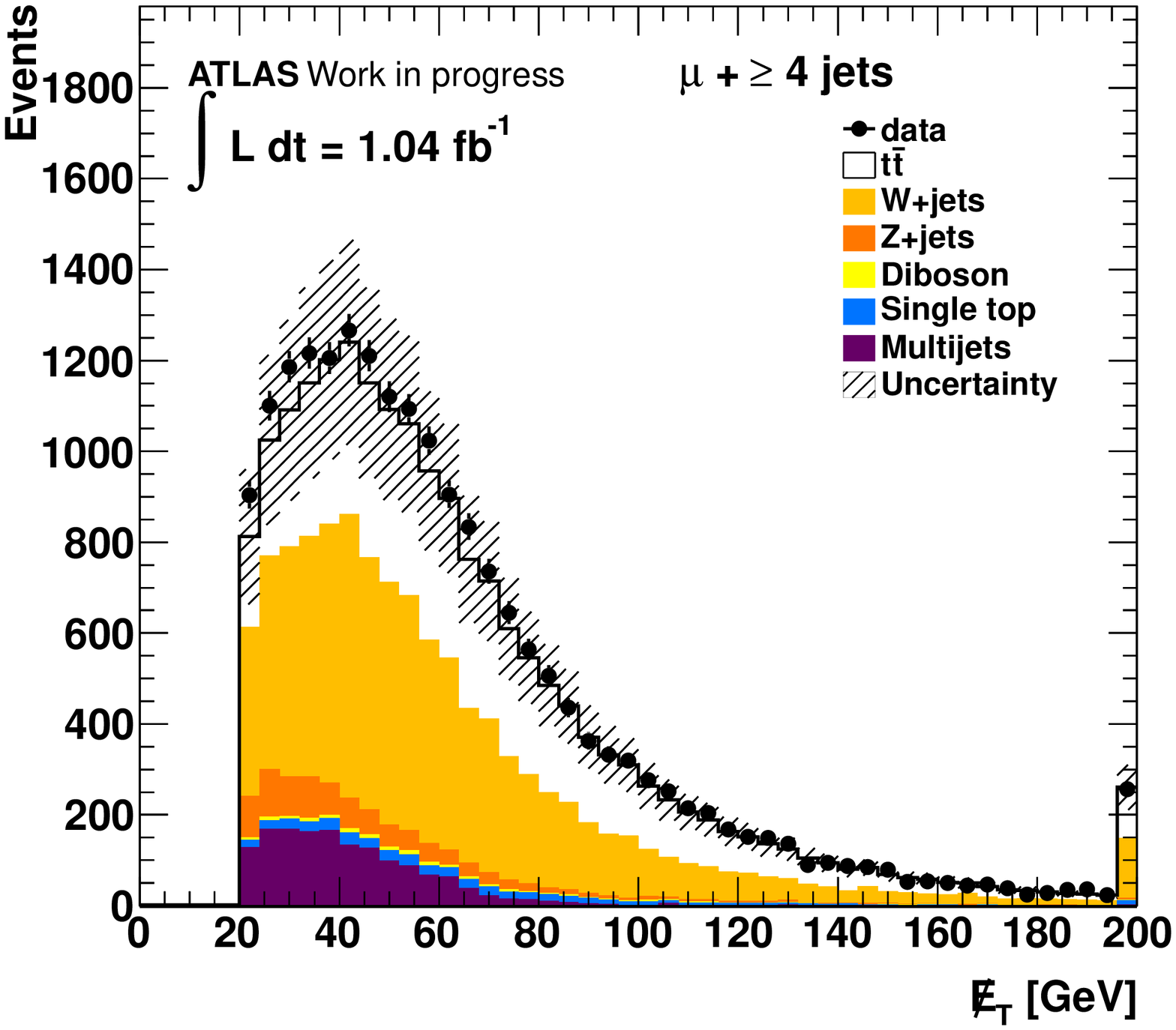}
    
    \vspace{-0.2 cm}
    \caption{Control plots for the muon+jets channel prior to the requirement of at least one $b$ tagged jet. From the top left to the bottom right, the transverse momentum $p_{\rm{T}}$, the pseudorapidity $\eta$ and the azimuthal angle $\phi$ of the selected muon are shown. Additional plots show the transverse momentum of the leading jet ($p_{\rm{T}}(j_1)$), the $W$ transverse mass $m_{\rm{T}}(W)$ and the transverse missing energy \met. Uncertainties are statistical and for $W$+jets also include systematic uncertainties on normalisation. For the QCD multijet background, a conservative 100\,\% systematic uncertainty was assumed. In addition, the uncertainties on luminosity, jet energy scale and \ttbar~cross-section are shown.}
    \label{controlplots_muons_pretag}
  \end{center}
\end{figure}

\begin{figure}[!htbp]
  \begin{center}
    \includegraphics[width=\plotwidth]{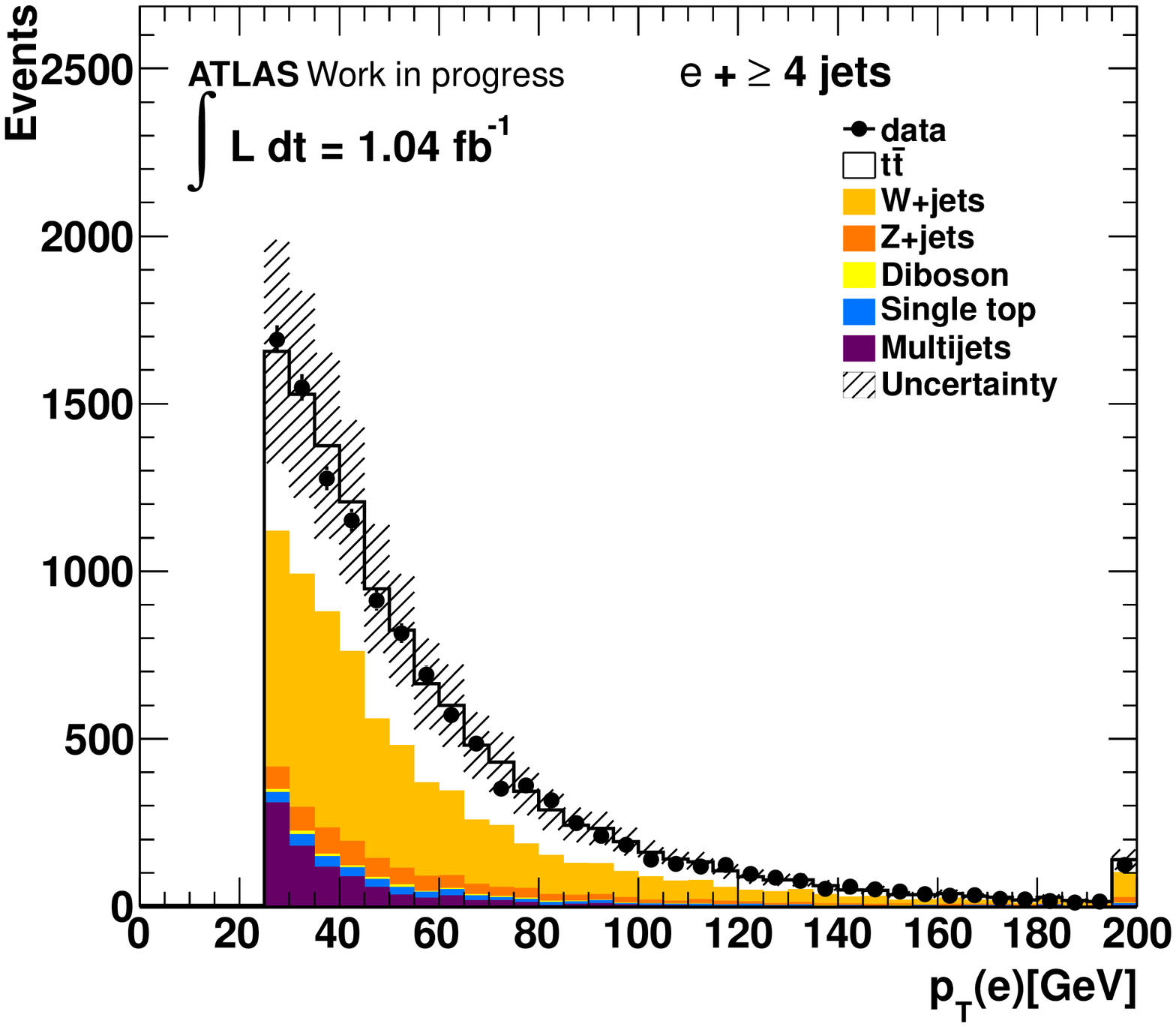}
    \quad\quad
    \includegraphics[width=\plotwidth]{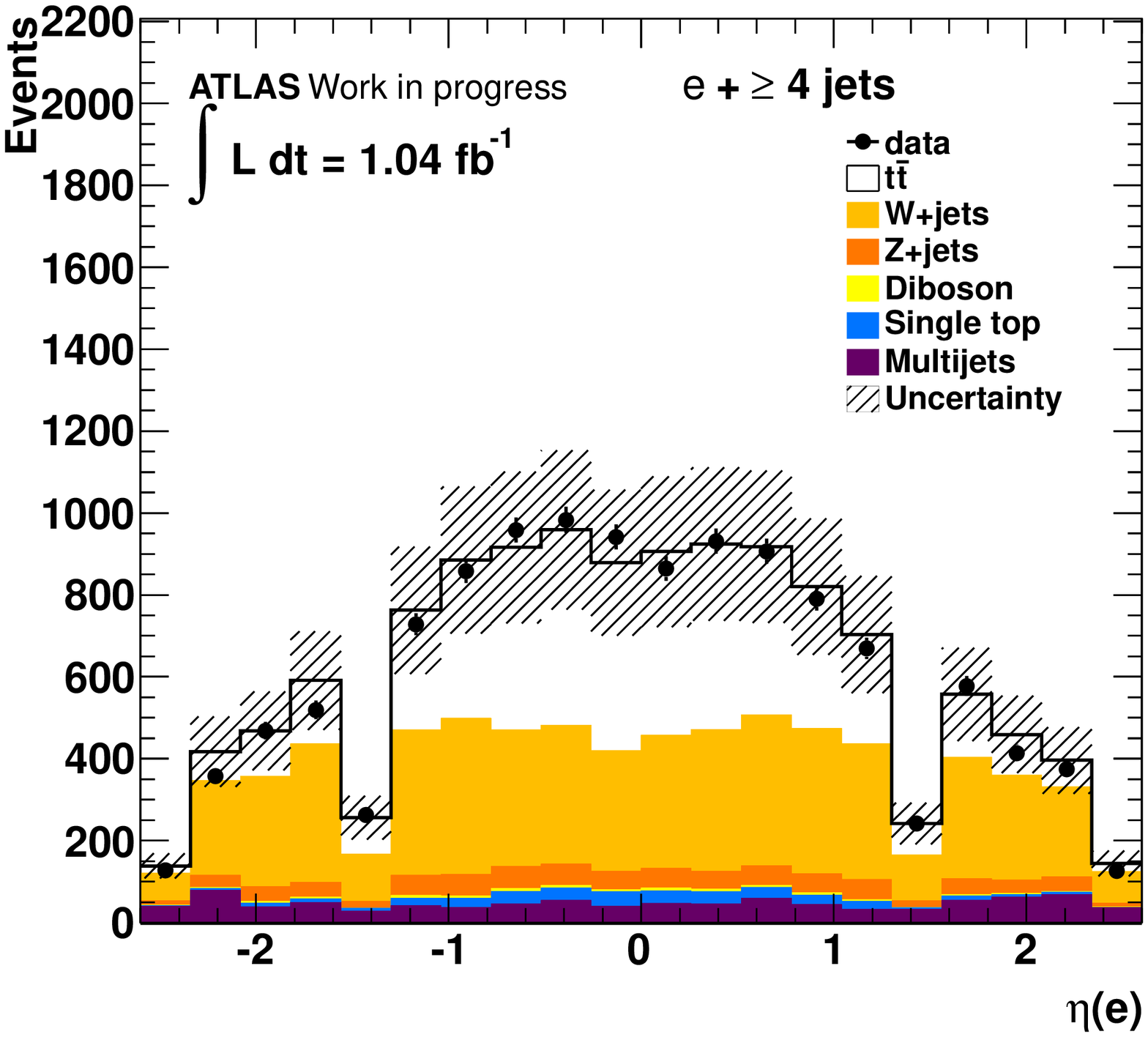}
    \includegraphics[width=\plotwidth]{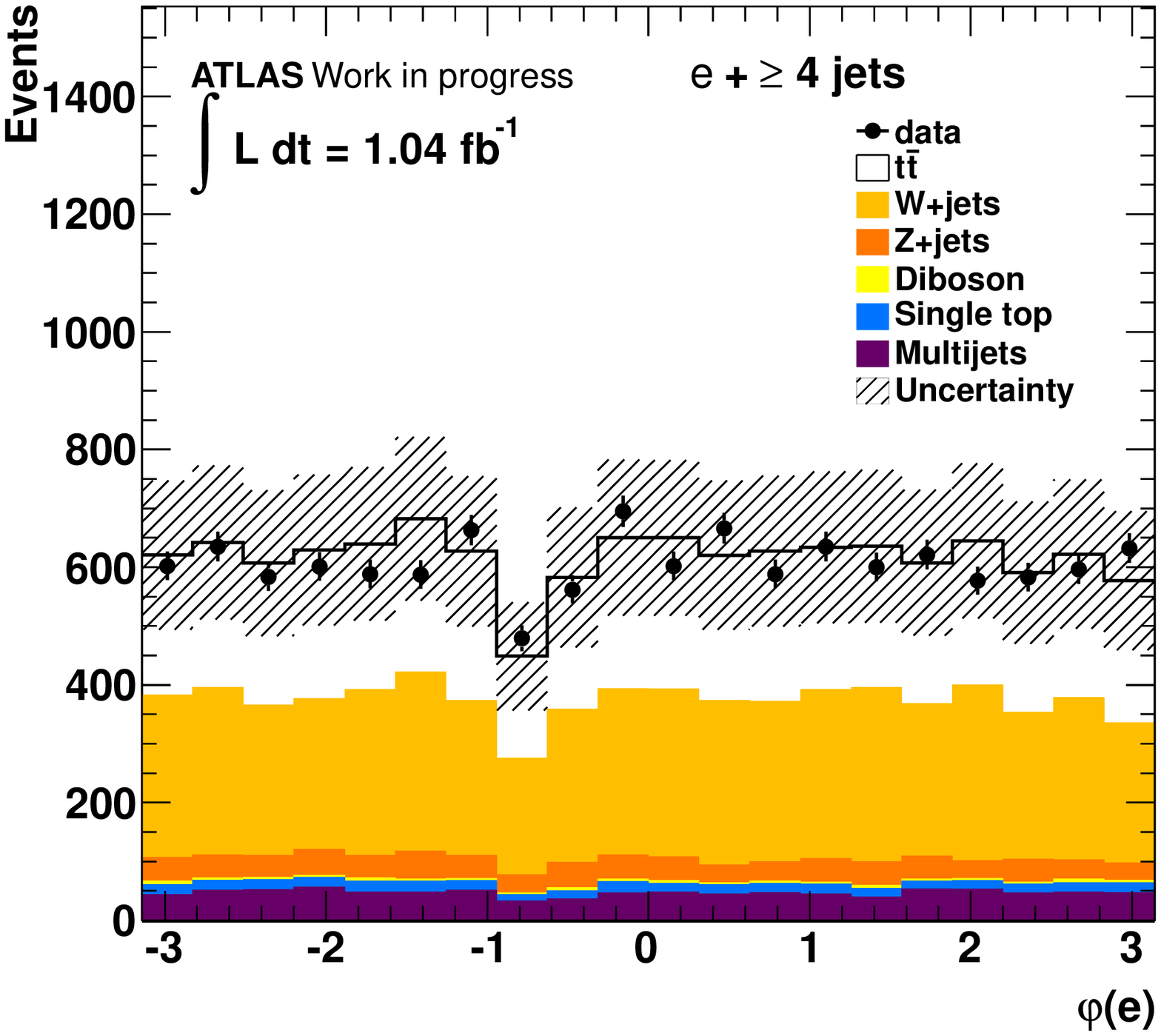}
    \quad\quad
    \includegraphics[width=\plotwidth]{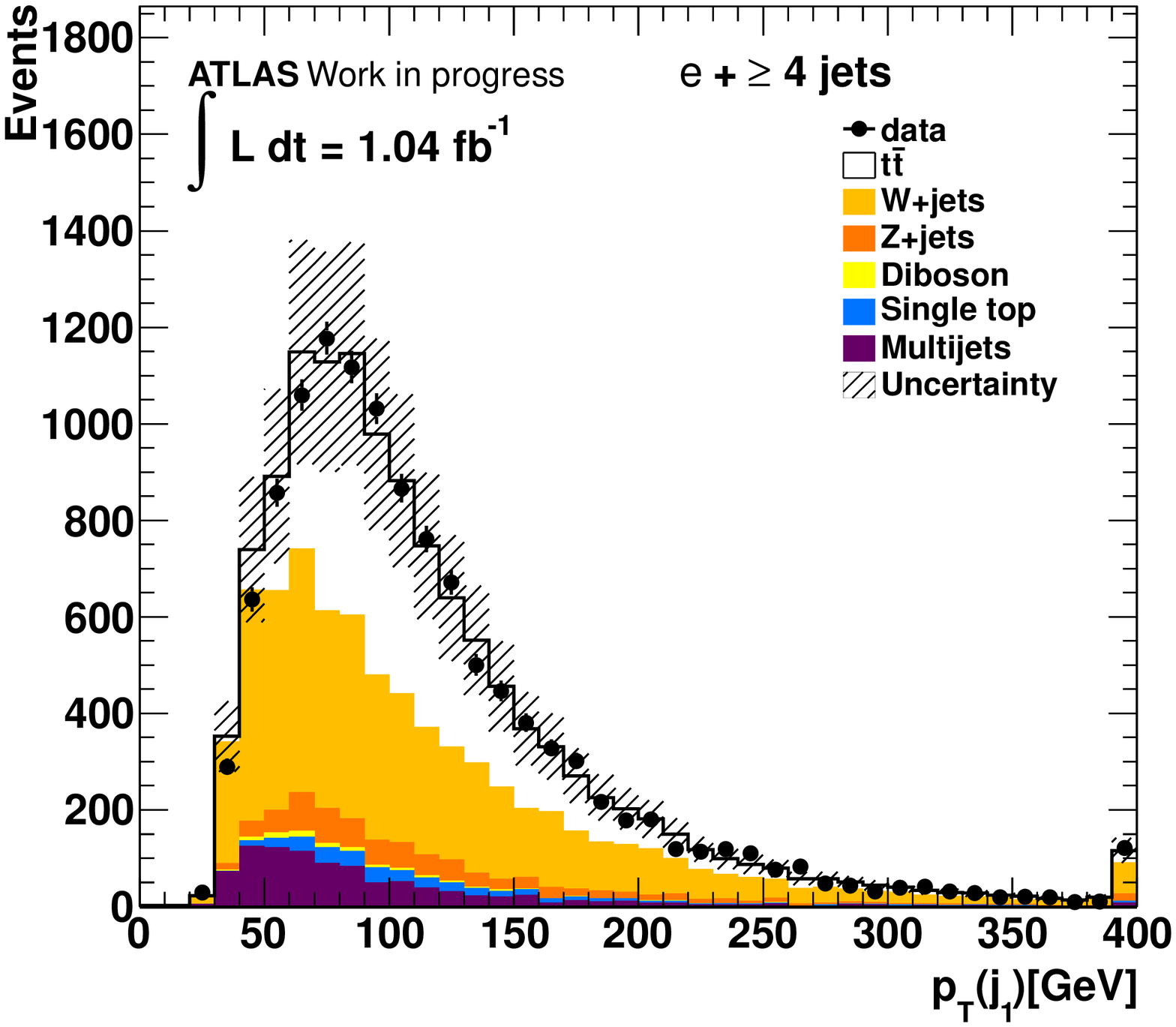}
    \includegraphics[width=\plotwidth]{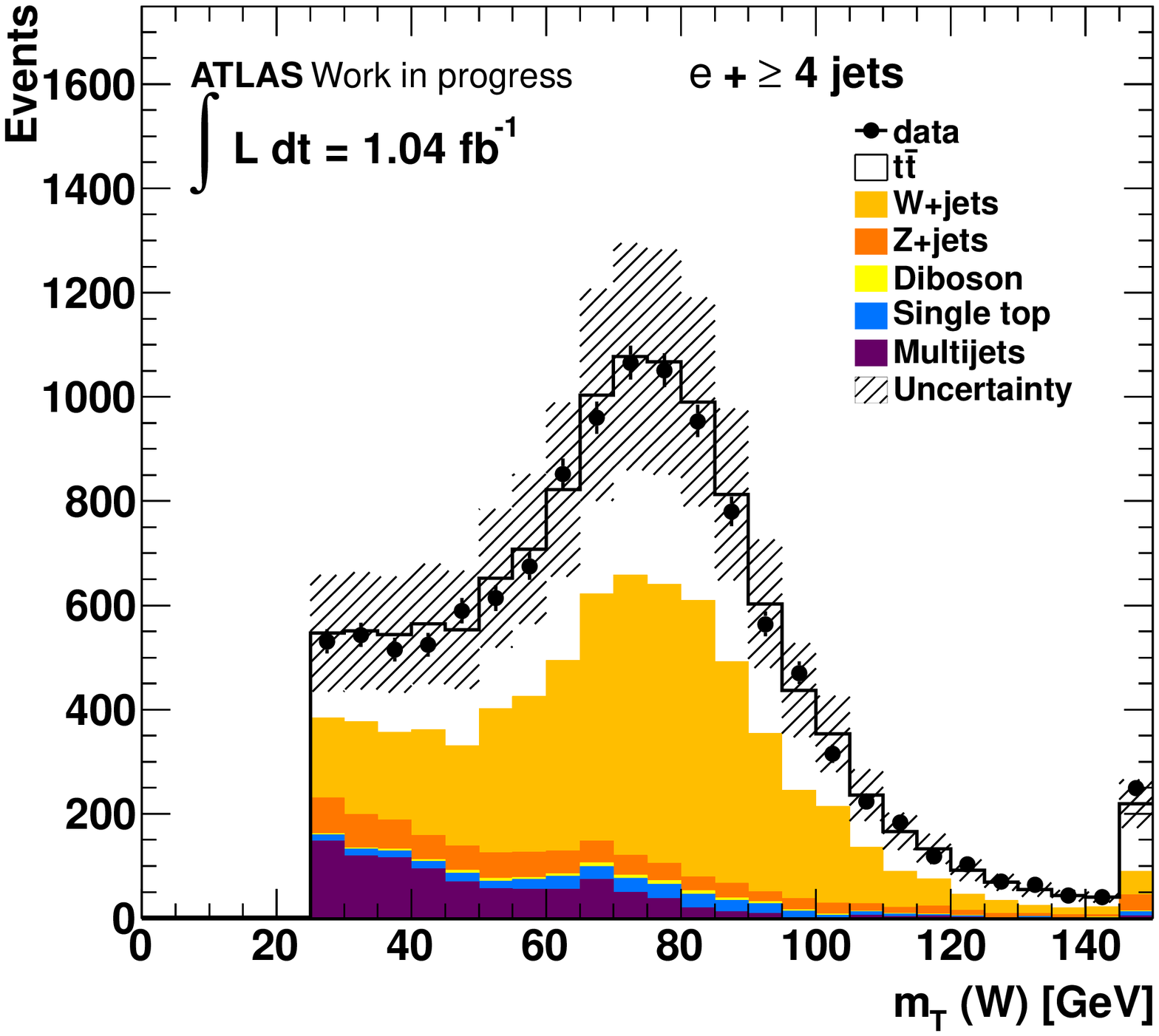}
    \quad\quad
    \includegraphics[width=\plotwidth]{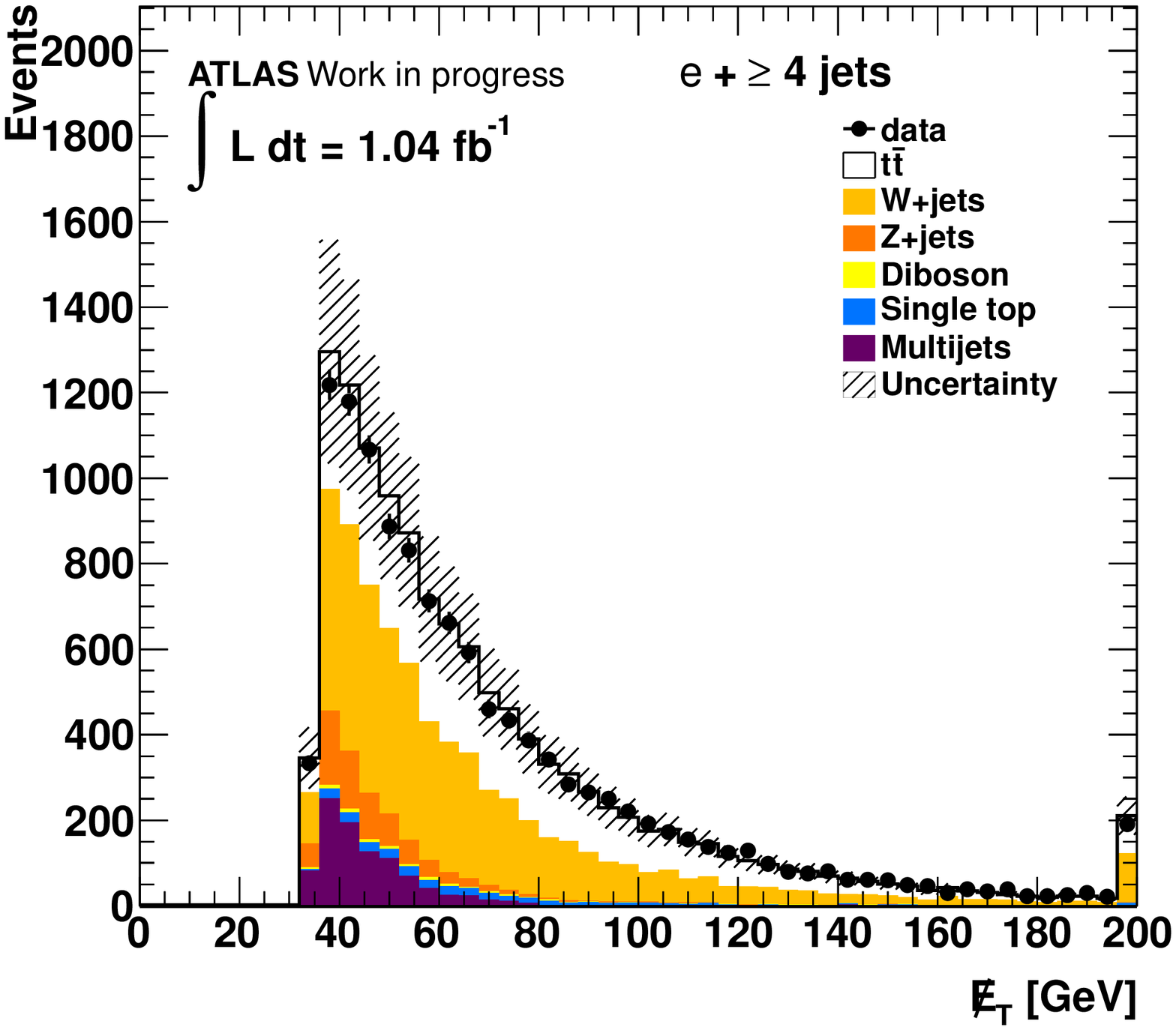}
    
    \vspace{-0.2 cm}
    \caption{Control plots for the electron+jets channel prior to the requirement of at least one $b$ tagged jet. From the top left to the bottom right, the transverse momentum $p_{\rm{T}}$, the pseudorapidity $\eta$ and the azimuthal angle $\phi$ of the selected electron are shown. Additional plots show the transverse momentum of the leading jet ($p_{\rm{T}}(j_1)$), the $W$ transverse mass $m_{\rm{T}}(W)$ and the transverse missing energy \met. Uncertainties are statistical and for $W$+jets also include systematic uncertainties on normalisation. For the QCD multijet background, a conservative 100\,\% systematic uncertainty was assumed. In addition, the uncertainties on luminosity, jet energy scale and \ttbar~cross-section are shown.}
    \label{controlplots_electrons_pretag}
  \end{center}
\end{figure}

\begin{figure}[!htbp]
  \begin{center}
    \includegraphics[width=\plotwidth]{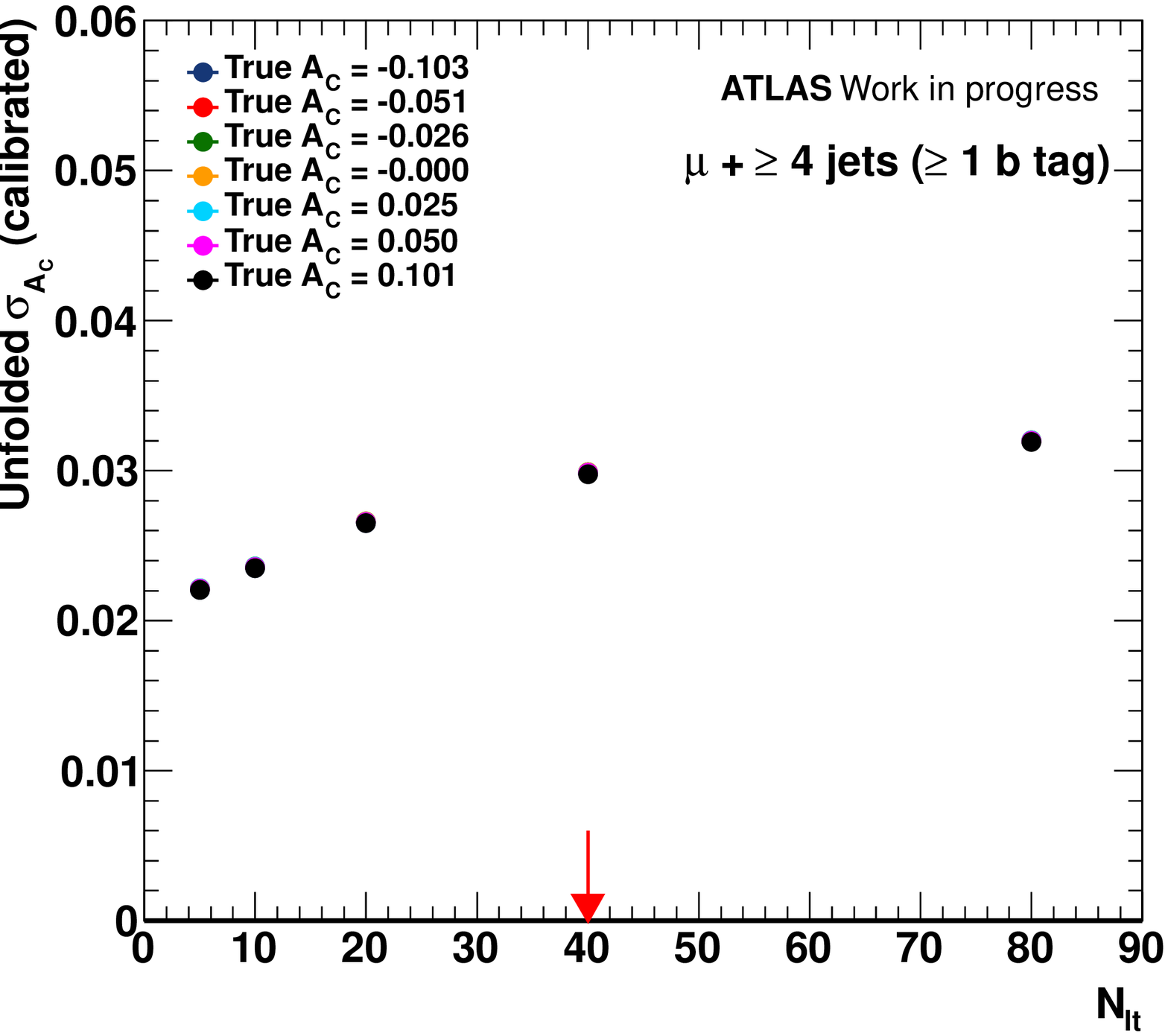}
    \quad\quad
    \includegraphics[width=\plotwidth]{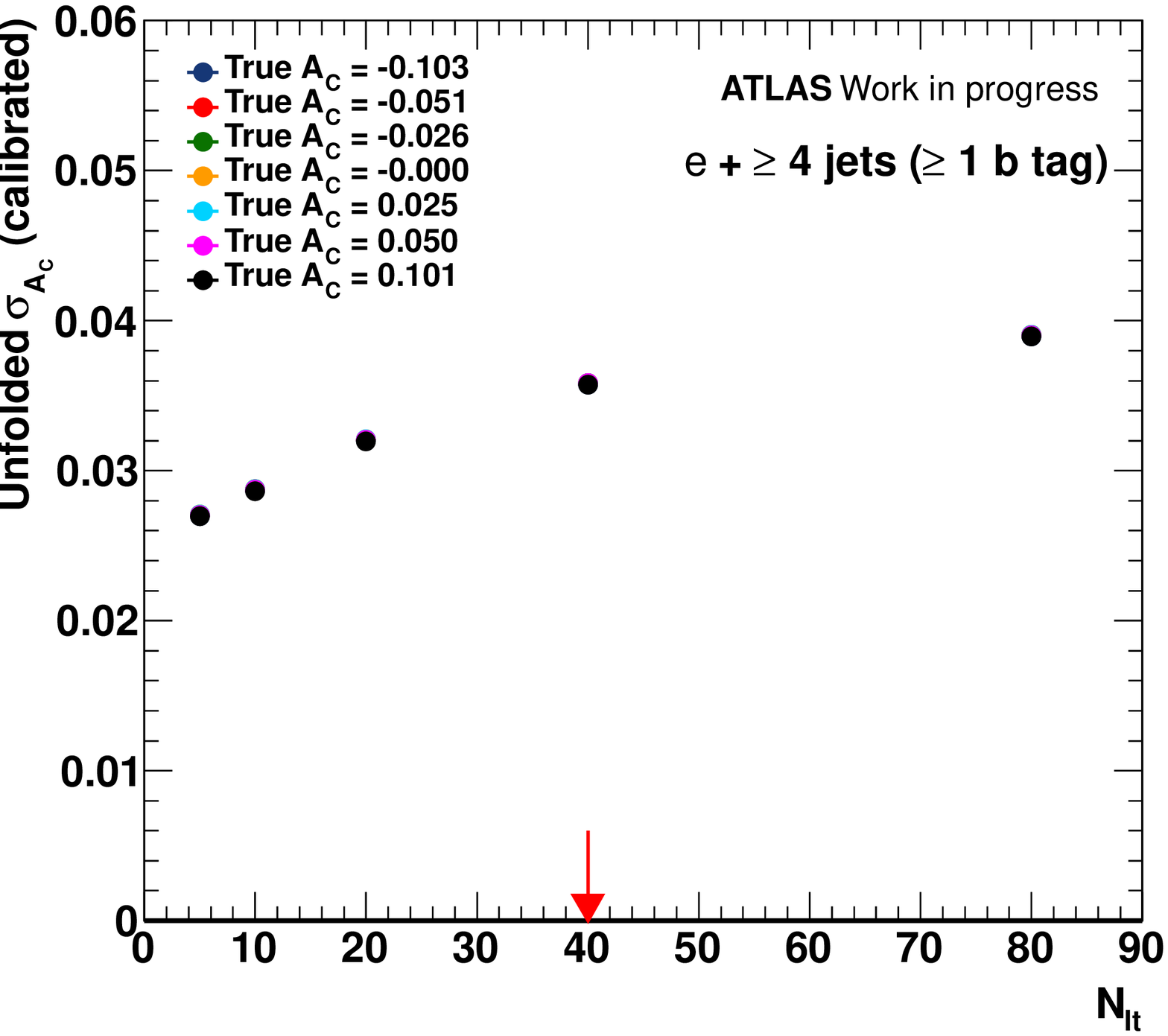}
    \includegraphics[width=\plotwidth]{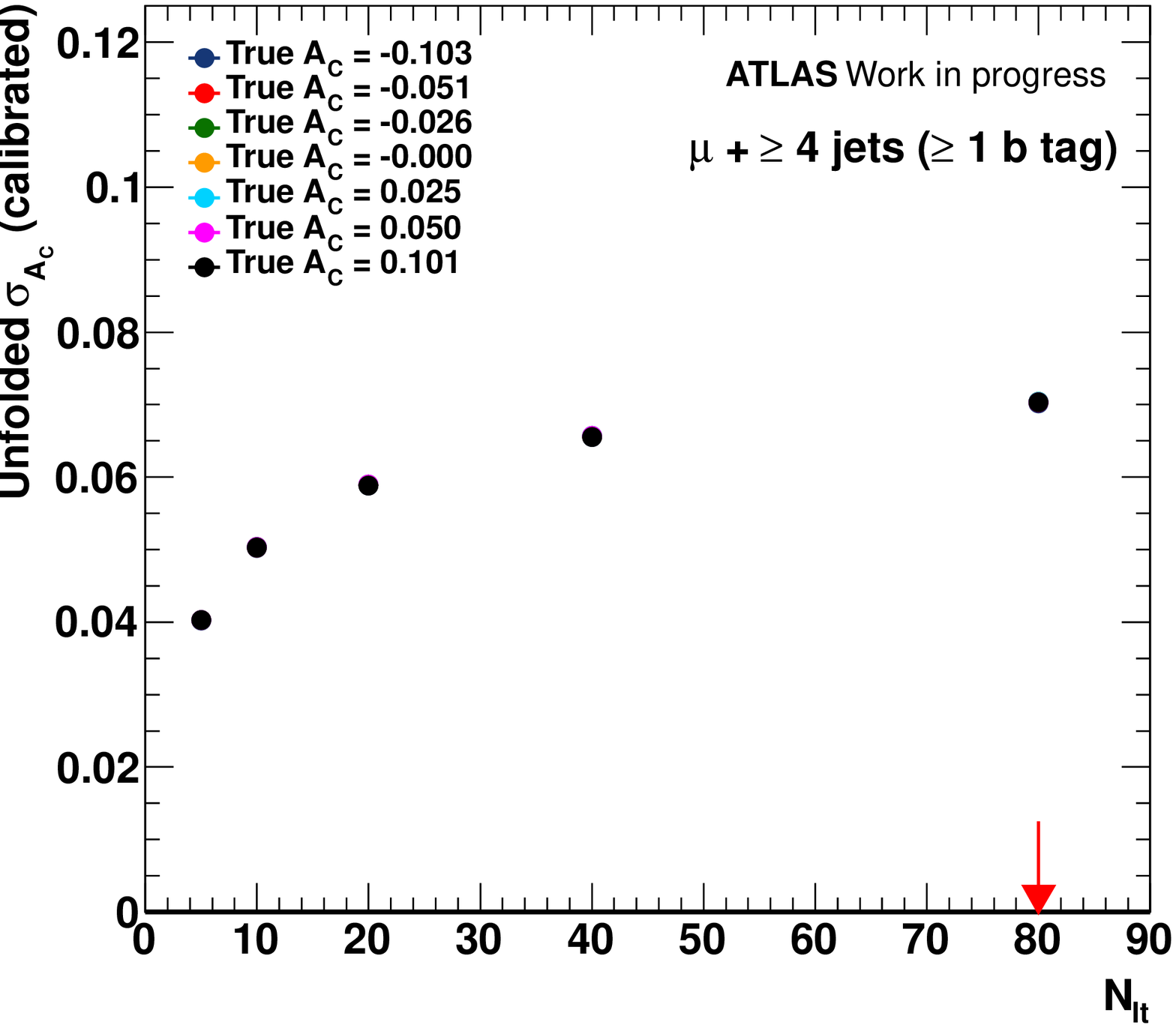}
    \quad\quad
    \includegraphics[width=\plotwidth]{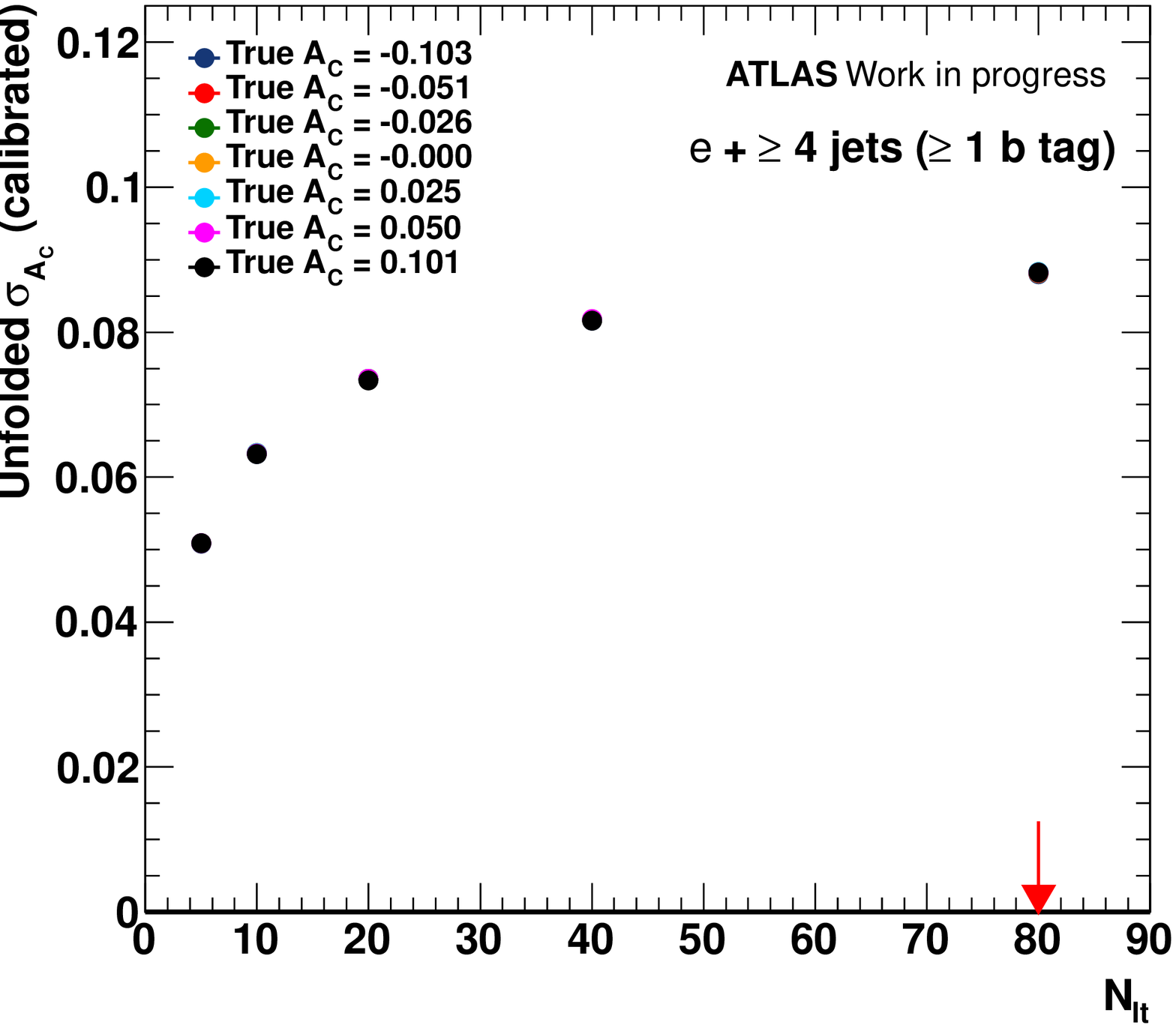}
    \includegraphics[width=\plotwidth]{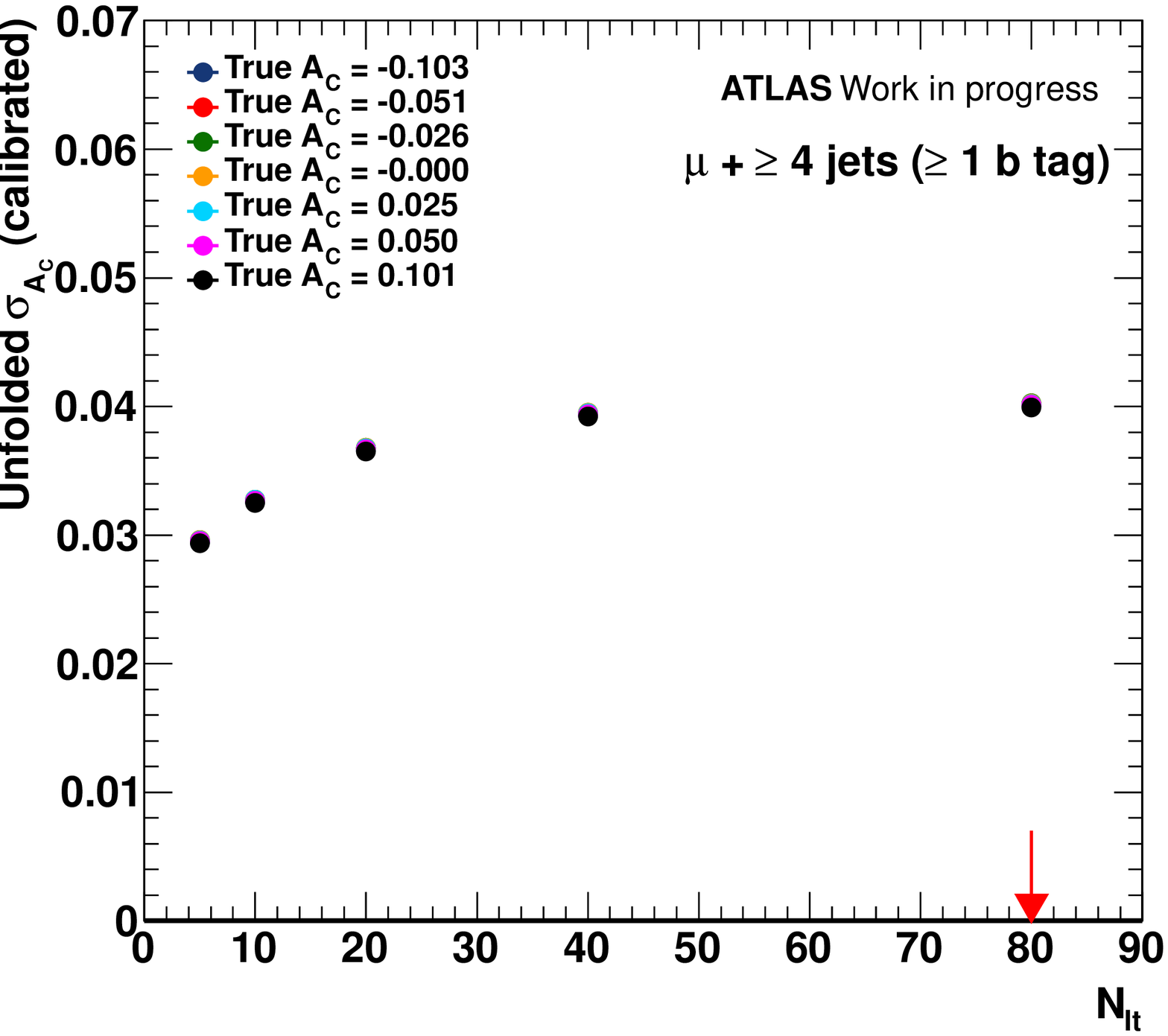}
    \quad\quad
    \includegraphics[width=\plotwidth]{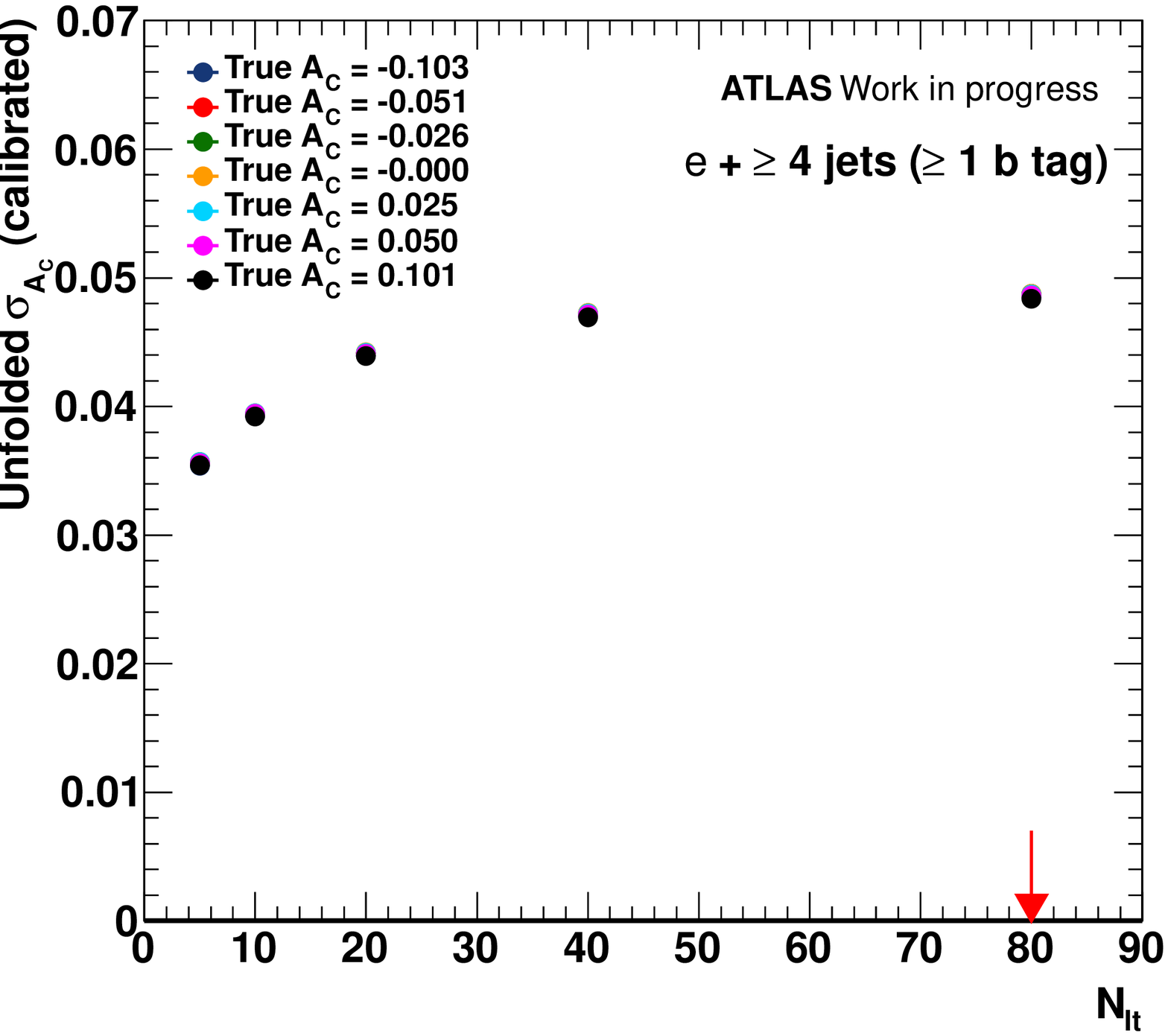}
    
    \vspace{-0.2 cm}
    \caption{Expected statistical uncertainties on the unfolded value of $A_C^{\text{unf}}$ as a function of the regularisation parameter $N_{\text{It}}$ for different injected true asymmetries in the observable $A_C^{\text{true}}$ for both the muon+jets channel (left) and electron+jets channel (right). The top row shows the respective distribution for the inclusive measurement, while the lower rows show the corresponding distributions for $M_{t\bar{t}} < 450$\,GeV and $M_{t\bar{t}} > 450$\,GeV, respectively. For the simultaneous unfolding in $|y_t| - |y_{\bar{t}}|$ and $M_{t \bar{t}}$, a cut on the event reconstruction likelihood $\log{L}$ was applied to improve the $M_{t\bar{t}}$ resolution of the selected events. The arrows indicate the chosen values for $N_{\text{It}}$.}
    \label{fig:unfolding:calibuncert}
  \end{center}
\end{figure}

\begin{figure}[!htbp]
  \begin{center}
    \includegraphics[width=\plotwidth]{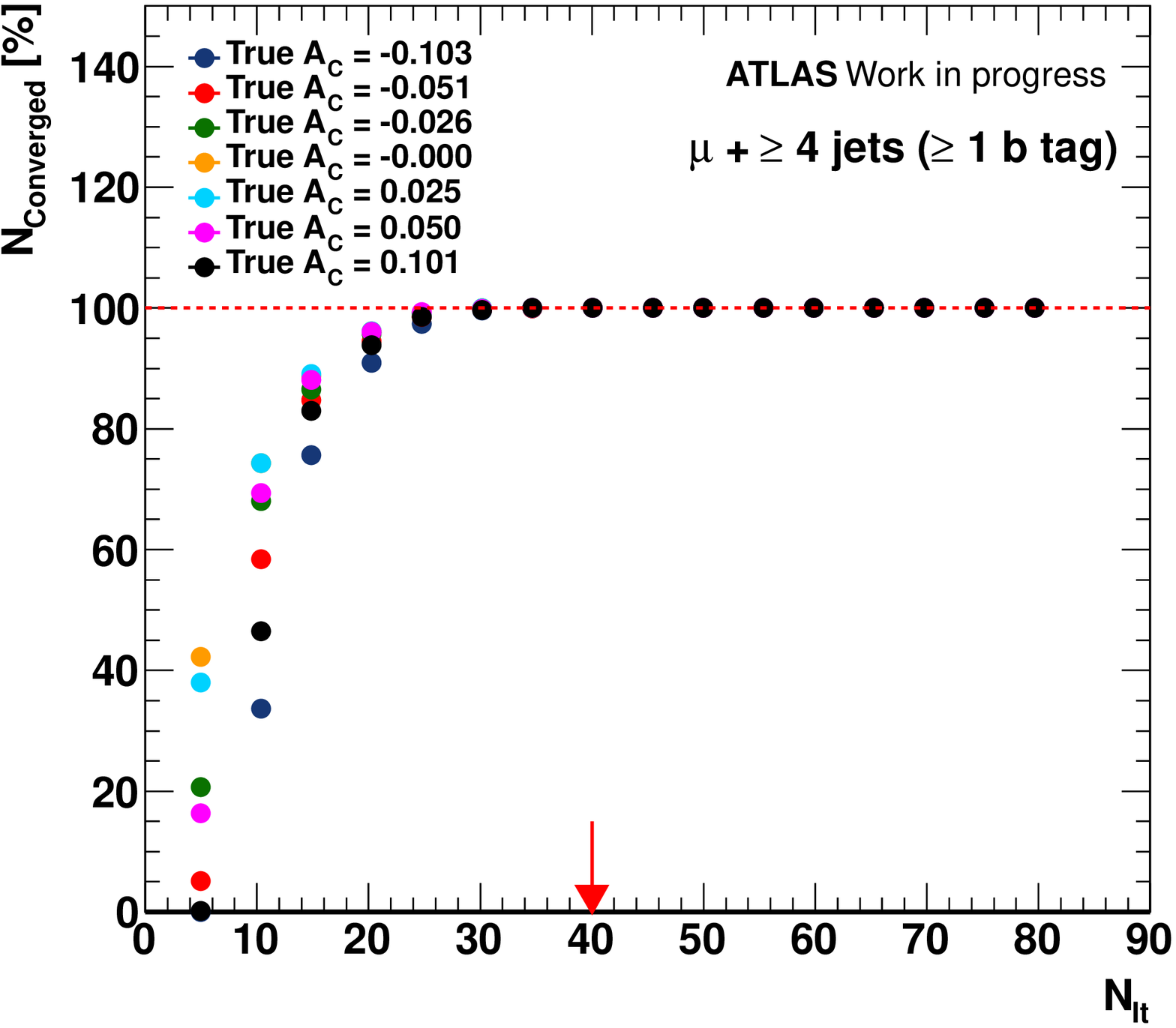}
    \quad\quad
    \includegraphics[width=\plotwidth]{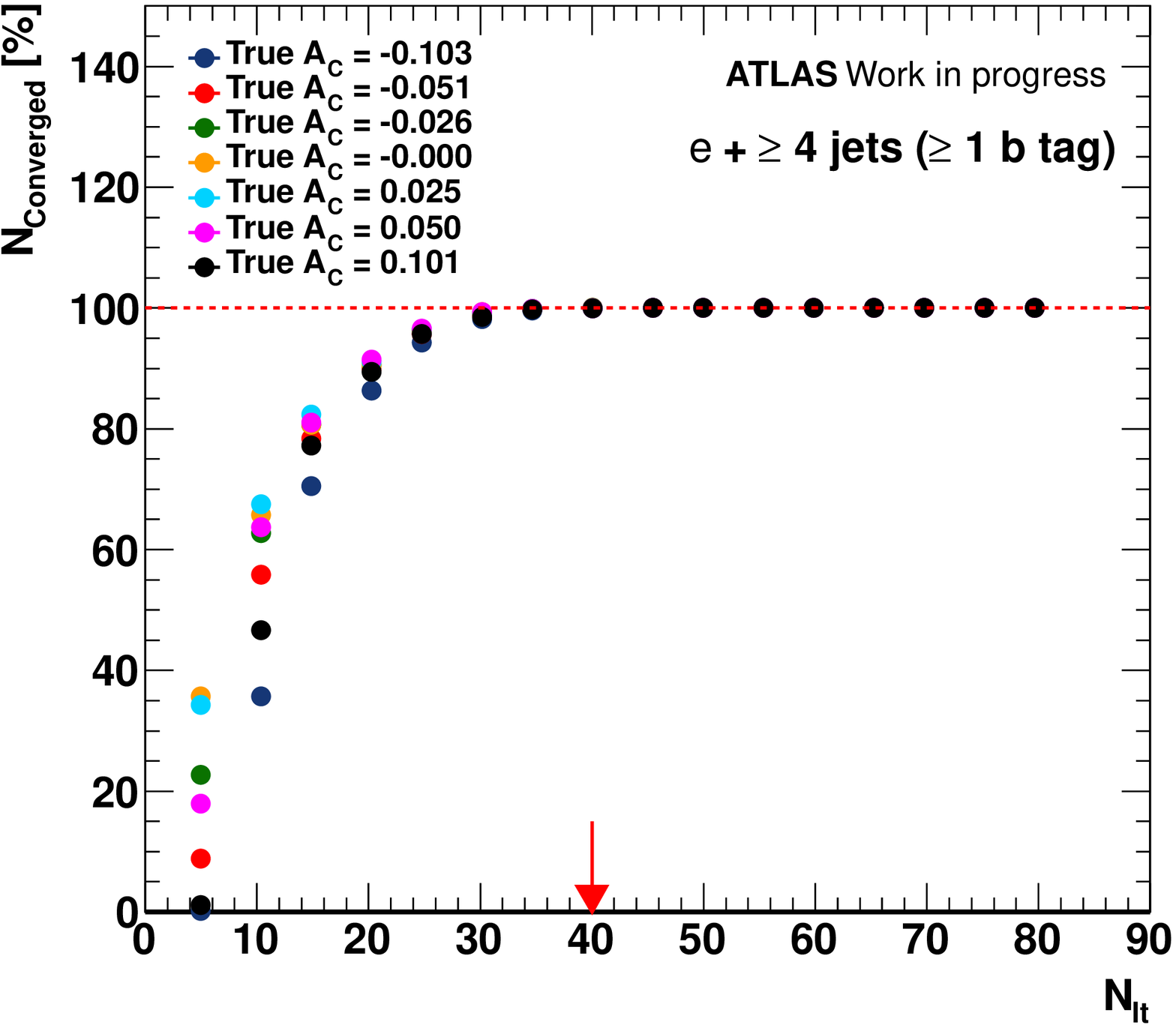}
    \includegraphics[width=\plotwidth]{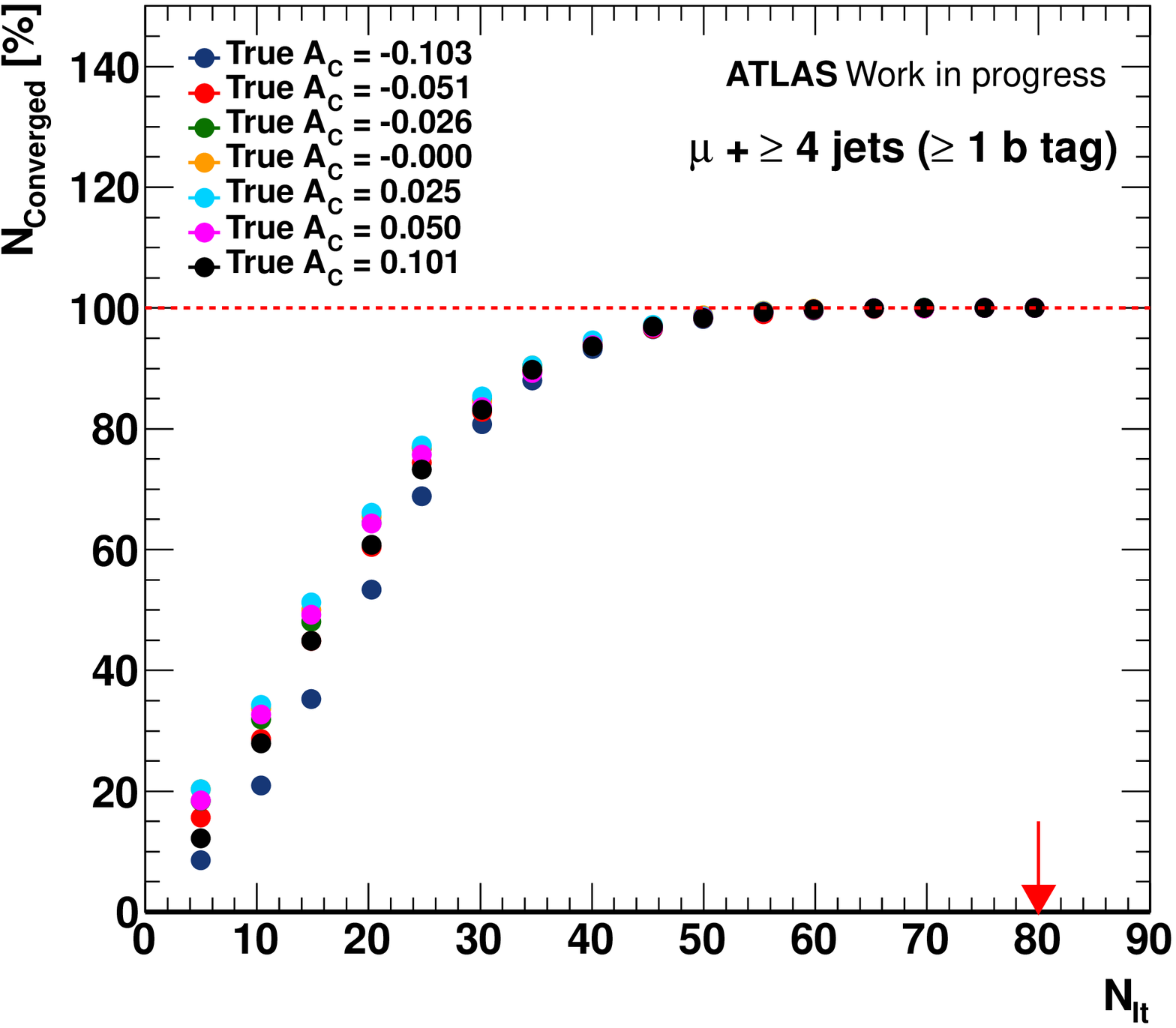}
    \quad\quad
    \includegraphics[width=\plotwidth]{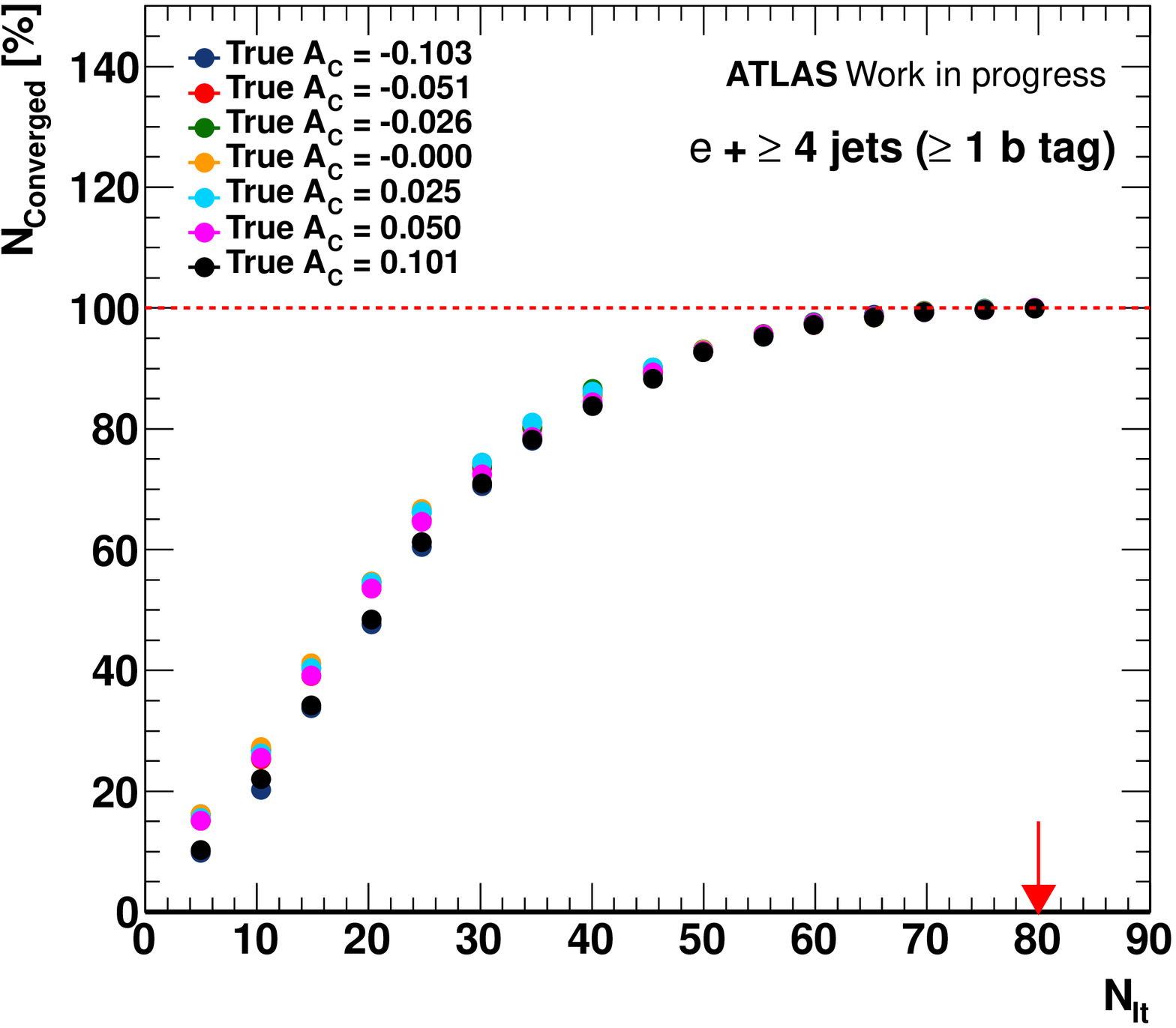}
    \includegraphics[width=\plotwidth]{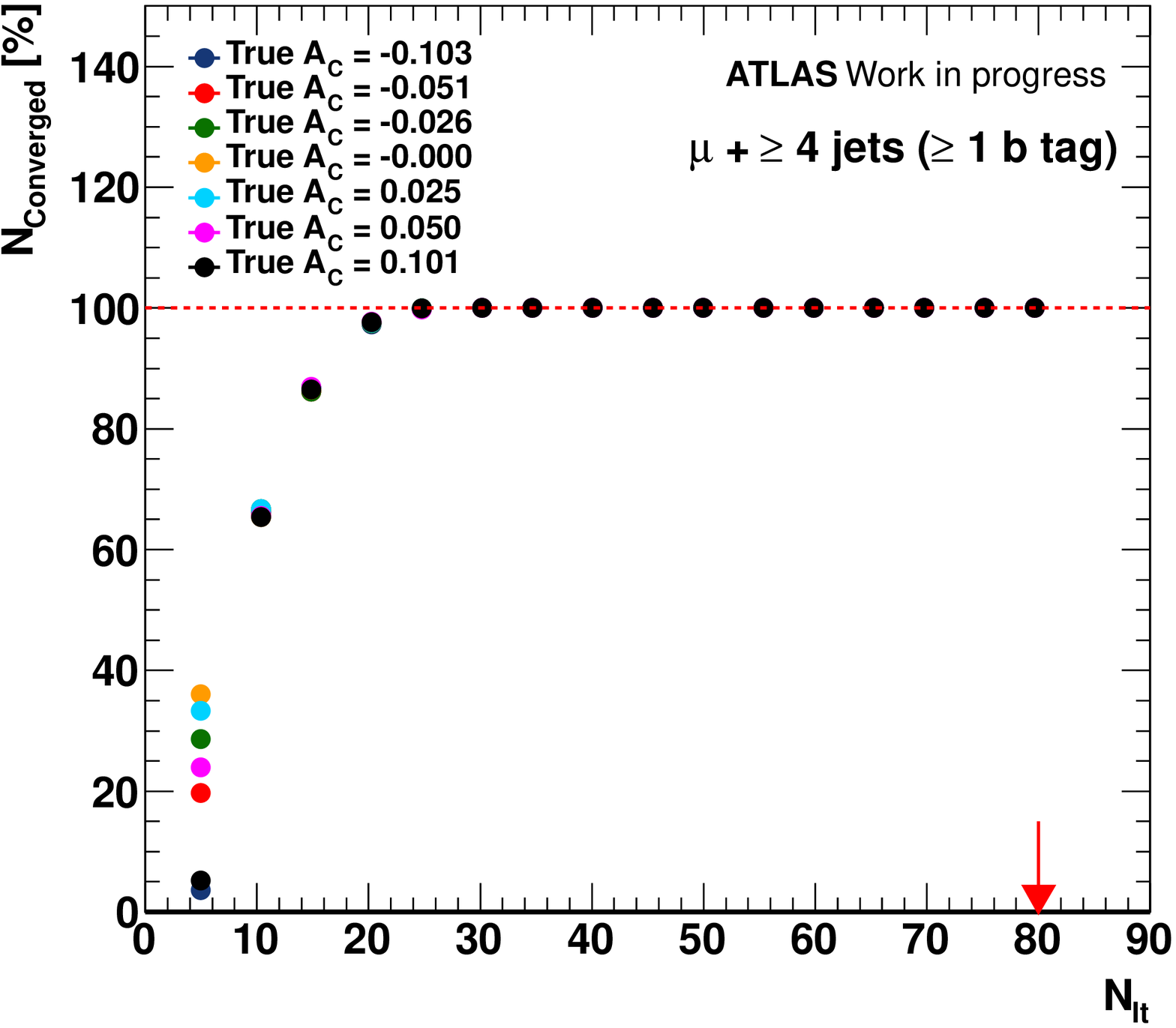}
    \quad\quad
    \includegraphics[width=\plotwidth]{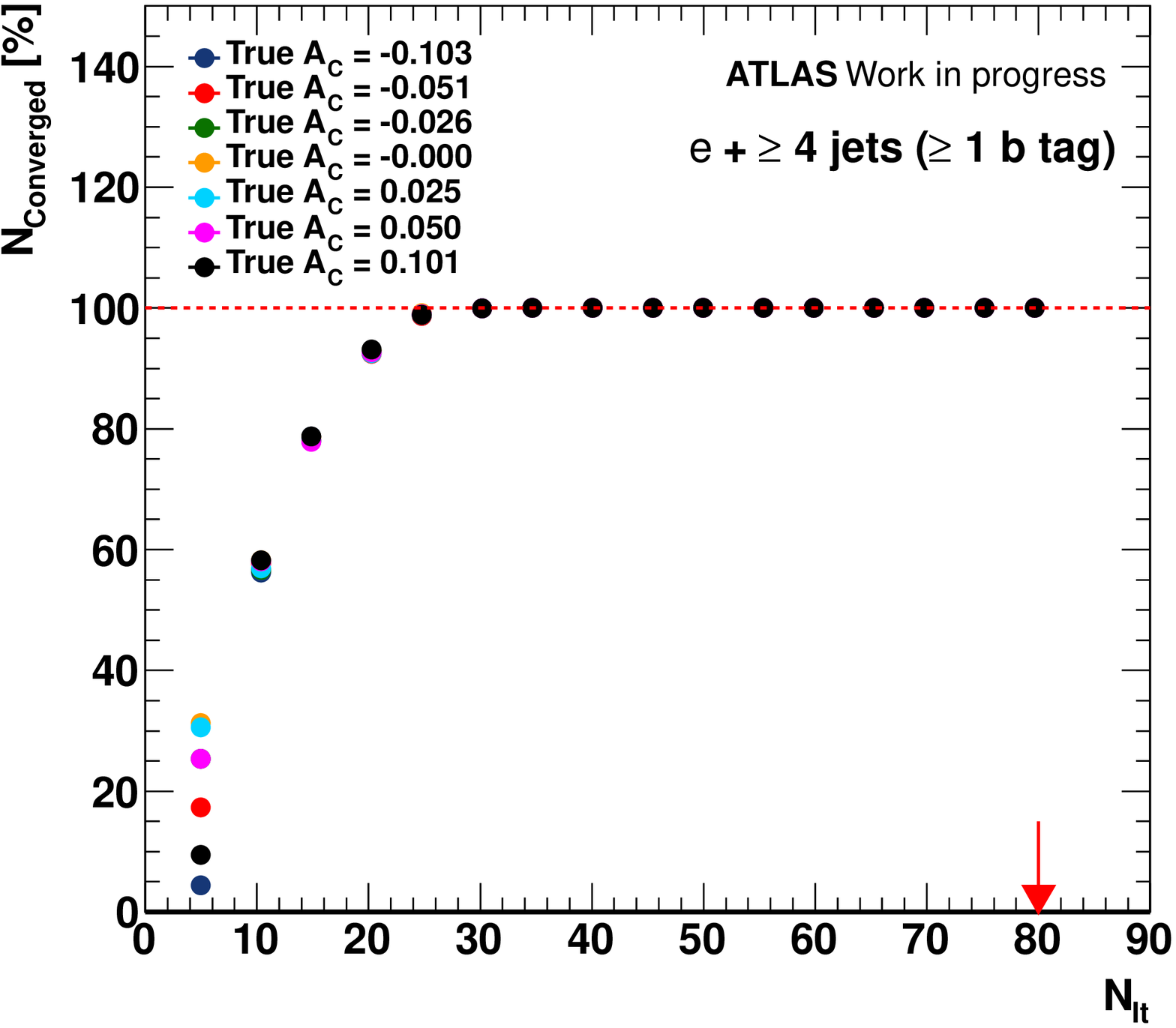}
    
    \vspace{-0.2 cm}
    \caption{Distributions showing the convergence behaviour of ensembles of pseudodata as a function of regularisation for different injected true asymmetries $A_C^{\text{true}}$ for both the muon+jets channel (left) and electron+jets channel (right), parametrised by the number of iterations $N_{\text{It}}$ in the unfolding procedure. The relative amount of ensembles which have reached a convergent state at given $N_{\text{It}}$ (100\,\% being indicated by the red dashed line) are shown. The top row shows the respective distributions for the inclusive measurement, while the lower rows show the corresponding distributions for $M_{t\bar{t}} < 450$\,GeV and $M_{t\bar{t}} > 450$\,GeV, respectively. For the simultaneous unfolding in $|y_t| - |y_{\bar{t}}|$ and $M_{t \bar{t}}$, a cut on the event reconstruction likelihood $\log{L}$ was applied to improve the $M_{t\bar{t}}$ resolution of the selected events. The arrows indicate the chosen values for $N_{\text{It}}$.}
    \label{fig:conv1}
  \end{center}
\end{figure}

\begin{figure}[!htbp]
  \begin{center}
    \includegraphics[width=\plotwidth]{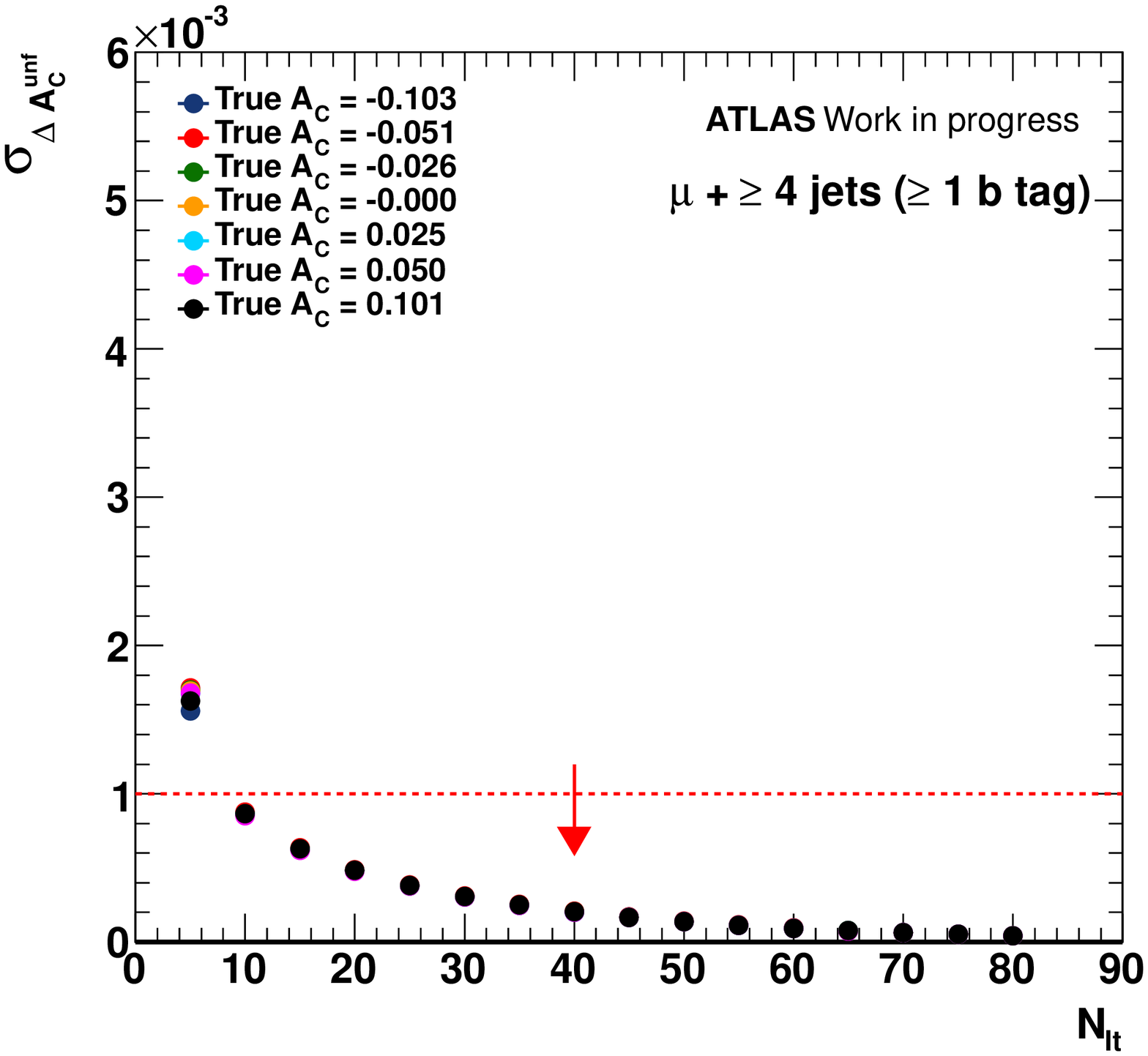}
    \quad\quad
    \includegraphics[width=\plotwidth]{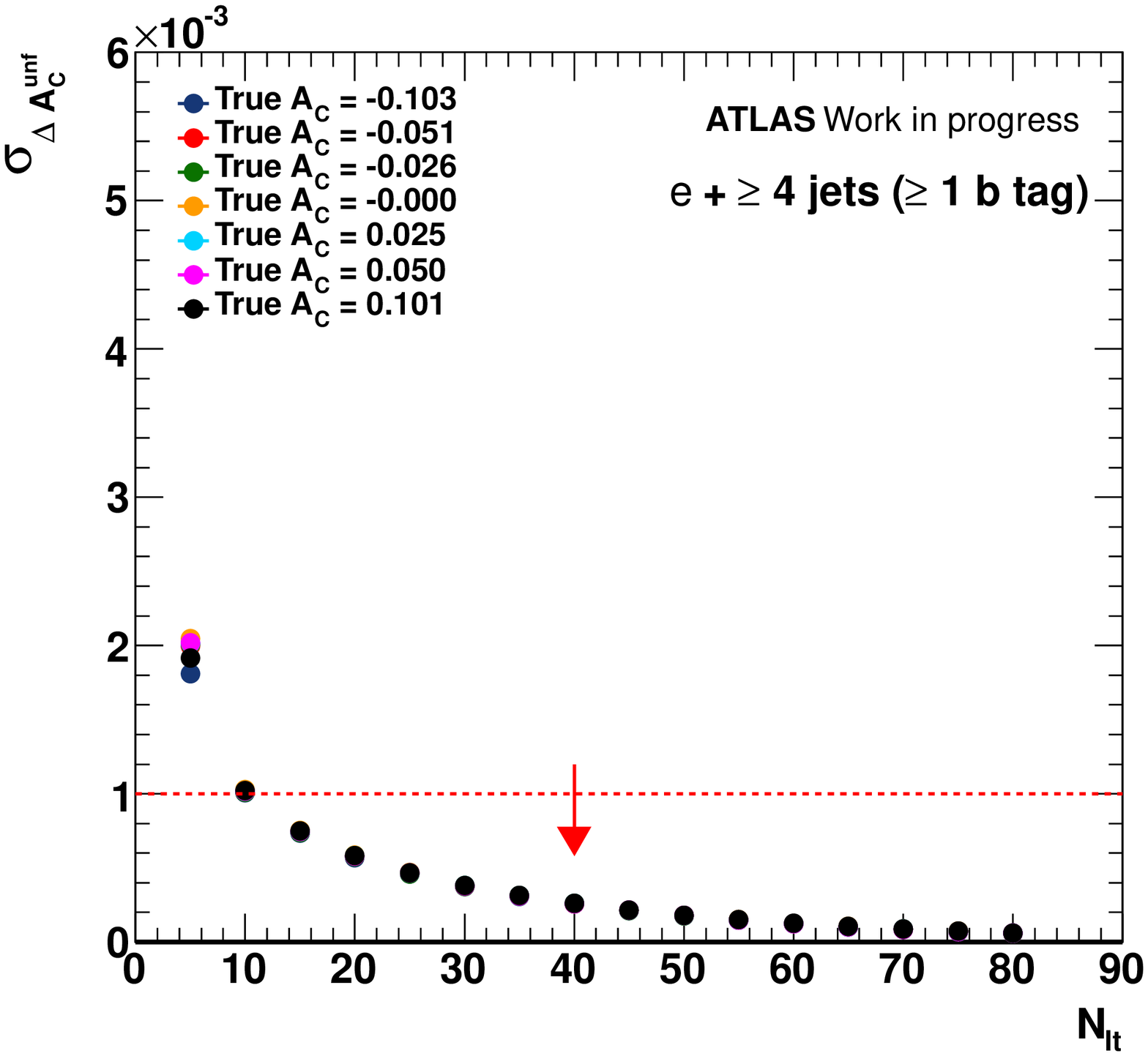}
    \includegraphics[width=\plotwidth]{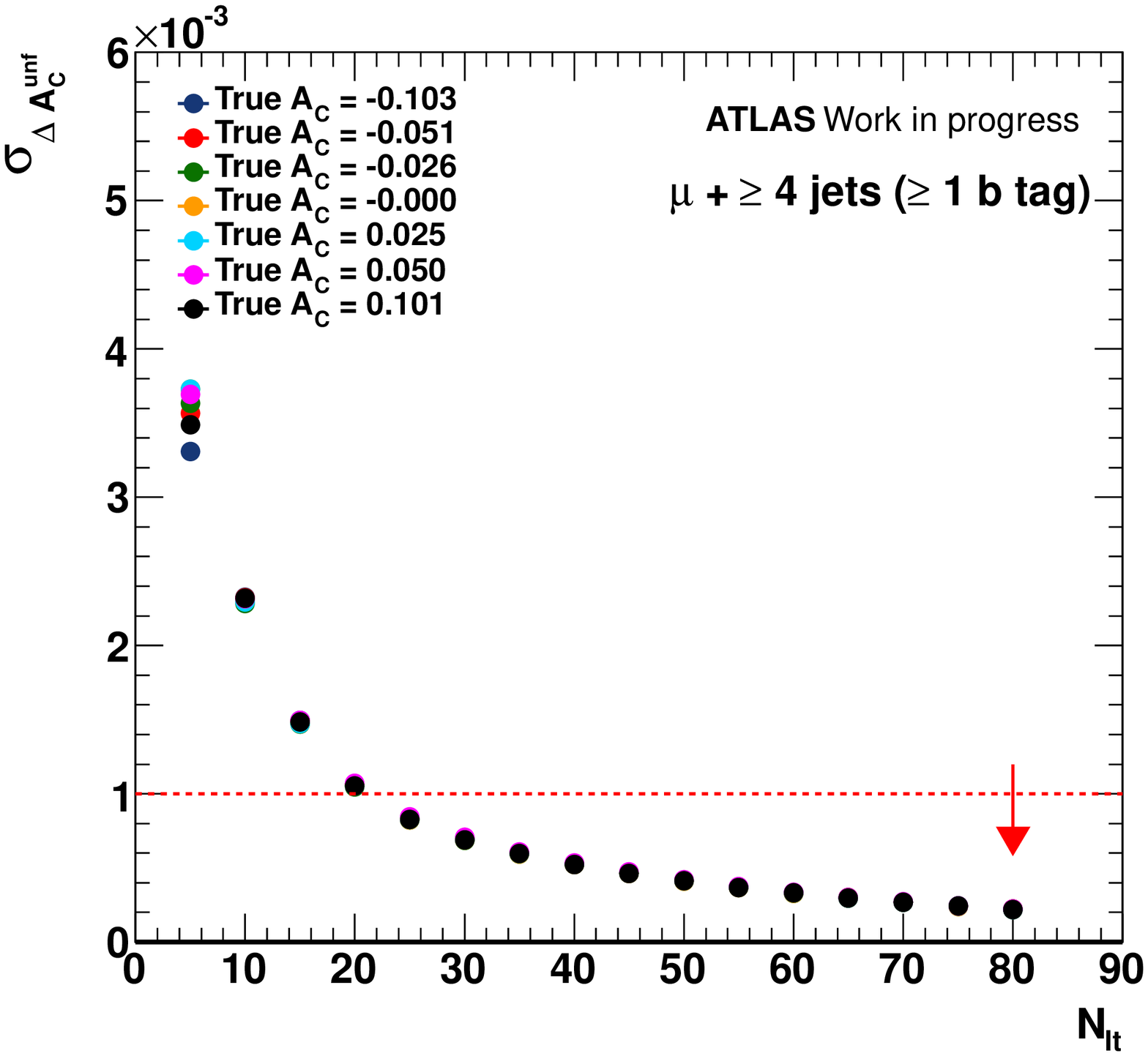}
    \quad\quad
    \includegraphics[width=\plotwidth]{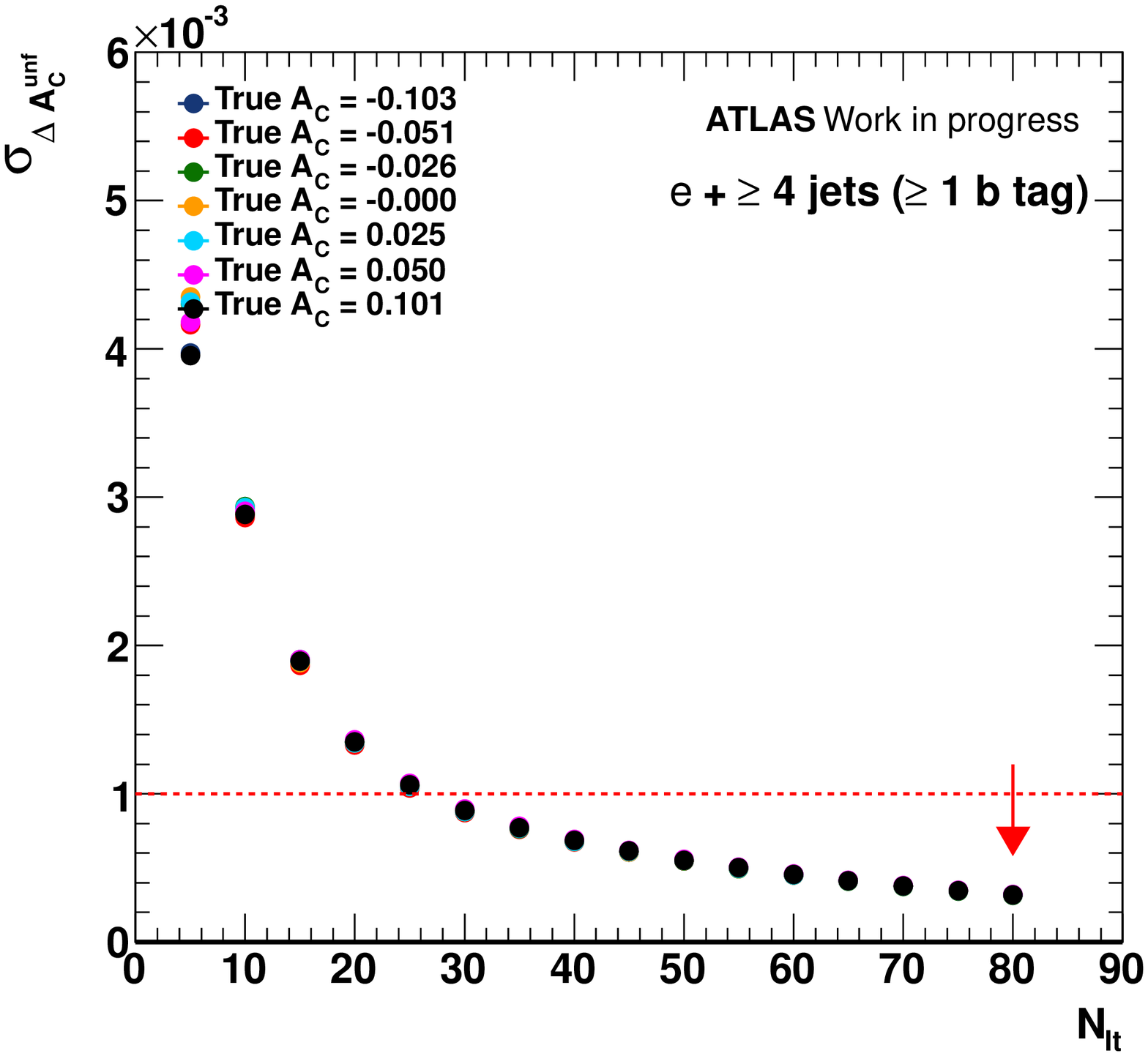}
    \includegraphics[width=\plotwidth]{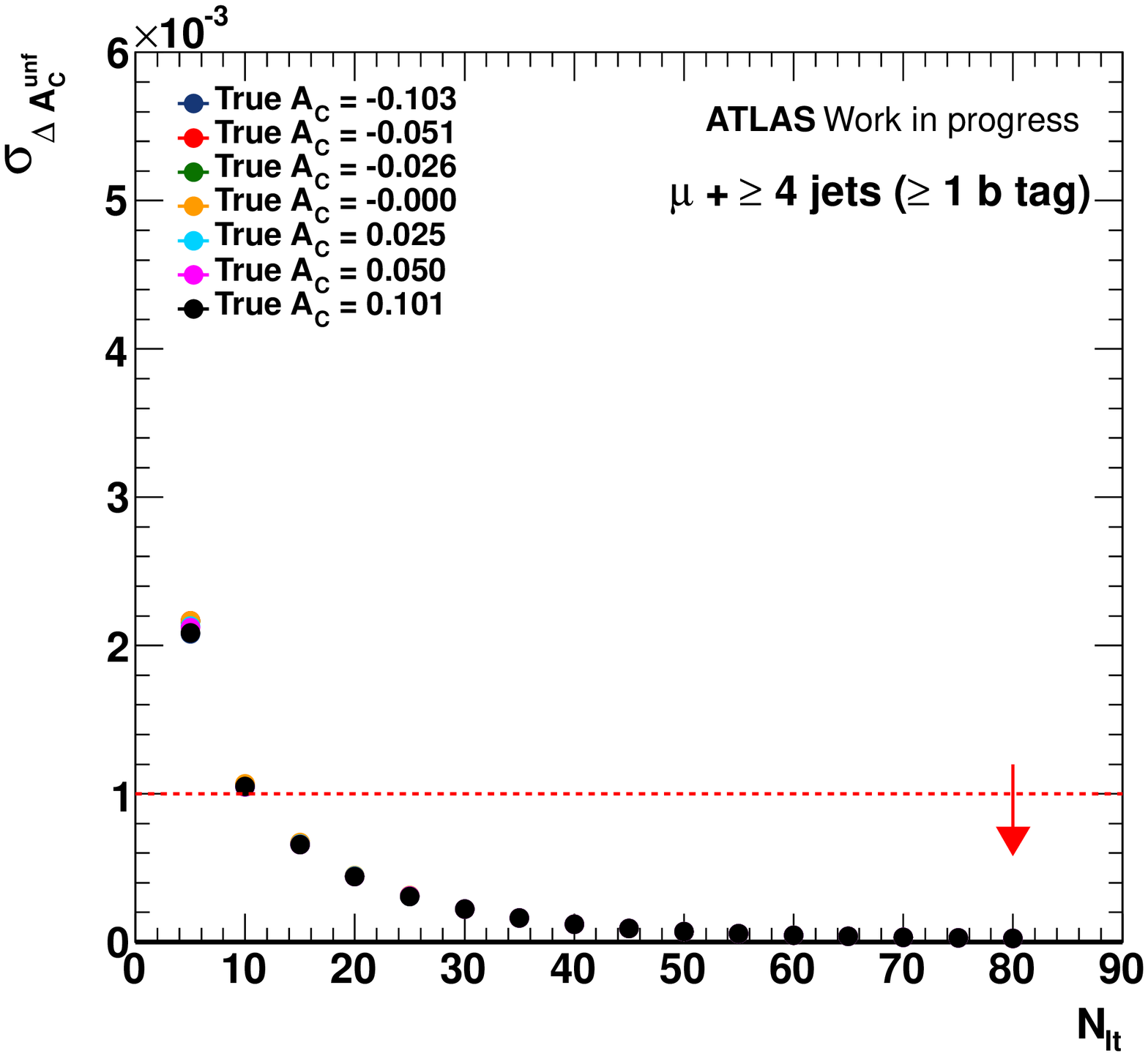}
    \quad\quad
    \includegraphics[width=\plotwidth]{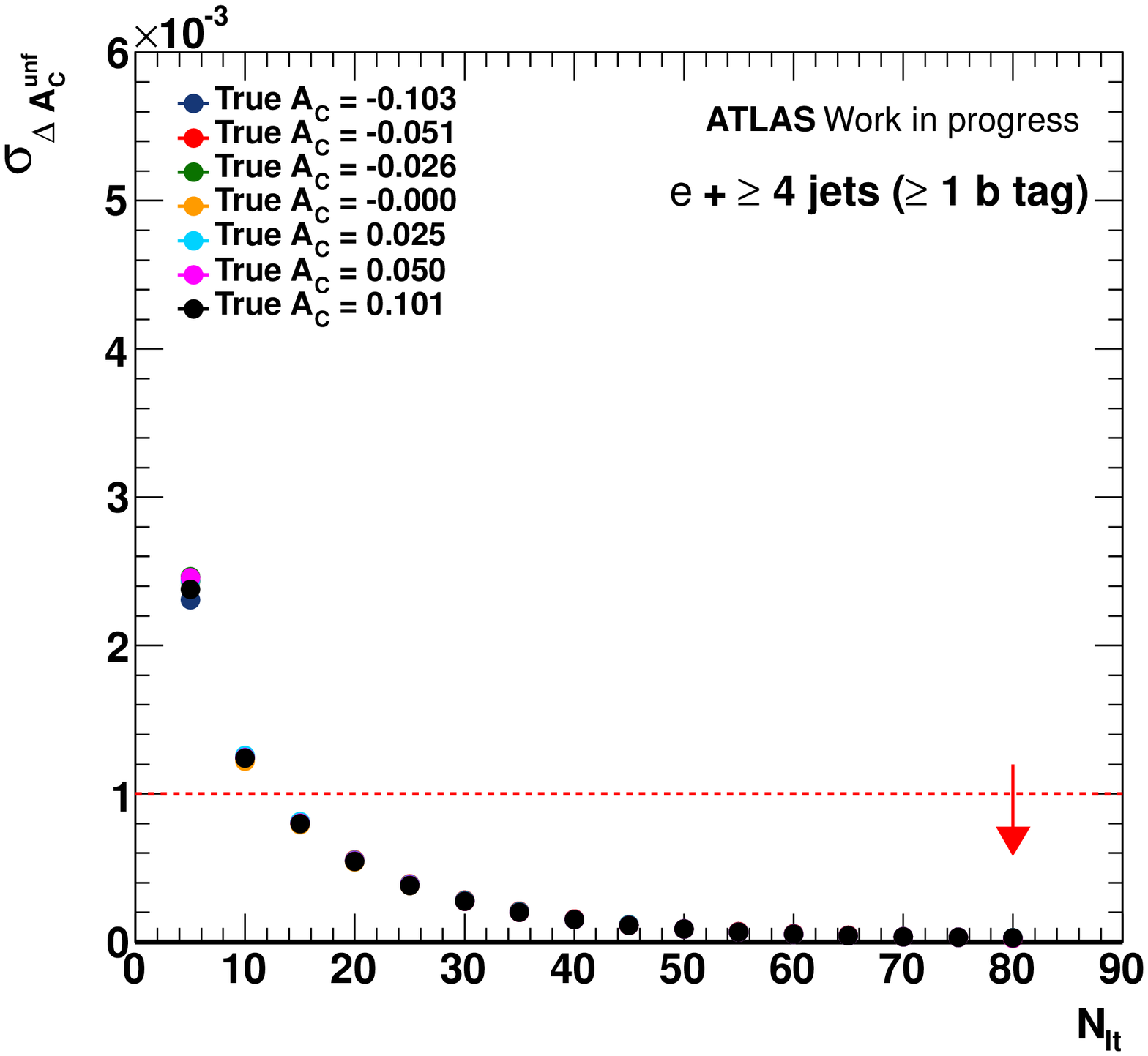}
    
    \vspace{-0.2 cm}
    \caption{Distributions showing the standard deviation of the variable $\Delta A_C^{\text{unf}}$ for ensembles of pseudodata as a function of regularisation for both the muon+jets channel (left) and electron+jets channel (right), parametrised by the number of iterations $N_{\text{It}}$ in the unfolding procedure. The applied convergence criterion of 0.001 is indicated by the red dashed line. The top row shows the respective dependencies for the inclusive measurement, while the lower rows show the corresponding dependencies for $M_{t\bar{t}} < 450$\,GeV and $M_{t\bar{t}} > 450$\,GeV, respectively. For the simultaneous unfolding in $|y_t| - |y_{\bar{t}}|$ and $M_{t \bar{t}}$, a cut on the event reconstruction likelihood $\log{L}$ was applied to improve the $M_{t\bar{t}}$ resolution of the selected events. The arrows indicate the chosen values for $N_{\text{It}}$.}
    \label{fig:conv2}
  \end{center}
\end{figure}

\begin{figure}[!htbp]
  \begin{center}
    \includegraphics[width=\plotwidth]{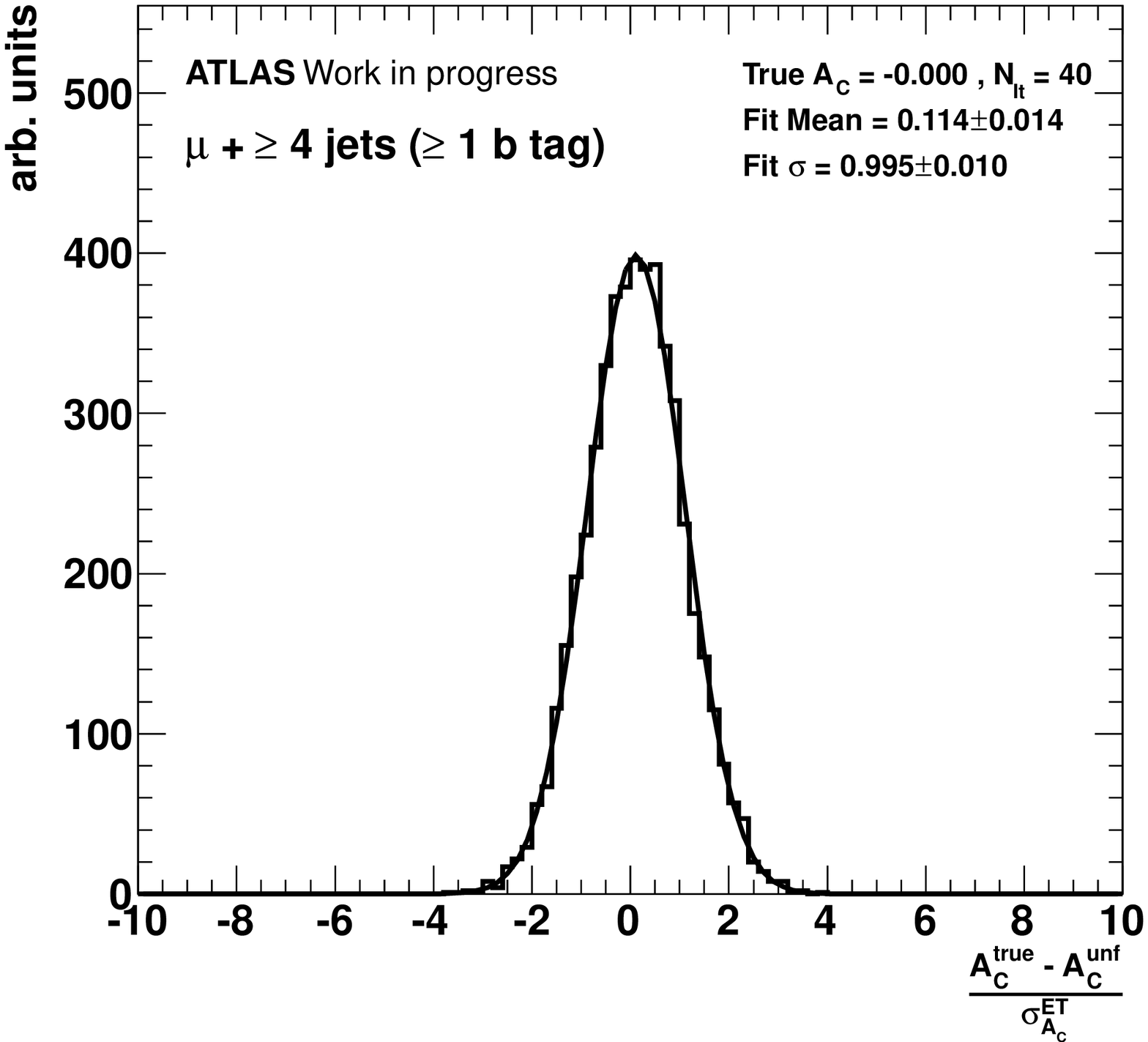}
    \quad\quad
    \includegraphics[width=\plotwidth]{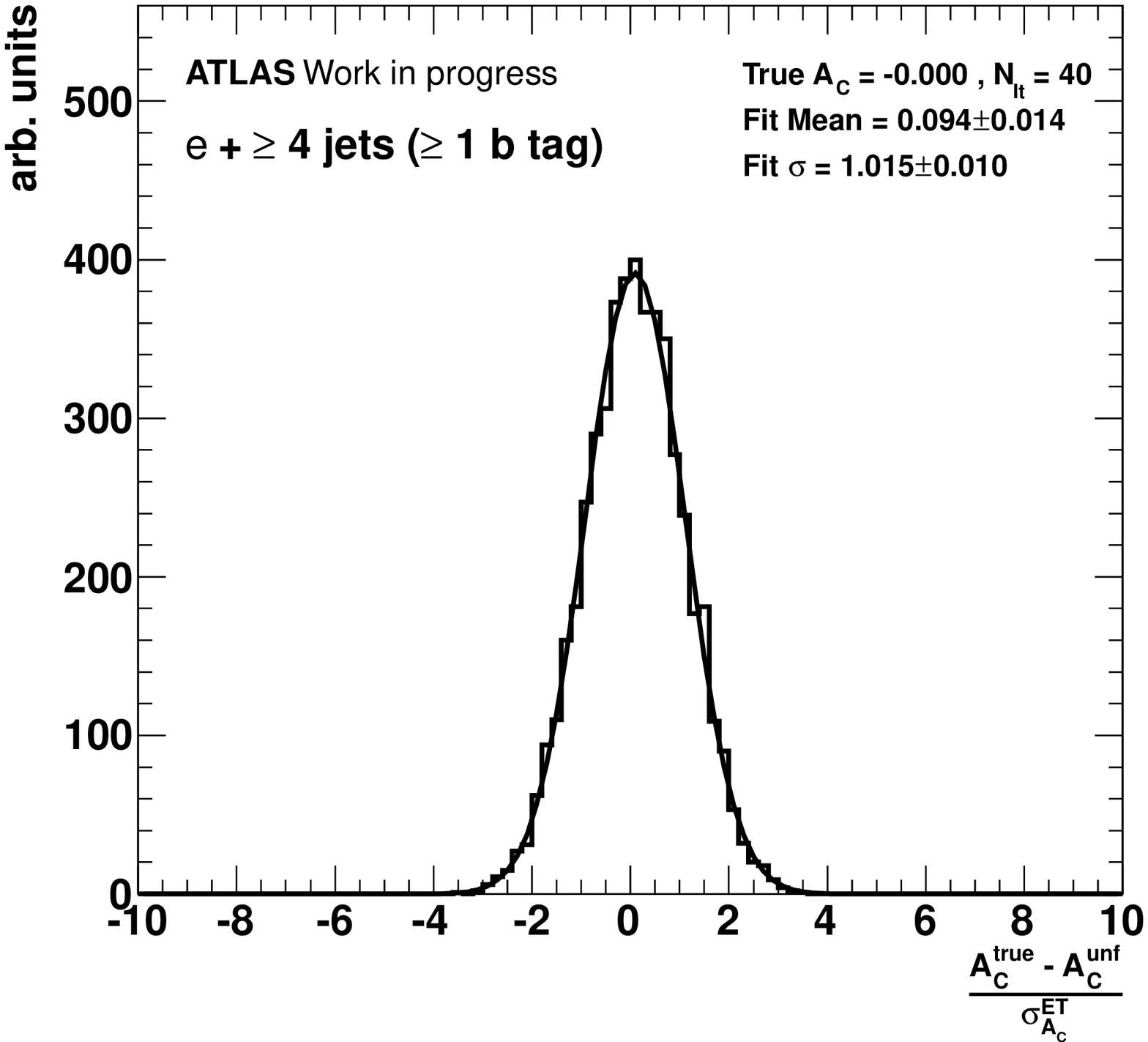}
    \includegraphics[width=\plotwidth]{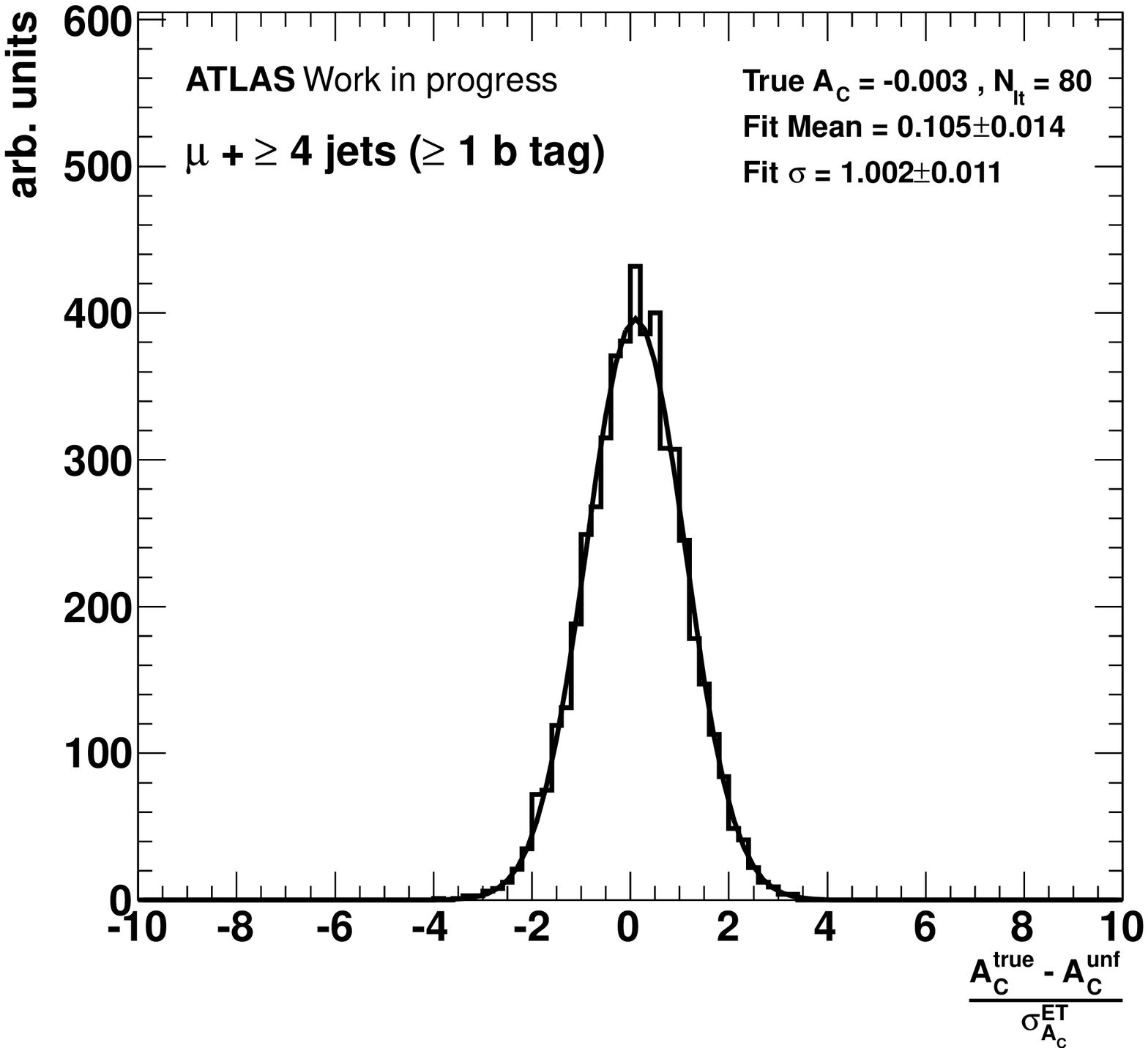}
    \quad\quad
    \includegraphics[width=\plotwidth]{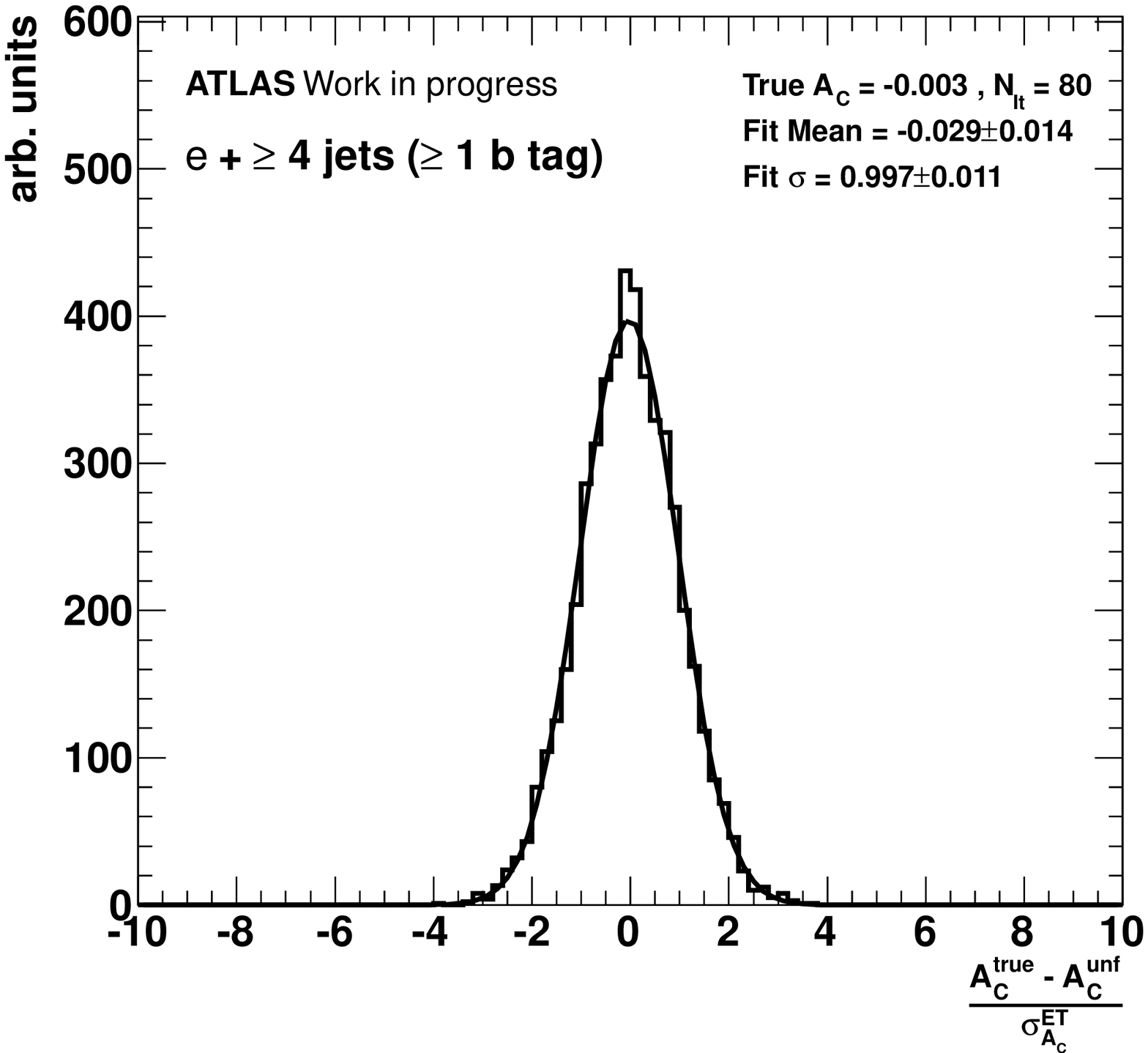}
    \includegraphics[width=\plotwidth]{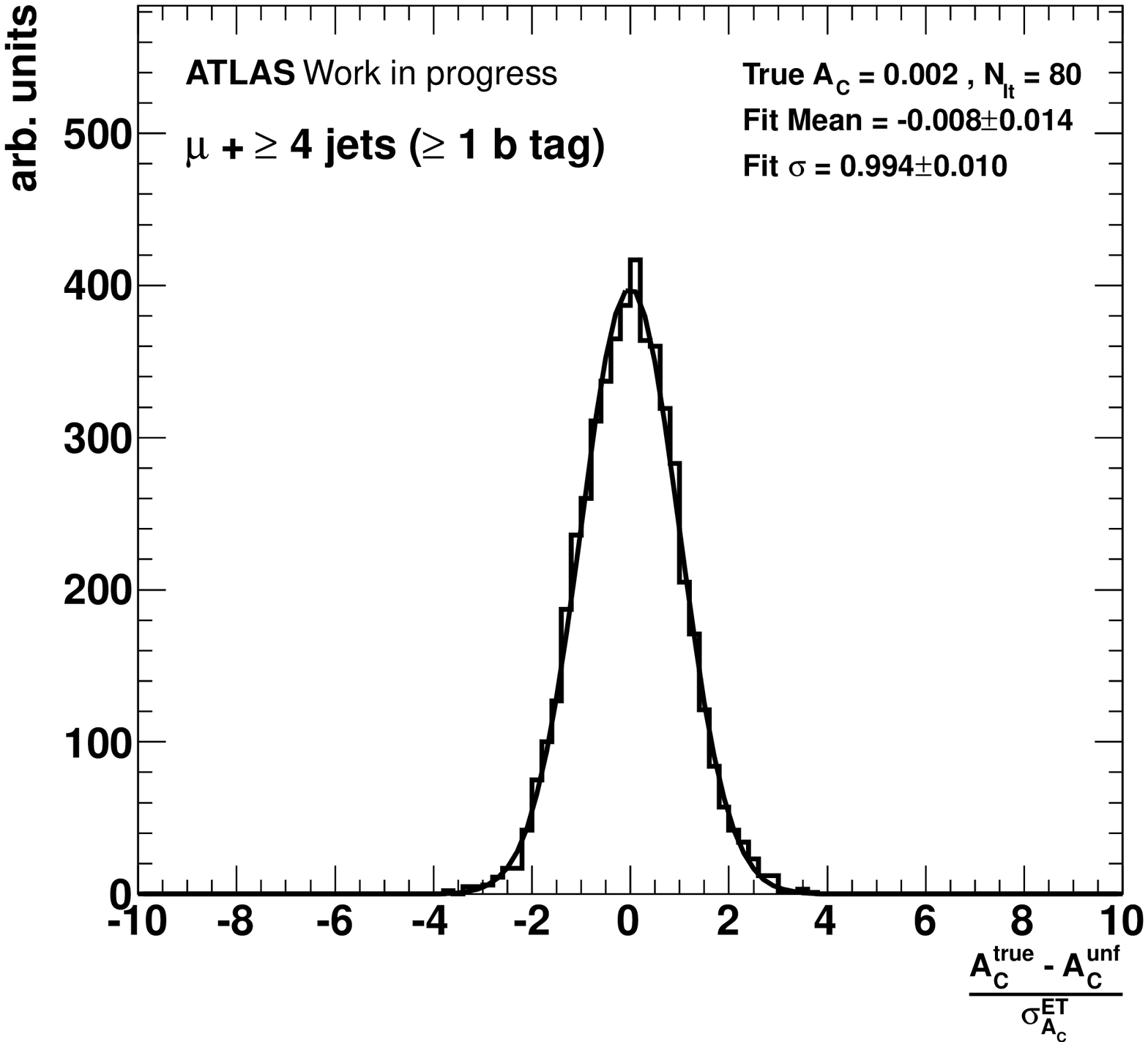}
    \quad\quad
    \includegraphics[width=\plotwidth]{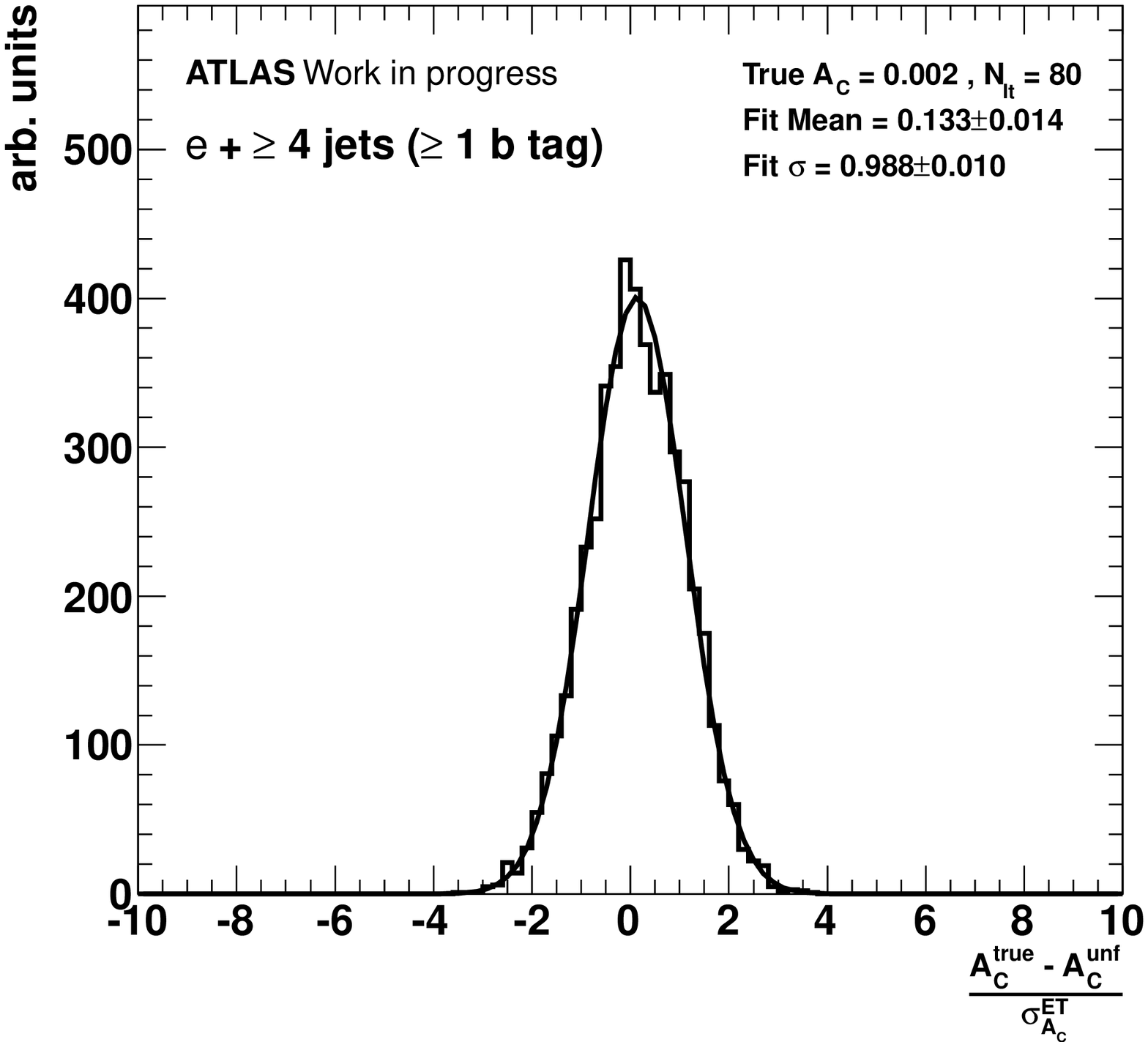}
    
    \vspace{-0.2 cm}
    \caption{Pull distributions showing the difference between true and unfolded asymmetries, normalised to the unfolding respective uncertainty, obtained from pseudoexperiments for both the muon+jets channel (left) and electron+jets channel (right), parametrised by the number of iterations $N_{\text{It}}$ in the unfolding procedure for a true asymmetry of $A_C^{\text{true}} = 0.0$. The top row shows the respective dependencies for the inclusive measurement, while the lower rows show the corresponding dependencies for $M_{t\bar{t}} < 450$\,GeV and $M_{t\bar{t}} > 450$\,GeV, respectively. For the simultaneous unfolding in $|y_t| - |y_{\bar{t}}|$ and $M_{t \bar{t}}$, a cut on the event reconstruction likelihood $\log{L}$ was applied to improve the $M_{t\bar{t}}$ resolution of the selected events.}
    \label{fig:pull_dist}
  \end{center}
\end{figure}

\begin{figure}[!htbp]
  \begin{center}
    \includegraphics[width=\plotwidth]{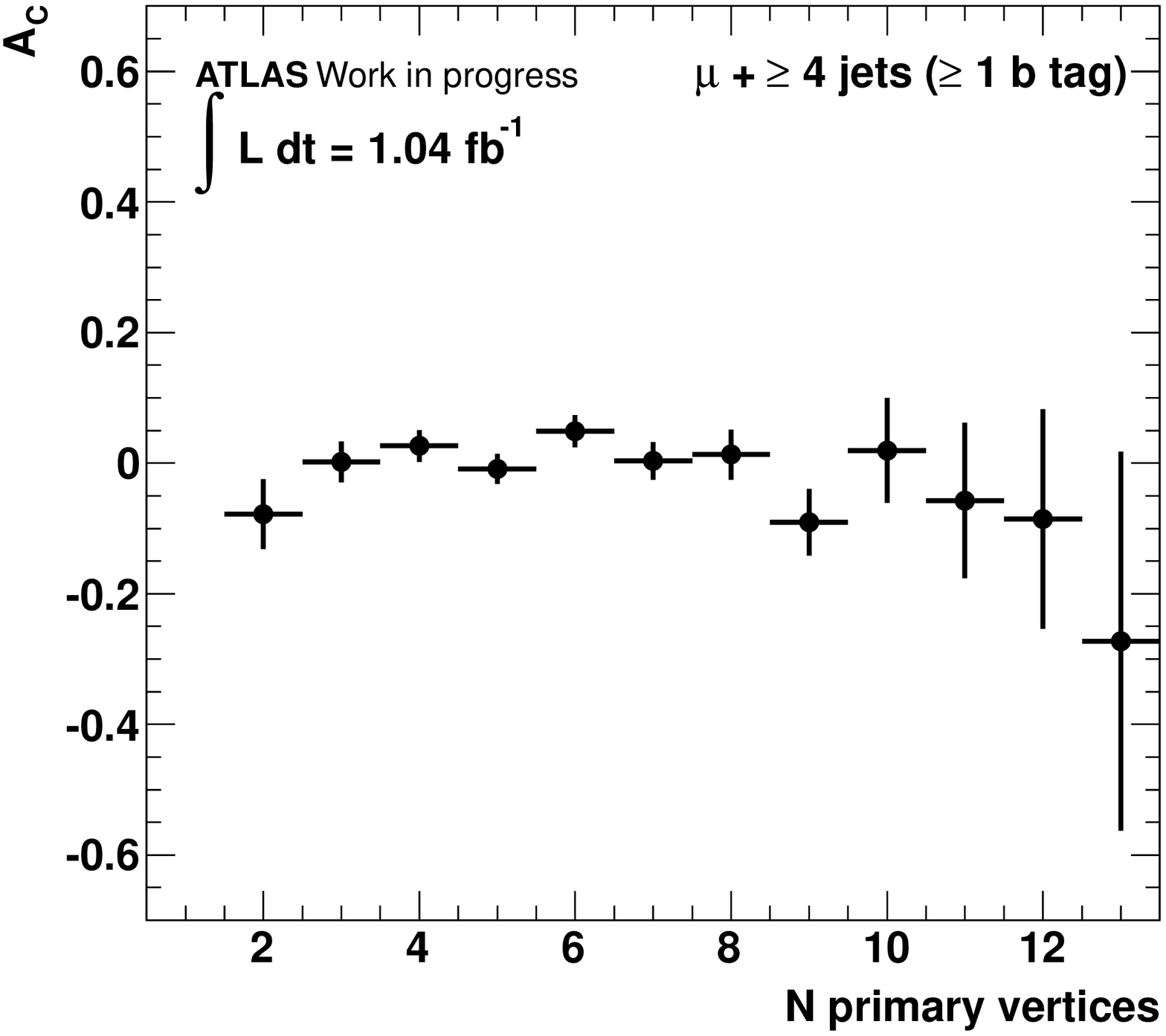}
    \quad\quad
    \includegraphics[width=\plotwidth]{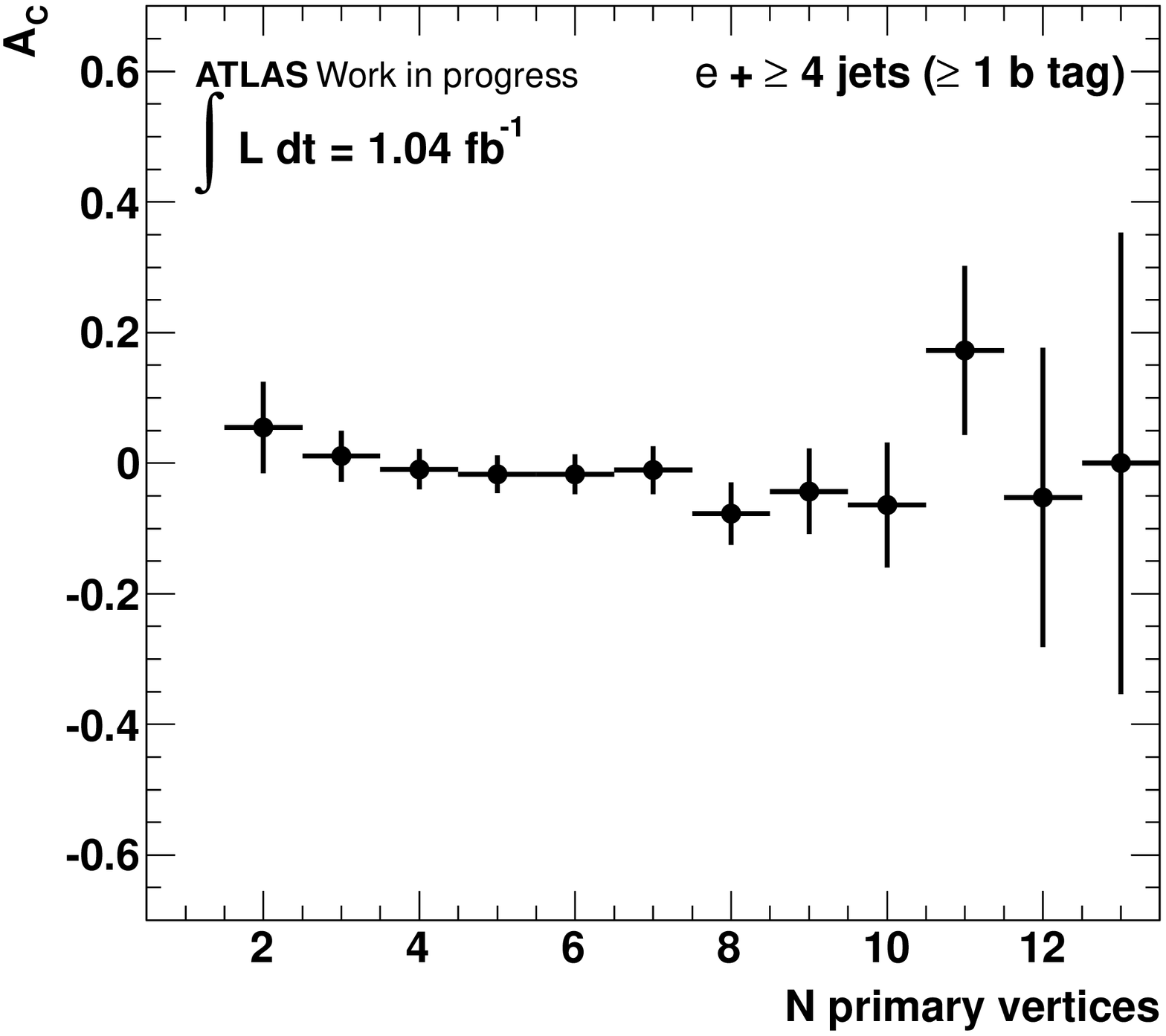}
    \includegraphics[width=\plotwidth]{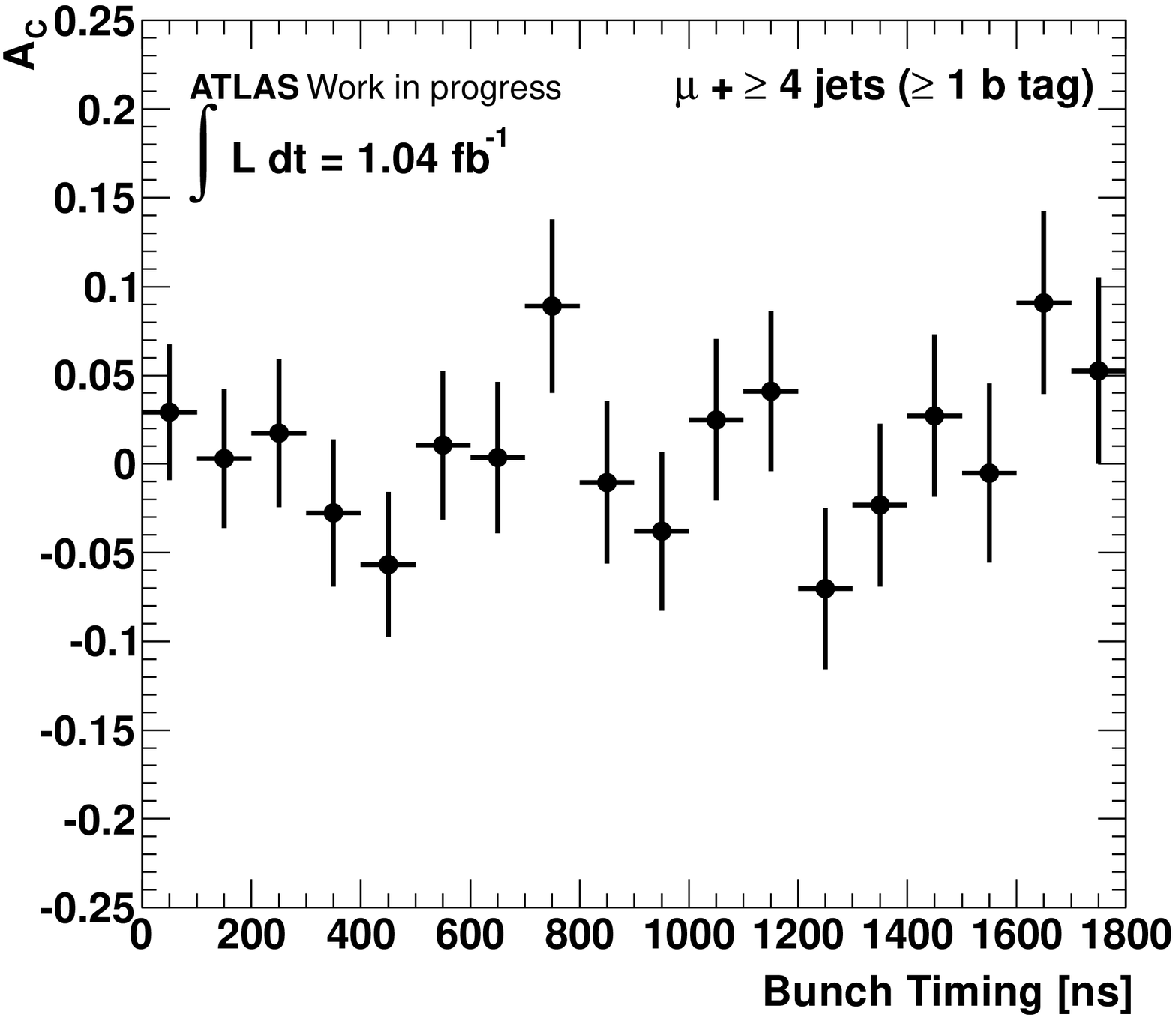}
    \quad\quad
    \includegraphics[width=\plotwidth]{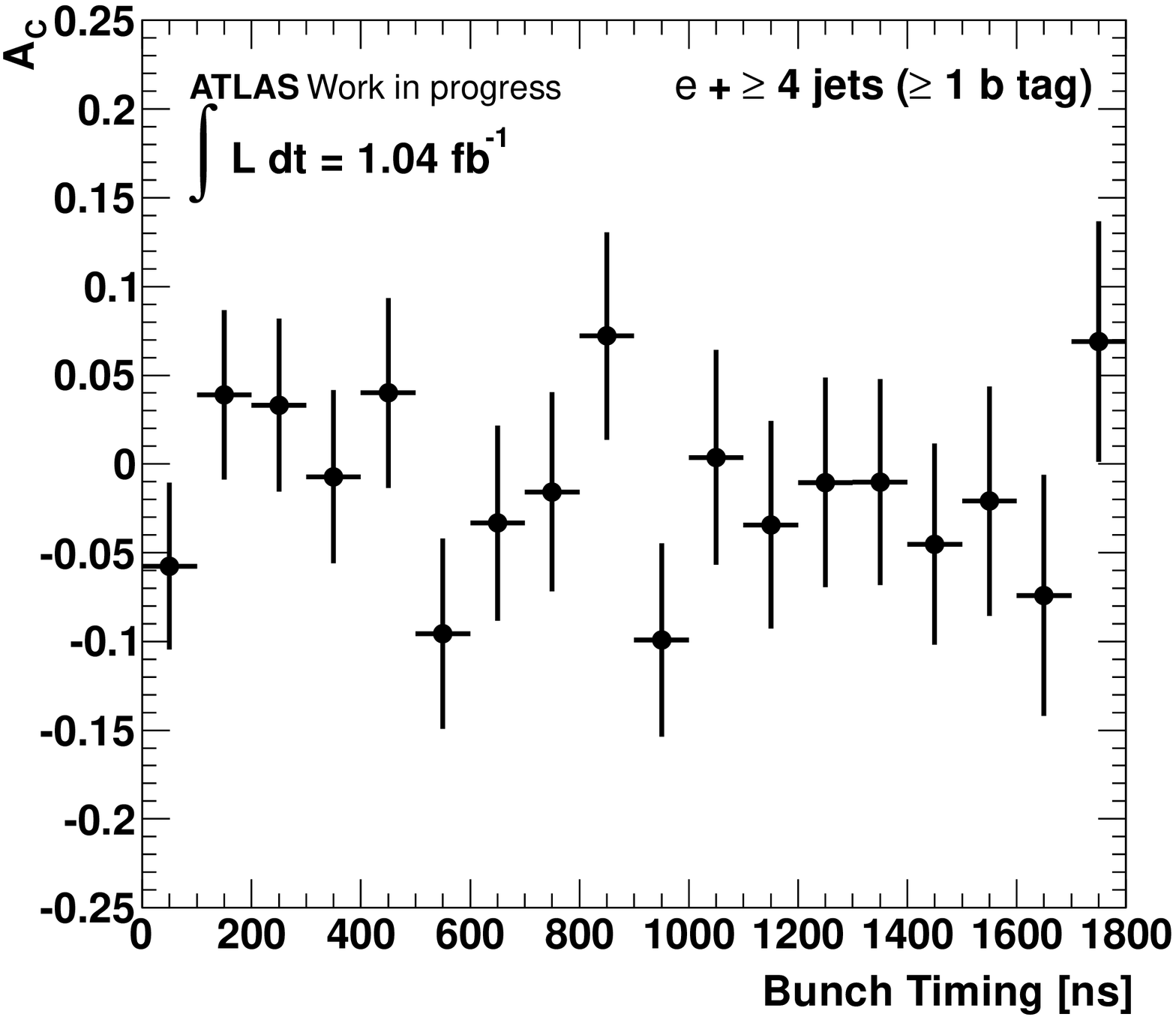}
    
    \vspace{-0.2 cm}
    \caption{Measured charge asymmetry $A_C^{\text{data}}$ before background subtraction and unfolding for both the muon+jets channel (left) and electron+jets channel (right) as a function of the number of primary vertices in the respective events (top row) and the bunch position (bottom row). Uncertainties are statistical only.}
    \label{fig:pile-up}
  \end{center}
\end{figure}
\clearpage

\chapter{Systematic Uncertainties}
\label{App:Systematics}
\begin{table}[htbp]
\begin{center}
{\small
\begin{tabular}{|l|r|r|}
\cline{2-3}\multicolumn{1}{c|}{} & \multicolumn{ 2}{c|}{Absolute systematic uncertainty} \\
\cline{2-3}\multicolumn{1}{c|}{} & Muon Channel & Electron Channel \\
\hline
QCD multijet  & $<$ 0.001 & 0.006 \\ 
\hline
Jet energy scale  & 0.005 & 0.004 \\ 
$b$ tag jet energy scale  & $<$ 0.001 & 0.001 \\ 
Pile-up jet energy scale  & $<$ 0.001 & $<$ 0.001 \\ 
\hline
Jet reco efficiency  & 0.001 & 0.002 \\ 
Jet energy resolution  & 0.004 & 0.001 \\ 
\hline
Muon efficiencies  & 0.001 & (n.a.) \\ 
Muon scales / resolution  & $<$ 0.001 & $<$ 0.001 \\ 
Electron efficiencies  & (n.a.) & 0.001 \\ 
Electron scales / resolution  & $<$ 0.001 & 0.001 \\ 
\hline
$b$ tag scale factors  & 0.002 & 0.003 \\ 
LAr hole uncertainty  & 0.002 & 0.001 \\ 
\hline
$W$+jets normalisation  & 0.003 & 0.004 \\ 
$W$+jets shape  & $<$ 0.001 & 0.002 \\ 
$Z$+jets normalisation  & $<$ 0.001 & $<$ 0.001 \\ 
$Z$+jets shape  & $<$ 0.001 & $<$ 0.001 \\ 
\hline
Single top  & $<$ 0.001 & $<$ 0.001 \\ 
Diboson  & $<$ 0.001 & $<$ 0.001 \\ 
Charge mis-identification  & $<$ 0.001 & $<$ 0.001 \\ 
$b$ tag charge  & 0.001 & 0.001 \\ 
\hline
Luminosity   & 0.001 & 0.001 \\
\hline
\hline
Combined & 0.008 & 0.010 \\
\hline
\end{tabular}
}
\end{center}
\caption{List of all systematic uncertainties taken into account for the measurement of the top charge asymmetry before unfolding.}
\label{Tab:SystematicsMeas}
\end{table}

\begin{table}[htbp]
\begin{center}
{\small
\begin{tabular}{|l|r|r|r|r|}
\cline{2-5}\multicolumn{1}{c|}{} & \multicolumn{ 4}{c|}{Absolute systematic uncertainty} \\ \cline{2-5}\multicolumn{1}{c|}{} & \multicolumn{ 2}{c|}{Muon Channel} & \multicolumn{ 2}{c|}{Electron Channel} \\ 
\cline{2-5}\multicolumn{1}{c|}{} & $M_{t\bar{t}} < 450\,\text{GeV}$ & $M_{t\bar{t}} > 450\,\text{GeV}$ & $M_{t\bar{t}} < 450\,\text{GeV}$ & $M_{t\bar{t}} > 450\,\text{GeV}$ \\
\hline
QCD multijet  & $<$ 0.001 & $<$ 0.001 & 0.013 & 0.013 \\ 
\hline
Jet energy scale  & 0.003 & 0.004 & 0.008 & 0.006 \\ 
$b$ tag jet energy scale  & $<$ 0.001 & $<$ 0.001 & 0.001 & 0.001 \\ 
Pile-up jet energy scale  & $<$ 0.001 & $<$ 0.001 & $<$ 0.001 & $<$ 0.001 \\ 
\hline
Jet reco efficiency  & $<$ 0.001 & 0.001 & 0.003 & 0.001 \\ 
Jet energy resolution  & 0.006 & 0.003 & 0.009 & 0.003 \\ 
\hline
Muon efficiencies  & $<$ 0.001 & 0.001 & (n.a.) & (n.a.) \\ 
Muon scales / resolution  & $<$ 0.001 & 0.001 & $<$ 0.001 & $<$ 0.001 \\ 
Electron efficiencies  & (n.a.) & (n.a.) & 0.001 & $<$ 0.001 \\ 
Electron scales / resolution  & $<$ 0.001 & $<$ 0.001 & 0.001 & 0.001 \\ 
\hline
$b$ tag scale factors  & $<$ 0.001 & 0.001 & 0.003 & 0.002 \\ 
LAr hole uncertainty  & 0.001 & 0.001 & 0.002 & 0.001 \\ 
\hline
$W$+jets normalisation  & 0.001 & 0.004 & 0.004 & 0.003 \\ 
$W$+jets shape  & $<$ 0.001 & $<$ 0.001 & $<$ 0.001 & 0.001 \\ 
$Z$+jets normalisation  & $<$ 0.001 & $<$ 0.001 & 0.001 & $<$ 0.001 \\ 
$Z$+jets shape  & $<$ 0.001 & $<$ 0.001 & $<$ 0.001 & $<$ 0.001 \\ 
\hline
Single top  & $<$ 0.001 & $<$ 0.001 & 0.001 & $<$ 0.001 \\ 
Diboson  & $<$ 0.001 & $<$ 0.001 & $<$ 0.001 & $<$ 0.001 \\ 
Charge mis-identification  & $<$ 0.001 & $<$ 0.001 & $<$ 0.001 & $<$ 0.001 \\ 
$b$ tag charge  & 0.001 & 0.001 & 0.001 & 0.001 \\ 
\hline
Luminosity  & $<$ 0.001 & 0.001 & 0.001 & $<$ 0.001 \\
\hline
\hline
Combined & 0.007 & 0.007 & 0.019 & 0.015 \\
\hline
\end{tabular}
}
\end{center}
\caption{List of all systematic uncertainties taken into account for the measurement of the top charge asymmetry before unfolding.}
\label{Tab:SystematicsMeas2D}
\end{table}
\clearpage

\chapter{SVD Unfolding}
\label{App:SVDUnfolding}
As an additional cross-check, the inclusive unfolding has been performed with the SVD unfolding procedure (for details, refer to \mbox{Chapter \ref{Unfolding}}), using otherwise identical analysis parameters. The same binning and the same response matrix as determined for the Bayesian iterative unfolding (c.f. \mbox{Chapter \ref{chap:results:unfolding}}) has been used.

A closure test has been performed to verify that the SVD unfolding approach can be used to recover an arbitrary asymmetry present in the true distribution. Similar to the Bayesian iterative unfolding, pseudoexperiments using statistically independent sets of pseudodata corresponding to the statistics expected in the used data sample were performed. Ensemble tests were conducted to confirm the linearity of the unfolding response in the true value of the chosen charge asymmetry observable, $A_C^{\text{true}}$, and the dependency of the obtained results on the regularisation of the unfolding procedure was studied.

The average unfolded value of $A_C^{\text{unf}}$ obtained from the respective sets of pseudodata as a function of the injected true value of $A_C^{\text{true}}$ can be found in \mbox{Figure \ref{fig:unfolding:calib:SVD}} for different regularisation strengths corresponding to 
choices of the regularisation parameter of $\tau = 2$, 3, 4, 5 and 6. Since $\tau$ corresponds directly to a fraction of the number of bins, it can not exceed the number of bins chosen for the unfolding by construction\cite{Hocker:1995kb}.
\begin{figure}[!htbp]
  \begin{center}
    \includegraphics[width=\plotwidth]{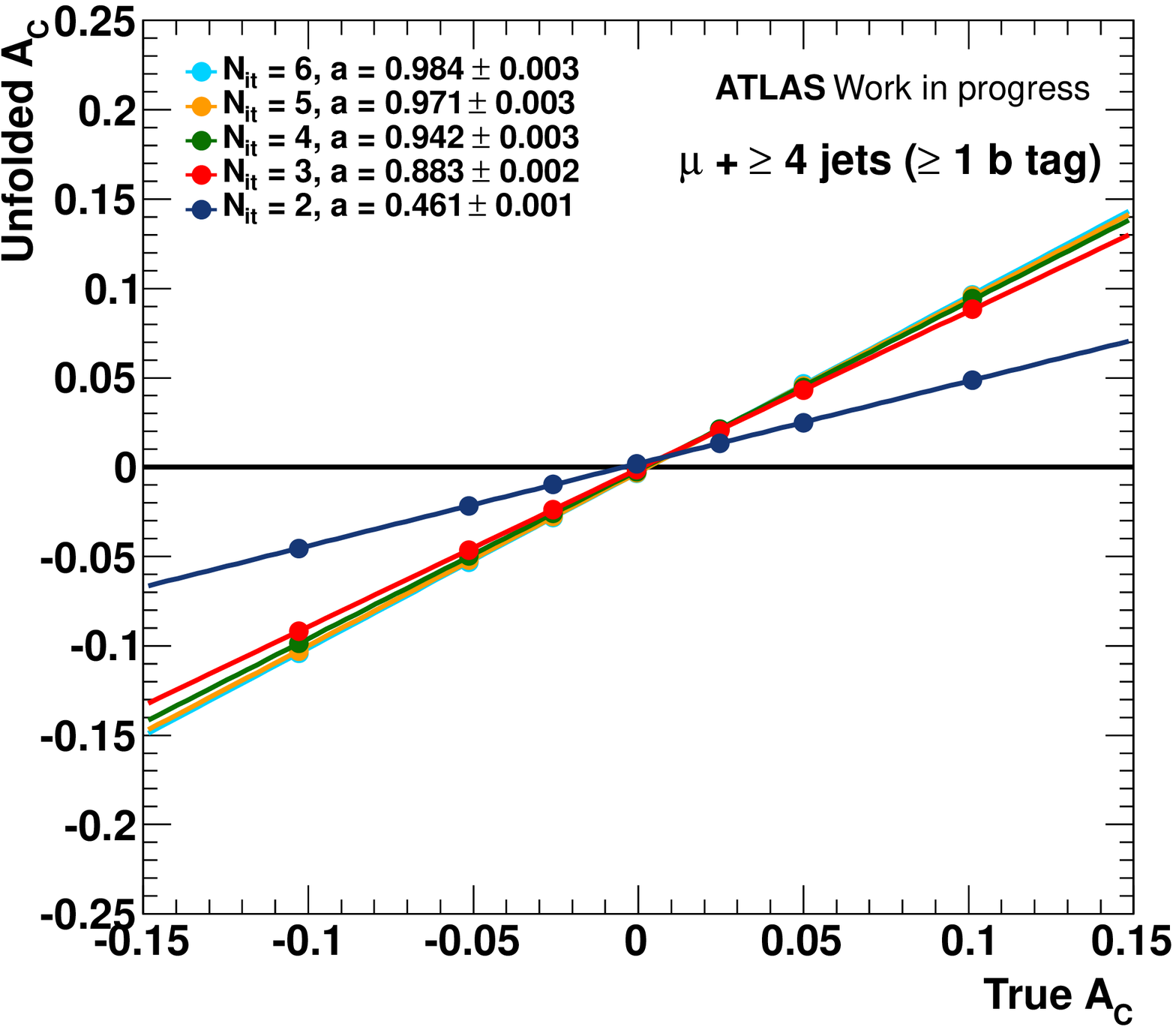}
    \quad\quad
    \includegraphics[width=\plotwidth]{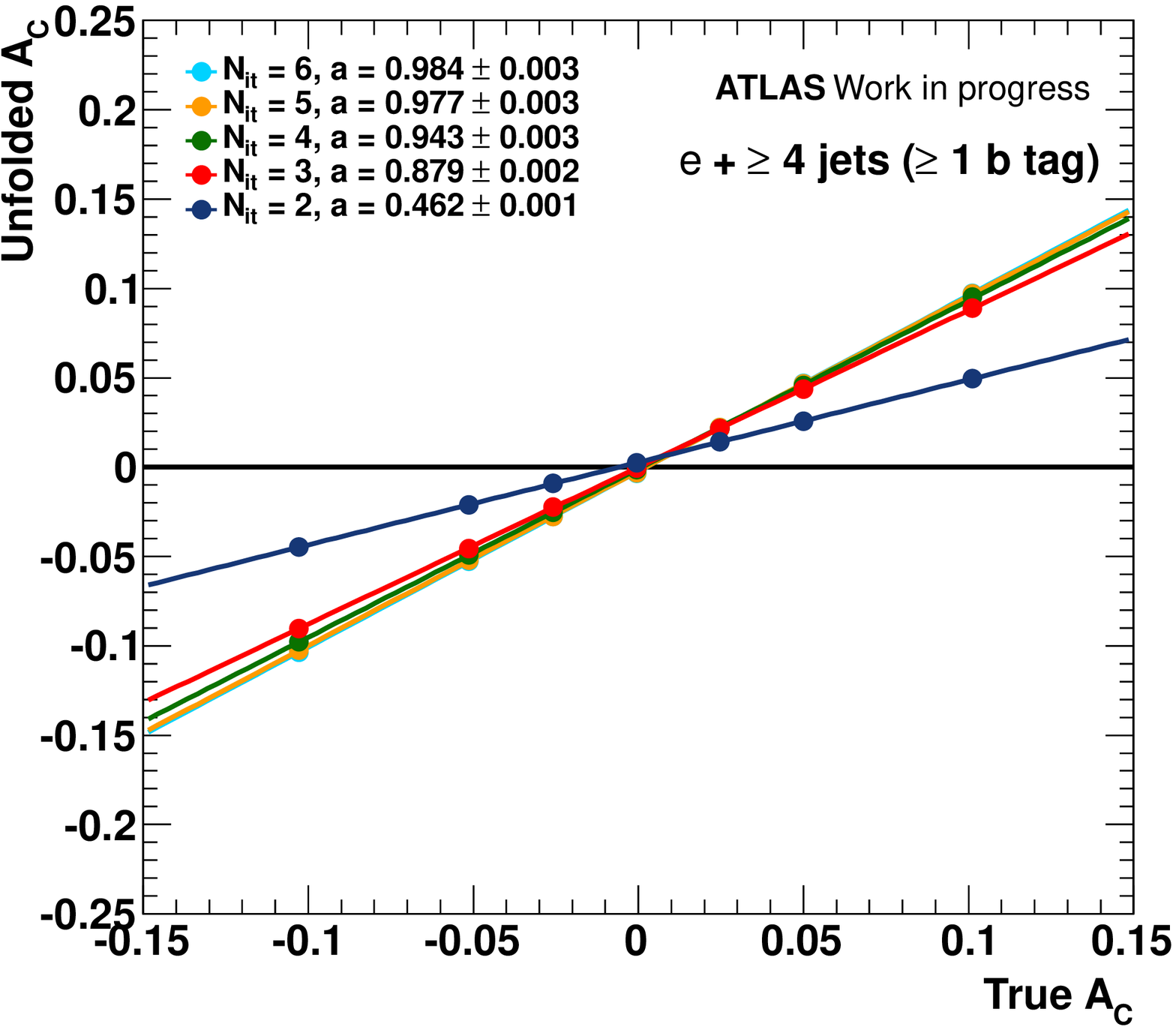}
    
    \vspace{-0.2 cm}
    \caption{The obtained inclusive asymmetry $A_C^{\text{unf}}$ after SVD unfolding as a function of injected true asymmetry $A_C^{\text{true}}$ for different regularisation parameters $\tau$ for both the muon+jets channel (left) and electron+jets channel (right).}
    \label{fig:unfolding:calib:SVD}
  \end{center}
\end{figure}
A straight line fit has been performed as described in \mbox{Chapter \ref{chap:results:unfolding}}. As can be seen in the respective fits, a slope close to one can be achieved in all cases for $\tau = 6$, indicating a proper correspondence of the unfolded asymmetry to the respective true value, independent of the strength of the injected asymmetry.

Since the SVD unfolding does not rely on an iterative procedure, no convergence criterion was imposed to choose a proper regularisation. Instead, the parameter $\tau$ was chosen such that the remaining unfolding bias was minimised, i.e. such that the slope parameter $a$ of the linear fit in the calibration was as close to unity as possible. As can be seen in \mbox{Figure \ref{fig:unfolding:calib:SVD}}, the optimal choice was given by $\tau = 6$ for both the muon+jets and electron+jets channel. 

Additional closure tests have been performed as described in \mbox{Chapter \ref{chap:results:unfolding}}, using ensembles of pseudodata to quantify any remaining bias from the unfolding at the chosen regularisation strengths. A corresponding systematic uncertainty was assigned to the unfolded result accordingly from the residuals of the respective pull distributions (c.f. \mbox{Chapter \ref{Systematics}}).

The unfolded integrated asymmetries $A_C^{\text{unf}}$ for the inclusive unfolding, using $\tau = 6$ in the SVD unfolding process, were determined to be
\begin{equation*}
  A_C^{\text{unf}} = -0.005 \pm 0.034\,\text{(stat.)} \pm 0.024\,\text{(syst.)}
\end{equation*}
in the muon channel and
\begin{equation*}
  A_C^{\text{unf}} = -0.056 \pm 0.043\,\text{(stat.)} \pm 0.029\,\text{(syst.)}
\end{equation*}
in the electron channel, respectively. In addition, the obtained distributions after unfolding can be found in \mbox{Figure \ref{fig:unfolding:data:SVD}}.
\begin{figure}[!htbp]
  \begin{center}
    \includegraphics[width=\plotwidth]{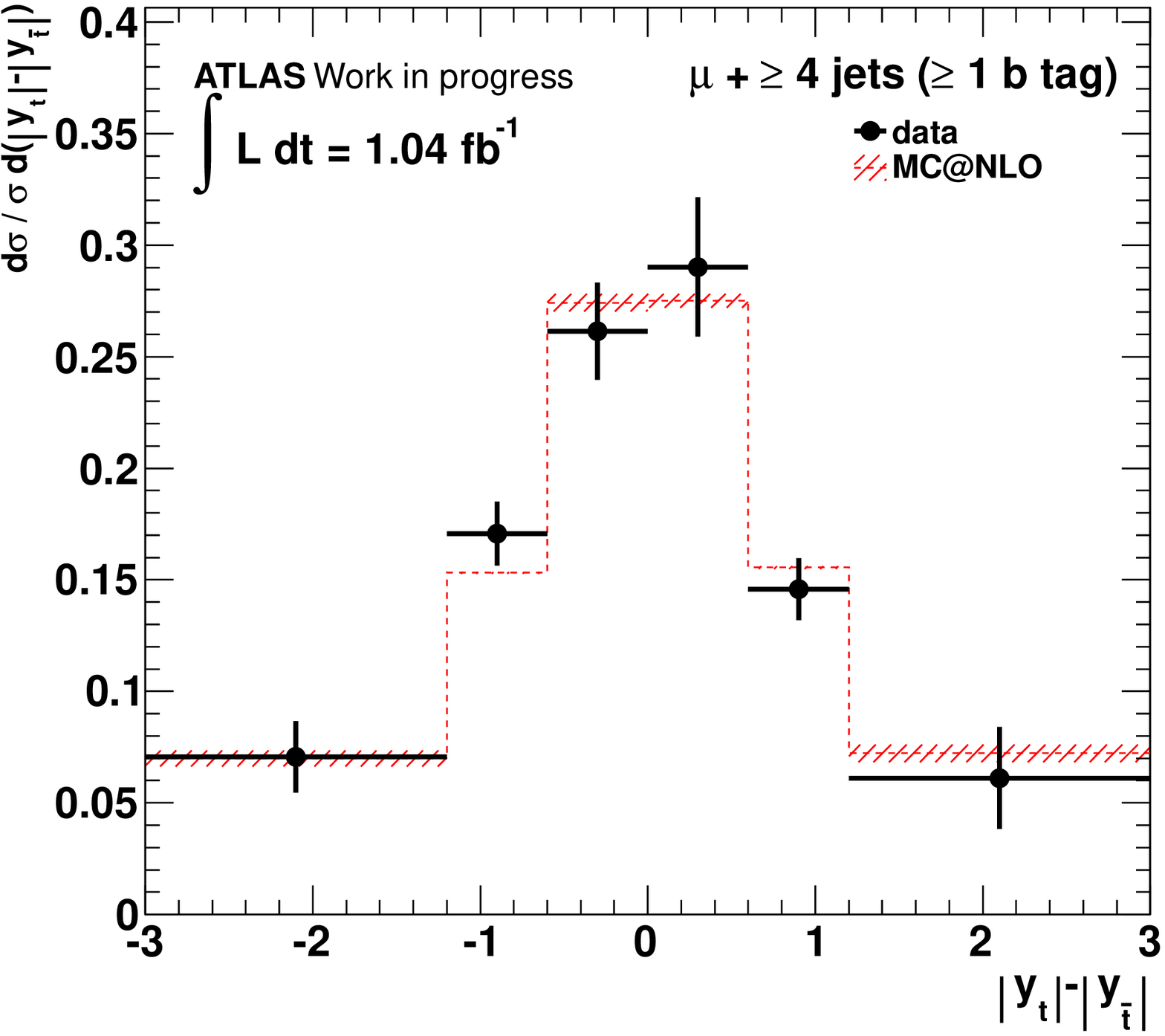}
    \quad\quad
    \includegraphics[width=\plotwidth]{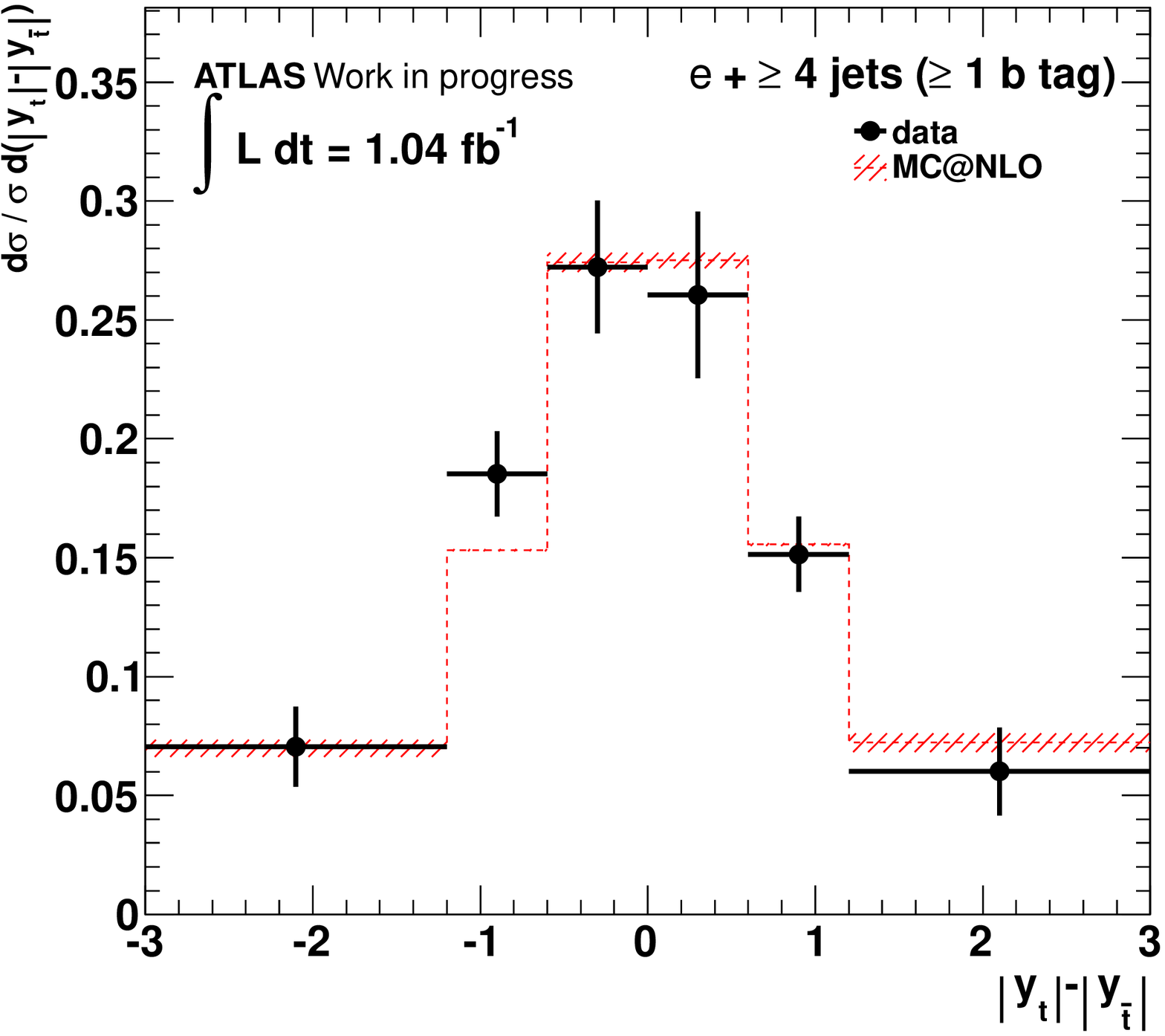}
    
    \vspace{-0.2 cm}
    \caption{Unfolded distribution of $|y_t|-|y_{\bar{t}}|$ using an SVD unfolding procedure, normalised to unity for both the muon+jets channel (left) and electron+jets channel (right). The uncertainties include both statistical and systematic shape components.}
    \label{fig:unfolding:data:SVD}
  \end{center}
\end{figure}

A summarised list of all systematics and their contribution to the overall systematic uncertainty can be found in \mbox{Table \ref{Tab:SystematicsUnf:SVD}}.
\begin{table}[htbp]
\begin{center}
{\small
\begin{tabular}{|l|r|r|}
\cline{2-3}\multicolumn{1}{c|}{} & \multicolumn{ 2}{c|}{Absolute systematic uncertainty} \\
\cline{2-3}\multicolumn{1}{c|}{} & Muon Channel & Electron Channel \\
\hline
QCD multijet  & 0.002 & 0.012 \\ 
\hline
Jet energy scale  & 0.007 & 0.013 \\ 
$b$ tag jet energy scale  & $<$ 0.001 & $<$ 0.001 \\ 
Pile-up jet energy scale  & 0.001 & 0.002 \\ 
\hline
Jet reco efficiency  & 0.003 & 0.001 \\ 
Jet energy resolution  & 0.008 & 0.004 \\ 
\hline
Muon efficiencies  & 0.001 & (n.a.) \\ 
Muon scales / resolution  & 0.001 & $<$ 0.001 \\ 
Electron efficiencies  & (n.a.) & 0.001 \\ 
Electron scales / resolution  & $<$ 0.001 & 0.002 \\ 
\hline
$b$ tag scale factors  & 0.003 & 0.005 \\ 
PDF uncertainty  & $<$ 0.001 & $<$ 0.001 \\ 
LAr hole uncertainty  & 0.004 & 0.001 \\ 
\hline
ISR and FSR  & 0.011 & 0.008 \\ 
\ttbar~modelling  & (0.008) 0.011 & 0.011 \\ 
Parton shower / fragmentation  & (0.001) 0.010 & 0.010 \\ 
Top mass  & 0.006 & (0.002) 0.006 \\ 
\hline
$W$+jets normalisation  & 0.005 & 0.008 \\ 
$W$+jets shape  & $<$ 0.001 & 0.004 \\ 
$Z$+jets normalisation  & $<$ 0.001 & $<$ 0.001 \\ 
$Z$+jets shape  & 0.001 & 0.005 \\ 
\hline
Single top  & $<$ 0.001 & $<$ 0.001 \\ 
Diboson  & $<$ 0.001 & $<$ 0.001 \\ 
Charge mis-identification  & $<$ 0.001 & $<$ 0.001 \\ 
$b$ tag charge  & 0.001 & 0.001 \\ 
\hline
MC statistics  & 0.004 & 0.006 \\
Unfolding bias  & 0.001 & 0.004 \\
\hline
Luminosity   & 0.001 & 0.001 \\
\hline
\hline
Combined & 0.024 & 0.029 \\
\hline
\end{tabular}
}
\end{center}
\caption{List of all systematic uncertainties taken into account for the unfolding procedure in the measurement of the top charge asymmetry. The numbers in brackets denote the uncertainties before using the largest uncertainty of both channels as conservative estimate.}
\label{Tab:SystematicsUnf:SVD}
\end{table}
All systematic uncertainties based on a replacement of the unfolding matrix which were of the same order of magnitude or lower than the respective MC statistics uncertainty could not be resolved to full extent due to the inherent fluctuations from limited statistics in the response matrix. In those cases, the largest of the systematic uncertainties in both channels was used for the final systematic uncertainty on the unfolded charge asymmetry. The resulting combined systematic uncertainties were 0.024 in the muon+jets channel and 0.029 in the electron+jets channel for the inclusive unfolding.

The \textsc{Blue} method was used to combine the measurements after performing the SVD unfolding in the muon+jets and electron+jets channel. The same assumptions about correlations as described in \mbox{Chapter \ref{chap:results:comb}} were used. A combined value of
\begin{equation*}
A_C^{\text{unf}} = -0.024 \pm 0.027\,\text{(stat.)} \pm 0.024\,\text{(syst.)}
\end{equation*}
was obtained, where the relative weight of the muon+jets channel result was 64.3\,\%.

The obtained results were compatible with the results from the Bayesian unfolding shown in \mbox{Chapter \ref{chap:results:unfolding}}.
\clearpage

\chapter{Additional Studies of Unfolding and Systematics}
\label{App:AltApproach}
As described in \mbox{Chapter \ref{chap:results:unfolding}}, the unfolding procedure employed used a strong regularisation in order to achieve a convergent state and to reduce the remaining bias after unfolding. However, this implies large statistical uncertainties on the obtained results. Furthermore, as described in \mbox{Chapter \ref{Systematics}}, the QCD multijet background contribution systematic uncertainty was conservatively assumed to be 100\,\% and no explicit shape uncertainty was included despite the availability of more advanced estimates.

As an additional cross-check, the unfolding procedure has been repeated, discarding the explicit requirement of convergence of the Bayesian iterative procedure. Instead, smaller values of $N_{\text{It}}$ were used, implying lower statistical uncertainties, and the expected remaining bias of the unfolded value $A_C^{\text{unf}}$ was extracted from pseudoexperiments and corrected for. This, however, implies larger assumptions about the underlying physics since the remaining bias is determined and corrected for based on the observable $A_C$ only. Consequently, no fully model-independent calibration can be performed and a bias not covered by the uncertainties can remain. Furthermore, the described approach does not allow for the extraction of the full unfolded distribution and the associated bin-by-bin uncertainties since the calibration is performed with respect to the integrated asymmetry observable $A_C$ only. 

A closure test has been performed to calibrate the unfolded asymmetry $A_C^{\text{unf}}$ with respect to the true asymmetry $A_C^{\text{true}}$. Similar to the procedure described in \mbox{Chapter \ref{chap:results:unfolding}}, pseudoexperiments using statistically independent sets of pseudodata corresponding to the statistics expected in the used data sample (after background subtraction) were created based on a Poissonian fluctuation of the respective reconstructed distributions of $|y_t| - |y_{\bar{t}}|$. Ensemble tests were performed to confirm the linearity of the unfolding response in the true value of the chosen charge asymmetry observable, $A_C^{\text{true}}$, and the dependence of the obtained results on the regularisation of the unfolding procedure was studied.

The average value of $A_C^{\text{unf}}$ obtained from the respective sets of pseudodata as a function of the injected true asymmetry, $A_C^{\text{true}}$, can be found for both the inclusive unfolding and the simultaneous unfolding in $|y_t| - |y_{\bar{t}}|$ and $M_{t \bar{t}}$ in \mbox{Figure \ref{fig:unfolding:calib:alt}} for different regularisation strengths, using $N_{\text{It}} = 2$, 4, 6, 7, 8 and 10 iterations in the Bayesian unfolding.
\begin{figure}[!htbp]
  \begin{center}
    \includegraphics[width=\plotwidth]{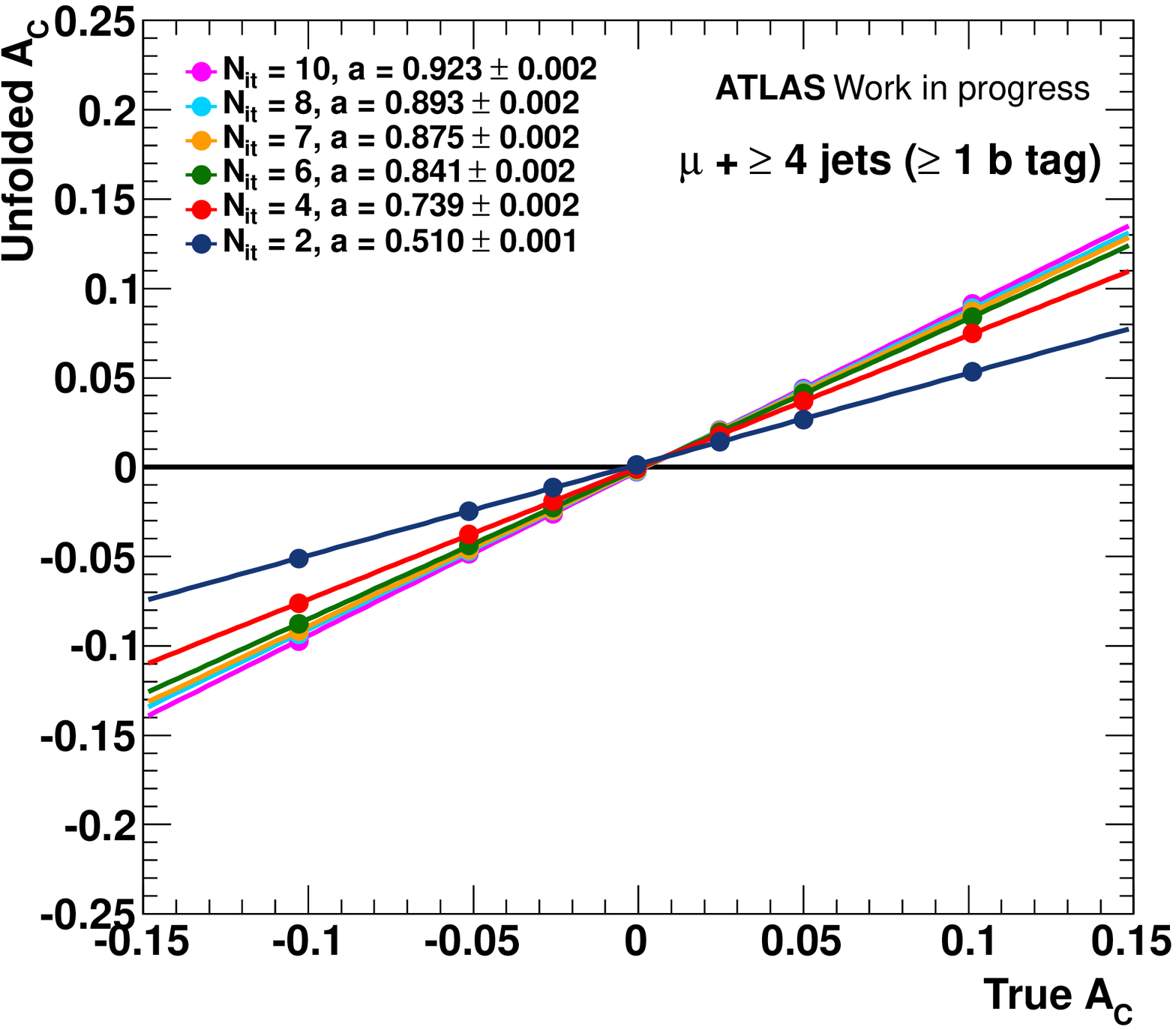}
    \quad\quad
    \includegraphics[width=\plotwidth]{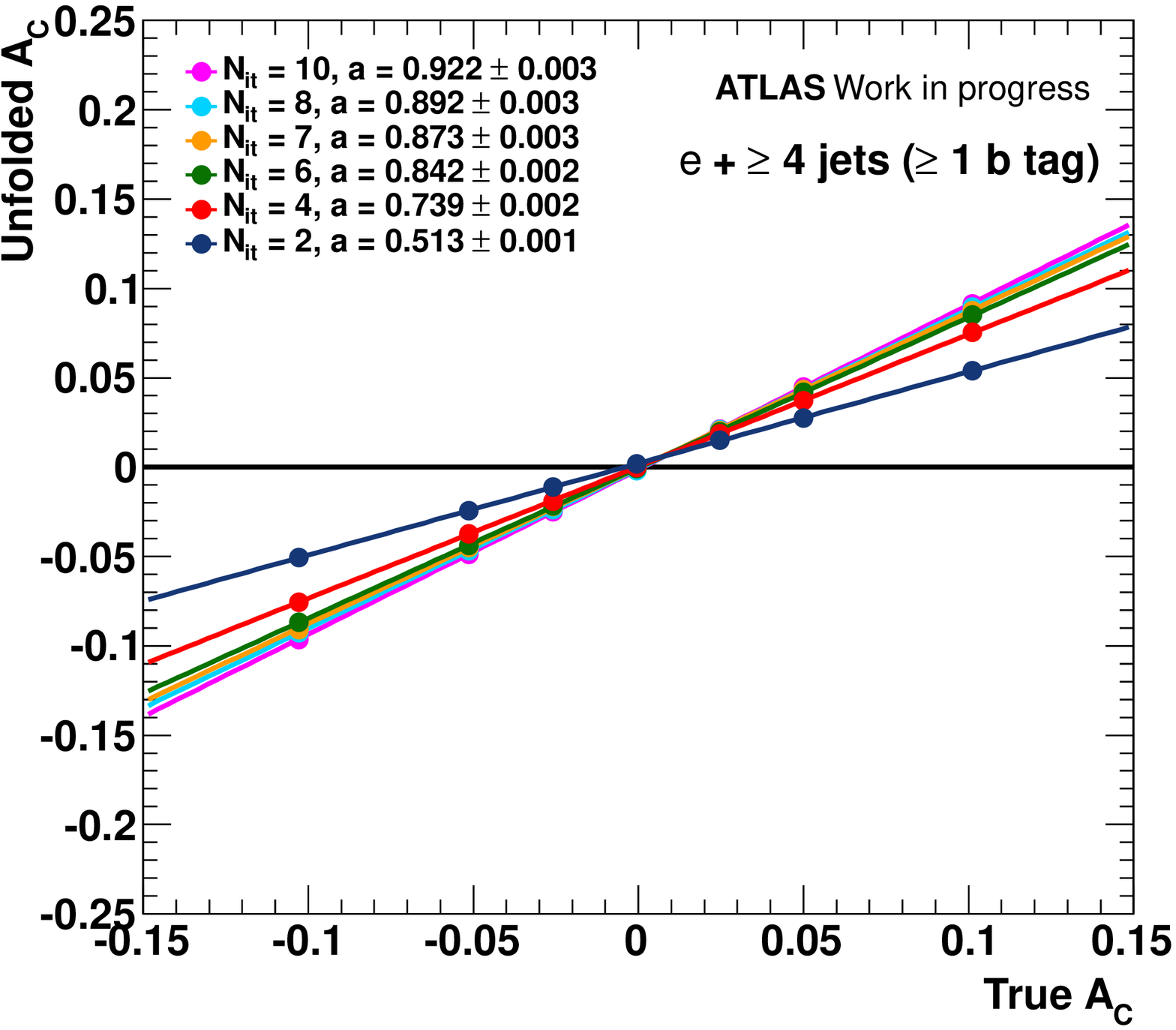}
    \includegraphics[width=\plotwidth]{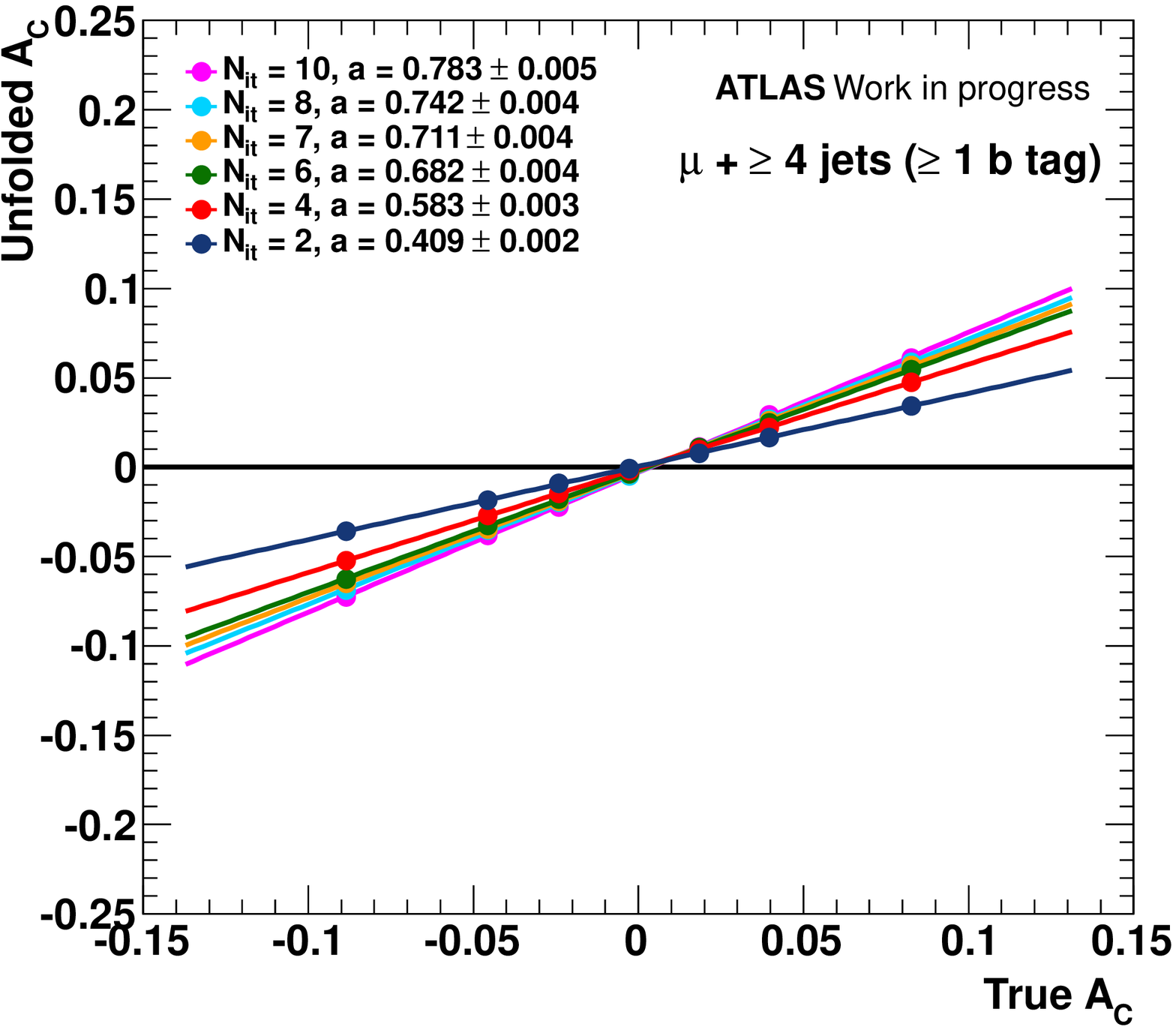}
    \quad\quad
    \includegraphics[width=\plotwidth]{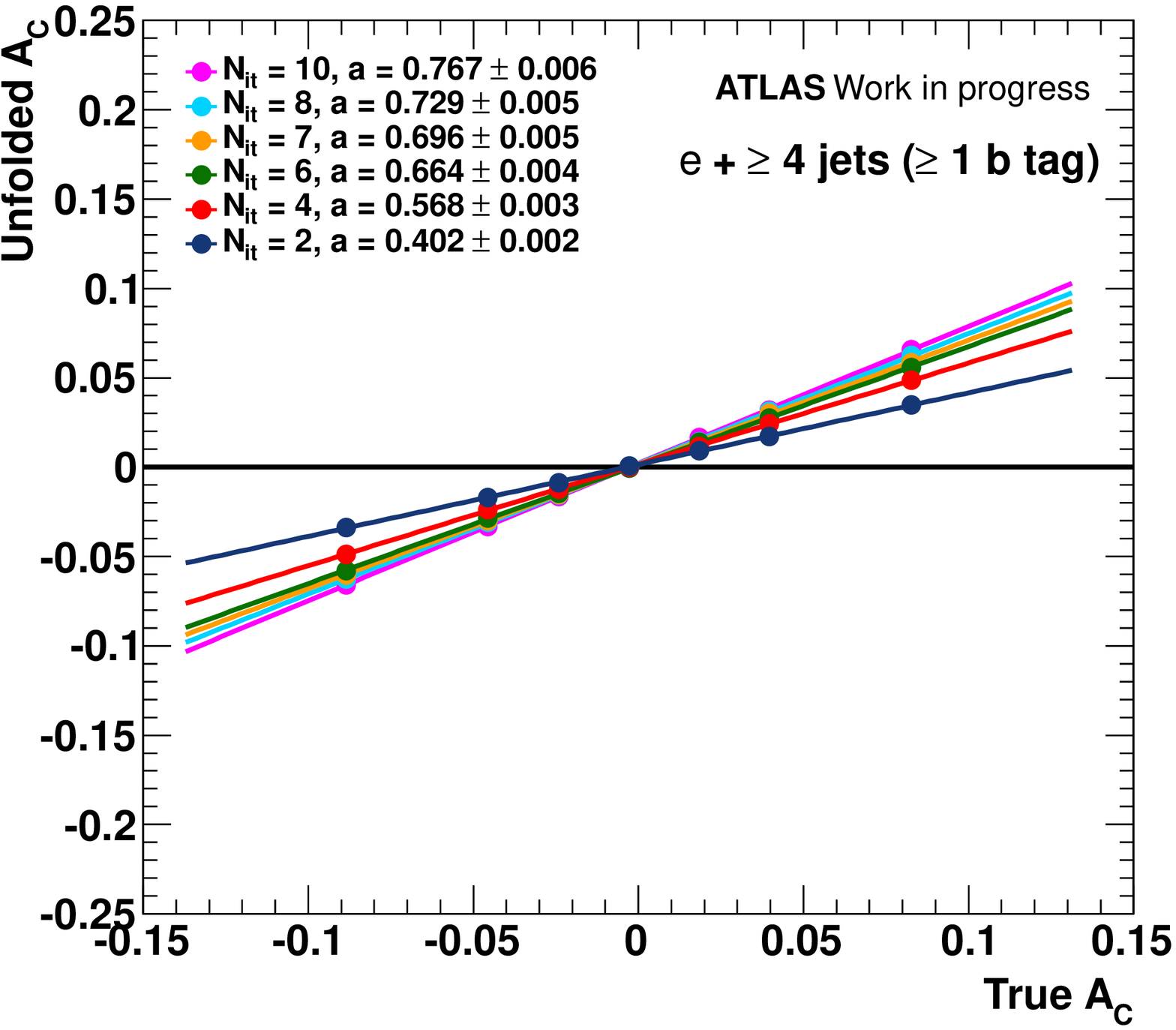}
    \includegraphics[width=\plotwidth]{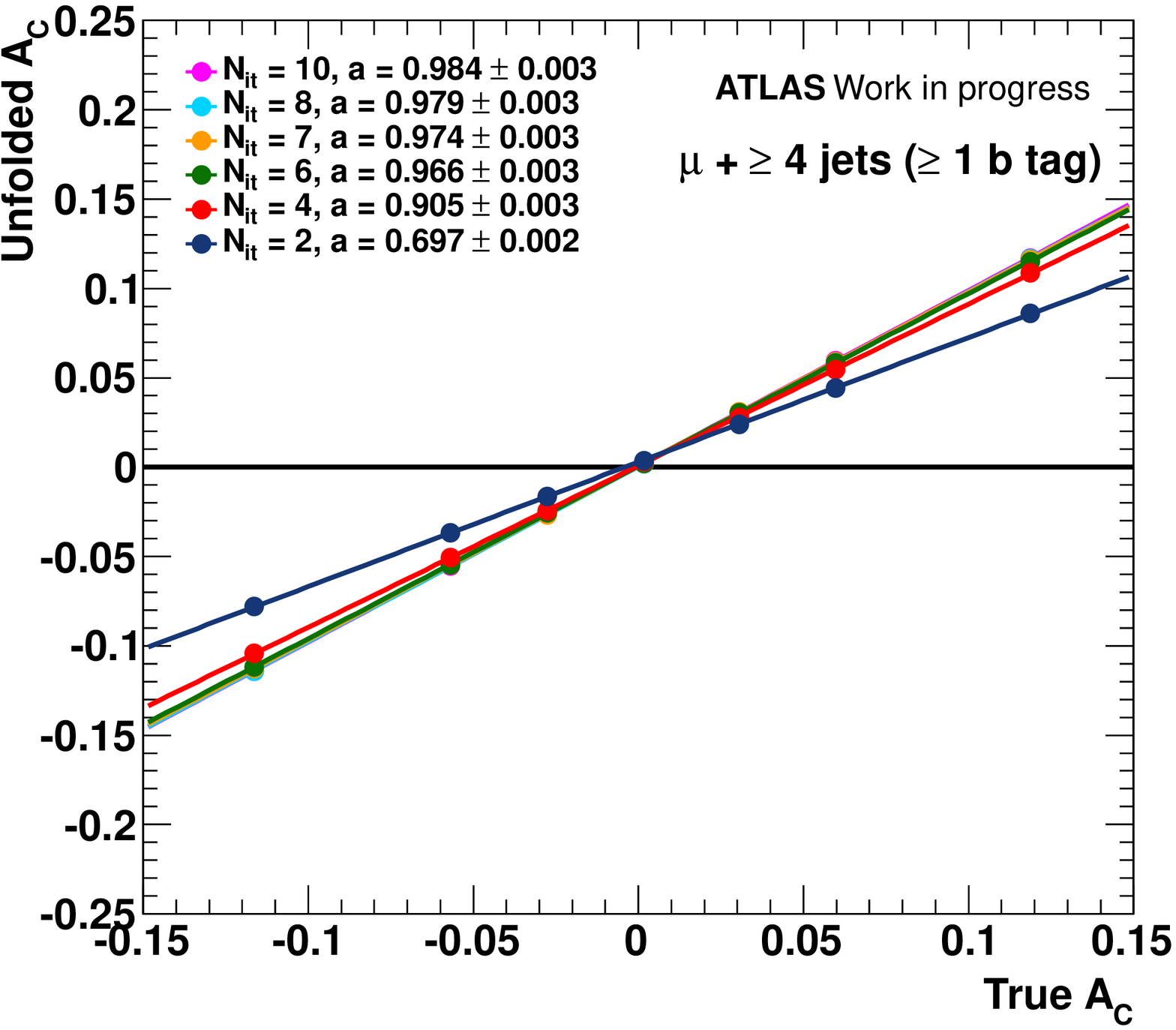}
    \quad\quad
    \includegraphics[width=\plotwidth]{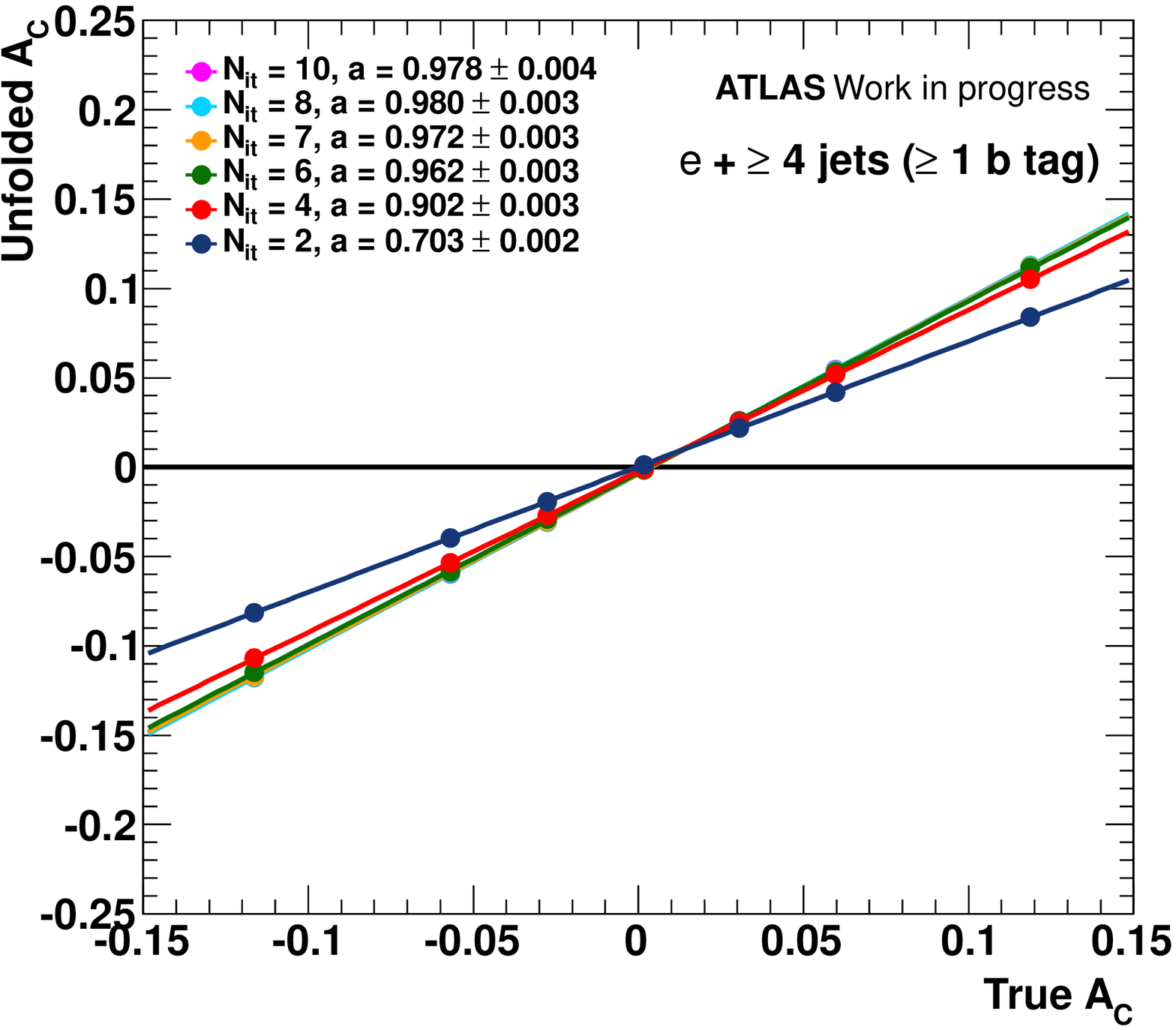}
    
    \vspace{-0.2 cm}
    \caption{The obtained overall inclusive asymmetry after unfolding, $A_C^{\text{unf}}$, as a function of injected true asymmetry $A_C^{\text{true}}$ for different regularisation parameters $N_{\text{It}}$ for both the muon+jets channel (left) and electron+jets channel (right). The top row shows the respective distribution for the inclusive measurement, while the lower rows show the corresponding distributions for $M_{t\bar{t}} < 450$\,GeV and $M_{t\bar{t}} > 450$\,GeV, respectively. For the simultaneous unfolding in $|y_t| - |y_{\bar{t}}|$ and $M_{t \bar{t}}$, a cut on the event reconstruction likelihood $\log{L}$ was applied to improve the $M_{t\bar{t}}$ resolution of the selected events.}
    \label{fig:unfolding:calib:alt}
  \end{center}
\end{figure}
A straight line fit has been performed as described in \mbox{Chapter \ref{chap:results:unfolding}}. As can be seen in the respective fits, the slopes differ from one, in particular for $M_{t\bar{t}} < 450$.

In addition, the respective expected statistical uncertainty on $A_C^{\text{unf}}$ is shown in \mbox{Figure \ref{fig:unfolding:calibuncert:alt}} as a function of $N_{\text{It}}$, taking into account the parameters for the slope and offset $a$ and $b$, respectively, from the straight line fits of the calibration curves. 
\begin{figure}[!htbp]
  \begin{center}
    \includegraphics[width=\plotwidth]{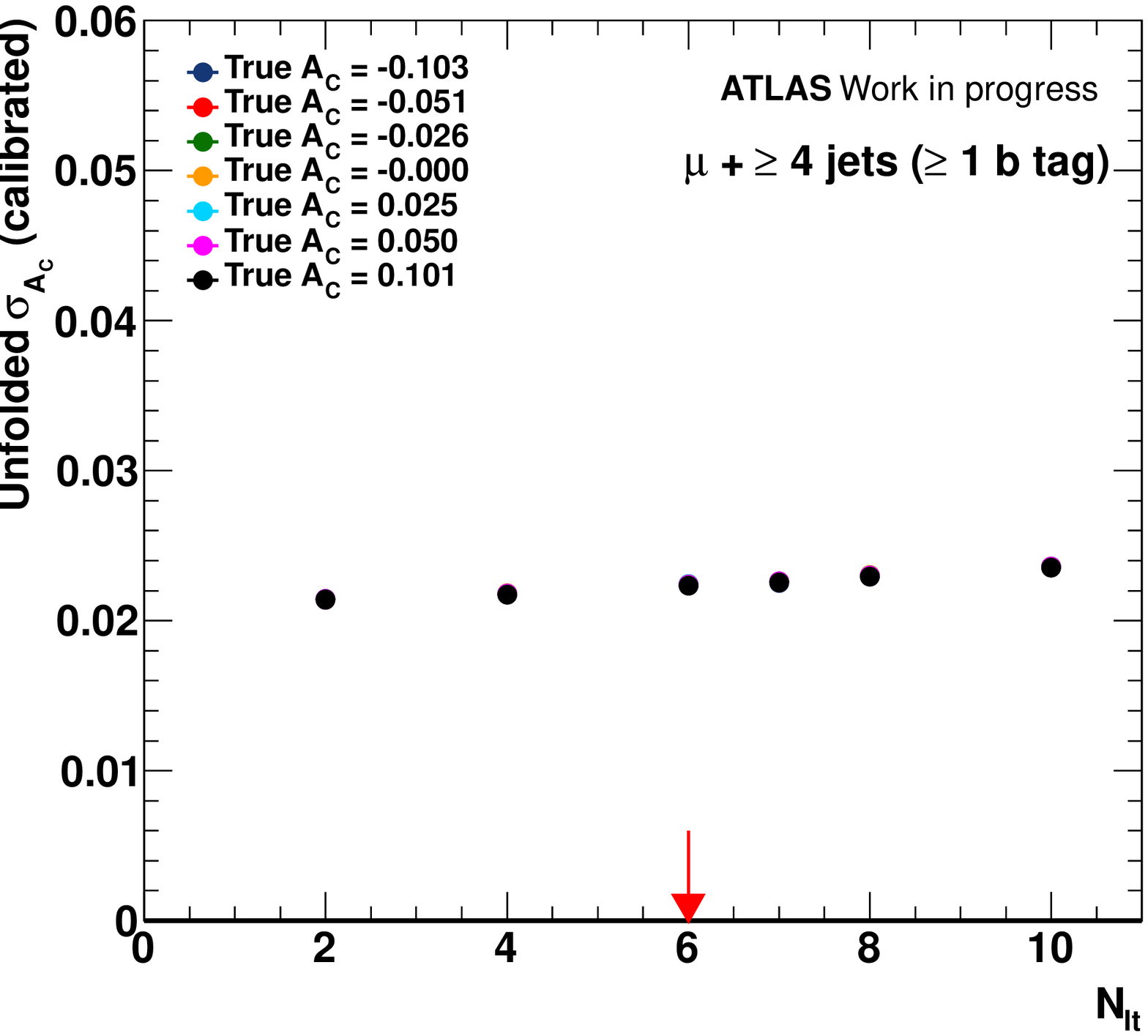}
    \quad\quad
    \includegraphics[width=\plotwidth]{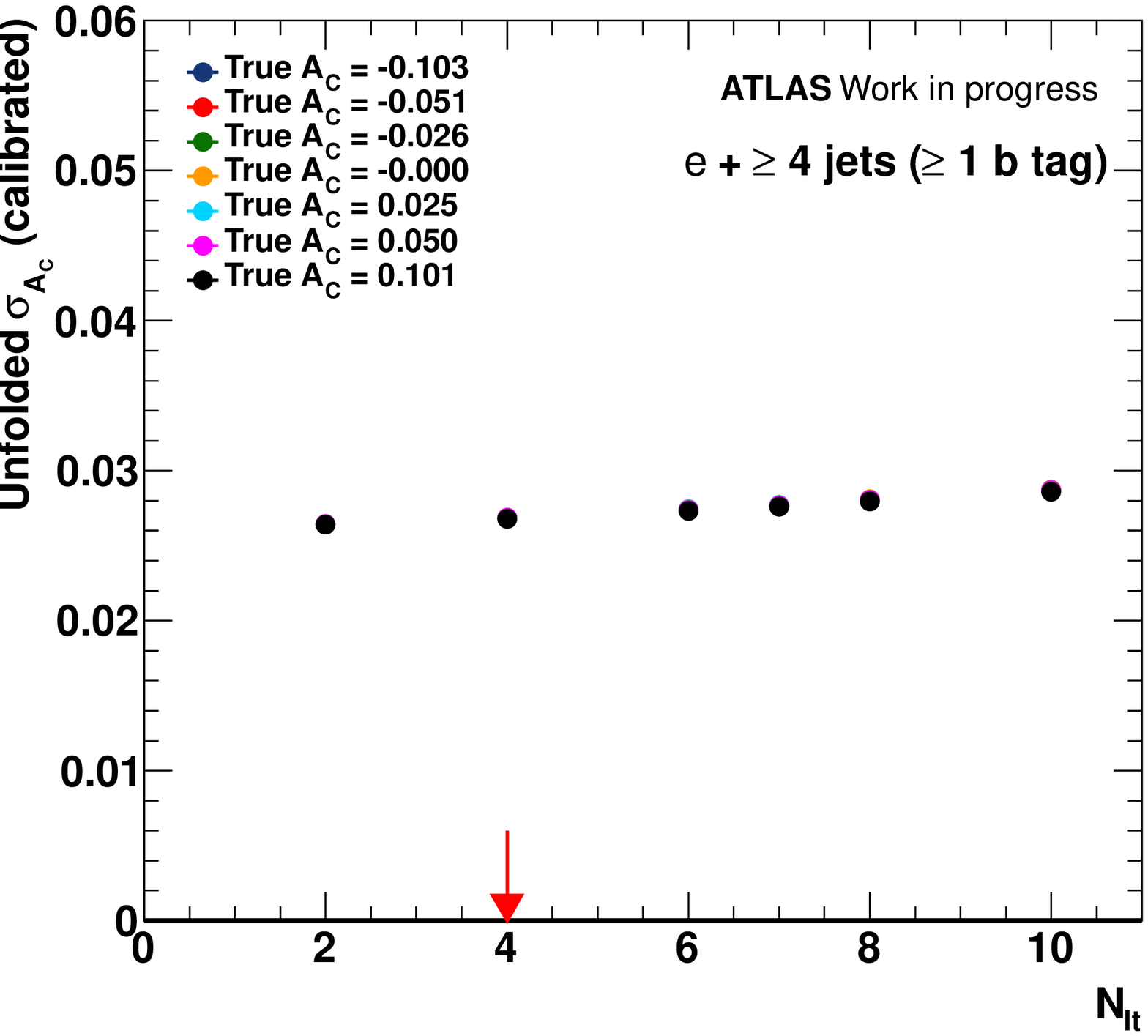}
    \includegraphics[width=\plotwidth]{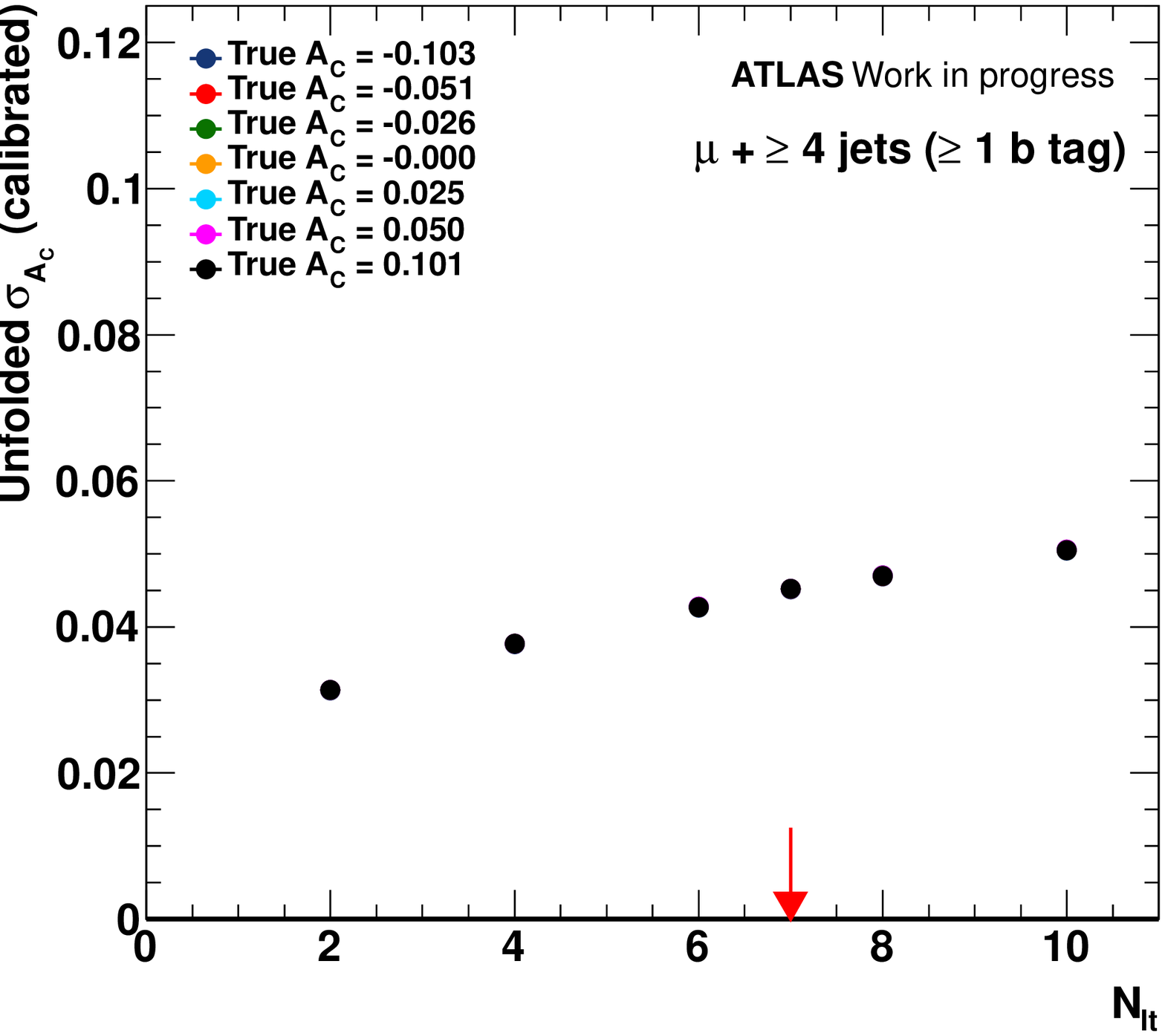}
    \quad\quad
    \includegraphics[width=\plotwidth]{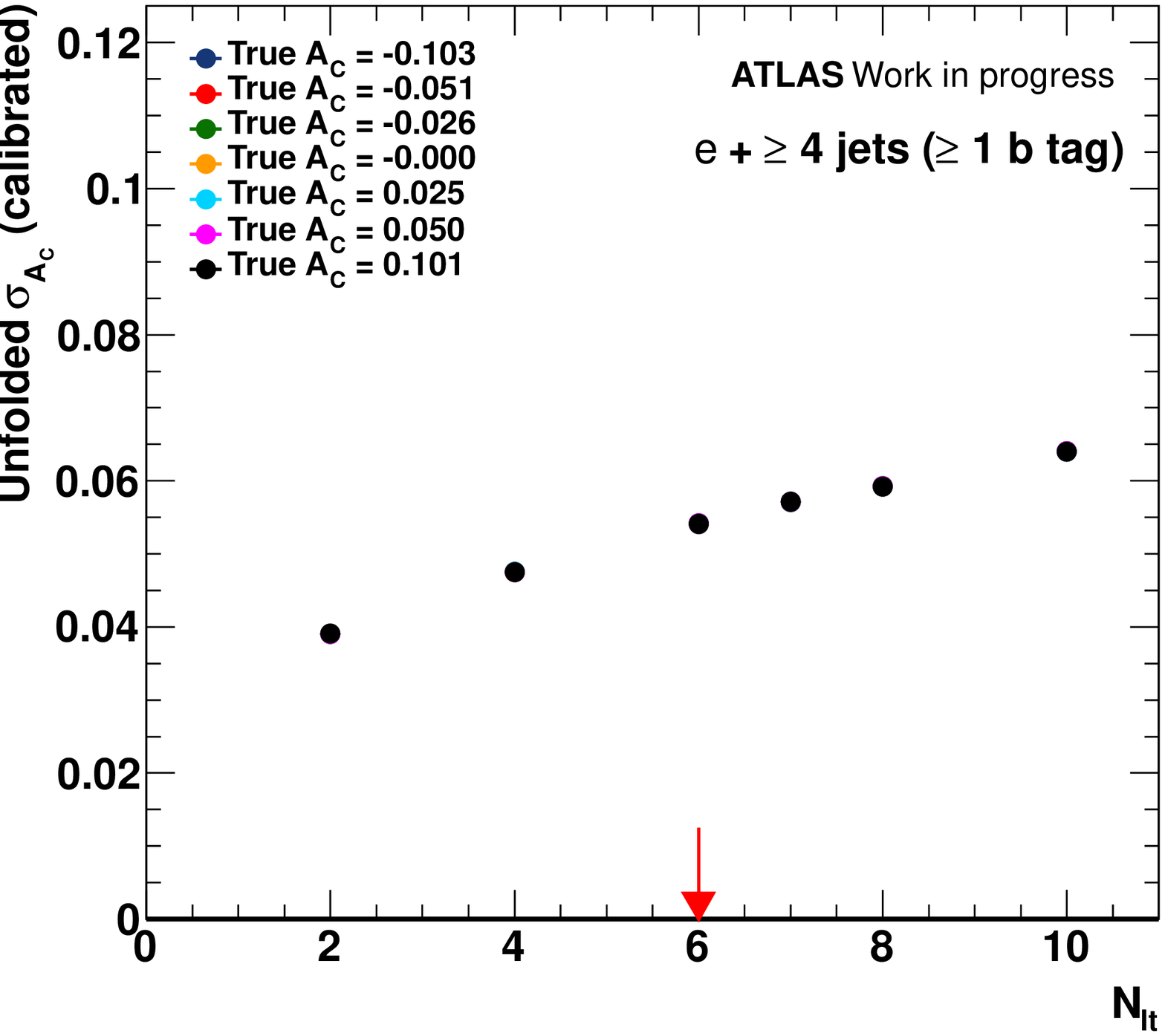}
    \includegraphics[width=\plotwidth]{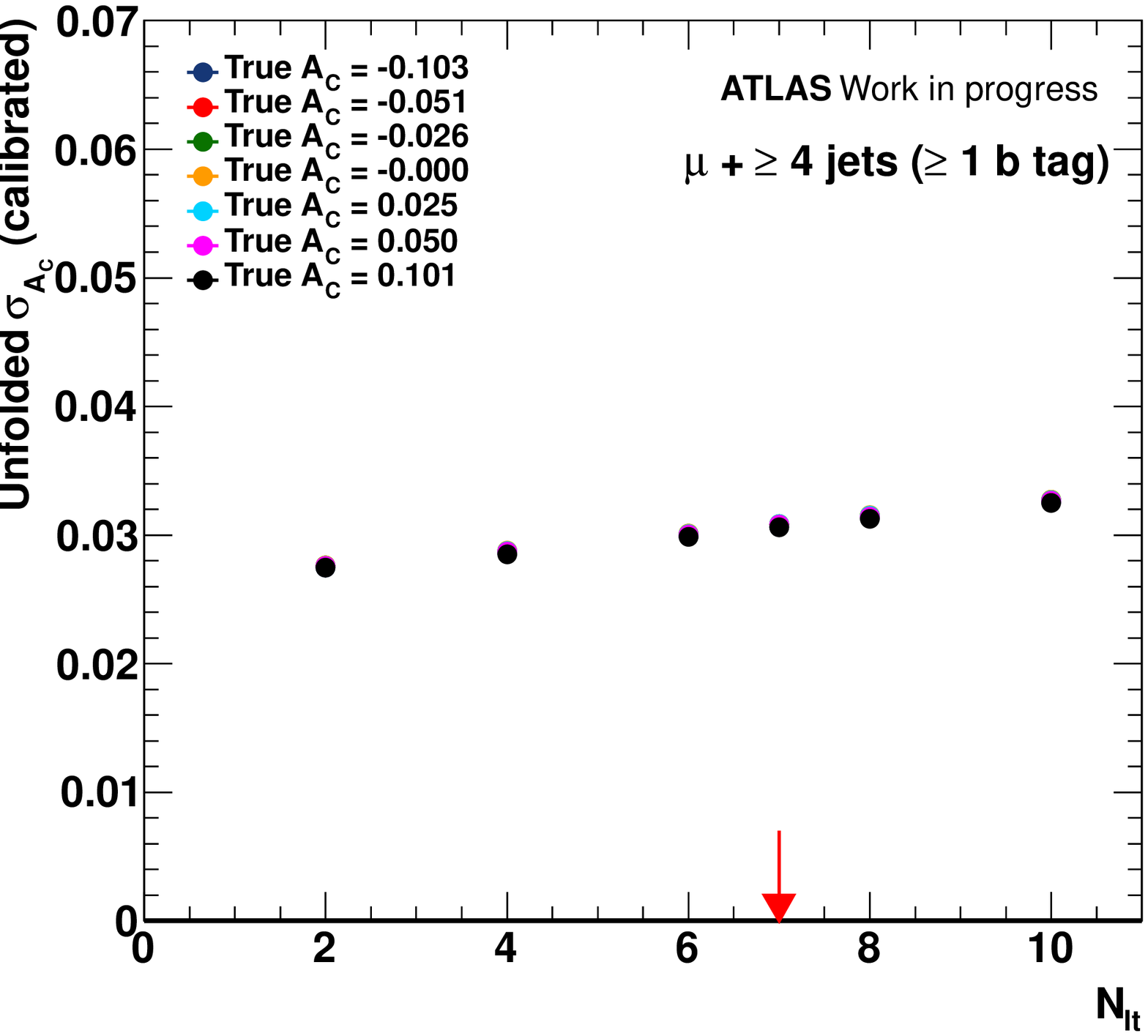}
    \quad\quad
    \includegraphics[width=\plotwidth]{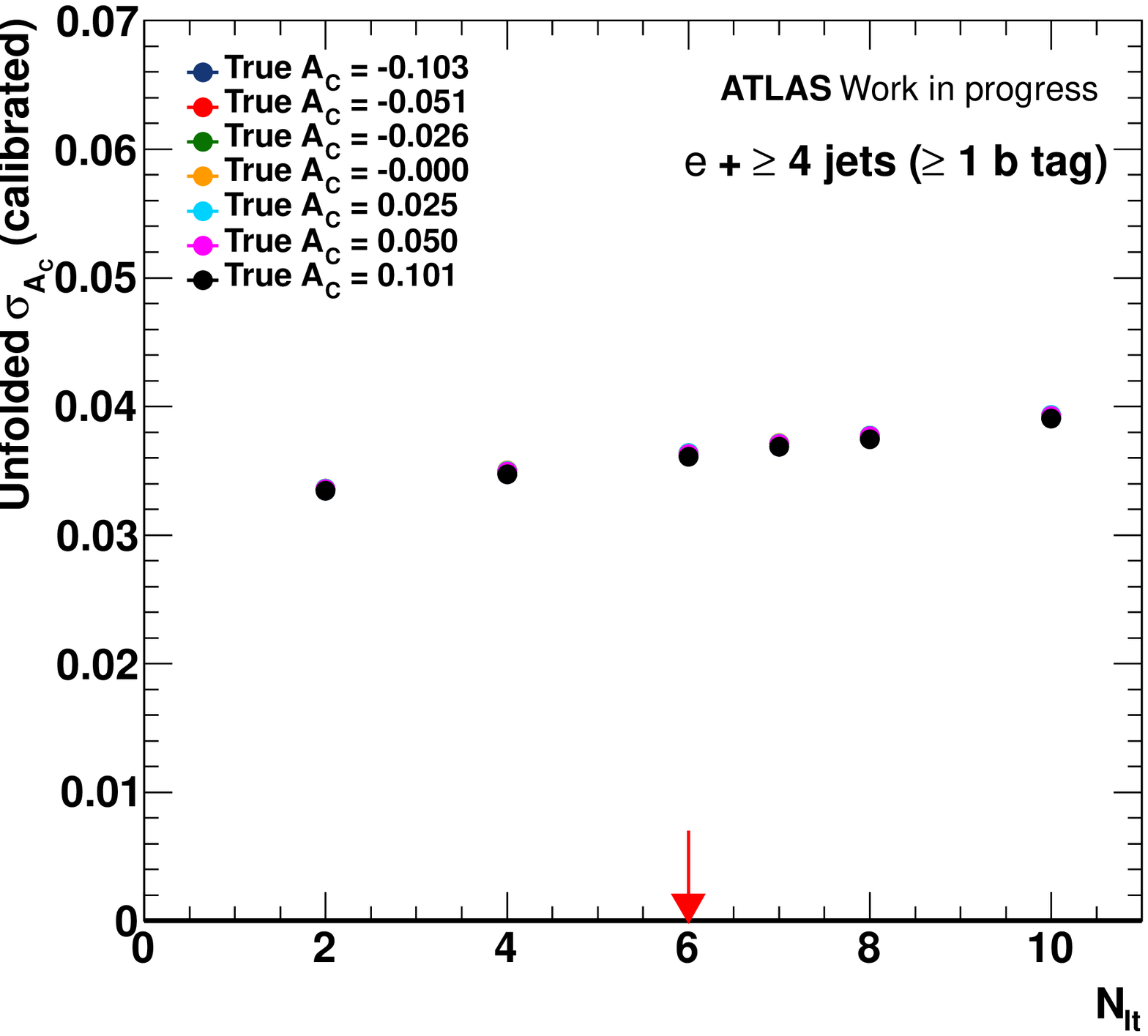}
    
    \vspace{-0.2 cm}
    \caption{Expected statistical uncertainties on $A_C^{\text{unf}}$ as a function of the regularisation parameter $N_{\text{It}}$ for different injected true asymmetries $A_C^{\text{true}}$ for both the muon+jets channel (left) and electron+jets channel (right). The top row shows the respective distribution for the inclusive measurement, while the lower rows show the corresponding distributions for $M_{t\bar{t}} < 450$\,GeV and $M_{t\bar{t}} > 450$\,GeV, respectively. For the simultaneous unfolding in $|y_t| - |y_{\bar{t}}|$ and $M_{t \bar{t}}$, a cut on the event reconstruction likelihood $\log{L}$ was applied to improve the $M_{t\bar{t}}$ resolution of the selected events. The arrows indicate the chosen values for $N_{\text{It}}$.}
    \label{fig:unfolding:calibuncert:alt}
  \end{center}
\end{figure}
For such small changes in $N_{\text{It}}$, the expected statistical uncertainty after correction increases only slightly.

The number of iterations for the unfolding of the used data set was chosen such that the statistical uncertainty was minimised, while requiring that any injected asymmetry in the tested range was recoverable within the expected statistical uncertainties. This approach, however, led to a remaining bias of the unfolded asymmetry $A_C^{\text{unf}}$ with respect to the true asymmetry $A_C^{\text{true}}$. Consequently, this bias was corrected for by taking into account the slope and offset parameters $a$ and $b$ of the straight line fit, respectively, to obtain a corrected value of the unfolded asymmetry, $A_C^{\text{unf,corr}}$, given by
\begin{equation}
A_C^{\text{unf,corr}} = \frac{A_C^{\text{unf}} - b}{a}.
\end{equation}
Consequently, the corresponding statistical and systematic uncertainties after correction, $\sigma_{A_C}^{\text{unf,corr}}$, can be determined by
\begin{equation}
\sigma_{A_C}^{\text{unf,corr}} = \frac{\sigma_{A_C}^{\text{unf}}}{a}.
\end{equation}
Following the described requirements, regularisation parameters of $N_{\text{It}} = 6$ for the muon+jets and $N_{\text{It}} = 4$ in the electron+jets channel were chosen for the inclusive unfolding, while for the simultaneous unfolding in $|y_t| - |y_{\bar{t}}|$ and $M_{t \bar{t}}$, $N_{\text{It}} = 7$ and $N_{\text{It}} = 6$ were chosen for the muon+jets and electron+jets channel, respectively. The corresponding slope and offset parameters extracted from the straight line fit in the calibration can be found in \mbox{Table \ref{tab:calibfactors:alt}} for completeness.
\begin{table}[h!tdp]
\begin{center}
{\small
\begin{tabular}{|l|c|c|c|c|c|c|}
\hline
           \multicolumn{1}{|c|}{} & \multicolumn{3}{c|}{muon+jets channel} & \multicolumn{3}{c|}{electron+jets channel} \\
\cline{2-7}\multicolumn{1}{|c|}{Unfolding} & $N_{\text{It}}$ & $a$ (slope) & $b$ (offset) & $N_{\text{It}}$ & $a$ (slope) & $b$ (offset) \\
\hline\hline
Inclusive & 6 & \phantom{-}0.841 $\pm$ 0.002 & -0.0009 $\pm$ 0.0001 & 4 & \phantom{-}0.739 $\pm$ 0.002 & \phantom{-}0.0004 $\pm$ 0.0001 \\
\hline
$M_{t\bar{t}} < 450\,\text{GeV}$ & 7 & \phantom{-}0.711 $\pm$ 0.004 & -0.0021 $\pm$ 0.0002 & 6 & \phantom{-}0.664 $\pm$ 0.004 & \phantom{-}0.0012 $\pm$ 0.0002 \\
\hline
$M_{t\bar{t}} > 450\,\text{GeV}$ & 7 & \phantom{-}0.974 $\pm$ 0.003 & \phantom{-}0.0007 $\pm$ 0.0002 & 6 & \phantom{-}0.902 $\pm$ 0.003 & -0.0032 $\pm$ 0.0002 \\
\hline
\end{tabular}
}
\caption{Slopes and offsets from the linear fit in the unfolding calibration. The parameters were obtained for a linear fit of the average unfolded value of $A_C^{\text{unf}}$ as a function of the true value $A_C^{\text{true}}$, obtained from sets of pseudoexperiments. For the simultaneous unfolding in $|y_t| - |y_{\bar{t}}|$ and $M_{t \bar{t}}$, a cut on the event reconstruction likelihood $\log{L}$ was applied to improve the $M_{t\bar{t}}$ resolution of the selected events.}
\label{tab:calibfactors:alt}
\end{center}
\end{table}

The unfolded integrated asymmetries in $A_C^{\text{unf}}$ are shown in \mbox{Table \ref{tab:unfolding:alt}} for both the inclusive unfolding and for the simultaneous unfolding in $|y_t| - |y_{\bar{t}}|$ and $M_{t\bar{t}}$, using the chosen regularisation strength in the unfolding process.
\begin{table}[!htbp]
\begin{center}
\begin{tabular}{|l|c|c|}
\hline
Unfolding                  & Observed                                                     & Predicted (\mcnlo) \\
\hline
\hline
\multicolumn{3}{|c|}{Muon+jets Channel} \\
\hline
Inclusive                  & $-0.020 \pm 0.027\,\text{(stat.)} \pm 0.022\,\text{(syst.)}$ & $0.0056 \pm 0.0003\,\text{(stat.)}$ \\
$M_{t\bar{t}} < 450$\,GeV  & $-0.025 \pm 0.054\,\text{(stat.)} \pm 0.034\,\text{(syst.)}$ & $0.0024 \pm 0.0004\,\text{(stat.)}$ \\
$M_{t\bar{t}} > 450$\,GeV  & $-0.015 \pm 0.035\,\text{(stat.)} \pm 0.021\,\text{(syst.)}$ & $0.0086 \pm 0.0004\,\text{(stat.)}$ \\
\hline
\hline
\multicolumn{3}{|c|}{Electron+jets Channel} \\
\hline
Inclusive                  & $-0.063 \pm 0.035\,\text{(stat.)} \pm 0.029\,\text{(syst.)}$ & $0.0056 \pm 0.0003\,\text{(stat.)}$ \\
$M_{t\bar{t}} < 450$\,GeV  & $-0.122 \pm 0.069\,\text{(stat.)} \pm 0.046\,\text{(syst.)}$ & $0.0024 \pm 0.0004\,\text{(stat.)}$ \\
$M_{t\bar{t}} > 450$\,GeV  & $-0.046 \pm 0.046\,\text{(stat.)} \pm 0.034\,\text{(syst.)}$ & $0.0086 \pm 0.0004\,\text{(stat.)}$ \\
\hline
\end{tabular}
\caption{Unfolded values of the charge asymmetry observable $A_C^{\text{unf}}$ for the muon+jets and electron+jets channel. The results for the inclusive measurement and the respective results for the simultaneous unfolding in $|y_t| - |y_{\bar{t}}|$ and $M_{t \bar{t}}$ for $M_{t \bar{t}} < 450$\,GeV and $M_{t \bar{t}} > 450$\,GeV, taking into account the correction for unfolding bias, are shown. For the simultaneous unfolding, a cut on the event reconstruction likelihood $\log{L}$ was applied to improve the $M_{t\bar{t}}$ resolution of the selected events. Furthermore, the respective \mcnlo~predictions are shown.}
\label{tab:unfolding:alt}
\end{center}
\end{table}

A summarised list of all systematics and their contribution to the overall systematic uncertainties can be found in \mbox{Table \ref{Tab:SystematicsUnf:alt}} for the inclusive unfolding and in \mbox{Table \ref{Tab:SystematicsUnf2D:alt}} for the simultaneous unfolding in $|y_t| - |y_{\bar{t}}|$ and $M_{t\bar{t}}$.
\begin{table}[htbp]
\begin{center}
{\small
\begin{tabular}{|l|r|r|}
\cline{2-3}\multicolumn{1}{c|}{} & \multicolumn{ 2}{c|}{Absolute systematic uncertainty} \\
\cline{2-3}\multicolumn{1}{c|}{} & Muon Channel & Electron Channel \\
\hline
QCD multijet  & 0.005 & 0.011 \\ 
\hline
Jet energy scale  & 0.008 & 0.013 \\ 
$b$ tag jet energy scale  & $<$ 0.001 & $<$ 0.001 \\ 
Pile-up jet energy scale  & 0.001 & 0.002 \\ 
\hline
Jet reco efficiency  & 0.002 & 0.001 \\ 
Jet energy resolution  & 0.009 & 0.003 \\ 
\hline
Muon efficiencies  & 0.002 & (n.a.) \\ 
Muon scales / resolution  & 0.001 & 0.001 \\ 
Electron efficiencies  & (n.a.) & 0.001 \\ 
Electron scales / resolution  & 0.001 & 0.002 \\ 
\hline
$b$ tag scale factors  & 0.004 & 0.006 \\ 
PDF uncertainty  & $<$ 0.001 & $<$ 0.001 \\ 
LAr hole uncertainty  & 0.003 & $<$ 0.001 \\ 
\hline
ISR and FSR  & 0.005 & 0.013 \\ 
\ttbar~modelling  & (0.002) 0.007 & 0.007 \\ 
Parton shower / fragmentation  & (0.004) 0.011 & 0.011 \\ 
Top mass  & 0.004 & 0.002 \\ 
\hline
$W$+jets normalisation  & 0.007 & 0.010 \\ 
$W$+jets shape  & $<$ 0.001 & 0.005 \\ 
$Z$+jets normalisation  & $<$ 0.001 & $<$ 0.001 \\ 
$Z$+jets shape  & 0.002 & 0.003 \\ 
\hline
Single top  & $<$ 0.001 & $<$ 0.001 \\ 
Diboson  & $<$ 0.001 & $<$ 0.001 \\ 
Charge mis-identification  & $<$ 0.001 & $<$ 0.001 \\ 
$b$ tag charge  & 0.001 & 0.001 \\ 
\hline
MC statistics  & 0.004 & 0.005 \\
\hline
Luminosity   & 0.002 & 0.002 \\
\hline
\hline
Combined & 0.022 & 0.029 \\
\hline
\end{tabular}
}
\end{center}
\caption{List of all systematic uncertainties taken into account for the unfolding procedure in the measurement of the top charge asymmetry. The numbers in brackets denote the uncertainties before using the larger uncertainty of both channels as conservative estimate.}
\label{Tab:SystematicsUnf:alt}
\end{table}

\begin{table}[htbp]
\begin{center}
{\small
\begin{tabular}{|l|r|r|r|r|}
\cline{2-5}\multicolumn{1}{c|}{} & \multicolumn{ 4}{c|}{Absolute systematic uncertainty} \\ \cline{2-5}\multicolumn{1}{c|}{} & \multicolumn{ 2}{c|}{Muon Channel} & \multicolumn{ 2}{c|}{Electron Channel} \\ 
\cline{2-5}\multicolumn{1}{c|}{} & $M_{t\bar{t}} < 450\,\text{GeV}$ & $M_{t\bar{t}} > 450\,\text{GeV}$ & $M_{t\bar{t}} < 450\,\text{GeV}$ & $M_{t\bar{t}} > 450\,\text{GeV}$ \\
\hline
QCD multijet  & 0.010 & 0.002 & 0.021 & 0.020 \\ 
\hline
Jet energy scale  & 0.007 & 0.009 & 0.021 & 0.011 \\ 
$b$ tag jet energy scale  & 0.001 & 0.001 & 0.001 & 0.002 \\ 
Pile-up jet energy scale  & 0.001 & 0.001 & 0.002 & 0.001 \\ 
\hline
Jet reco efficiency  & $<$ 0.001 & 0.001 & 0.007 & 0.001 \\ 
Jet energy resolution  & 0.009 & 0.005 & 0.011 & 0.007 \\ 
\hline
Muon efficiencies  & 0.001 & 0.001 & (n.a.) & (n.a.) \\ 
Muon scales / resolution  & 0.002 & 0.001 & $<$ 0.001 & 0.001 \\ 
Electron efficiencies  & (n.a.) & (n.a.) & 0.001 & 0.001 \\ 
Electron scales / resolution  & $<$ 0.001 & 0.001 & 0.005 & 0.003 \\ 
\hline
$b$ tag scale factors  & 0.001 & 0.003 & 0.007 & 0.004 \\ 
PDF uncertainty  & 0.001 & 0.001 & 0.001 & 0.002 \\ 
LAr hole uncertainty  & 0.001 & 0.001 & 0.002 & 0.002 \\ 
\hline
ISR and FSR  & 0.021 & 0.007 & 0.024 & 0.018 \\ 
\ttbar~modelling  & 0.016 & (0.004) 0.016 & 0.005 & (0.003) 0.005 \\ 
Parton shower / fragm.  & (0.003) 0.006 & 0.006 & (0.001) 0.012 & 0.012 \\ 
Top mass  & 0.010 & (0.003) 0.010 & 0.003 & 0.005 \\ 
\hline
$W$+jets normalisation  & 0.003 & 0.006 & 0.009 & 0.006 \\ 
$W$+jets shape  & $<$ 0.001 & $<$ 0.001 & 0.001 & 0.001 \\ 
$Z$+jets normalisation  & $<$ 0.001 & $<$ 0.001 & 0.001 & $<$ 0.001 \\ 
$Z$+jets shape  & 0.007 & $<$ 0.001 & 0.013 & 0.003 \\ 
\hline
Single top  & $<$ 0.001 & $<$ 0.001 & 0.001 & $<$ 0.001 \\ 
Diboson  & $<$ 0.001 & $<$ 0.001 & $<$ 0.001 & $<$ 0.001 \\ 
Charge mis-identification  & $<$ 0.001 & $<$ 0.001 & $<$ 0.001 & $<$ 0.001 \\ 
$b$ tag charge  & 0.001 & 0.001 & 0.001 & 0.001 \\ 
\hline
MC statistics  & 0.008 & 0.005 & 0.010 & 0.007 \\
\hline
Luminosity  & 0.001 & 0.001 & 0.002 & 0.001 \\
\hline
\hline
Combined & 0.034 & 0.021 & 0.046 & 0.034 \\
\hline
\end{tabular}
}
\end{center}
\caption{List of all systematic uncertainties taken into account for the unfolding procedure in the measurement of the top charge asymmetry. The numbers in brackets denote the uncertainties before using the larger uncertainty of both channels as conservative estimate.}
\label{Tab:SystematicsUnf2D:alt}
\end{table}

All systematic uncertainties based on a replacement of the unfolding matrix which were of the same order of magnitude or lower than the respective MC statistics uncertainty could not be resolved to full extent due to the inherent fluctuations from limited statistics in the response matrix. Hence, for those cases, the largest of the systematic uncertainties in both channels was used for the final systematic uncertainty on the unfolded charge asymmetry. The resulting combined systematic uncertainties were 0.022 in the muon+jets channel and 0.029 in the electron+jets channel for the inclusive unfolding. For the simultaneous unfolding in $|y_t| - |y_{\bar{t}}|$ and $M_{t\bar{t}}$ the resulting total systematics were 0.034 ($M_{t\bar{t}} < 450\,\text{GeV}$) and 0.021 ($M_{t\bar{t}} > 450\,\text{GeV}$) in the muon+jets channel, and 0.046 ($M_{t\bar{t}} < 450\,\text{GeV}$) and 0.034 ($M_{t\bar{t}} > 450\,\text{GeV}$) in the electron+jets channel, respectively.

The \textsc{Blue} method was used to combine the measurements after performing the unfolding for the described alternative approach in the muon+jets and electron+jets channel, taking into account the respective systematic uncertainties and the associated correlations. The same assumptions about correlations as described in \mbox{Chapter \ref{chap:results:comb}} were used and a combined value of
\begin{equation*}
A_C^{\text{unf}} = -0.035 \pm 0.021\,\text{(stat.)} \pm 0.021\,\text{(syst.)}
\end{equation*}
for the inclusive unfolding was obtained, where the relative weight of the muon+jets channel result was 65.7\,\%. The combination of the results for the simultaneous unfolding in $|y_t| - |y_{\bar{t}}|$ and $M_{t \bar{t}}$ yielded
\begin{eqnarray*}
A_C^{\text{unf}} (M_{t\bar{t}} < 450\,\text{GeV}) & = & -0.058 \pm 0.043\,\text{(stat.)} \pm 0.035\,\text{(syst.)},\\
A_C^{\text{unf}} (M_{t\bar{t}} > 450\,\text{GeV}) & = & -0.025 \pm 0.028\,\text{(stat.)} \pm 0.022\,\text{(syst.)},
\end{eqnarray*}
where the relative weight of the muon+jets channel was 65.6\,\% and 69.4\,\%, respectively.

As expected, the resulting statistical uncertainties are significantly lower than the ones obtained in the procedure described in \mbox{Chapter \ref{Results}}. This reduction is due to the smaller number of iterations used. The obtained central values from the two different approaches were compatible within the respective statistical uncertainties. 

The systematic uncertainties in the individual channels were of the same order for both approaches for the inclusive unfolding. Larger deviations were observed for the simultaneous unfolding in $|y_t| - |y_{\bar{t}}|$ and $M_{t\bar{t}}$, where the statistical component in the evaluation of the systematic uncertainties was larger. These statistical fluctuations are expected to be reduced by using a smaller amount of iterations in the unfolding procedure. This effect can be seen in particular for the (dominant) contributions involving the replacement of the response matrix with samples with a smaller number of events. For most of these cases, the obtained systematic uncertainties are significantly lower than the ones obtained in the procedure described in \mbox{Chapter \ref{Results}}.

Consequently, the unfolding procedure used in the main part of this analysis, despite being more conservative, is expected to be significantly more stable, in particular due to the reduced model dependency and the requirement of a convergent unfolding process.

In addition, the conservative systematic uncertainties assumed for the QCD multijet normalisation in the muon+jets channel have been replaced by a combined normalisation and shape systematic uncertainty as described in \mbox{Chapter \ref{chap:QCDMu}}. All other parameters of the analysis have not been changed with respect to the nominal procedure described in \mbox{Chapter \ref{chap:results:unfolding}}. This approach yielded uncertainties on the measurement which were significantly lower than for the assumption of a 100\,\% normalisation uncertainty, as expected. Uncertainties of 0.0007 for the inclusive unfolding, and 0.002 ($M_{t\bar{t}} < 450\,\text{GeV}$) and 0.001 ($M_{t\bar{t}} > 450\,\text{GeV}$) for the simultaneous unfolding in $|y_t| - |y_{\bar{t}}|$ and $M_{t\bar{t}}$ were obtained (compared to 0.0011, 0.018 and 0.004, respectively, as shown in \mbox{Chapter \ref{Results}}).
\clearpage

\chapter{$\mathbf{b}$ Tag Weighting Control Plots}
\label{AppBTagW}
It was verified that the direct application of $b$ tag efficiencies to distributions obtained without the requirement of at least one $b$ tagged jet yields effective distributions comparable to the ones obtained with the full event selection. A comparison was performed for the distributions of $|y_t| - |y_{\bar{t}}|$ both for the background contributions (c.f. \mbox{Figure \ref{fig:BTagWBG}}) only and for the data distribution after background subtraction (c.f. \mbox{Figure \ref{fig:BTagWSIG}}). The distributions are in excellent agreement within the shown statistical uncertainties.
\begin{figure}[h!tb]
  \begin{centering}
    \mbox{
      \includegraphics[width=\plotwidth]{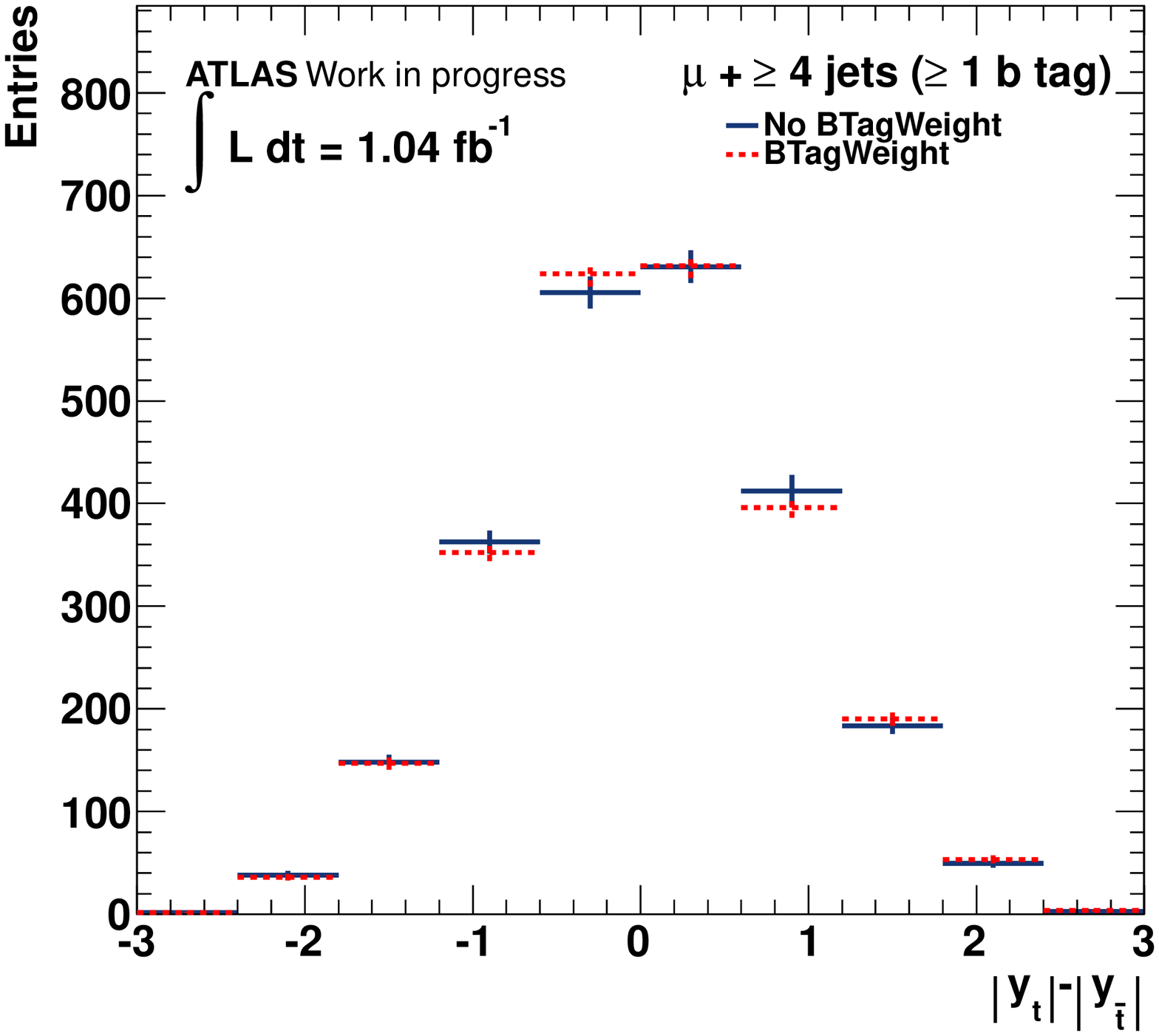}
      \quad\quad
      \includegraphics[width=\plotwidth]{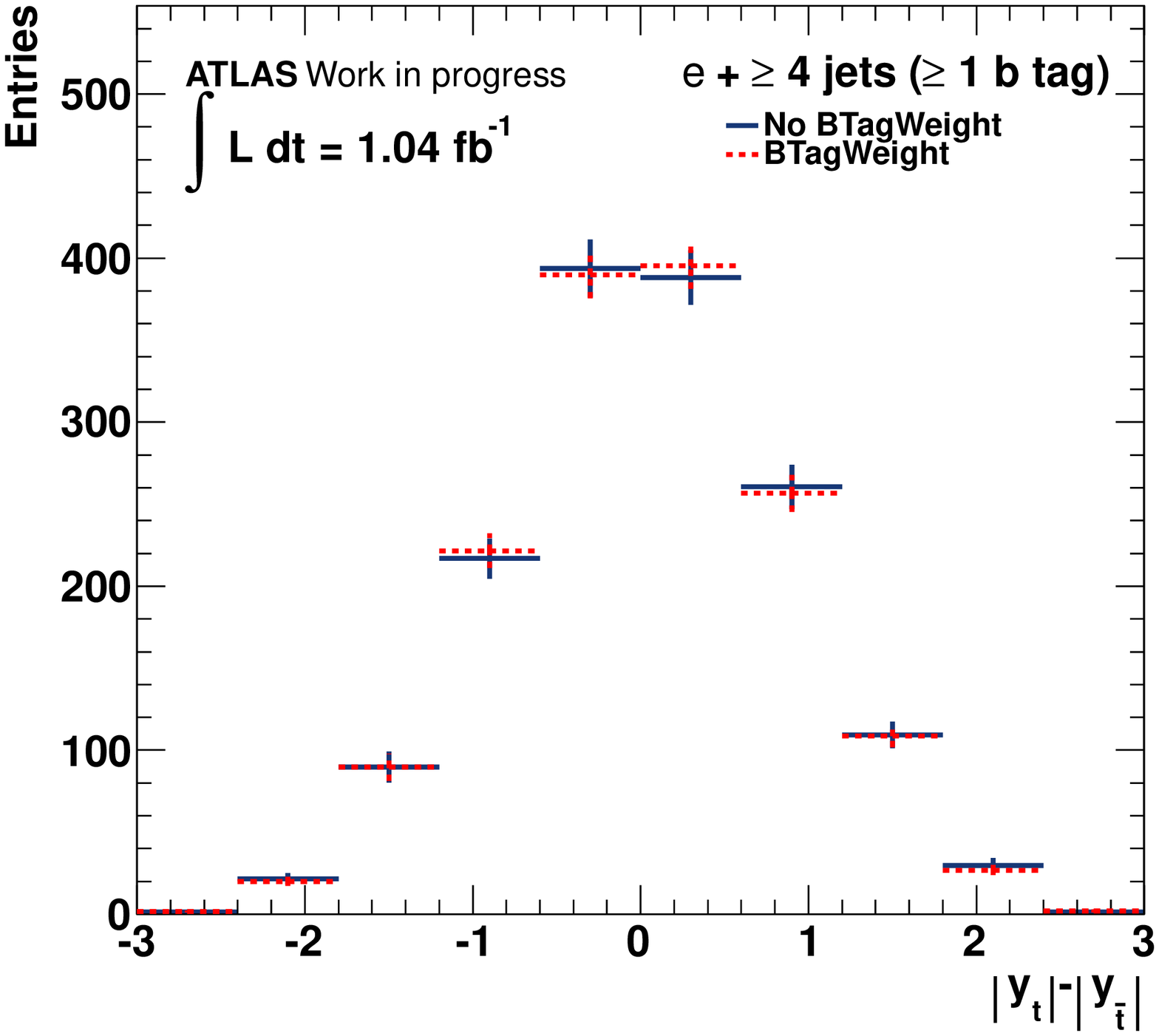}
    }
    
    \vspace{-0.2 cm}
    \caption[\quad $b$ tag weighting for background distribution]{Comparison of the distribution for $|y_t| - |y_{\bar{t}}|$ for the Monte Carlo background contribution for the full event selection (blue) and without $b$ tag requirement, but with $b$ tag efficiencies applied directly in a reweighting approach (red, dashed). The bin-by-bin statistical uncertainty is reduced by up to 25\,\%.}
    \label{fig:BTagWBG}
  \end{centering}
\end{figure}
\begin{figure}[h!tb]
  \begin{centering}
    \mbox{
      \includegraphics[width=\plotwidth]{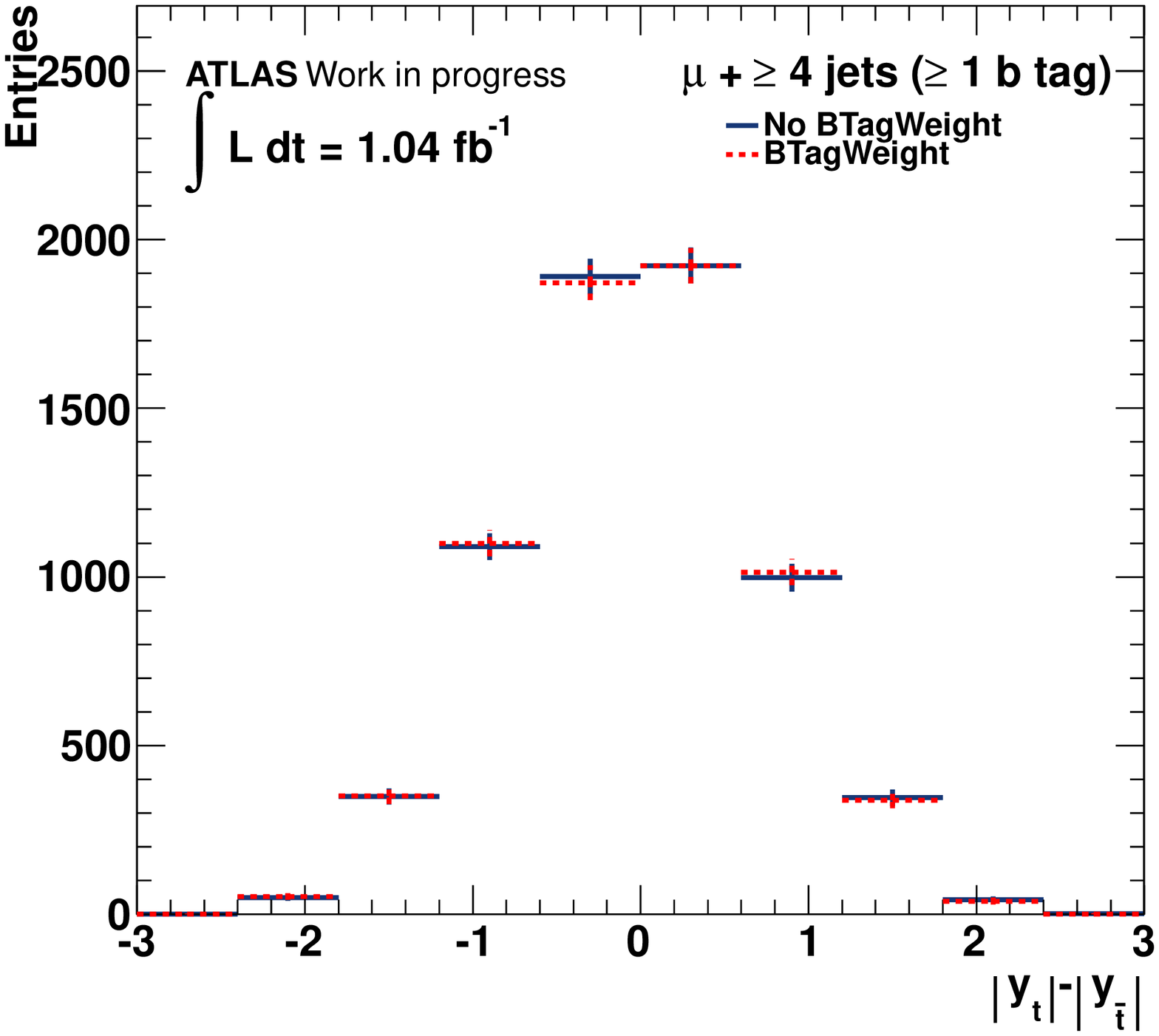}
      \quad\quad
      \includegraphics[width=\plotwidth]{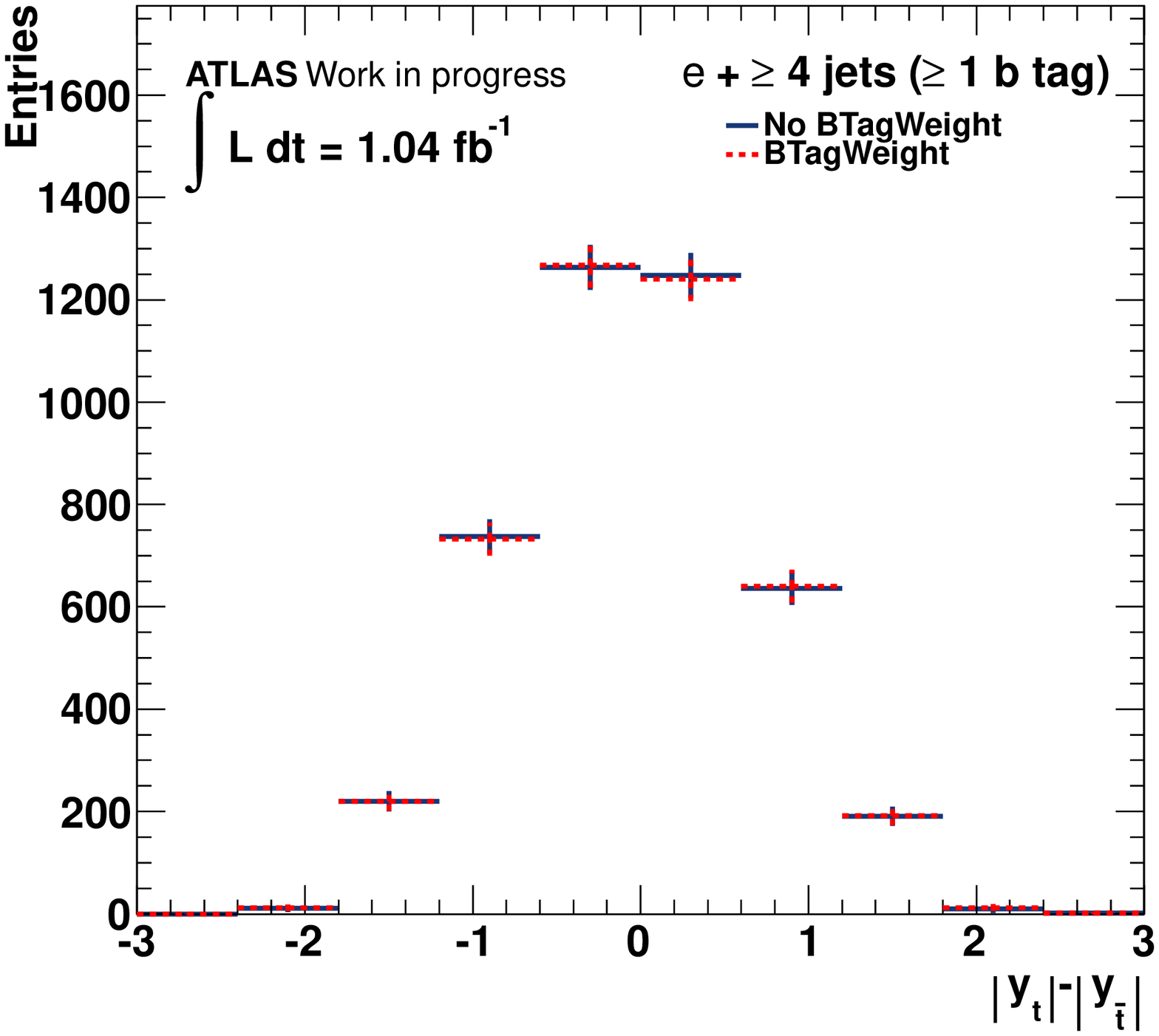}
    }
    
    \vspace{-0.2 cm}
    \caption[\quad $b$ tag weighting for signal distribution]{Comparison of the distribution for $|y_t| - |y_{\bar{t}}|$ for data after background subtraction for the full event selection (blue) and without $b$ tag requirement, but with $b$ tag efficiencies applied directly in a reweighting approach to the subtracted background contribution (red, dashed).}
    \label{fig:BTagWSIG}
  \end{centering}
\end{figure}
\clearpage

\end{appendix}

  \endgroup
  
  \clearpage
  \addcontentsline{toc}{chapter}{Bibliography}
  \bibliographystyle{habil}
  \bibliography{Thesis}{}
  
  \chapter*{Acknowledgements}
\addcontentsline{toc}{chapter}{Acknowledgements} 
Foremost, I would like to express my sincere gratitude to Prof. Dr. Arnulf Quadt and Dr. Carsten Hensel, who have been my advisers during my whole thesis and the accompanying studies at the $2^{\text{nd}}$ Institute of Physics.

I also like to thank Carsten and Kevin for the constant support and for continuously providing constructive advice, and most importantly, for always having an open ear.

Sincere thanks are extended to everyone involved in the extensive proofreading process of my thesis draft, Adam, Erik, Kerim, and especially Kevin, who spent a lot of his time (and most likely several red pencils) correcting this document at various draft stages. The many suggestions I received from all of them were a great help.

Thanks also go to Frederic, Kerim and Umberto for the extremely productive work we did together on the summer conference note, the paper and the two internal ATLAS notes on the top quark charge asymmetry. In addition, Frederic kindly provided me with his Fortran implementation of the \textsc{Blue} method, for which I would like to thank him in particular.

A special thank goes to all of my colleagues and friends at the $2^{\text{nd}}$ Institute of Physics, especially Adam, Elisabeth, Matthias and Philipp for providing diversion from work whenever I was in need.

I cannot finish without extending a very personal thank you to Martina for constantly supporting and motivating me, in particular during the last year, with her always finding the right words to cheer me up. Furthermore, it was her who helped keeping away {\em real life} problems from me during my time of working on this analysis and writing my thesis.

Finally, I would like to thank my parents for providing me with the opportunity to be where I am. Without them, none of this would have been possible. Their support and encouragement during school and my studies have been priceless.

\vspace*{2cm}

\begin{flushright}
\textsc{Science. It works, bitches.}\\
{\small -- \textit{Randall Munroe}, xkcd \#54}
\end{flushright}

\begin{flushright}
\textsc{Five card stud, nothing wild. And the sky is the limit.}\\
{\small -- \textit{Capt. Jean-Luc Picard}, U.S.S. Enterprise}
\end{flushright}

\end{document}